\documentclass[a4paper,12pt,twoside]{book}
\pdfoutput=1
\usepackage{general_cite}
\usepackage{caption}
\usepackage{bm}       
\usepackage{setspace}
\usepackage{graphicx}
\usepackage{amssymb}
\usepackage{amsbsy}   
\usepackage{amsmath}
\usepackage{url}
\usepackage{epsfig, natbib, mydefs}
\usepackage{aastex_hack}
\usepackage{deluxetable}
\usepackage{longtable}
\usepackage{fancyhdr}
\usepackage[avantgarde]{quotchap}
\usepackage[calcwidth]{titlesec}
\usepackage{rotating}

\fancypagestyle{plain}{
\fancyhf{}
\fancyfoot[CO]{\thepage}

}

\makeatletter
\def\cleardoublepage{\clearpage\if@twoside \ifodd\c@page\else
	\hbox{}
	\vspace*{\fill}
	\thispagestyle{empty}
	\newpage
	\if@twocolumn\hbox{}\newpage\fi\fi\fi}
\makeatother

\renewcommand\chapterheadstartvskip{\vspace*{-5\baselineskip}}

\titleformat{\section}[hang]{\sffamily\bfseries}
{\Large\thesection}{12pt}{\Large}[{\titlerule[0.5pt]}]

\begin{document}

\frontmatter
\begin{titlepage}
\begin{center}

{\textsc{\Large\bf 
	Studies of Radio Galaxies and Starburst Galaxies using Wide-field, High Spatial Resolution Radio Imaging
}} \\

\vspace{4cm}

{\Large\bf Emil Lenc} \\

\vspace{2cm}

\centerline{\psfig{figure=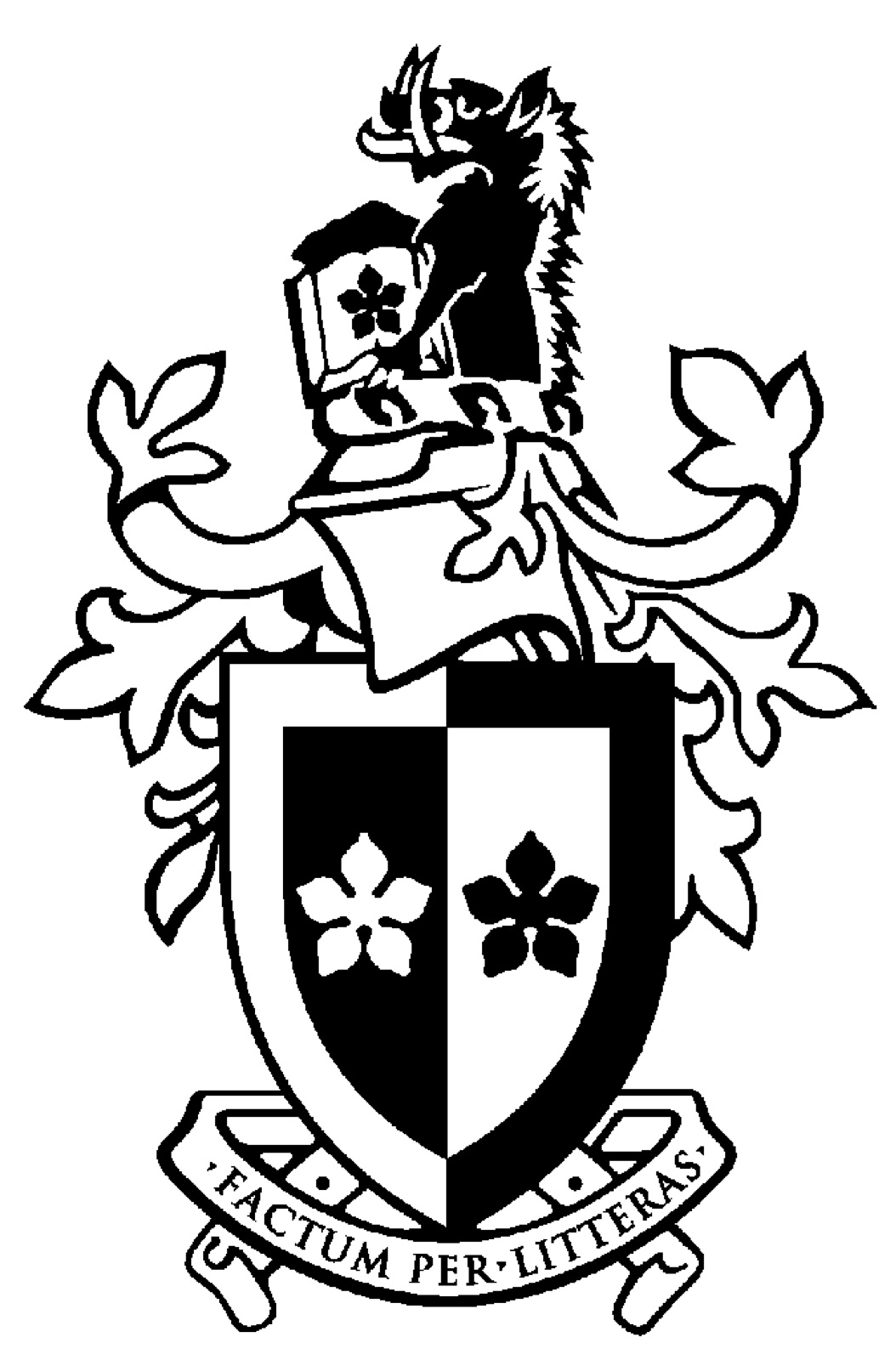,width=4cm}}

\vspace{2cm}

{\it A Dissertation \\
Presented in fulfillment of the requirements\\
for the degree of\\
Doctor of Philosophy\\
at Swinburne University Of Technology\\
}

\vspace{1cm}
{\it February 2009}

\end{center} 
\end{titlepage}


\chapter*{Abstract}

This thesis reports on the application of new wide-field Very Long Baseline Interferometry (VLBI) imaging techniques using real data for the first time. These techniques are used to target three specific science areas: (i) a sub-parsec-scale study of compact radio sources in nearby starburst galaxies, (ii) a study of jet interactions in active radio galaxies, and (iii) an unbiased study of the sub-arcsecond, 90 cm sky.

Six local southern starburst galaxies are surveyed on sub-parsec-scales using wide-field VLBI techniques. Compact radio sources are detected in two of the most prominent galaxies, NGC 253 and NGC 4945. Modelling of the compact source spectra reveal the majority have steep spectra, associated with supernova remnants, and are significantly free-free absorbed by a dense ionised screen. Limits on the supernova rate and star formation rate in these galaxies are estimated based on source fading, source population modelling, and on source counts and sizes. No or few compact radio sources are detected in the less prominent galaxies, presumably as a result of reduced star formation and/or star formation in sparse environments that result in weak and short-lived supernovae and remnants.

The hot spots and interaction regions of three active radio galaxies are studied at parsec--scales, for the first time, using wide-field VLBI imaging. The resulting images have provided the most detailed views of these regions to date. In two of the target sources, PKS $0518-458$ (Pictor A) and PKS $0521-365$, the hot spot emission is resolved into a set of compact components. The emission mechanisms in the hot spots are discussed based on their morphology and additional multi-wavelength data.

Two overlapping 28 deg$^2$ regions are surveyed in detail using wide-field VLBI techniques in the first systematic (and non-biased), deep, high resolution survey of the low frequency sky. This represents a field of view two orders of magnitude greater than anything previously attempted in a single pointing with VLBI. A total of 27 sources were detected as far as $2\arcdeg$ from the phase centre. The results of the survey suggest that new low frequency telescopes, such as LOFAR and SKA, should detect many compact radio sources and that plans to extend these arrays to baselines of several thousand kilometres are warranted.

\chapter*{Acknowledgments}

Having reached the conclusion of this grand and exciting adventure, I would like to sincerely thank and acknowledge those that have supported me over the past four years.

Firstly, I wish to thank my wife Wendy for the great sacrifices she has endured over these years to enable me to follow my new field of research. This work would not have been possible without her love and support. I also wish to apologise to my son Arin, who came into the world half-way through my PhD thesis, for not being able to provide the amount of attention he deserved from his loving father.

Special thanks go to my PhD thesis supervisor, Prof. Steven Tingay. On our first meeting he handed me a radio image from his recent research. On the face of it, the image appeared to contain two apparently nondescript ``blobs'' on an otherwise blank piece of paper. He then explained at length, with great enthusiasm, how these were probably supernova remnants and how they could be used to gain a better understanding of the inner workings of the host galaxy. His enthusiasm and constant stream of ideas encouraged me to pursue research under his guidance. His help and support have ever since been above and beyond the call of duty and I am greatly indebted to him. I also wish to thank Prof. Matthew Bailes, Director of the Centre for Astrophysics and Supercomputing at Swinburne University of Technology, for his support and for providing me with the opportunity to study at the centre. Thanks also go to the staff of Swinburne Astronomy Online (SAO), particularly Sarah Maddison and Glen Mackie (who also acted as co-ordinating supervisor in my final year), for re-kindling my fascination with astronomy.

I would like to thank my collaborators for helping take my research to even higher levels. Special thanks go to Mike Garrett, Olaf Wucknitz and James Anderson of the Joint Institute for VLBI in Europe (JIVE) for their support, hospitality and friendship during my international placement at JIVE. I would also like to thank Gianfranco Brunetti and Marco Bondi from the Italian National Institute for Astrophysics (INAF) in Bologna, Italy for their contribution to the X-ray modelling in chapter \ref{chap:pictora}. Finally, thanks also go to Michael Dahlem, Alessandro Marconi, Niruj R. Mohan, Juergen Ott, Rick Perley, Tim Roberts, Nick Schurch and Kim A. Weaver for generously providing access to their data for use in Figures \ref{fig:figmultiwav}, \ref{fig:p3figmultiwav} and \ref{fig:p4fig3}.

A great deal of support has also been provided by staff at the Australia Telescope National Facility (ATNF). There are too many to thank individually but I would particularly like to thank Anastasios K. Tzioumis for his support as an ATNF co-supervisor and for advancing the capabilities of the Long Baseline Array (LBA) to newer and greater heights. I would also like to thank Tim Cornwell for his valuable advise and help with wide-field and high-dynamic range imaging problems. Special thanks also go to Robert Braun and Ray Norris for their support in my transition from doctoral to post-doctoral life at the ATNF.

Finally, this work was supported financially by a Swinburne University of Technology Chancellor's Research Scholarship and a CSIRO top-up scholarship. The work conducted in Chapter \ref{chap:lfwfvlbi} was supported by the European Community's Sixth Framework Marie Curie Research Training Network Programme, Contract No.\linebreak[4]\mbox{MRTN-CT-2004-505183} ``ANGLES''.

The data presented in this work were obtained using the Australia Telescope (ATCA), the LBA, the National Radio Astronomy Observatory (NRAO) Very Long Baseline Array (VLBA) and the European VLBI Network (EVN). The ATCA and LBA are funded by the Australian Commonwealth Government for operation as a national facility managed by the CSIRO. The NRAO is a facility of the National Science Foundation operated under cooperative agreement by Associated Universities, Inc. The EVN is a joint facility of European, Chinese, South African and other radio astronomy institutes funded by their national research councils.

This research has made use of the NASA/IPAC Extragalactic Database (NED), which is operated by the Jet Propulsion Laboratory, California Institute of Technology, under contract with the National Aeronautics and Space Administration. Some of the data presented in this thesis were obtained from the Multimission Archive at the Space Telescope Science Institute (MAST). STScI is operated by the Association of Universities for Research in Astronomy, Inc., under NASA contract NAS5-26555. Support for MAST for non-HST data is provided by the NASA Office of Space Science via grant NAG5-7584 and by other grants and contracts.

\chapter*{Declaration}

This thesis contains no material that has been accepted for the award of any other degree or diploma. To the best of my knowledge, this thesis  contains no material previously published or written by another author, except where due reference is made in the text of the thesis. All work presented is primarily that of the author, except where indicated below.

{\bf Chapter \ref{chap:ngc253}:} The initial calibration of the NGC 253 VLBI data were performed by Steven Tingay (Curtin University of Technology).

{\bf Chapter \ref{chap:ngc4945}:} The interpretation of apparent jet in NGC 4945 is by Steven Tingay (Curtin University of Technology).

{\bf Chapter \ref{chap:pictora}:} A substantial proportion of the hot spot comparison was performed by Steven Tingay (Curtin University of Technology). The modelling of X-ray emission was performed by Gianfranco Brunetti and Marco Bondi (Italian National Institute for Astrophysics).

{\bf Chapter \ref{chap:lfwfvlbi}:} The initial calibration of narrow-band field was performed by Olaf Wucknitz (Joint Institute for VLBI in Europe - JIVE). The ionospheric modelling software was developed by James M. Anderson (JIVE).

\vspace{5cm}
\noindent
Emil Lenc

\vspace{5mm}
\noindent
27 August, 2008

\tableofcontents
\listoffigures
\listoftables

\mainmatter

\linespread{1.0}
\normalsize
\begin{savequote}[20pc]
\sffamily
There is a theory which states that if ever anybody discovers exactly what the Universe is for and why it is here, it will instantly disappear and be replaced by something even more bizarre and inexplicable. There is another theory which states that this has already happened.
\qauthor{Douglas Adams}
\end{savequote}

\chapter{Introduction}
\label{chap:intro}

\linespread{1.3}
\normalsize

\section{Motivation}
\label{sec:intro.motivation}

Traditionally, Very Long Baseline Interferometry (VLBI) has been a narrow field-of-view technique, generally concerned with bright, compact sources subtending significantly less than one arcsecond on the sky.  This situation has suited the computational facilities available until recently, since VLBI data-sets in their unaveraged form can require significant amounts of disk space and computing power in order to form images. When VLBI data-sets are averaged as a function of frequency and/or time, the undistorted field-of-view is reduced, alleviating the computing issues but retaining the generally very small required field of view.

However, with current computational facilities, it is now feasible to work with VLBI data-sets in their unaveraged state, allowing images to be made over much wider fields of view, 10's of arcseconds, arcminutes or even degrees (if the VLBI correlator can output data with a fine enough frequency and temporal resolution).  In addition, a great deal of progress has been made recently on wide-field imaging algorithms.  In particular, the $w$-projection technique \citep{cor03}, which removes image distortions due to the effects of non-coplanar arrays, is a significant improvement over previous methods such as faceting \citep{Perley:1999p10576}.  $W$-projection is also computationally efficient, providing an order magnitude improvement over previous methods.

Due to advances such as these, VLBI wide-field imaging opens up new science possibilities for radio astronomy.  In particular, since many VLBI arrays use large antennas, wide-field, sensitive, and high angular resolution observations are possible.  The use of hard disk recording of data, as opposed to the traditional use of tapes, enables greater bandwidths to be recorded resulting in a factor of $\sim2$ times improvement in sensitivity. Furthermore, software correlation, such as that now used routinely at the Australian Long Baseline Array \citep{Deller:2007p10545}, supports finer frequency and temporal correlator outputs thus increasing the available field-of-view.

The wide-field VLBI imaging techniques used in this research are being tested with real data for the first time. The $w$-projection technique, for example, has not yet been adopted by the wider astronomical community as it has only recently been introduced. Yet the advantages it provides in terms of image quality, available field-of-view, and performance will become critical for the success of instruments such as the SKA which are being designed for wide-field, high sensitivity and high resolution observations. Utilising these techniques on real astronomical sources, together with the high data rates now available with LBA hard disk recording and the finer frequency and temporal outputs available with the LBA software correlator, will demonstrate the first steps towards SKA data reduction.

There are three areas of science interest which will be targeted with wide-field VLBI observations as part of this thesis. The first is the study of compact radio sources in local starburst galaxies. This study is a continuation of the pioneering work of \citet{Pedlar:1999p3534} and \citet{Tingay:2004p778} and aims to survey the local population of starbursts visible from the Southern Hemisphere. The second science aim is to study jet interaction regions in southern AGN. This work will build on the initial efforts of \citet{Young:2005p5449} and will aim to survey bright interaction regions in southern AGN exhibiting complex morphologies. The final science aim will test the application of wide-field VLBI in an ambitious wide-field survey at 90 cm. This will be a pioneering effort to survey a large fraction of the primary beam in a single VLBI observation. Each of the science aims will be discussed in more detail in the following subsections.

\subsection{Starburst Galaxies}
\label{sec:introsb}
Starburst galaxies, such as the prominent southern starburst NGC 253, host large numbers of massive stars that quickly evolve to form supernovae \citep{Engelbracht:1998p874}. In NGC 253 the supernovae drive activity in the nuclear region and produce strong winds out of the disk of the galaxy that are visible at X-ray and H$\alpha$ wavelengths \citep{Weaver:2002p844, Strickland:2002p797}. The accumulation of gas that drives the star formation can be caused by interactions between galaxies. For example the merger of gas rich spiral galaxies can result in the accumulation of 60\% of the gas within the inner 100 pc of the merger product \citep{Barnes:1996p10524}.  Gas can also be funnelled into the nuclear region of a galaxy by dynamical processes associated with bar instabilities \citep{Koribalski:1995p891}.

A significant fraction of the available gas for star formation will be exhausted within a single dynamical time-scale. For example, by fitting an initial mass function (IMF) to NGC 253 with an evolutionary synthesis model, \citet{Engelbracht:1998p874} found that the galaxy had been undergoing rapid massive star formation for 20-30 million years and that it is now in a late phase of its starburst activity. Their emission-line spectrum models from a late phase starburst are also consistent with the observed optical spectrum. 

A dramatic spectral flattening has been observed at radio frequencies in nuclear starburst regions of galaxies such as NGC 253 \citep{Carilli:1996p10536} and is typical of free-free absorption by highly ionised gas. Similar results were obtained by observing supernova remnants in M82 over a range of radio wavelengths \citep{McDonald:2002p3330}. \citet{McDonald:2002p3330} detected 30 supernova remnants in M82 and modelled their spectra with a simple power law, a free-free absorbed power law, and a synchrotron self-absorbed spectrum. They found that free-free absorbed power law models fit the sources best with a free-free emission measure toward supernova remnants of $\sim10^{7}$ cm$^{-6}$ pc.

Starbursts are therefore the result of processes acting on a wide range of scales and themselves drive energetic activity on a wide range of scales.  By direct observation of the supernova remnants in starburst regions at radio wavelengths (to mitigate against severe extinction at optical wavelengths and provide very high angular resolution) it is possible to investigate the remnants themselves, using them to reconstruct the supernova and star formation history of the starburst.  Further, it is possible to use radio observations to investigate the ionised gaseous environment of the starburst.  Such observations provide a link between large-scale dynamical effects in the galaxies, activity in the star forming region itself, and the energetic phenomena that are in turn driven by the starburst. 

The combination of wide fields of view, high sensitivity, and high resolution at radio wavelengths enable the study of populations of faint, compact sources over wide fields.  For example, the supernova remnants in nearby starburst galaxies can be detected and identified at low radio frequencies, where they are brightest.  This is impossible using short baseline interferometers such as the VLA or ATCA.  In some of the first wide-field imaging attempts made using the Australian Long Baseline Array (LBA), \citet{Tingay:2004p778} observed NGC 253, a southern starburst galaxy, at 1.4 GHz, approximately matching the angular resolution of previous VLA observations at higher frequencies (up to 22 GHz: \citet{Ulvestad:1997p907}).  \citet{Tingay:2004p778} was therefore able to extend the measured radio spectra of individual supernova remnants in the starburst down to 1.4 GHz, where free-free absorption due to the ionised interstellar medium of NGC 253 causes a sharp turnover in the spectra of the remnants.  The free-free optical depth as a function of line-of-sight through NGC 253 was therefore determined, and the inferred electron densities and temperatures compared to values derived from radio recombination line observations \citep{Mohan:2002p792}. In addition to low frequency observations, high frequency VLA and/or ATCA observations can be used to measure the spectral indices of the compact radio sources and distinguish flat spectrum HII regions from steep spectrum supernova remnants \citep{Ulvestad:1997p907,McDonald:2002p3330}.

Observations of supernova remnants in a galaxy can also be used to place limits on the supernova rate of the galaxy, particularly where multi-epoch observations are made. There are at least three approaches that can be used to measure the supernova rate:
\begin{itemize}
\item An estimation based on source counts, source sizes and an estimated expansion rate for the supernova remnants \citep{Antonucci:1988p10521, Pedlar:1999p3534}.
\item Statistical modelling using a hypothetical population of supernovae and supernova remnants that have Poisson distributed occurrence times \citep{Ulvestad:1991p1008}.
\item An estimation based on observed source fading between epochs \citep{Ulvestad:1997p907}.
\end{itemize}

\begin{figure}[ht]
\epsscale{0.5}
\plotone{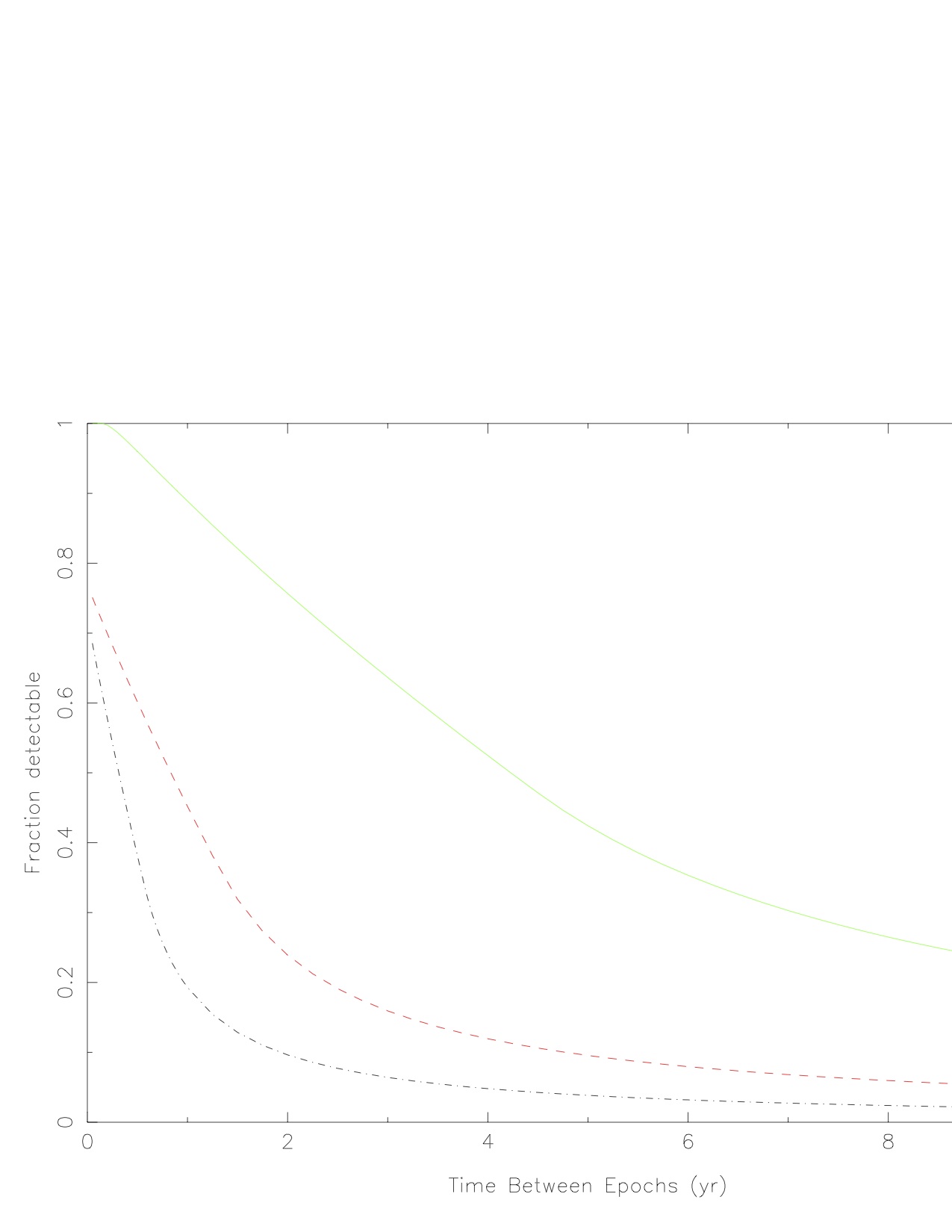}
\caption[Proportion of type II supernova remnants detectable as a function of the time.]{Proportion of type II supernova remnants detectable as a function of the time between epochs for the 5 GHz VLA observations (lower curve) of Ulvestad \& Antonucci and Mohan, 2.3 GHz LBA observations with tape-based recording (middle curve) and new 2.3 GHz LBA observations with hard disk recording and software correlation (top curve: corresponding to the observations proposed here). Following the same assumptions of \citet{Ulvestad:1991p1008}, the population of type II supernovae used to derive the above plots are assumed to have a uniformly distributed peak luminosity between 5-20 times that of Cas A, have a light curve that approximately follows that of SN 1980k \citep{Weiler:1986p9393} and are Poisson distributed in time.}
\label{fig:figdetect}
\end{figure}

Previous attempts to measure the supernova rate in starburst galaxies, such as NGC 253, using the VLA have been hindered by the diffuse emission of these galaxies and the lack of spectral indices for some of the sources, thus complicating the identification and separation of supernova remnants from thermal sources. Furthermore, at least in NGC 253, the size estimates were limited by poor image resolution and none of the remnants have yet been observed to expand or fade, making age estimates difficult. Finally, the statistical methods used by \citet{Ulvestad:1991p1008,Ulvestad:1994p1041,Ulvestad:1997p907} to model the effect of non-detections did not take into consideration revised distances to the remnants, the differing sensitivity limits of the observations or the effects of free-free absorption. When such effects are simulated, it turns out that the VLA is in fact a poor instrument for detecting supernova remnants (see Figure \ref{fig:figdetect}) with a detection rate of only 13\% for a period of 1.5 years between epochs and as low as 5\% for a period of 4 years between epochs. The poor performance of the VLA is caused by an effective reduction in the sensitivity limit as a result of confusion between compact radio sources and the diffuse emission of the galaxy.

Observations at 2.3 GHz with the LBA (Long Baseline Array) can achieve greater sensitivity and resolution than the VLA at 5 GHz, see Figure \ref{fig:figdetect}. At LBA resolutions the diffuse galactic emission is completely resolved, thereby allowing a more straightforward view of the individual supernova remnants and HII regions. By processing the VLBI data-sets in their unaveraged form with bandwidth synthesis and wide-field imaging algorithms, such as the $w$-projection technique of \citet{cor03} (see Section \S~\ref{sec:wfvlbi.wfeffects}) it is also possible to image a wide field-of-view. Further increases in sensitivity can be gained by recording directly to hard disk (Figure \ref{fig:figdetect}) and utilising the LBA software correlator \citep{Deller:2007p10545} to correlate the data. The increased sensitivity and resolution combined with a wide field-of-view can aid in estimating the supernova rate in starburst galaxies.

Using the high sensitivity and high resolution available with the LBA, a selection of southern starbursts will be studied using wide-field imaging techniques. The selection criteria for these sources is described in Section \S~\ref{sec:selection.sb}. This will be the first time the starburst region of many of these galaxies has been observed. My aim is to measure the spectra of compact radio sources in these galaxies and model them using techniques similar to those pioneered by \citet{McDonald:2002p3330} and \citet{Tingay:2004p778}. By measuring the free-free opacity of each of the sources the distribution of ionised gas in the galaxy can be determined. This can be compared with multi-wavelength data to understand the larger structures of the galaxy. The high resolution of the LBA will resolve structure in some of the supernova remnants and enable estimates of the supernova rate to be determined. Second epoch observations, where available, can be used to further refine these estimates.

\subsection{Jet Interactions}
\label{sec:introagn}
Interactions between relativistic jets and their environments are clearly a major factor in the evolution of powerful radio galaxies.  The morphologies of typical FR-II radio galaxies are dominated by the termination of the relativistic jets from the nucleus in roughly symmetric, compact and bright hot spots where the jets interact with the intergalactic medium. A nearby Southern Hemisphere example of this type of interaction is the radio galaxy Pictor A \citep{Perley:1997p6689}.  At the opposite end of the size and age scale of radio galaxies, interactions between relativistic jets and a dense nuclear environment dominate the morphologies of Gigahertz Peaked Spectrum (GPS) radio galaxies \citep{ODea:1998p10573}.  Jet interactions can also affect the morphologies of radio galaxies via entrainment, the process by which a relativistic jet gradually entrains material from the galactic environment, causing the jet to decelerate and expand.  This process is thought to explain the appearance of FR-I type radio galaxies such as Centaurus A e.g. \citet{Bicknell:1994p10527}.

An unusual type of jet interaction can be produced when a jet, which previously existed in a steady state, is disrupted by material that enters the path of jet propagation.  This can occur if the jet direction varies as a function of time, perhaps via precession, as thought to occur in a number of AGN (e.g. PKS B2152$-$699, \citet{Young:2005p5449}; 3C 273, \citet{Savolainen:2006p28354}; and BL Lac, \citet{Stirling:2003p28368}), or if material is introduced into the galaxy by a merger event (e.g. 3C 120, \citet{Gomez:2006p28376}).  In such cases, radio galaxies often posses a normal jet and lobe structure on one side of the nucleus and a disrupted morphology on the opposite side of the nucleus.  The disrupted morphology presents itself as a remnant lobe with a bright radio hot spot at the site of interaction.

A study of the jet interactions allows us to probe the conditions in the relativistic jets.  High resolution multi-wavelength observations from radio to X-rays allow us to estimate the energy budget, making it possible to estimate the energy expended in the interaction, by examining the emission over a range of wavelengths.  Knowledge of the emission mechanisms and their efficiencies allow estimates of the energy transfer between jet and environment and therefore parameters such as the jet kinetic luminosity or power \citep[e.g.][]{Punsly:2001p10577}.  Estimates of jet kinetic power from an examination of jet interaction regions are potentially useful in studying the connection between jet power and the accretion rate onto the super-massive black holes.  \citet{Wang:2004p9708} claim an inverse relationship between accretion rate as a fraction of the Eddington limit and jet power.  However, \citet{Punsly:2005p10552}, find that jet power is unrelated to accretion rate, evidenced by measurements of the extended radio power in quasars, blazars, and radio galaxies and optical spectra of their nuclei. Thus observations of jet interactions provide a new and independent method to estimate jet power.

Models of the X-ray emission mechanisms in radio galaxy hot spots are relatively uncertain, with non-thermal models reliant on estimates of the photon spectral index of the electron population that is scattering low energy photons, either the radio photons themselves (synchrotron self-Compton emission; SSC) or external photons (external Compton emission; EC).  In the case of SSC, uncertainties in the size of the radio emission region \citep[e.g.][]{Young:2001p9220} are also a problem for models.  High resolution radio observations can be used to better measure the size, shape, flux density, and spectrum of the compact radio emission in the hot spots and in the process remove the observational uncertainties from the models. 

\begin{figure}[ht]
\epsscale{0.6}
\plotone{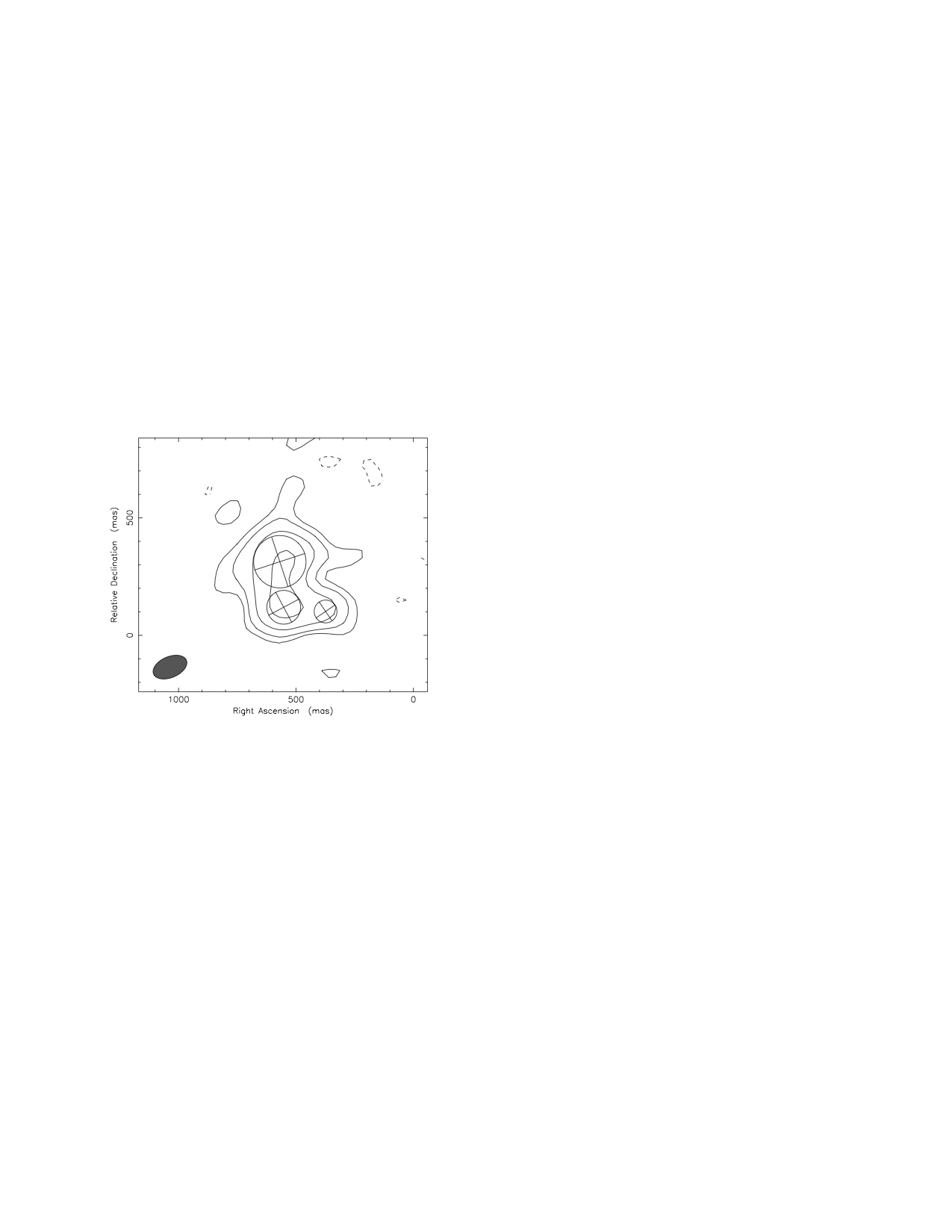}
\caption[Australian LBA image of the southern lobe hot spot of PKS B2152$-$699 at 1.4 GHz.]{Southern lobe hot spot of PKS B2152$-$699 as imaged at 1.4 GHz using the Australian VLBI array \citep{Young:2005p5449}. Radio contours are at $-10$,10,20,40 and 80\% of the peak flux density of 17.2 mJy beam$^{-1}$. The beam is $152\times90$ mas at a position angle of $-67^{\circ}$. The model components fit to the data are shown superimposed on the image.}
\label{fig:fig2152}            
\end{figure}

Wide-field, sensitive, and high resolution VLBI observations allow compact features in radio galaxies, other than the nucleus, to be imaged, for example the hot spots in the radio lobes that terminate the relativistic jets, or other hot spots due to jet/cloud interactions. \citet{Young:2005p5449} have used LBA observations of PKS B2152$-$699 to detect the southern lobe hot spot in this radio galaxy at 1.4 GHz, producing the highest angular resolution image of a radio galaxy hot spot to date (Figure \ref{fig:fig2152}).  In contrast, the northern jet in PKS 2152$-$699 undergoes a heavy interaction with a cloud of gas in the outskirts of the host galaxy.  \citet{Young:2005p5449} combined the wide-field LBA data, \emph{Chandra} X-ray imaging data, and \emph{HST} optical data to form a high resolution multi-wavelength view of PKS B2152$-$699 and concluded that the morphology of the galaxy, including the complex jet/cloud interaction region, can be explained by twin jets, precessing at a rate of a few times $10^{-4}$ degrees per year. Their LBA observations did not detect, with enough significance, the jet-interaction with the extra-nuclear cloud of the northern hot spot - it appeared in the image as a 3$\sigma$ feature and was therefore not a robust detection.

I aim to use the high sensitivity and high resolution available with the LBA to study the hot spots and jet-cloud interaction regions of a selection of southern radio galaxies (see Chapter \ref{chap:selection}) using wide-field VLBI imaging techniques. The high resolution data will be used with \emph{Chandra}, \emph{Spitzer} and \emph{HST} data, where available, to accurately model the X-ray emission from the radio galaxies using the techniques pioneered by \citet{Young:2005p5449}. The high sensitivity in combination with the high resolution available with the LBA may also be sufficient to resolve the shock regions for the first time.

\subsection{Wide-field VLBI surveys}
\label{sec:introsurvey}

The next generation of low frequency instruments, such as the LOw Frequency ARray (LOFAR), the European expansion of LOFAR (E-LOFAR) and the Square-Kilometre Array (SKA) are ambitious facilities that will face many technical and data processing challenges. Furthermore, the low frequency sky that these instruments will be surveying is not very well known and even less is known at the sub-arcsecond scale; this poses potential risks with regards to the calibration strategies envisaged for these instruments. Previous VLBI snapshot surveys at wavelengths around 90 cm only targeted the brightest of sources and were plagued by poor sensitivity, radio interference and limited coherence times. Furthermore, the potential for wide-field imaging was limited by the poor spectral and temporal resolution of early generation hardware correlators, the available data storage and computing performance at the time. As a result, images of only a few tens of sources have been published to date and little is known about the sub-arcsecond and sub-Jansky population of low-frequency sources.

With recent improvements in storage and computing facilities, wide-field processing algorithms and VLBI correlators, it is now feasible to image wide fields of view at VLBI resolution. I aim to perform a VLBI survey of the entire Full-Width Half Maximum (FWHM) primary beam of a 25 m VLBA antenna for an observation of a target field at 90 cm. The approach will build upon a deep VLBI survey of a $36\arcmin$ wide field performed by \citet{Garrett:2005p10555} at 20 cm in which a central bright source was used as an in-beam calibrator. Other field sources may be used to refine the calibration and thus correct for ionospheric effects that plague low frequency observations. This approach will permit the imaging of many potential target sources simultaneously by taking advantage of the full sensitivity of the observation across the entire field of view (28 deg$^{2}$). The results of this survey will provide an important indication of what may be seen by future low-frequency instruments such as the Low Frequency Array (LOFAR), European LOFAR (E-LOFAR) and the Square Kilometre Array (SKA).

\section{Outline of thesis}
\label{sec:intro.thesis}

In chapter \ref{chap:wfvlbi}, a brief theoretical introduction to the fundamentals of aperture synthesis is given. Particular emphasis is given to the factors that affect the available field-of-view and how these can be overcome.

In chapter \ref{chap:selection}, sources are defined and selected for the three areas of science to be investigated with wide-field VLBI techniques: starbursts, AGN jet interactions, and a low frequency wide-field survey.

In chapters \ref{chap:ngc253} to \ref{chap:starbursts}, observations of the selected sample of southern starburst galaxies are described. In chapter \ref{chap:ngc253}, the first epoch VLBI observation of NGC 253 at 2.3 GHz is described and related to both previous VLA observations and 1.4 GHz LBA observations of the source. In chapter \ref{chap:ngc4945}, ATCA $17-23$ GHz observations and the first and second epoch VLBI observations of NGC 4945 at 2.3 GHz are described. Finally, in chapter \ref{chap:starbursts}, the ATCA $17-23$ GHz and 2.3 GHz VLBI observations of the remaining starburst galaxies are described. These starburst observations have resulted in the highest resolution studies of these galaxies.

In chapters \ref{chap:pictora} and \ref{chap:jets}, observations of the selected sample of AGN hot spots are described. In chapter \ref{chap:pictora}, the first VLBI observations of the north-west and south-east hot spots of Pictor A at 1.6 GHz and 2.3 GHz are described and compared to VLA, \emph{Chandra} and \emph{Spitzer} observations of the same source. A multi frequency analysis is used to determine the emission mechanism in the hot spots. In chapter \ref{chap:jets}, observations of the remaining jet-interaction sources of the selected sample, PKS $0344-345$ and PKS $0521-365$, are described.

In chapter \ref{chap:lfwfvlbi}, an observation of the first low frequency, wide-field VLBI survey is described. The survey is the widest VLBI survey performed; achieving a survey area of 28 deg$^{2}$.

Finally, in chapter \ref{chap:conclusion}, a summary of all of the wide-field VLBI results is given as well as future directions for new science, follow-up work and improved processing algorithms.

\linespread{1.0}
\normalsize
\begin{savequote}[20pc]
\sffamily
Every person takes the limits of their own field of vision for the limits of the world.
\qauthor{Arthur Schopenhauer}
\end{savequote}

\chapter{An Introduction to Wide-field VLBI Imaging}
\label{chap:wfvlbi}

\linespread{1.3}
\normalsize

\section{Introduction}
\label{sec:wfvlbi.intro}

To help in understanding the limitations on the field of view in VLBI imaging and how these may be overcome, a brief introduction to aperture synthesis will be given. The intention here is not to explain all of the technical details of radio interferometry but rather to highlight the assumptions, in traditional VLBI, that affect the overall field of view and how these limitations can be overcome. Furthermore, the role of the atmosphere in radio observations will not be discussed here; where it plays a role in limiting the quality of data it will be discussed in the observation section of the subsequent chapters. For a deeper discussion of radio interferometry, we refer the reader to standard texts on the subject e.g. \citet{Taylor:1999p26878} and \citet{Thompson:2001p21070}.

\section{Fundamentals of Aperture Synthesis}
\label{sec:wfvlbi.apsynthesis}

\begin{figure}[ht]
\epsscale{0.5}
\plotone{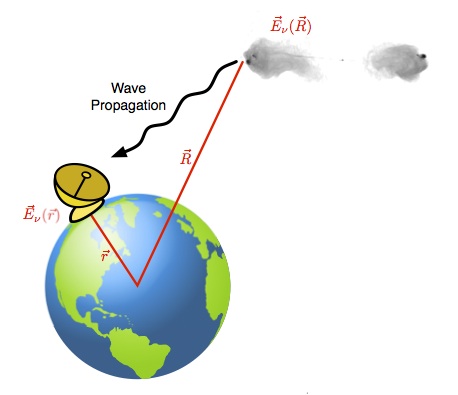}
\caption[Propagation of an electromagnetic wave.]{Propagation of an electromagnetic wave $\vec{E}_{\nu}(\vec{r})$ from an astrophysical source at $\vec{R}$, to a radio antenna at $\vec{r}$.}
\label{fig:figantgeom}
\end{figure}

For simplicity, consider a quasi-monochromatic component $\vec{E}_{\nu}(\vec{R})$ of a time-varying field $\vec{E}(\vec{R},t)$ emitted from an astrophysical radio source at position $\vec{R}$ and observed from an Earth-based antenna at position $\vec{r}$, see Figure \ref{fig:figantgeom}. The linearity of Maxwell's equations allows each of the complex components of the signal, $\vec{E}_{\nu}(\vec{R})$, to be superposed at the observer:

\begin{equation}
\vec{E}_{\nu}(\vec{r})=\int \int \int P_\nu(\vec{R},\vec{r})\vec{E}_{\nu}(\vec{R})dx dy dz.
\label{eq:wfvlbi.eq1}
\end{equation}

The propagator, $P_\nu(\vec{R},\vec{r})$, describes how the electric field at $\vec{R}$ affects the electric field at $\vec{r}$. For simplicity, if electromagnetic radiation is considered to be a scalar field, thus ignoring all polarisation phenomena, and that the emitting source is at infinite distance then we only need to consider the surface brightness of the source ${\cal E}_{\nu}(\vec{R})$. Furthermore, we can assume that the electromagnetic wave propagates through empty space, thus the propagator in Equation \ref{eq:wfvlbi.eq1} takes on a simple form shown in Equation \ref{eq:wfvlbi.eq2}.

\begin{equation}
E_{\nu}(\vec{r})=\int {\cal E}_{\nu}(\vec{R})\frac{e^{2\pi i \nu |\vec{R}-\vec{r}|/c}}{|\vec{R}-\vec{r}|}dS.
\label{eq:wfvlbi.eq2}
\end{equation}

Here $E_{\nu}(\vec{r})$ is the quantity observed by a radio telescope and $dS$ is the element of surface area on the celestial sphere.

An interferometer is a device that measures the spatial coherence function $V_{\nu}(\vec{r_{1}},\vec{r_{2}})$, which is the correlation of the field at points $\vec{r_{1}}$ and $\vec{r_{2}}$ and is defined as $V_{\nu}(\vec{r_{1}},\vec{r_{2}})=\langle E_{\nu}(\vec{r_{1}})E^{*}_{\nu}(\vec{r_{2}}) \rangle$. Substituting Equation \ref{eq:wfvlbi.eq2} for $E_{\nu}(\vec{r})$ and using the simplifying assumptions that the radiation from the source is spatially incoherent and that small terms of order $|\vec{r}/\vec{R}|$ can be neglected as a result of the large distance to the emitting source, the spatial coherence function simplifies to

\begin{equation}
V_{\nu}(\vec{r_{1}},\vec{r_{2}})\approx\int I_\nu(\vec{s})e^{ -2 \pi i \nu\vec{s}\cdot(\vec{r}_1-\vec{r}_2)/c}d\Omega,
\label{eq:wfvlbi.eq3}
\end{equation}

where $\vec{s}$ is the unit vector pointing towards the source and $d\Omega$ is the surface element of the celestial sphere. $I_\nu$ is the spatial intensity distribution of electromagnetic radiation produced by an astronomical object at a particular frequency.

An appropriate co-ordinate system can be chosen for the interferometer baselines such that the $u$ and $v$ axes form a plane, often referred to as the $(u,v)$ plane, perpendicular to the line of sight with the $w$ axis along the line of sight \citep{Clark:1999p18936}. The co-ordinates in $(u,v,w)$ space are measured in units of wavelengths. The sky co-ordinates, in the image plane, are the direction cosines $l$ and $m$. For an array of Earth-based radio telescopes, the baseline vector measured in $(u,v,w)$-space will change orientation, with respect to the line of site of the target source, as the Earth rotates on its axes. Consequently, over the course of many hours, many points in the $(u,v,w)$ can be sampled by each of the baselines taking part in the observation. Using this co-ordinate system, the general response of a two-element interferometer to spatially incoherent radiation from a distant source can be written as:

\begin{equation}
V_\nu(u,v,w)=\iint I_\nu(l,m)\frac{e^{-2\pi i[ul+vm+w(\sqrt{1-l^{2}-m^{2}}-1)]}}{\sqrt{1-l^{2}-m^{2}}}dl\,dm.
\label{eq:wfvlbi.eq4}
\end{equation}

In traditional VLBI it is assumed that only a small field-of-view will be imaged and so the third term in the exponential can be ignored. This is an important assumption that will be revisited in Section \S~\ref{sec:wfvlbi.nce}. The simplification leads to the form: 

\begin{equation}
V_\nu(u,v)=\iint I_\nu(l,m)e^{-2\pi i(ul+vm)}dl\,dm
\label{eq:wfvlbi.eq5}
\end{equation}

This can be readily inverted, using the Fourier Transform, to recover the true brightness distribution of the radio source:

\begin{equation}
I_\nu(l,m)=\iint V_\nu(u,v)e^{2\pi i(ul+vm)}du\,dv
\end{equation}

In practice, it is not possible to directly use the inverse Fourier transform to determine the brightness distribution from the visibilities, as the spatial coherence function, $V_{\nu}(u,v)$, is only known for a discrete subset of the $(u,v)$ plane. As a result, a sampling function, $S(u,v)$ must be introduced:

\begin{equation}
I^{D}_{\nu}(l,m)=\iint S(u,v)V_\nu(u,v)e^{2\pi i(ul+vm)}du\,dv.
\end{equation}

$I^{D}_{\nu}$ is referred to as the \emph{dirty image} and is related to the true brightness distribution, $I_{\nu}$, of the radio source (using the convolution theorem for Fourier transforms) as follows:

\begin{equation}
I^{D}_{\nu}=I_{\nu}*B,
\end{equation}

Where $B$ is the point spread function corresponding to the sampling function $S(u,v)$ and is often referred to as the \emph{dirty beam},

\begin{equation}
B(l,m)=\iint S(u,v)e^{2\pi i(ul+vm)}du\,dv.
\end{equation}

The true brightness distribution, $I_{\nu}$, can then be recovered from the dirty image, $I_{\nu}^{D}$, using the dirty beam, $B$, through the process of deconvolution.

\section{Wide field effects}
\label{sec:wfvlbi.wfeffects}
\subsection{Bandwidth Smearing}

\begin{figure}[p]
\mbox{
\epsscale{0.44}
\plotone{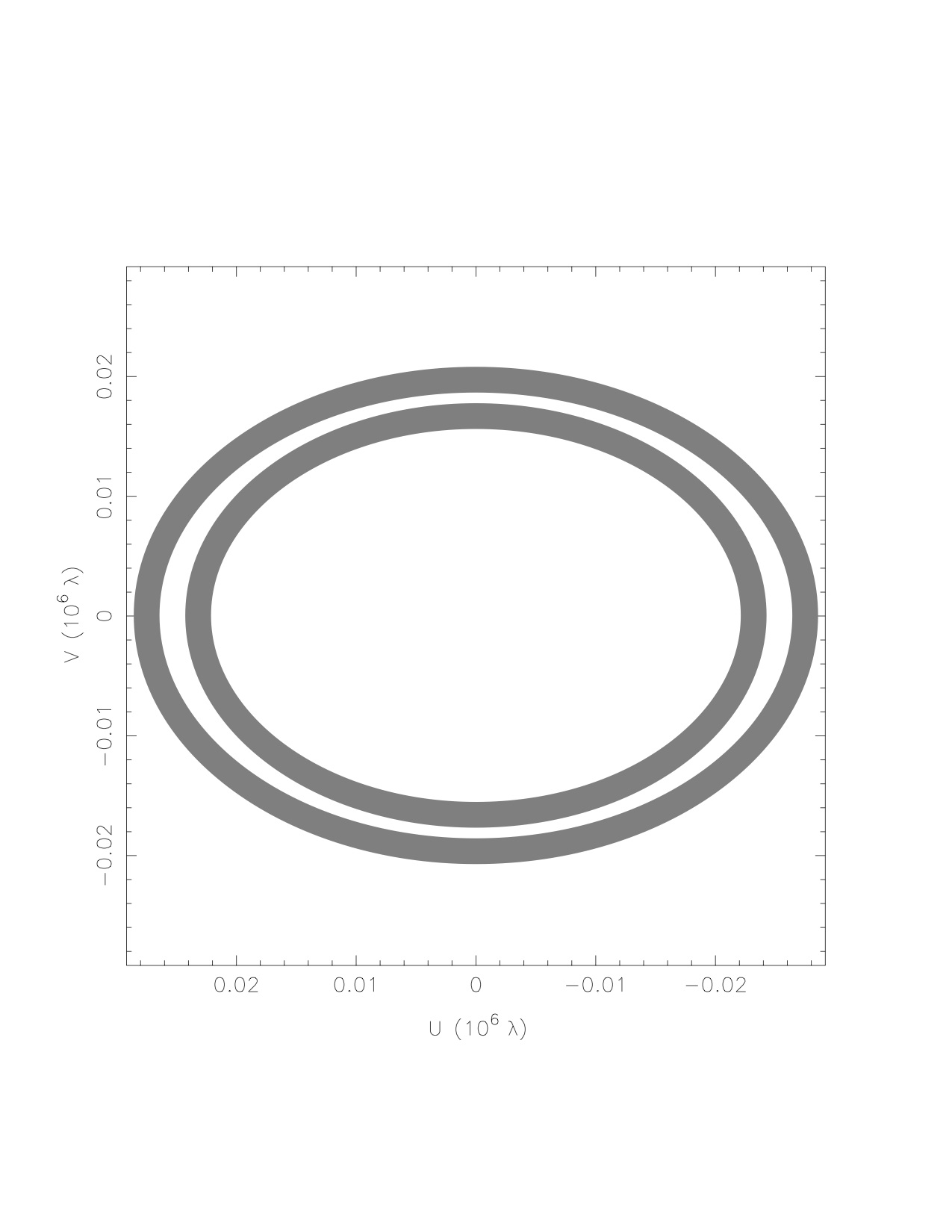} \quad
\epsscale{0.48}
\plotone{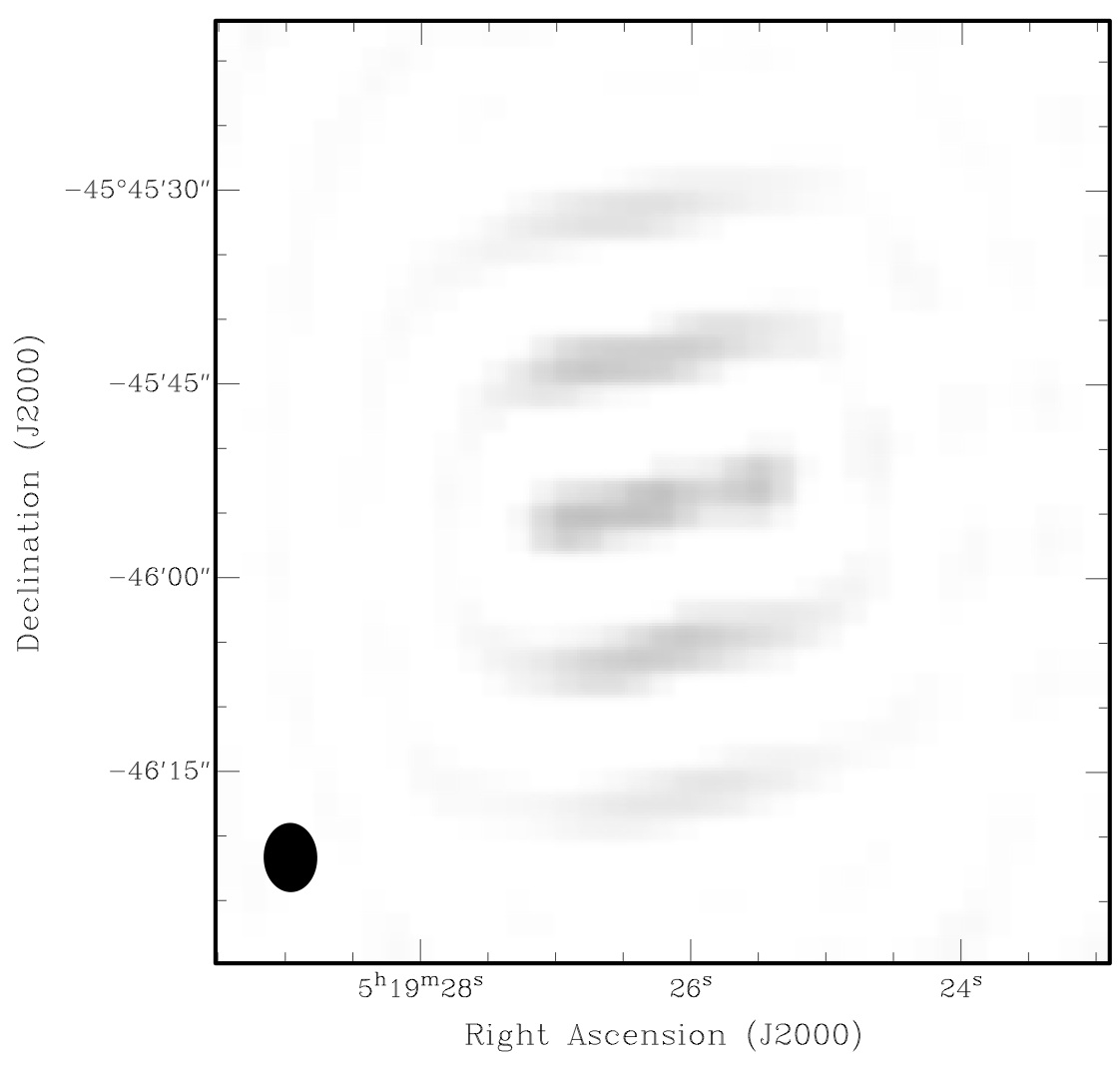}
}
\mbox{
\epsscale{0.44}
\plotone{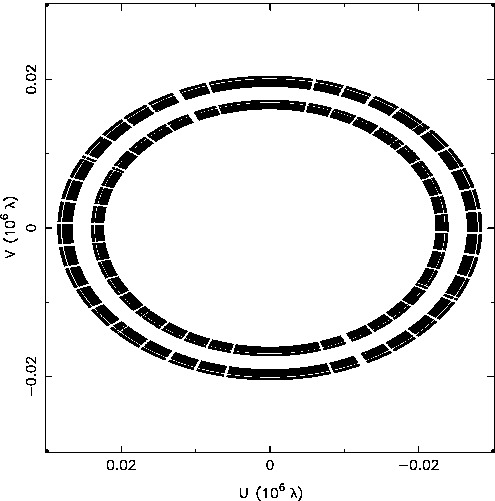} \quad
\epsscale{0.48}
\plotone{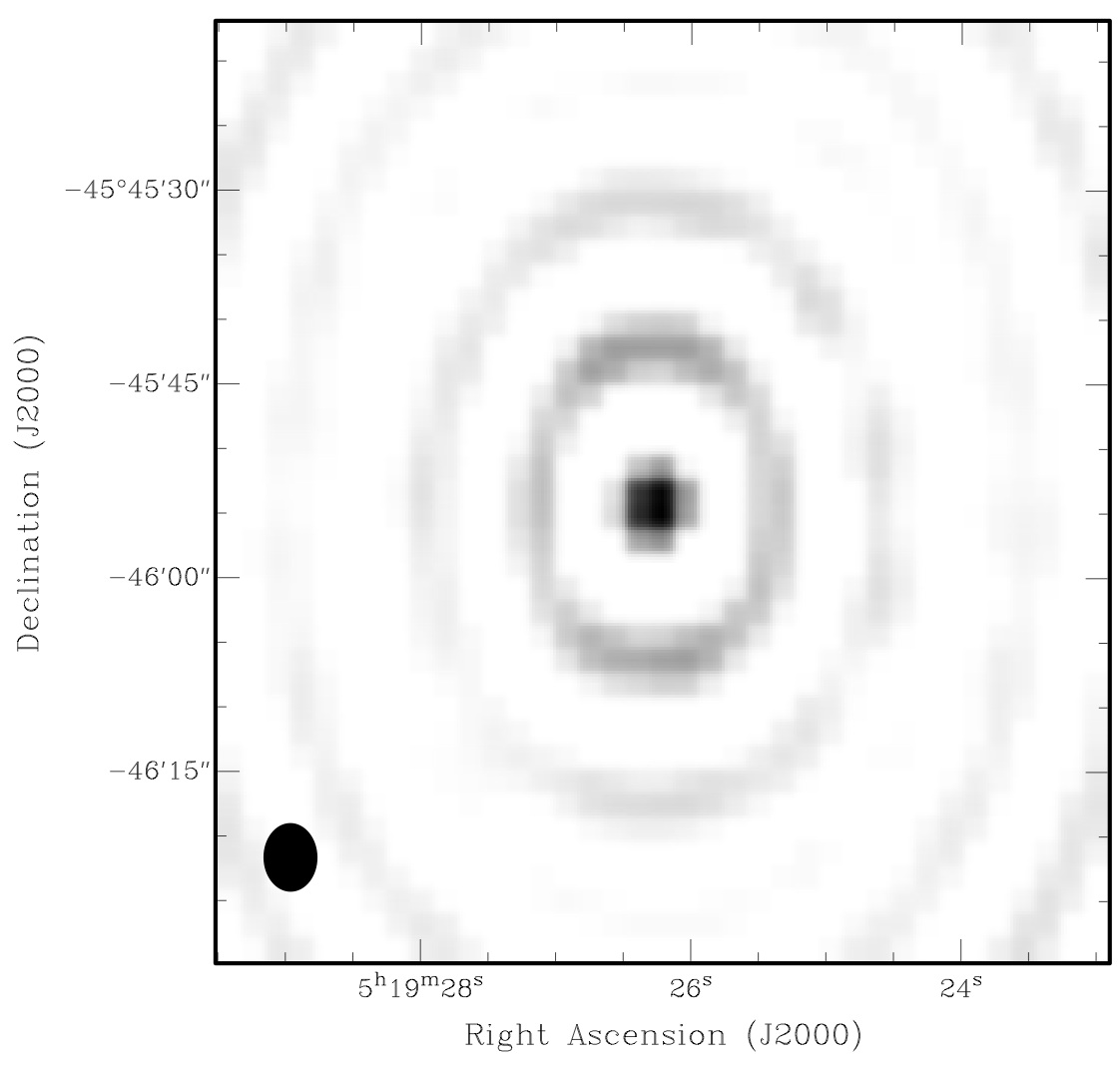}
}
\caption[Effects of bandwidth smearing.]{The effects of bandwidth smearing. Top left: $(u,v)$ visibilities based on the centre frequency of a single channel with a significant bandwidth (represented by the thickness of the grey curve). Top right: The resulting dirty image of a point source approximately $4\arcmin$ from the phase centre of a 20 cm observation. Bottom left: The same channel bandwidth is now subdivided into sub-channels each with only one sixteenth of the original bandwidth. Bottom right: the resulting dirty image of the same point source using the smaller sub-channels and multi-frequency synthesis techniques.}
\label{fig:figbwsmearing}
\end{figure}

The equations in Section \S~\ref{sec:wfvlbi.apsynthesis} assume monochromatic signals and a monochromatic intensity distribution. In practice, the receiver passbands and the correlation process result in one or more channels of finite width, $\Delta\nu$. However, the visibility data are treated as if they are measured at a single central frequency of $\nu_{0}$. This point actually represents an averaging across the bandwidth, $\Delta\nu$, of the channel. If the degree of averaging is significant then this may lead to visible distortion in the image plane in the form of radial smearing, see Figure \ref{fig:figbwsmearing}. The effect is a form of chromatic aberration and is commonly referred to by radio astronomers as bandwidth smearing. Bandwidth smearing and time-average smearing (which will be covered further in the following section) can be explained intuitively as averaging over high frequency variations in the visibilities in the $(u,v)$ plane in orthogonal directions: radially for bandwidth averaging and azimuthally for time averaging.

Assuming a square bandpass, with no tapering and square $(u,v)$ coverage, the width of the radial distortion of the observed source caused by bandwidth smearing increases as $\Delta\nu\theta_{0}$, whereas the peak intensity decreases as $1/(\Delta\nu\theta_{0})$ \citep{Bridle:1999p10564}. Thus, the severity of smearing increases with distance from the phase centre, $\theta_{0}$, of the observation and with the bandwidth, $\Delta\nu$. In general, the smearing process preserves the integrated flux of the source and may be recovered through deconvolution \citep{Bridle:1999p10564}. The source structure, however, can not be recovered as averaging across the band destroys this information. Furthermore, sensitivity in the wider field is reduced as a result of the reduced peak intensity. For wide-field radio astronomy this poses a problem, on the one hand a large bandwidth is desired to improve sensitivity but on the other it restricts the field that may be imaged.

The simplest strategy for reducing the effect of bandwidth smearing is to correlate wide-band data into a large number of narrow-band spectral channels and calculate the $(u,v)$ locations for each channel separately. This technique, referred to as multi-frequency synthesis, can dramatically reduce the effects of bandwidth smearing and as an added bonus can also be used to increase the overall $(u,v)$ coverage of an observation, Figure \ref{fig:figbwsmearing} illustrates the improvements in both the $(u,v)$ and image plane. One drawback of this technique is that to increase the total number of channels from one to $n$, $n$ times more data must be stored and processed - thus added pressure is placed on storage and processing resources. Secondly, correlators, such as those used by the LBA \citep{Cannon:1997p10533}, EVN \citep{vanLangevelde:2004p9712} and VLBA \citep{Benson:1995p26884}, have traditionally been implemented in hardware, with the restricted nature of their designs often limiting the spectral resolution of output visibilities. This has changed recently with the introduction of software correlators and their mainstream use for VLBI at facilities such as the LBA \citep{Deller:2007p10545} and the VLBA, their flexible design allows the generation of visibilities with fine spectral resolution. The only restrictions that remain are those of data storage and processing.

\subsection{Time Average Smearing}

\begin{figure}[ht]
\mbox{
\epsscale{0.44}
\plotone{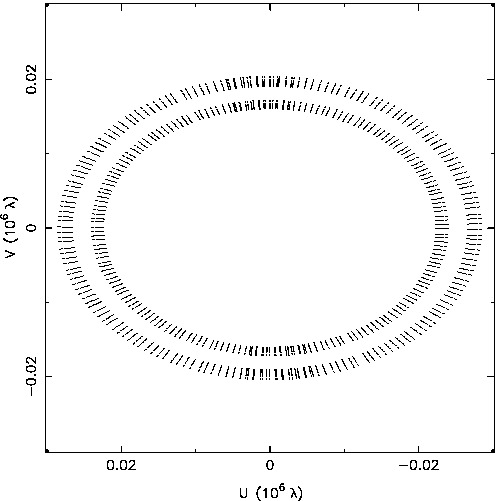} \quad
\epsscale{0.48}
\plotone{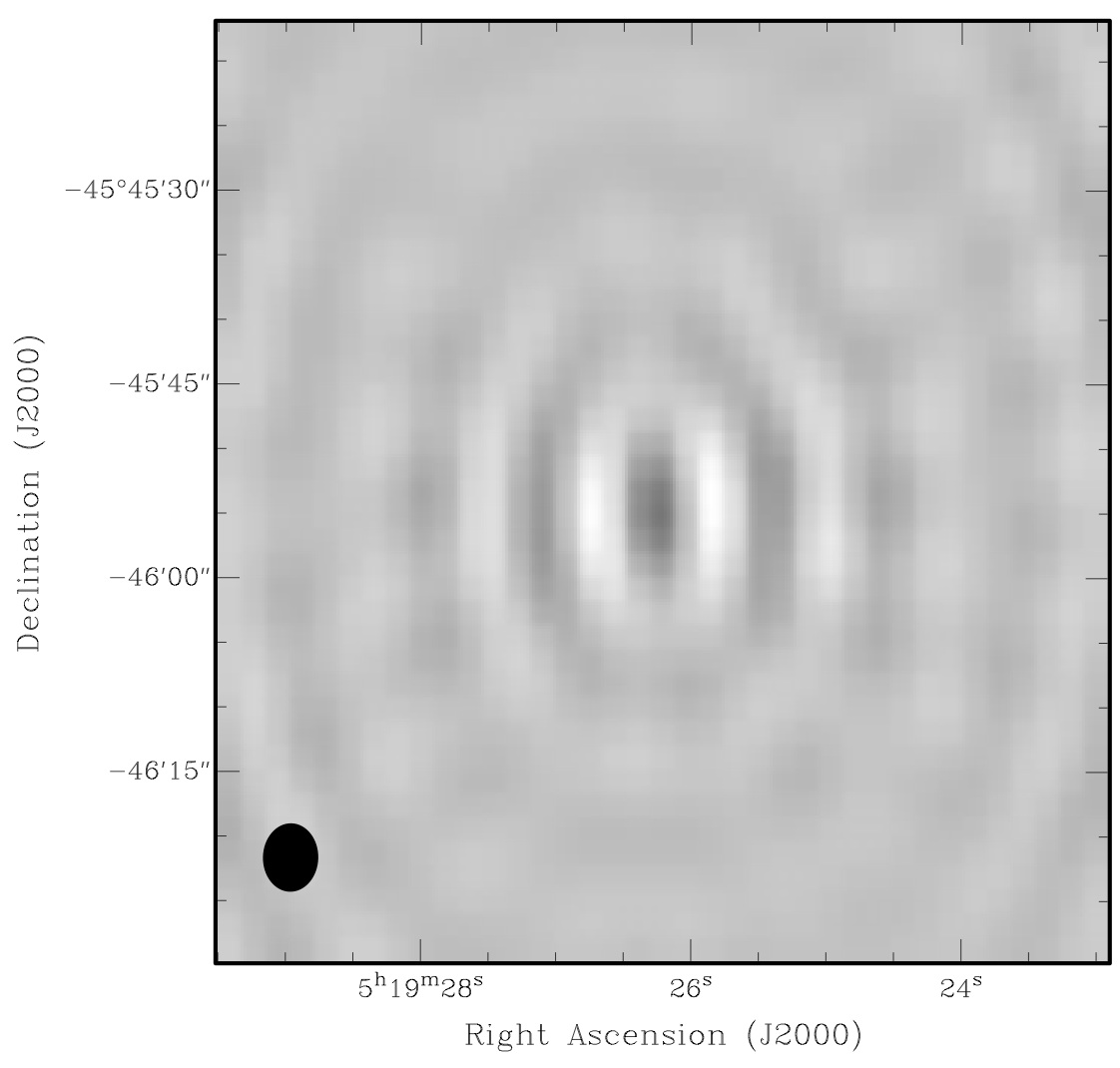}
}
\caption[Effects of time-average smearing.]{The effects of time-average smearing. Left: $(u,v)$ visibilities averaged in time by a factor sixteen greater than those in Figure \ref{fig:figbwsmearing}. Right: The resulting dirty image of a point source approximately $4\arcmin$ from the phase centre of a 20 cm observation.}
\label{fig:figtasmearing}
\end{figure}

The averaging of visibility data over time is another cause of image smearing and is referred to as time-average smearing. Visibility data output from a correlator are averaged over a specified time, $\tau_{a}$, and assigned a $(u,v)$ value that corresponds to the midpoints of the averaging period \citep{Bridle:1999p10564}. However, during this period the baseline vector would not have remained constant. The rotation of the Earth would have caused the vector to rotate through $\omega_E\Delta\tau$, where $\omega_E$ is the angular velocity of the earth. Where this rotation is significant, the $(u,v)$ points are effectively under-sampled. The effect on the image plane is complex and depends greatly on the location of the source in the Celestial Sphere. However, the general effect is a loss of peak amplitude. Furthermore, the magnitude of this loss increases with $\tau_{a}$ and with the distance of the source from the phase centre, $\theta_{0}$, and so is also an important effect to consider in wide-field imaging. Figure \ref{fig:figtasmearing} illustrates the effect of time-average smearing in which the $(u,v)$ data are averaged over a period sixteen times greater than those in Figure \ref{fig:figbwsmearing}.

The solution to time-average smearing is similar to that for bandwidth smearing. The correlator needs to generate visibilities with fine enough temporal resolution i.e. small $\tau_{a}$. As in the case for bandwidth smearing, the drawbacks are similar. Traditional correlators are limited in the temporal resolution with which they can output visibilities (generally of the order of $\sim1$ second) and secondly, for each halving of $\tau_{a}$ there is a subsequent doubling in visibility data. Fortunately, the flexibility of software correlators can at least alleviate the first of these issues with their ability to output data at any desired temporal resolution within fundamental limits i.e. high time resolution imposes limits on the frequency resolution and vice versa.

\subsection{Non-coplanar Effects}
\label{sec:wfvlbi.nce}

The general response of a two-element interferometer to spatially coherent radiation from a distant source, was defined in Equation \ref{eq:wfvlbi.eq4}. This form can be exactly inverted using a simple two-dimensional Fourier transform in situations where the interferometer array is aligned in the East-West direction and thus all visibilities lie on a plane in $(u,v,w)$ space \citep{Perley:1999p10576} - this, however, is only possible with small arrays and is not possible with Earth-based VLBI in a useful manner. A second possibility, as discussed in Section \S~\ref{sec:wfvlbi.apsynthesis}, is to limit the field-of-view, in which case Equation \ref{eq:wfvlbi.eq4} simplifies to the approximation defined in Equation \ref{eq:wfvlbi.eq5}, a form that is readily inverted by a two-dimensional Fourier transform. However, this method results in serious limits on the field of view possible with VLBI.

\begin{figure}[ht]
\epsscale{0.6}
\plotone{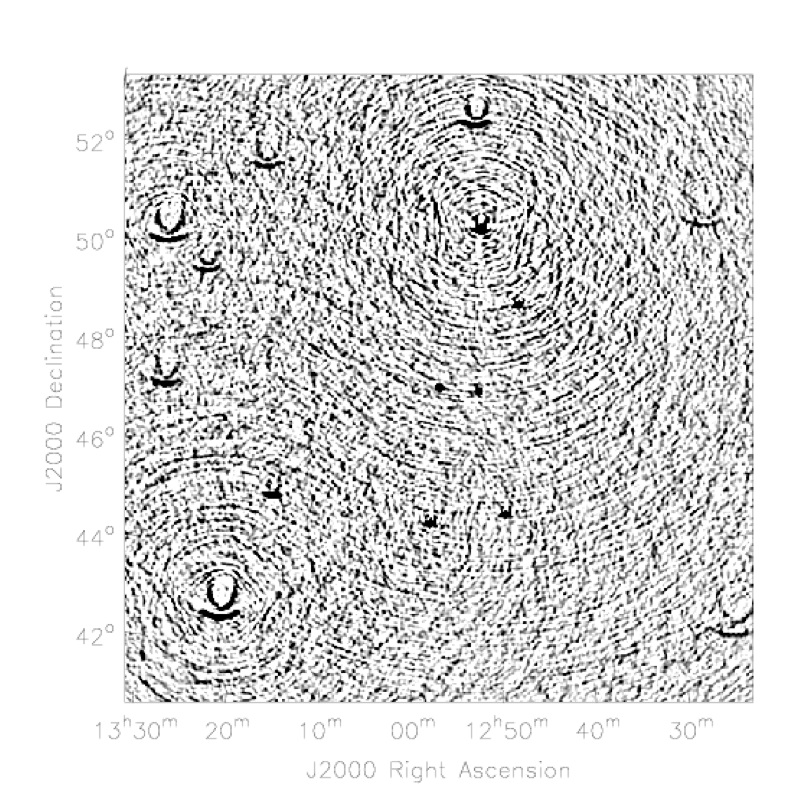}
\caption[Non-coplanar effects.]{The dirty image of a simulated 74 MHz field in which non-coplanar effects have not been taken into consideration \citep{cor03}.}
\label{fig:fignce}
\end{figure}

A lack of consideration of non-coplanar effects results in position-dependent phase-shifts across the field and a variation in size structures sampled by a given baseline pair of antennas across the field \citep{cor03}. The scale of the effect is approximately proportional to $B\lambda/D^{2}$, so it is most troublesome for longer wavelength observations ($\lambda$), long baselines ($B$) and arrays with small diameter antennas ($D$). The effect also increases with distance from the phase centre. Figure \ref{fig:fignce} shows the result of non-coplanar effects on a 74 MHz VLA observation of a simulated field.

Non-coplanar effects can be alleviated using a number of different techniques. In a non-coplanar array the measured visibilities lie on a surface within three-dimensional space rather than on a single two-dimensional plane. As such, the most intuitive solution to the problem is to move up to a three dimensional Fourier transform \citep{Perley:1999p10576,cor03}. There are two main drawbacks to this approach: firstly, adding a third dimension greatly increases the processing required to transform the visibilities; and secondly, the resulting image cube is mostly devoid of emission since the sky brightness is constrained to a spherical surface within the cube.

The most commonly used approach for dealing with non-coplanar effects is the use of image-plane facets \citep{Perley:1999p10576}. The field-of-view is divided into a grid of $N\times N$ facets, each is imaged separately and then reconciled at a later stage of the deconvolution process to generate the final image. By reducing the size of each facet the severity of non-coplanar effects can be minimised. Unfortunately, the performance of the faceting algorithm significantly degrades with $N$, approximately $O(N^{2})$, and is typically $30-100$ times slower than the standard two-dimensional FFT approach.

In VLBI imaging, at least with the simple arrays we have available currently, much of the image plane is in fact devoid of emission since such observations are particularly sensitive to small compact objects and not to large-scale emission. So in circumstances in which the positions of a small number of compact sources are known, based on a previous wide-field survey at low resolution, only small regions around each of the source centres need to be imaged. In this process, the visibility data must be phase shifted to the centre of each source to be imaged. As long as the number of sources doesn't exceed the number of facets required to image the entire field, great savings in processing time can be achieved. As an alternative for extremely wide-fields, it is also feasible to re-correlate the raw observed data such that visibilities are generated for the correlation centre of each source of interest (or at least a group of sources if they can be easily imaged within a single field). The added benefit of this approach is that the effects of bandwidth and time-average smearing can be reduced, since the source is now at the phase centre, this results in smaller data-sets as lower temporal and spectral resolution can be used to generate the visibilities.

In cases where insufficient a priori information is available for a field or where transient sources are the target of interest, it is desirable to image the entire field of view. While faceting is one means of performing such a task it has been previously mentioned that the algorithm is rather processor intensive. Another alternative is the recently developed $w$-projection algorithm \citep{cor03}. The algorithm works by projecting $w$ out of the visibilities after which a simple two-dimensional Fourier transform can be used to invert them into the image plane. For the same dynamic range, $w$-projection provides approximately an order of magnitude improvement in processing time over faceting. The $w$-projection algorithm, however, only works optimally when the entire image can be transformed in computer memory, as opposed to faceting in which small facets can be processed individually. This requirement can limit image sizes to the order of $10^4\times10^4$ pixels based on typical memory availability for computing workstations today. Nonetheless, the faceting approach can be used in conjunction with $w$-projection to generate even larger fields of view.

\section{Application of wide-field VLBI}
\label{sec:wfvlbi.wfapplication}

\begin{figure}[p]
\epsscale{0.7}
\plotone{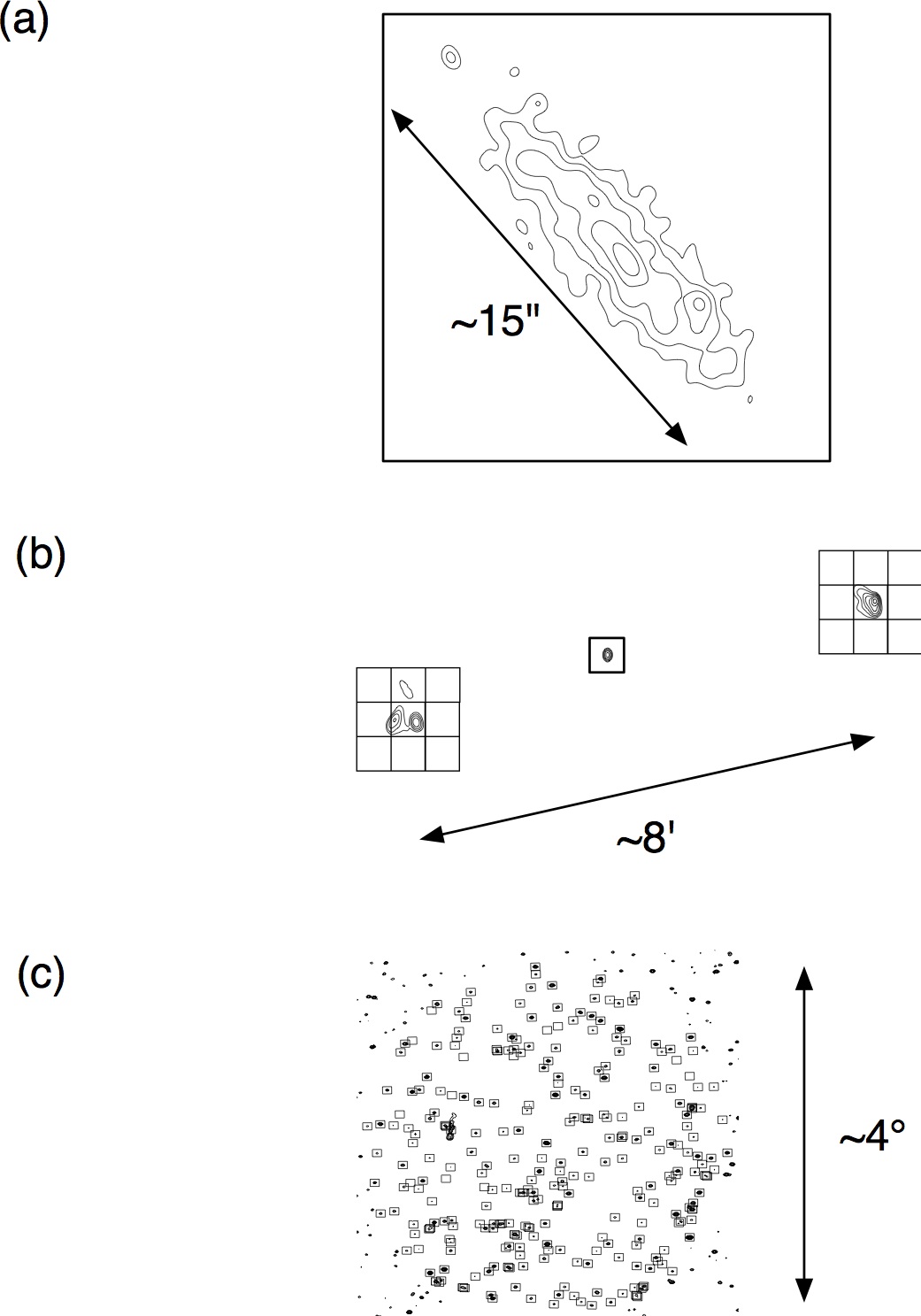}
\caption[Application of wide-field VLBI.]{(a) A nearby prototypical nuclear starburst can be imaged with a single field with $w$-projection or using standard two-dimensional fourier techniques. The entire region needs to be imaged to enable the detection of new supernova events and/or remnants. (b) The regions around radio AGN hot spots can be surveyed using faceting or $w$-projection techniques; the AGN nucleus can be used for in-beam calibration. The remaining regions in between can be safely ignored as they are unlikely to contain significant emission. (c) For large surveys, where a priori information is available for source positions, sources can be targeted individually using the phase-centre shifting approach.}
\label{fig:figwfapp}
\end{figure}

The three prime areas of scientific focus, as outlined in Chapter \ref{chap:intro}, all apply and test different techniques for achieving wide fields of view. The three methods are shown graphically in Figure \ref{fig:figwfapp}.

For the study of starburst galaxies it will be necessary to search for faint and often extended sources, typically supernovae and supernova remnants. While the position of some of these sources may be known a priori, one aspect of the science is to search for new sources as they appear. As a result, it is necessary to image the entire field of interest, see Figure \ref{fig:figwfapp}(a). In the case of nearby nuclear starburst galaxies, this region is typically of the order of $0.25-1.0\arcmin$. At full LBA resolution, this would require an image of the order of $10^4\times10^4$ pixels, a size that is feasibly produced with either a faceted approach or with $w$-projection. In observations of NGC 253, Chapter \ref{chap:ngc253}, the $w$-projection approach is utilised to search the wider field of this source. In NGC 4945, Chapter \ref{chap:ngc4945}, a simple two-dimensional Fourier transform approach is used owing to the smaller size of the emission region. For the remaining starbursts, a simple two-dimensional Fourier transform approach is utilised for the main area of interest, however, phase-centre shifting is employed to image other regions that may also be of interest i.e. where starburst activity is assumed to occur in more than one region.

In the study of AGN jet interactions, Chapters \ref{chap:pictora} and \ref{chap:jets}, the bright cores of these galaxies are used as in-beam calibrators. The hot spots can be quite distant from the AGN nucleus and so phase-centre shifting is utilised to image the hot spots and/or interaction regions. The in-between regions are not imaged as they are unlikely to contain any significant emission. In the case of hot spots with extended emission, a grid of facets is used to image the entire region containing this emission (see Figure \ref{fig:figwfapp}(b)).

Finally, for a wide-field, low-frequency VLBI survey, described in Chapter \ref{chap:lfwfvlbi}, a priori source positions are used to target each source of interest separately (see Figure \ref{fig:figwfapp}(c)). Where available, in-beam calibration is utilised for initial calibration of the field. Then, for each source, a phase-centre shift is applied to place the source at the centre of the field to be imaged, the resulting visibilities are subsequently averaged in time and frequency, and finally imaged over a small region for analysis. The process is repeated for each of the target sources. This technique allows for a large number ($\sim1000$) of sources to be imaged over a very large field of view ($\sim20-30$ square degrees).

In general, confusion from peripheral sources on the edge of the primary beam is generally not as problematic with VLBI as it is in lower angular resolution arrays such as the ATCA and VLA. However, as an added precaution, bright sources (such as an in-beam calibrator) are modelled for (after calibration) and subtracted from the calibrated $(u,v)$ visibilities to minimise the effects of confusion.

\linespread{1.0}
\normalsize
\begin{savequote}[20pc]
\sffamily
Space is big. You just won't believe how vastly, hugely, mind-bogglingly big it is. I mean, you may think it's a long way down the road to the chemist's, but that's just peanuts to space.
\qauthor{Douglas Adams}
\end{savequote}

\chapter{Source definition and selection}
\label{chap:selection}

\linespread{1.3}
\normalsize

\section{Introduction}
\label{sec:selection.intro}

The three different studies that will be performed using wide-field techniques are aimed at very different types of science and require different selection strategies to maximise the success of each study. For the southern starburst and jet-interaction investigations, many of the sources in the southern sky are not well studied even at arc-second resolution, so extra attention is required to select sources that are appropriate for study with VLBI. Whereas, for the low frequency wide-field survey, the selection of individual sources is not so much of interest but rather the nature of the low frequency radio sky at VLBI resolution. The selection criteria used to define the sources for each science aim is described separately in the following sections.  

\section{Starbursts}
\label{sec:selection.sb}

The main target in the observations of the starburst galaxies are supernova remnants, which are expected as a result of high rates of massive star formation. Extra-galactic supernova remnants are generally weak at radio wavelengths. For example, at a distance of approximately 4 Mpc, typical remnants have integrated flux densities of a few to a few tens of milli-Jansky's at 20 cm wavelengths \citep{McDonald:2002p3330, Tingay:2004p778}. At VLBI resolution, for distances up to about 8 Mpc, these sources can be resolved, possibly reducing the peak brightness below the detection threshold of the observation. While these sources may be unresolved at distances above 8 Mpc, their flux density naturally reduces by the inverse square of the distance and so once again the source may fall below the detection threshold. So, to maximise the probability of detecting supernova remnants only local galaxies, up to a distance of 8 Mpc (corresponding to recession velocities less than approximately 600 km s$^{-1}$), have been considered.

To further maximise detections using the Long Baseline Array (LBA), only galaxies with 1.4 GHz flux densities greater than 10 mJy, and with declinations less than -20$^{\circ}$ were considered.  A good indicator of active star formation is the far-infrared luminosity, so to ensure that the galaxies were in fact starburst galaxies, only candidate galaxies with a far-infrared luminosity of greater than $5\times10^{8}$ L$_{\sun}$ were selected. Finally, only candidate galaxies that appear to have populations of compact radio sources with individual flux densities greater than 1 mJy were considered.  A search in NED and IRAS \citep{Sanders:2003p7880} catalogues revealed six galaxies that met all of the selection criteria: M83, NGC 55; NGC 253; NGC 1313; NGC 4945; and NGC 5253. The proximity of these galaxies allows them to be studied at sub-parsec spatial resolutions using the LBA. The southern declination of some of these sources (NGC 55, NGC 1313 and NGC 4945 in particular) places them out of reach of instruments such as the VLA and VLBA and so have been poorly studied at radio wavelengths. The selection of galaxies also contains a good mix of edge-on and face-on galaxies providing an opportunity to compare and contrast the ionised structure of galaxies with different orientations.

\section{Jet interactions}
\label{sec:selection.agn}

\begin{figure}[ht]
\epsscale{0.4}
\begin{center}
\mbox{
\plotone{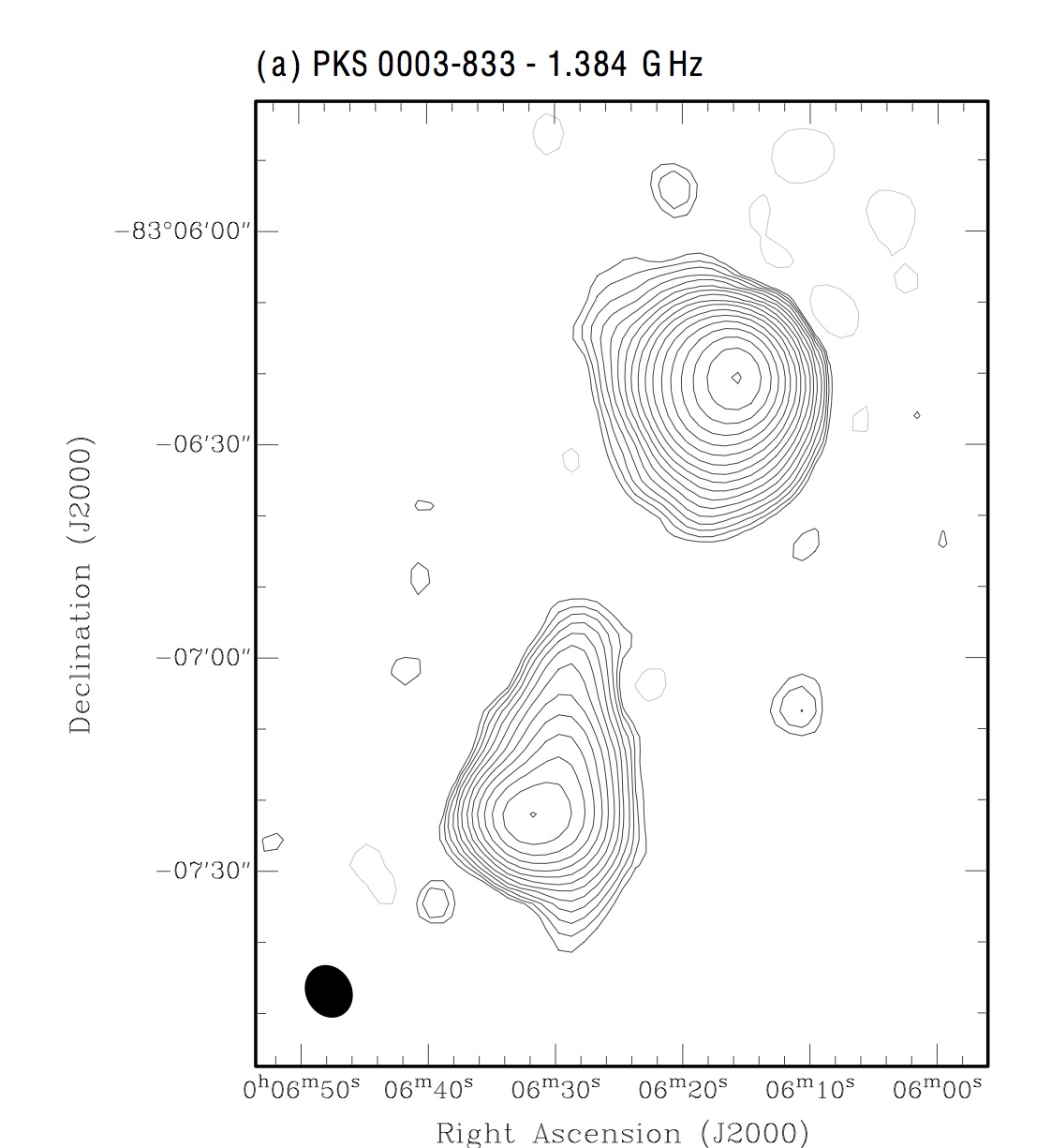} \quad
\plotone{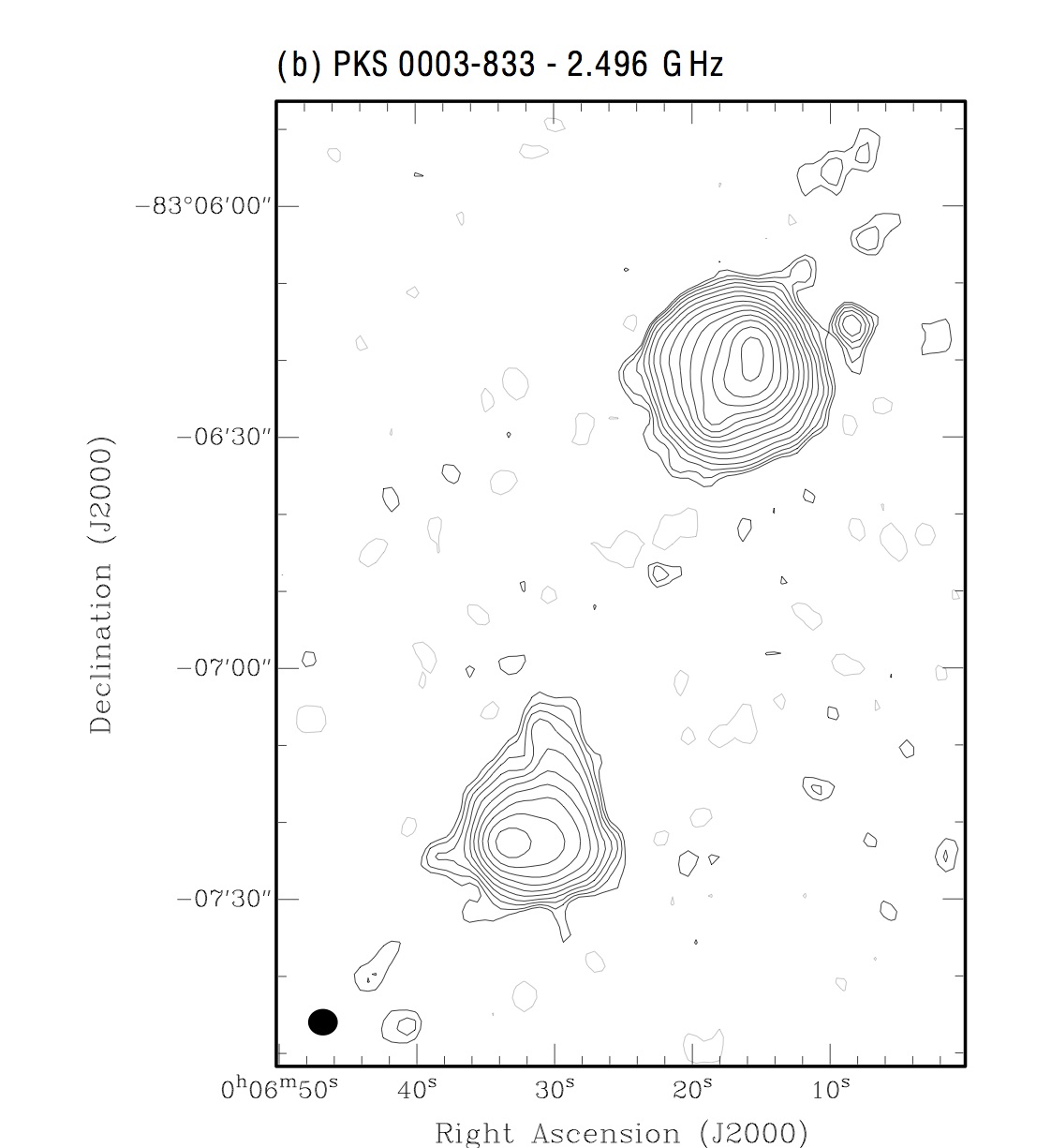}
}
\mbox{
\plotone{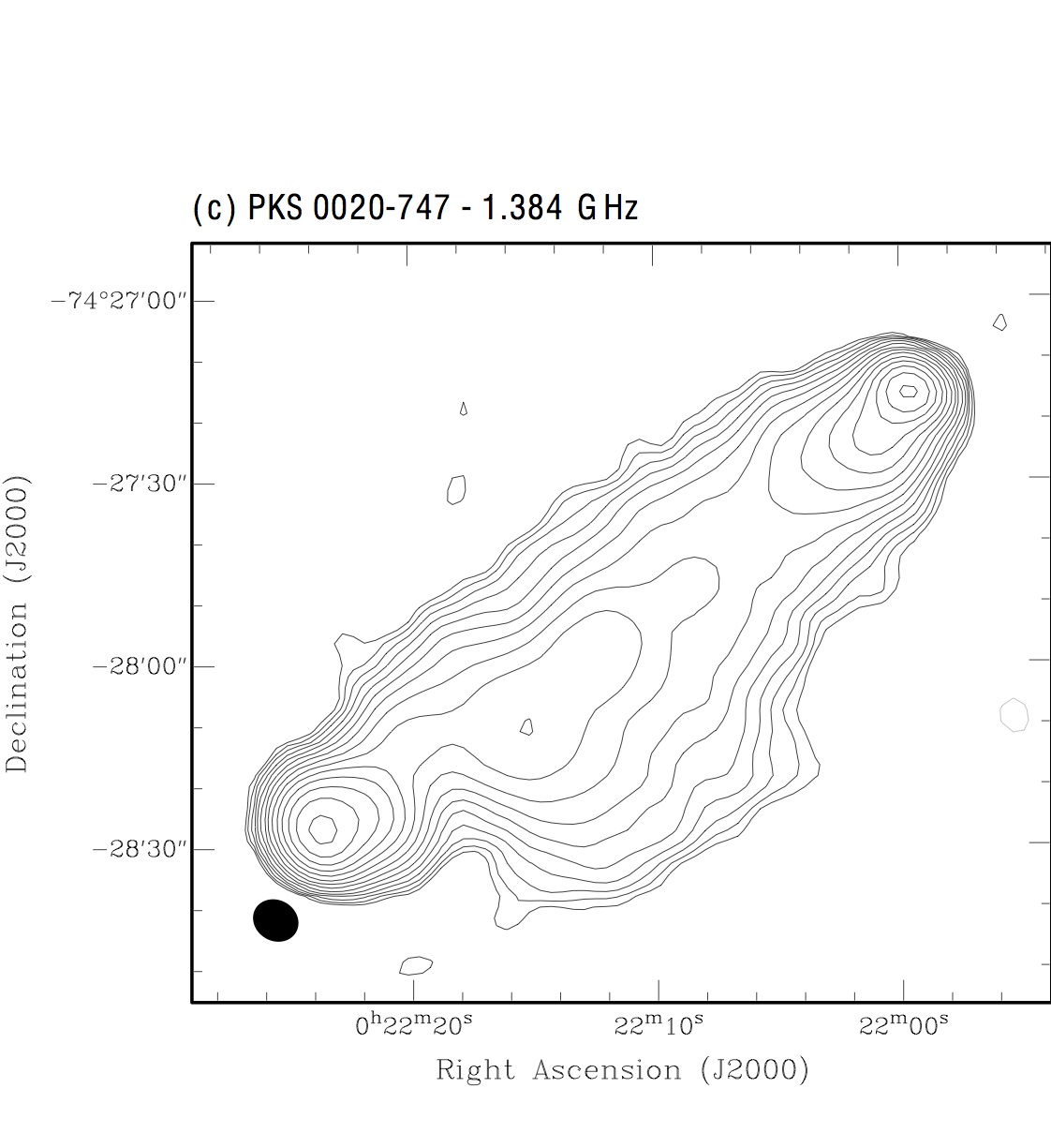} \quad
\plotone{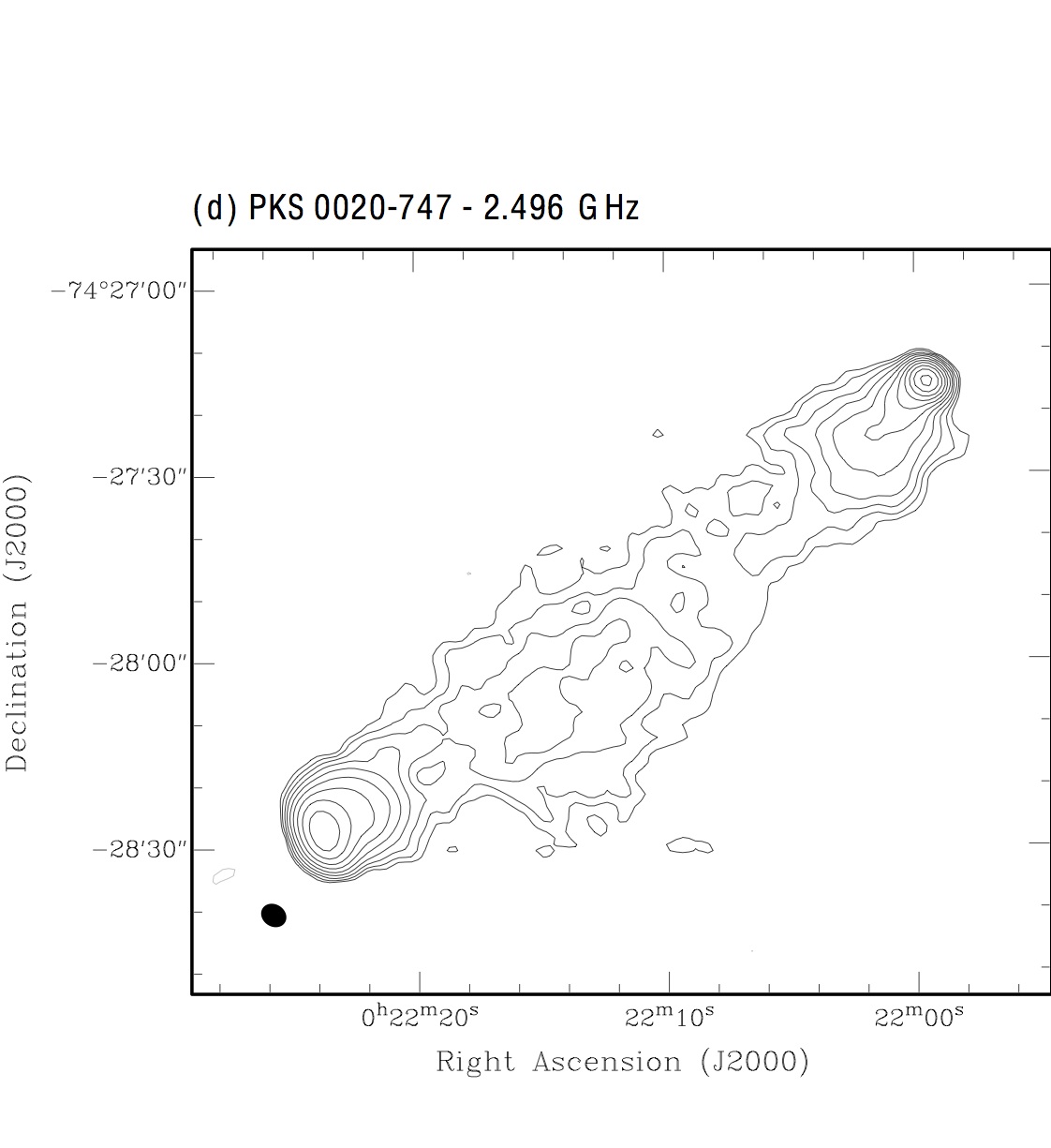}
}
\caption[Naturally weighted ATCA images of AGN.]{Naturally weighted ATCA images of AGN sources. Contours are drawn at $\pm2^{\frac{1}{2}}, \pm2^{1}, \pm2^{\frac{3}{2}}, \cdots$ times the $3\sigma$ rms noise. Restoring beam and rms image noise for all images can be found in Table \ref{tab:tabselAGN}.}
\label{fig:figself3}            
\end{center}
\end{figure}

\begin{figure}[ht]
\epsscale{0.7}
\plotone{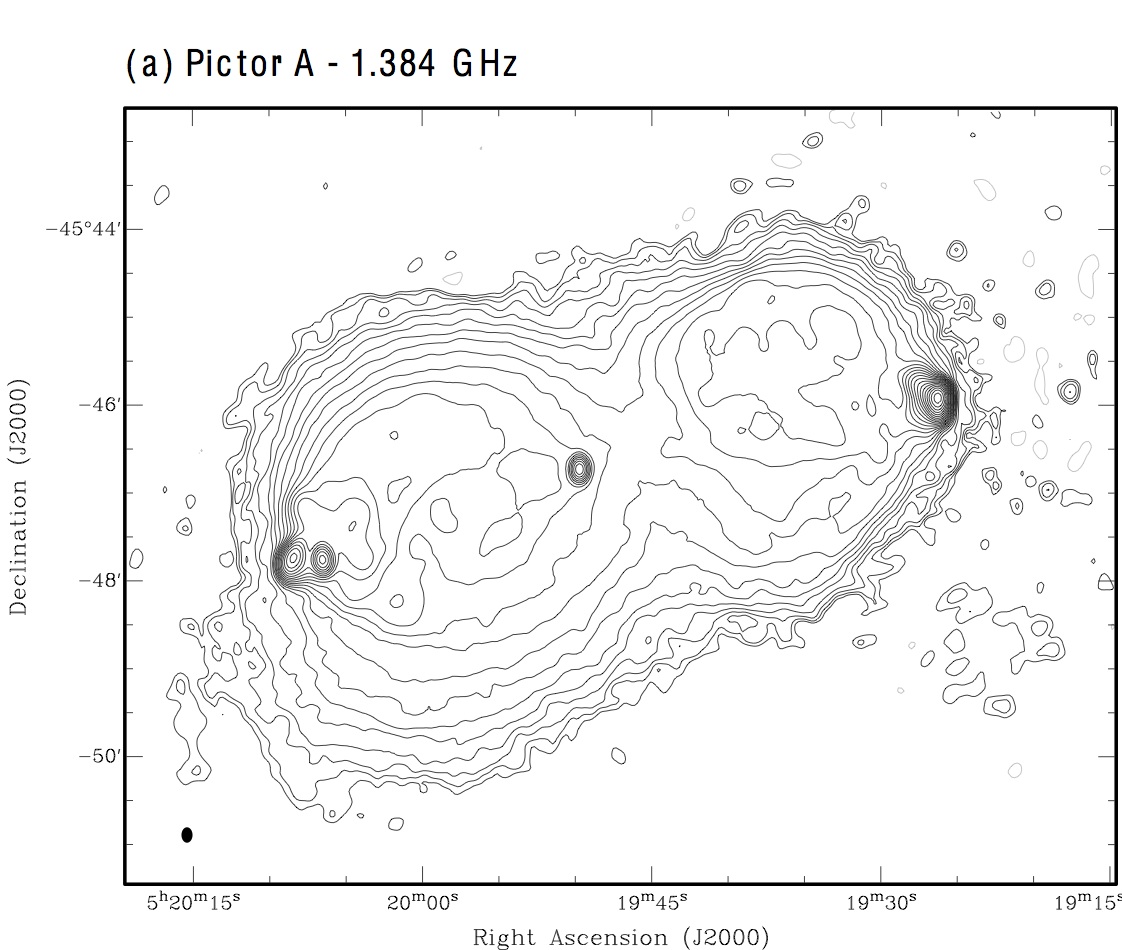}
\plotone{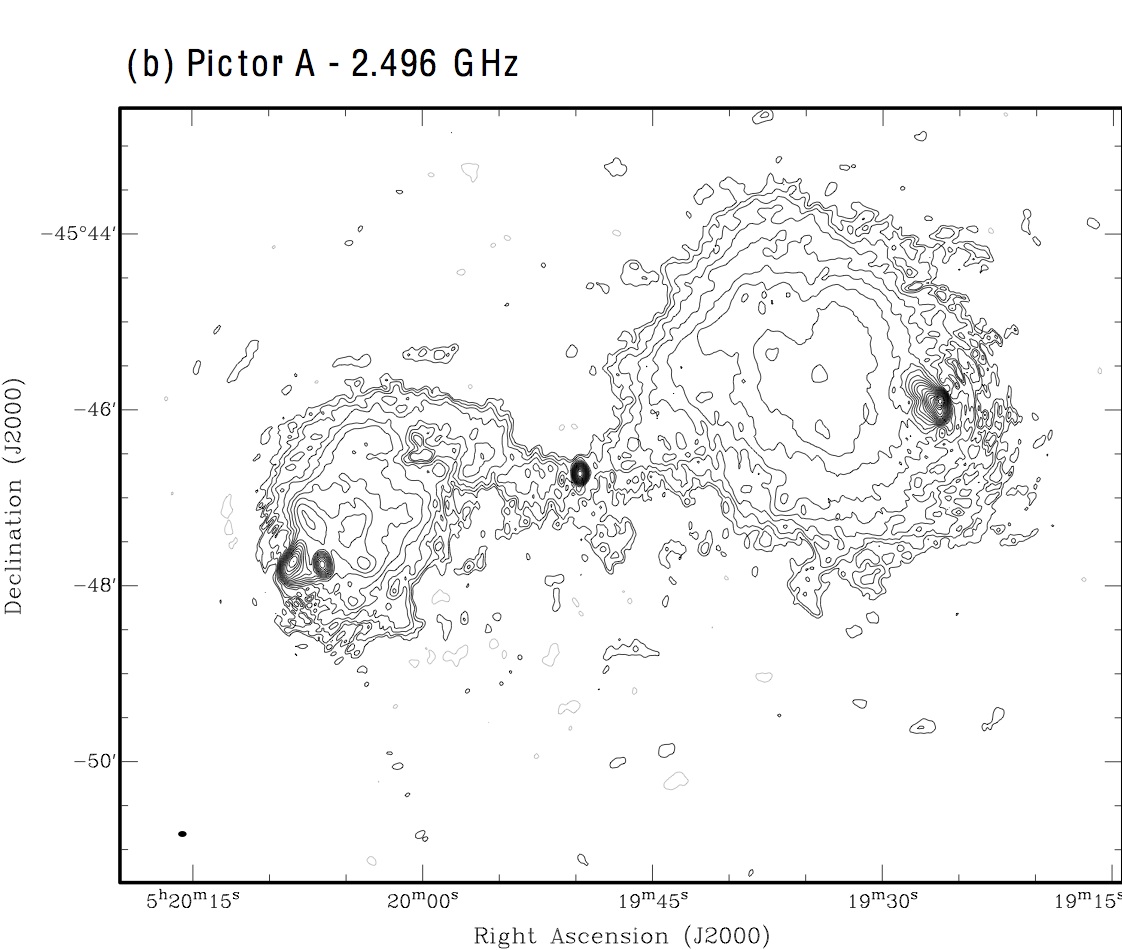}
\caption[Naturally weighted 20 cm and 13 cm ATCA images of Pictor A.]{Naturally weighted ATCA images of Pictor A. Contours are drawn at $\pm2^{\frac{1}{2}}, \pm2^{1}, \pm2^{\frac{3}{2}}, \cdots$ times the $3\sigma$ rms noise. Restoring beam and rms image noise for all images can be found in Table \ref{tab:tabselAGN}.}
\label{fig:figselpica}            
\end{figure}

The ideal source for imaging an AGN jet interaction region at VLBI resolution would require the nucleus to be bright and unresolved and the interaction region to be bright and compact. Under these conditions the nucleus can be used as an in-beam calibrator and if the hot spot is sufficiently bright it will still be detectable even if it is partially resolved. To find candidate sources, images from a 843 MHz Molonglo Observatory Synthesis Telescope (MOST) southern ($\delta<-30\arcdeg$) sky survey of bright ($>0.4$ Jy) and extended ($>30\arcsec$) sources \citep{Jones:1992p681} were examined for evidence of unusual morphology. From a total of 193 sources, 34 galaxies were selected. The MOST images had insufficient resolution ($\sim44\arcsec$) to determine if the interaction regions were sufficiently compact, so the selected sources were subsequently observed with the Australia Telescope Compact Array (ATCA) at 1.4 GHz and 2.5 GHz. The observations for 33 of the sources consisted of several snap-shots taken with two different 6 km configurations of the ATCA between October and November 2000. Observations of the low red-shift radio galaxy PKS $0518-458$ (Pictor A) consisted of a single deep 14 hour observation of the source. A summary of the observations is listed in Table \ref{tab:seltabobs}. Images resulting from the 1.4 GHz and 2.5 GHz observations are shown in Figures \ref{fig:figself3} and \ref{fig:figselpica}. The red-shift for each source, where known, and map statistics are listed in Table \ref{tab:tabselAGN}.

\begin{table}[ht]
\begin{center}
{ \normalsize
\begin{tabular}{lcccccccc} \hline \hline
Frequency & Date                            & Config.\tablenotemark{a}  \\
(MHz)     &                                 &          \\ \hline \hline
1384.0    & 10/11 OCT 2000                  & 6A       \\
2496.0    & \nodata                         & \nodata  \\
\hline
1384.0    & 31 OCT 2000                     & 6C       \\
2496.0    & \nodata                         & \nodata  \\
\hline
1384.0    & 02/03 NOV 2000                  & 6C       \\
2496.0    & \nodata                         & \nodata  \\
\hline
1384.0    & 27/28 AUG 2001\tablenotemark{b} & 6B       \\
2496.0    & \nodata                         & \nodata  \\
\hline
4800.0    & 27/28 JUN 2005\tablenotemark{c} & 6B       \\
8640.0    & \nodata                         & \nodata  \\
18000.0   & \nodata                         & \nodata  \\
18128.0   & \nodata                         & \nodata  \\
\hline
18000.0   & 21 MAR 2006\tablenotemark{d}    & 6C       \\
18128.0   & \nodata                         & \nodata  \\ \hline
\tablenotetext{a}{All ATCA configurations are east-west arrays with a maximum baseline of 6 km. The 6A configuration has antennas at 628 m, 1500 m, 2587 m, 2923 m and 5939 m west of the eastern-most antenna. The 6B configuration has antennas at 949 m, 2219 m, 2755 m, 2969 m and 5969 m west of the eastern-most antenna. The 6C configuration has antennas at 153 m, 1730 m, 2143 m, 2786 m, 6000 m west of the eastern-most antenna. }
\tablenotetext{b}{Observation of Pictor A (PKS $0518-458$) only.}
\tablenotetext{c}{Observations of PKS $0344-345$, PKS $0703-595$, PKS $1637-771$ and PKS $2152-699$ only.}
\tablenotetext{d}{Observations of PKS $2152-699$ only.}
\end{tabular}
\caption{Summary of AGN observations.}
\label{tab:seltabobs}
}
\end{center}
\end{table}

\begin{figure}[ht]
\epsscale{0.55}
\plotone{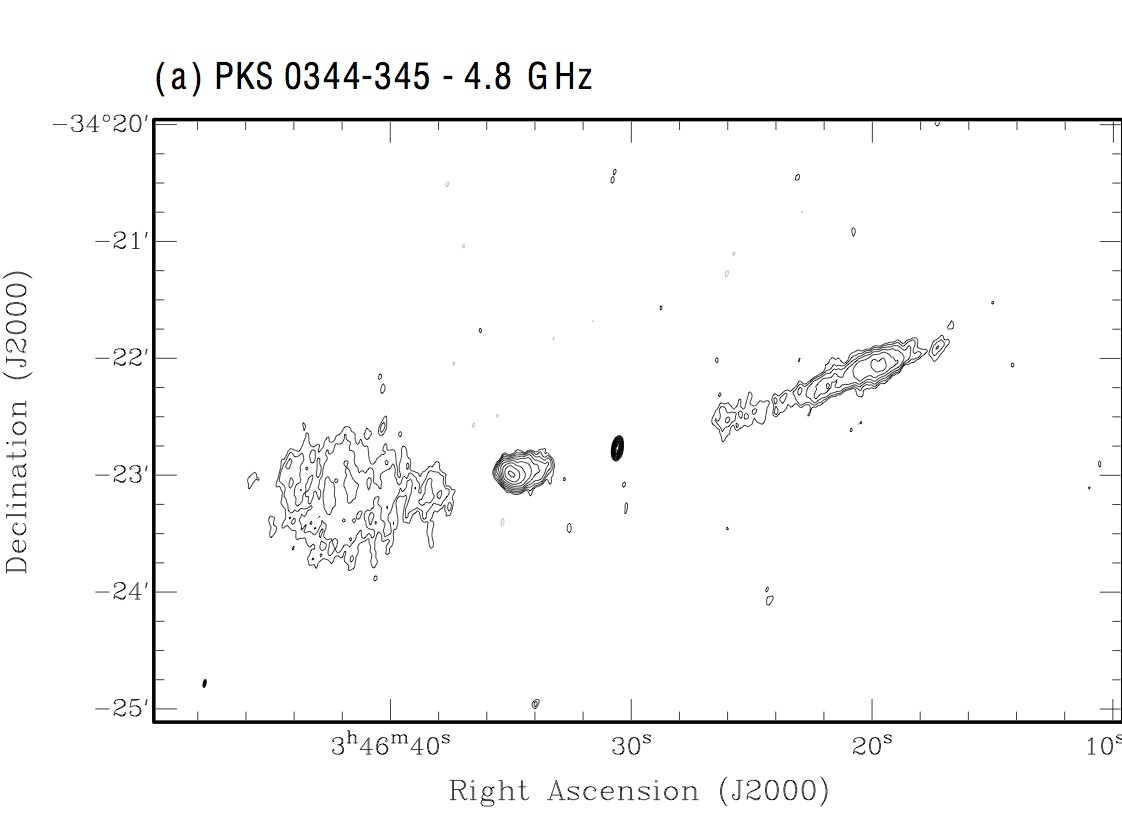}
\plotone{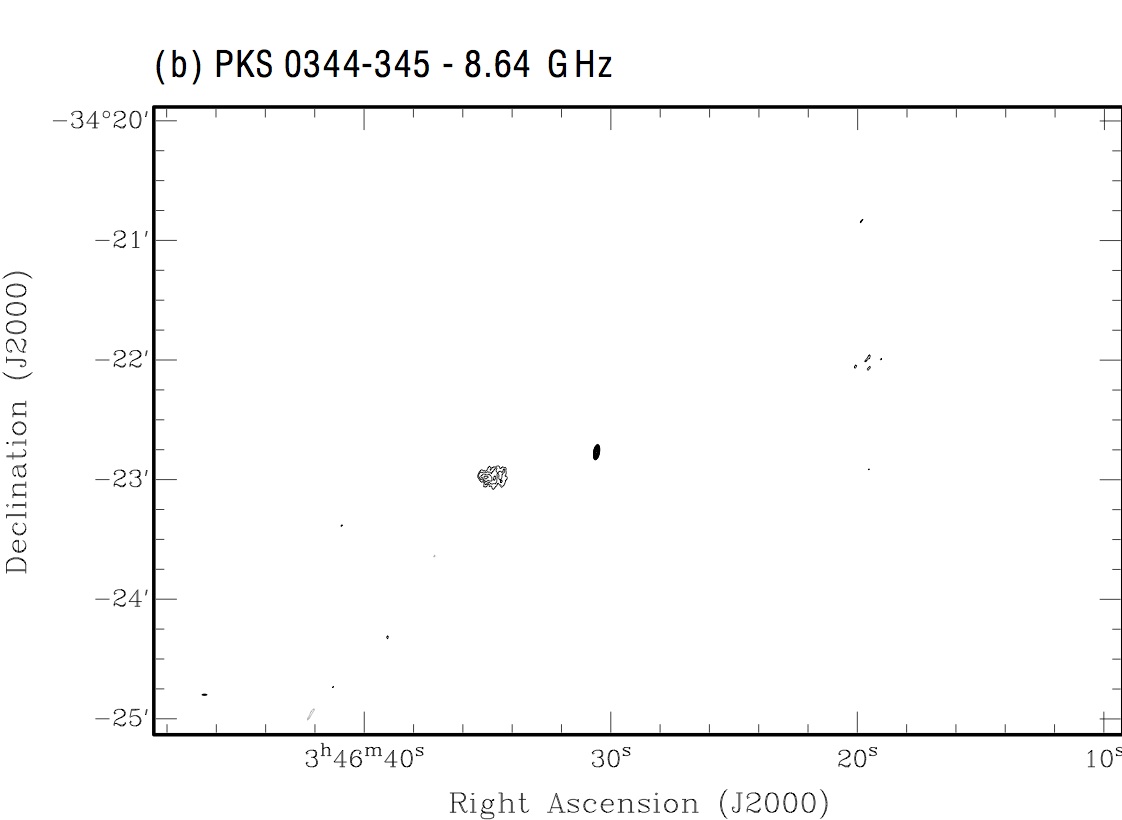}
\plotone{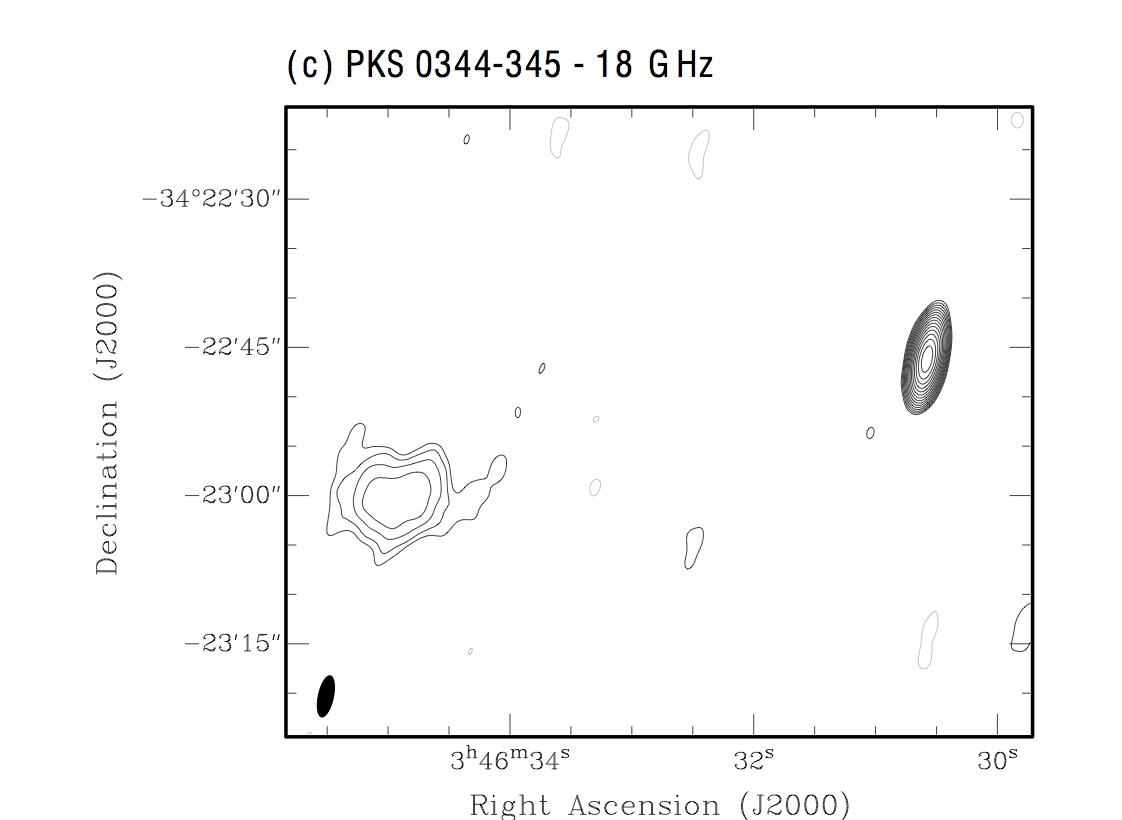}
\caption[Naturally weighted ATCA images of PKS $0344-345$.]{Naturally weighted ATCA images of PKS $0344-345$. Contours are drawn at $\pm2^{\frac{1}{2}}, \pm2^{1}, \pm2^{\frac{3}{2}}, \cdots$ times the $3\sigma$ rms noise. Restoring beam and rms image noise for all images can be found in Table \ref{tab:tabselAGN}.}
\label{fig:figsel0344}            
\end{figure}

For each of the sources in the sample, a search was performed to detect compact components (e.g. a knot or cloud) along the jet line between the core and the termination hotspot. Such components are likely to be associated with the interaction of the jet with a dense region within the galaxy, the intergalactic medium, or another galaxy. From the resulting images, five of the galaxies were found to contain excellent candidate jet-cloud or jet-galaxy interactions: PKS $0344-345$, PKS $0518-458$, PKS $0703-595$, PKS $1637-771$ and PKS $2152-699$. Further observations of these sources were made at higher frequencies with the ATCA to test the compactness of the interaction regions, see Table \ref{tab:seltabobs}. Images of PKS $0344-345$, PKS $0703-595$ and PKS $1637-771$ were made at 4.8, 8.64 and 18 GHz and are shown in Figures \ref{fig:figsel0344}, \ref{fig:figsel0703} and \ref{fig:figsel1637}, respectively. PKS $2152-699$ had been previously imaged at 4.7 GHz and 8.6 GHz \citep{Fosbury:1998p7267}, so an additional 18 GHz image of this source was made to complement the existing data, see Figure \ref{fig:figsel2152}. Similarly, PKS $0518-458$ had been previously imaged at 4.8 GHz, 8.4 GHz and 15 GHz using the VLA \citep{Perley:1997p6689}, so no further images were required of this source. In PKS $0703-595$ and PKS $1637-771$ the interaction regions and hot spots were completely resolved out in the highest resolution images. In PKS $0518-458$ and PKS $2152-699$, the interaction regions remained compact even in the highest resolution images. In PKS $0344-345$ the interaction region was still detectable at 18 GHz but was partially resolved. It is possible that this source may still be detectable with VLBI at low frequencies where the steep spectral index of the interaction region helps. Its apparent interaction with neighbouring galaxies also makes this source of particular interest.

\begin{sidewaysfigure}[ht]
\epsscale{1.0}
\plotone{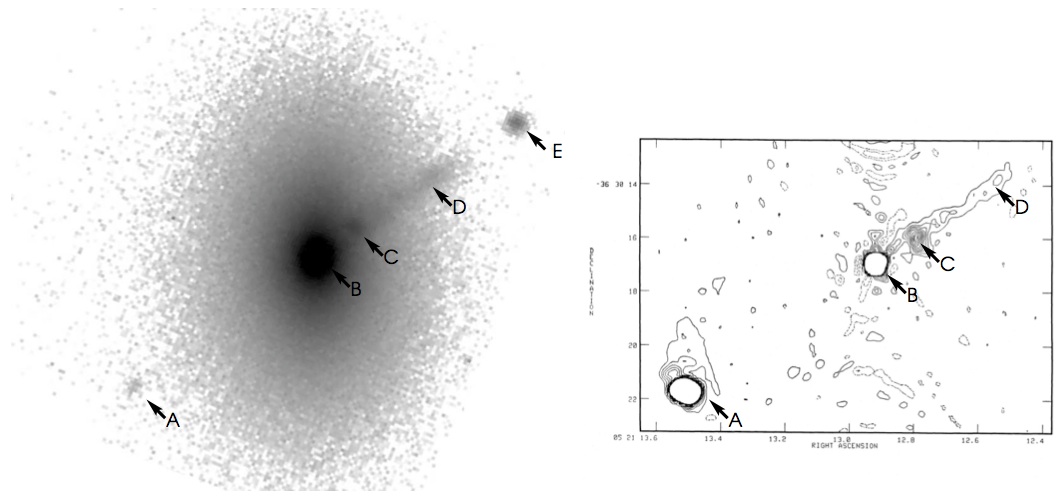}
\caption[J-band and 15 GHz images of PKS $0521-365$.]{Left, Magellan infrared J-band image of PKS $0521-365$. Right, total intensity map of PKS $0521-365$ at 15 GHz, using the VLA in A/B hybrid configuration \citep{Keel:1986p10557}. Radio contours range from -30 to 30 mJy beam$^{-1}$ in 3 mJy beam$^{-1}$ steps, peak flux density is 2.0418 Jy beam$^{-1}$. Radio restoring beam is $\sim300$ mas. Right ascension and declination in B1950 coordinates.}
\label{fig:fig0521}
\end{sidewaysfigure}

\begin{sidewaysfigure}[ht]
\epsscale{0.3}
\mbox{
\plotone{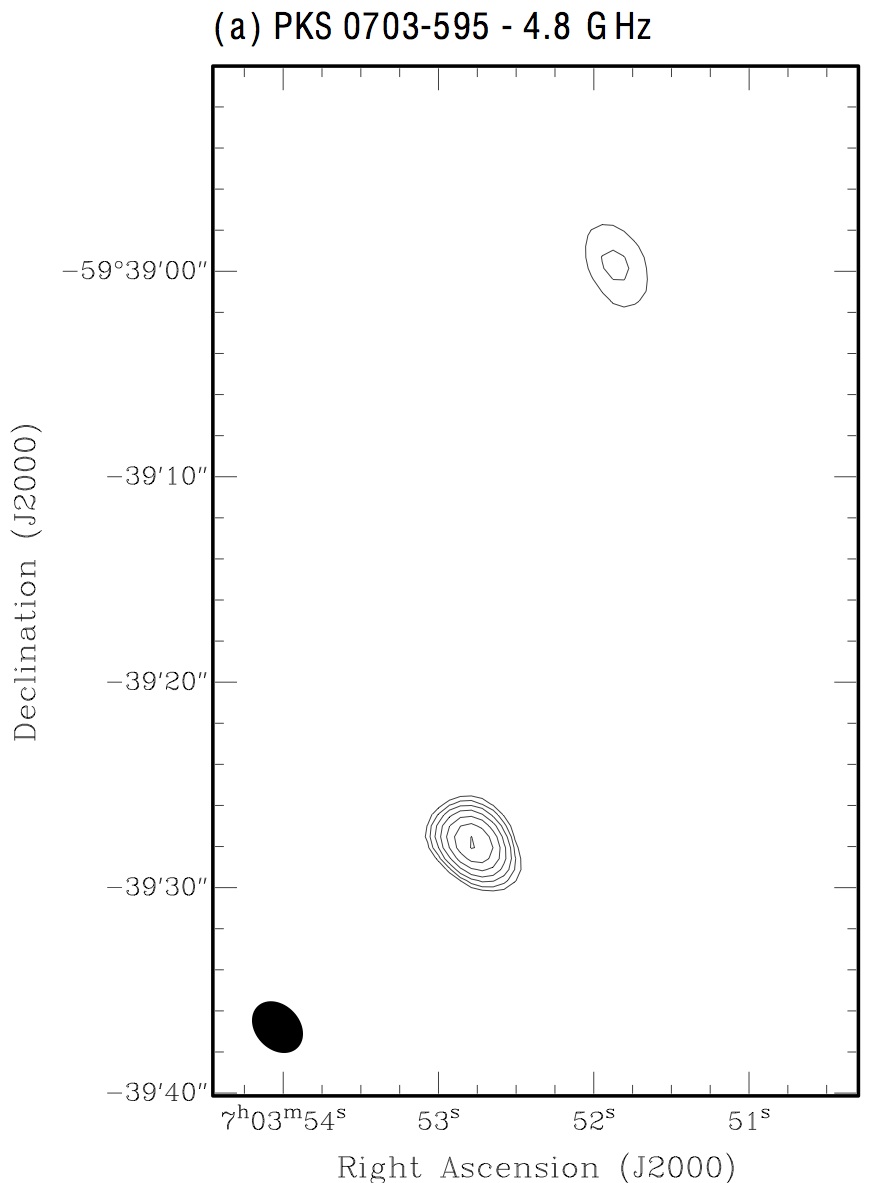} \quad
\plotone{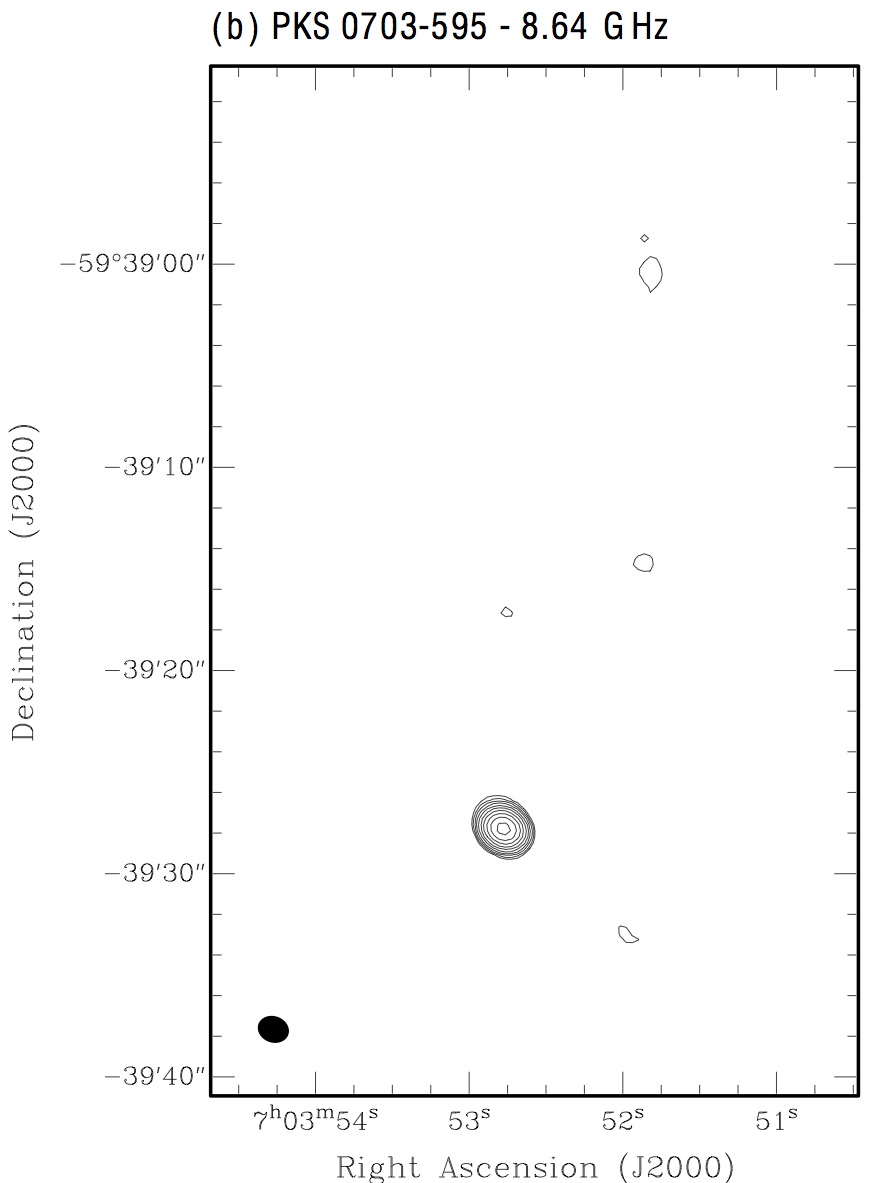} \quad
\plotone{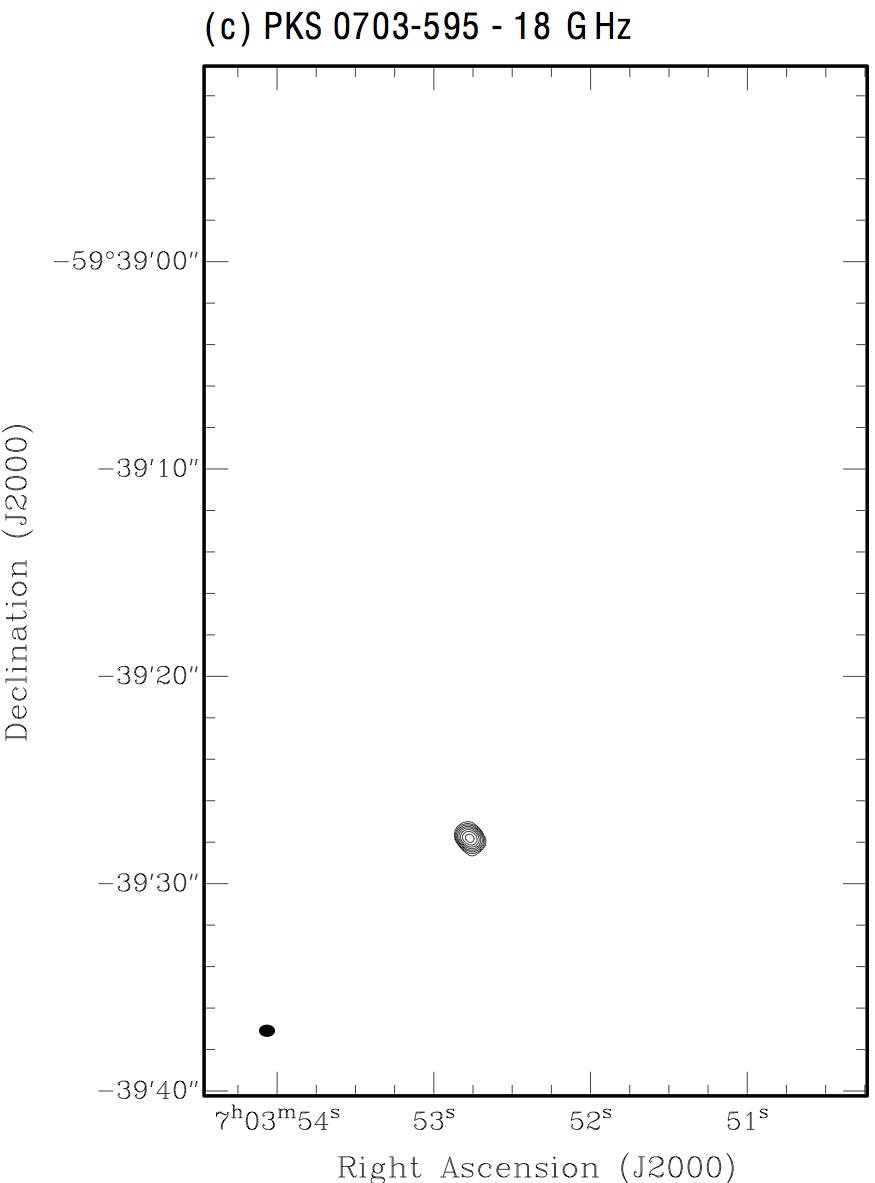}
}
\caption[Naturally weighted ATCA images of PKS $0703-595$.]{Naturally weighted ATCA images of PKS $0703-595$. Contours are drawn at $\pm2^{\frac{1}{2}}, \pm2^{1}, \pm2^{\frac{3}{2}}, \cdots$ times the $3\sigma$ rms noise. Restoring beam and rms image noise for all images can be found in Table \ref{tab:tabselAGN}.}
\label{fig:figsel0703}            
\end{sidewaysfigure}

\begin{sidewaysfigure}[ht]
\epsscale{0.3}
\mbox{
\plotone{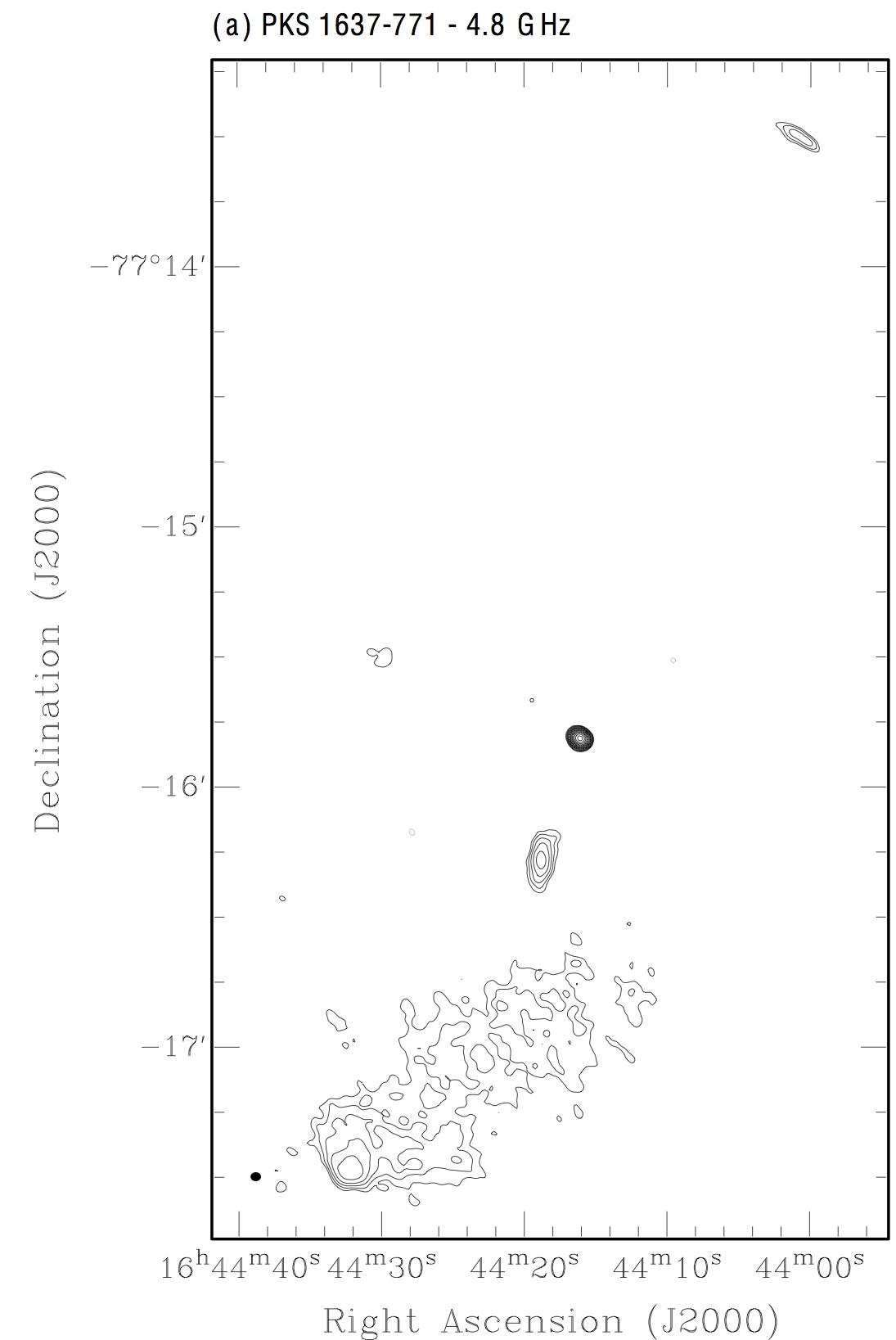} \quad
\plotone{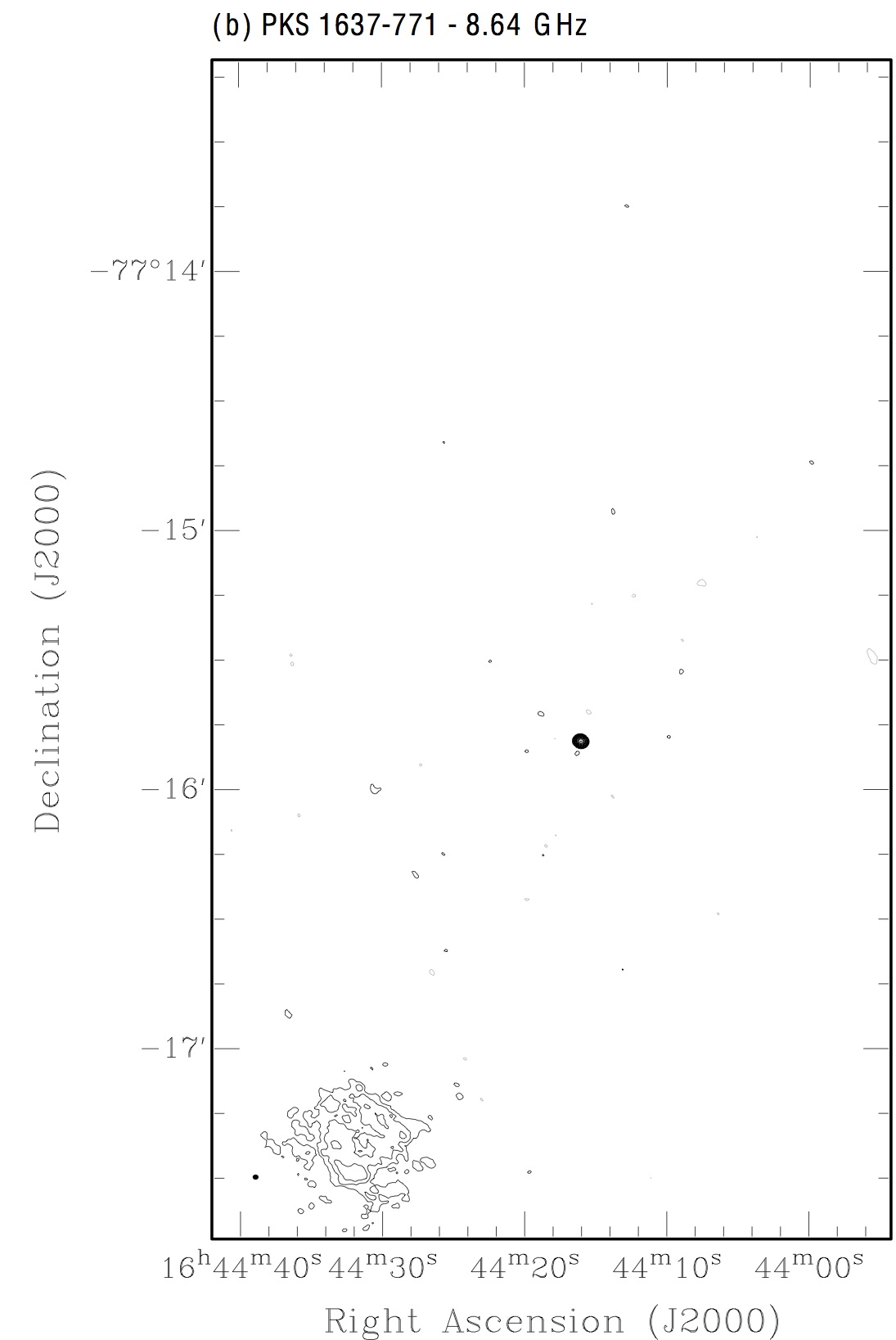} \quad
\plotone{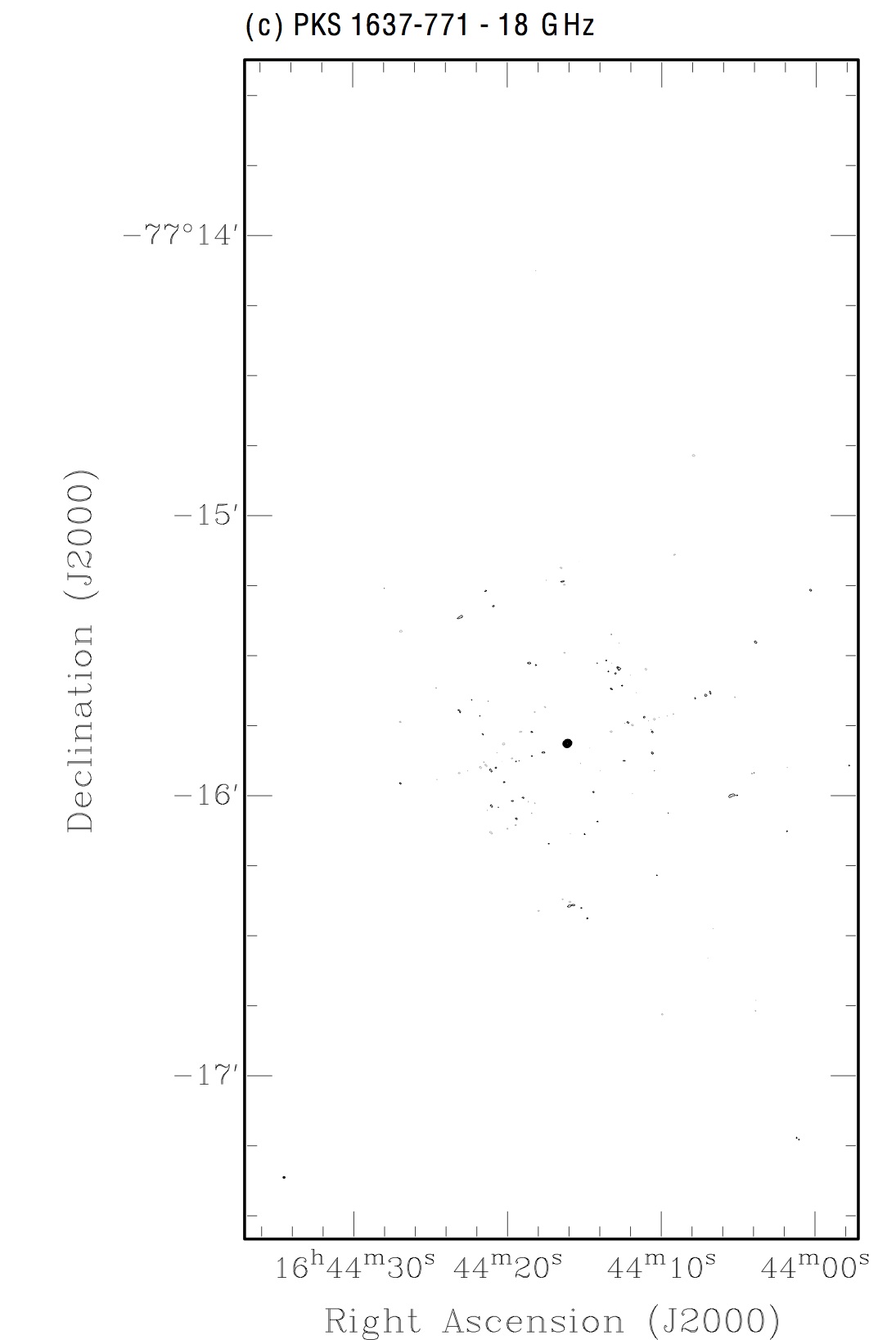}
}
\caption[Naturally weighted ATCA images of PKS $1637-771$.]{Naturally weighted ATCA images of PKS $1637-771$. Contours are drawn at $\pm2^{\frac{1}{2}}, \pm2^{1}, \pm2^{\frac{3}{2}}, \cdots$ times the $3\sigma$ rms noise. Restoring beam and rms image noise for all images can be found in Table \ref{tab:tabselAGN}.}
\label{fig:figsel1637}            
\end{sidewaysfigure}

\begin{figure}[ht]
\epsscale{0.7}
\plotone{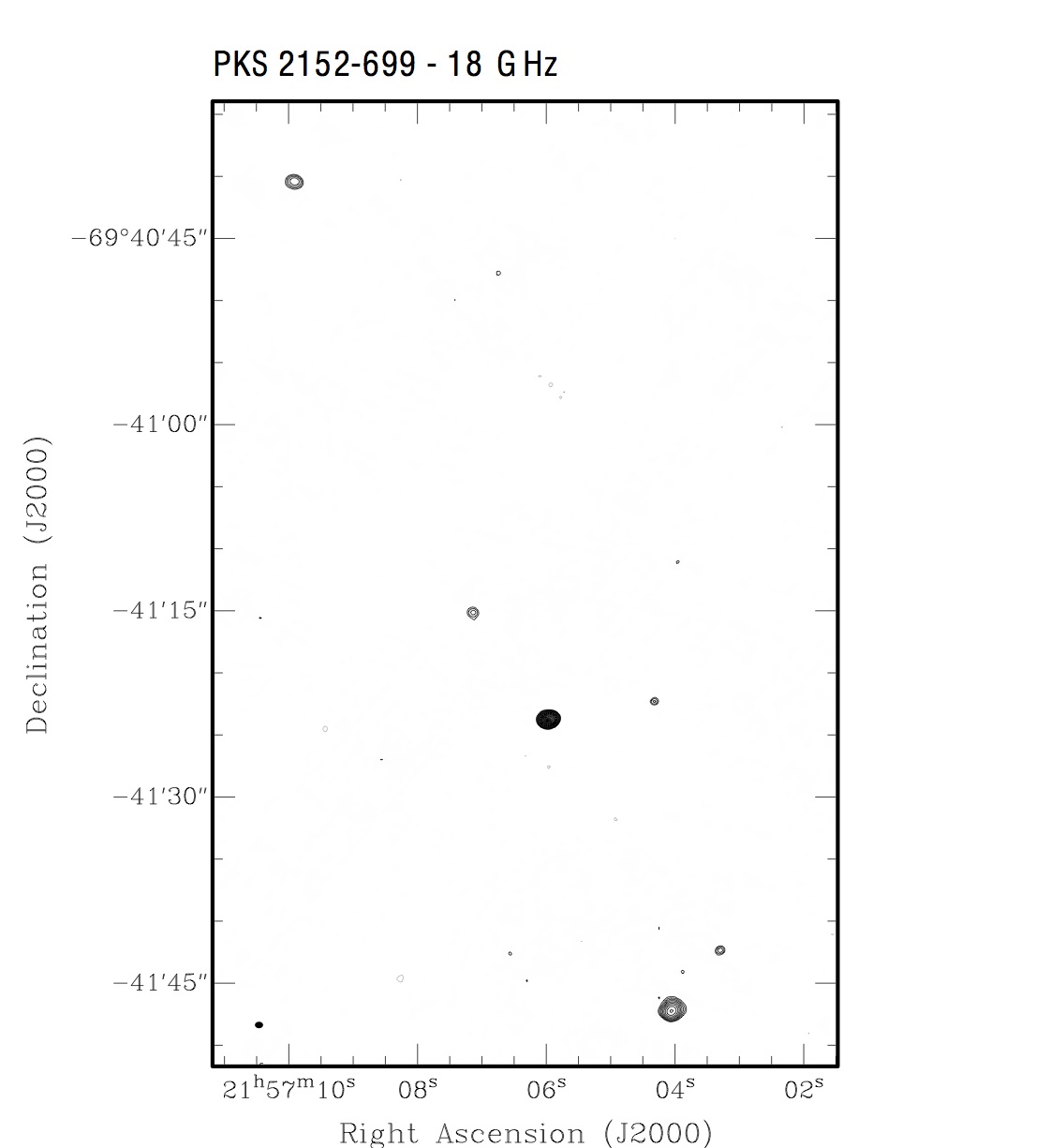}
\caption[Naturally weighted ATCA image of PKS $2162-699$ at 18 GHz.]{Naturally weighted ATCA image of PKS $2152-699$. Contours are drawn at $\pm2^{\frac{1}{2}}, \pm2^{1}, \pm2^{\frac{3}{2}}, \cdots$ times the $3\sigma$ rms noise. Restoring beam and rms image noise for all images can be found in Table \ref{tab:tabselAGN}.}
\label{fig:figsel2152}            
\end{figure}

The galaxy PKS $0521-365$ was not listed in the \citet{Jones:1992p681} survey, most likely because of its smaller size ($<30\arcsec$), but shows evidence for a jet interaction in 15 GHz VLA images of the source, see Figure \ref{fig:fig0521}. Our PKS $0521-365$ Magellan J-band images, also shown in Figure \ref{fig:fig0521}, reveal a hot spot to the south-east (A), the nucleus (B), a knot (C) in jet (D) and an apparent hot spot to the north-west (E).

After the selection process the final list of candidate sources was narrowed down to four sources: PKS $0344-345$, PKS $0518-458$, PKS $0521-365$, PKS $2152-699$. Unfortunately, due to reduced availability of one of the most sensitive elements of the LBA, the 70 m NASA Deep Space Network (DSN) antenna at Tidbinbilla, observations of PKS $2152-699$ could not be scheduled within the time frame of the research presented here.

\clearpage
\section{Low frequency wide-field survey}

To examine the potential of wide-field VLBI in survey applications for upcoming low-frequency instruments such as the Low Frequency Array (LOFAR), European LOFAR (E-LOFAR) and the Square Kilometre Array (SKA), an appropriate low-frequency observation of a test field using a long baseline array and correlated to minimise smearing effects is required. The target of the observation is not particularly important, rather it is the entire field and what it contains that is of interest. Such an experiment can in fact lever off an existing low frequency observation, effectively piggy-backing off the observation, assuming that the raw recorded data is available and that it can be re-correlated with high frequency and temporal resolution.

Such an observation was made on 11 November 2005 for the gravitation lens B$0218+357$. The observation was a 90 cm VLBI observation using all ten NRAO Very Long Baseline Array (VLBA) antennas, the Westerbork Synthesis Radio Telescope (WSRT) as a phased array and the Jodrell Bank, 76 m Lovell Telescope (JB). The primary aim of the observation was to investigate, in detail, the propagation effects in the lensing galaxy and the substructure in the lens. While the data was correlated in narrow-field mode for the primary aim, it was also correlated in wide-field mode to serve as a wide-field test observation to study the faint source population at 90 cm.

\clearpage
\epsscale{0.4}
\begin{center}
\mbox{
\plotone{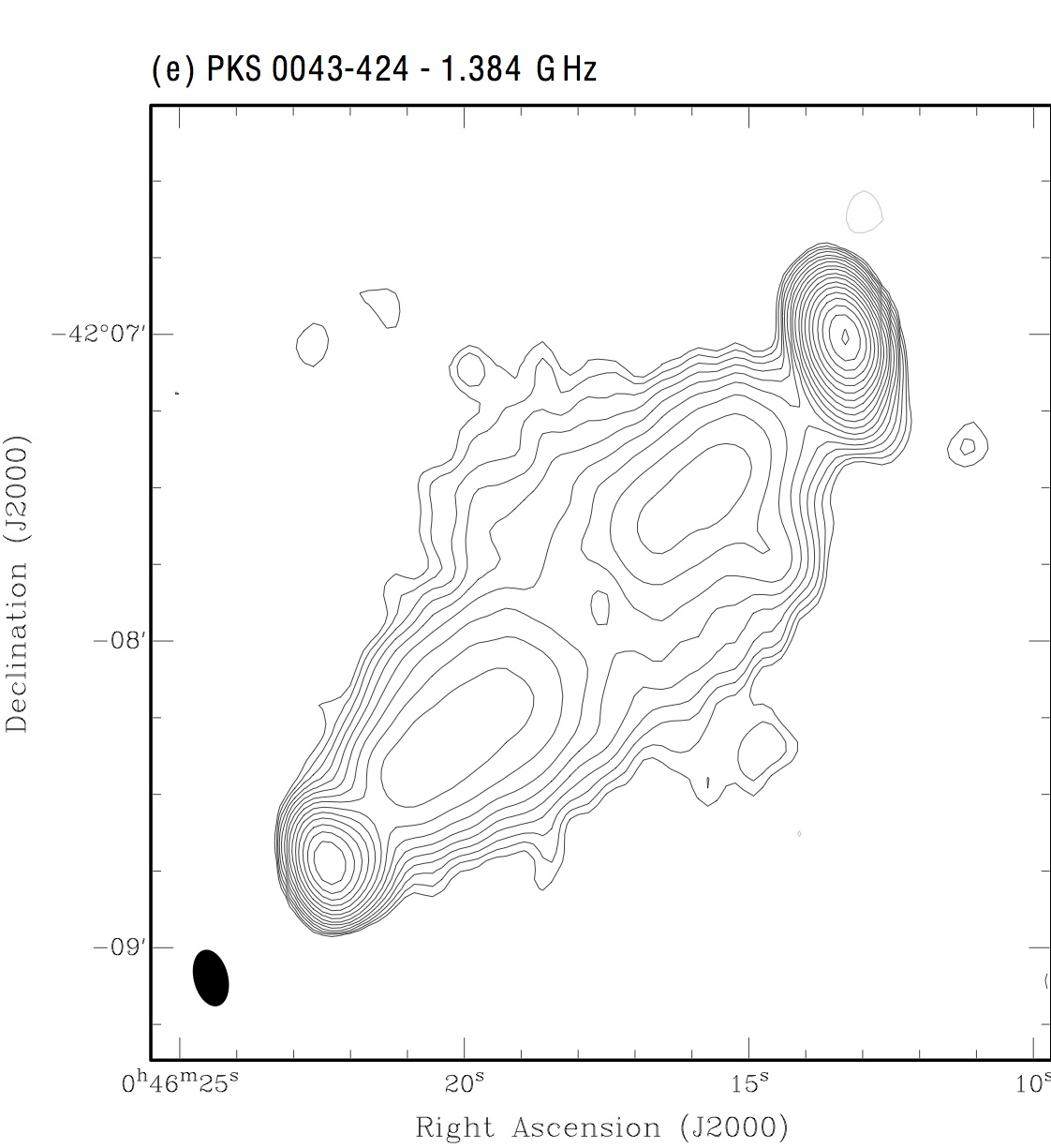} \quad
\plotone{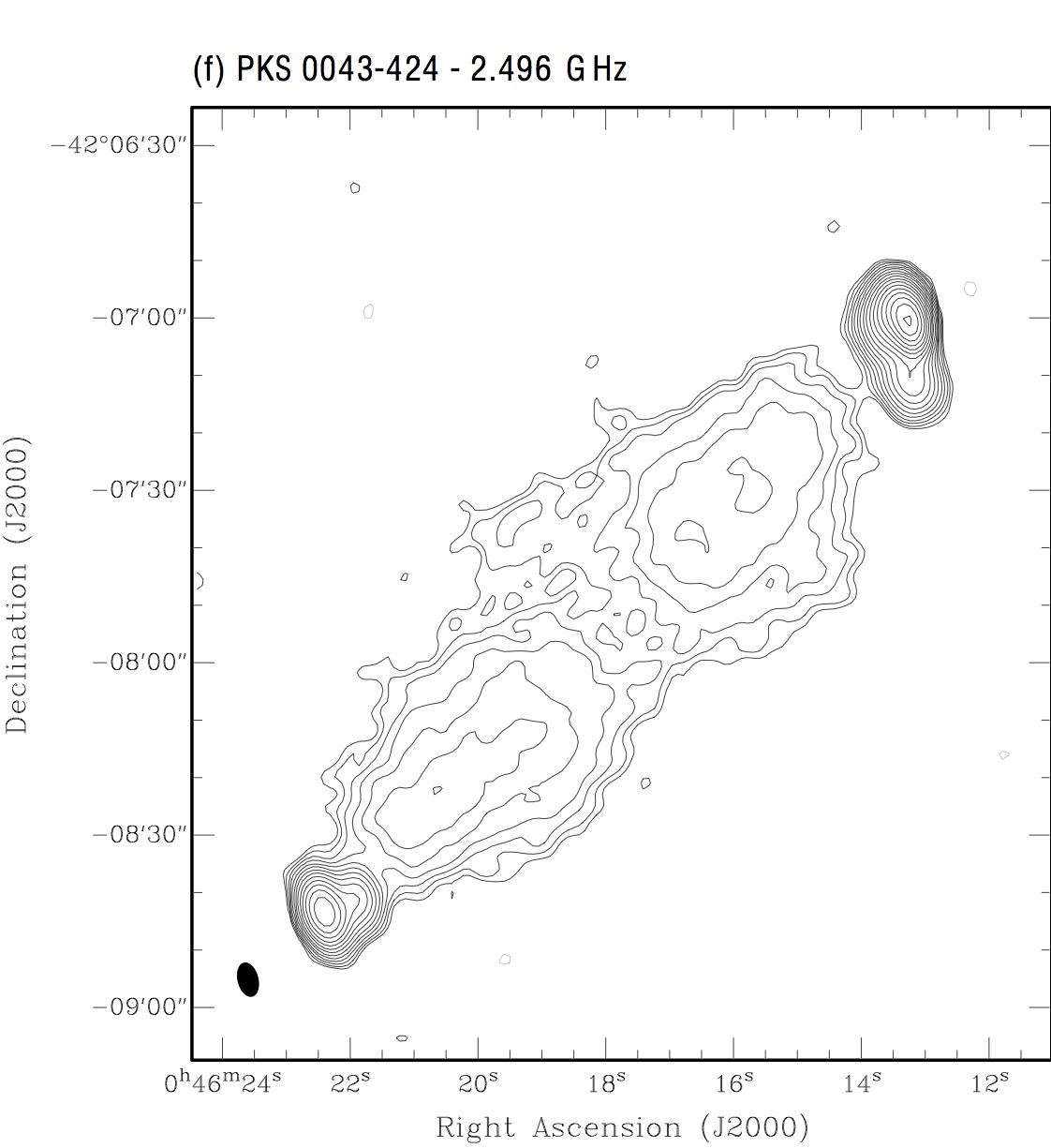}
}\\[5mm]
\mbox{
\plotone{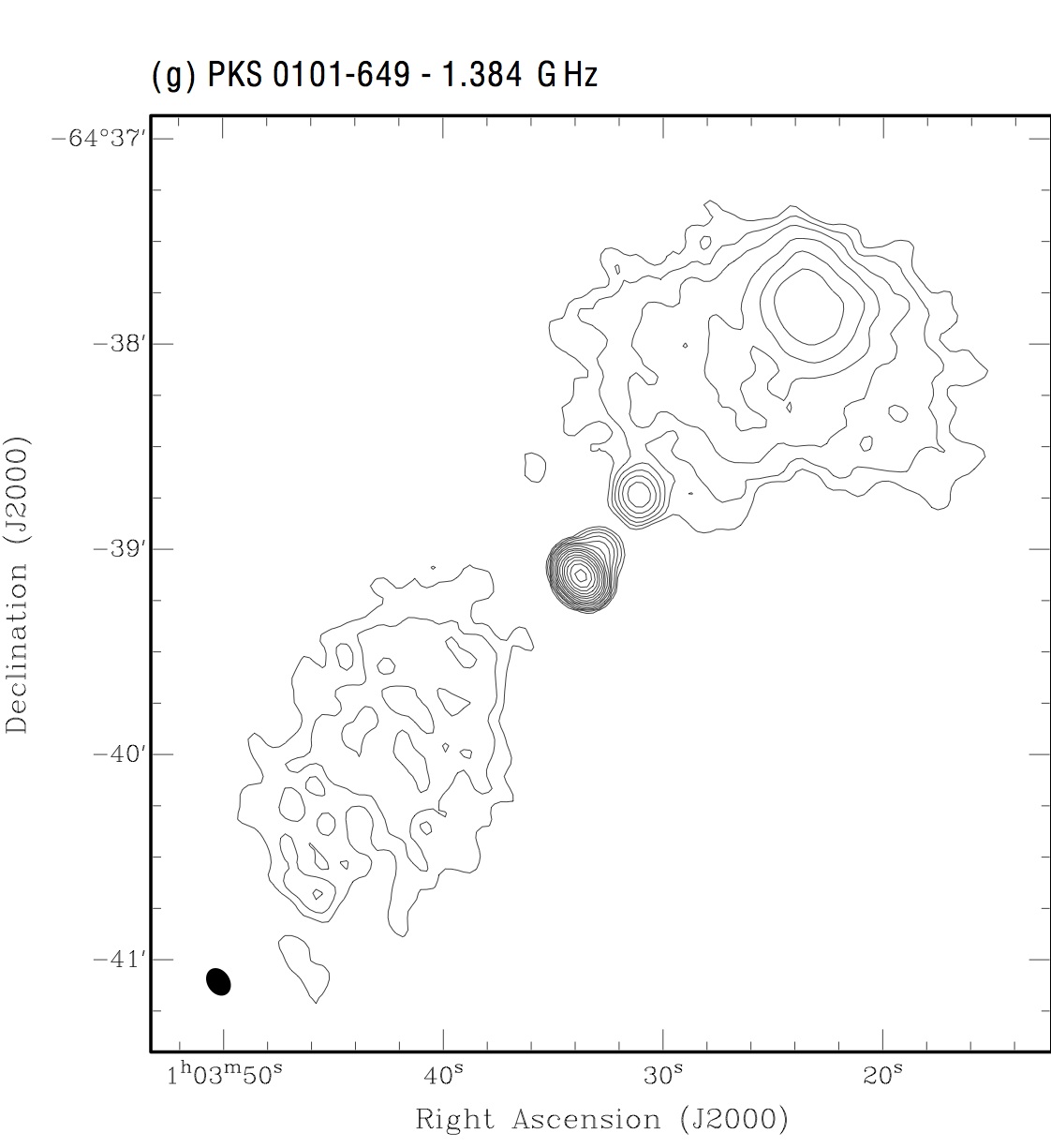} \quad
\plotone{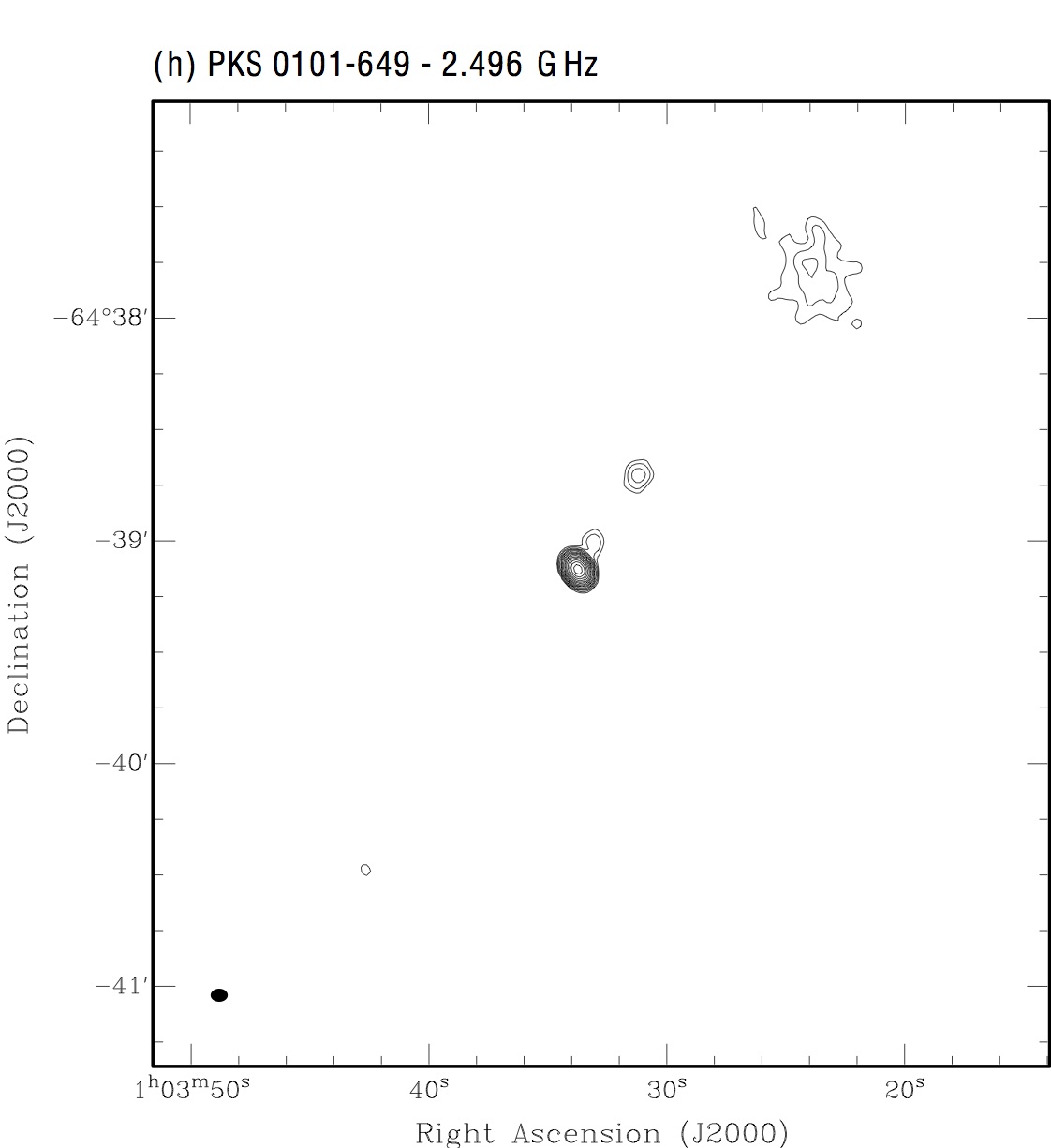}
}\\[5mm]
\mbox{
\plotone{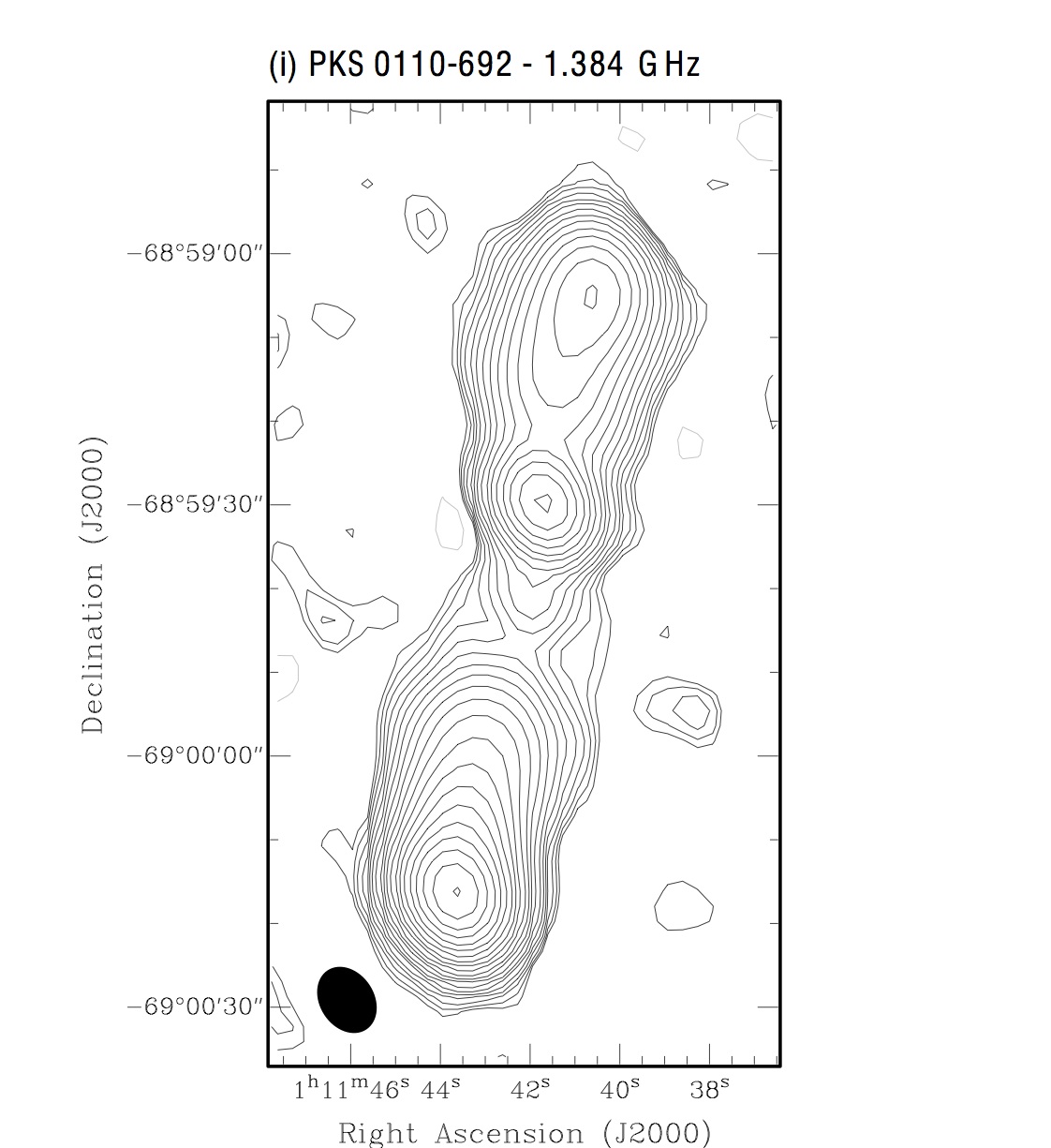} \quad
\plotone{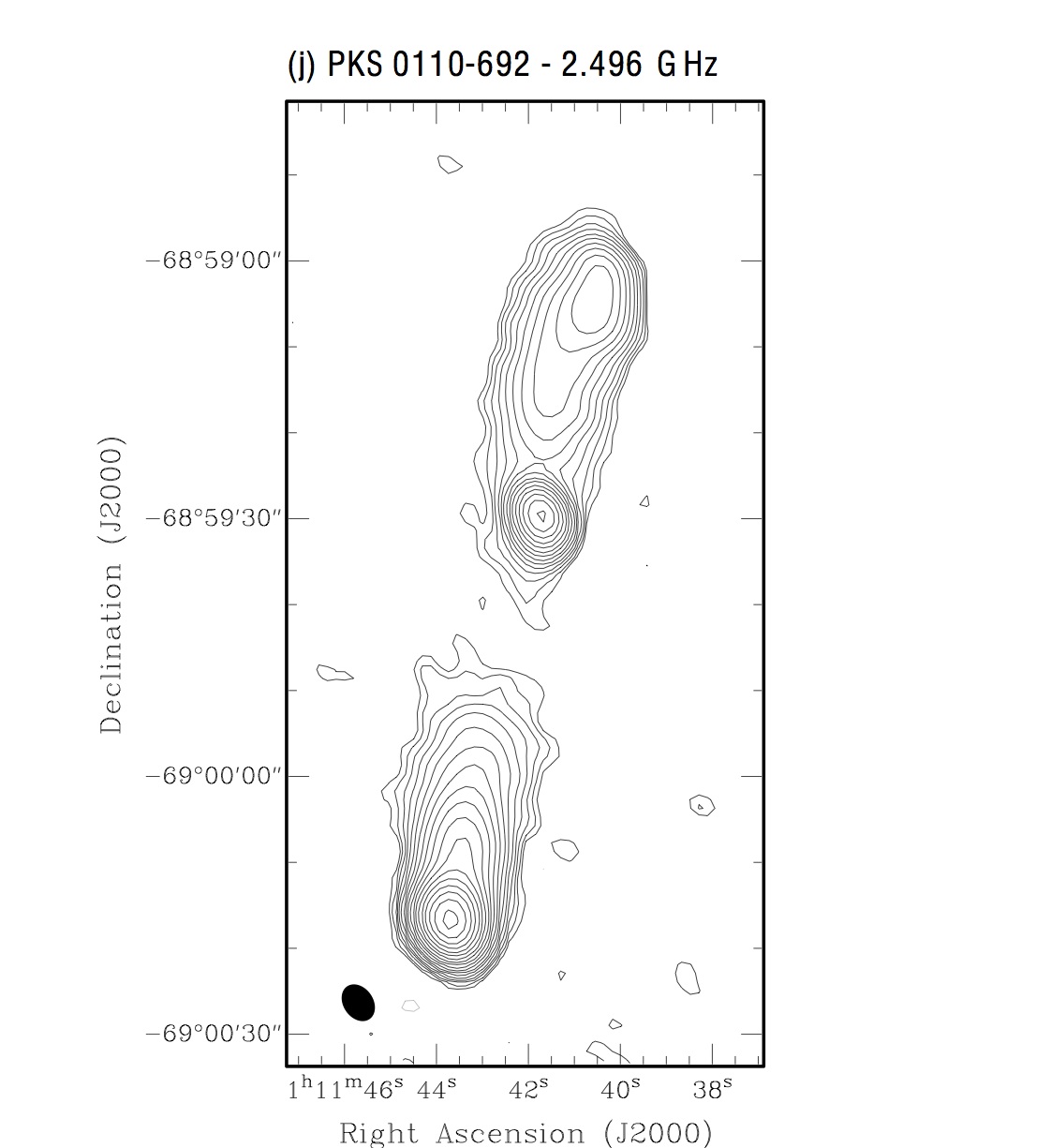}
}\\[5mm]
{Fig. 3.1. --- Continued}
\end{center}
\clearpage
\epsscale{0.4}
\begin{center}
\mbox{
\plotone{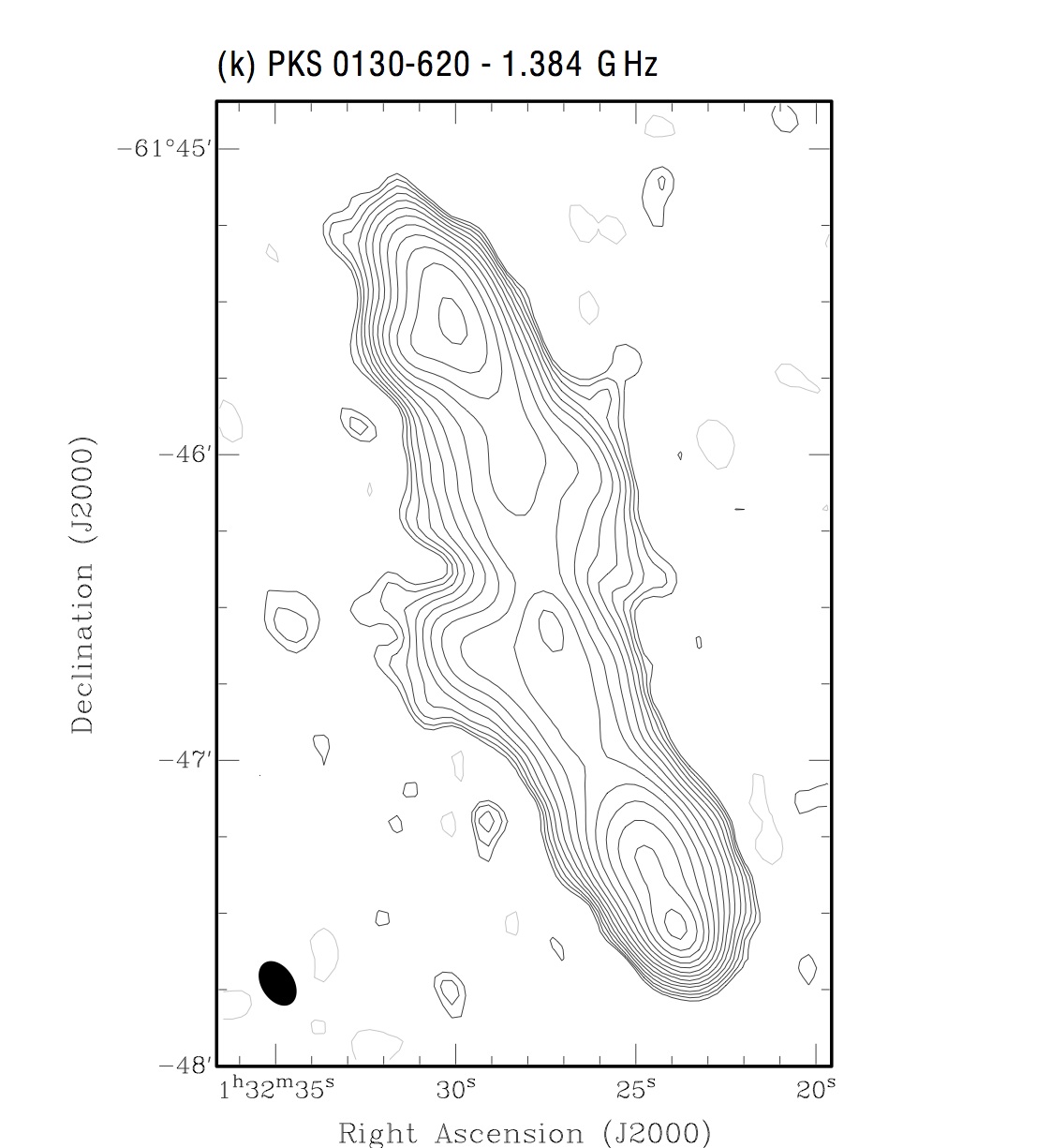} \quad
\plotone{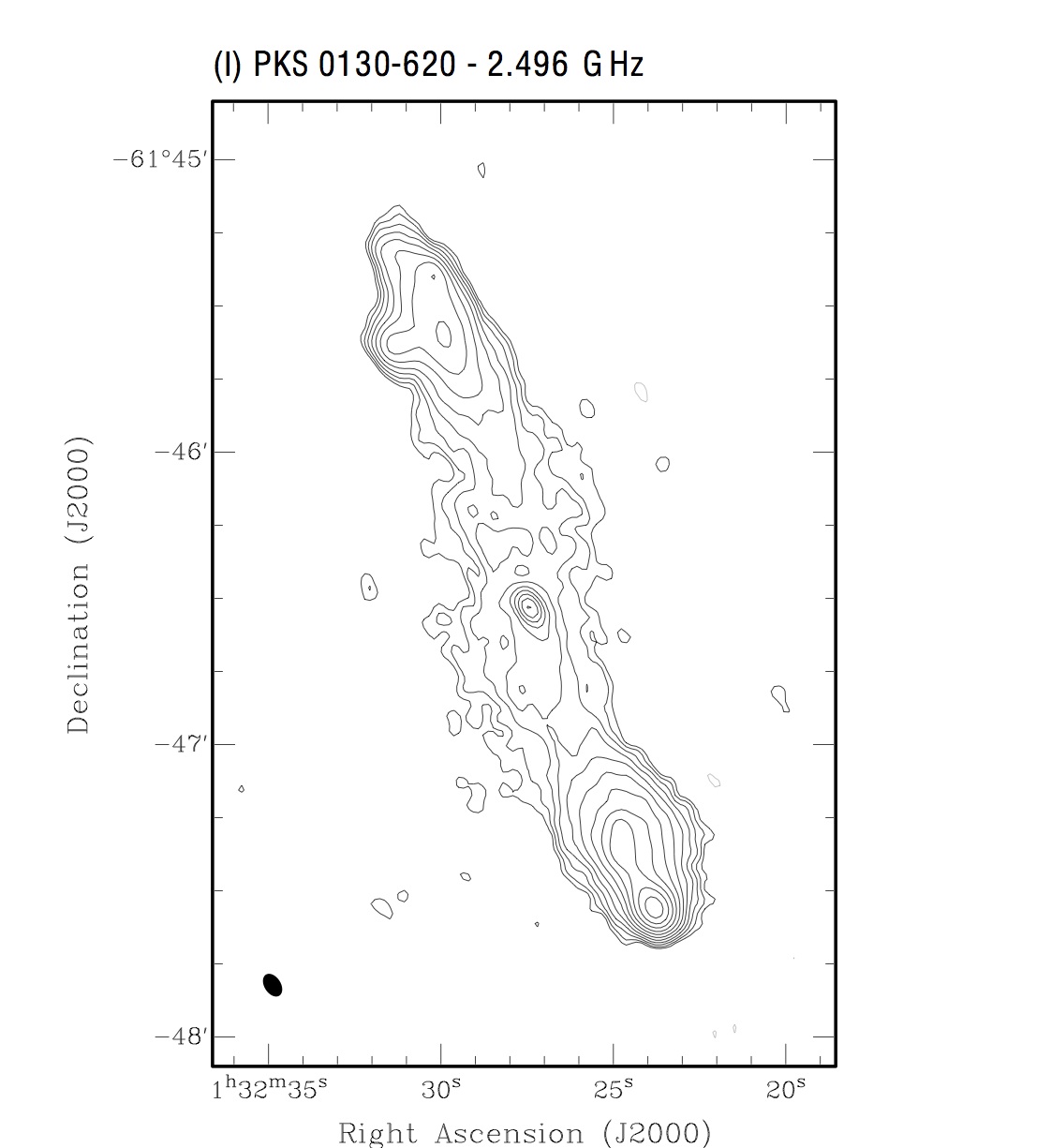}
}\\[5mm]
\mbox{
\plotone{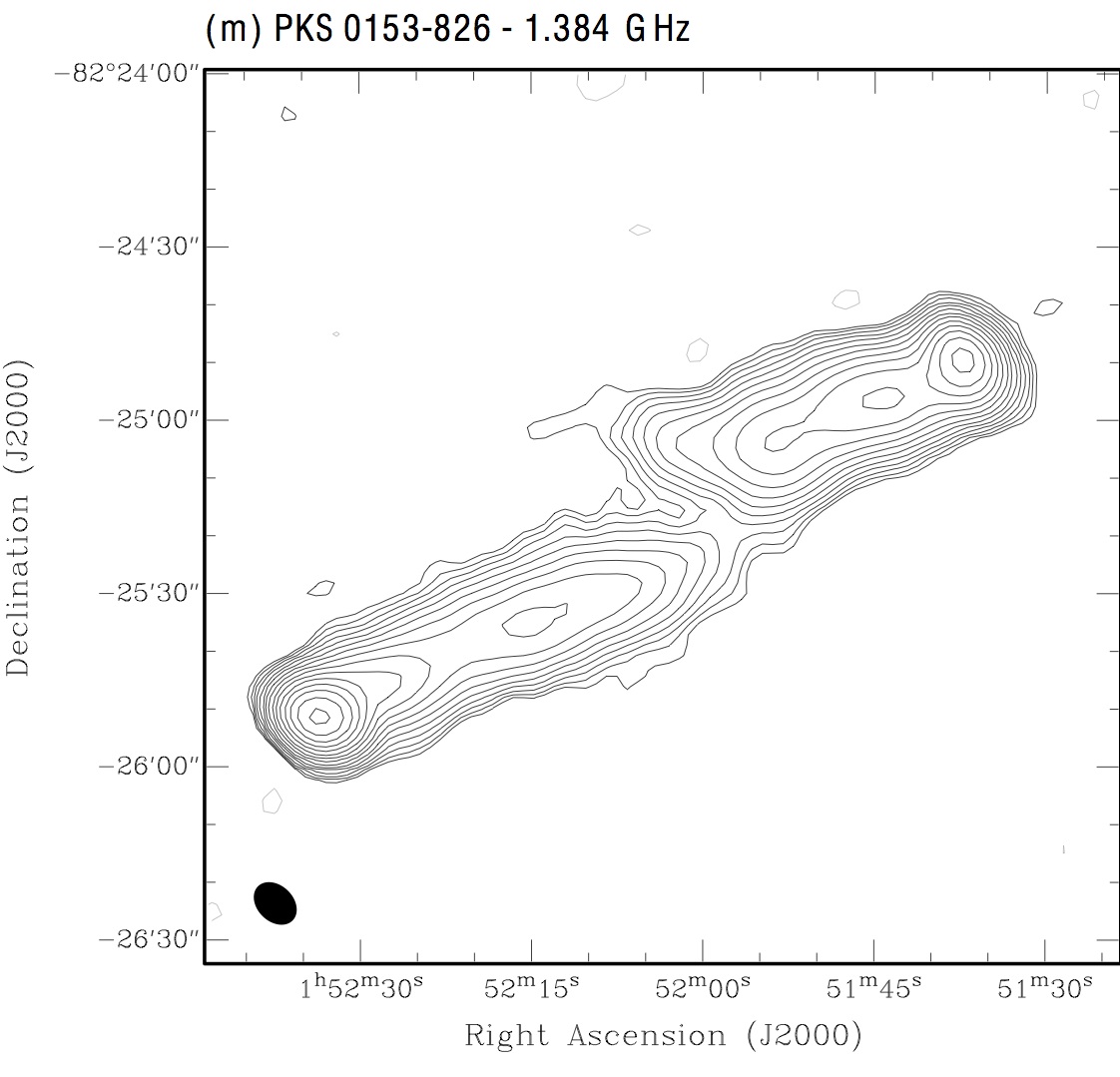} \quad
\epsscale{0.45}
\plotone{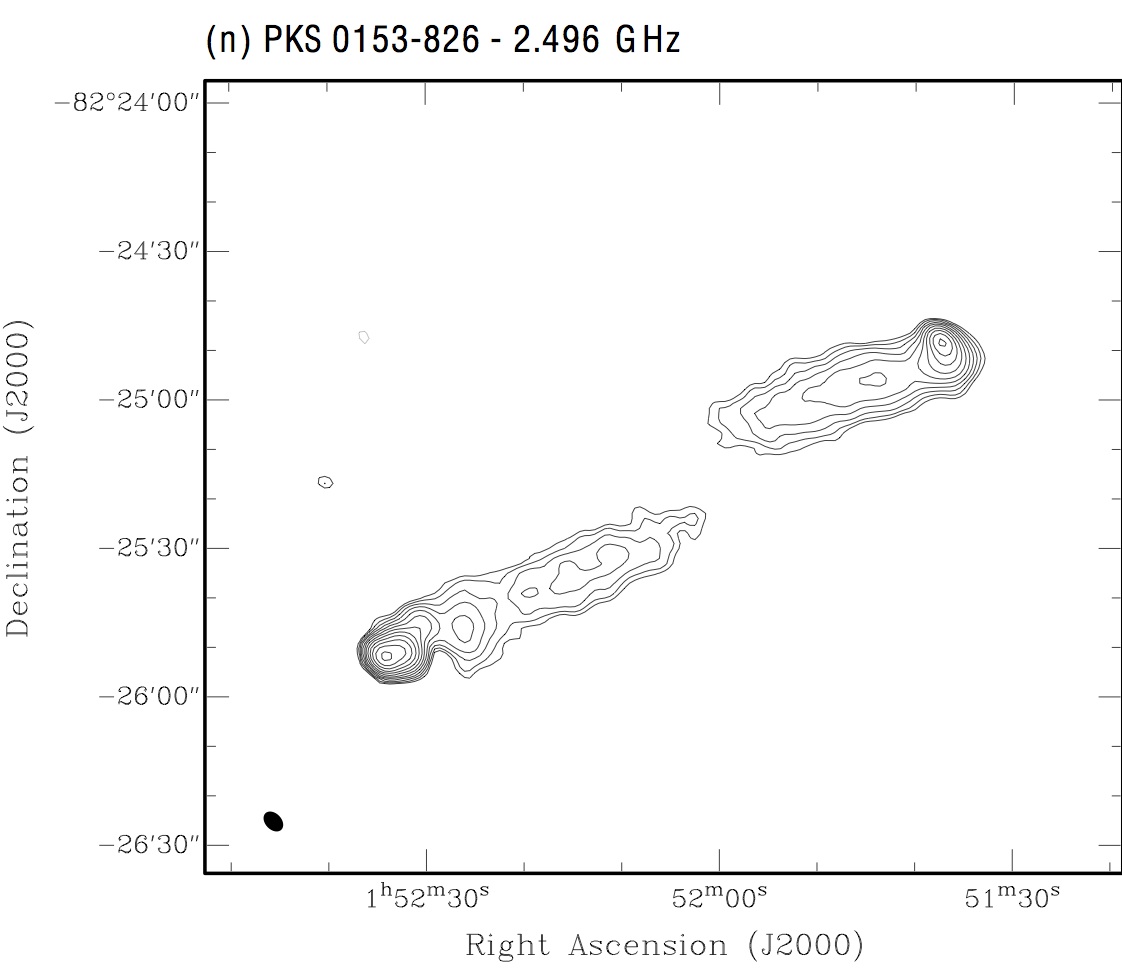}
}\\[5mm]
\mbox{
\epsscale{0.4}
\plotone{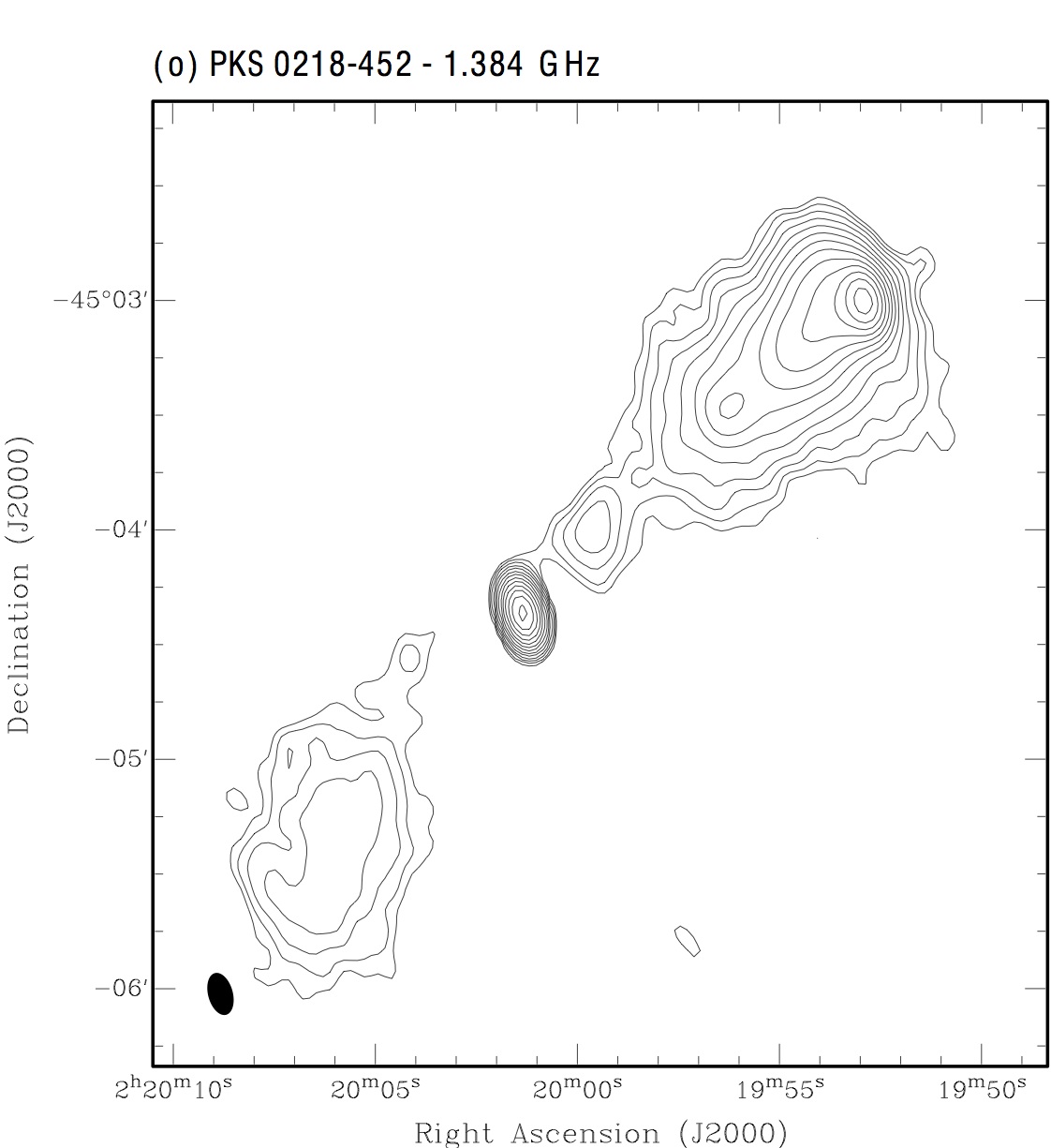} \quad
\plotone{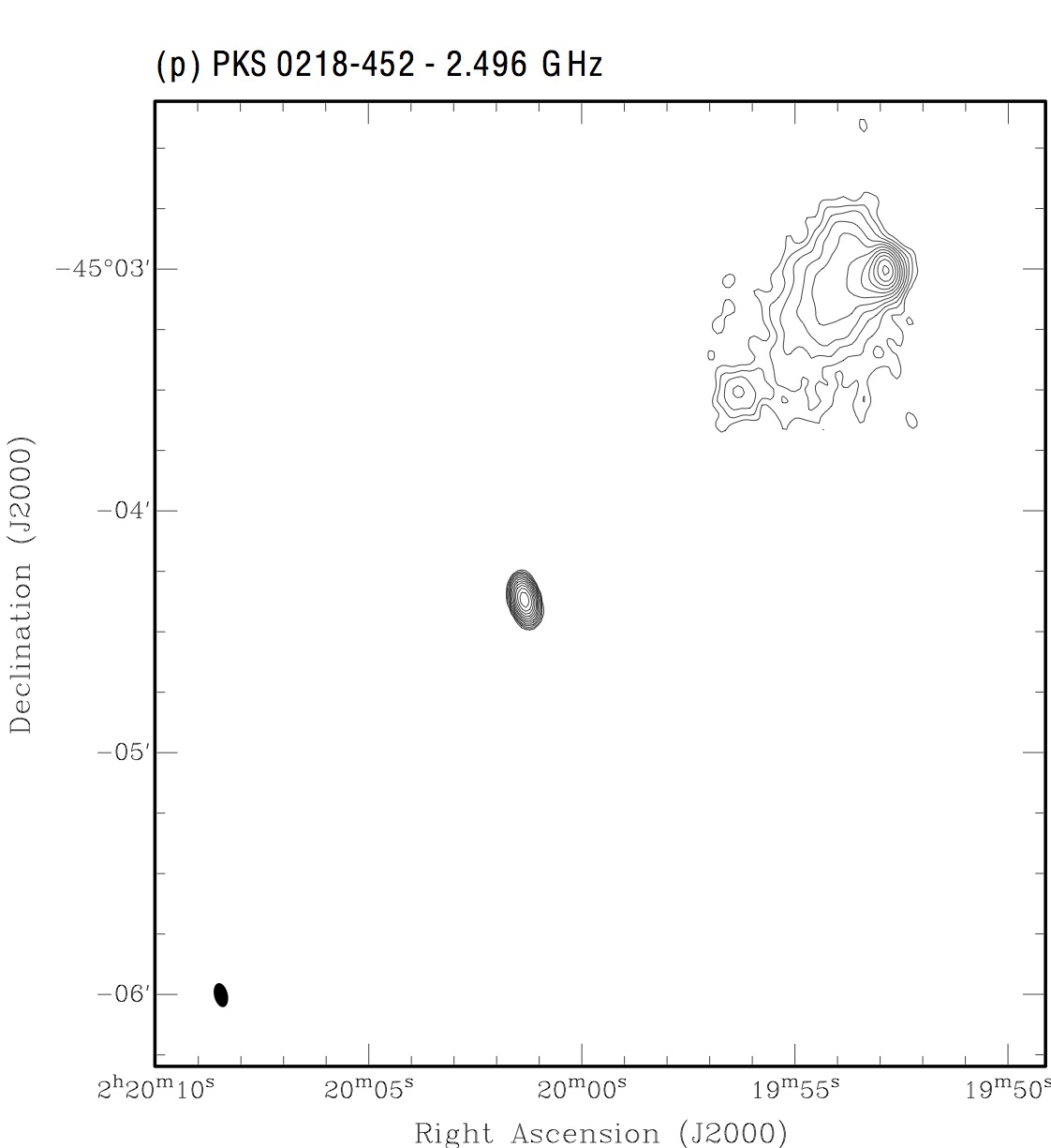}
}\\[5mm]
{Fig. 3.1. --- Continued}
\end{center}
\clearpage
\begin{center}
\mbox{
\plotone{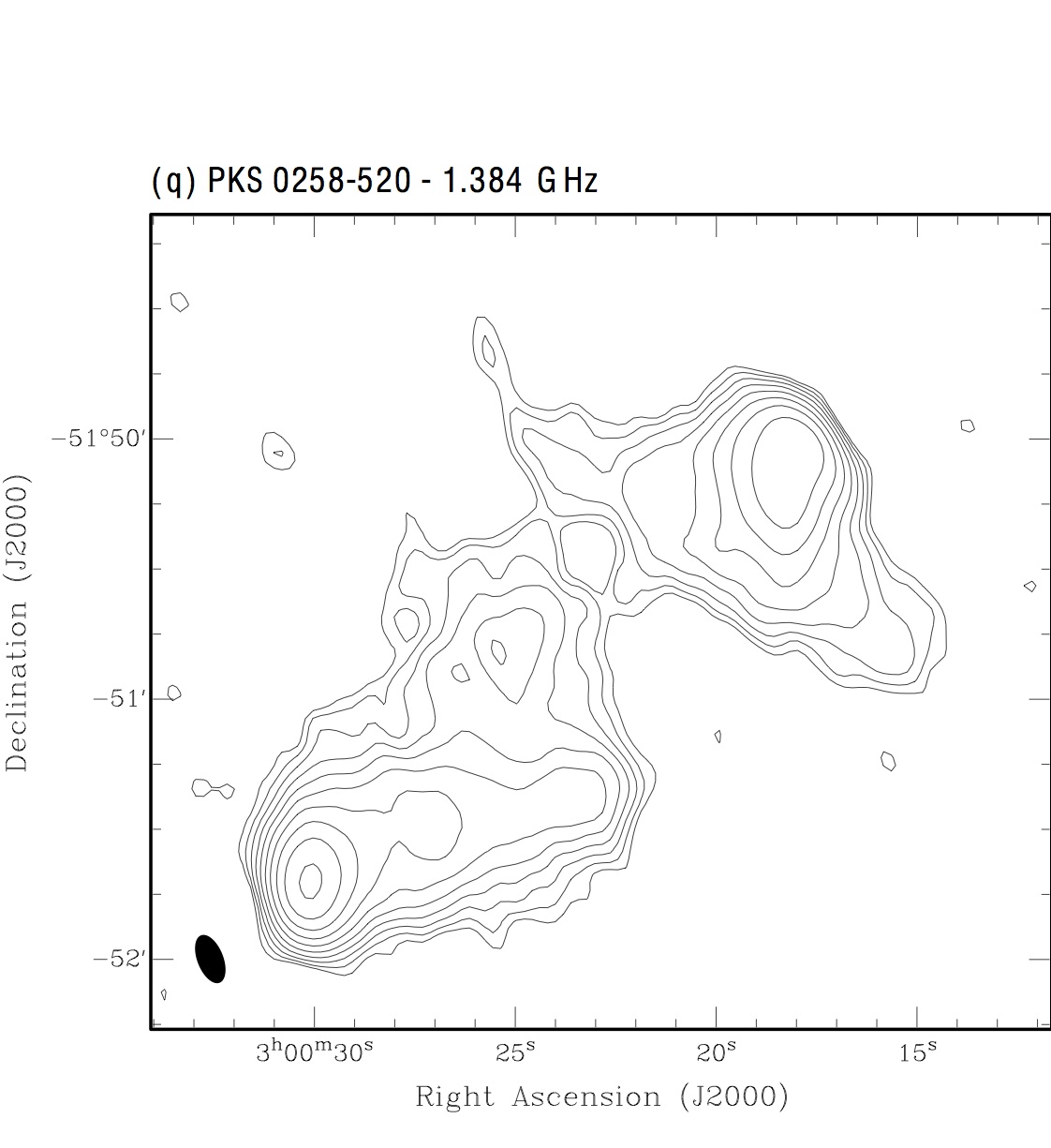} \quad
\plotone{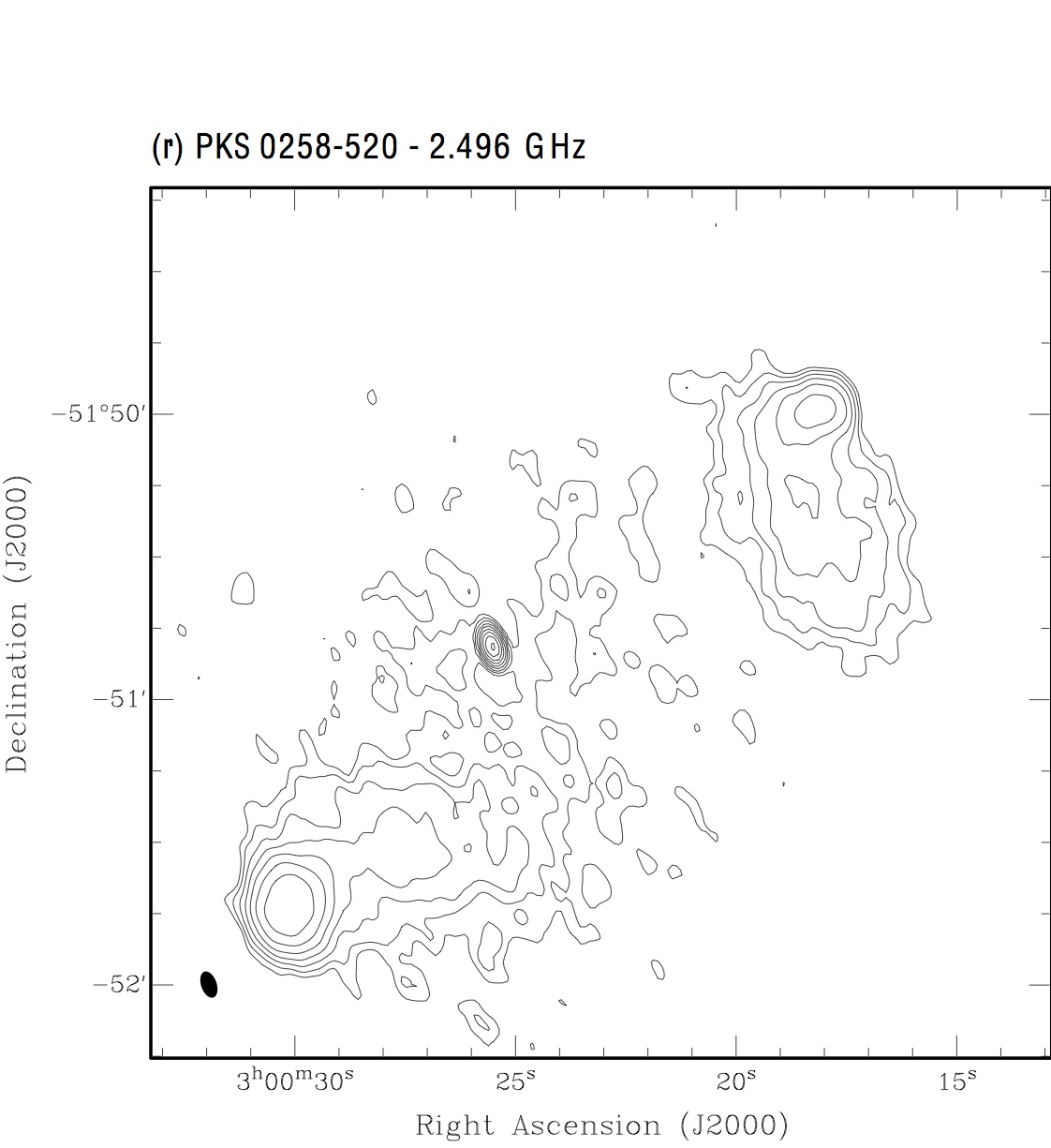}
}\\[5mm]
\mbox{
\plotone{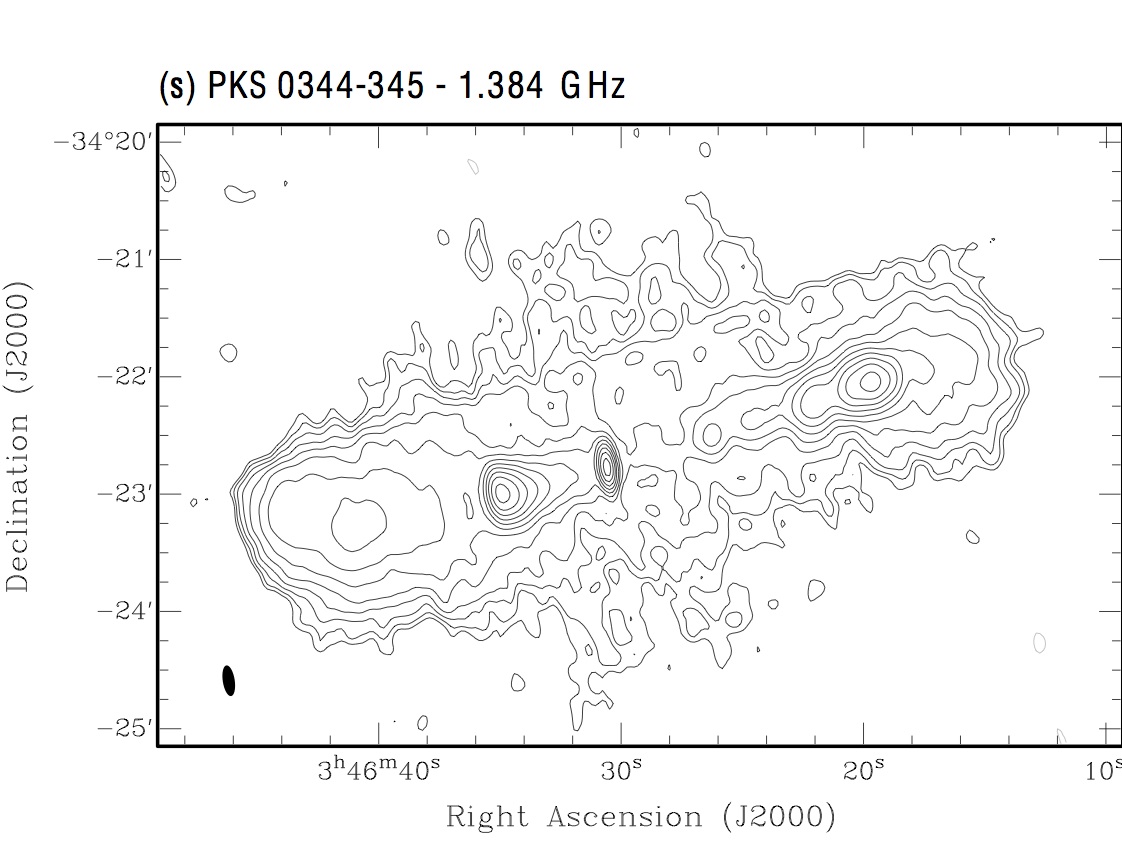} \quad
\plotone{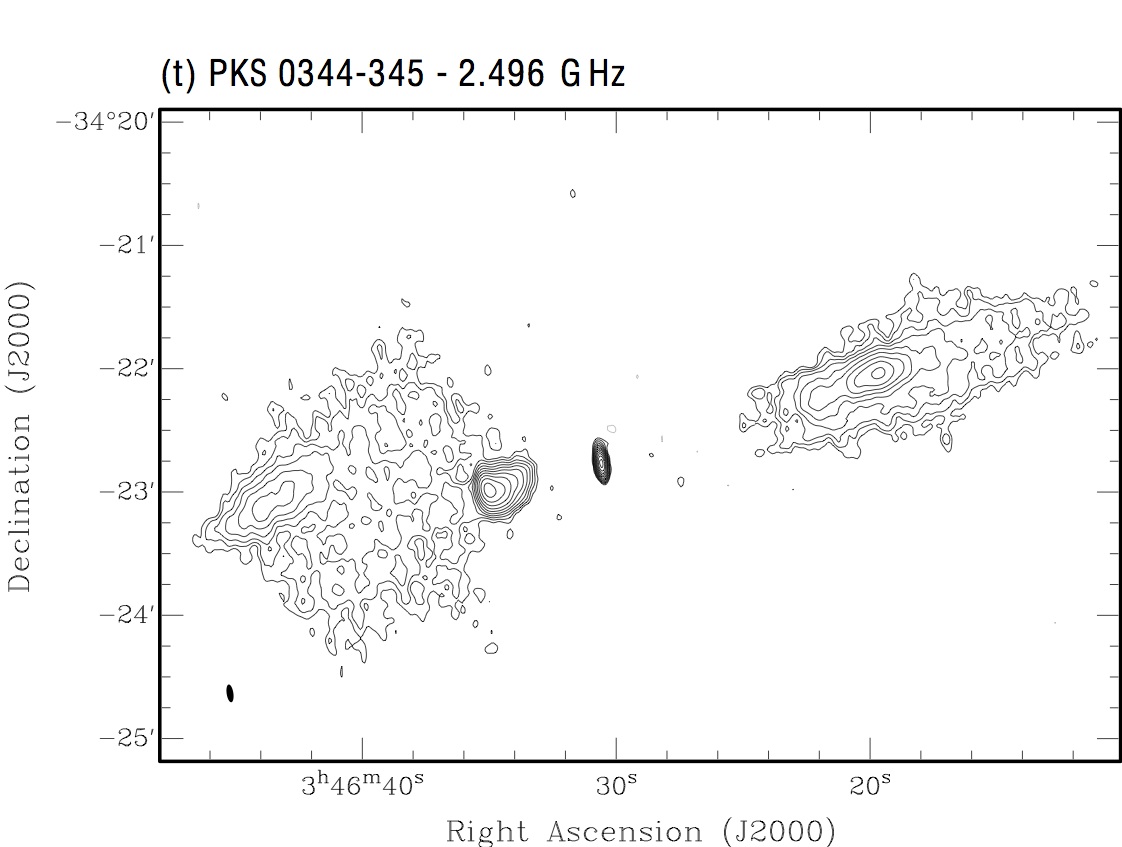}
}\\[5mm]
\mbox{
\plotone{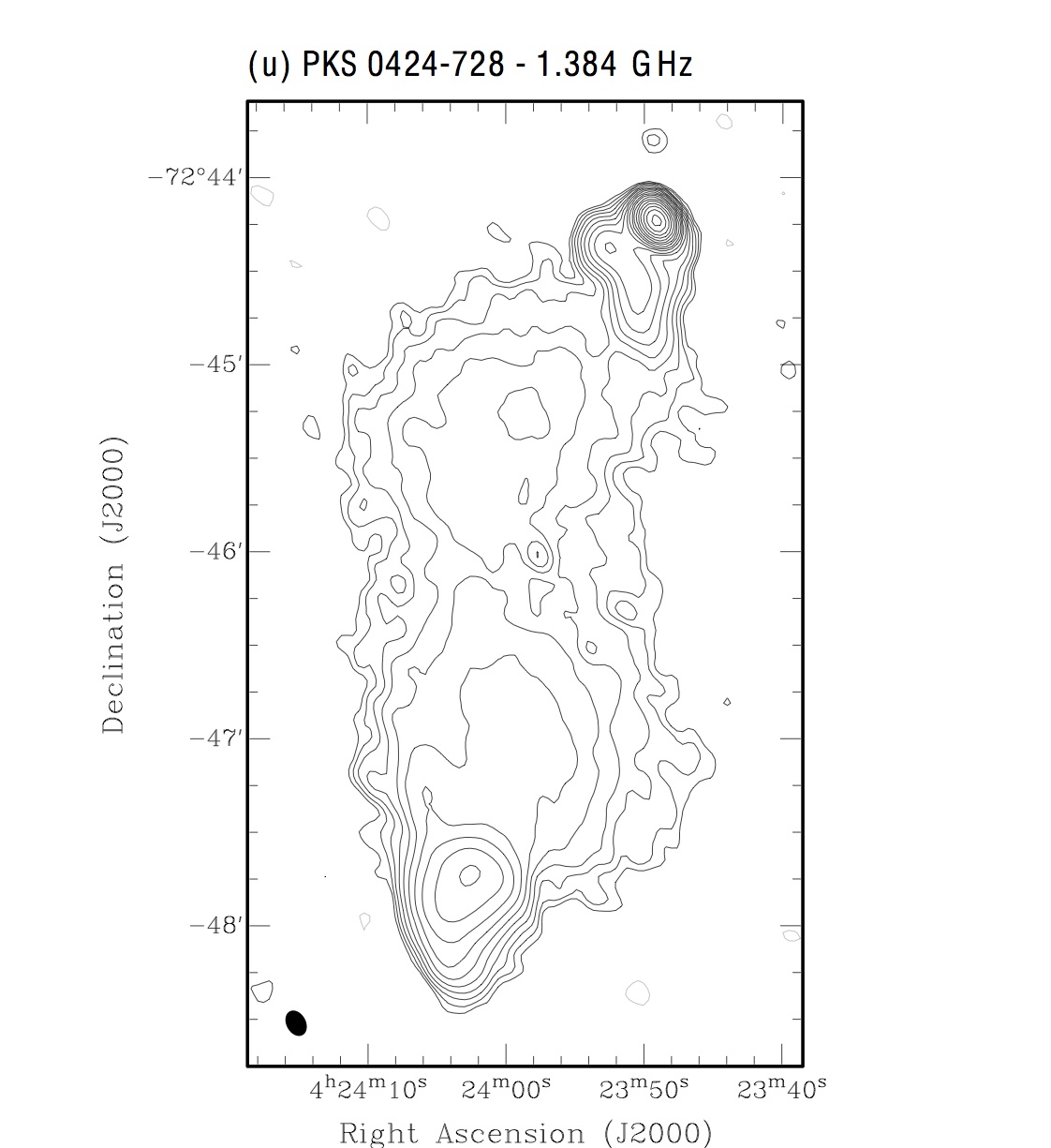} \quad
\plotone{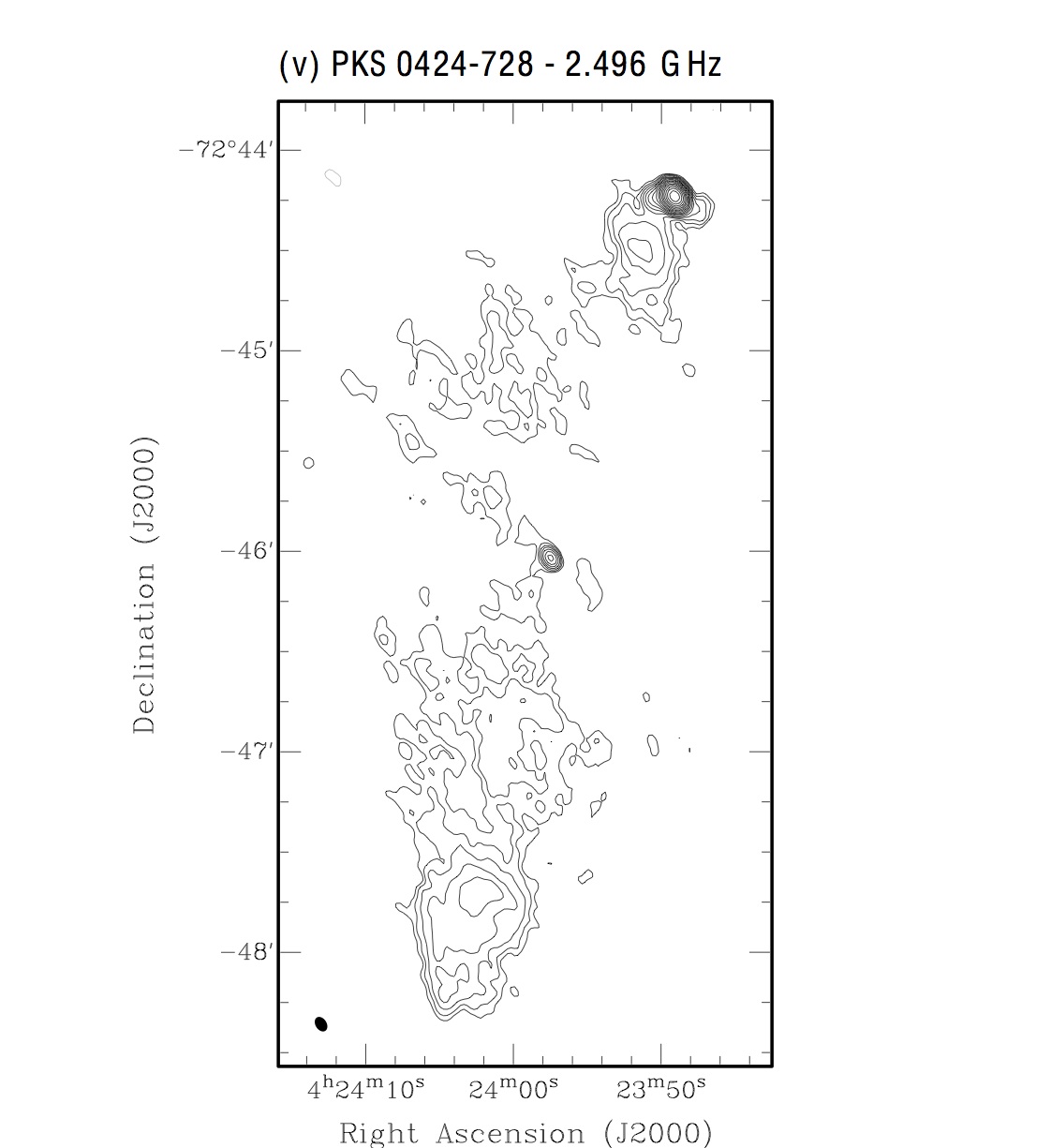}
}\\[5mm]
{Fig. 3.1. --- Continued}
\end{center}
\clearpage
\begin{center}
\mbox{
\plotone{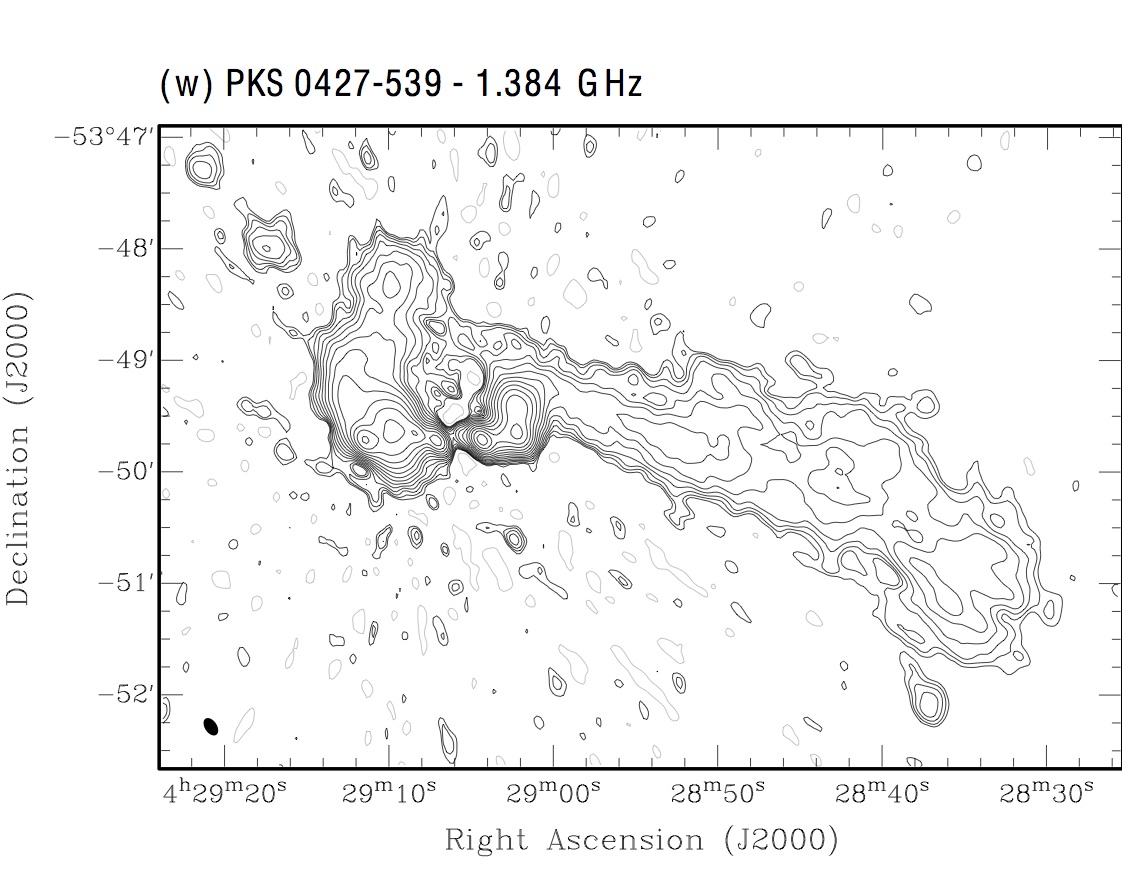} \quad
\plotone{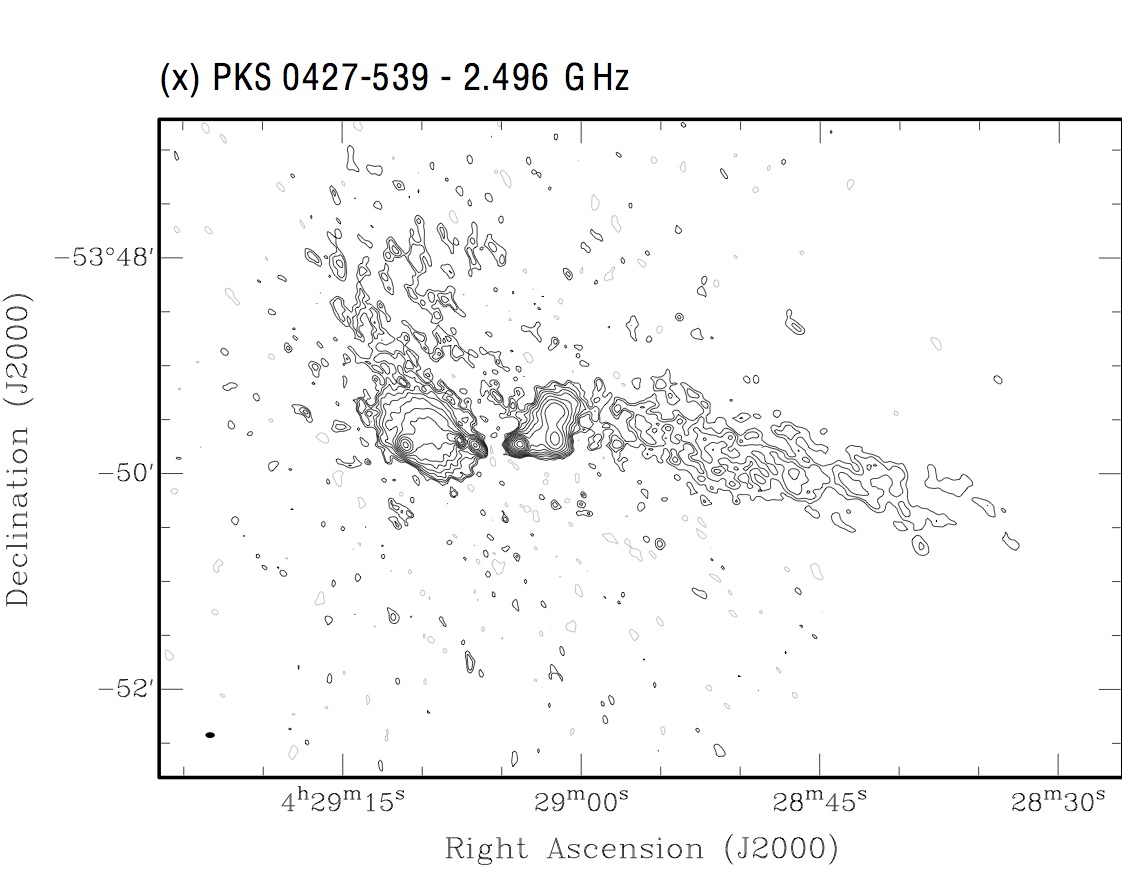}
}\\[5mm]
\mbox{
\plotone{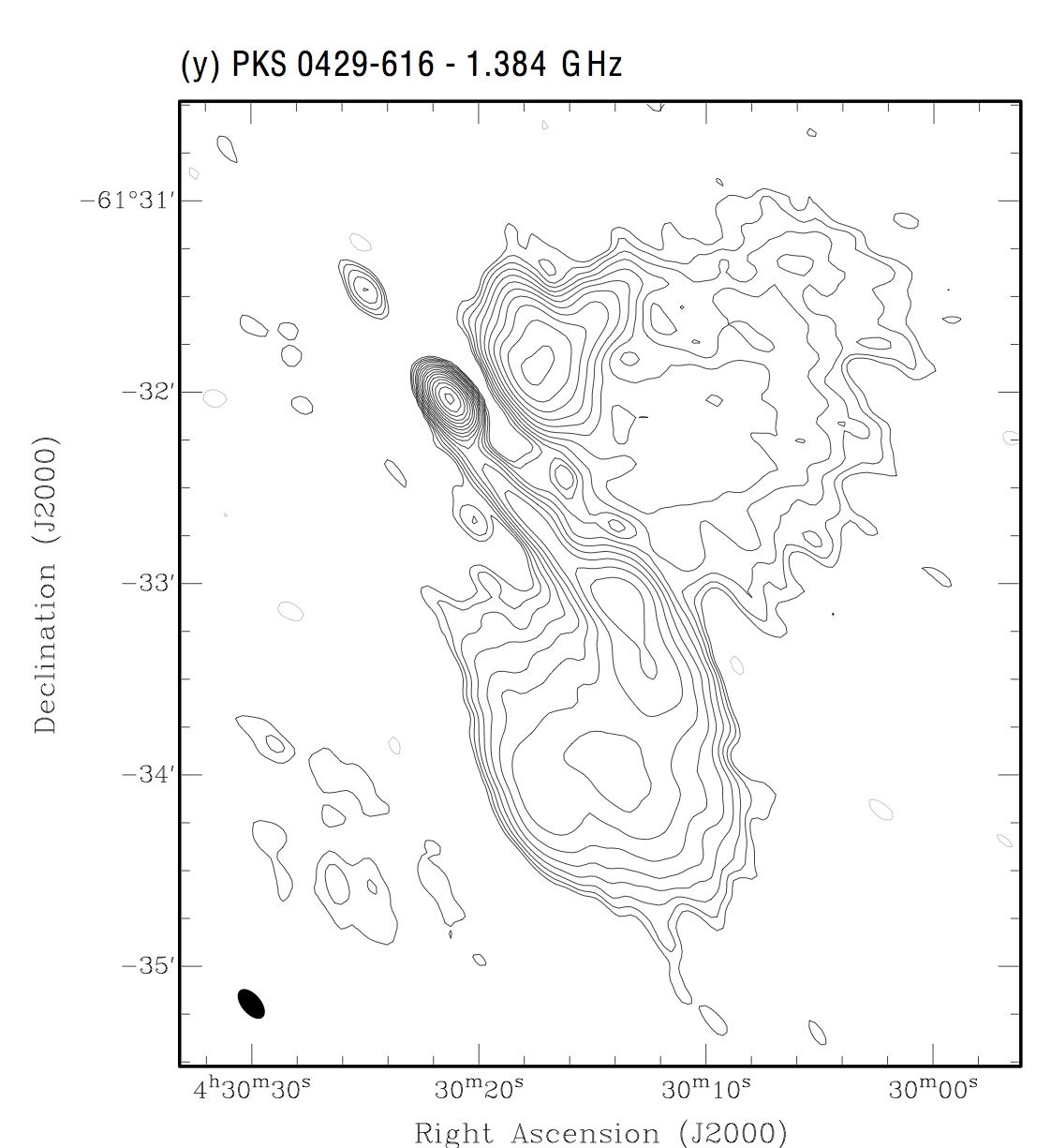} \quad
\plotone{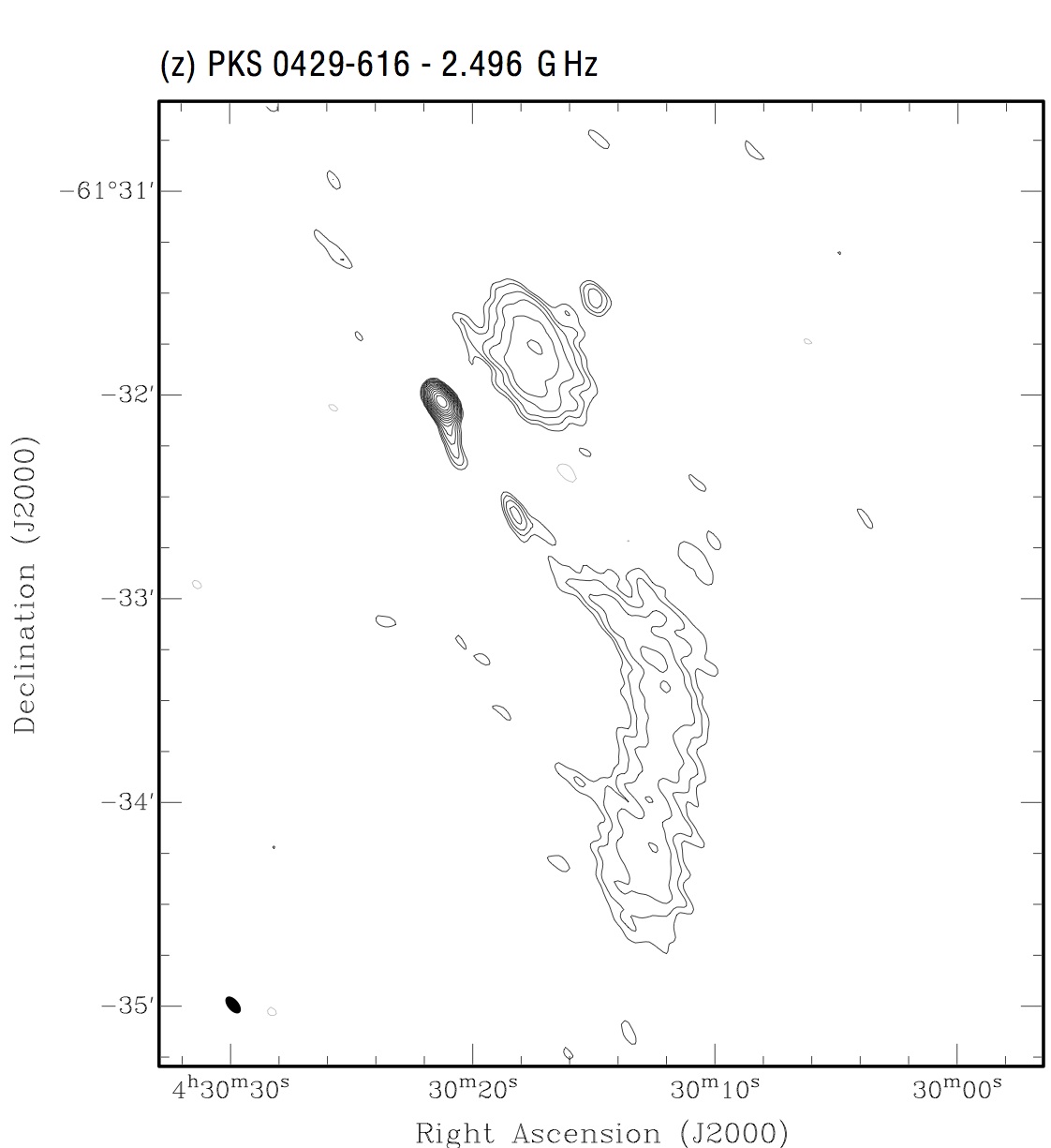}
}\\[5mm]
\mbox{
\plotone{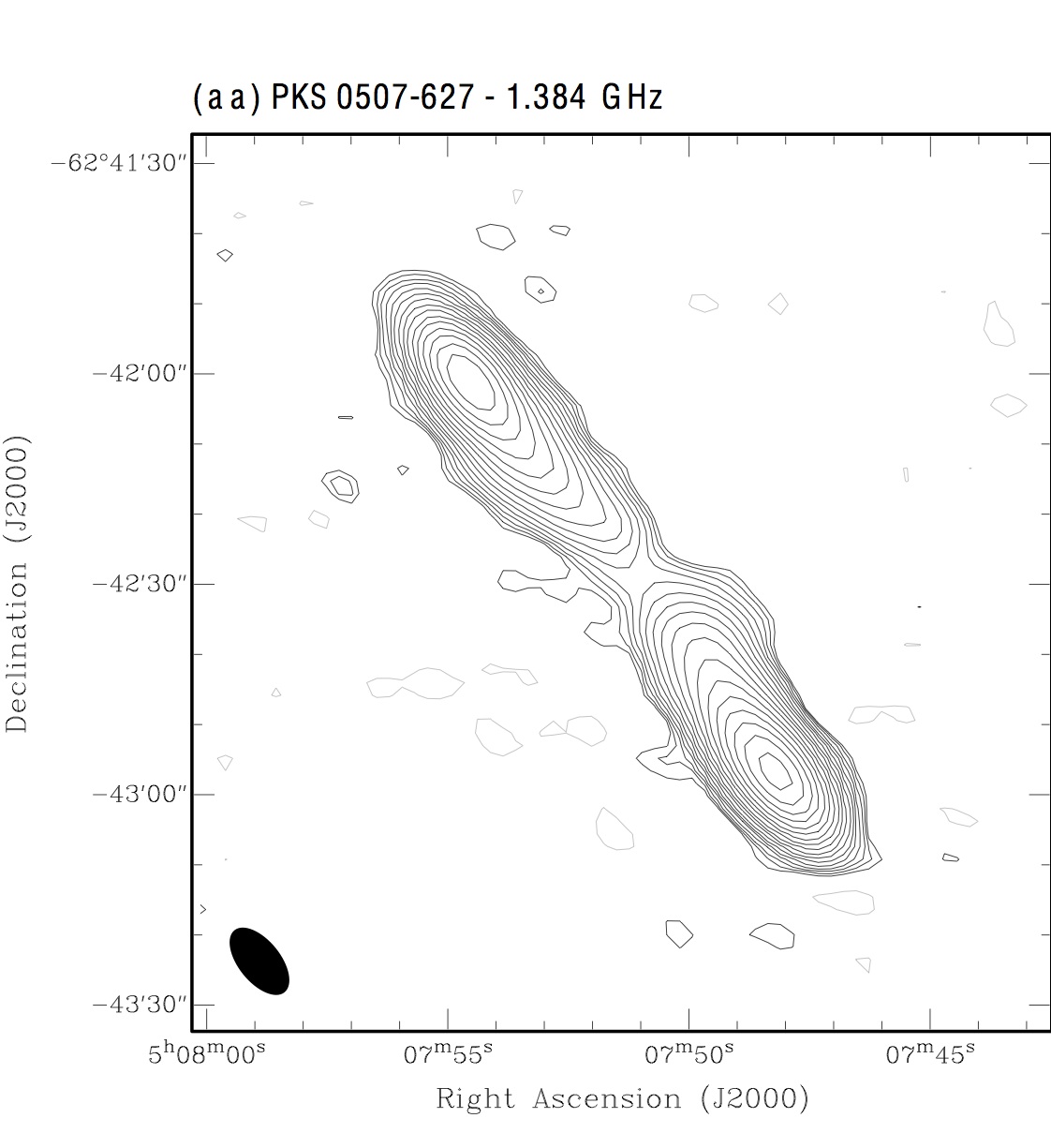} \quad
\plotone{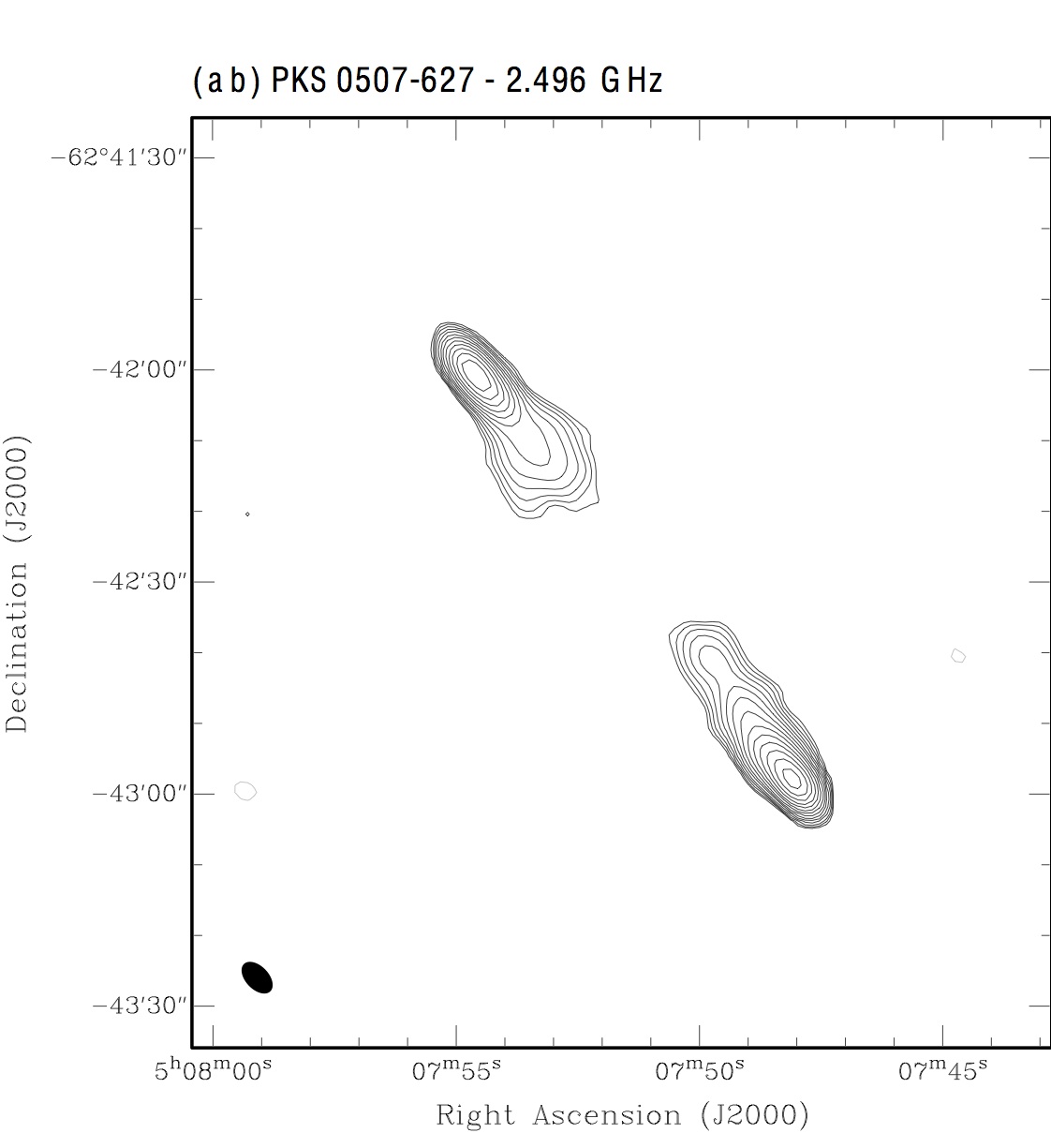}
}\\[5mm]
{Fig. 3.1. --- Continued}
\end{center}
\clearpage
\begin{center}
\mbox{
\plotone{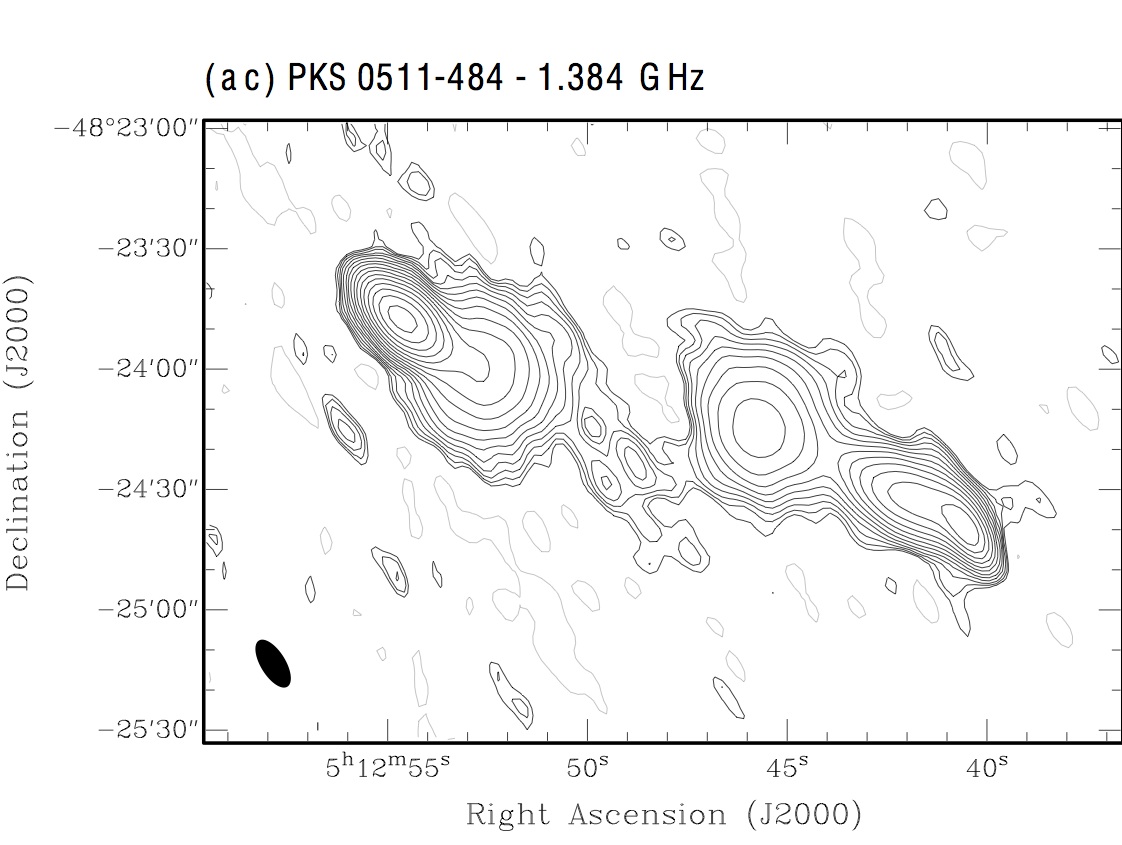} \quad
\plotone{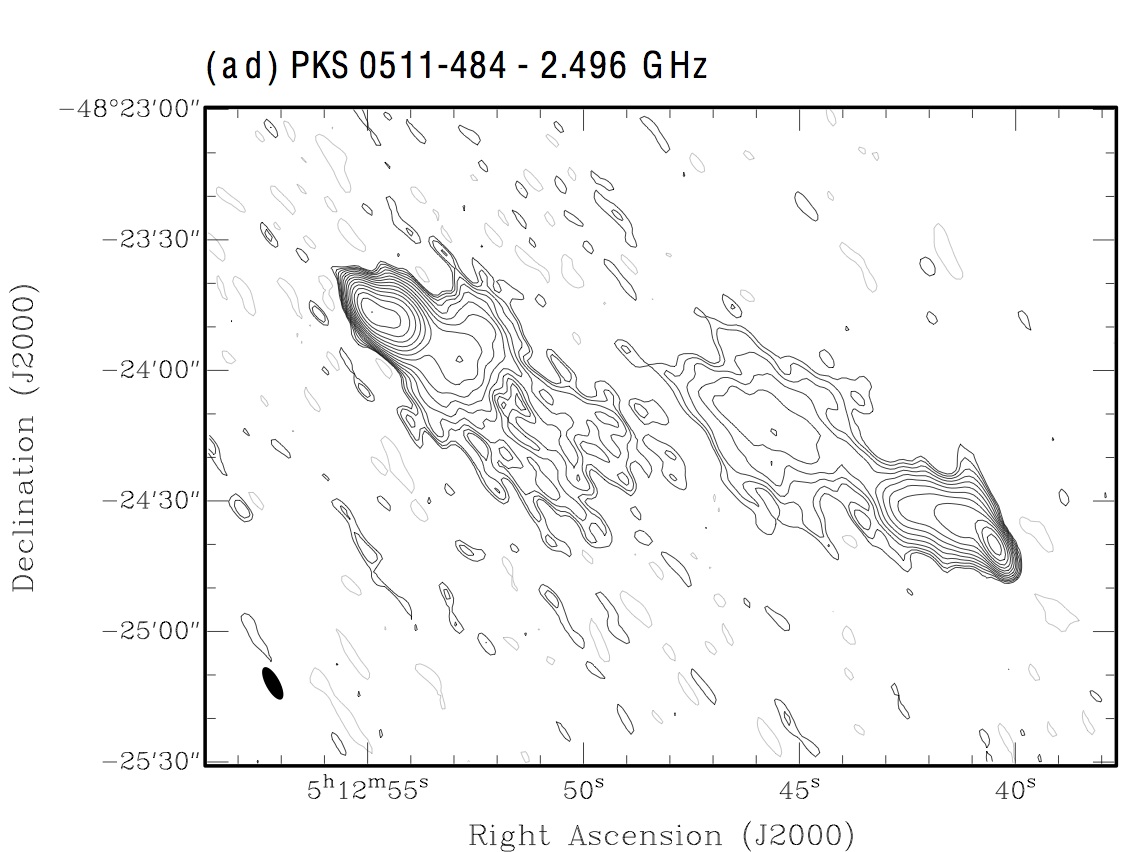}
}\\[5mm]
\mbox{
\plotone{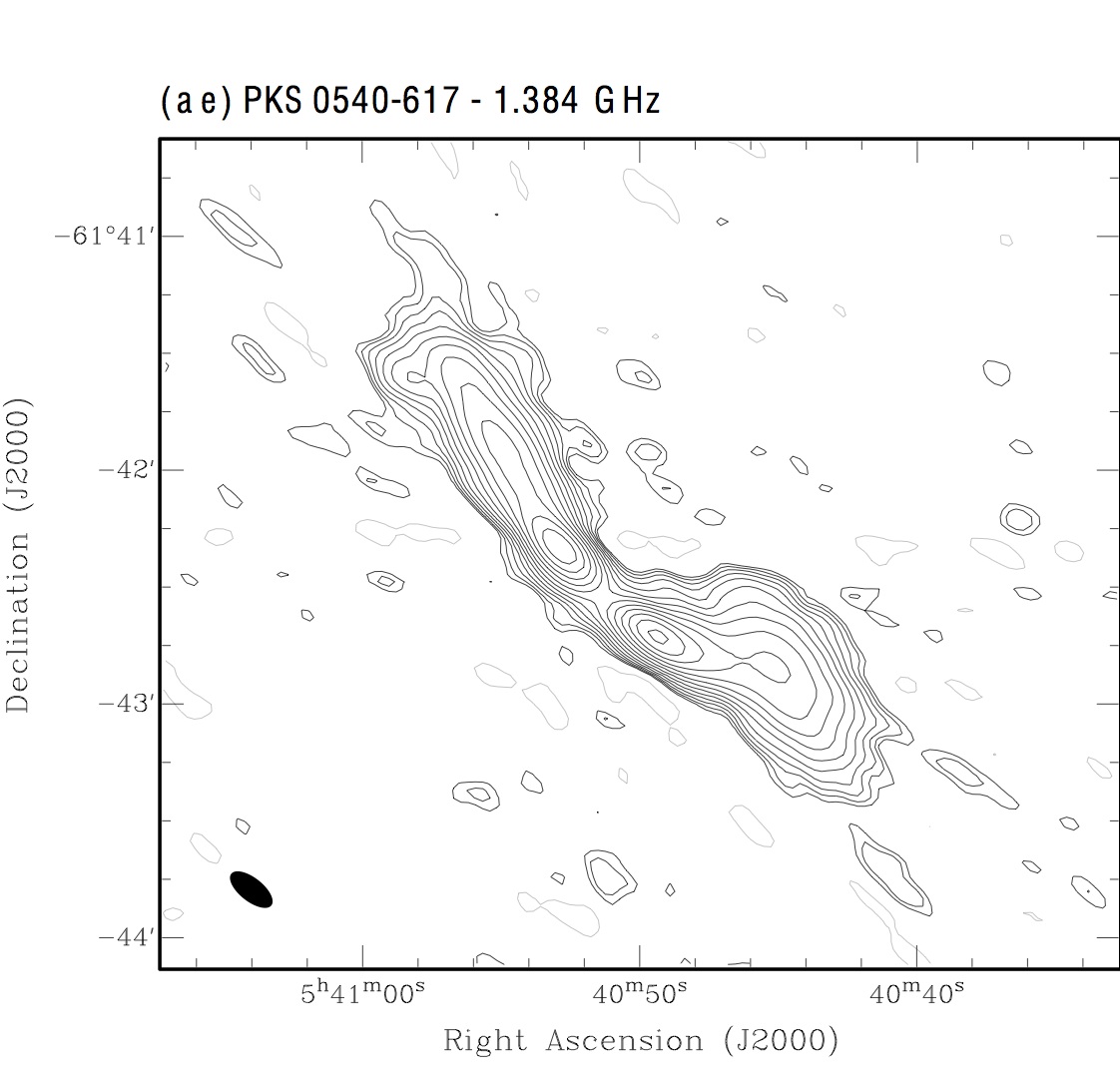} \quad
\plotone{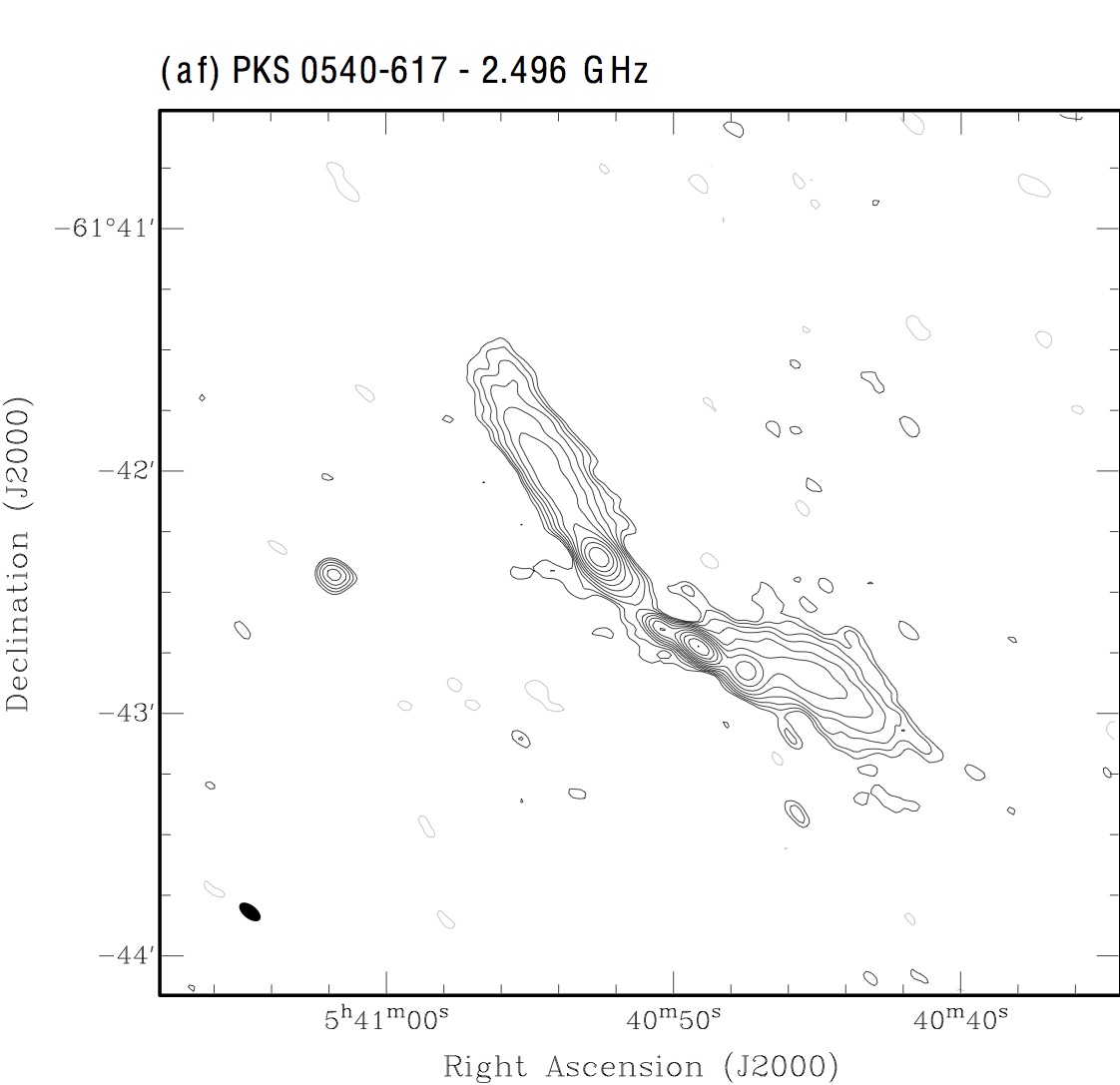}
}\\[5mm]
\mbox{
\plotone{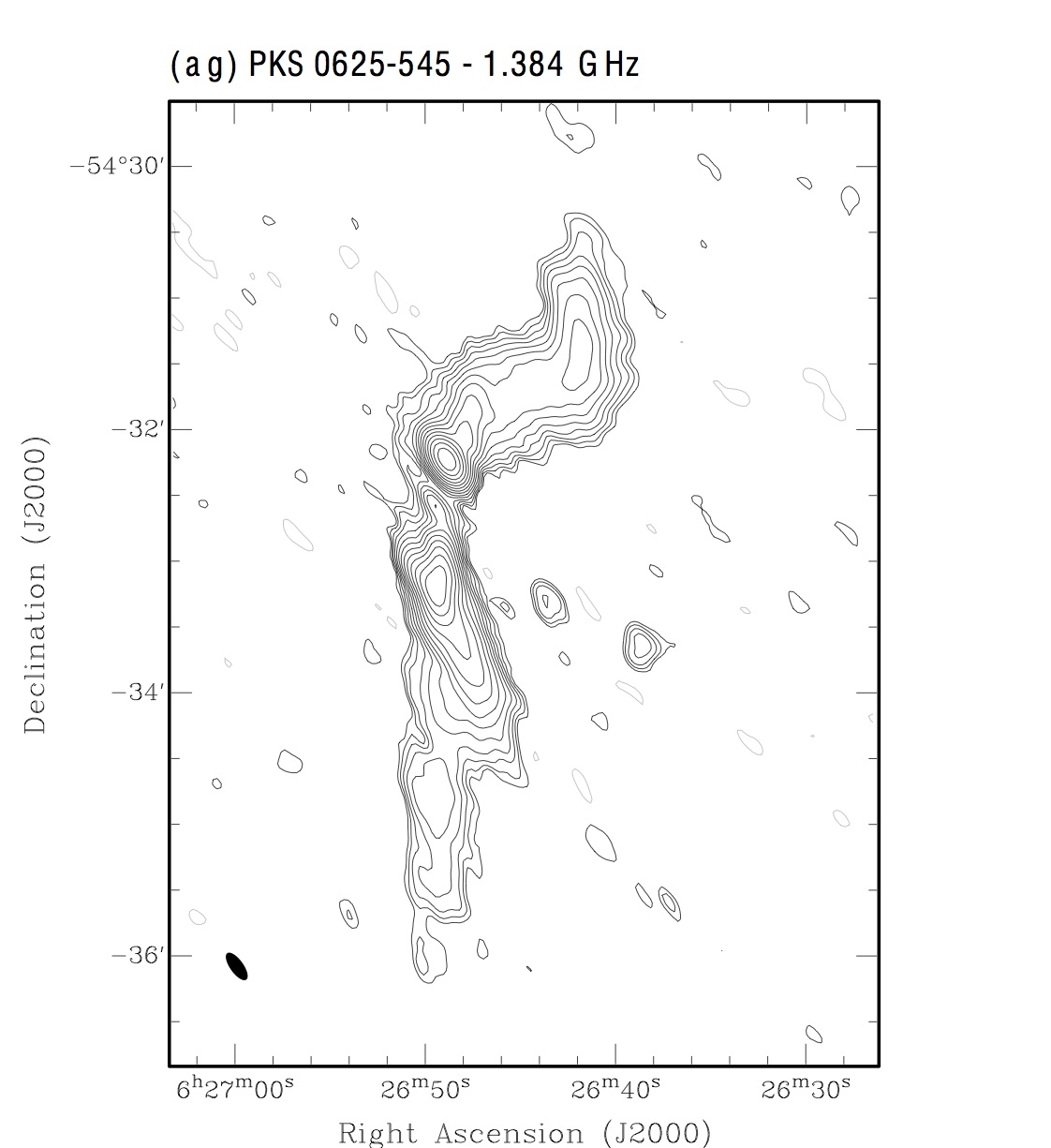} \quad
\plotone{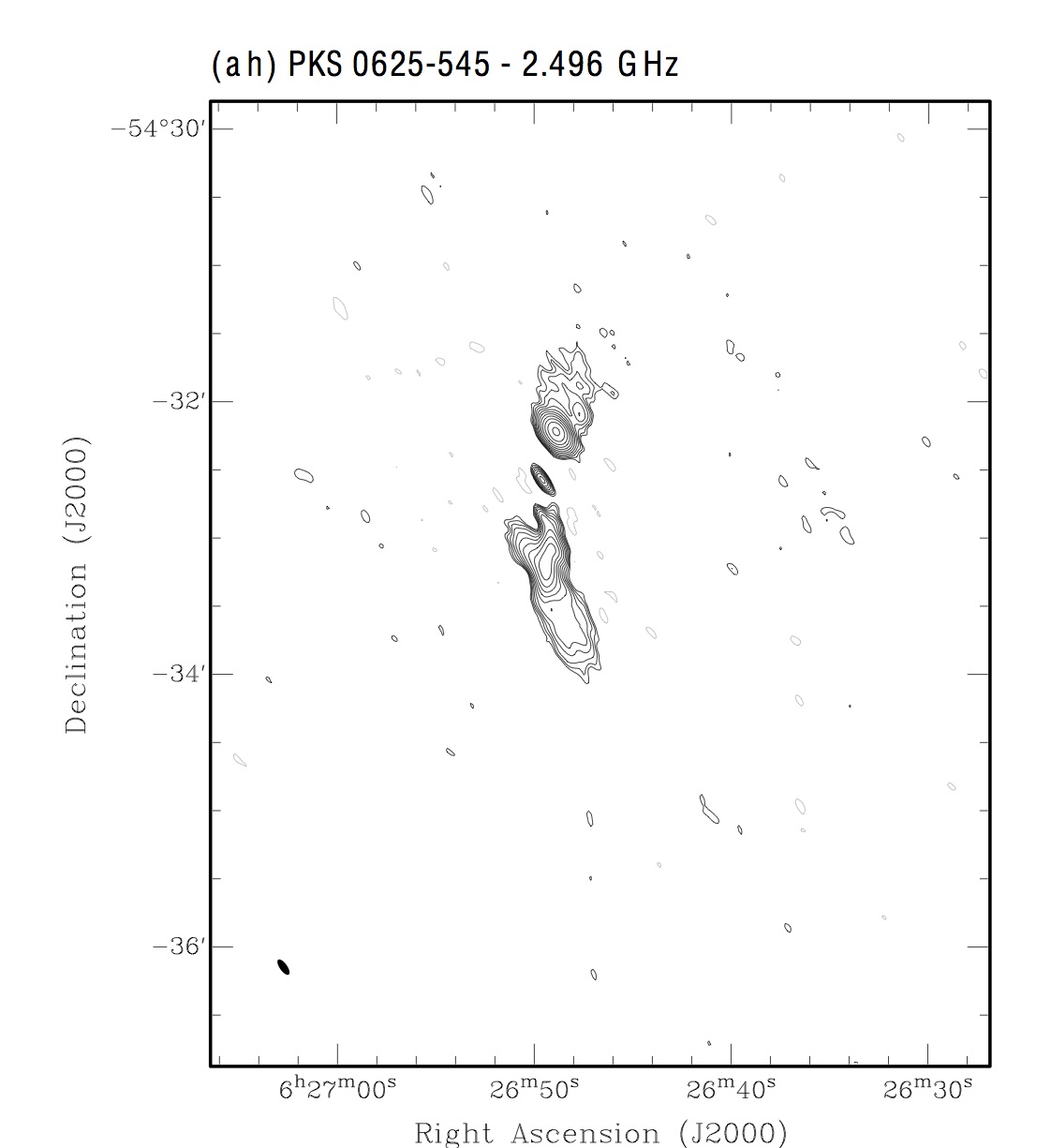}
}\\[5mm]
{Fig. 3.1. --- Continued}
\end{center}
\clearpage
\begin{center}
\mbox{
\plotone{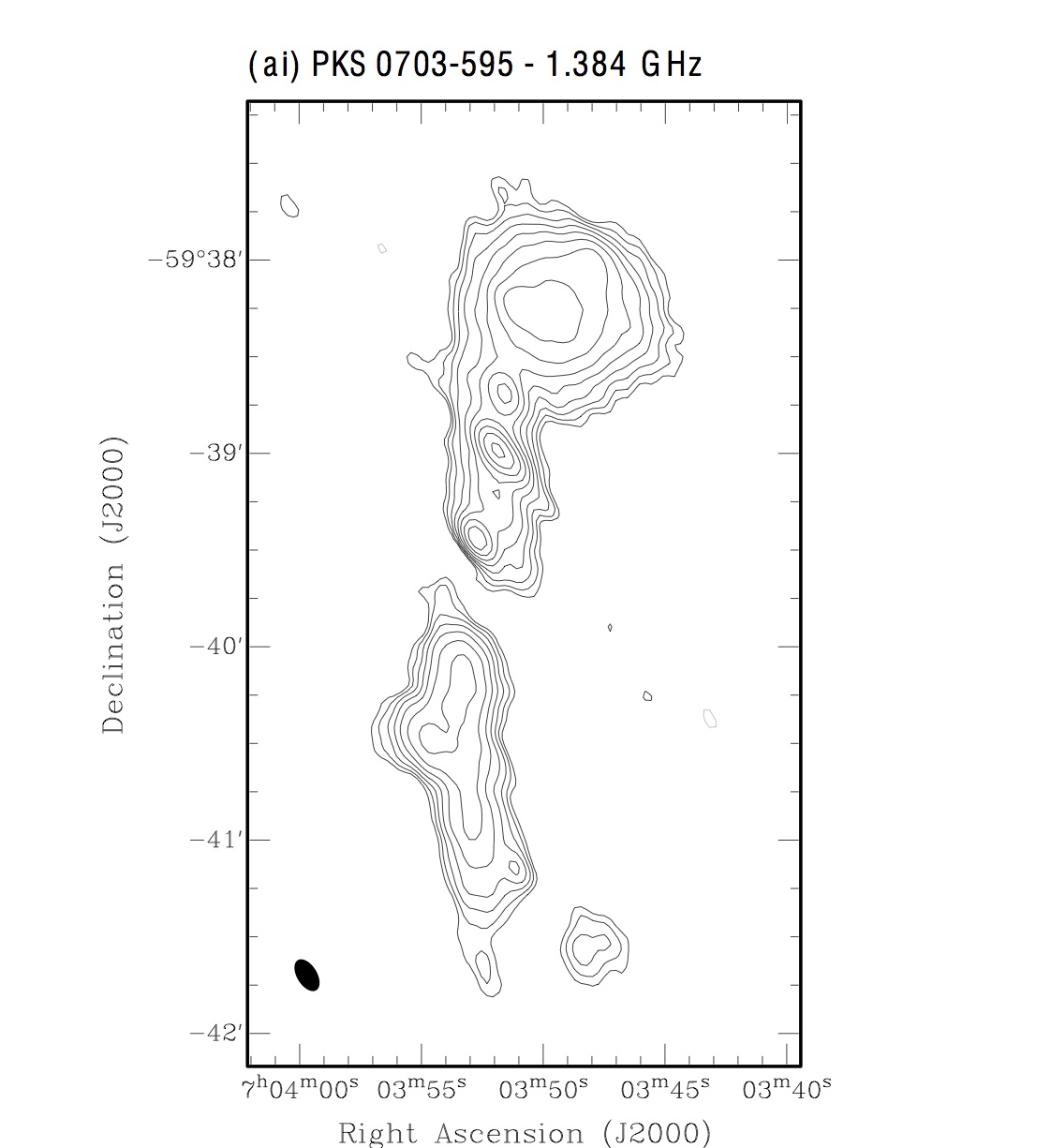} \quad
\plotone{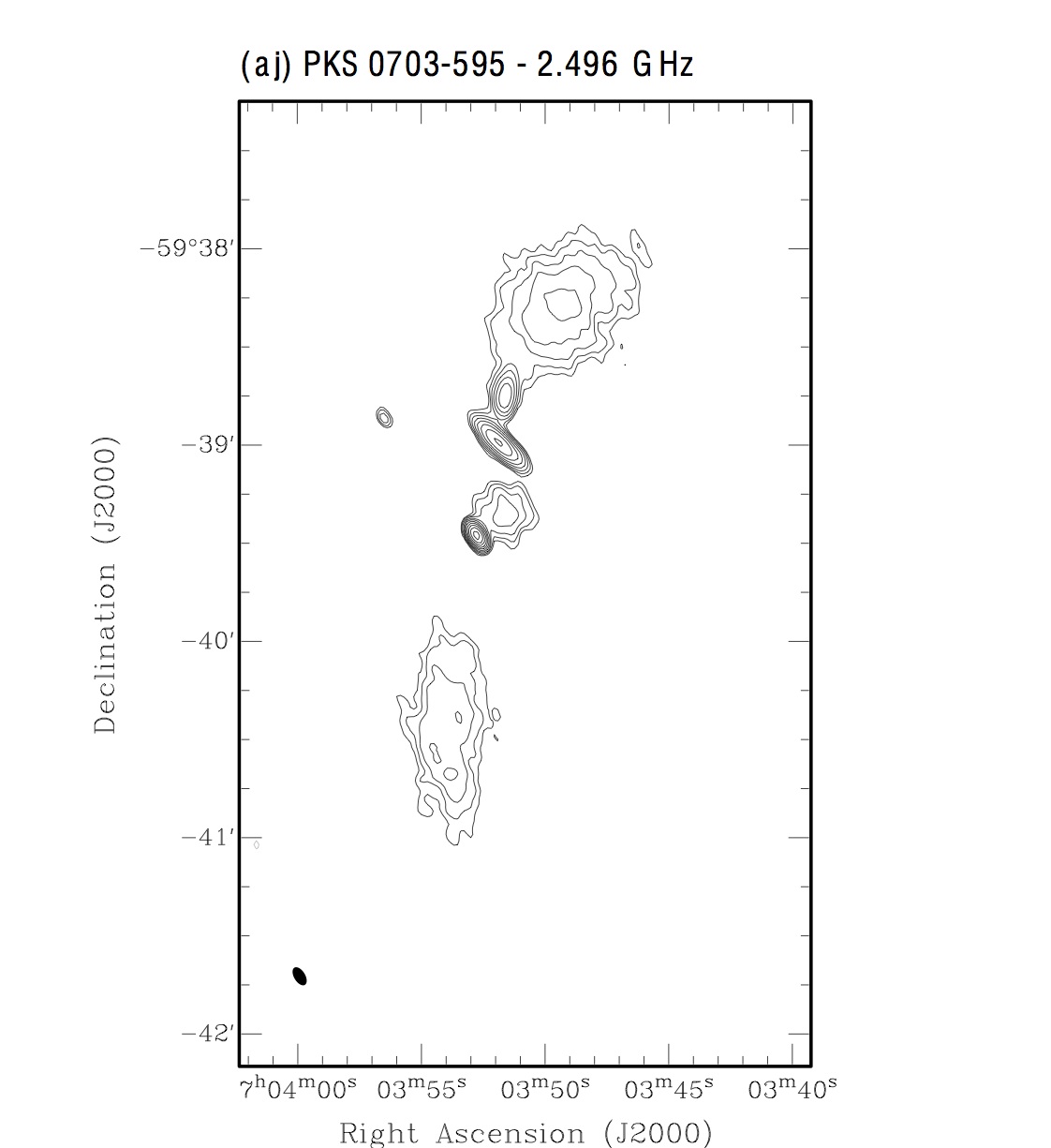}
}\\[5mm]
\mbox{
\plotone{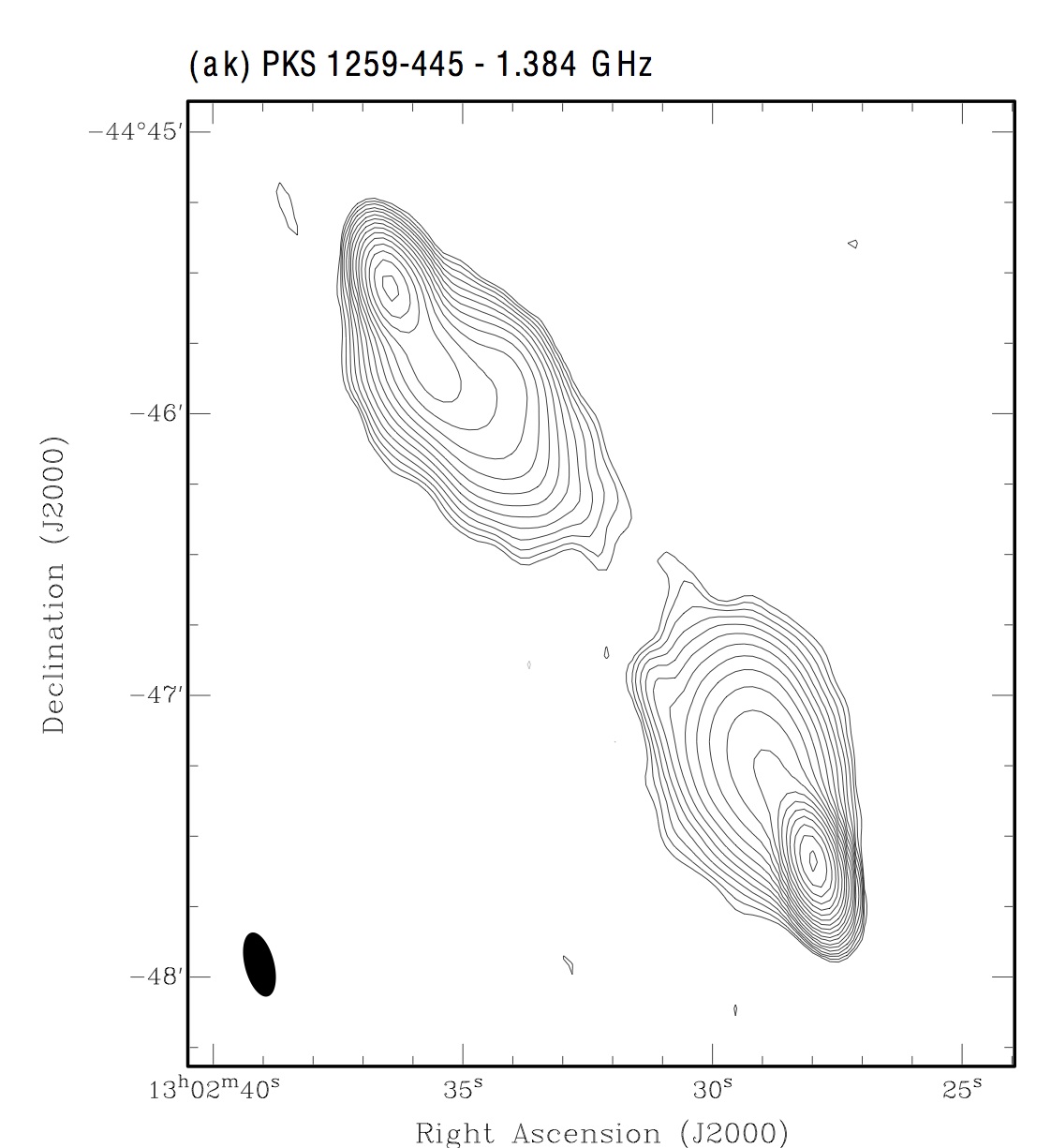} \quad
\plotone{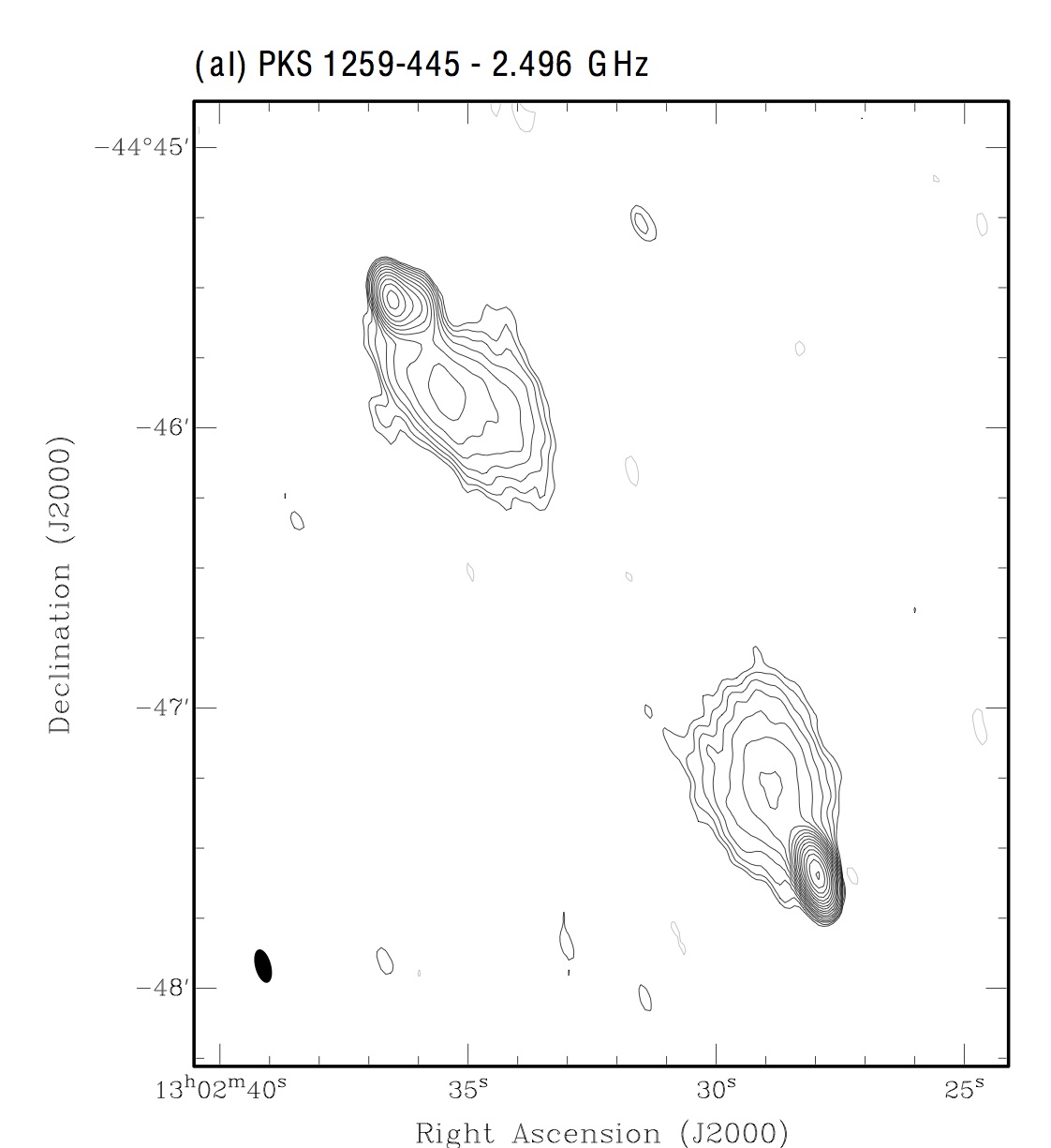}
}\\[5mm]
\mbox{
\plotone{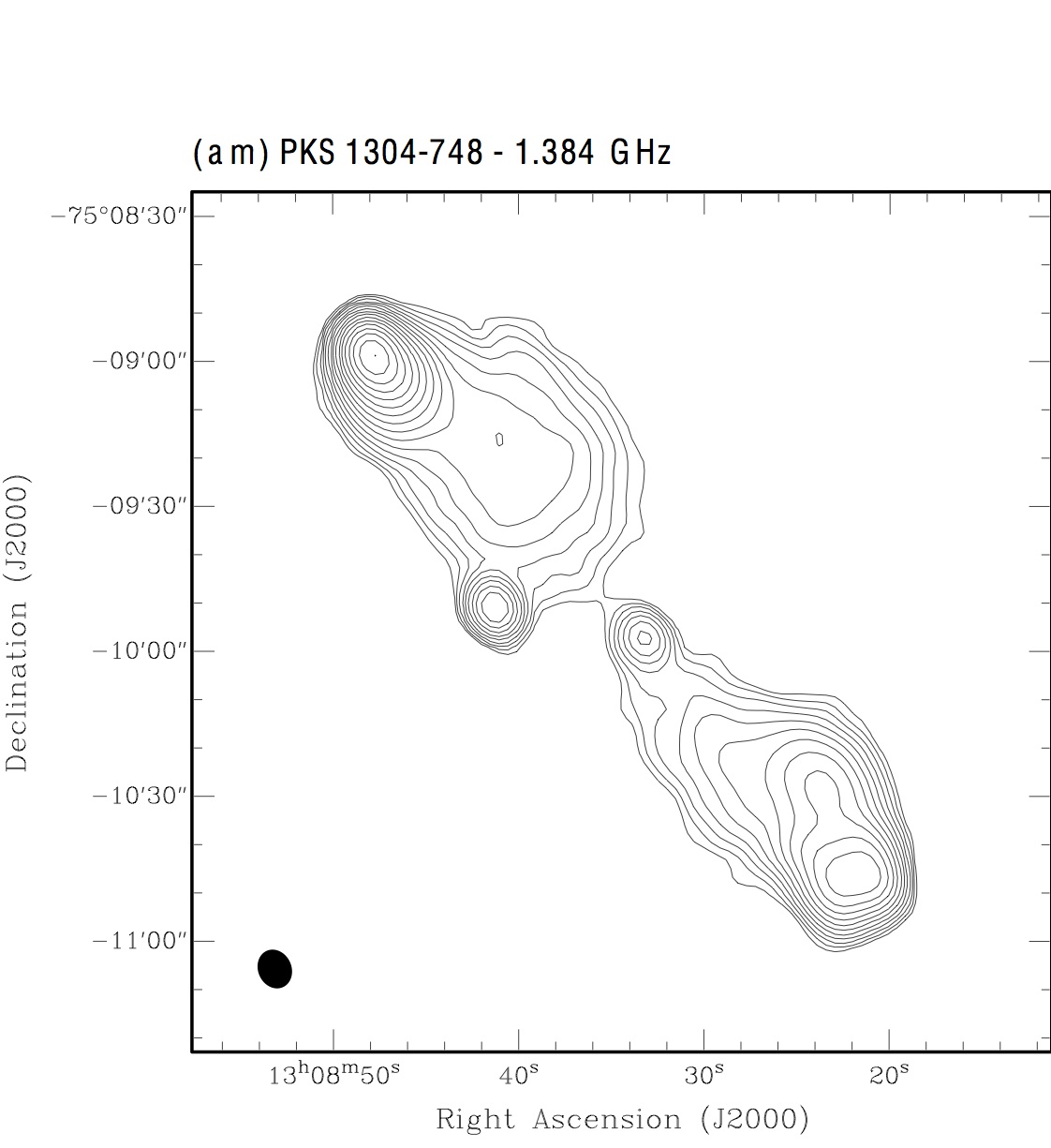} \quad
\plotone{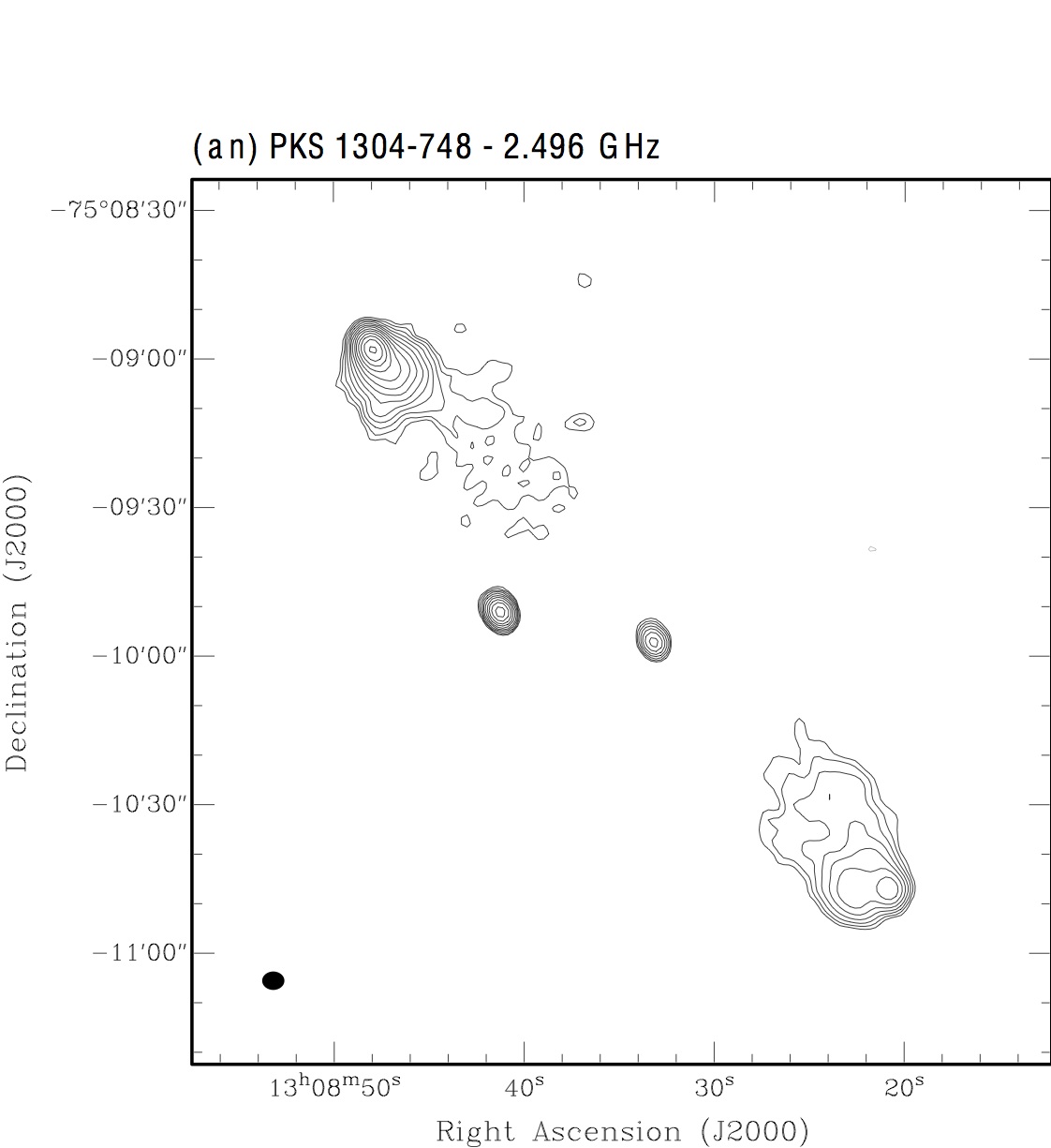}
}\\[5mm]
{Fig. 3.1. --- Continued}
\end{center}
\clearpage
\begin{center}
\mbox{
\plotone{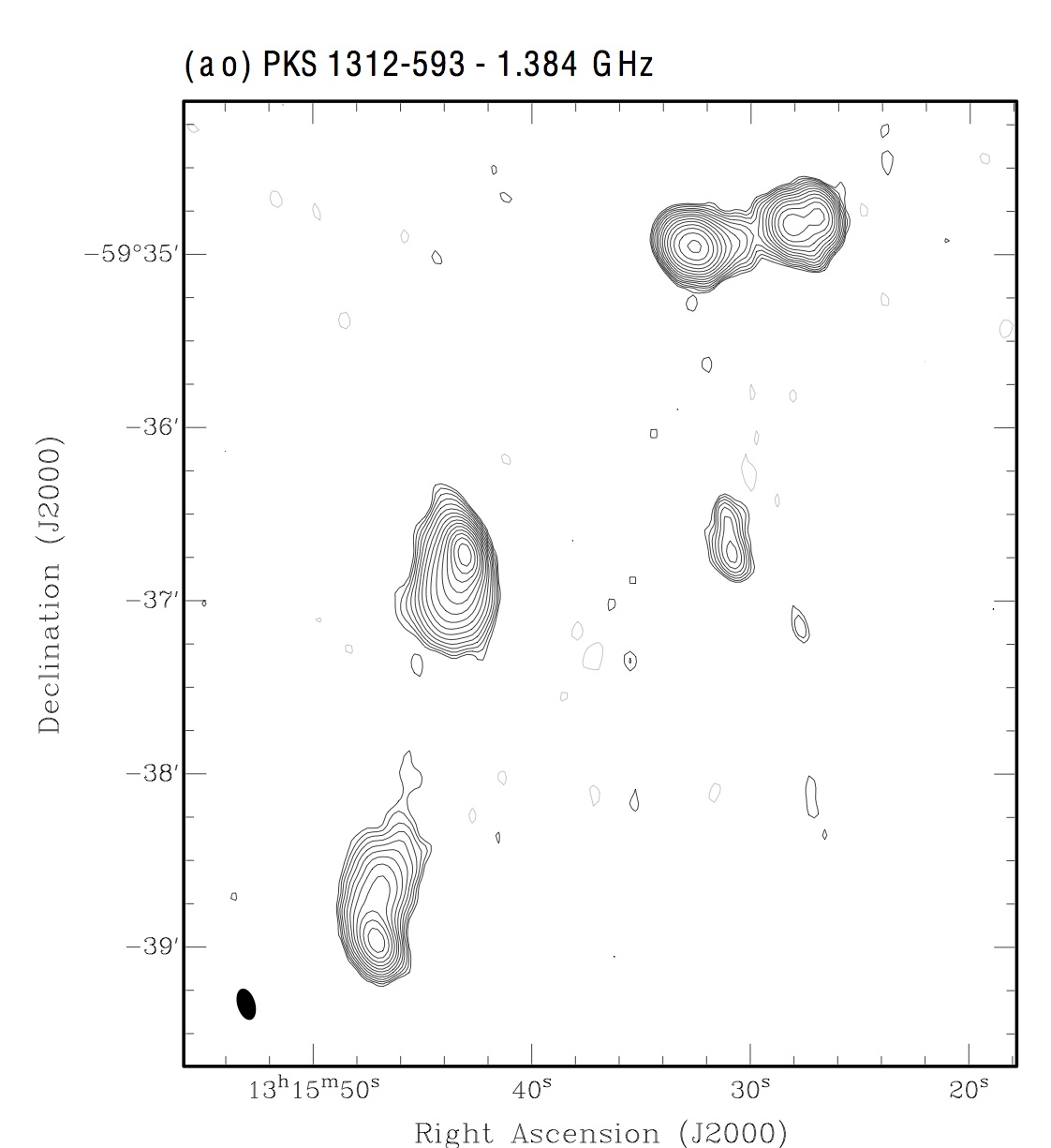} \quad
\plotone{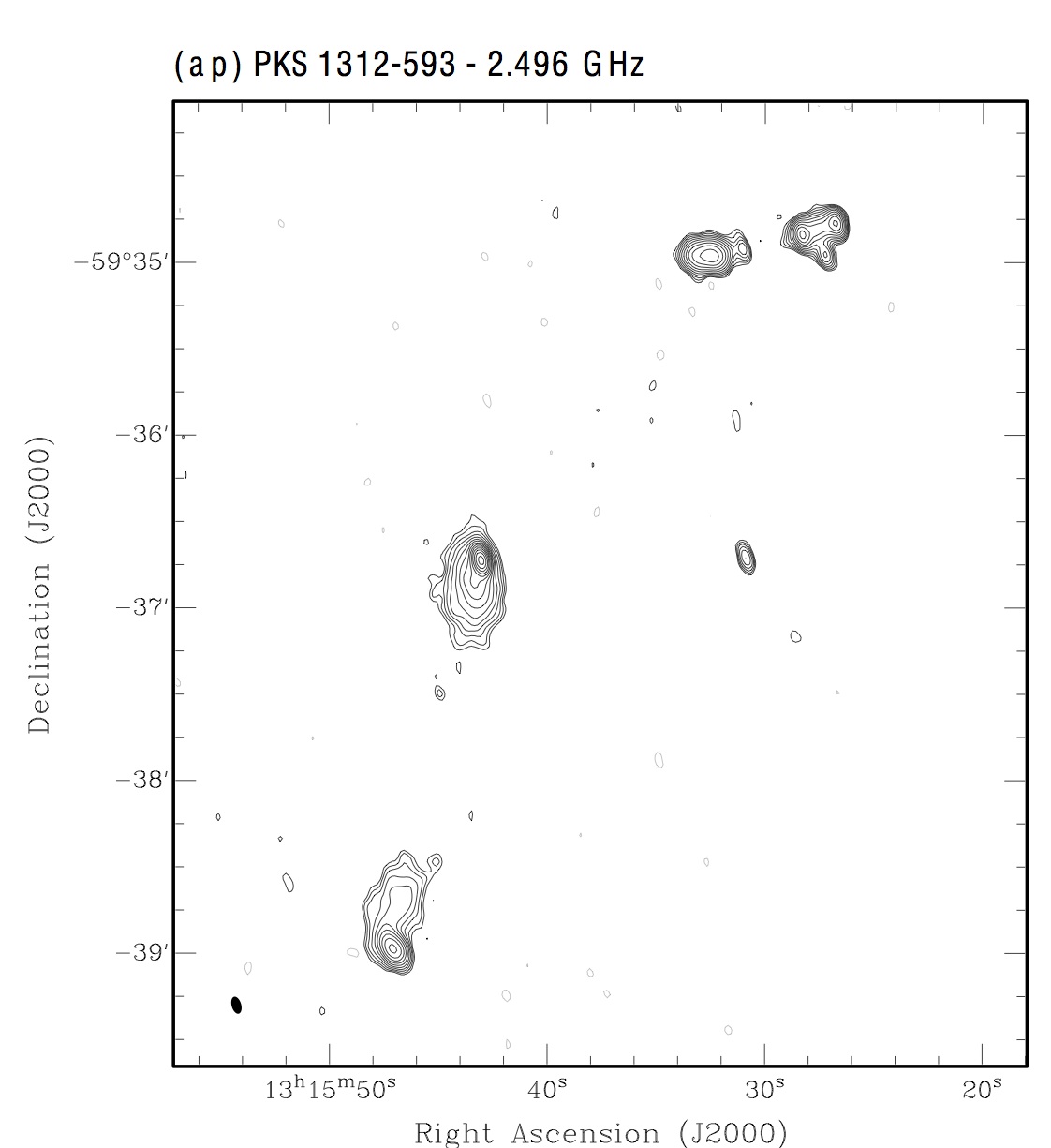}
}\\[5mm]
\mbox{
\plotone{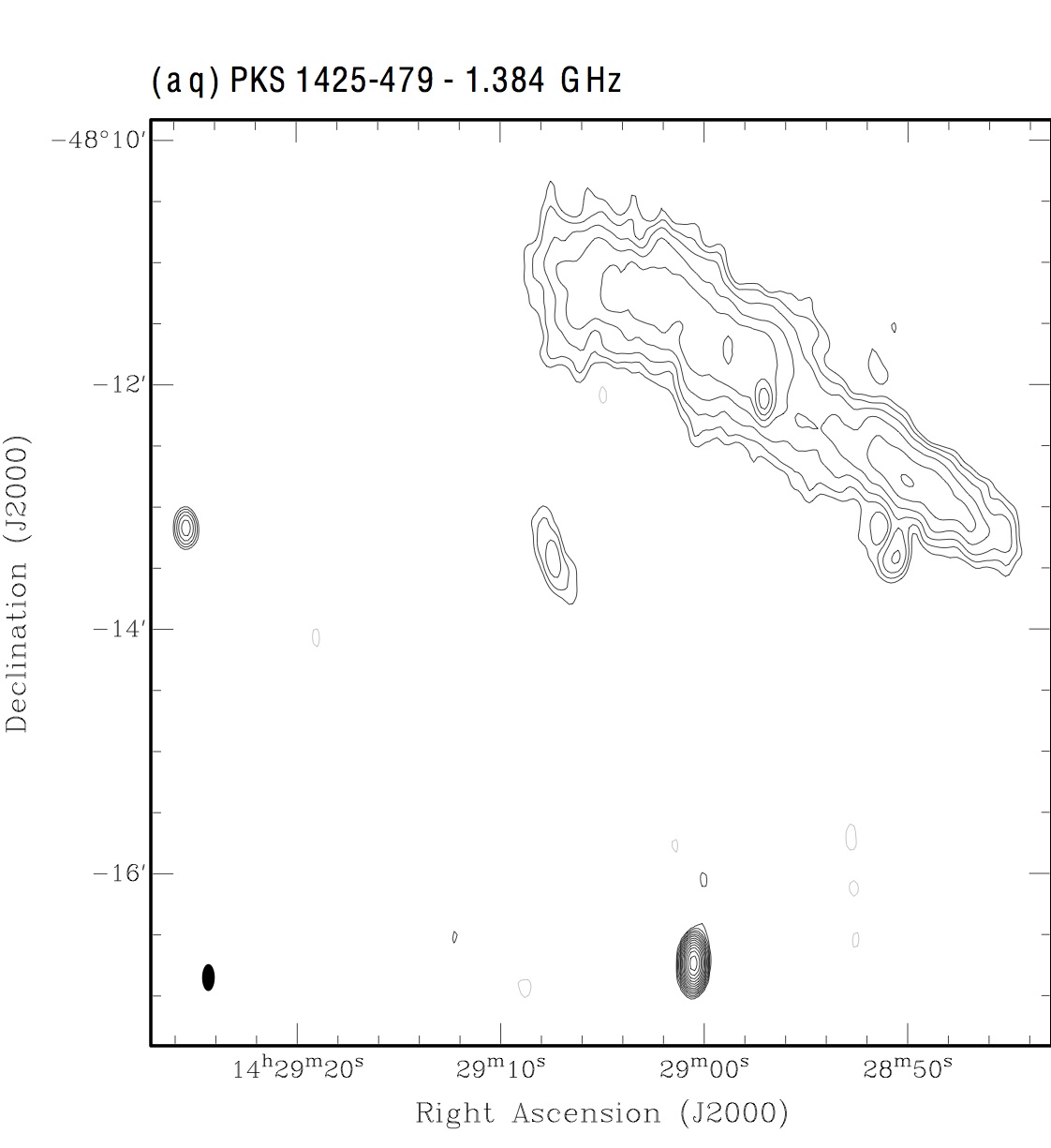} \quad
\plotone{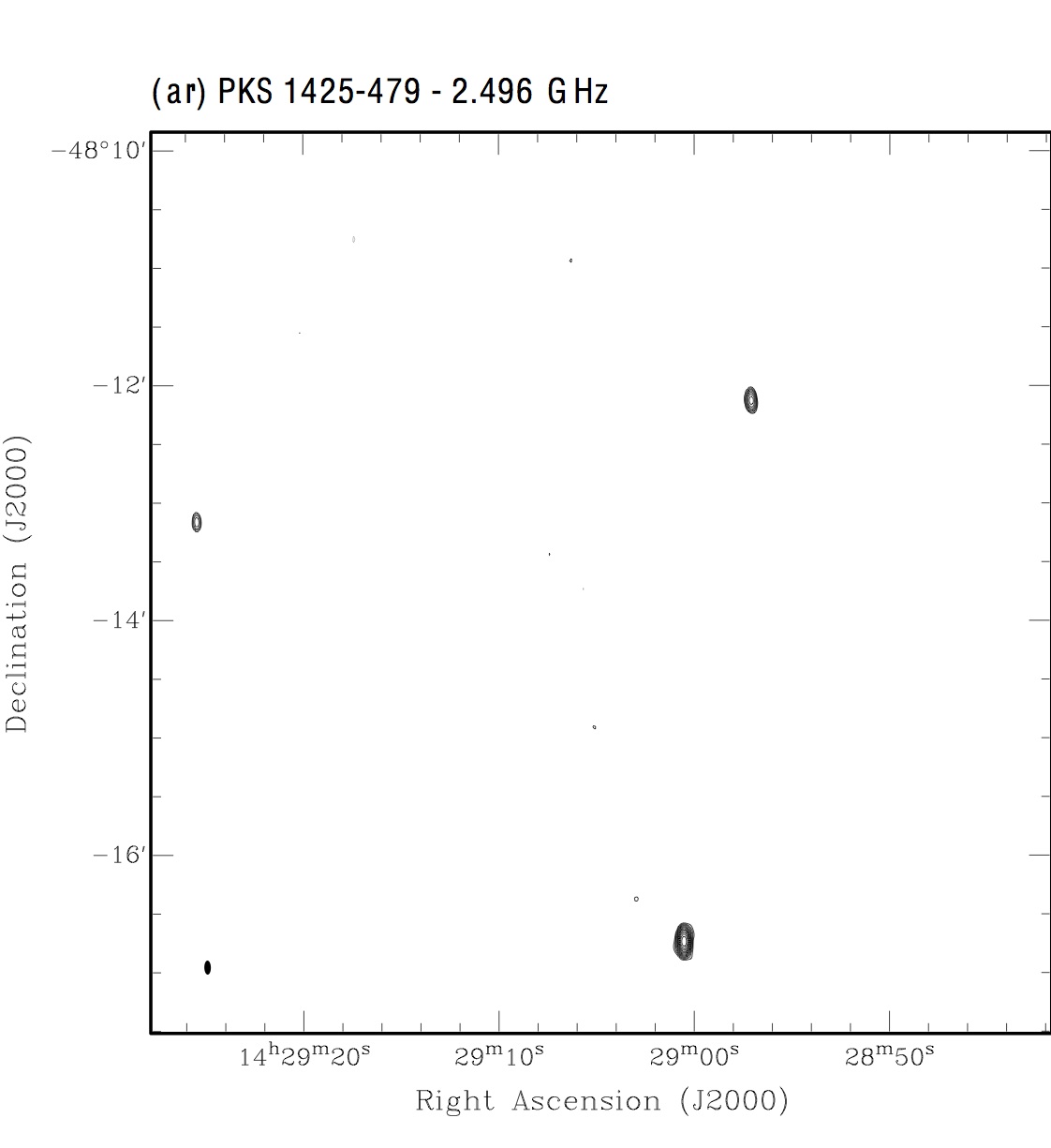}
}\\[5mm]
\mbox{
\plotone{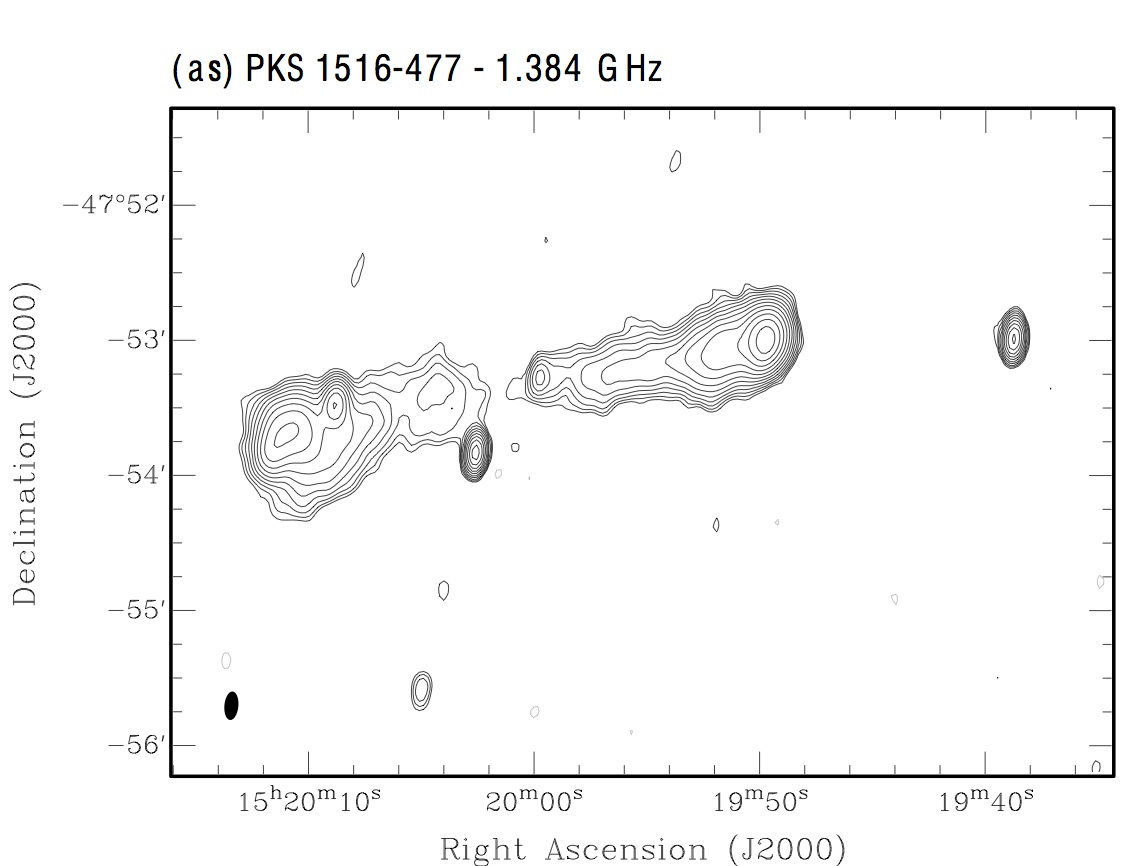} \quad
\plotone{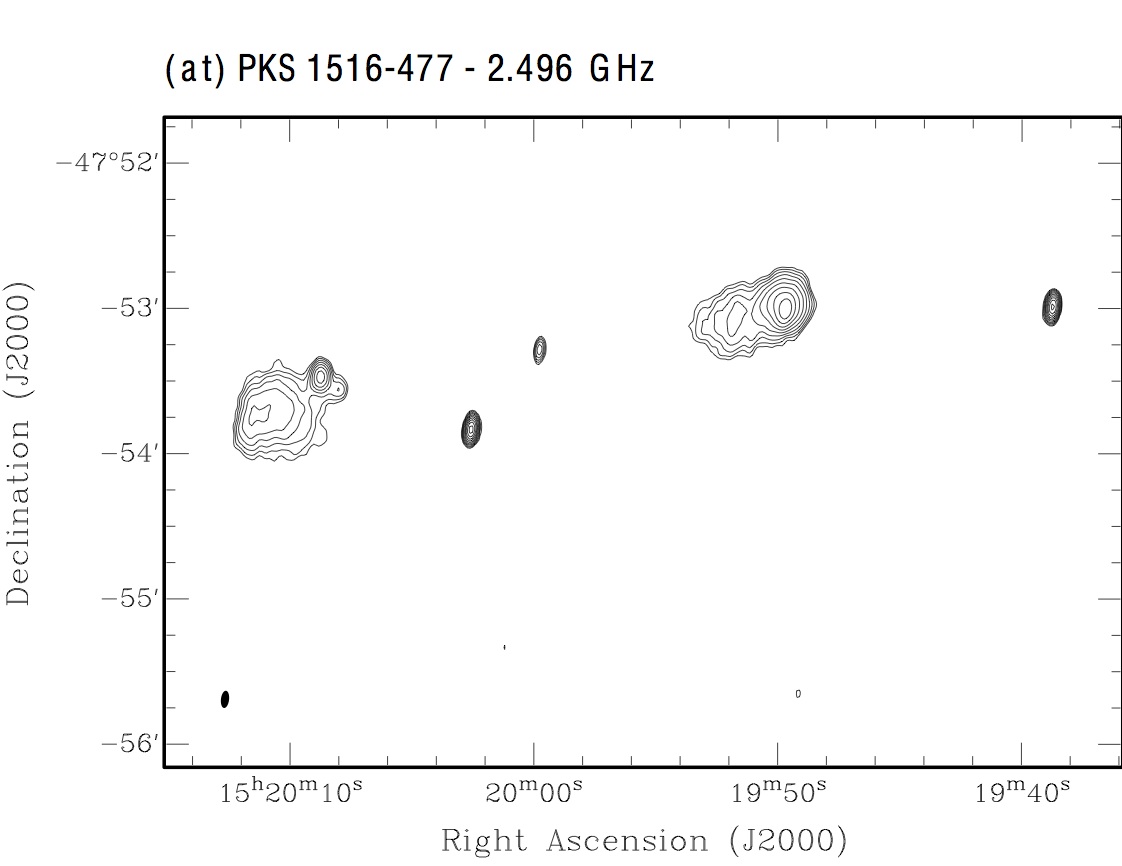}
}\\[5mm]
{Fig. 3.1. --- Continued}
\end{center}
\clearpage
\begin{center}
\mbox{
\plotone{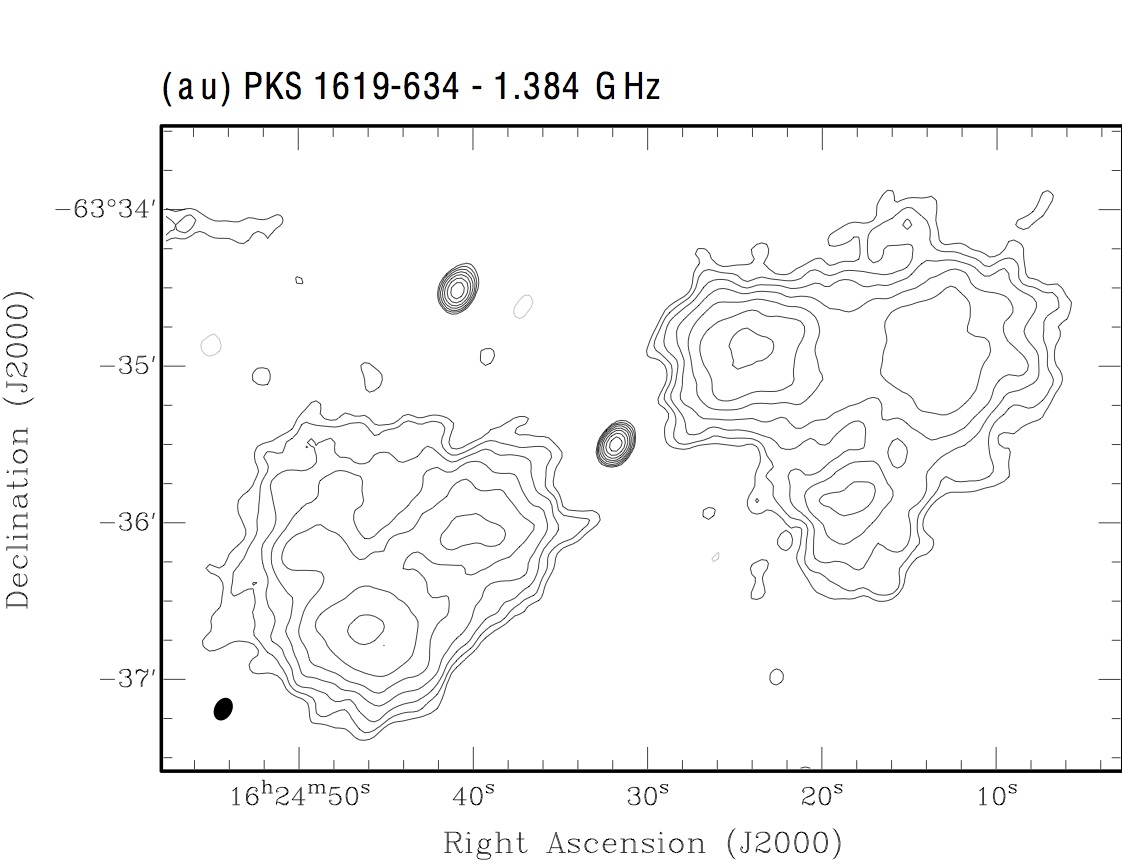} \quad
\plotone{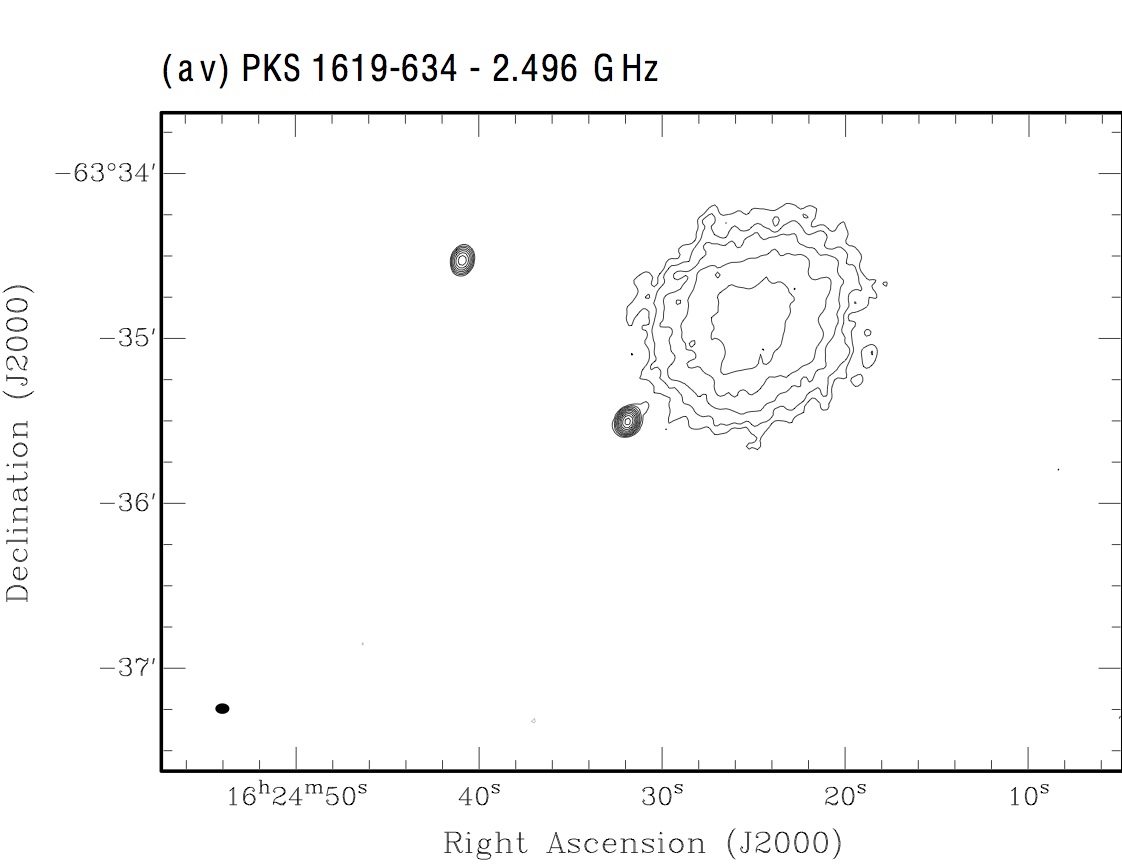}
}\\[5mm]
\mbox{
\plotone{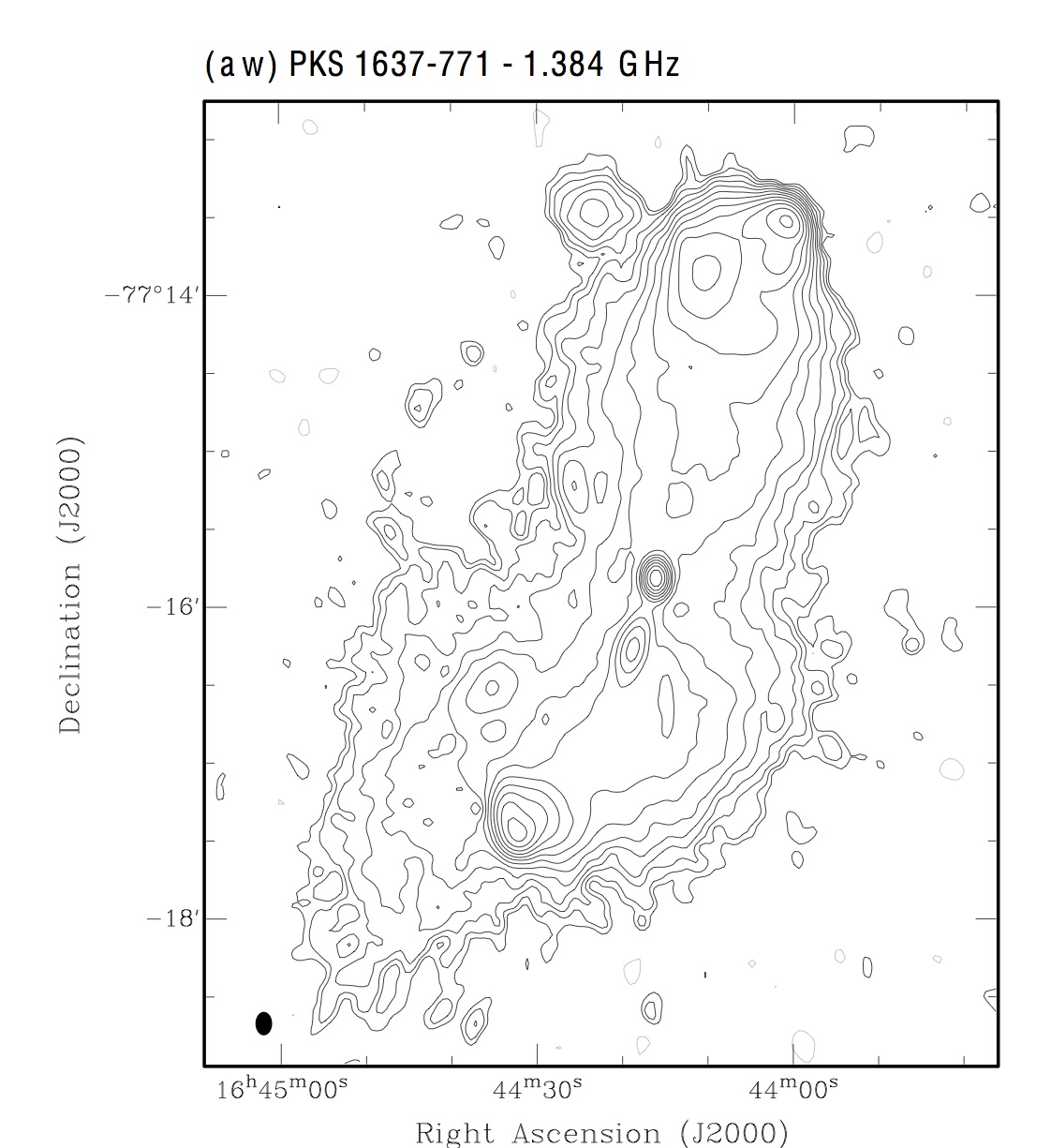} \quad
\plotone{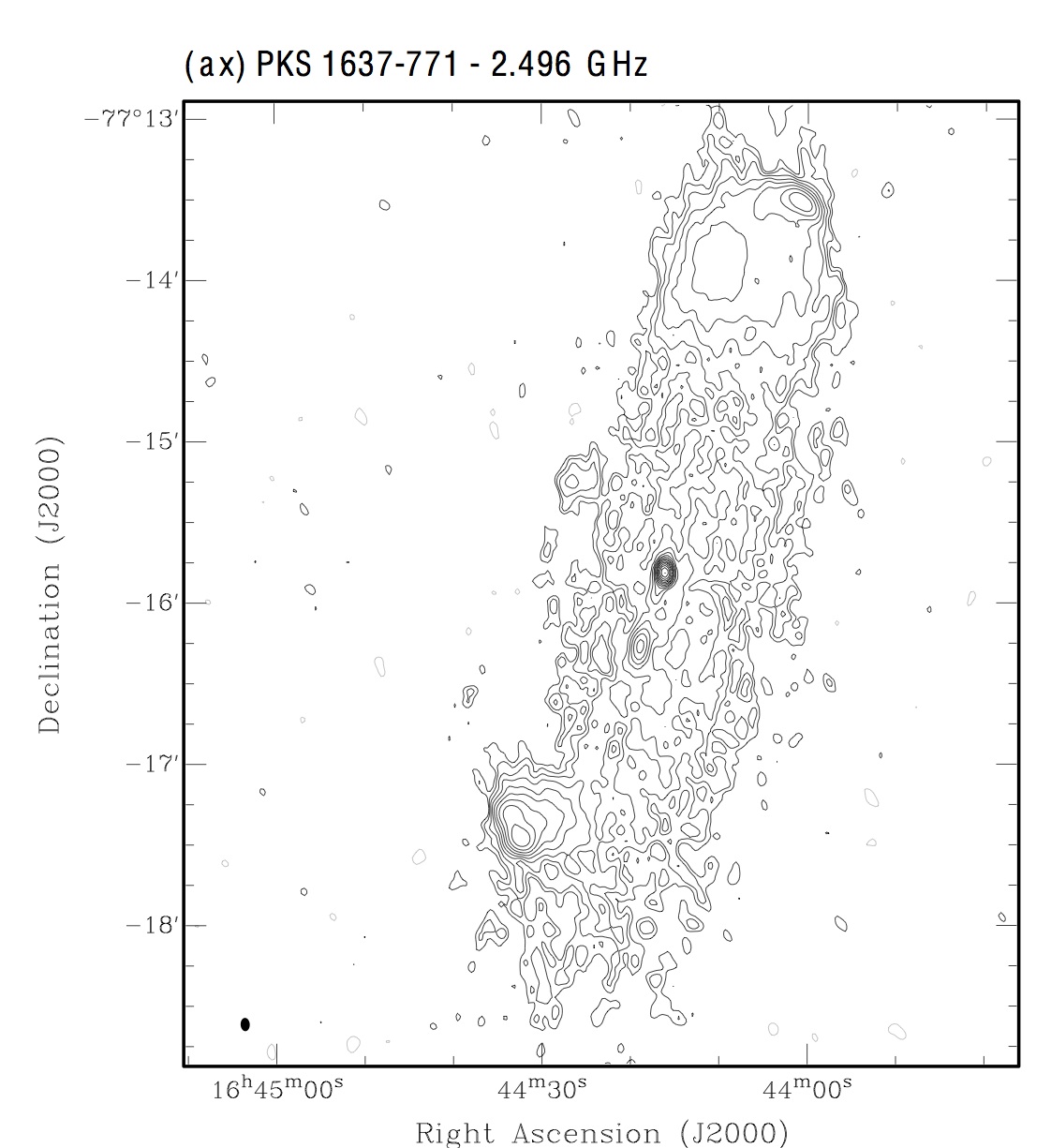}
}\\[5mm]
\mbox{
\plotone{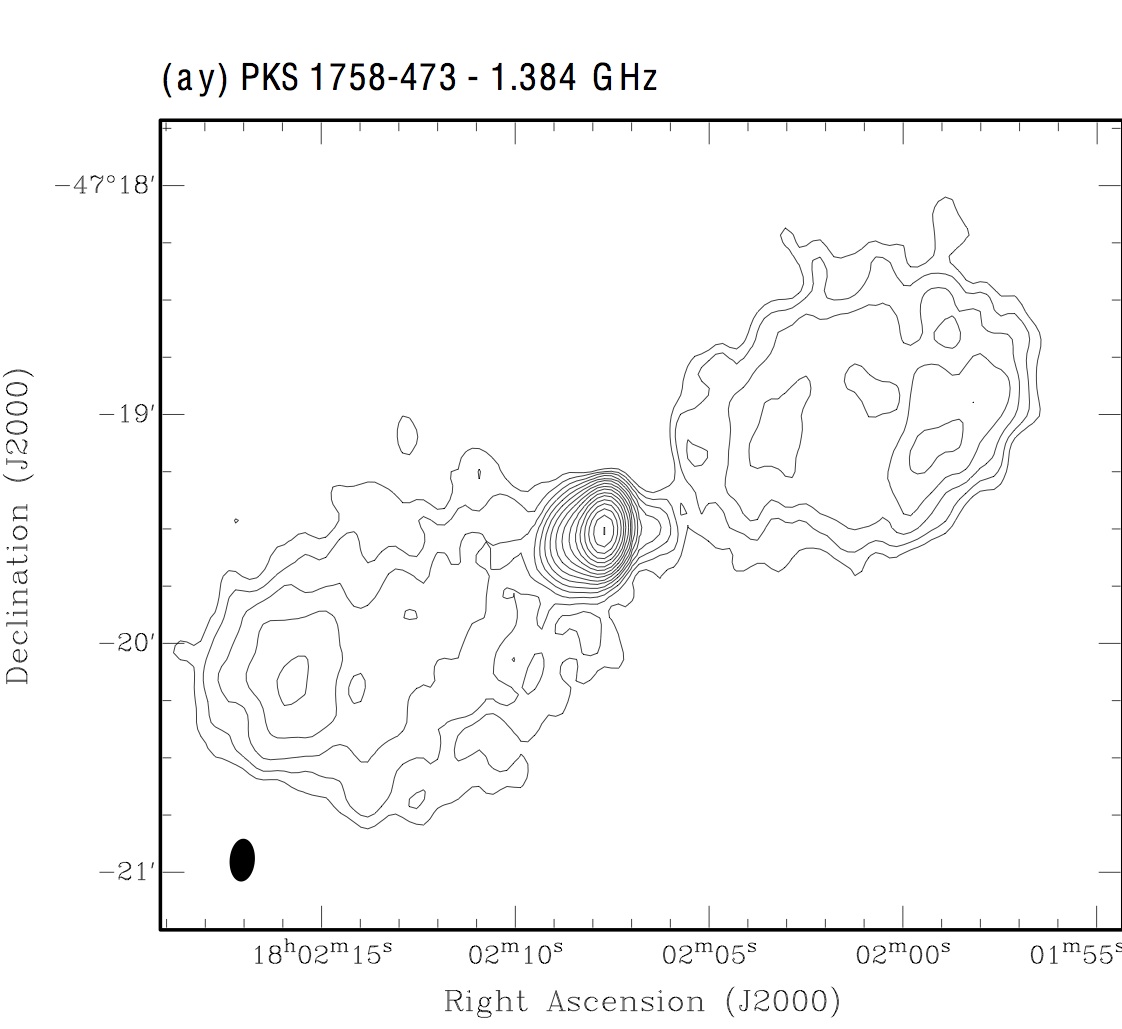} \quad
\plotone{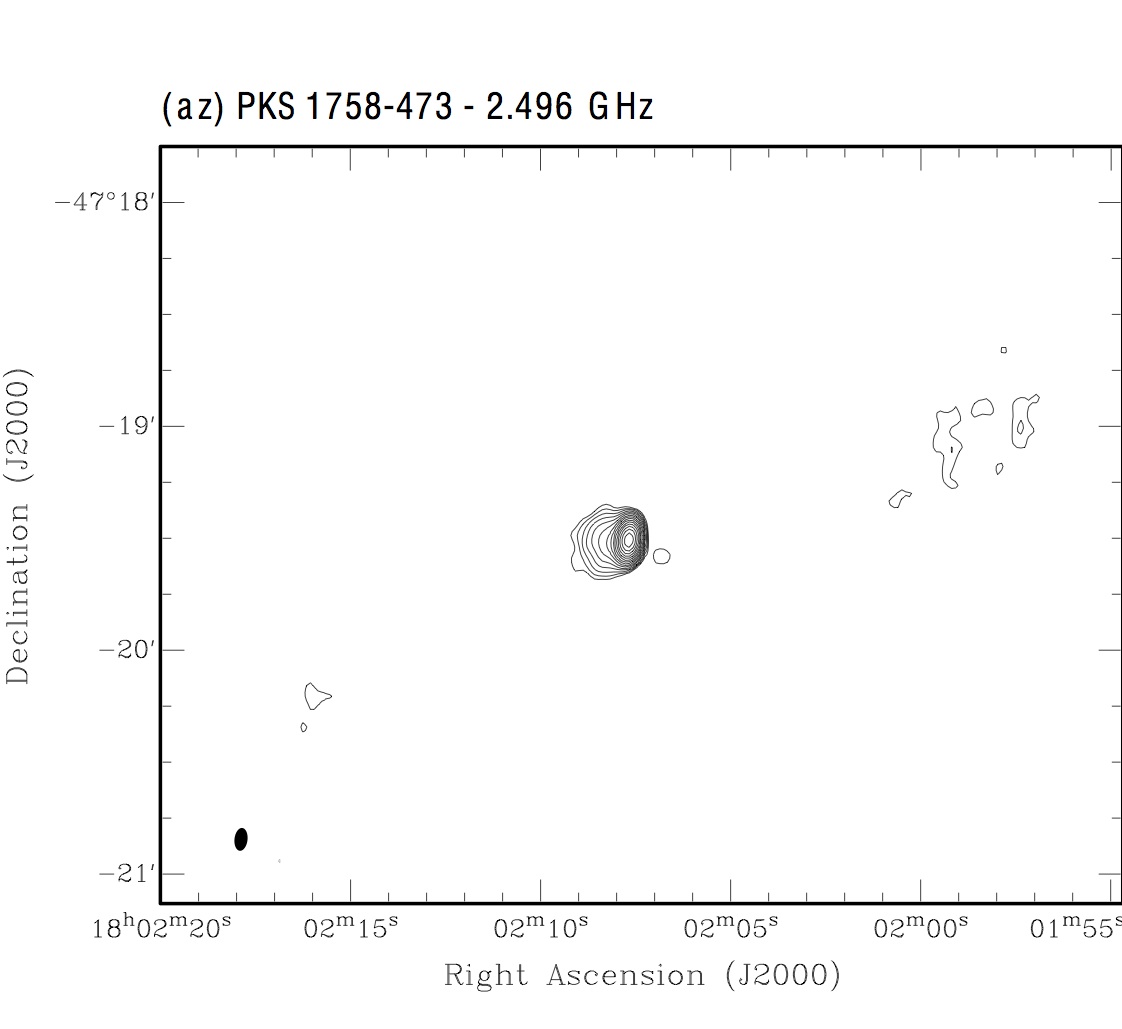}
}\\[5mm]
{Fig. 3.1. --- Continued}
\end{center}
\clearpage
\begin{center}
\mbox{
\plotone{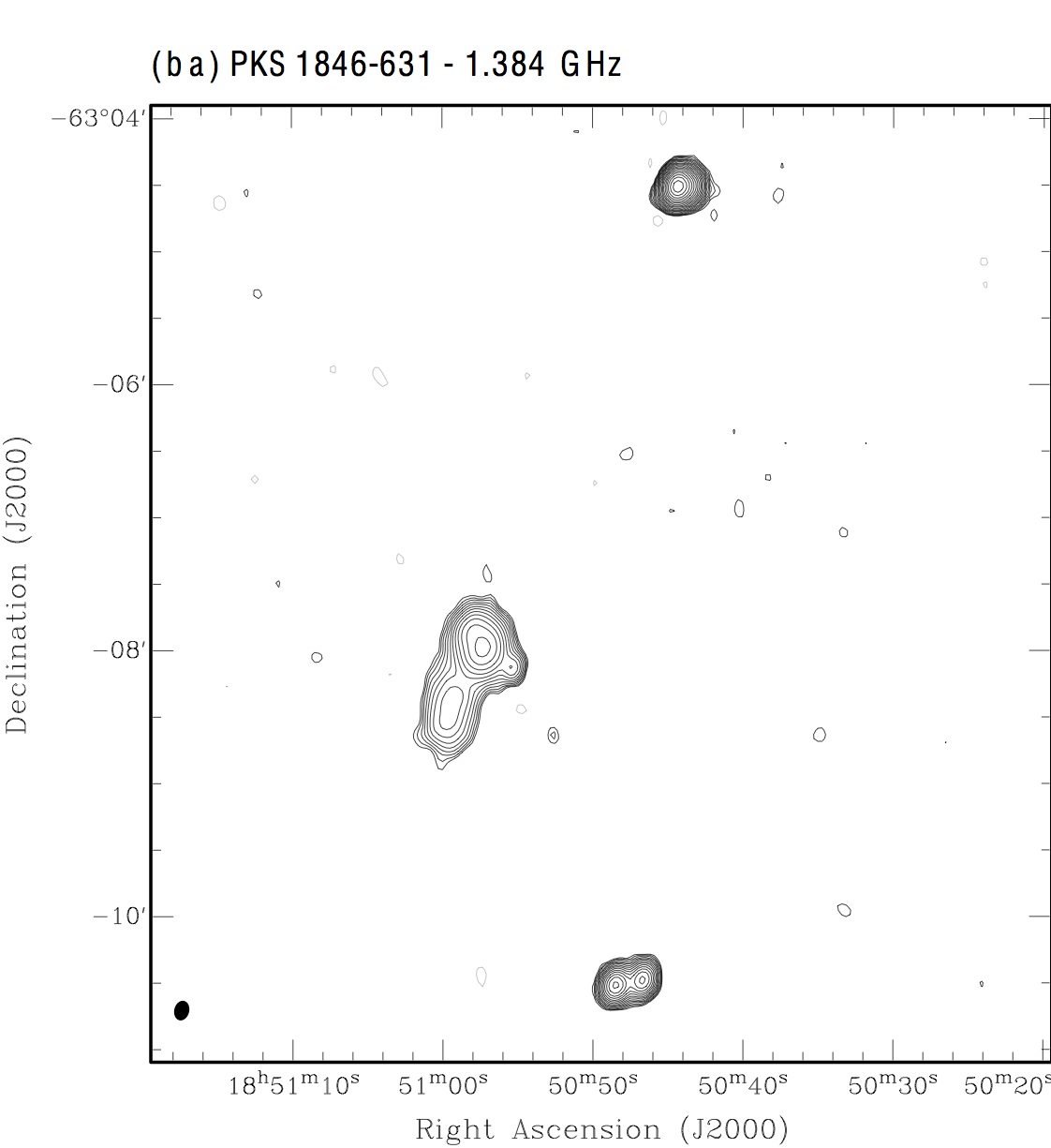} \quad
\plotone{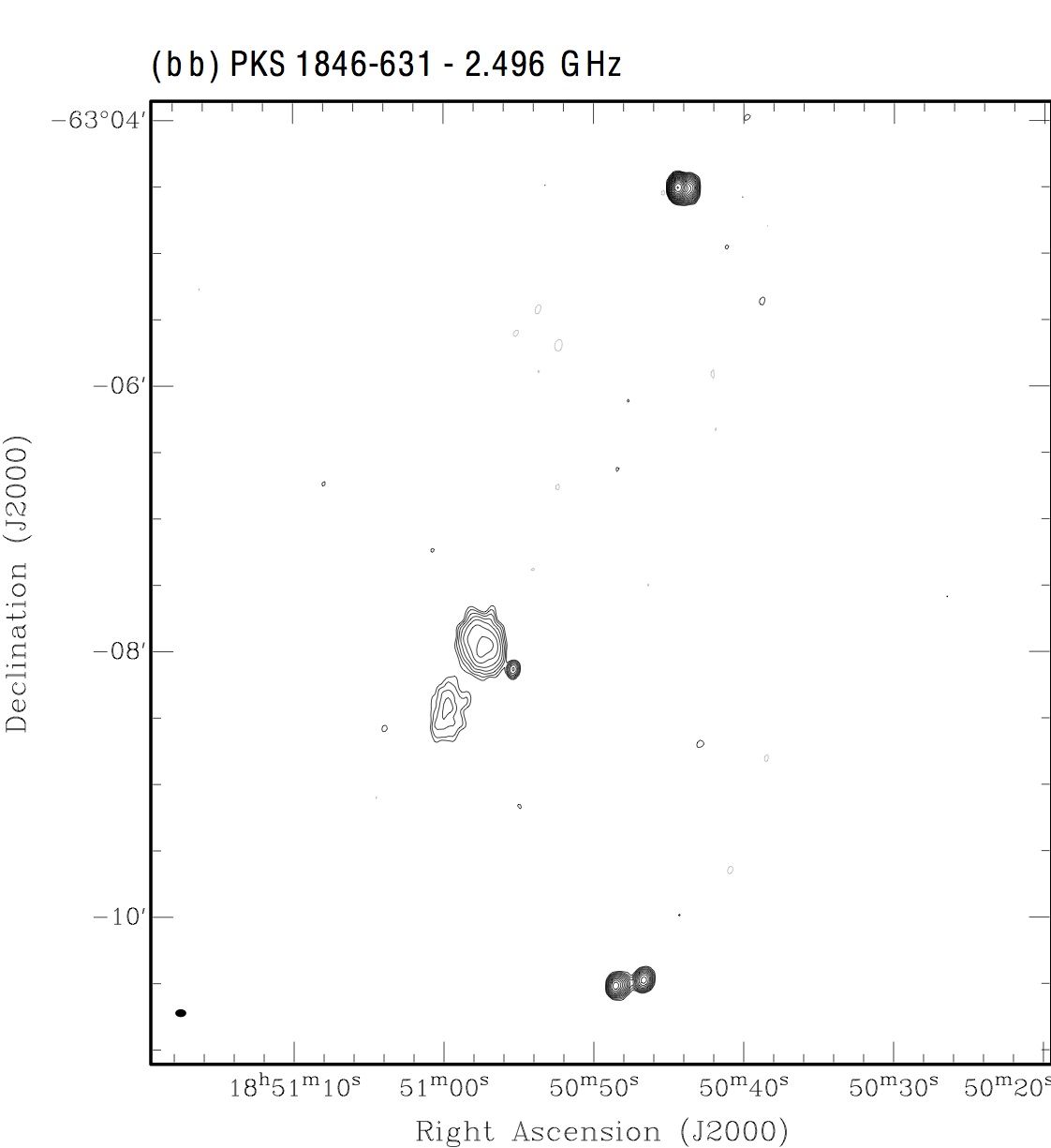}
}\\[5mm]
\mbox{
\plotone{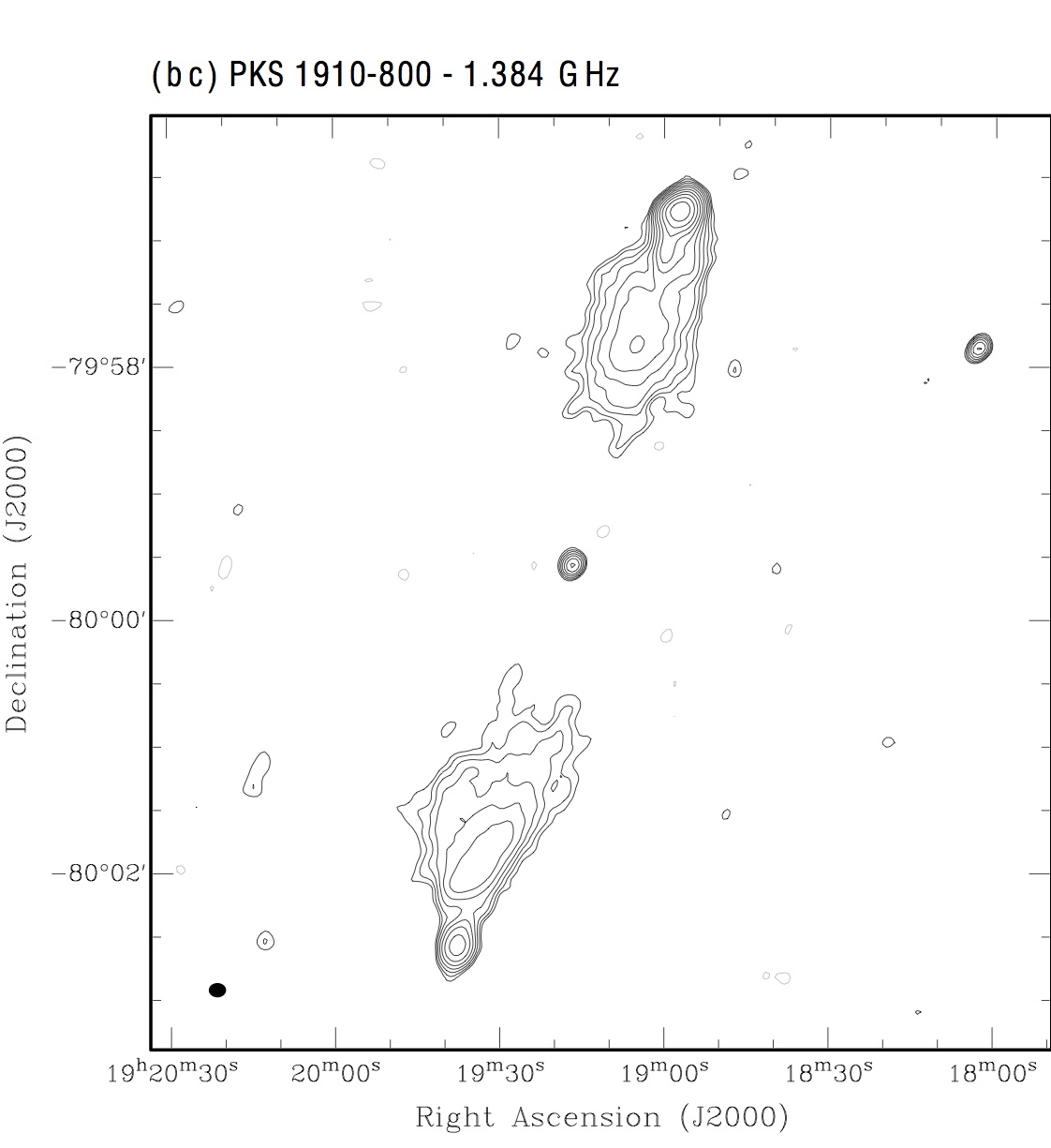} \quad
\plotone{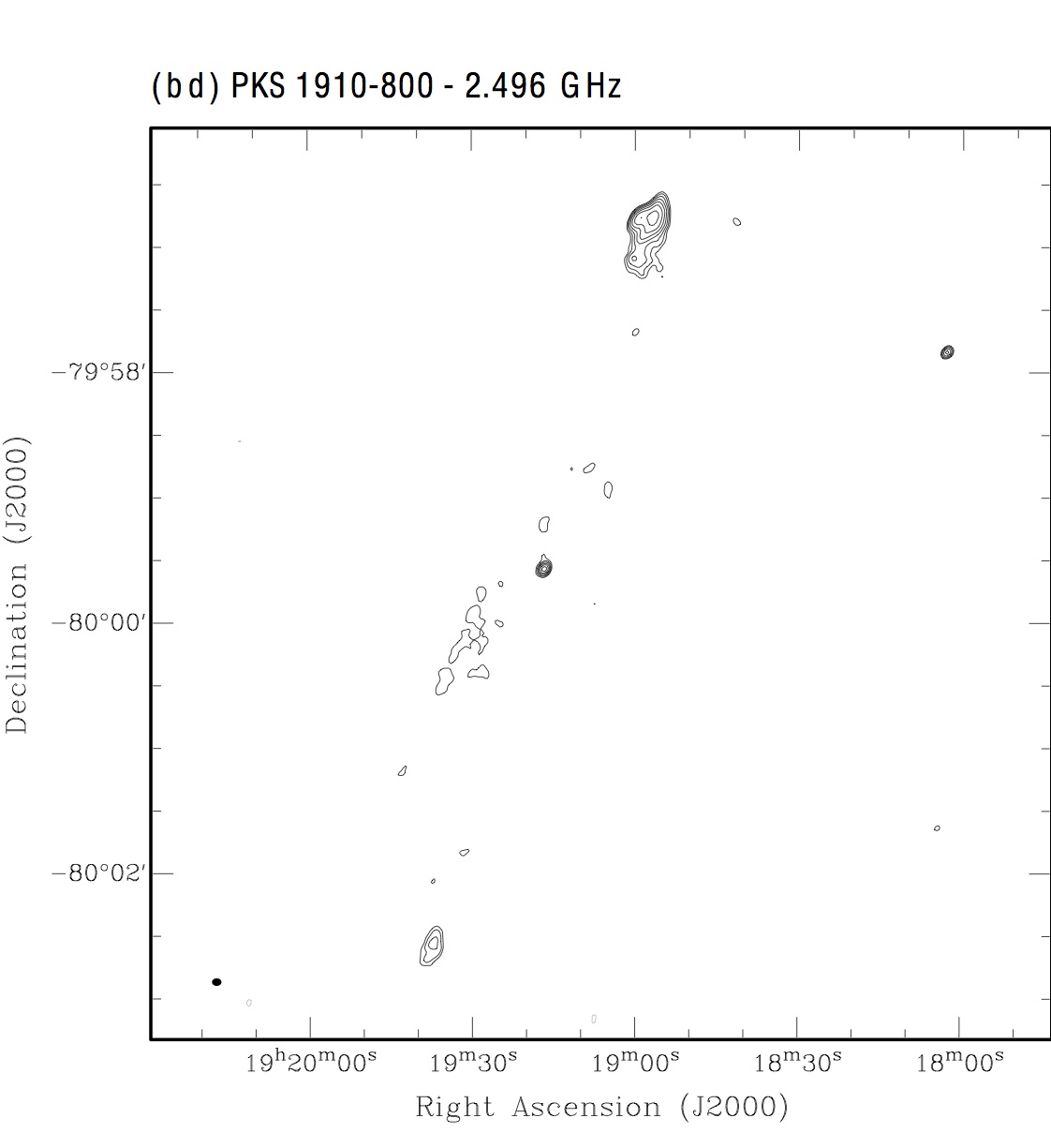}
}\\[5mm]
\mbox{
\plotone{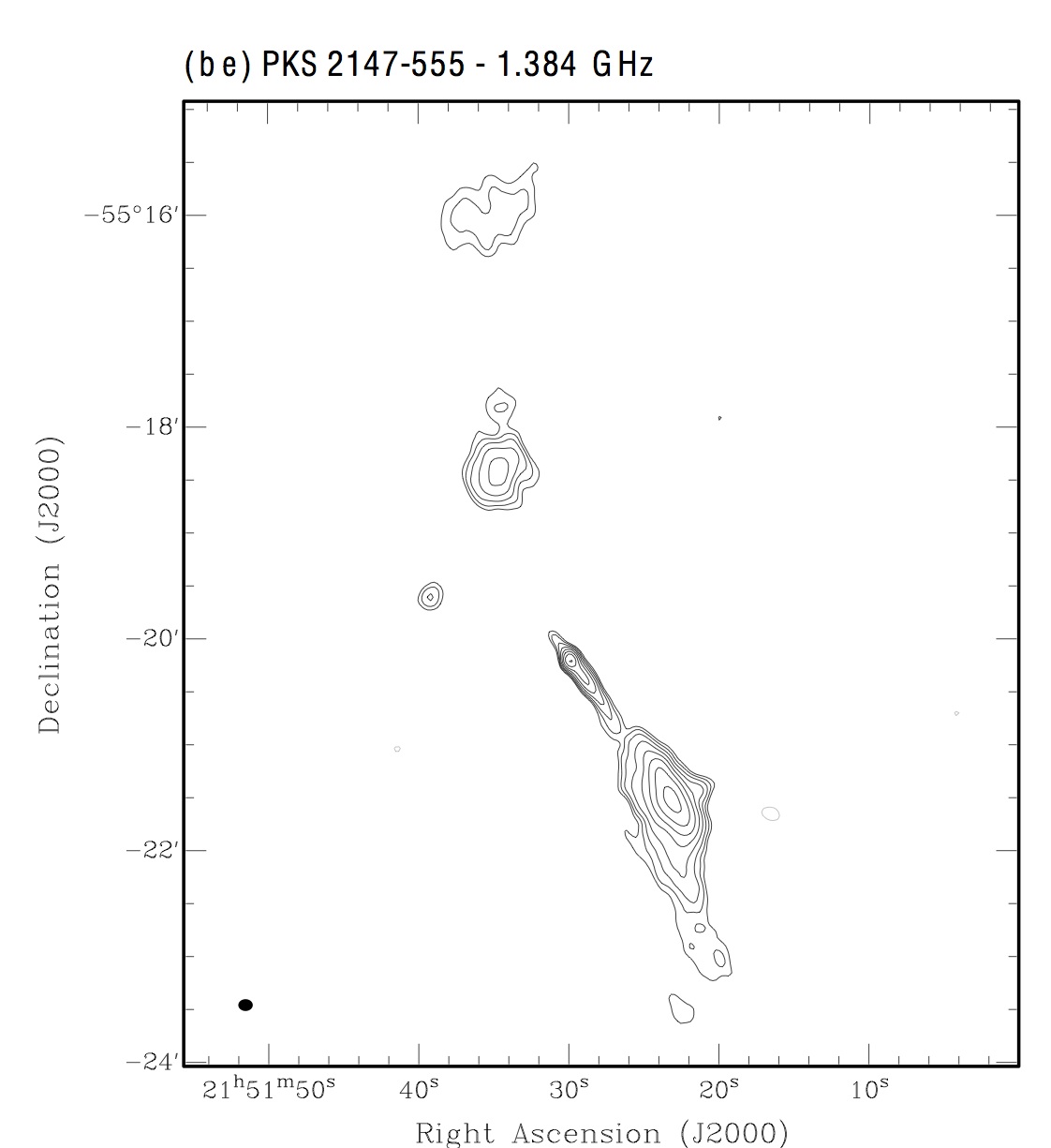} \quad
\plotone{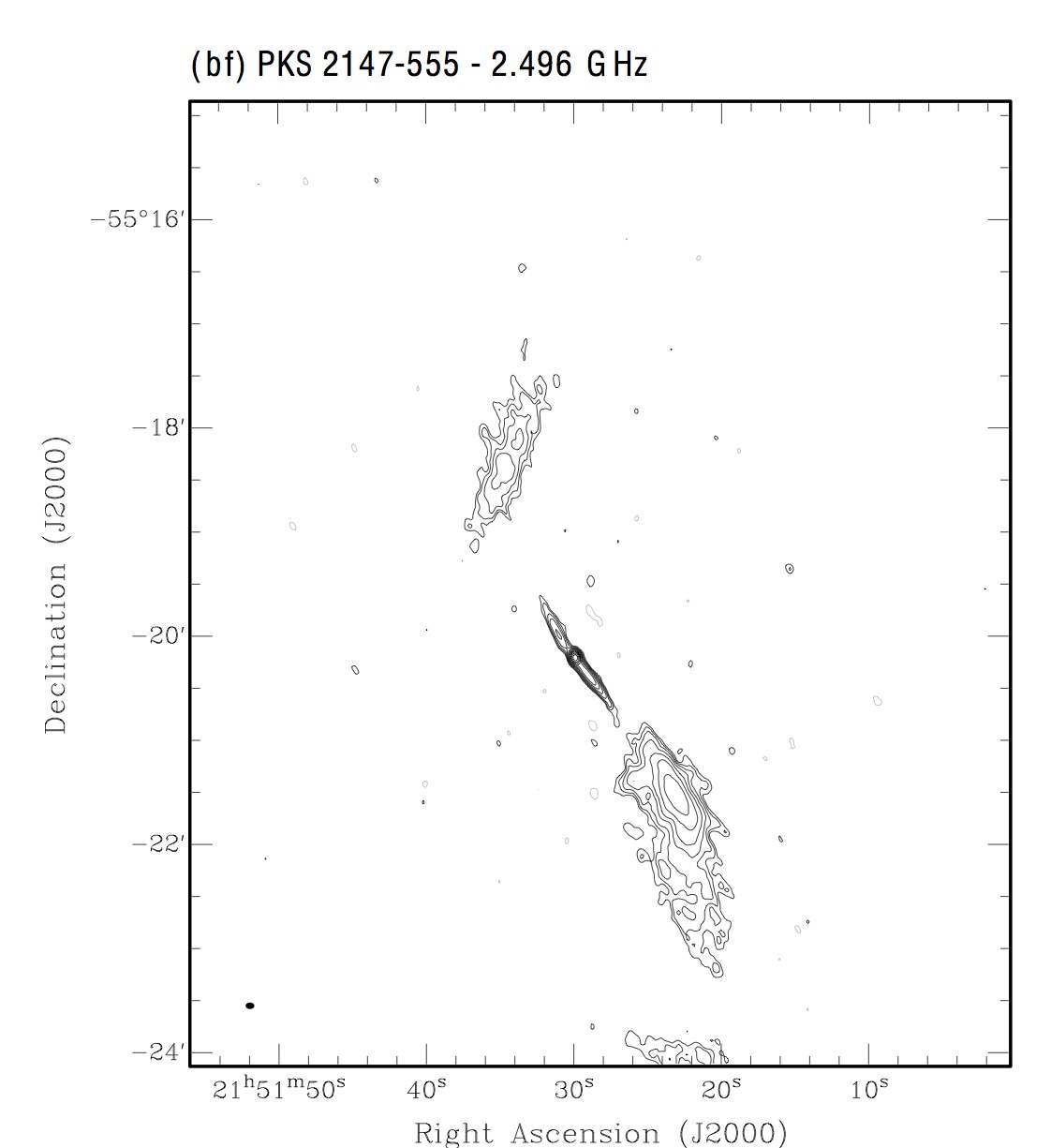}
}\\[5mm]
{Fig. 3.1. --- Continued}
\end{center}
\clearpage
\begin{center}
\mbox{
\plotone{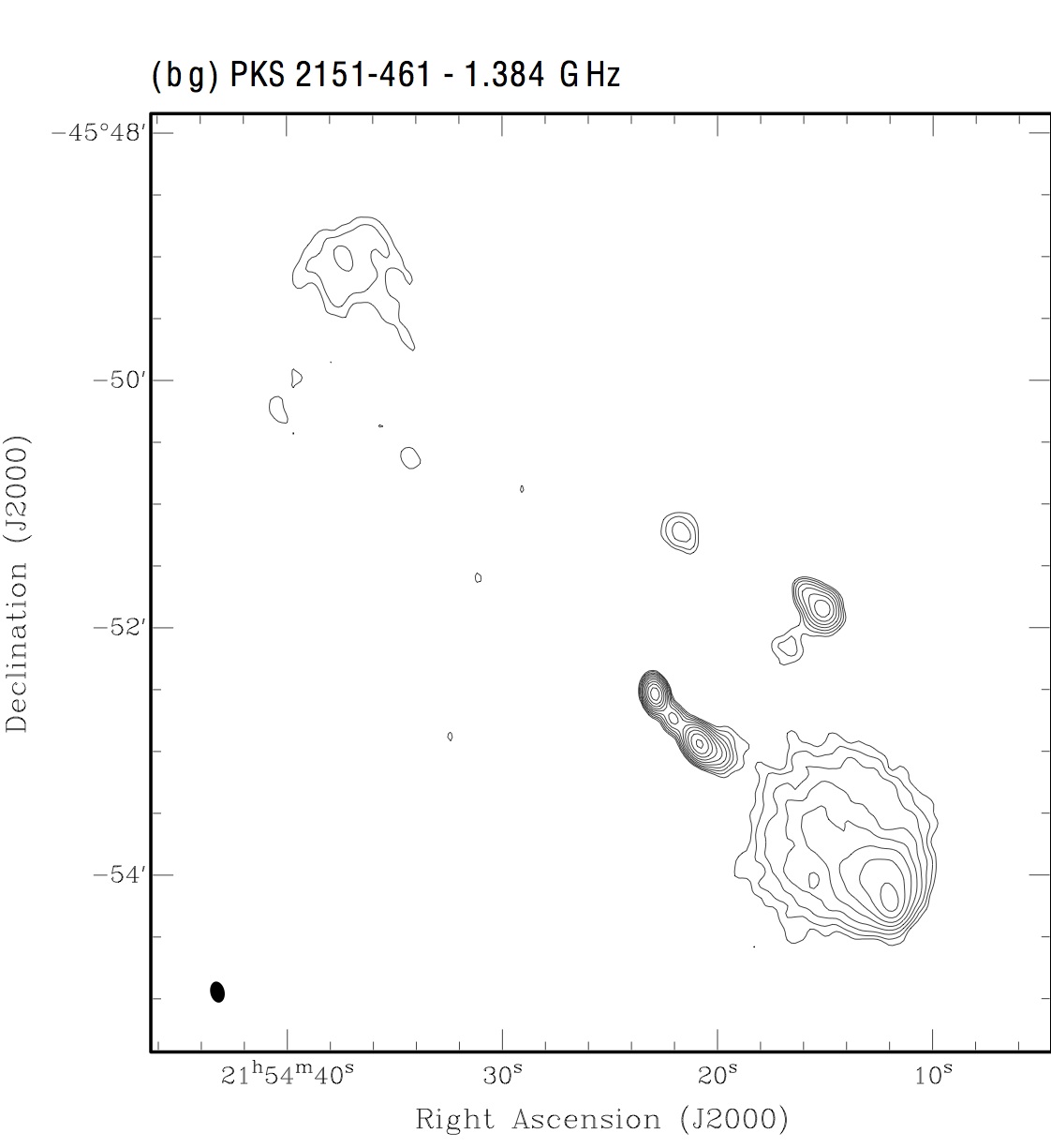} \quad
\plotone{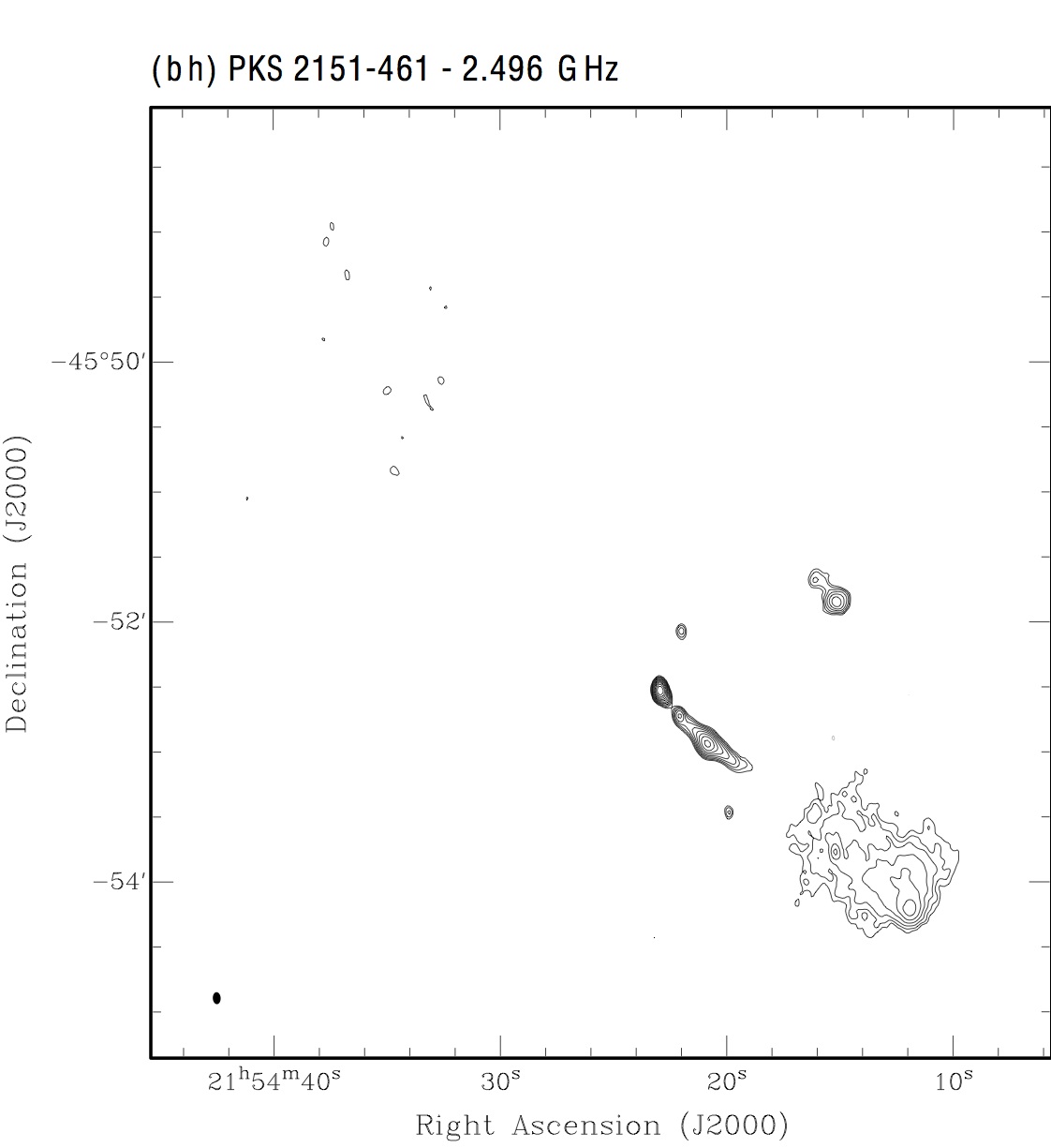}
}\\[5mm]
\mbox{
\plotone{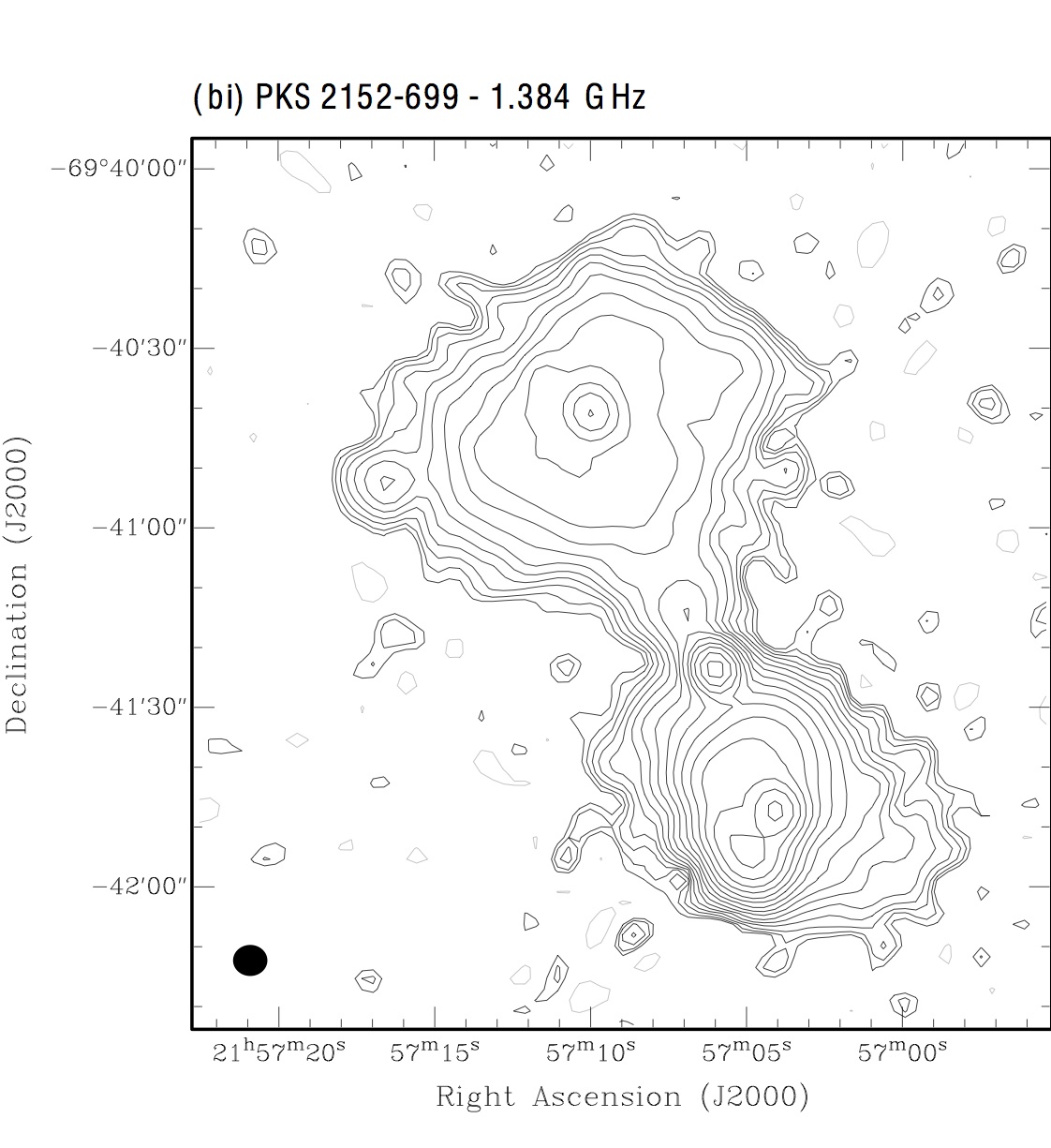} \quad
\plotone{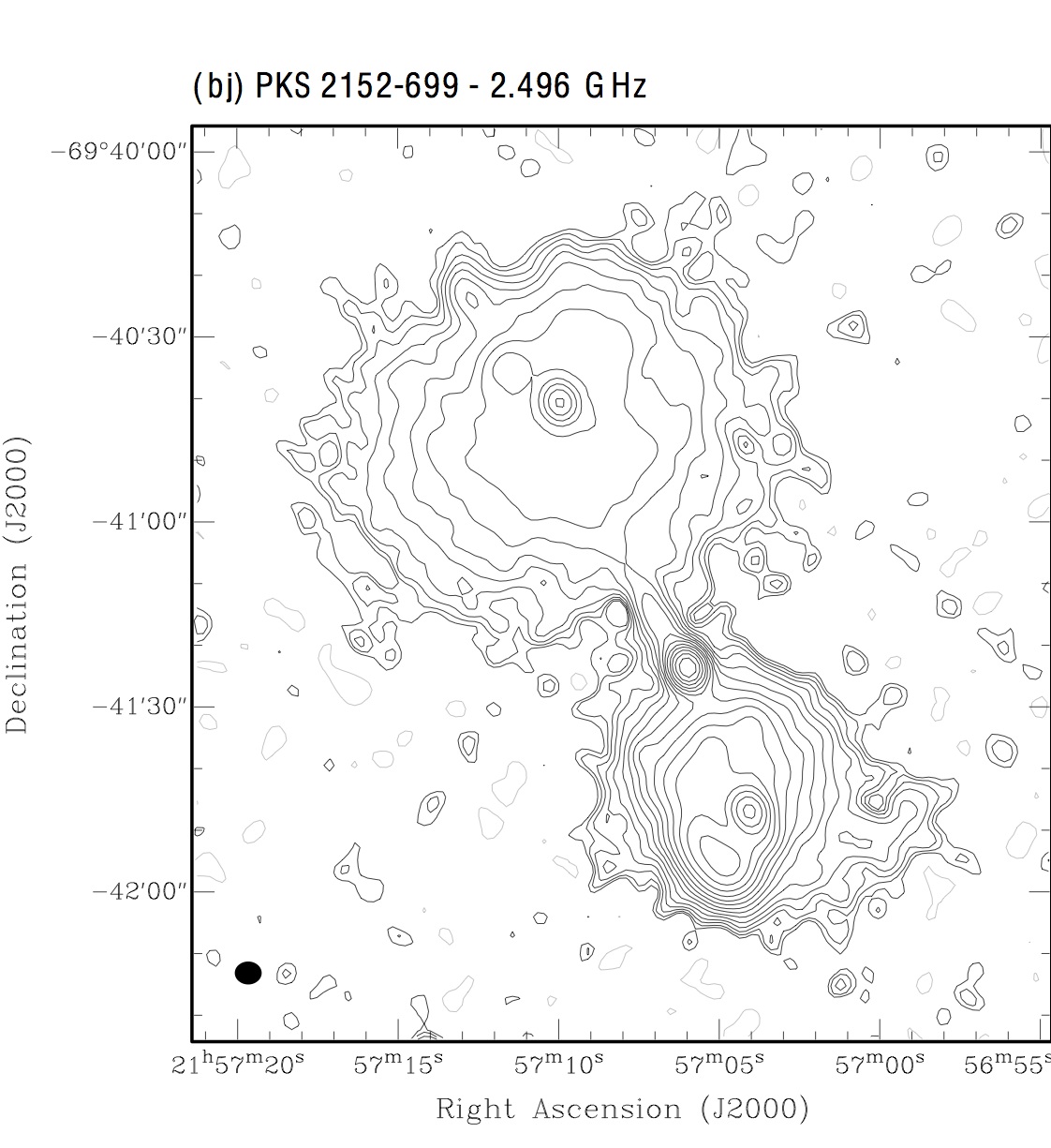}
}\\[5mm]
\mbox{
\plotone{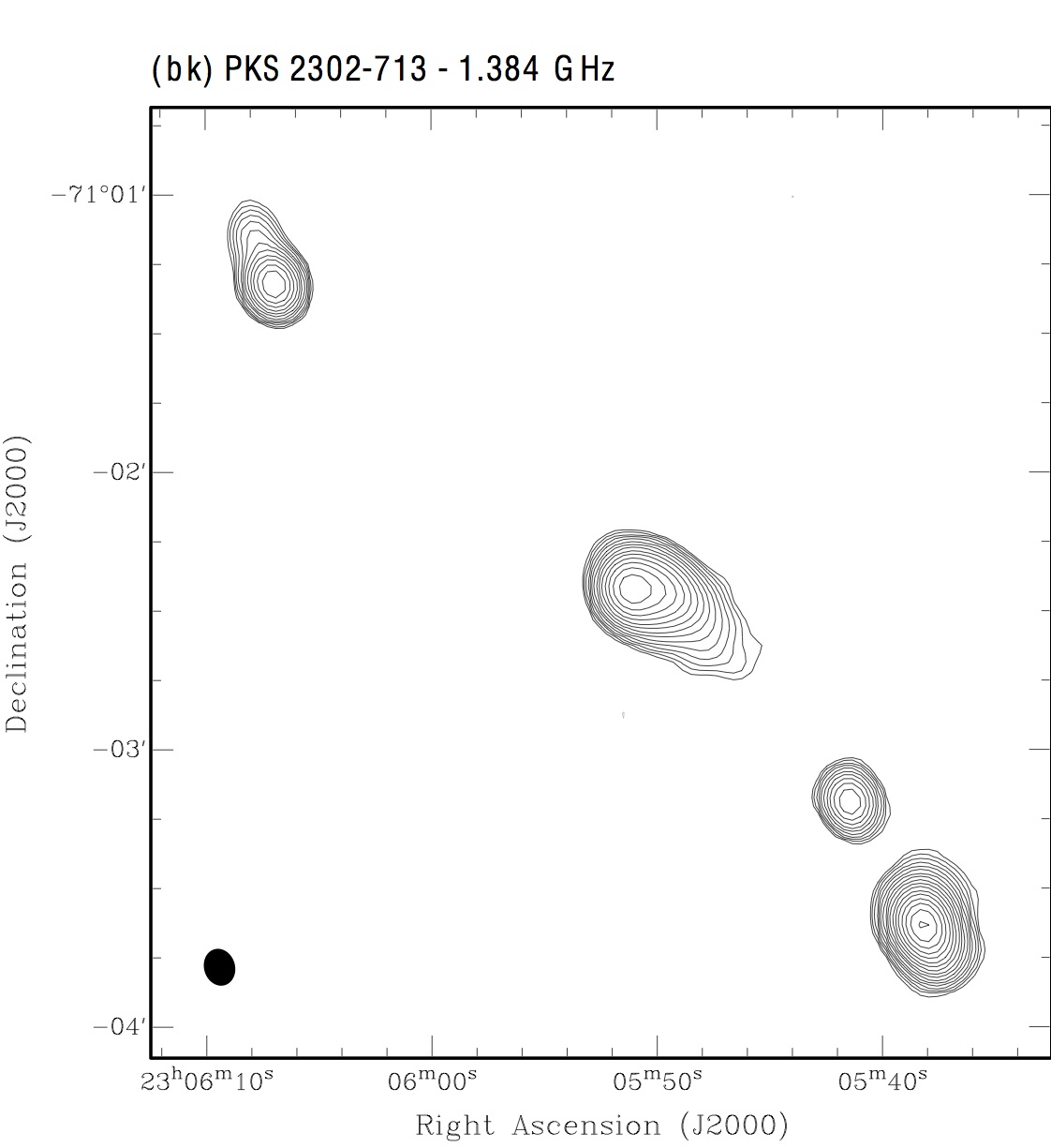} \quad
\plotone{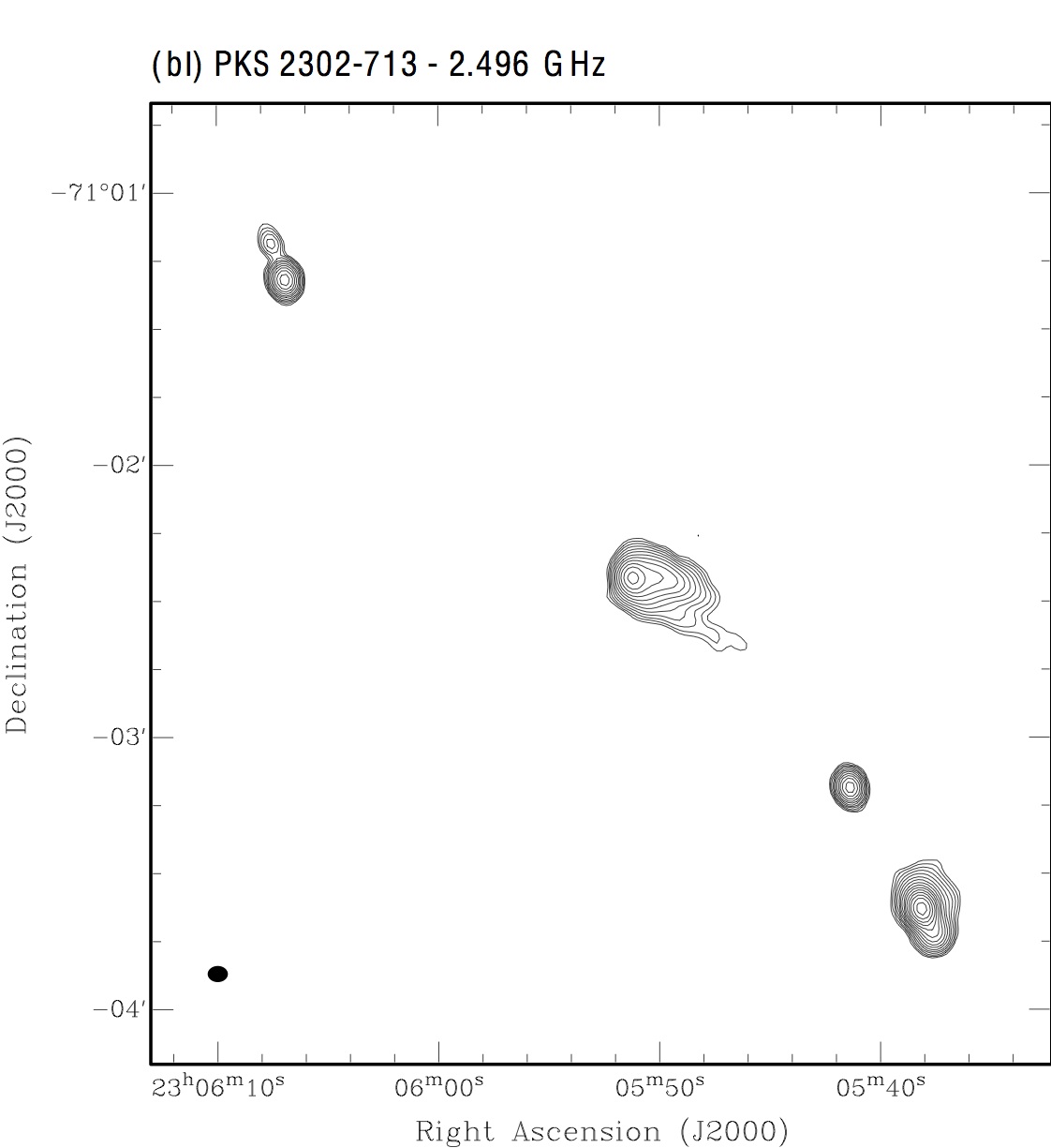}
}\\[5mm]
{Fig. 3.1. --- Continued}
\end{center}

\clearpage

\clearpage

\clearpage

\clearpage

\renewcommand{\thefootnote}{\alph{footnote}}
\scriptsize
\begin{center}
\begin{longtable}{llllccccc}
\caption[Map statistics for southern AGN survey images.]{Map statistics for southern AGN survey images.}
\label{tab:tabselAGN}
 \\
\hline \hline \\[-2ex]
   Figure & Source & z & $\nu$ & Beam     & Beam PA   & $\sigma$          & $S_{P}$         & $S_{I}$ \\
          &        &   & (GHz)     & (arcsec) & (deg)     & (mJy beam$^{-1}$) & (Jy beam$^{-1}$) & (Jy)           \\[0.5ex] \hline
   \\[-1.8ex]
\endfirsthead

\multicolumn{9}{c}{{\tablename} \thetable{} -- Continued} \\[0.5ex]
\hline \hline \\[-2ex]
Figure & Source & z & $\nu$ & Beam     & Beam PA   & $\sigma$          & $S_{P}$         & $S_{I}$ \\
       &        &   & (GHz)     & (arcsec)    & (deg)     & (mJy beam$^{-1}$) & (Jy beam$^{-1}$) & (Jy)           \\[0.5ex] \hline
   \\[-1.8ex]
\endhead

\multicolumn{9}{l}{{Continued on Next Page\ldots}} \\
\endfoot

\\[-1.8ex] \hline \hline
\endlastfoot

\ref{fig:figself3}(a)   & PKS $0003-833$ & 0.32      & 1.4  &  $7.7\times6.5$  & 28  & 0.527 & 0.588   & 2.20   \\ 
\ref{fig:figself3}(b)   & PKS $0003-833$ & 0.32      & 2.5  &  $3.9\times3.5$  & 33  & 0.329 & 0.147   & 1.30   \\
\ref{fig:figself3}(c)   & PKS $0020-747$ & \nodata   & 1.4  &  $7.7\times6.7$  & 59  & 0.145 & 0.0626  & 0.753  \\ 
\ref{fig:figself3}(d)   & PKS $0020-747$ & \nodata   & 2.5  &  $4.3\times3.6$  & 56  & 0.150 & 0.0228  & 0.380  \\
\ref{fig:figself3}(e)   & PKS $0043-424$ & 0.116     & 1.4  & $11.3\times6.7$  & 14  & 1.403 & 2.27    & 7.94   \\ 
\ref{fig:figself3}(f)   & PKS $0043-424$ & 0.116     & 2.5  &  $6.1\times3.7$  & 14  & 0.930 & 1.06    & 5.20   \\
\ref{fig:figself3}(g)   & PKS $0101-649$ & \nodata   & 1.4  &  $8.6\times6.6$  & 33  & 0.240 & 0.153   & 0.725  \\ 
\ref{fig:figself3}(h)   & PKS $0101-649$ & \nodata   & 2.5  &  $4.6\times3.5$  & 31  & 0.255 & 0.171   & 0.510  \\
\ref{fig:figself3}(i)   & PKS $0110-692$ & \nodata   & 1.4  &  $8.5\times6.5$  & 33  & 0.269 & 0.854   & 1.88   \\ 
\ref{fig:figself3}(j)   & PKS $0110-692$ & \nodata   & 2.5  &  $4.6\times3.5$  & 35  & 0.207 & 0.360   & 1.09   \\
\ref{fig:figself3}(k)   & PKS $0130-620$ & \nodata   & 1.4  &  $9.6\times6.4$  & 33  & 0.193 & 0.0841  & 1.02   \\ 
\ref{fig:figself3}(l)   & PKS $0130-620$ & \nodata   & 2.5  &  $5.1\times3.4$  & 33  & 0.160 & 0.0273  & 0.650  \\
\ref{fig:figself3}(m)   & PKS $0153-826$ & \nodata   & 1.4  &  $8.4\times6.2$  & 46  & 0.153 & 0.0917  & 0.545  \\ 
\ref{fig:figself3}(n)   & PKS $0153-826$ & \nodata   & 2.5  &  $4.6\times3.3$  & 44  & 0.167 & 0.0352  & 0.326  \\
\ref{fig:figself3}(o)   & PKS $0218-452$ & 0.162145  & 1.4  & $11.3\times6.5$  & 15  & 0.165 & 0.0543  & 0.474  \\ 
\ref{fig:figself3}(p)   & PKS $0218-452$ & 0.162145  & 2.5  &  $6.1\times3.4$  & 13  & 0.162 & 0.0407  & 0.243  \\
\ref{fig:figself3}(q)   & PKS $0258-520$ & 0.07299   & 1.4  & $11.7\times6.0$  & 21  & 0.189 & 0.0194  & 0.517  \\ 
\ref{fig:figself3}(r)   & PKS $0258-520$ & 0.07299   & 2.5  &  $5.7\times3.3$  & 19  & 0.123 & 0.00895 & 0.587  \\
\ref{fig:figself3}(s)   & PKS $0344-345$ & 0.053804  & 1.4  & $15.9\times6.1$  & 9   & 0.350 & 0.0790  & 3.30   \\ 
\ref{fig:figself3}(t)   & PKS $0344-345$ & 0.053804  & 2.5  &  $8.8\times3.3$  & 9   & 0.232 & 0.0528  & 1.97   \\
\ref{fig:figself3}(u)   & PKS $0424-728$ & 0.1921    & 1.4  &  $8.7\times6.2$  & 29  & 0.200 & 0.352   & 1.70   \\ 
\ref{fig:figself3}(v)   & PKS $0424-728$ & 0.1921    & 2.5  &  $4.7\times3.3$  & 33  & 0.188 & 0.197   & 1.13   \\
\ref{fig:figself3}(w)   & PKS $0427-539$ & 0.0412    & 1.4  & $10.5\times6.5$  & 34  & 0.377 & 0.176   & 5.46   \\ 
\ref{fig:figself3}(x)   & PKS $0427-539$ & 0.0412    & 2.5  &  $5.3\times3.2$  & 32  & 0.331 & 0.0781  & 3.63   \\
\ref{fig:figself3}(y)   & PKS $0429-616$ & \nodata   & 1.4  & $11.1\times6.1$  & 40  & 0.247 & 0.213   & 1.94   \\ 
\ref{fig:figself3}(z)   & PKS $0429-616$ & \nodata   & 2.5  &  $5.9\times3.3$  & 40  & 0.332 & 0.180   & 0.919  \\
\ref{fig:figself3}(aa)  & PKS $0507-627$ & \nodata   & 1.4  & $11.4\times6.0$  & 39  & 0.368 & 0.465   & 1.19   \\ 
\ref{fig:figself3}(ab)  & PKS $0507-627$ & \nodata   & 2.5  &  $5.4\times3.3$  & 44  & 0.439 & 0.204   & 0.674  \\
\ref{fig:figself3}(ac)  & PKS $0511-484$ & 0.3063    & 1.4  & $13.5\times6.2$  & 32  & 0.685 & 0.628   & 2.72   \\ 
\ref{fig:figself3}(ad)  & PKS $0511-484$ & 0.3063    & 2.5  &  $8.4\times3.2$  & 29  & 0.372 & 0.203   & 1.62   \\
\ref{fig:figself3}(ae)  & PKS $0540-617$ & \nodata   & 1.4  & $13.1\times6.0$  & 52  & 0.291 & 0.0864  & 0.818  \\ 
\ref{fig:figself3}(af)  & PKS $0540-617$ & \nodata   & 2.5  &  $6.2\times3.3$  & 52  & 0.222 & 0.0432  & 0.523  \\
\ref{fig:figself3}(ag)  & PKS $0625-545$ & 0.051742  & 1.4  & $14.9\times6.1$  & 36  & 0.533 & 0.277   & 3.09   \\ 
\ref{fig:figself3}(ah)  & PKS $0625-545$ & 0.051742  & 2.5  &  $8.0\times3.3$  & 35  & 0.439 & 0.0911  & 1.03   \\
\ref{fig:figself3}(ai)  & PKS $0703-595$ & \nodata   & 1.4  & $11.0\times6.2$  & 32  & 0.174 & 0.0185  & 0.446  \\ 
\ref{fig:figself3}(aj)  & PKS $0703-595$ & \nodata   & 2.5  &  $6.2\times3.4$  & 32  & 0.162 & 0.0181  & 0.305  \\
\ref{fig:figself3}(ak)  & PKS $1259-445$ & 0.6       & 1.4  & $14.0\times6.3$  & 14  & 0.256 & 0.602   & 1.62   \\ 
\ref{fig:figself3}(al)  & PKS $1259-445$ & 0.6       & 2.5  &  $7.4\times3.4$  & 15  & 0.282 & 0.320   & 0.927  \\
\ref{fig:figself3}(am)  & PKS $1304-748$ & \nodata   & 1.4  &  $8.2\times6.9$  & 24  & 0.153 & 0.118   & 0.529  \\ 
\ref{fig:figself3}(an)  & PKS $1304-748$ & \nodata   & 2.5  &  $4.5\times3.7$  & 28  & 0.175 & 0.0503  & 0.329  \\
\ref{fig:figself3}(ao)  & PKS $1312-593$ & \nodata   & 1.4  & $11.2\times6.2$  & 17  & 0.214 & 0.149   & 0.902  \\ 
\ref{fig:figself3}(ap)  & PKS $1312-593$ & \nodata   & 2.5  &  $6.1\times3.4$  & 16  & 0.184 & 0.0842  & 0.517  \\
\ref{fig:figself3}(aq)  & PKS $1425-479$ & 0.11      & 1.4  & $13.2\times6.2$  & 0   & 0.305 & 0.102   & 0.897  \\ 
\ref{fig:figself3}(ar)  & PKS $1425-479$ & 0.11      & 2.5  &  $7.2\times3.3$  & 1   & 0.211 & 0.0543  & 0.0652 \\
\ref{fig:figself3}(as)  & PKS $1516-477$ & \nodata   & 1.4  & $12.7\times6.2$  & -5  & 0.203 & 0.0508  & 0.718  \\ 
\ref{fig:figself3}(at)  & PKS $1516-477$ & \nodata   & 2.5  &  $7.2\times3.4$  & -7  & 0.170 & 0.0183  & 0.332  \\
\ref{fig:figself3}(au)  & PKS $1619-634$ & \nodata   & 1.4  &  $9.2\times6.7$  & -28 & 0.167 & 0.0104  & 0.963  \\ 
\ref{fig:figself3}(av)  & PKS $1619-634$ & \nodata   & 2.5  &  $5.1\times3.8$  & -19 & 0.142 & 0.0158  & 0.479  \\
\ref{fig:figself3}(aw)  & PKS $1637-771$ & 0.0427    & 1.4  &  $9.0\times6.4$  & 0   & 0.333 & 0.174   & 6.90   \\ 
\ref{fig:figself3}(ax)  & PKS $1637-771$ & 0.0427    & 2.5  &  $4.9\times3.4$  & 3   & 0.334 & 0.182   & 4.81   \\
\ref{fig:figself3}(ay)  & PKS $1758-473$ & 0.122864  & 1.4  & $11.3\times6.6$  & -5  & 0.209 & 0.162   & 0.567  \\ 
\ref{fig:figself3}(az)  & PKS $1758-473$ & 0.122864  & 2.5  &  $6.2\times3.5$  & -7  & 0.217 & 0.154   & 0.323  \\
\ref{fig:figself3}(ba)  & PKS $1846-631$ & \nodata   & 1.4  &  $9.0\times6.8$  & -15 & 0.244 & 0.677   & 1.38   \\ 
\ref{fig:figself3}(bb)  & PKS $1846-631$ & \nodata   & 2.5  &  $5.0\times3.5$  & -9  & 0.244 & 0.392   & 0.824  \\
\ref{fig:figself3}(bc)  & PKS $1910-800$ & 0.346     & 1.4  &  $8.2\times6.8$  & -41 & 0.255 & 0.0453  & 0.872  \\ 
\ref{fig:figself3}(bd)  & PKS $1910-800$ & 0.346     & 2.5  &  $4.6\times3.6$  & -42 & 0.238 & 0.00941 & 0.504  \\
\ref{fig:figself3}(be)  & PKS $2147-555$ & 0.146     & 1.4  &  $8.2\times6.7$  & -15 & 0.676 & 0.0371  & 1.42   \\ 
\ref{fig:figself3}(bf)  & PKS $2147-555$ & 0.146     & 2.5  &  $5.1\times3.7$  & 10  & 0.212 & 0.0396  & 1.05   \\
\ref{fig:figself3}(bg)  & PKS $2151-461$ & 0.038783  & 1.4  & $10.4\times6.9$  & 13  & 0.329 & 0.0389  & 0.929  \\ 
\ref{fig:figself3}(bh)  & PKS $2151-461$ & 0.038783  & 2.5  &  $5.6\times3.6$  & 3   & 0.161 & 0.0308  & 0.606  \\
\ref{fig:figself3}(bi)  & PKS $2152-699$ & 0.028273  & 1.4  &  $5.7\times5.2$  & 26  & 2.089 & 1.77    & 27.0   \\ 
\ref{fig:figself3}(bj)  & PKS $2152-699$ & 0.028273  & 2.5  &  $4.4\times3.7$  & 19  & 1.133 & 0.975   & 18.1   \\
\ref{fig:figself3}(bk)  & PKS $2302-713$ & 0.384     & 1.4  &  $8.1\times6.7$  & 16  & 0.186 & 0.211   & 0.573  \\ 
\ref{fig:figself3}(bl)  & PKS $2302-713$ & 0.384     & 2.5  &  $4.5\times3.6$  & 11  & 0.156 & 0.104   & 0.413  \\ 
\ref{fig:figselpica}(a) & PKS $0518-458$ & 0.035058  & 1.4  &  $10.6\times7.6$ & 0   & 0.435 & 6.73    & 76.8   \\ 
\ref{fig:figselpica}(b) & PKS $0518-458$ & 0.035058  & 2.5  &  $5.8\times4.0$  & 0   & 0.450 & 3.37    & 41.5   \\
\ref{fig:figsel0344}(a) & PKS $0344-345$ & 0.053804  & 4.8  &  $4.7\times2.0$  & -11 & 0.100 & 0.069   & 0.746  \\ 
\ref{fig:figsel0344}(b) & PKS $0344-345$ & 0.053804  & 8.64 &  $3.0\times1.1$  & -9  & 0.160 & 0.068   & 0.174  \\
\ref{fig:figsel0344}(c) & PKS $0344-345$ & 0.053804  & 18.0 &  $4.4\times1.6$  & -12 & 0.102 & 0.072   & 0.089  \\
\ref{fig:figsel0703}(a) & PKS $0703-595$ & \nodata   & 4.8  &  $2.8\times2.2$  & 44  & 0.500 & 0.017   & 0.049  \\ 
\ref{fig:figsel0703}(b) & PKS $0703-595$ & \nodata   & 8.64 &  $1.6\times1.3$  & 70  & 0.190 & 0.021   & 0.021  \\
\ref{fig:figsel0703}(c) & PKS $0703-595$ & \nodata   & 18.0 &  $0.8\times0.6$  & 72  & 0.280 & 0.017   & 0.020  \\
\ref{fig:figsel1637}(a) & PKS $1637-771$ & 0.0427    & 4.8  &  $2.4\times2.1$  & 59  & 0.290 & 0.184   & 1.39   \\ 
\ref{fig:figsel1637}(b) & PKS $1637-771$ & 0.0427    & 8.64 &  $1.4\times1.2$  & 76  & 0.220 & 0.242   & 1.09   \\
\ref{fig:figsel1637}(c) & PKS $1637-771$ & 0.0427    & 18.0 &  $0.7\times0.6$  & -61 & 0.230 & 0.435   & 0.450  \\
\ref{fig:figsel2152}    & PKS $2152-699$ & 0.028273  & 18.0 &  $0.5\times0.7$  & -79 & 0.590 & 0.767   & 1.31   \\
\end{longtable}
\end{center}
\normalsize

\linespread{1.0}
\normalsize
\begin{savequote}[15pc]
\sffamily
Night hides a world,\\
but reveals a Universe.
\qauthor{Iranian Proverb}
\end{savequote}

\chapter{The Compact Radio Source Population of NGC 253}
\label{chap:ngc253}
\begin{center}
{\it Adapted from:}

E. Lenc \& S.J. Tingay

Astronomical Journal, 132, 133 (2006)
\end{center}
\small
Wide-field, very long baseline interferometry (VLBI) observations of the nearby starburst galaxy NGC 253, obtained with the Australian Long Baseline Array (LBA), have produced a 2.3 GHz image with a maximum angular resolution of 15 mas (0.3 pc). Six sources were detected, all corresponding to sources identified in higher frequency ($>5$ GHz) VLA images. One of the sources, supernova remnant 5.48$-$43.3, is resolved into a shell-like structure approximately 90 mas (1.7 pc) in diameter. From these data and additional data from the literature, the spectra of 20 compact radio sources in NGC 253 were modelled and found to be consistent with free-free absorbed power laws. Broadly, the free-free opacity is highest toward the nucleus but varies significantly throughout the nuclear region ($\tau_0\sim 1->20$), implying that the overall structure of the ionised medium is clumpy. Of the 20 sources, eight have flat intrinsic spectra associated with thermal radio emission and the remaining 12 have steep intrinsic spectra, associated with synchrotron emission from supernova remnants. A supernova rate upper limit of 2.4 yr$^{-1}$ is determined for the inner 320 pc region of the galaxy at the 95\% confidence level, based on the lack of detection of new sources in observations spanning almost 17 years and a simple model for the evolution of supernova remnants. A supernova rate of $>0.14 (v/ 10^{4})$ yr$^{-1}$ is implied from estimates of supernova remnant source counts, sizes and expansion rates, where $v$ is the radial expansion velocity of the supernova remnant in km s$^{-1}$. A star formation rate of $3.4 (v/10^{4}) < SFR(M\geq5M_{\Sun})<59$ M$_{\Sun}$ yr$^{-1}$ has been estimated directly from the supernova rate limits and is of the same order of magnitude as rates determined from integrated FIR and radio luminosities.
\clearpage
\linespread{1.3}
\normalsize
\section{Introduction}
\label{sec:p1introduction}
Starburst galaxies are those galaxies containing regions (usually associated with the nucleus) currently undergoing high rates of star formation, much higher than seen in galaxies like our own and such that a significant fraction of the available gas for star formation will be exhausted within a single dynamical timescale \citep{Lehnert:1996p10560}.  A starburst is triggered by the rapid accumulation of gas in a small volume, causing large numbers of massive stars to form.  The accumulation of gas can be caused by interactions between galaxies, for example the merger of gas rich spiral galaxies can result in the accumulation of 60\% of the gas within the inner 100 pc of the merger product \citep{Barnes:1996p10524}.  Gas can be funnelled into the nuclear region of a galaxy by dynamical processes associated with bar instabilities \citep{Combes:2000p10567}.

Once formed, the starburst hosts large numbers of massive stars that ionise their environment and evolve quickly to form supernovae \citep{Engelbracht:1998p874}.  The supernova remnants can drive activity in the nuclear region and can produce strong winds out of the disk of the galaxy \citep{Doane:1993p10547}.

Starbursts are therefore the result of processes acting on a wide range of scales and themselves drive energetic activity on a wide range of scales.  By direct observation of the supernova remnants in starburst regions at radio wavelengths (to mitigate against severe extinction at optical wavelengths and provide very high angular resolution) it is possible to investigate the individual remnants and use them to reconstruct the supernova and star formation history of the starburst.  Furthermore, it is possible to use radio observations to investigate the ionised gaseous environment of the starburst.  Such observations provide a link between large-scale dynamical effects in the galaxies, activity in the star forming region itself, and the energetic phenomena in turn driven by the starburst. 

NGC 253 is a prominent nearly edge-on starburst galaxy in the Southern Hemisphere ($\delta\sim-25^{\circ}$), its proximity making it an ideal target for high spatial resolution observations. Previous studies of NGC 253 have assumed a distance of 2.5 Mpc \citep{Turner:1985p1103}, based on the estimate of \citet{deVaucouleurs:1978p10546}. Both this and the revised estimate of 2.58$\pm$0.7 Mpc \citep{Puche:1988p10553} were derived from the Tully-Fisher relation. More recently \citet{Karachentsev:2003p10562} determined the magnitude of the tip of the red giant branch from Hubble Space Telescope/WFPC2 images and arrived at a more reliable distance estimate of 3.94$\pm$0.5 Mpc. In this paper this new distance estimate is adopted.

In many respects NGC 253 is a twin of it's Northern Hemisphere counterpart M82. Both are of similar size, are nearly edge-on, barred, and have similar infrared spectra and luminosities. \cite{Engelbracht:1998p874} confirmed the existence of a large-scale bar $\sim$7 kpc in diameter crossing a circumnuclear ring of similar size in a de-reddened K-band image of NGC 253. Using a combination of optical, infrared and millimetre observations, \citet{Forbes:2000p950} concluded that a ring of young star clusters $\sim$80 pc in diameter defined the inner edge of a cold gas torus centred on the assumed nucleus. The torus is fed by gas channelled in by the large-scale bar and deposited at the Inner Lindblad Resonance \citep{Arnaboldi:1995p10523}.  Australia Telescope Compact Array (ATCA) images of para-NH$_{3}$ (1,1) and (2,2) and ortho-NH$_{3}$ (3,3) and (6,6) were used to probe the temperature of the nuclear regions and highlight some of the inner structure of the torus where the temperatures are greatest \citep{Ott:2005p785}. The full extent of the torus is seen more clearly in CO and [\ion{C}{1}] emission line maps which trace cool gas and dust \citep{Israel:1995p905}, these observations showing that the torus is clumpy and appears to have a diameter of $\sim900$ pc.

Associated with the nuclear region of the galaxy is gas producing strong optical emission lines such as H$\alpha$, [\ion{S}{2}], and [\ion{O}{3}] \citep{Forbes:2000p950,Carral:1994p10535}.  The environment is heavily ionised and the emission line region has a complex morphology \citep{Forbes:2000p950}.  Also associated with the nuclear region, and having its origin in the inner starburst region, is a powerful wind that has formed a bipolar nuclear outflow cone perpendicular to the disk of the galaxy. The outflow has been investigated by \citet{Heckman:1990p2260} and \citet{Schulz:1992p1036} using H$\alpha$ spectral line measurements but is seen more clearly in X-ray observations by Chandra \citep{Weaver:2002p844, Strickland:2002p797}.  \ion{H}{1} plumes are also seen bordering the H$\alpha$ and X-ray halo emission and are also believed to be associated with the inner starburst region and active star formation in the disk \citep{Boomsma:2005p10530}. It is thought that the wind is matter blown from the central starburst region by the ensemble of massive stars and supernova remnants within the inner 80 pc.

\begin{figure}[ht]
\epsscale{1.0}
\plotone{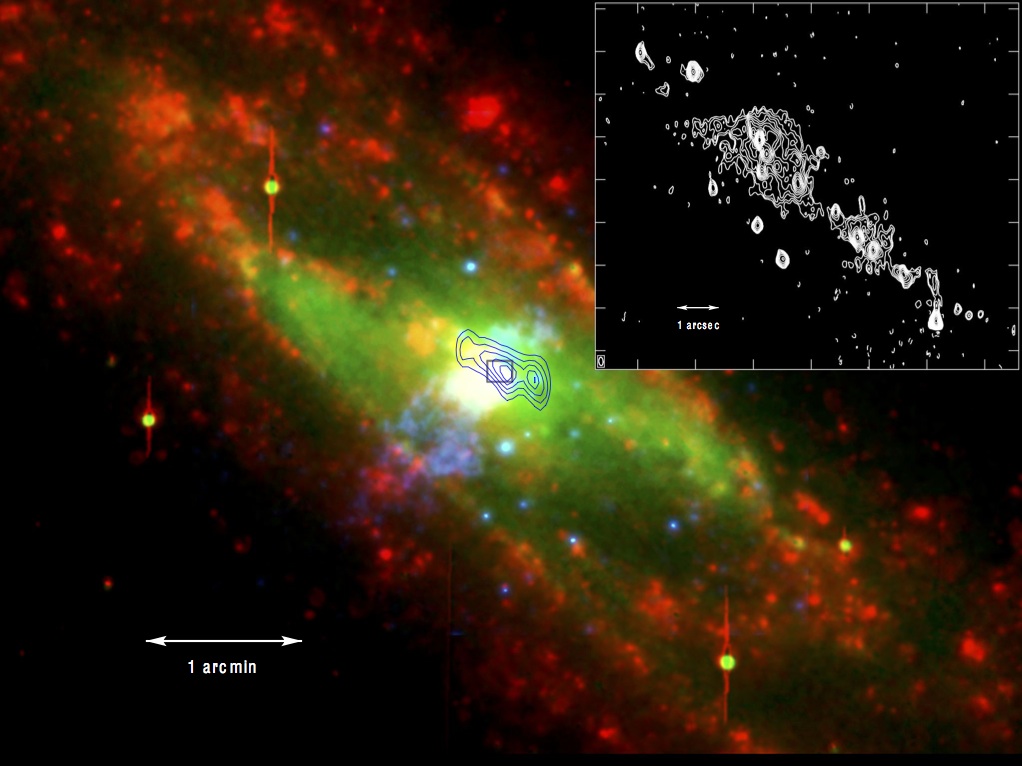}
\caption[A three-colour composite, multi-wavelength view of NGC 253.]{A three-colour composite, multi-wavelength view of NGC 253. Red, green and blue indicate H$\alpha$ (courtesy of M. Dahlem), K$-$band 2MASS IR \citep{Jarrett:2003p8453}, and Chandra soft X-ray \citep{Weaver:2002p844} respectively. The Owens Valley Radio Interferometer CO emission line contour map of the inner 3 kpc is shown in blue with contours at $19\%$, $38\%$, $57\%$, $76\%$ and $87\%$ of the peak (courtesy of M. Dahlem and F. Walter and first published by \citet{Ott:2005p785}). Inset: VLA A configuration 2 cm image of central boxed region, contours are logarithmic intervals of $2^{1/2}$, beginning at 0.225 mJy beam$^{-1}$ (U\&A97).}
\label{fig:figmultiwav}            
\end{figure}

Figure \ref{fig:figmultiwav} presents images of the galaxy at various wavelengths, on different spatial scales, to illustrate of the relationships between the galaxy bar (IR), the molecular torus (CO), the inner starburst region (radio), and the wind eminating from that region (H$\alpha$ and X-ray).  The supernova remnants that are a primary subject of this paper are concentrated within the inner 80 pc of the galaxy but are also distributed widely outside the inner starburst region. 

At radio wavelengths, \citet[hereafter U\&A97]{Ulvestad:1997p907} imaged NGC 253 with the VLA at frequencies between 1.5 GHz and 23 GHz over a period of eight years and identified 64 individual compact sources within 320 pc of the nucleus. The identification of sources at lower frequencies ($<10$ GHz) was hindered by confusion and the diffuse emission of the galaxy. Spectral indices, with an error less than 0.4, could therefore only be determined for 17 compact sources, mainly those isolated sources outside the nuclear region. In the absence of any new sources or any significant fading of existing sources over the course of their observations, U\&A97 derived a supernova rate of no more than 0.3 yr$^{-1}$. At 23 GHz the assumed nucleus (5.78-39.4) is the strongest source and is unresolved. Approximately half of the 17 detected compact sources with measured spectral indices were shown by U\&A97 to be associated with \ion{H}{2} regions. The remainder are most likely to be supernova remnants.

In the first attempt to resolve the inner starburst region of NGC 253 at low radio frequencies, \citet{Tingay:2004p778} used the Australian Long Baseline Array (LBA) in the first multi-baseline VLBI observation of NGC 253. At 1.4 GHz the observation approximately matched the angular resolution of the VLA at 23 GHz, thus overcoming confusion problems. Two sources were detected, neither of which corresponded to the assumed nucleus. \citet{Tingay:2004p778} showed that the radio emission from compact sources in NGC 253 is partially absorbed by ionised gas with free-free optical depths ranging between $\tau_{0}=2.5$ and $\tau_{0}>8$ ($\tau_{0}$ denotes the free-free optical depth at 1 GHz).  Further evidence for free-free absorption in NGC 253 comes from measurements of the nuclear starburst region between 300 MHz and 1.5 GHz by \citet{Carilli:1996p10536}, showing a dramatic spectral flattening at low frequencies. The modelling of radio-recombination line emission from the nuclear region of NGC 253 suggests that the lines originate from a gas with an electron density of between $\sim7\times10^{3}$ and $\sim1.7\times10^{4}$ cm$^{-3}$ and with a temperature of $7.5\times10^{3}$ K, distributed in a spherical structure of uniform density and a diameter of 3$-$4 pc \citep{Mohan:2002p792}. \citet{Tingay:2004p778} found that these results were consistent with a high degree ($\tau_{0} \sim 36$) of free-free absorption towards the assumed nucleus, 5.79$-$39.0.

Similar results for the supernova remnants in M82 have been obtained by \citet{Pedlar:1999p3534} and \citet{McDonald:2002p3330}.  In M82 30 supernova remnants have been detected from a population of 46 compact radio sources \citep{McDonald:2002p3330} and several resolved \citep{Pedlar:1999p3534, Muxlow:2005p3176}.  \citet{McDonald:2002p3330} found a free-free optical depth toward supernova remnants of $\tau_{0}\sim3$.

Motivated by the strong indicators of free-free absorption in NGC 253, we have embarked on a program of VLBI observations of this galaxy and other prominant Southern Hemisphere starburst galaxies. This paper reports results from a new observation of the nuclear region of NGC 253 at 2.3 GHz. The resulting data have better resolution and sensitivity than the 1.4 GHz observations  of \citet{Tingay:2004p778} and we have used them to investigate in more detail the free-free absorbed spectra of supernova remnants and \ion{H}{2} regions in the NGC 253 starburst (\S~\ref{sec:p1ffmodel}). In particular we are interested in probing the structure of the ionised environment of the supernova remnants and the relationship between the ionised medium and the other components of the galaxy (\S~\ref{sec:p1spectra}), further constraining the supernova rate in NGC 253 using previously published images and our new VLBI images (\S~\ref{sec:p1snrate}) and re-evaluating estimates of the star formation rate based on our results (\S~\ref{sec:p1sfrate}).

A subsidiary focus of this project is to achieve high sensitivity, high fidelity, high resolution, and computationally efficient imaging over wide fields of view (at least compared to traditional VLBI imaging techniques).  As such, the associated exploration of the relevant techniques is a small first step toward the much larger task of imaging with the next generation of large radio telescopes, for example the Square Kilometre Array (SKA: \citet{Hall:2005p10570, Carilli:2004p10534}).  The SKA will use baselines ranging up to approximately 3000 km and cover fields of view greater than one square degree instantaneously at frequencies of around 1 GHz.  The computational load required for this type of imaging task is vast and will drive the development of novel imaging algorithms, using techniques such as $w$-projection \citep{cor03} and the use of super-computing facilities.  Investigating the performance of these techniques now, under the most challenging current observing conditions, is therefore a useful activity.

%
%
%
\section{Observations, data reduction, and results}

\subsection{Observations and data reduction}
\begin{figure}[ht]
\epsscale{0.5}
\plotone{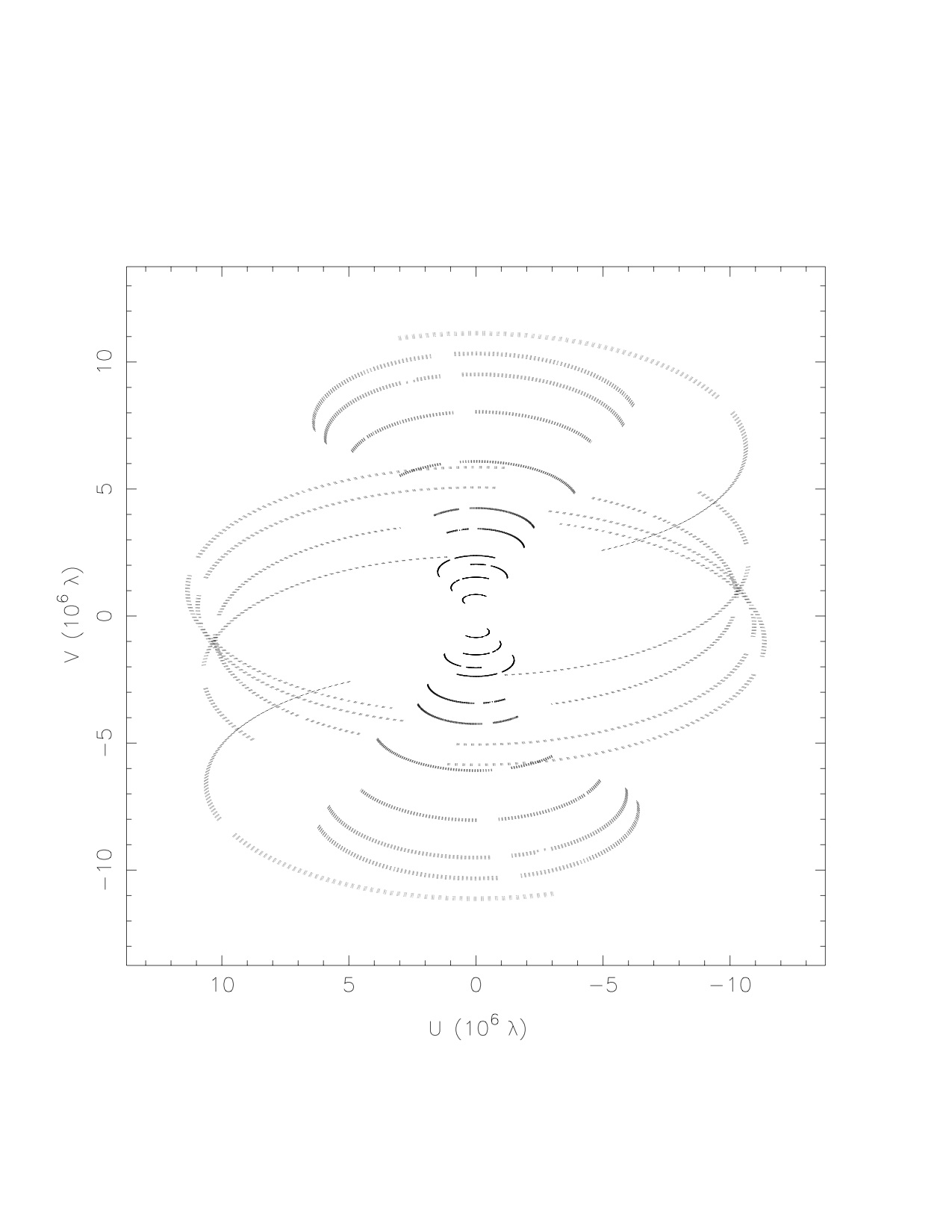}
\caption{Spatial frequency ($u,v$) coverage for NGC 253 at a frequency of 2.3 GHz.}
\label{fig:figuvcov}            
\end{figure}

A VLBI observation of NGC 253 was made on 16/17 April, 2004 (19:00 - 07:00 UT) using a number of Australian radio telescopes: the 70 m NASA Deep Space Network (DSN) antenna at Tidbinbilla; the 64 m antenna of the Australia Telescope National Facility (ATNF) near Parkes; 5 $\times$ 22 m antennas of the ATNF Australia Telescope Compact Array (ATCA) near Narrabri used as a phased array; the ATNF Mopra 22 m antenna near Coonabarabran; the University of Tasmania's 26 m antenna near Hobart; and the University of Tasmania's 30 m antenna near Ceduna.  The observation utilised the S2 recording system \citep{Cannon:1997p10533} to record 2 $\times$ 16 MHz bands (digitally filtered 2-bit samples) in the frequency ranges: 2252 - 2268 MHz and 2268 - 2284 MHz.  Both bands were upper side band and right circular polarisation.  The spatial frequency ($u,v$) coverage for the observation is shown in Figure \ref{fig:figuvcov}.

During the observation, three minute scans of NGC 253 ($\alpha = 00$:47:33.178; $\delta = -25$:17:17.060 [J2000]) were scheduled, alternating with three minute scans of a nearby phase reference calibration source, PKS J0038$-$2459 ($\alpha = 00$:38:14.735493; $\delta = -24$:59:02.23519 [J2000]).

The recorded data were correlated using the ATNF Long Baseline Array (LBA) processor at ATNF headquarters in Sydney \citep{Wilson:1992p9290}. The data were correlated using an integration time of 2 seconds and with 32 frequency channels across each 16 MHz band (channel widths of 0.5 MHz).

The correlated data were imported into the AIPS\footnote{The Astronomical Image Processing System (AIPS) was developed and is maintained by the National Radio Astronomy Observatory, which is operated by Associated Universities, Inc., under co-operative agreement with the National Science Foundation} package for initial processing. The data for the phase reference source were fringe-fit (AIPS task FRING) using a one minute solution interval, finding independent solutions for each of the two 16 MHz bands.  The delay and phase solutions for the phase reference source were examined and averaged over each three minute calibrator scan, following editing of bad solutions, before being applied to both the phase reference source and NGC 253.  Further flagging of the data was undertaken via application of a flag file that reflected the times during which each of the antennas were expected to be slewing, or time ranges that contained known bad data.  Finally, data from the first 30 seconds of each scan from baselines involving the ATCA or Parkes were flagged, to eliminate known corruption of the data at the start of each scan at these two telescopes.

During correlation, nominal (constant) system temperatures (in Jy) for each antenna were applied to the correlation coefficients in order to roughly calibrate the visibility amplitudes (mainly to ensure roughly correct weights during fringe-fitting).  Following fringe-fitting, the nominal calibration was refined by collecting and applying the antenna system temperatures (in K) measured during the observation, along with the most recently measured gain (in Jy/K) for each antenna.  Further refinements to the amplitude calibration were derived from observations of a strong and compact radio source (PKS B1921$-$293) made immediately before the NGC 253 observations. PKS B1921$-$293 was observed using the same array as NGC 253, at the same frequencies and bandwidths, and simultaneously the PKS B1921$-$293 data were recorded at the ATCA. PKS B1921$-$293 is unresolved at this frequency and on these VLBI baselines, therefore the flux measured at the ATCA could be used to check and refine the amplitude calibration for the VLBI data.  The PKS B1921$-$293 data were also used to derive a bandpass calibration in the MIRIAD package \citep{Sault:1995p10582}, using task MFCAL, which was applied to the data for NGC 253 and the phase reference source.  The edge channels of each band were edited from the data-set (3 channels from the lower edge and 2 channels from the upper edge of each 32 channel band).

\begin{figure}[ht]
\plotone{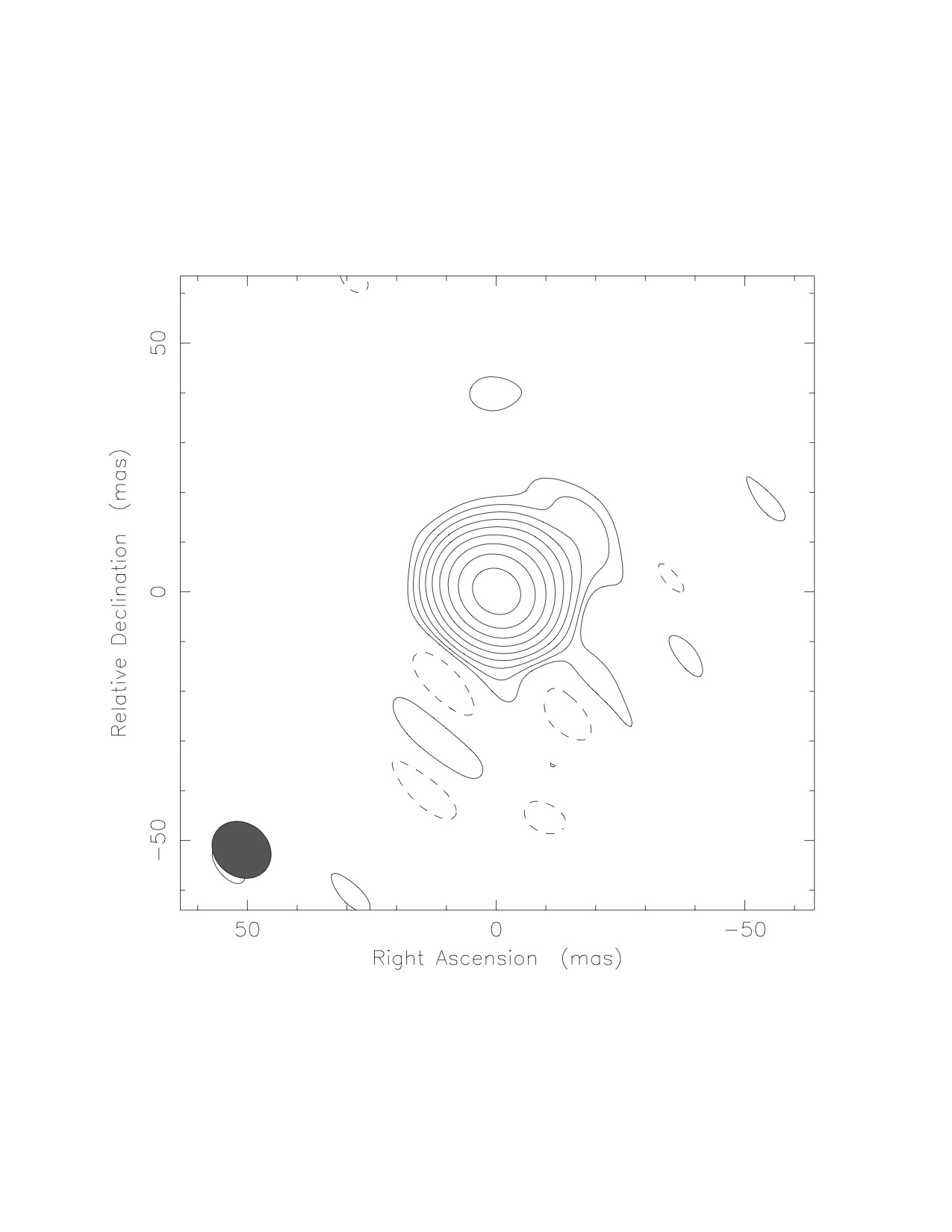}
\caption[Australian LBA image of the phase calibrator PKS J0038$-$2459 at a frequency of 2.3 GHz.]{Australian LBA image of the source used for phase calibration, PKS J0038$-$2459 at a frequency of 2.3 GHz. The image has an RMS noise level of approximately 0.25 mJy/beam. The peak is 361 mJy/beam and contours are at -0.25\%, 0.25\%, 0.5\%, 1\%, 2\%, 4\%, 8\%, 16\%, 32\% and 64\% of the peak. The beam size is approximately 13 $\times$ 11 mas at a position angle of $51^{\circ}$}
\label{fig:figj0038}            
\end{figure}

In DIFMAP \citep{Shepherd:1994p10583} the data for the phase reference calibrator were vector averaged over a 30 second period and then imaged using standard imaging techniques (de-convolution and self-calibration of both phase and amplitude).  The resulting image (Figure \ref{fig:figj0038}) shows that PKS J0038$-$2459 is highly compact, with no significant structure on these baselines at this frequency. During self-calibration, amplitude corrections of less than 10\% were noted and applied to the NGC 253 data in a second iteration of the amplitude calibration in AIPS.  For the Tidbinbilla antenna, no system temperatures were measured, so a constant value as a function of time was initially applied. To refine the Tidbinbilla calibration a gain-elevation curve was derived in MIRIAD using the task MFCAL. This curve was applied to the data-set to complete the amplitude calibration procedure.

\begin{figure}[ht]
\epsscale{0.7}
\centerline{\includegraphics[angle=270,width=\textwidth]{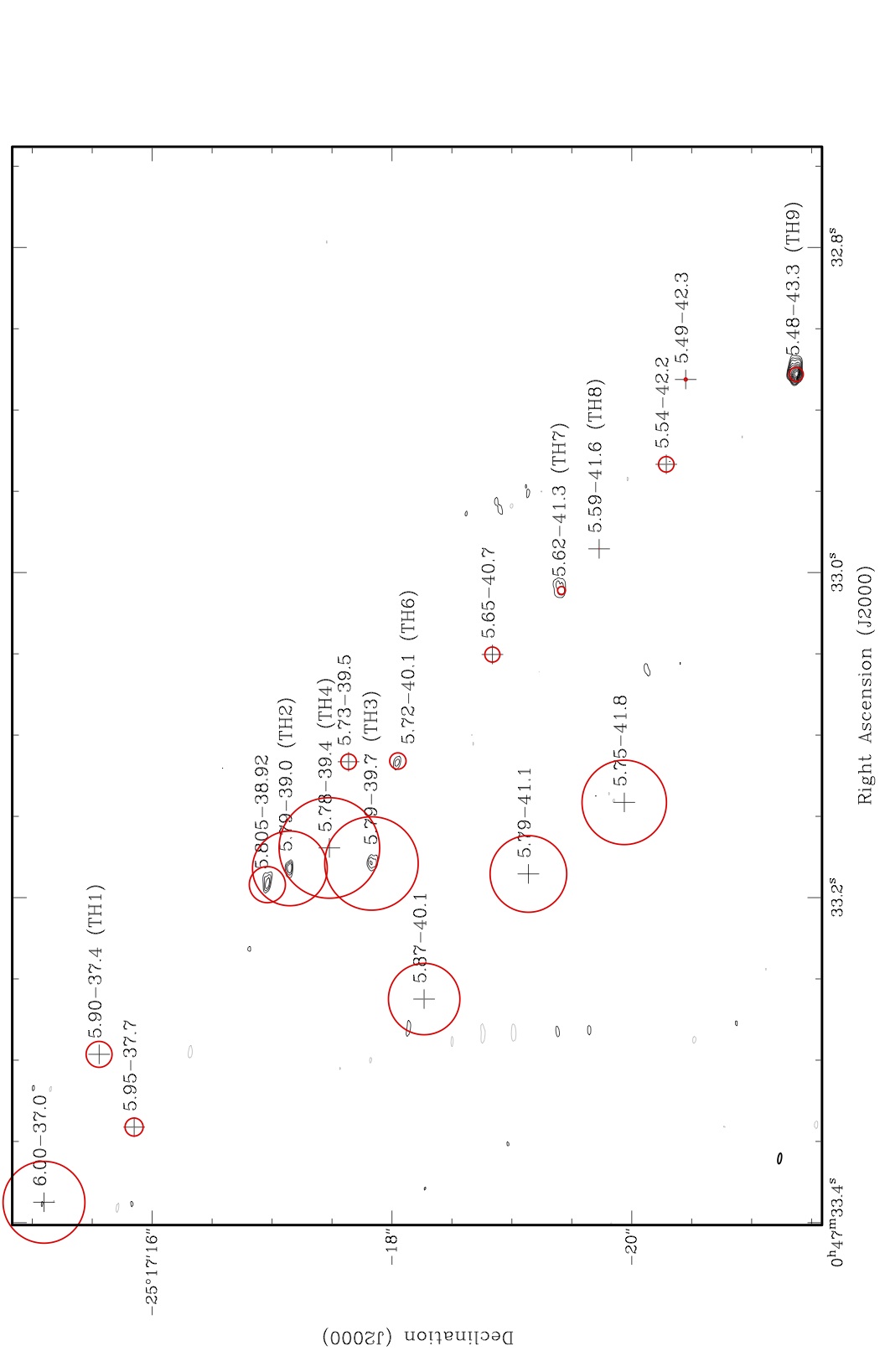}}
\caption[Australian LBA image of NGC 253 at a frequency of 2.3 GHz.]{Australian LBA image of NGC 253 at a frequency of 2.3 GHz. The image has an RMS noise level of approximately 0.24 mJy/beam. The peak is 8.56 mJy/beam and contours are at -10\%, 10\%, 15\%, 20\%, 30\%, 40\%, 50\%, 60\%, 70\%, 80\%, 90\% and 100\% of the peak. The beam size is approximately 90 $\times$ 30 mas at a position angle of $81^{\circ}$. The overlayed circles give an indication of the degree of absorption in the vicinity of the source with the diameter of the circle being directly proportional to $\tau_{0}$ (or the lower limit on $\tau_{0}$). All sources are labelled using U\&A97 notation, \citet{Turner:1985p1103} sources are labelled within parentheses and non-detections are marked with a cross. }
\label{fig:figngc253}            
\end{figure}

Once amplitude calibrated, with the delay and phase solutions from fringe-fitting applied, and the bandpass calibration applied, the data for both phase reference source and NGC 253 were exported as FITS files.  Both data-sets were examined in DIFMAP and some editing was undertaken. Imaging of the NGC 253 data took place in AIPS++\footnote{AIPS++ (Astronomical Information Processing System) was originally developed by an international consortium of observatories (ASTRON, ATNF, JBO, NCSA, NRAO) as a toolkit for the analysis/reduction of radio astronomical data. Development now continues based on project/instrument needs.}, using the w-projection algorithm. W-projection allows wide fields to be imaged, accounting for the w-term in such a way that imaging artefacts are at far lower levels than for other more traditional VLBI wide-field imaging techniques such as faceting \citep{Perley:1999p10576}. Following a number of trial imaging runs, the parameters chosen for imaging were as follows: cell size = 11 mas; weighting = briggs and weighting robustness = 2. The chosen weighting scheme minimised noise at the expense of resolution but provided a beam shape that enabled sources to be identified more readily with the available spatial frequency coverage. Figure \ref{fig:figngc253} shows the image of NGC 253 using the data-set described above but excluding the data from the Hobart and Ceduna antennas (the longest baselines).

\begin{figure}[ht]
\plotone{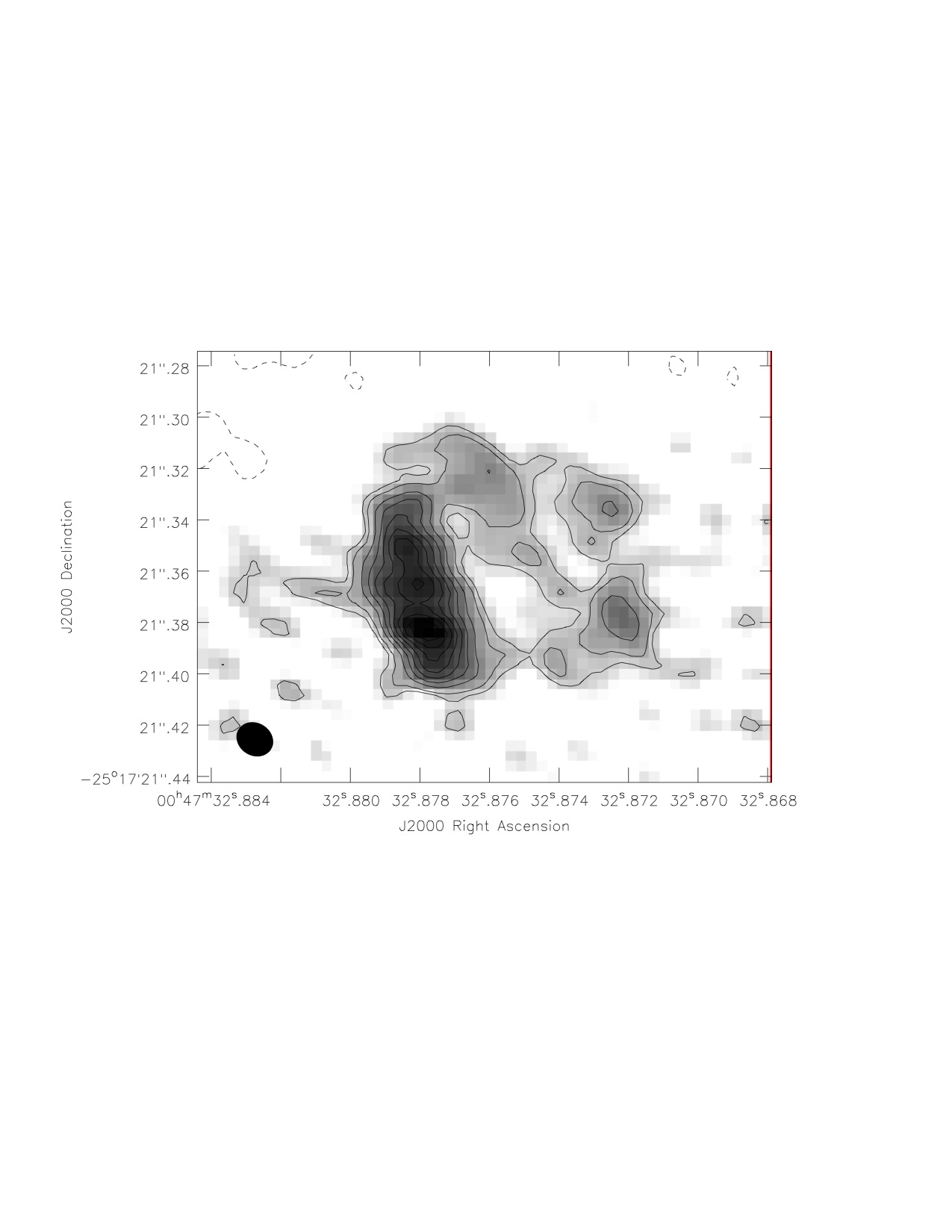}
\caption[Australian LBA image of the compact source 5.48$-$43.3 in NGC 253 at a frequency of 2.3 GHz.]{Australian LBA image of the compact source 5.48$-$43.3 in NGC 253 at a frequency of 2.3 GHz. The image has an RMS noise level of approximately 0.16 mJy/beam. The peak is 2.42mJy/beam and contours are at -15\%, 15\%, 20\%, 30\%, 40\%, 50\%, 60\%, 70\%, 80\%, 90\% and 100\% of the peak. The beam size is approximately 15 $\times$ 13 mas at a position angle of $58^{\circ}$.}
\label{fig:figsnr}            
\end{figure}

In Figure \ref{fig:figngc253}, a number of compact radio sources have clearly been detected.  The one sigma RMS noise in Figure \ref{fig:figngc253} is 0.24 mJy/beam, a factor of 4 greater than the theoretically predicted thermal noise value\footnote{Estimated with the ATNF VLBI sensitivity calculator: \url{http://www.atnf.csiro.au/vlbi/calculator/}} for this array, bandwidth and observation time.  As there were no sufficiently bright sources to perform self-calibration, the higher noise level may be attributed to residual phase errors on time-scales less than the 3 minute calibrator-target duty cycle. Amplitude errors and limited spatial frequency coverage are also likely to have been contributing factors. We take a 5 sigma detection limit in this image of 1.2 mJy/beam. Figure \ref{fig:figsnr} is an image of 5.48$-$43.3 (bottom right corner in Figure \ref{fig:figngc253}) made from the full VLBI data-set (including the Hobart and Ceduna data) and with similar imaging parameters used to produce \ref{fig:figngc253}, except that a smaller cell size of 4 mas was chosen.

\subsection{Identification of Sources}
\label{sec:p1sourceid}
A list of the sources detected above the 5 sigma detection threshold in the new VLBI data is given in Table \ref{tab:p1tabsources}. Flux density errors of $\pm10$\% are listed due to uncertainties in the absolute flux density scale for Southern Hemisphere VLBI \citep{rey94}.  The total flux density for each source was determined by summing its CLEAN model components in the image plane and an estimate of the source size, for all but 5.48$-$43.3, was obtained by fitting a Gaussian source to the CLEAN model components. As can be seen from the list, the brightest of the detected sources (5.48$-$43.3) is well resolved into a shell at this resolution. The shell diameter of 5.48$-$43.3 was measured by taking a cross-cut through the centre and measuring the distance between shell maxima. 

A comparison of the source positions with previous high resolution imaging of NGC 253 shows that each of the detected sources can be readily identified. Table \ref{tab:p1tabsources} lists identifications with 1.3 cm and 2 cm sources found by U\&A97 with VLA observations. Where available, identifications with 2 cm sources found by \citet{Turner:1985p1103}, also with the VLA, are specified with the ``TH$n$" notation originally adopted by U\&A97. All other compact radio sources in the NGC 253 starburst detected by U\&A97 at 23 GHz lie below our 1.2 mJy/beam detection limit at 2.3 GHz. \citet{Tingay:2004p778} sources (A) and (B), detected at 1.4 GHz with the LBA, are also identified with the U\&A97 sources 5.48$-$43.3 and 5.62$-$41.3 respectively but are not listed in the table.

\begin{table}[ht]
\begin{center}
{ \scriptsize
\begin{tabular}{llclllllll} \hline \hline
\multicolumn{2}{c}{Position\tablenotemark{a}}             & & \multicolumn{2}{c}{Identification} &      &      & \multicolumn{3}{c}{Source Size} \\
                      \cline{1-2}                                      \cline{4-5}                                         \cline{8-10}
$\alpha$(J2000) & $\delta$(J2000) & &                      &                          &  Peak\tablenotemark{d}      & $S_{2.3}$\tablenotemark{e} & Major & Minor & P.A. \\
00:47                  & -25:17                & & U\&A97\tablenotemark{b}            & T\&H85\tablenotemark{c}                  & (mJy/beam)      &  (mJy) & (mas) & (mas) & (deg) \\ \hline \hline
32.878                & 21.372               & & 5.48$-$43.3     & TH9        & 8.6$\pm0.9$      & 32$\pm3$      & $90\pm7$ & $90\pm7$ & 0 \\
33.011                & 19.414               & & 5.62$-$41.3     & TH7        & 1.7$\pm0.2$      & 8.8$\pm0.9$  & $130\pm15$ & $90\pm15$ & 98 \\
33.116                & 18.050               & & 5.72$-$40.1     & TH6        & 1.5$\pm0.2$      & 3.8$\pm0.4$  & $60\pm15$ & $50\pm15$ & 47 \\
33.179                & 17.830               & & 5.79$-$39.7     & TH3        & 1.8$\pm0.2$      & 6.2$\pm0.6$  & $50\pm15$ & $60\pm15$ & 128 \\
33.182                & 17.148               & & 5.79$-$39.0     & TH2        & 2.4$\pm0.2$      & 5.7$\pm0.6$  & $80\pm15$ & $60\pm15$ & 141 \\
33.192                & 16.961               & & 5.805$-$38.92 & $\cdots$ & 2.1$\pm0.2$      & 6.8$\pm0.7$  & $130\pm15$ & $50\pm15$ & 78 \\ \hline
\tablenotetext{a}{The measured 2.3 GHz LBA source positions.}
\tablenotetext{b}{Source identifications with 1.3 cm and 2 cm sources found by U\&A97.}
\tablenotetext{c}{Source identifications with 2 cm sources found by \citet{Turner:1985p1103} using the ``TH$n$" designation of U\&A97.}
\tablenotetext{d}{The measured peak flux density for the sources at 2.3 GHz.}
\tablenotetext{e}{The measured flux density for the sources at 2.3 GHz.}
\end{tabular}
\caption{Compact sources detected in NGC 253 with 2.3 GHz VLBI.}
\label{tab:p1tabsources}
}
\end{center}
\end{table}

We find a position difference of 71$\pm$53 mas between our J2000 source positions and the precessed B1950 positions of the 23 GHz sources from U\&A97. These differences reduce to 51$\pm$24 mas if the weakest source, 5.72$-$40.1, is excluded from the sample. We also find an offset of 94$\pm$13 mas between the 1.4 GHz LBA source positions of \citet{Tingay:2004p778} and our 2.3 GHz positions. A contribution to the difference between our derived positions and the positions listed by U\&A97 comes from the uncertainty in the phase calibrator positions used at the VLA during the NGC 253 observations of U\&A97. For example, the calibrator 0008$-$264, used by U\&A97 for their 1.3 cm, 2.0 cm, 3.6 cm, and 6.0 cm observations of 1987, 1989 and 1991, had an uncertainty of 150 mas at the time of its use as a calibrator. The calibrator used for our VLBI observations, J0038$-$2459, is used as part of the ICRF and has a quoted uncertainty of $\sim0.7$ mas \citep{Beasley:2002p10526}. Our observations further benefited from a smaller beam size and an improved spatial frequency coverage compared to the 1.4 GHz LBA observation. We estimate our astrometric errors, based on half the minor axis of the beam, to be $\pm7$ mas for 5.48$-$43.3 and $\pm15$ mas for the remaining sources. 

\section{Discussion}

\subsection{Radio spectra and free-free absorption modelling of the compact sources}
\label{sec:p1ffmodel}
U\&A97 measured the spectral indices of 23 compact radio sources in NGC 253 using VLA observations between 5 GHz and 23 GHz (17 with spectral index errors of less than 0.4). Assuming that the spectral indices apply down to 2.3 GHz and that the sources are not resolved out by the higher resolution available with the LBA, 18 of these sources should be visible above the 5 $\sigma$ detection limit of our 2.3 GHz LBA observations. These sources are listed in Table \ref{tab:p1tabflux} with flux density measurements from VLA observations at 5 GHz, 8.3 GHz, 15 GHz and 23 GHz (U\&A97), together with LBA observations of two sources (5.48$-$43.3 and 5.62$-$41.3) at 1.4 GHz and flux density upper limits of 1.8 mJy at 1.4 GHz for the remaining sources \citep{Tingay:2004p778}.

\begin{table}[p]
\begin{center}
\begin{tabular}{lrrrrrr} \hline \hline
      & \multicolumn{2}{c}{LBA\tablenotemark{b} (mJy)} & \multicolumn{4}{c}{VLA\tablenotemark{c} (mJy)}                  \\
              \cline{2-3}                 \cline{4-7}
VLA ID\tablenotemark{a}      & ~~~~$S_{1.4}$  & ~~~~$S_{2.3}$  & ~~~~$S_{5}$    & ~~~~$S_{8.3}$  & ~~~~$S_{15}$  & ~~~~$S_{23}$  \\ \hline \hline 
4.81$-$43.6 &  $<$1.8    & $<$1.2     &  1.39       & $\cdots$ & 0.37       & $\cdots$ \\
5.48$-$43.3 &  16           & 32         &  27.10     & 20.49      & 12.51     &  9.79     \\
5.49$-$42.3 &  $<$1.8    & $<$1.2     &  $\cdots$ & 1.12       & $\cdots$ &  0.83     \\
5.54$-$42.2 &  $<$1.8    & $<$1.2     &  2.82       & 2.73        &  3.42      &  4.48     \\
5.59$-$41.6 &  $<$1.8    & $<$1.2     &  $\cdots$ & 2.97       & $\cdots$ &  5.72     \\
5.62$-$41.3 &  6             & 8.8           &  9.77       & 7.29        & 7.47       &  5.84     \\
5.65$-$40.7 &  $<$1.8    & $<$1.2     &  $\cdots$ & 1.52       & $\cdots$ &  1.23     \\
5.72$-$40.1 &  $<$1.8    & 3.8           &  7.74       & 6.53        & 7.82       &  7.88     \\
5.73$-$39.5 &  $<$1.8    & $<$1.2     &  $\cdots$ & 2.42       & $\cdots$ &  2.75     \\
5.75$-$41.8 &  $<$1.8    & $<$1.2     &  6.98       & 4.83        & 3.13       &  2.57     \\
5.78$-$39.4 &  $<$1.8    & $<$1.2     &  $\cdots$ & 16.86     & $\cdots$ &  9.69     \\
5.79$-$39.0 &  $<$1.8    & 5.7           &  38.64     & 47.95      &  40.32    &  35.79    \\
5.79$-$39.7 &  $<$1.8    & 6.2           &  $\cdots$ & $\cdots$ & $\sim$4.1 &  $\sim$1.6 \\
5.79$-$41.1 &  $<$1.8    & $<$1.2     &  4.73       & 3.17        &  1.94      &  1.00     \\
5.805$-$38.92 & $<$1.8 & 6.8           &  $\cdots$ & $\cdots$ & $\sim$3.0 &  $\sim$1.9 \\
5.87$-$40.1 &  $<$1.8    & $<$1.2     &  3.77       & 2.11        &  1.36      &  0.72     \\
5.90$-$37.4 &  $<$1.8    & $<$1.2     &  3.97       & 3.96        &  5.89      &  6.72     \\
5.95$-$37.7 &  $<$1.8    & $<$1.2     &  1.20       & $\cdots$  &  0.57      & $\cdots$ \\
6.00$-$37.0 &  $<$1.8    & $<$1.2     &  5.85       & 3.48         &  2.43      &  1.68     \\
6.40$-$37.1 &  $<$1.8    & $<$1.2     &  2.87       & $\cdots$  &  1.25      & $\cdots$ \\ \hline
\tablenotetext{a}{``VLA ID" is the identification from U\&A97.}
\tablenotetext{b}{$S_{1.4}$ is the measured flux density from the LBA \citep{Tingay:2004p778}. $S_{2.3}$ is the measured 2.3 GHz flux density from this work. Upper limits on the flux density at 1.4 GHz and 2.3 GHz are provided for sources that have not been detected.}
\tablenotetext{c}{$S_{5}$, $S_{8.3}$, $S_{15}$ and $S_{23}$ are the measured 5 GHz, 8.3 GHz, 15 GHz and 23 GHz flux densities from the VLA respectively (U\&A97).}
\end{tabular}
\caption{Flux densities of sources detected in NGC 253.}
\label{tab:p1tabflux}
\end{center}
\end{table}

Comparing this list with the sources in Table \ref{tab:p1tabsources} reveals that only 4 sources (5.48$-$43.3, 5.62$-$41.3, 5.72$-$40.1 and 5.79$-$39.0) were detected with the LBA at 2.3 GHz. The 2.3 GHz flux densities for all 18 sources are listed in Table \ref{tab:p1tabflux} with upper limits of 1.2 mJy for non-detections. A further two sources, 5.805$-$38.92 and 5.97$-$39.7, were detected with the LBA at 2.3 GHz and also by U\&A97 using the 15 GHz VLA A and 23 GHz VLA (A+B) configurations, and are also listed in Table \ref{tab:p1tabflux}. However, due to mismatches in resolution these source fluxes are only considered as approximate measurements by U\&A97. As a check to ensure that our measured fluxes at 2.3 GHz were not an underestimate as a result of higher resolution, total flux density measurements were compared between the 6 telescope LBA (Parkes, Narrabri, Mopra, Tidbinbilla, Hobart and Ceduna) observations and the 4 telescope array (excluding the longer baselines associated with Ceduna and Hobart) observations and were found to be the same within the measurement errors. Furthermore, U\&A97 found no significant variability in individual radio sources in NGC 253 over a period of 8 years, so it can be assumed that these measured fluxes, at multiple wavelengths, over multiple epochs, can be used to construct the spectra. 

\citet{Tingay:2004p778} argued that free-free absorption was the cause for the sharp downturn observed in the spectra of the compact radio sources at 1.4 GHz. At 2.3 GHz the effect of free-free absorption is not as great as at 1.4 GHz, so the detection of two sources at 1.4 GHz and a further four sources at 2.3 GHz appears to support this argument. The non-detection of the remaining 14 sources thus suggests that the effect of free-free absorption is still significant at 2.3 GHz. To test this quantitatively, an analysis similar to that undertaken by \citet{McDonald:2002p3330} for M82 was performed.

\citet{McDonald:2002p3330} constructed the spectra of 20 compact radio sources in M82 from existing and new observations between 1.42 GHz and 15 GHz using MERLIN and the VLA. The spectra were investigated using three models as described by the equations

\begin{equation}
\label{eq:plaw}
S(\nu)=S_{0}\nu^{\alpha},
\end{equation}
\begin{equation}
\label{eq:ff}
S(\nu)=S_{0}\nu^{\alpha}e^{-\tau(\nu)},
\end{equation}
\begin{equation}
\label{eq:sa}
S(\nu)=S_{0}\nu^{2}(1-e^{-\tau(\nu)}),
\end{equation}
where
\begin{equation}
\tau(\nu)=\tau_{0}\nu^{-2.1}.
\end{equation}

Equation \ref{eq:plaw} represents a simple power-law spectrum, Equation \ref{eq:ff} a free-free absorbed power-law spectrum, and Equation \ref{eq:sa} a self-absorbed bremsstrahlung (free-free absorbed) spectrum. In these expressions $\alpha$ is the optically thin intrinsic spectral index, $\tau_{0}$ is the free-free optical depth at 1 GHz and $S_{0}$ is the intrinsic flux density of the source at 1 GHz. The combination of high frequency VLA data and 1.4 GHz LBA data, together with the detection of six sources at 2.3 GHz, and the more stringent upper limit for non-detections at 2.3 GHz, allow us to use these models to constrain the free-free parameters of the 20 compact radio sources listed in Table \ref{tab:p1tabflux}.

The spectrum of each source was tested against each of the three models described above using a reduced$-\chi^{2}$ criterion to determine the best fit. For all sources, the free-free absorbed power-law model produced a significantly better fit than the self-absorbed and power-law models, for example giving a mean reduced$-\chi^{2}$ fit of 0.8, 23, and 40 for models 2, 3 and 1 respectively. Upper limits of 1.8 mJy and 1.2 mJy were used where there were non-detections from the LBA observations at 1.4 GHz and 2.3 GHz, respectively. 

\begin{figure}[p]
\epsscale{0.8}
\plotone{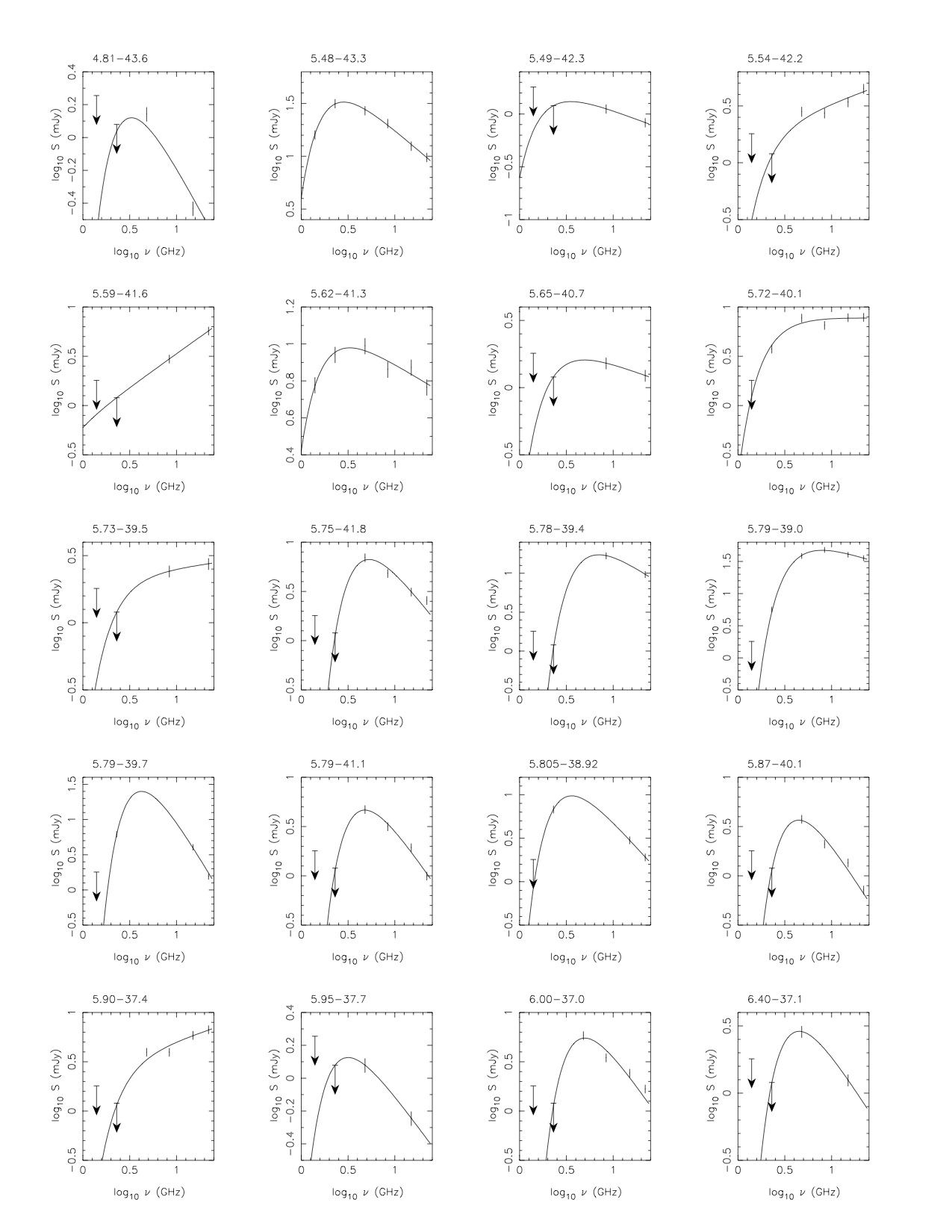}
\caption[Measured flux densities and free-free absorption models for 20 sources in NGC 253.]{Measured flux densities (points with error bars) and free-free absorption models (solid line) for 20 U\&A97 sources. The error bars are $\pm$10\% of the measured flux densities. 3$\sigma$ and 5$\sigma$ flux density upper limits are shown for sources not detected at 1.4 GHz (1.8 mJy) and 2.3 GHz (1.2 mJy) respectively.}
\label{fig:figff}            
\end{figure}

The resulting free parameters $S_{0}$, $\tau_{0}$ and $\alpha$ from the model fits are listed in Table \ref{tab:p1tabff}. The free-free absorbed power law model for each source is shown against the corresponding measured spectrum in Figure \ref{fig:figff}. The data shown in Table \ref{tab:p1tabff} and Figure \ref{fig:figff} confirm the free-free absorption interpretation of \citet{Tingay:2004p778}. In particular, the new 2.3 GHz data for 5.48$-$43.3 and 5.62$-$41.3 complete spectra that are very well fit by free-free absorbed power laws and are close to those adopted by \citet{Tingay:2004p778}. Furthermore, the results for the six sources listed in Table 2 of \citet{Tingay:2004p778} are consistent with the more stringently constrained free-free absorbed power law models listed in Table \ref{tab:p1tabff}.

Using similar criteria to that used by U\&A97, of the 20 sources listed in Table \ref{tab:p1tabff}, 9 have flat intrinsic power law spectra ($\alpha > -0.4$) indicative of \ion{H}{2} regions dominated by thermal radio emission (T), the 11 remaining sources have steep intrinsic spectra (S), as normally associated with supernova remnants \citep{McDonald:2002p3330}. The approximately equal number of thermal versus non-thermal sources is consistent with the findings of U\&A97 in NGC 253, however a slightly larger proportion of non-thermal sources is present in M82 \citep{McDonald:2002p3330}.

\subsection{Comparison with multi-wavelength data-sets}
\label{sec:p1spectra}

When the modelled free-free opacity is illustrated against the source location (Figure \ref{fig:figngc253}) it is clear that it does not vary smoothly across the central region of the galaxy, but is rather clumpy. Since free-free absorption is associated with ionised gas, it is worthwhile investigating tracers of ionised gas at other wavelengths in an attempt to understand the distribution of gas in these regions.

\emph{HST} images of NGC 253 at optical wavelengths provide the closest match in resolution to those obtained with the LBA and VLA. Of particular interest is the H$\alpha$ emission line as it traces diffuse ionised gas and is prominent in pre and post-main sequence stars. The \ion{S}{3} line has been used together with the H$\alpha$ line to highlight regions with photo-ionization from \ion{H}{2} regions and OB stars \citep{Forbes:2000p950}. \emph{HST} H$\alpha$ and [\ion{S}{3}] emission line images were made from archival \emph{HST} data and the images shifted so that the source co-ordinates were aligned with those listed by \citet{Forbes:2000p950}. The two images were then correlated against the free-free opacities modelled at the compact source locations. As many sources only had lower limits for the free-free optical depth, simple statistical techniques to test the relationship between measured intensity and the free-free opacity would not fully utilise all of the data available. Instead, 'survival analysis' methods were employed to take into consideration the lower limits when performing linear regression tests on the data.  The survival analysis was performed on the data using tasks in the ASURV package \citep{Lavalley:1992p10571}. The resulting correlation tests indicated that no trend was apparent between the H$\alpha$ or [\ion{S}{3}] emission line intensity and the modelled free-free optical depth for any of the sources. The lack of correlation is most likely due to the high levels of extinction that are apparent at optical wavelengths and are characteristic of starburst galaxies \citep{Engelbracht:1998p874}. This extinction makes properties deduced from optical measurements unrepresentative \citep[e.g.][]{Rieke:1980p1072, Puxley:1991p3674}.

\begin{table}[ht]
\begin{center}
\begin{tabular}{lrrrrl} \hline \hline
Source	&  $S_{0}$ (mJy) &   $\alpha$	&  $\tau_{0}$	& Type \\ \hline \hline
4.81$-$43.6 & 7.78		& -1.06		& $>$6.2     & S \\
5.48$-$43.3 & 104.7		& -0.77		& 3.3	   & S \\
5.49$-$42.3 & 2.34		& -0.33		& $>$2.2     & T    \\ 
5.54$-$42.2 & 1.66		& 0.30		& $>$3.6     & T    \\
5.59$-$41.6 & 0.70		& 0.68		& $>$0.16   & T    \\ 
5.62$-$41.3 & 16.0		& -0.31		& 1.8	   & T    \\
5.65$-$40.7 & 2.73		& -0.26		& $>$3.5     & T    \\ 
5.72$-$40.1 & 8.02		& -0.01		& 3.8	   & T    \\
5.73$-$39.5 & 2.05		& 0.10		& $>$3.6   & T    \\ 
5.75$-$41.8 & 93.2		& -1.23		& $>$20	& S \\
5.78$-$39.4 & 125.0		& -0.82		& $>$23	& S \\
5.79$-$39.0 & 155.9		& -0.47		& 17		& S \\
5.79$-$39.7 & 1694.0        & -2.21		& 22		& S \\
5.79$-$41.1 & 79.6		& -1.39		& $>$18	& S \\
5.805$-$38.92 & 78.9	& -1.20		& 8.4	& S    \\
5.87$-$40.1 & 71.7		& -1.51		& $>$17	& S \\
5.90$-$37.4 & 2.52		& 0.32		& $>$6.0   & T    \\
5.95$-$37.7 & 4.74		& -0.78		& $>$4.2   & S \\
6.00$-$37.0 & 95.8		& -1.37		& $>$19	& S \\
6.40$-$37.1 & 25.1		& -1.09		& $>$12	& S \\ \hline
\end{tabular}
\caption[Parameters of the free-free absorption models for all detected sources in NGC 253]{Parameters of the free-free absorption models for all compact sources, as discussed in the text. Thermal sources are shown with type T and non-thermal synchrotron sources are shown as type S.}
\label{tab:p1tabff}
\end{center}
\end{table}

To overcome the problems associated with extinction it is necessary to observe at wavelengths, such as X-ray and radio, that can penetrate gas and dust. Radio observations in particular also need to be toward the high frequency end of the spectrum to avoid the effects of free-free absorption. Australia Telescope Compact Array (ATCA) images of the Ammonia \citep{Ott:2005p785} and HCN emission lines \citep{ott04} and \emph{Chandra} hard X-ray images \citep{Weaver:2002p844} were obtained and compared against the modelled free-free optical depth at each of the source locations. Ammonia (NH$_{3}$) is a good temperature indicator whereas HCN is a good tracer for high density molecular gas ($>10^{4}$ cm$^{-3}$). The hard X-ray observations show regions of photo-ionization by massive young stars or SNR-driven shocks. All three images were analysed using a survival analysis against the modelled free-free optical depth with no clear trend evident. Here the lack of correlation is most likely due to the mismatches in resolution. In comparison to the 2.3 GHz data, the ATCA and \emph{Chandra} images were of substantially lower resolution and so were not able to resolve sufficient detail to provide a useful comparison with the VLBI data.

High resolution radio recombination line (RRL) observations with the VLA provide a reasonable match in resolution to the VLBI observations. Furthermore, the high frequency H92$\alpha$ (8.3094 GHz) and H75$\alpha$ (15.2815 GHz) lines are not adversely affected by the effects of free-free absorption and can be used to estimate temperatures, densities, and kinematics \citep{Lobanov:2005p10572}. The hydrogen radio recombination line (RRL), which occurs in thermal gas, is insensitive to dust extinction and can be used to probe highly obscured star formation.  \citet{Mohan:2005p784} have imaged RRL emission in NGC 253 at arcsecond resolution and detected 8 sources. By modelling the emission, \citet{Mohan:2005p784} provide size and electron density estimates for these sources. 

When the RRL source associated with the core is overlaid with the core position in the 2.3 GHz LBA continuum map, 6 of the sources modelled for free-free absorption in Table \ref{tab:p1tabff} are found to be located near 5 of the RRL sources. By assuming that the sources are centrally located within a uniform density sphere of ionised gas of temperature 7500 K, the expected free-free opacities have been estimated, the results listed in Table \ref{tab:p1tabrrlff}.

\begin{table}[ht]
\begin{center}
{ \footnotesize
\begin{tabular}{llccccccc} \hline \hline
                        &                      &                   &  \multicolumn{2}{c}{$n_{e}$}                     & \multicolumn{2}{c}{$l$} & \multicolumn{2}{c}{$\tau_{0}$} \\
                                                                          \cline{4-5}                                                    \cline{6-7}                        \cline{8-9}
RRL                 & VLA              &                   & low                          & high                         & low   & high   & low            & high           \\
Source             & Source         & $\tau_{0}$ & (10$^{3}$ cm$^{-3}$) & (10$^{3}$ cm$^{-3}$)  & (pc)  & (pc)    &                  &                  \\ \hline \hline
5.49$-$42.3     & 5.49$-$42.3 & $>$2.2       & 4                             & 10                            & 2      & 3.5     & 7.8             & 85             \\
5.59$-$41.6     & 5.59$-$41.6 & $>$0.16     & 2                             & 20                            & 1      & 5        & 1.0             &  490          \\
5.72$-$40.1     & 5.72$-$40.1 & 3.8             & 2                             & 10                            & 1.4   & 2.7     & 1.4             &   66           \\ 
5.75$-$39.9     & $\cdots$       & $\cdots$    & 2                             & 10                            & 1.7   & 2.9     & 1.7             & 71             \\
5.753$-$38.95 & $\cdots$       & $\cdots$    & 2                             & 10                            & 1.8   & 5        & 4.9             &  44            \\ 
5.773$-$39.54 & 5.78$-$39.4 & $>$23        & 0.9                          & 6                              & 2.6   & 6.4     & 1.3             & 32             \\
5.777$-$38.98 & $\cdots$       & $\cdots$    & 8                             & 10                            & 2      & 2.5     & 31              & 61             \\ 
5.795$-$39.05 & 5.805$-$38.92 & 8.4         & 7                             & 10                            & 2.5   & 2.5     & 30              & 61             \\
5.795$-$39.05 & 5.79$-$39.0 & 17              & 7                             & 10                            & 2.5   & 2.5     & 30              & 61             \\ \hline
\end{tabular}
\caption{Comparison of free-free opacity determined from source spectrum modelling and from radio recombination line modelling.}
\label{tab:p1tabrrlff}
}
\end{center}
\end{table}

Four of the VLA sources (5.49$-$42.3, 5.59$-$41.6, 5.72$-$40.1 and 5.78$-$39.4) have free-free opacities that are within the limits imposed by the RRL models (labelled as 'low' and 'high' in Table \ref{tab:p1tabrrlff}). The first three sources are identified as thermal regions. However 5.78$-$39.4 is non-thermal ($\alpha\sim-0.8$). The close proximity of 5.78$-$39.4 to the assumed core 5.79$-$39.0, where dense ionised gas is expected, together with the observed RRL emission, suggests that it may be a supernova remnant embedded in or behind a \ion{H}{2} region. 

The two VLA sources associated with the 5.795$-$39.05 RRL source, the assumed core (5.79$-$39.0) and a nearby non-thermal source (5.805$-$38.92) have free-free optical depths that fall well below the limits imposed by the RRL models. This suggests that the sources may not be centrally located within the gas distribution. If the sources are in front of or to the side of the high density region then they may exhibit the lower free-free optical depth that is inferred.

Many of the other VLA sources exhibit a significant degree of free-free absorption but do not have any associated RRL emission. Most of these sources are non-thermal and so may not necessarily have a RRL component. However, there are five thermal sources that do not have RRL sources associated with them. Four of these are weak and may fall below the sensitivity limit of the RRL observation. The remaining source has a combination of thermal and non-thermal emission (5.62$-$41.3) and it is possible that the thermal component is not strong enough to be detected within the sensitivity limits of the RRL observation. Interestingly, 5.62$-$41.3 does lie within a beam width of the RRL emission associated with 5.59$-$41.6 and were it associated with that emission it would have a free-free optical depth that is consistent with that expected from the source parameters modelled by \citet{Mohan:2005p784} from the RRL emission.

\subsection{Comments on individual compact radio sources}

\subsubsection{Resolved SNR 5.48$-$43.3}

Source 5.48$-$43.3 dominates observations at 1.4 GHz and 2.3 GHz. In our high resolution image (Figure \ref{fig:figsnr}) 5.48$-$43.3 appears to be resolved into a shell of approximately 90 mas in diameter corresponding to an extent of approximately 1.7 pc. If we assume an average expansion velocity of $v=10,000$ km s$^{-1}$ relative to the shell centre, similar to that measured for the supernova remnant 43.31+592 in M82 \citep{Pedlar:1999p3534}, this implies an age of $\sim 80 (10^{4}/v)$ years for this remnant. The eastern side of the remnant appears 2 to 3 times brighter than the western side. This may be the result of interaction with denser interstellar material in that direction. The circular symmetry of the remnant may indicate that the interaction has only recently begun or that the density of the ISM is only marginally greater.

With a total flux of 32 mJy at 2.3 GHz, 5.48$-$43.3 is approximately half the size and 25 times brighter than Cas A would be at this frequency if it were at the same distance. Furthermore, with the assumed expansion velocity, 5.48$-$43.3 is one quarter the estimated age of Cas A. 

\subsubsection{Assumed Core 5.79$-$39.0}

The measured size of 5.79$-$39.0 in the 2.3 GHz LBA image is approximately 80$\times$60 mas, consistent with the lower limit of 30 mas measured with the Parkes-Tidbinbilla Interferometer \citep{Sadler:1995p10581} and the upper limit of $\sim100$ mas measured by U\&A97 with 8 GHz VLA observations. 5.79$-$39.0 is not detected in the full resolution LBA image at 2.3 GHz as would be expected given its size and flux density.

5.79$-$39.0 is the strongest source in 5$-$23 GHz VLA images but is not visible in 1.4 GHz LBA observations \citep{Tingay:2004p778}. In our 2.3 GHz LBA image the source is detected but is weaker than expected based on its high frequency spectral index, suggesting that there is significant absorption at lower frequencies. Our source modelling (\S\ref{sec:p1ffmodel}) shows that the source is best fit with a free-free absorbed power law model with optical depth $\tau_{0}\sim17$. Evidence of free-free absorption in the nuclear region has also been found by \citet{Reynolds:1983p1062}, \citet{Carilli:1996p10536} and \citet{Tingay:2004p778}.

\subsubsection{Resolved SNRs 5.62$-$41.3 and 5.79$-$39.7}

The source 5.62$-$41.3 lies in a region that has the highest HCN/CO ratio and thus the highest average density ($\sim10^{5}$ cm$^{-3}$) of any region studied by \citet{Paglione:1995p946}. This is one of only two sources detected at 1.4 GHz with the LBA \citep{Tingay:2004p778}. U\&A97 suggest that the source is a mix of supernova remnants and \ion{H}{2} regions. In our 2.3 GHz LBA observations the source is extended ($130\times85$ mas) in the low resolution image (Figure \ref{fig:figngc253}) and is not detected in the full resolution image. It does not appear to be greatly affected by free-free absorption and so may lie in front of the dense gas detected by \citet{Paglione:1995p946}. Its modelled spectral index of $-0.31$ is at the low end of that measured by U\&A97.

Source 5.79$-$39.7 is the smallest supernova remnant in our 2.3 GHz LBA image but is not detected in the higher resolution image. It also appears to be in a region with significant levels of ionised gas as it is heavily affected by free-free absorption.

\subsubsection{The \ion{H}{2} region 5.72$-$40.1}
Source 5.72$-$40.1 is a flat spectrum source which has been observed to vary in its flux significantly at 5 GHz and 8.3 GHz from VLA observations (U\&A97). Our measured size of $60\times50$ mas is significantly smaller than the size of $230\times120$ mas measured at 23 GHz and $200\times90$ mas measured at 15 GHz using the VLA (U\&A97). It also has a position offset that is not as consistent as the other sources when compared against VLA positions. These two facts suggest that it may have structure that varies significantly with frequency. This source is associated with RRL emission that is consistent with a free-free absorbed \ion{H}{2} region.

\subsection{The supernova rate in NGC 253}
\label{sec:p1snrate}
With new observations made since those of U\&A97 and with an improved distance determination to NGC 253 \citep{Karachentsev:2003p10562} it is worth reinvestigating the limits that may be placed on the supernova rate, discussed in detail by U\&A97.  The increased distance estimate to NGC 253 suggests that earlier spatial measures were underestimated by a factor of $\sim1.6$. Similarly, luminosity and energy estimates would have been underestimated by a factor of $\sim2.5$. The overall effect on the population of observed supernova remnants in NGC 253 is that they are not only older but also more luminous than previously estimated.

An upper limit to the supernova rate can be made by assuming that all of the FIR luminosity is reprocessed supernova energy, that is, all energy released by supernovae is eventually thermalized. Using this assumption \citet{Antonucci:1988p10521} determined an upper limit of 3 yr$^{-1}$ for the supernova rate. Adjusting the original FIR luminosity measured by \citet{Telesco:1980p9718} for a distance of 3.94 Mpc gives a total FIR luminosity of $3.8\times10^{10} L_{\Sun}$ or $\sim1.4\times10^{44}$ ergs s$^{-1}$ for NGC 253. Dividing this by the canonical $10^{51}$ ergs of total energy output per supernova results in a maximum rate of $\sim4.5$ supernovae per year. This method has been shown to grossly overestimate supernova rates since FIR emission also results from stellar heating of dust and can be contaminated by the presence of an AGN \citep{Condon:1992p10540}.

A lower limit on the supernova rate can be made based on the number of detected remnants, their size and an assumed expansion rate. \citet{Ulvestad:1994p1041} assumed that of order 100 compact sources were supernova remnants with upper size limits of 2$-$5 parsecs based on a distance of 2.5 Mpc to NGC 253. Furthermore, assuming an expansion at 5,000 km s$^{-1}$ (rate of expansion of the FWHM of a Gaussian model), \citet{Ulvestad:1994p1041} arrived at a supernova rate of 0.1$-$0.25 yr$^{-1}$. As only $\sim50\%$ of the compact radio sources in NGC 253 are associated with supernova remnants (U\&A97) and the revised source sizes are now of order 3$-$8 parsecs, a supernova rate of $>0.14-0.36$ is expected given an assumed expansion rate of 10,000 km s$^{-1}$. Since the mean expansion rate of the supernova remnants in NGC 253 is unclear, we can rewrite the limit as $\nu_{SN}>0.14 (v/10^{4})$ yr$^{-1}$, where $v$ is the shell radial expansion velocity in km s$^{-1}$. This is of the same order of magnitude as the supernova rate determined for M82, using similar methods, $\sim0.07$ yr$^{-1}$ \citep{Pedlar:2003p3309}.

Having observed NGC 253 at a second epoch, \citet{Ulvestad:1991p1008} found that no new sources stronger than 3 mJy at 5 GHz were detected after a period of 18 months. They statistically modelled the effect of such non-detections on the supernova rate. Their model assumed a hypothetical population of supernovae that peaked at 5 GHz 100 days after their optical maxima, had a 5 GHz peak flux that was with equal probability either 5, 10, 15, or  20 times that of Cas. A and decayed as the $-$0.7 power of time. \citet{Ulvestad:1991p1008} determined that approximately $2/3$ of supernovae that occur during the 18 month period between epochs should be detectable. Further assuming that the occurrence of supernovae in NGC 253 are Poisson distributed, \citet{Ulvestad:1991p1008} determined with 95\% confidence that the upper limit of the supernova rate was 3.0 yr$^{-1}$. The same model was used to determine an upper limit of $\sim1.4$ yr$^{-1}$ at the end of a third epoch \citep{Ulvestad:1994p1041} and $\sim0.3$ yr$^{-1}$ at the end of the fourth epoch (U\&A97).

Since the fourth epoch two further high resolution radio observations of NGC 253 have been made. The first was four years later by \citet{Mohan:2005p784} at 5.0 GHz and the second was the observation made 4.8 years later again at 2.3 GHz as described in this paper. Although the \citet{Mohan:2005p784} image can be directly compared to the historical data at the same frequency, the observations at 2.3 GHz require extra consideration of the nature of supernova remnants at lower frequencies, the differing sensitivity limits,  and the effects of free-free absorption. Regardless of these differences, what is clear is that no new sources are evident in either the \citet{Mohan:2005p784} images or our 2.3 GHz image.

To place limits on the supernova rate based on these observations, we have developed a new model, based in principle on that of \citet{Ulvestad:1991p1008}. At 5 GHz the hypothetical supernova for this model is identical to the one described in \citet{Ulvestad:1991p1008}, however at 2.3 GHz it is assumed that the supernova remnant will peak 200 days after the optical peak, albeit at the same flux density as at 5 GHz - this approximately follows the SN 1980k light curve determined by \citet{Weiler:1986p9393}.

A Monte-Carlo simulation was used to drive the model by randomly choosing a uniformly distributed peak luminosity between 5 and 20 times that of Cas A., which was then scaled to flux density for a given distance to NGC 253 and also adjusted for free-free absorption using Equation \ref{eq:ff}. A Poisson distributed random number was then used to define how many supernovae, given a specified supernova rate, would occur between two epochs and at which uniformly distributed times these supernovae would occur. Given the peak flux density and timing of the supernova, a test was made to determine if each supernova remnant could be detected when observed at the subsequent epoch given the sensitivity limit for the observation. By executing the Monte-Carlo simulation over 10,000 iterations, the proportion of supernova remnants detected at the end of each epoch was determined. The simulation was seeded with a supernova rate of 0.1 yr$^{-1}$ across all epochs, the resulting output giving the confidence level for a non-detection at each epoch at that rate. Linear interpolation was then used to determine a new rate that would drive the simulation towards a 95\% confidence level and the simulation was repeated until the 95\% level was reached.

Table \ref{tab:p1tabsnrate} shows the results of three separate tests with the Monte-Carlo simulation. In each test, a run was made for each epoch to place a limit on the supernova rate at the time of that observation. Also, given that a supernova remnant can peak and fade beneath the sensitivity limit between two epochs, the proportion of supernova remnants ($\beta_{SN}$) that can actually be detected at the end of each epoch was also noted.

\begin{table}[ht]
\begin{center}
{ \small
\begin{tabular}{llllllllll} \hline \hline
           &               &                      &  & \multicolumn{2}{c}{Test 1\tablenotemark{a}} & \multicolumn{2}{c}{Test 2\tablenotemark{b}} & \multicolumn{2}{c}{Test 3\tablenotemark{c}} \\
                                                                          \cline{5-6}                                                    \cline{7-8}                        \cline{9-10}
Epoch & Time    & $\nu_{obs}$  & Sensitivity   & $\beta_{SN}$ & $\nu_{SN}$   & $\beta_{SN}$ & $\nu_{SN}$ & $\beta_{SN}$    & $\nu_{SN}$  \\ 
           & (yr)       & (GHz)            & (mJy)          &                   & (yr$^{-1}$) &  & (yr$^{-1}$) &  & (yr$^{-1}$) \\ \hline \hline
1         & $\cdots$& 5.0               & 3.0               & $\cdots$    & $\cdots$ & $\cdots$& $\cdots$& $\cdots$& $\cdots$ \\  
2         & 1.5        & 5.0                & 3.0              & 0.828         & $<2.4$     & 0.220         & $<9.1$   & 0.128          & $<16$       \\
3         & 2.5        & 5.0                & 3.0              & 0.678         & $<1.0$     & 0.132         & $<4.5$   & 0.077          & $<7.8$      \\
4         & 4.0        & 5.0                & 3.0              & 0.477         & $<0.62$   & 0.083         & $<3.0$   & 0.048          & $<5.2$      \\
5         & 4.0        & 5.0                & 3.0              & 0.477         & $<0.44$   & 0.083         & $<2.3$   & 0.048          & $<3.9$      \\
6         & 4.8        & 2.3                & 1.2              & 0.993         & $<0.26$   & 0.697         & $<0.64$ & 0.100          & $<2.4$      \\  \hline
\tablenotetext{a}{Monte Carlo test run with a distance of 2.5 Mpc.}
\tablenotetext{b}{Monte Carlo test run with a distance of 3.94 Mpc.}
\tablenotetext{c}{Monte Carlo test run with a distance of 3.94 Mpc and a free-free opacity of $\tau_{0}=6.0$.}
\end{tabular}
\caption[NGC 253 supernova rate upper limit based on Monte Carlo simulations.]{The supernova rate upper limit based on Monte Carlo simulations run over six observing epochs. The time between epochs, the observing frequency and sensitivity of the observation are listed. At the end of each epoch the proportion of supernova remnants detected ($\beta_{SN}$) in that epoch is listed together with an estimate of the supernova rate upper limit based on all observations prior to and including that epoch. }
\label{tab:p1tabsnrate}
}
\end{center}
\end{table}

In test 1 the distance to NGC 253 was set to 2.5 Mpc and no free-free absorption was assumed in order to provide a direct comparison to the \citet{Ulvestad:1991p1008, Ulvestad:1994p1041, Ulvestad:1997p907} results.  At epoch 2, the first test resulted in a supernova rate of $<2.4$ yr$^{-1}$ compared to $<3.0$ yr$^{-1}$ obtained by \citet{Ulvestad:1991p1008}. This difference can be attributed to the proportion detected ($\sim83$\%) being greater than initially anticipated with a simpler model ($\sim67$\%). The proportion detected at epoch 3 ($\sim68$\%) is similar to that predicted by \citet{Ulvestad:1991p1008} however the cumulative effect of a greater rate of detection at the second epoch results in a supernova rate of only $<1.0$ yr$^{-1}$ compared to $<1.4$ yr$^{-1}$ obtained by \citet{Ulvestad:1994p1041}. At the fourth epoch a lower proportion of detections are predicted ($\sim48$\%) as a result of an increased number of supernova remnants fading below the sensitivity limits over the longer period between observations, this results in a supernova rate of $<0.62$ yr$^{-1}$ which is almost twice the maximum predicted by U\&A97. Given further observations by \citet{Mohan:2005p784} and those presented in this paper it is possible to add two further epochs to the test which in the absence of further detections results in a further lowering of the maximum allowable supernova rate to $<0.44$ yr$^{-1}$ and $<0.26$ yr$^{-1}$ for epochs 5 and 6 respectively. The results of our simulations show that the model originally used by \citet{Ulvestad:1991p1008} does not accurately determine the supernova rate when the time between epochs exceeds approximately 2 years, under the assumptions used.

In test 2 the Monte-Carlo simulation was repeated for NGC 253 at the revised distance of 3.94 Mpc. The increased distance results in more supernova remnants falling below the sensitivity limits of the observations. For observations at 5 GHz this results in the possibility of only 8$-$22\% of all supernovae occurring between epochs being detected. The overall effect is to increase the upper limit of the supernova rate to 3.0 yr$^{-1}$ at the end of epoch 4 and to 0.6 yr$^{-1}$ at the end of epoch 6.

It has been shown in \S~\ref{sec:p1spectra} that the supernova remnants in NGC 253 appear to be situated in or behind a screen of ionised gas. In test 3 the effects of free-free absorption on the upper limit of the supernova rate are taken into consideration with the Monte-Carlo simulation. For this test a median value of $\tau_{0}=6$ was taken based on free-free modelling in \S~\ref{sec:p1spectra}. The effect of free-free absorption at 5 GHz is not greatly significant but is sufficient to drop the proportion of detections to between 5\% and 13\%. At 2.3 GHz the effect of free-free absorption is more significant with the proportion of supernova remnants that can be detected dropping from 70\% to 10\%. Overall, this results in upper limits on the supernova rate of 5.2 yr$^{-1}$ at epoch 4 and 2.4 yr$^{-1}$ at epoch 6.

The overall supernova rate upper limit of 2.4 yr$^{-1}$ is below the upper limit determined from FIR emission, however it is more than an order of magnitude greater than that determined from source counts and sizes alone. U\&A97 observed no significant source fading over a period of 8 years and on this basis they could further limit the supernova rate to no greater than 0.3$-$0.6 yr$^{-1}$. The apparent lack of significant variability on time-scales of many years has also been observed for a majority of compact radio sources in M82 by \citet{Kronberg:2000p10558}. It is possible that the dense environment around the nuclear regions of NGC 253 and M82 inhibit the expansion and fading of the remnants. If this is the case then the remnants may be older than they appear and this could reduce the supernova rate further. It is also possible that we are only seeing the tip of the iceberg, there may be a large background supernova rate but we only observe the rare bright events.

To improve the supernova rate estimates further it would be useful to refine the Monte-Carlo simulator to include a more realistic probability distribution for the supernova remnant peaks and make better use of the \citet{Weiler:1986p9393} radio supernova light curves. A reduction in the upper limit of the supernova rate can be achieved by increasing the probability of detecting new supernovae as they occur. This may be achieved with frequent high sensitivity observations of NGC 253. Furthermore, through multi-epoch observations, it should be possible to measure the rate of expansion of the brighter more resolved remnants and so allow for age determination at a greater level of accuracy and subsequently improved estimates of the supernova rate. 

\subsection{The star formation rate in NGC 253}
\label{sec:p1sfrate}

The star-formation rate (SFR) of a star-forming galaxy is directly proportional to its radio luminosity $L_{\nu}$ at wavelength $\nu$ \citep{Condon:1992p10540, Haarsma:2000p10556}:
\begin{equation}
\left( \frac{SFR(M\geq5M_{\Sun})}{M_{\Sun}\mathrm{yr}^{-1}} \right)=Q \left \{ \frac{ \frac{L_{\nu}}{\mathrm{W Hz^{-1}}}}{\left[ 5.3\times10^{21} \left( \frac{\nu}{\mathrm{GHz}}\right) ^{-0.8} + 5.5\times10^{20} \left( \frac{\nu}{\mathrm{GHz}} \right) ^ {-0.1} \right]} \right \}.
\end{equation}
\citet{Condon:1992p10540} derives this relation purely from radio considerations by calculating the contribution of synchrotron radio emission from SN remnants and of thermal emission from \ion{H}{2} regions to the observed radio luminosity. The factor $Q$ accounts for the mass of all stars in the interval $0.1-100 M_{\Sun}$ and has a value of 8.8 if a Saltpeter IMF ($\gamma=2.5$) is assumed - this IMF is assumed for the remainder of this section. \citet{Ott:2005p785} used a measure of the 24 GHz continuum emission in NGC 253 to derive a SFR which, when adjusted for a distance of 3.94 Mpc, gives a SFR of 10.4$\pm$1 M$_{\Sun}$ yr$^{-1}$.

A large proportion of the bolometric luminosity of a galaxy is absorbed by interstellar dust and re-emitted in thermal IR. As a result FIR emission is a good tracer for young stellar populations. By modelling the total energy emission of a massive star and assuming that the contribution of dust heating by old stars is negligable, \citet{Condon:1992p10540} determined the following relation between the star-formation rate of a galaxy and the FIR luminosity $L_{IR}$:

\begin{equation}
\label{eq:sfr2lfir}
\left( \frac{SFR(M\geq5M_{\Sun})}{M_{\Sun}\mathrm{yr}^{-1}} \right)=9.1\times10^{-11}L_{FIR}/L_{\Sun}.
\end{equation}
Using \citet{Radovich:2001p10551} FIR measures for NGC 253, adjusted for a distance of 3.94 Mpc, gives a SFR of 1.8$-$2.8 M$_{\Sun}$ yr$^{-1}$ for the inner $\sim300$ pc nuclear region and 3.5$-$4.3 M$_{\Sun}$ yr$^{-1}$ for the entire galaxy.

Finally, the star formation rate can be determined from the supernova rate directly. For most galaxies the total radio non-thermal and thermal luminosities and the FIR/radio ratio is proportional to the average star formation rate for $M\geq5M_{\Sun}$ \citep{Condon:1992p10540}. Furthermore, as all stars more massive than $8M_{\Sun}$ become radio supernova, the star formation rate can be determined using the following relationship: 

\begin{equation}
\label{eq:sfr2snr}
\left( \frac{SFR(M\geq5M_{\Sun})}{M_{\Sun}\mathrm{yr}^{-1}}\right) \sim 24.4 \times \left[ \frac{\nu_{SN}}{\mathrm{yr}^{-1}}\right].
\end{equation}
Using the supernova rate limits of $0.14 (v/10^{4})<\nu_{SN}<2.4$ yr$^{-1}$ (\S~\ref{sec:p1snrate}) the limits on the star formation rate are $3.4 (v/10^{4}) < SFR(M\geq5M_{\Sun})<59$ M$_{\Sun}$ yr$^{-1}$. This is of the same order of magnitude as the nuclear SFR determined from FIR emissions and in agreement with that determined from 24 GHz continuum emissions alone. Using Equations \ref{eq:sfr2lfir} and \ref{eq:sfr2snr}, the supernova rate can be estimated from the FIR-based SFR estimate giving a value of 0.07$-$0.11 yr$^{-1}$. This figure is in line with measurements based on source sizes and suggests that the simple model used to determine the supernova rate from FIR emission in \S~\ref{sec:p1snrate} was grossly overestimating the supernova rate. It is encouraging that the three separate estimates of the SFR in NGC 253 all produce results that are only a factor of a few different. These estimates are similar to star-formation rates derived from the supernova rate and FIR emission in M82 which give values of $\sim1.8$ and $\sim2.0$ M$_{\Sun}$ yr$^{-1}$ respectively \citep{Pedlar:2003p3309}.

\section{Summary}

We have imaged NGC 253 at 2.3 GHz using the LBA to produce the highest resolution image of the nuclear starburst region of this galaxy to date. We find the following results:
\begin{itemize}
\item Six compact radio sources were detected and identified against higher frequency VLA observations (U\&A97). Two of these sources are also identified in a 1.4 GHz LBA image \citep{Tingay:2004p778}.
\item In the highest resolution image (13 mas beam), the supernova remnant\linebreak[4]\mbox{5.48$-$43.3} is resolved into a shell-like structure approximately 90 mas (1.7 pc) in diameter with the eastern limb substantially brighter than the west. Assuming an average radial expansion velocity of $v=10,000$ km s$^{-1}$, the remnant is estimated to be approximately $80 (10^{4}/v)$ years of age.
\item By combining flux density measures from 1.4 GHz LBA, 2.3 GHz LBA and high frequency VLA observations, with upper flux density limits for non-detections, the spectra for 20 compact radio sources were determined. The spectra of these sources were fit by a free-free absorbed power law model.
\item 12 of the 20 sources have steep intrinsic spectra normally associated with supernova remnants, the eight remaining sources have flat intrinsic power law spectra ($\alpha>-0.4$) indicative of \ion{H}{2} regions.
\item Based on the modelled free-free opacities of 20 sources, the morphology of the ionised medium in the central region of NGC 253 is complex and clumpy in nature.
\item Multi-wavelength comparisons of the free-free opacity against optical tracers of ionised gas, such as H$\alpha$ and [\ion{S}{3}] emission line images, show no significant correlation. The lack of correlation here is most likely due to the characteristically high levels of interstellar extinction associated with starbursts.
\item Multi-wavelength comparisons of the free-free opacity against non-optical tracers of ionised gas, such as X-ray and ammonia and HCN emission line images, also failed to show any significant correlation. The lack of correlation here is attributed to the  substantially lower resolution of these images compared to the LBA images.
\item A comparison with radio recombination line images show that four of the modelled sources have free-free optical depths expected by RRL models.
\item A supernova rate upper limit of 2.4 yr$^{-1}$ in the inner 320 pc region of NGC 253 was derived from the absence of any new sources, taking into consideration the improved distance measure to the galaxy, a median free-free opacity and the sensitivity limits of 6 observations over a period of 16.8 years. A supernova rate of $>0.14 (v/ 10^{4})$ yr$^{-1}$ has been estimated based on an estimate of supernova remnant source counts, their sizes and their expansion rates.
\item Both upper and lower limits on the supernova rate could be further constrained with more frequent, high sensitivity observations with the LBA or the VLBA.
\item A star formation rate of $3.4 (v/10^{4}) < SFR(M\geq5M_{\Sun})<59$ M$_{\Sun}$ yr$^{-1}$ has been estimated directly from supernova rate limits for the inner 320 pc region of the galaxy.
\end{itemize}

\linespread{1.0}
\normalsize
\begin{savequote}[20pc]
\sffamily
Something unknown is doing we don't know what.
\qauthor{Sir Arthur Eddington}
\end{savequote}

\chapter{The Compact Radio Source Population of NGC 4945}
\label{chap:ngc4945}
\begin{center}
{\it Adapted from:}

E. Lenc \& S.J. Tingay

Astronomical Journal, 137, 537 (2009)
\end{center}
\small
Wide-field, very long baseline interferometry (VLBI) observations of the nearby starburst galaxy NGC 4945, obtained with the Australian Long Baseline Array (LBA), have produced 2.3 GHz images over two epochs with a maximum angular resolution of 15 mas (0.3 pc). 15 sources were detected, 13 of which correspond to sources identified in higher frequency (3 cm and 12 mm) ATCA images. Four of the sources are resolved into shell-like structures ranging between 60 and 110 mas (1.1 to 2.1 pc) in diameter. From these data the spectra of 13 compact radio sources in NGC 4945 were modelled; nine were found to be consistent with free-free absorbed power laws and four with a simple power law spectrum. The free-free opacity is highest toward the nucleus but varies significantly throughout the nuclear region ($\tau_0\sim 6-23$), implying that the overall structure of the ionised medium is clumpy. Of the 13 sources, 10 have steep intrinsic spectra associated with synchrotron emission from supernova remnants, the remaining sources have flat intrinsic spectra which may be associated with thermal radio emission. A non-thermal source with a jet-like morphology is detected $\sim1\arcsec$ from the assumed location of the AGN. A type II supernova rate upper limit of 15.3 yr$^{-1}$ is determined for the inner 250 pc region of the galaxy at the 95\% confidence level, based on the lack of detection of new sources in observations spanning 1.9 years and a simple model for the evolution of supernova remnants. A type II supernova rate of $>0.1 (v/ 10^{4})$ yr$^{-1}$ is implied from estimates of supernova remnant source counts, sizes and expansion rates, where $v$ is the radial expansion velocity of the supernova remnant in km s$^{-1}$. A star formation rate of $2.4 (v/10^{4}) < SFR(M\geq5M_{\Sun})<370$ M$_{\Sun}$ yr$^{-1}$ has been estimated directly from the supernova rate limits and is of the same order of magnitude as rates determined from integrated FIR (1.5 M$_{\Sun}$ yr$^{-1}$) and radio luminosities ($14.4\pm1.4$ $(Q/8.8)$ M$_{\Sun}$ yr$^{-1}$). The supernova rates and star formation rates determined for NGC 4945 are comparable to those of NGC 253 and M82.

\clearpage
\linespread{1.3}
\normalsize
\section{Introduction}
\label{sec:p3introduction}
This paper is the second of a series of papers on the sub-parsec scale properties of local ($D<10$ Mpc), bright ($S_{1.4}>10$ mJy) southern ($\delta<-20\arcdeg$) starburst galaxies. Starburst galaxies are the result of processes acting on a wide range of scales and themselves drive energetic activity on a wide range of scales. In Paper I \citep{Lenc:2006p6695} of this series we discussed the properties of the nuclear starburst region of NGC 253. Direct observations of supernova remnants enabled an investigation of the star-formation and supernova history. Furthermore, high-resolution radio observations were used to map the ionized gaseous environment of the starburst region. These observations provide a link between the large-scale dynamical effects that feed the starburst region, the activity in the starburst region, and the energetic phenomenon that in turn is driven by the starburst. In the present work (Paper II of the series) we discuss the sub-parsec scale properties of NGC 4945 and implications on the star-formation and supernova rate of this galaxy.

NGC 4945 is a nearly edge-on spiral galaxy associated with the Centaurus group of galaxies at a declination of $\sim-40\arcdeg$ \citep{Webster:1979p9704}. The galaxy has been classified as SB(s)cd or SAB(s)cd and has an optical extent of 17 arcmin \citep{deVaucouleurs:1964p10569, Braatz:1997p10531}. Distance estimates for NGC 4945 have varied between 3.7 \citep{Mauersberger:1996p7843} and 8.1 Mpc \citep{Baan:1985p10435}. More recently, \citet{Karachentsev:2007p7477} have determined the magnitude of stars at the tip of the red giant branch using \emph{Hubble Space Telescope} (\emph{HST}) Advanced Camera for Surveys (ACS) images and arrived at a more reliable distance estimate of $3.82\pm0.31$ Mpc. This new distance estimate will be adopted throughout this paper and implies a spatial scale of $\sim19$ pc arcsec$^{-1}$.

While it is classified as a Seyfert 2 galaxy \citep{Braatz:1997p10531, Madejski:2000p9093}, it exhibits both starburst and Seyfert characteristics \citep[e.g.][]{Whiteoak:1986p9113}. The galaxy is one of the strongest and richest sources of extragalatic molecular lines with $\sim50$ line features detected to date \citep[e.g.][]{Henkel:1994p9082, Curran:2001p10543, Wang:2004p6109}. Evidence of an active galactic nucleus (AGN) comes from strong and variable hard X-ray emission \citep{Iwasawa:1993p6812}. H$_{2}$O megamaser emission has been observed from the nucleus \citep{Greenhill:1997p7509} and is assumed to be distributed in a $\sim3$ pc disk around the AGN.

Although the central region of NGC 4945 is highly obscured by gas and dust at optical wavelengths \citep{Madejski:2000p9093}, it is one of the three brightest IRAS point sources beyond the Magellanic Clouds \citep{Sanders:2003p7880}. Neutral hydrogen observations of NGC 4945 reveal the signature of a bar associated with the disk and can be traced out to a radius of $\sim7$ kpc \citep{Ott:2001p3998}. CO(2$-$1) observations suggest that the bar extends inwards to within $\sim100$ pc of the galaxy centre \citep{Chou:2007p10537}, where it may act to feed gas and dust into a fast-rotating disk of cold gas \citep{Dahlem:1993p10544, Mauersberger:1996p7843}. HNC(1-0) maps of the disk reveal a dense inner region \citep{HuntCunningham:2005p10542} that may fuel a circumnuclear starburst ring of similar extent observed by \citet{Marconi:2000p7671} with \emph{HST} PA$\alpha$ images. \citet{Chou:2007p10537} also detect a smaller-scale ($r<27$ pc), high density, kinetically de-coupled component in the nuclear region and suggest that it may be evidence of a circumnuclear torus around the AGN \citep{Antonucci:1993p10522}. 

Associated with the nuclear region, and having its origin in the inner starburst region, is a powerful wind that forms a cone-shaped plume perpendicular to the disk of the galaxy. The plume, extending $\sim500$ pc north-west, is clearly observed in H$\alpha$ spectral line observations \citep{Rossa:2003p10550} and \emph{Chandra} soft X-ray images \citep{Schurch:2002p7434}. Such plumes have been found in other nearby starburst galaxies \citep[e.g.][]{Strickland:2004p9721} and are believed to be driven by supernovae in the starburst region \citep{Doane:1993p10547,Strickland:2004p1157}.

The bar associated with the disk of the galaxy, a fast-rotating disk of cold gas, a starburst ring and the conical nuclear outflow are all characteristics that are observed in other nearby starburst galaxies such as NGC 253 \citep{Lenc:2006p6695} and M82 \citep{ForsterSchreiber:2000p8653}. These features are interpreted in terms of bar-induced gas dynamics and star formation. Even though NGC 4945, NGC 253 and M82 are associated with different galaxy groups and are situated in different locations in the celestial sphere, they share several further similarities. All three galaxies are nearly edge-on, all have distances close to 4 Mpc and all have similar infrared luminosities \citep{Rice:1988p7793}. It is these similarities that make them particularly interesting targets for study and comparison in other respects.

\begin{figure}[ht]
\epsscale{1.0}
\plotone{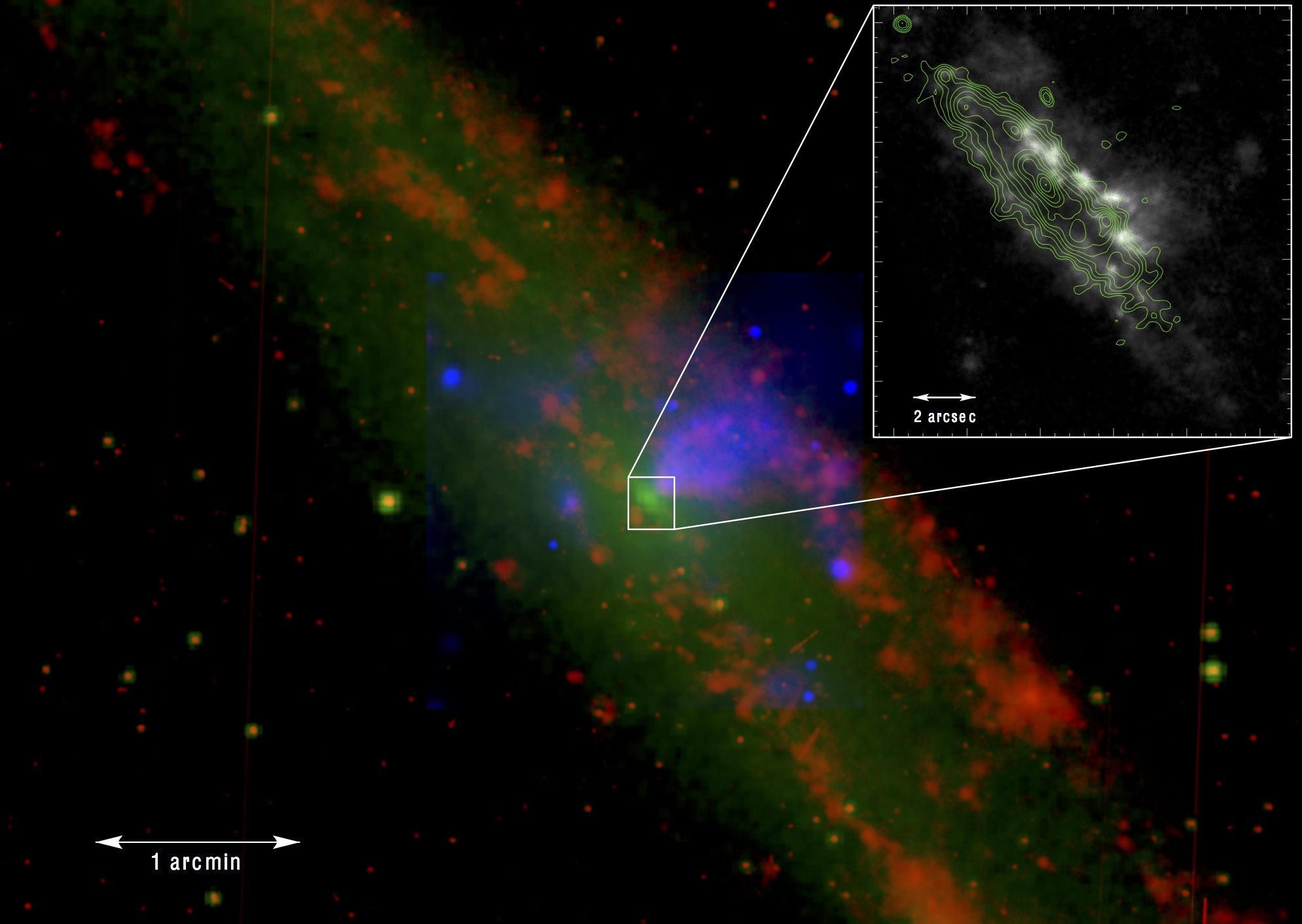}
\caption[A three-colour composite, multi-wavelength view of NGC 4945]{A three-colour composite, multi-wavelength view of NGC 4945. Red, green and blue indicate H$\alpha$ \citep{Rossa:2003p10550}, K$-$band Two Micron All Sky Survey IR \citep{Jarrett:2003p8453}, and \emph{Chandra} soft X-ray \citep{Schurch:2002p7434} respectively. Inset: \emph{HST} Pa$\alpha$ image \citep{Marconi:2000p7671} with ATCA 23 GHz contours overlaid, contours are logarithmic intervals of $2^{1/2}$, beginning at 1.02 mJy beam$^{-1}$.}
\label{fig:p3figmultiwav}            
\end{figure}

Figure \ref{fig:p3figmultiwav} presents images of NGC 4945 at various wavelengths, on different spatial scales, to illustrate the relationships between the distribution of cool stars (K-band), ionised hydrogen gas (H$\alpha$), the inner starburst region (radio and Pa$\alpha$), and the wind emanating from that region (H$\alpha$ and X-ray). The supernova remnants that are a primary subject of this paper are concentrated within the inner 250 pc of the galaxy. 

The observation of new supernovae and their associated remnants in the starburst regions of nearby galaxies is rare as these regions are highly obscured by gas and dust. Only a few have been observed in local starbursts and are generally revealed in X-rays where such events are more conspicuous and then followed up with multi-wavelength observations. In a few cases these can be traced back to an earlier optical or infrared outburst, the presumed supernova event, as with SN 1996cr in Circinus \citep{Bauer:2007p10525} and SN 1978K in NGC 1313 \citep{Smith:2007p7796, Ryder:1993p7804}. However, in most cases, where the remnants are embedded deep in the starburst region and only X-ray and radio observations can reveal their presence, they are discovered long after the event and their ages remain uncertain. Such a population of supernova remnants has been discovered in M82 \citep{Pedlar:1999p3534, McDonald:2002p3330} and NGC 253 \citep{Tingay:2004p778, Lenc:2006p6695} using VLBI observations.

Motivated by the existence of compact radio sources and strong indicators of free-free absorption in similar starburst galaxies, such as NGC 253 and M82, we have embarked on a program of VLBI observations of NGC 4945 and other prominant Southern Hemisphere starburst galaxies. This paper reports results from new observations of the nuclear region of NGC 4945 at a range of frequencies between 2.3 GHz and 23 GHz. The resulting data have better resolution and sensitivity than previous observations and we have used them to investigate in detail the free-free absorbed spectra of supernova remnants and \ion{H}{2} regions in the NGC 4945 starburst (\S~\ref{sec:p3ffmodel}). In particular we are interested in probing the structure of the ionised environment of the supernova remnants, constraining the supernova rate in NGC 4945 using our new ATCA and VLBI images (\S~\ref{sec:snrate}) and re-evaluating estimates of the star formation rate based on our results (\S~\ref{sec:sfrate}).

A subsidiary focus of this project is to achieve high sensitivity, high fidelity, high resolution, and computationally efficient imaging over wide fields of view (at least compared to traditional VLBI imaging techniques).  As such, the associated exploration of the relevant techniques is a small first step toward the much larger task of imaging with the next generation of large radio telescopes, for example the Square Kilometre Array (SKA: \citet{Hall:2005p10570, Carilli:2004p10534}).  The SKA will use baselines ranging up to approximately 3000 km and cover fields of view greater than one square degree instantaneously at frequencies of around 1 GHz.  The computational load required for this type of imaging task is vast and will drive the development of novel imaging and calibration algorithms and the use of super-computing facilities.  Investigating the performance of these techniques now, under the most challenging current observing conditions, is therefore a useful activity.

\section{Observations, correlation and data reduction}
\subsection{LBA Observations}
A VLBI observation of NGC 4945 was made on 13/14 May, 2005 using a number of the Long Baseline Array (LBA) telescopes: the 70 m NASA Deep Space Network (DSN) antenna at Tidbinbilla; the 64 m antenna of the Australia Telescope National Facility (ATNF) near Parkes; 5 $\times$ 22 m antennas of the ATNF Australia Telescope Compact Array (ATCA) near Narrabri used as a phased array; the ATNF Mopra 22 m antenna near Coonabarabran; the University of Tasmania's 26 m antenna near Hobart; and the University of Tasmania's 30 m antenna near Ceduna.  The observation utilised the S2 recording system \citep{Cannon:1997p10533} to record 2 $\times$ 16 MHz bands (digitally filtered 2-bit samples) in the frequency ranges: 2252 - 2268 MHz and 2268 - 2284 MHz.  Both bands were upper side band and right circular polarisation.

\begin{table}[ht]
\begin{center}
{ \tiny
\begin{tabular}{lcccccccc} \hline \hline
Observatory & Frequency & $\alpha$ & $\delta$ & Date & Config. & Duration & Bandwidth & $\Delta$t \\
            & (MHz)     & (J2000)  & (J2000)  &      &         & (h)      & (MHz)     & (s) \\ \hline \hline
ATCA    & 4800.0  & $13\rah5\ram28\fs00$ & $-49\arcdeg28\arcmin12\farcs00$ & 19/20 NOV 1993 & 6A      & 11  & 128 & 15 \\
\nodata & 8640.0  & \nodata              & \nodata                         & \nodata        & \nodata & 11  & 128 & 15 \\
\hline
LBA\tablenotemark{a} & 2252.0  & $13\rah5\ram27\fs50$ & $-49\arcdeg28\arcmin6\farcs00$  & 13/14 MAY 2005 & \nodata & 9   & 16  & 2  \\
\nodata & 2268.0  & \nodata              & \nodata                         & \nodata        & \nodata & 9   & 16  & 2  \\
\hline
ATCA    & 17000.0 & $13\rah5\ram27\fs50$ & $-49\arcdeg28\arcmin6\farcs00$  & 18 MAR 2006    & EW367   & 3 & 128 & 15 \\
\nodata & 19000.0 & \nodata              & \nodata                         & \nodata        & \nodata & 3 & 128 & 15 \\
\nodata & 21000.0 & \nodata              & \nodata                         & \nodata        & \nodata & 3.5 & 128 & 15 \\
\nodata & 23000.0 & \nodata              & \nodata                         & \nodata        & \nodata & 3.5 & 128 & 15 \\
\hline
ATCA    & 17000.0 & $13\rah5\ram27\fs50$ & $-49\arcdeg28\arcmin6\farcs00$  & 24/25 MAR 2006 & 6C      & 9 & 128 & 15 \\
\nodata & 19000.0 & \nodata              & \nodata                         & \nodata        & \nodata & 9 & 128 & 15 \\
\nodata & 21000.0 & \nodata              & \nodata                         & \nodata        & \nodata & 9 & 128 & 15 \\
\nodata & 23000.0 & \nodata              & \nodata                         & \nodata        & \nodata & 9 & 128 & 15 \\
\hline
LBA\tablenotemark{b} & 2269.0  & $13\rah5\ram27\fs50$ & $-49\arcdeg28\arcmin6\farcs00$  & 21 MAR 2007    & \nodata & 11   & 16  & 2 \\
\nodata & 2285.0  & \nodata              & \nodata                         & \nodata        & \nodata & 11   & 16  & 2 \\
\nodata & 2301.0  & \nodata              & \nodata                         & \nodata        & \nodata & 11   & 16 & 2 \\
\nodata & 2317.0  & \nodata              & \nodata                         & \nodata        & \nodata & 11   & 16 & 2 \\ \hline
\tablenotetext{a}{Australian LBA observation included 9 hours at ATCA and Mopra, Tidbinbilla, Hobart and Ceduna, and 6 hours at Parkes.}
\tablenotetext{b}{Australian LBA observation included 11 hours at ATCA and Mopra, Hobart and Ceduna, 10 hours at Parkes and 5 hours at Tidbinbilla.}
\end{tabular}
\caption{Summary of NGC 4945 observations.}
\label{tab:p3tabobs}
}
\end{center}
\end{table}

A second epoch LBA observation was made on 21 March, 2007 using the same set of radio telescopes except that only 3 $\times$ 22 m antennas of the ATCA were used as a phased array to maintain a wide field of view given the longer baseline configuration of the ATCA at that time. The second epoch observation utilised hard disk data recorders (Phillips et al. 2008, in preparation); allowing dual circular polarisation data across $4\times16$ MHz bands to be recorded at Parkes, ATCA, and Mopra, and $2\times16$ MHz bands at the remaining antennas. Observing parameters associated with each of the LBA observations are shown in Table \ref{tab:p3tabobs}. 

During each of the VLBI observations, three minute scans of NGC 4945 ($\alpha = 13\rah5\ram27\fs50$; $\delta = -49\arcdeg28\arcmin6\farcs00$ [J2000]) were scheduled, alternating with three minute scans of a nearby phase reference calibration source, J1237$-$5046 ($\alpha = $\linebreak[4]$12\rah37\ram15\fs239286$; $\delta = -50\arcdeg46\arcmin23\farcs17260$ [J2000]).

\subsection{LBA correlation}
The first epoch data were correlated using the ATNF Long Baseline Array (LBA) processor at ATNF headquarters in Sydney \citep{Wilson:1992p9290}. The data were correlated using an integration time of 2 seconds and with 32 frequency channels across each 16 MHz band (channel widths of 0.5 MHz) to limit the effects of bandwidth smearing (a form of chromatic aberration) and time-averaging smearing \citep{Cotton:1999p19452,Bridle:1999p10564}. The ATCA primary beam limits the field of view of this observation to a half-width half-maximum (HWHM) of $\sim45\arcsec$. At the HWHM point, bandwidth smearing and time-averaging smearing losses are estimated to be approximately 15\% and 4\%, respectively.

The second epoch data were correlated using the DiFX Software Correlator \citep{Deller:2007p10545}. The data were correlated using an integration time of 2 seconds and with 64 frequency channels across each 16 MHz band (channel widths of 0.25 MHz). The ATCA primary beam limits the field of view of this observation to a HWHM of $\sim1\arcmin$. At the HWHM point, bandwidth smearing and time-averaging smearing losses are estimated to be approximately 8\% and 7\%, respectively.

\subsection{LBA Data Reduction}
\label{sec:p3lbareduction}
The correlated data were imported into the AIPS\footnote{The Astronomical Image Processing System (AIPS) was developed and is maintained by the National Radio Astronomy Observatory, which is operated by Associated Universities, Inc., under co-operative agreement with the National Science Foundation} package for initial processing. The data for the phase reference source were fringe-fit (AIPS task FRING) using a one minute solution interval, finding independent solutions for each of the 16 MHz bands.  The delay and phase solutions for the phase reference source were examined and averaged over each three minute calibrator scan, following editing of bad solutions, before being applied to both the phase reference source and NGC 4945.  Further flagging of the data was undertaken via application of a flag file that reflected the times during which each of the antennas were expected to be slewing, or time ranges that contained known bad data.  Finally, data from the first 30 seconds of each scan from baselines involving the ATCA or Parkes were flagged, to eliminate known corruption of the data at the start of each scan at these two telescopes.

\begin{figure}[ht]
\epsscale{0.45}
\begin{center}
\mbox{
\plotone{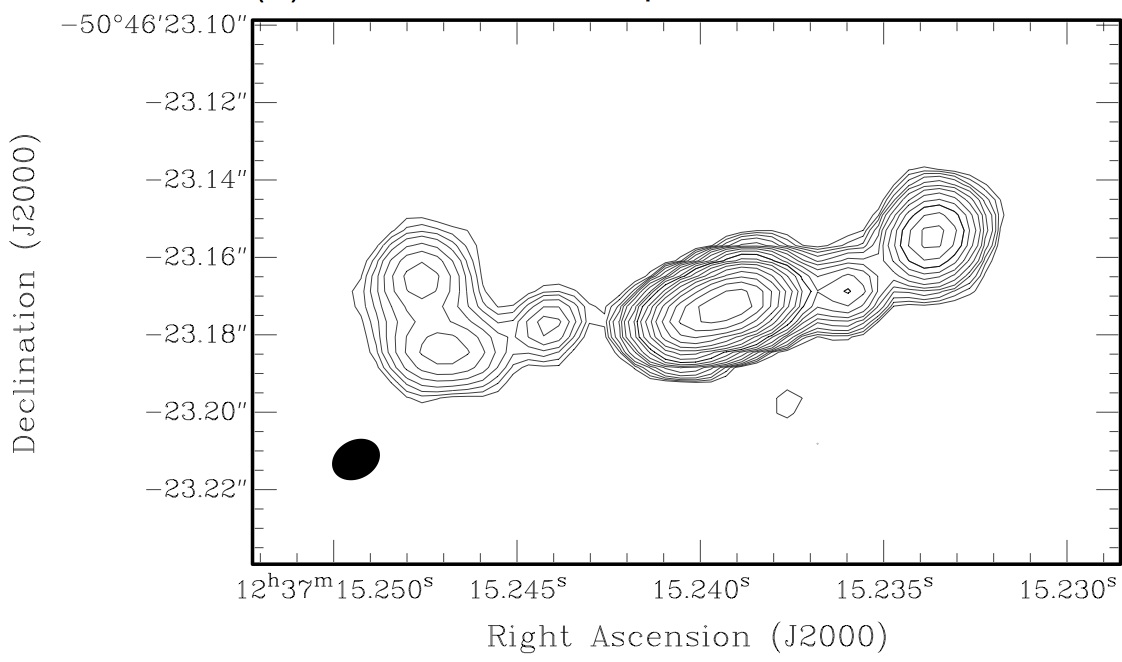} \quad
\plotone{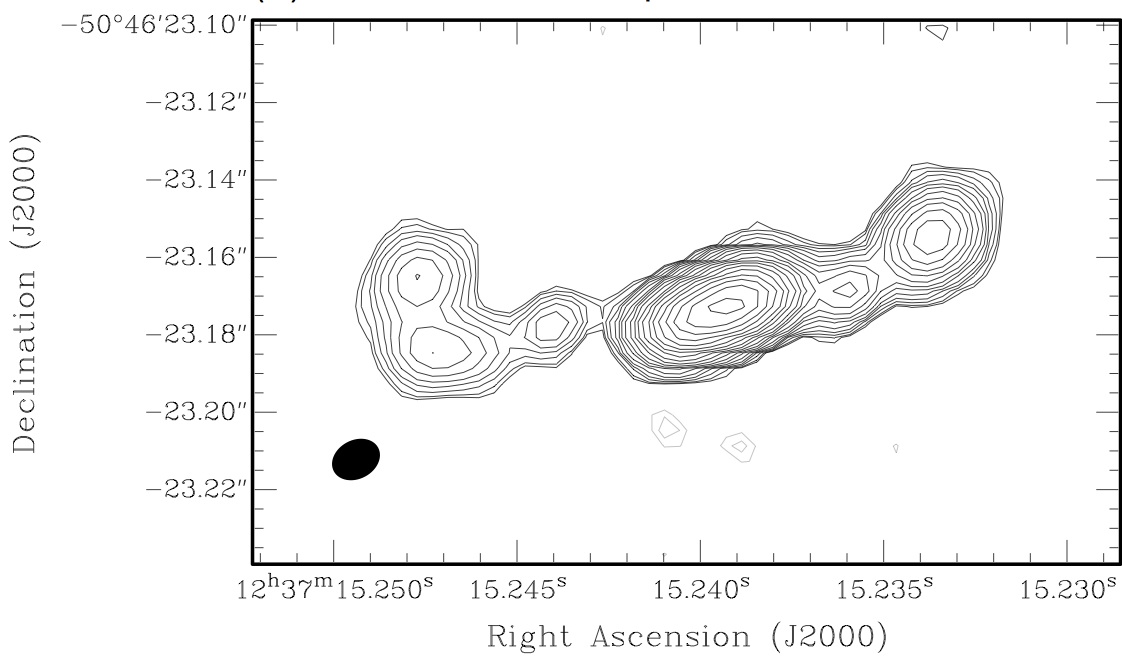}
}
\caption[Australian LBA image of the phase calibrator J1237$-$5046 at 2.3 GHz]{Uniformly-weighted total-power map of the source used for phase calibration, J1237-5046, as observed with the Australian LBA at 2.3 GHz over two epochs. Map statistics for the individual maps are shown in Table \ref{tab:p3tabimage}. Contours are drawn at $\pm2^{0}, \pm2^{\frac{1}{2}}, \pm2^{1}, \pm2^{\frac{3}{2}}, \cdots$ times the $3\sigma$ rms noise.}
\label{fig:p3figJ1237-5046}            
\end{center}
\end{figure}

During correlation, nominal (constant) system temperatures (in Jy) for each antenna were applied to the correlation coefficients in order to roughly calibrate the visibility amplitudes (mainly to ensure roughly correct weights during fringe-fitting).  Following fringe-fitting, the nominal calibration was refined by collecting and applying the antenna system temperatures (in K) measured during the observation, along with the most recently measured gain (in Jy/K) for each antenna. Further refinement in the calibration was complicated by the complex structure of the phase reference source J1237$-$5046. To account for this structure a new DIFMAP \citep{Shepherd:1994p10583} task, \emph{cordump}\footnote{The \emph{cordump} patch is available for DIFMAP at \url{http://astronomy.swin.edu.au/~elenc/DifmapPatches/}} \citep{Lenc:2006p32}, was developed to enable the transfer of all phase and amplitude corrections made in DIFMAP during the imaging process to an AIPS compatible solution table. The phase reference data were averaged in frequency and exported to DIFMAP where several iterations of modelling and self-calibration of both phases and amplitudes were performed. The resulting contour maps of the phase reference source are shown in Figure \ref{fig:p3figJ1237-5046} and have a measured one sigma RMS noise of 0.342 and 0.292 mJy beam$^{-1}$ in epochs 1 and 2, respectively. \emph{cordump} was then used to transfer the phase and amplitude corrections back to the unaveraged AIPS data-set.

Averaging in frequency was performed in AIPS to increase the channel widths to 2 MHz; this significantly reduced the data-set to a manageable size and improved imaging performance. While this also reduced the overall field of view, the starburst region of interest was unaffected by the averaging process. To facilitate imaging in DIFMAP, a ParselTongue\footnote{A Python scripting tool for AIPS. ParselTongue was developed in the context of the ALBUS project, which has benefited from research funding from the European Community's sixth Framework Programme under RadioNet R113CT 2003 5058187. ParselTongue is available for download at \url{http://www.radionet-eu.org/rnwiki/ParselTongue}} script was written to convert the averaged frequency channels into intermediate frequencies (IFs). This conversion allowed DIFMAP to treat the frequency channels independently in the $(u,v)$ plane rather than averaging them together, thus avoiding any further bandwidth smearing effects during the imaging process.

\begin{figure}[p]
\epsscale{0.55}
\begin{center}
\plotone{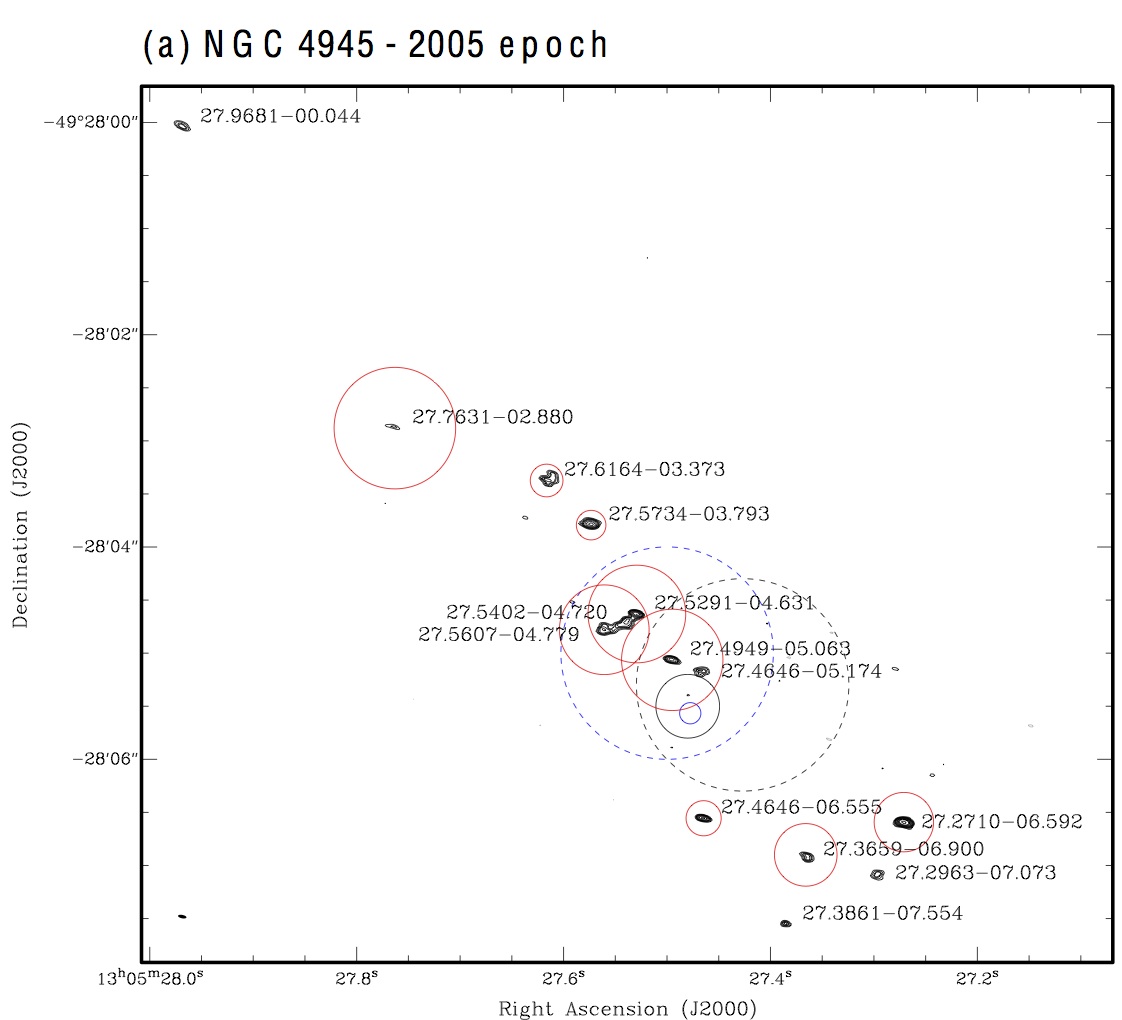}
\plotone{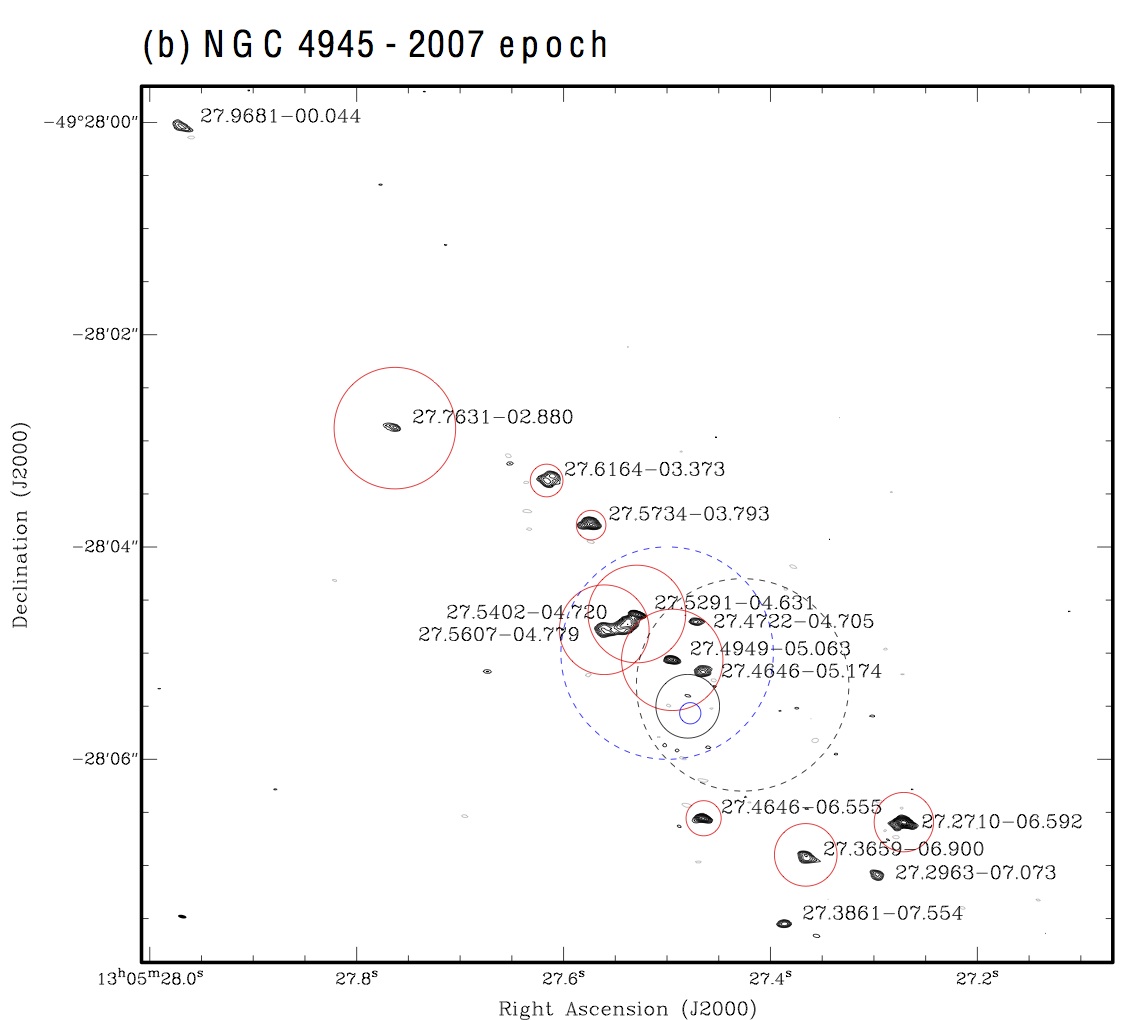}
\caption[Low-resolution Australian LBA image of NGC 4945 at 2.3 GHz]{Uniformly-weighted total-power map of NGC 4945 as observed with the Australian LBA at 2.3 GHz over two epochs. Map statistics for the individual maps are shown in Table \ref{tab:p3tabimage}. Contours are drawn at $\pm2^{\frac{1}{2}}, \pm2^{1}, \pm2^{\frac{3}{2}}, \cdots$ times the $3\sigma$ rms noise. The overlaid circles give an indication of the degree of absorption in the vicinity of the source, with the diameter of the circle being directly proportional to $\tau_{0}$. The dashed blue circle and dashed black circle indicate the positional error of the Chandra X-ray \citep{Schurch:2002p7434} and \emph{HST} K-band \citep{Marconi:2000p7671} peak, respectively. The solid blue circle and solid black circle mark the positional error of the H$_{2}$O megamaser \citep{Greenhill:1997p7509} and the centre of the HNC cloud \citep{HuntCunningham:2005p10542}.}
\label{fig:p3figLBANGC4945}            
\end{center}
\end{figure}

Initial imaging of NGC 4945 was performed at reduced resolution ($80\times32$ mas) by excluding data from the Hobart and Ceduna antennas (the longest baselines); uniform weighting was applied to maximise resolution at the expense of image noise. Several iterations of CLEAN and phase self-calibration were performed in DIFMAP. The resulting images are shown in Figure \ref{fig:p3figLBANGC4945} and achieve a one sigma RMS noise of 0.152 and 0.082 mJy beam$^{-1}$ for epoch 1 and 2, respectively. The image map parameters for both figures are shown in Table \ref{tab:p3tabimage}.

A second phase of imaging was repeated at full resolution ($16\times15$ mas) with data from all antennas. The imaging process followed the same procedure to that used in the initial imaging phase except that natural weighting was applied to provide a compromise between the increased resolution available with the full array and the available image noise. Separate postage-stamp images were made for both epochs for each of the sources which fell above the six sigma detection limit. The images from both epochs were then combined to produce a high dynamic range image map for each of the detected sources. The epoch 1, epoch 2 and combined image maps for each of the detected sources are shown in Figure \ref{fig:p3fighra} and the associated map statistics tabulated in Table \ref{tab:p3tabimage}. The epoch 1, epoch 2 and combined maps, achieve a one sigma RMS noise of 0.149, 0.110 and 0.075 mJy beam$^{-1}$, respectively. These figures compare favourably with the expected theoretically predicted thermal noise of $\sim0.120$, $\sim0.080$ and $\sim0.070$ mJy beam$^{-1}$, respectively\footnote{Estimated with the ATNF VLBI sensitivity calculator: \url{http://www.atnf.csiro.au/vlbi/calculator/}}.

Caution must be taken to avoid reading too much into the observed structure of some of the weaker sources at full resolution. It is possible that a number of these are showing signs of structural break-up during the deconvolution process as a result of limited $(u,v)$ coverage and sensitivity e.g. $27.5291-04.631$, $27.6164-03.373$ and $27.5734-03.793$. Nonetheless, the $6\sigma$ limit is very conservative and features above this level can be trusted with a high degree of confidence. Imaging of the brightest sources with uniform weighting applied (not shown here) resulted in a reduced signal to noise, as a consequence of a smaller restoring beam and increased image noise, when compared to the naturally weighted images. However the structures observed above the $6\sigma$ threshold remained consistent between the two sets of images.

\begin{table}[p]
\begin{center}
{ \scriptsize
\begin{tabular}{lcccccccc} \hline \hline
Figure & Source & Frequency & Synthesized Beam & $\sigma$ & Peak Flux & Integrated Flux \\
       &        & (GHz)     & (mas)            & (mJy beam$^{-1}$) & (mJy beam$^{-1}$) & (mJy) \\ \hline \hline
\ref{fig:p3figJ1237-5046}(a)  & J1237$-$5046   & 2.3 & $13\times10$     & 0.342 & 520  & 1100 \\
\ref{fig:p3figJ1237-5046}(b)  & J1237$-$5046   & 2.3 & $13\times10$     & 0.292 & 530  & 1100 \\
\ref{fig:p3figLBANGC4945}(a)  & NGC 4945       & 2.3 & $80\times32$     & 0.152 & 24  & 150 \\
\ref{fig:p3figLBANGC4945}(b)  & NGC 4945       & 2.3 & $80\times32$     & 0.085 & 23  & 160 \\
\ref{fig:p3figlr}(a)           & $27.5291-04.631$ & 2.3 & $80\times32$     & 0.152 & 7.3  & 30 \\
\ref{fig:p3figlr}(b)           & $27.5291-04.631$ & 2.3 & $80\times32$     & 0.085 & 6.8  & 36 \\
\ref{fig:p3figlr}(c)           & $27.6164-03.373$ & 2.3 & $80\times32$     & 0.152 & 2.1  & 8.5 \\
\ref{fig:p3figlr}(d)           & $27.6164-03.373$ & 2.3 & $80\times32$     & 0.085 & 1.8  & 8.2 \\
\ref{fig:p3figATCANGC4945}(a) & NGC 4945       & 4.8 & $2132\times1680$ & 0.520 & 580 & 2700  \\
\ref{fig:p3figATCANGC4945}(b) & NGC 4945       & 8.6 & $1155\times955$  & 0.202 & 170 & 1300  \\
\ref{fig:p3figATCANGC4945}(c) & NGC 4945       & 17  & $611\times510$   & 0.250 & 67 & 940  \\
\ref{fig:p3figATCANGC4945}(d) & NGC 4945       & 19  & $560\times443$   & 0.270 & 58 & 840  \\
\ref{fig:p3figATCANGC4945}(e) & NGC 4945       & 21  & $495\times399$   & 0.320 & 48 & 800  \\
\ref{fig:p3figATCANGC4945}(f) & NGC 4945       & 23  & $453\times363$   & 0.340 & 46 & 780  \\
\ref{fig:p3fighra}(a)          & $27.6164-03.373$ & 2.3 & $16\times15$     & 0.149 & 1.3 & 10 \\
\ref{fig:p3fighra}(b)          & $27.6164-03.373$ & 2.3 & $16\times15$     & 0.110 & 0.76 & 9.0 \\
\ref{fig:p3fighra}(c)          & $27.6164-03.373$ & 2.3 & $16\times15$     & 0.075 & 1.0 & 9.6 \\
\ref{fig:p3fighra}(d)          & $27.5734-03.793$ & 2.3 & $16\times15$     & 0.149 & 1.8 & 11 \\
\ref{fig:p3fighra}(e)          & $27.5734-03.793$ & 2.3 & $16\times15$     & 0.110 & 2.2 & 12 \\
\ref{fig:p3fighra}(f)          & $27.5734-03.793$ & 2.3 & $16\times15$     & 0.075 & 1.9 & 12 \\
\ref{fig:p3fighra}(g)          & $27.5291-04.631$ & 2.3 & $16\times15$     & 0.149 & 3.5 & 38 \\
\ref{fig:p3fighra}(h)          & $27.5291-04.631$ & 2.3 & $16\times15$     & 0.110 & 3.0 & 30 \\
\ref{fig:p3fighra}(i)          & $27.5291-04.631$ & 2.3 & $16\times15$     & 0.075 & 3.3 & 34 \\
\ref{fig:p3fighra}(j)          & $27.4949-05.063$ & 2.3 & $16\times15$     & 0.149 & 2.2 & 5.7 \\
\ref{fig:p3fighra}(k)          & $27.4949-05.063$ & 2.3 & $16\times15$     & 0.110 & 2.4 & 5.8 \\
\ref{fig:p3fighra}(l)          & $27.4949-05.063$ & 2.3 & $16\times15$     & 0.075 & 2.3 & 5.7 \\
\ref{fig:p3fighra}(m)          & $27.4646-06.555$ & 2.3 & $16\times15$     & 0.149 & 2.5 & 5.9 \\
\ref{fig:p3fighra}(n)          & $27.4646-06.555$ & 2.3 & $16\times15$     & 0.110 & 3.0 & 7.8 \\
\ref{fig:p3fighra}(o)          & $27.4646-06.555$ & 2.3 & $16\times15$     & 0.075 & 2.8 & 6.9 \\
\ref{fig:p3fighra}(p)          & $27.4646-05.174$ & 2.3 & $16\times15$     & 0.149 & 0.66 & 2.3 \\
\ref{fig:p3fighra}(q)          & $27.4646-05.174$ & 2.3 & $16\times15$     & 0.110 & 0.82 & 4.1 \\
\ref{fig:p3fighra}(r)          & $27.4646-05.174$ & 2.3 & $16\times15$     & 0.075 & 0.70 & 3.2 \\
\ref{fig:p3fighra}(s)          & $27.3861-07.554$ & 2.3 & $16\times15$     & 0.149 & 1.1 & 1.0 \\
\ref{fig:p3fighra}(t)          & $27.3861-07.554$ & 2.3 & $16\times15$     & 0.110 & 0.91 & 2.3 \\
\ref{fig:p3fighra}(u)          & $27.3861-07.554$ & 2.3 & $16\times15$     & 0.075 & 0.97 & 1.6 \\
\ref{fig:p3fighra}(v)          & $27.3659-06.900$ & 2.3 & $16\times15$     & 0.149 & 1.1 & 3.1 \\
\ref{fig:p3fighra}(w)          & $27.3659-06.900$ & 2.3 & $16\times15$     & 0.110 & 0.84 & 5.7 \\
\ref{fig:p3fighra}(x)          & $27.3659-06.900$ & 2.3 & $16\times15$     & 0.075 & 0.93 & 4.4 \\
\ref{fig:p3fighra}(y)          & $27.2710-06.592$ & 2.3 & $16\times15$     & 0.149 & 11 & 37 \\
\ref{fig:p3fighra}(z)          & $27.2710-06.592$ & 2.3 & $16\times15$     & 0.110 & 9.1 & 30 \\
\ref{fig:p3fighra}(aa)         & $27.2710-06.592$ & 2.3 & $16\times15$     & 0.075 & 9.9 & 33 \\ \hline
\end{tabular}
\caption{Map Statistics for NGC 4945 images.}
\label{tab:p3tabimage}
}
\end{center}
\end{table}

\subsection{ATCA Observations and Data Reduction}

Observations of NGC 4945 were carried out using all six antennas in the EW367 and 6C configurations of the Australia Telescope Compact Array (ATCA) on the 2006 March 18, 24 and 25. The EW367 configuration is compact with most baselines less than 367 m in length, the 6C configuration is for high resolution observations and has a spread of baselines up to 6 km in length. The flux scale was set using PKS B1934$-$638 and J1326$-$5256 was used as a phase calibrator. During each observing run, we alternated between two setups every 6.5 minutes. The first setup observed at 17 GHz and 19 GHz and the second setup at 21 GHz and 23 GHz. For intermediate frequencies of 4.8 GHz and 8.64 GHz, archive data were obtained from the Australia Telescope Online Archive. The data were observed with the 6A configuration of the ATCA, which has a spread of baselines up to 6 km in length, on 1993 November 19 and 20. The flux scale was set using PKS B1934$-$638 and PKS 1320$-$446 was used as a phase calibrator. A list of all ATCA observations reported in this paper and their associated observing parameters are tabulated in Table \ref{tab:p3tabobs}.

All ATCA data were initially calibrated using the MIRIAD software package \citep{Sault:1995p10582}. Subsequent calibration, deconvolution and imaging was performed with the DIFMAP \citep{Shepherd:1994p10583} software package and final images were created using the KARMA software package \citep{Gooch:1996p7263}. The resulting contour maps are shown in Figure \ref{fig:p3figATCANGC4945} with all associated beam parameters, image rms noise, peak and integrated flux densities tabulated in Table \ref{tab:p3tabimage}.

\section{Results}
\subsection{Identification of Sources}
\label{sec:sourceid}
\begin{figure}[ht]
\epsscale{0.4}
\begin{center}
\mbox{
\plotone{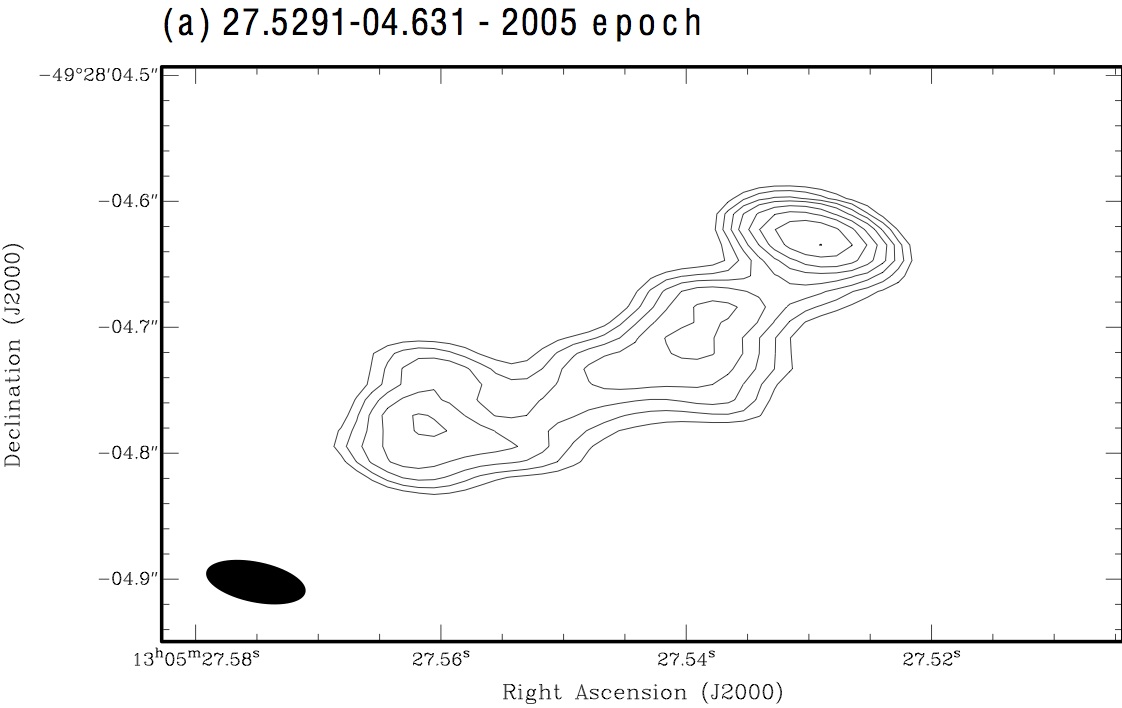} \quad
\plotone{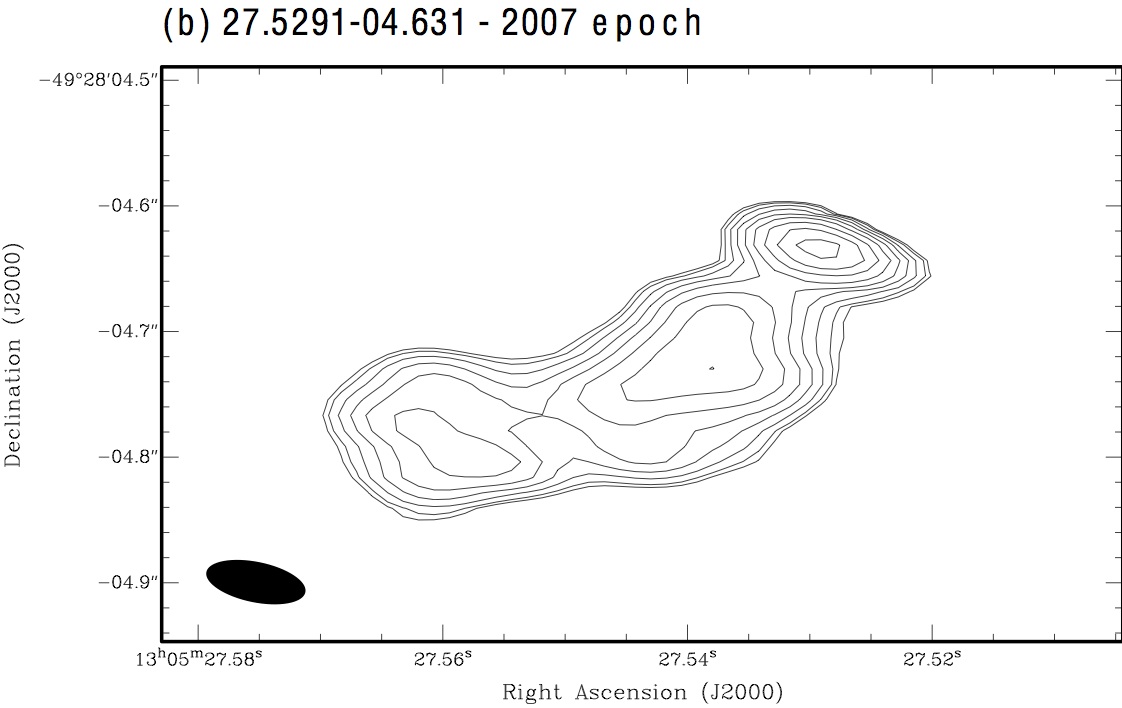}
}
\mbox{
\plotone{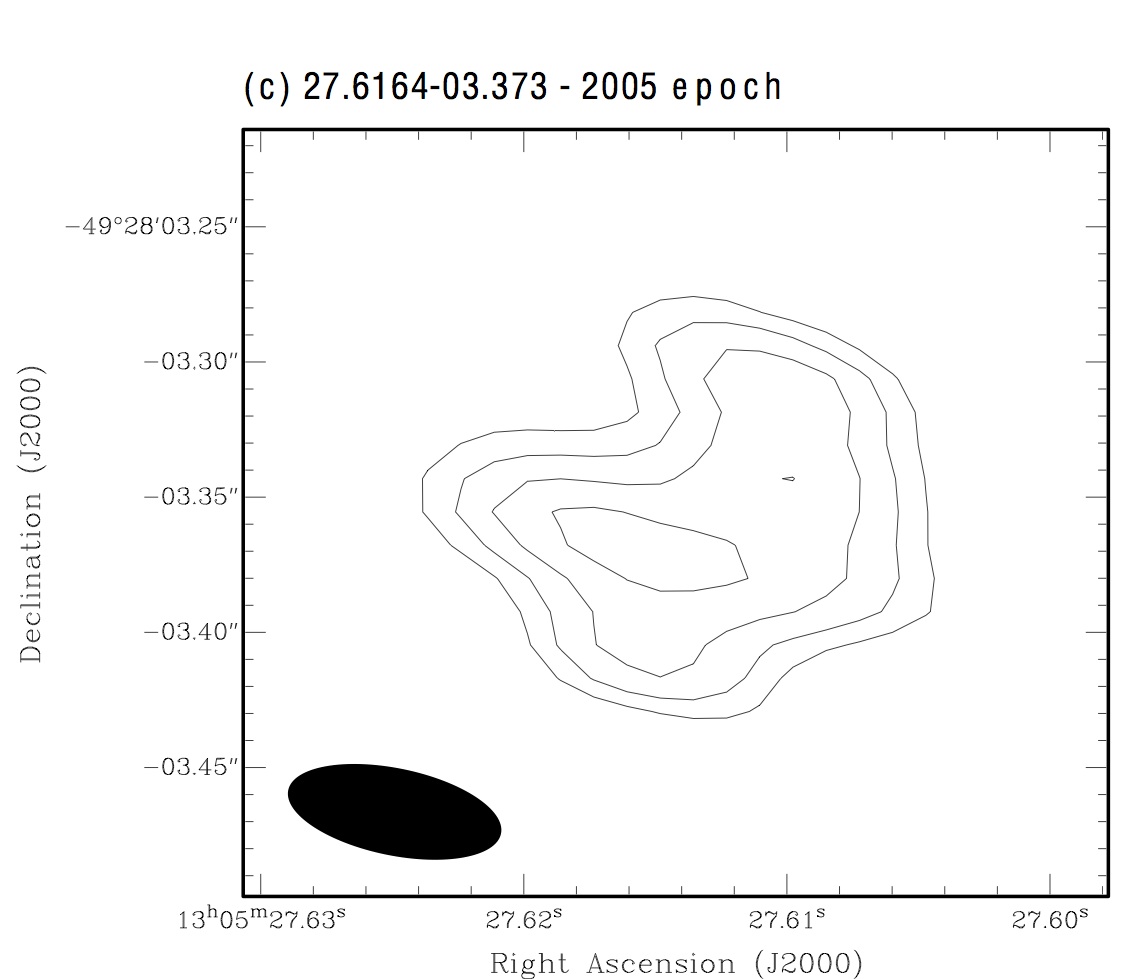} \quad
\plotone{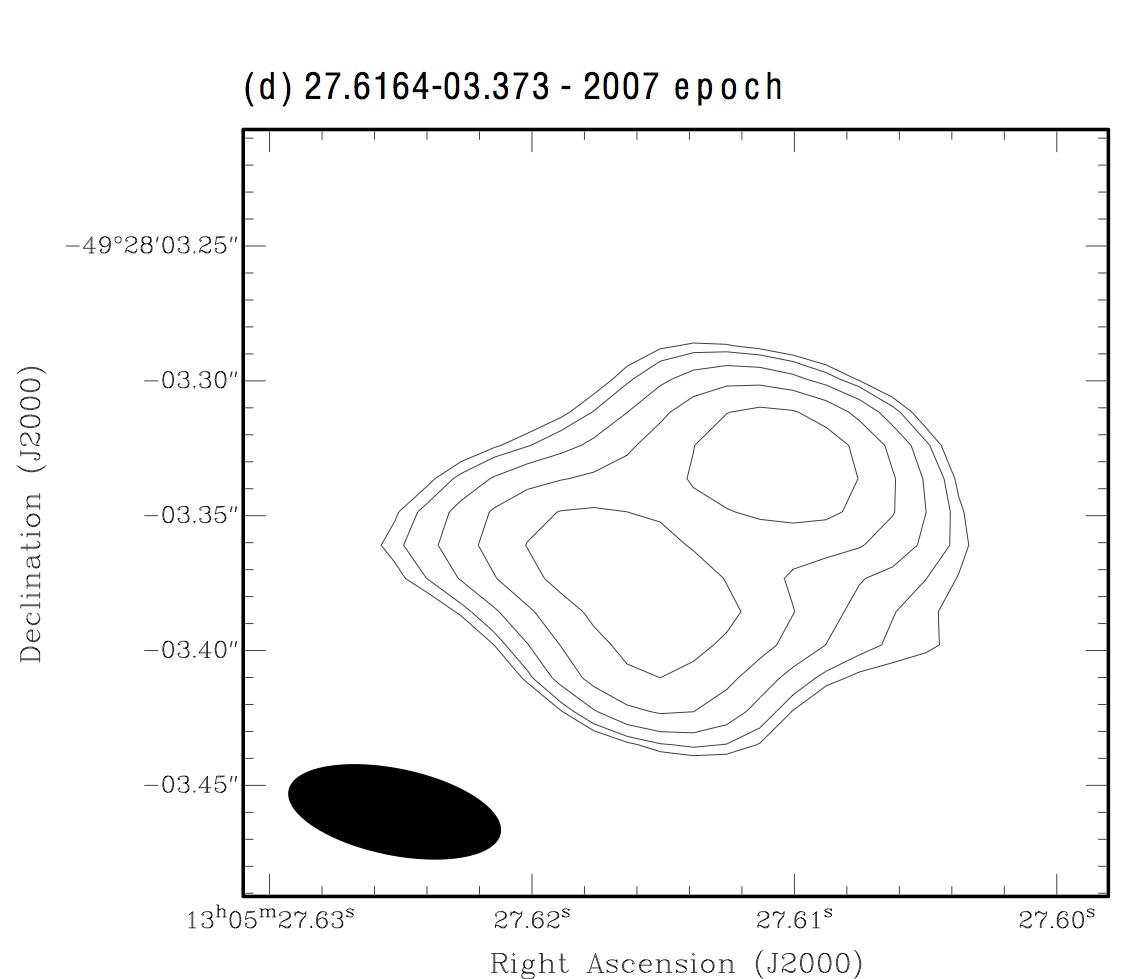}
}
\caption[Detailed image maps of 27.5291$-$04.631 and 27.6164$-$03.373 as observed with the Australian LBA at 2.3 GHz over two epochs]{Uniformly-weighted total-power maps of 27.5291$-$04.631 and 27.6164$-$03.373 as observed with the Australian LBA at 2.3 GHz over two epochs. Map statistics for the individual maps are shown in Table \ref{tab:p3tabimage}. Contours are drawn at $\pm2^{\frac{1}{2}}, \pm2^{1}, \pm2^{\frac{3}{2}}, \cdots$ times the $3\sigma$ rms noise.}
\label{fig:p3figlr}            
\end{center}
\end{figure}

\begin{figure}[p]
\epsscale{0.4}
\begin{center}
\mbox{
\plotone{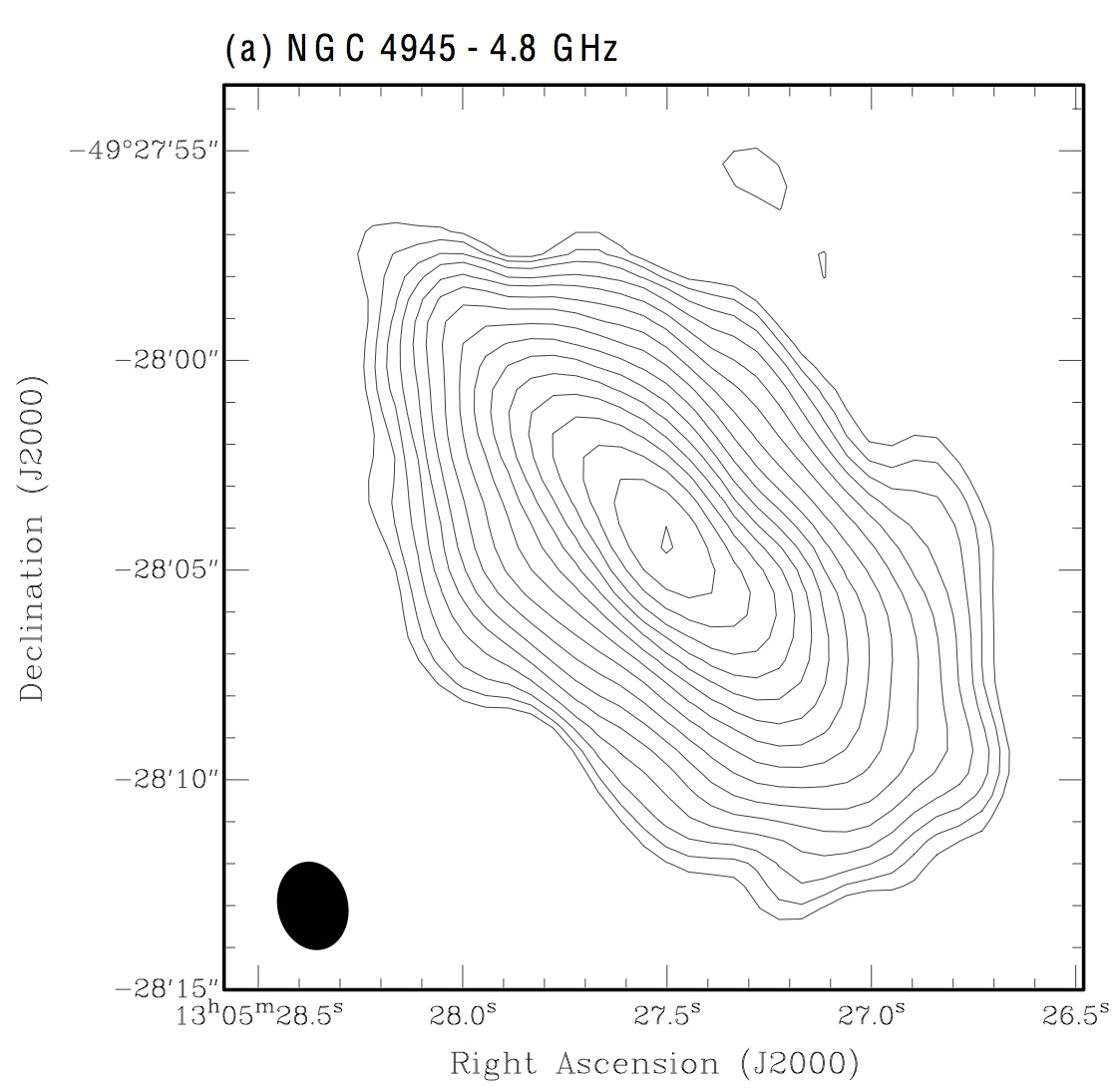} \quad
\plotone{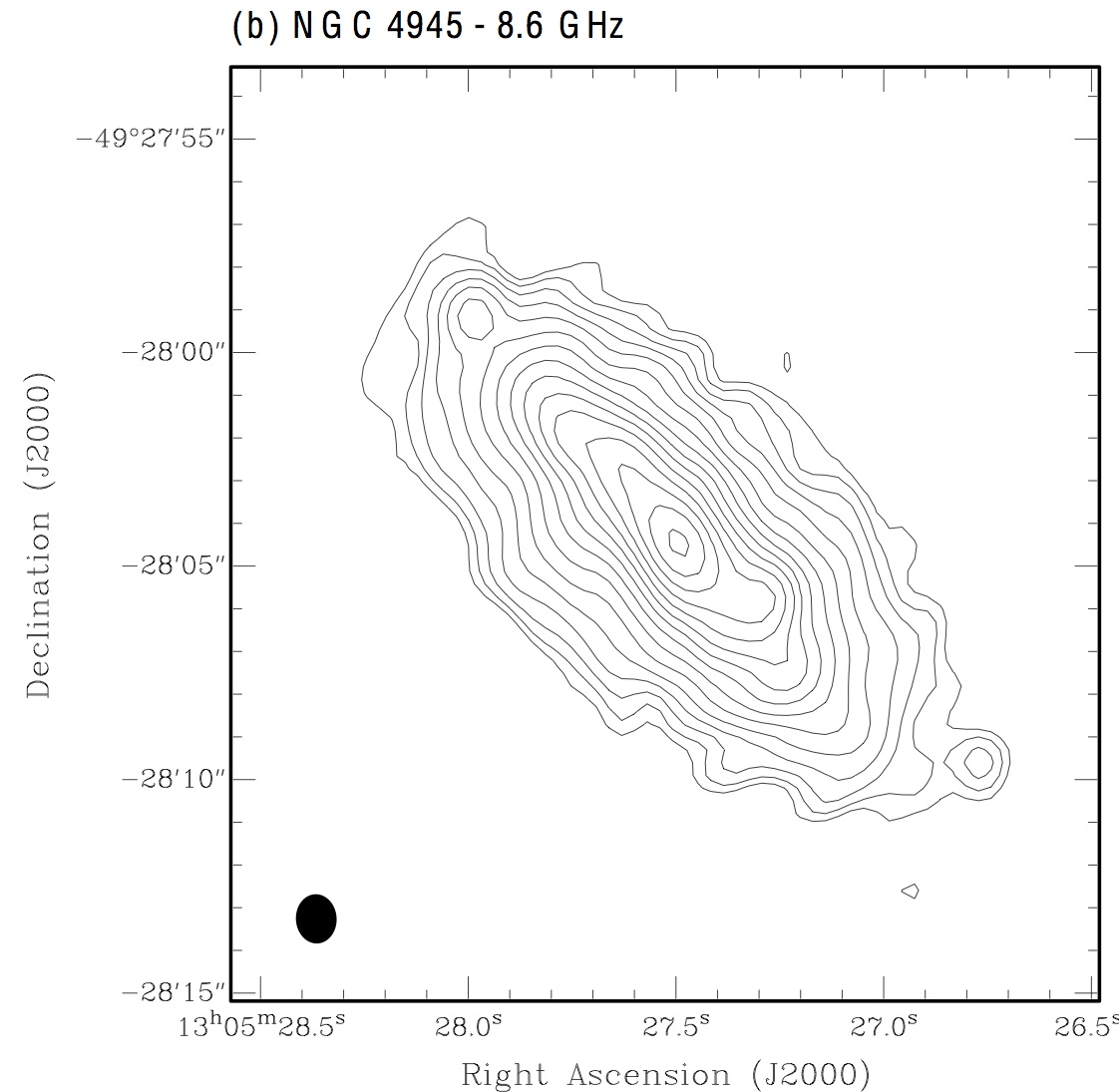}
}
\mbox{
\plotone{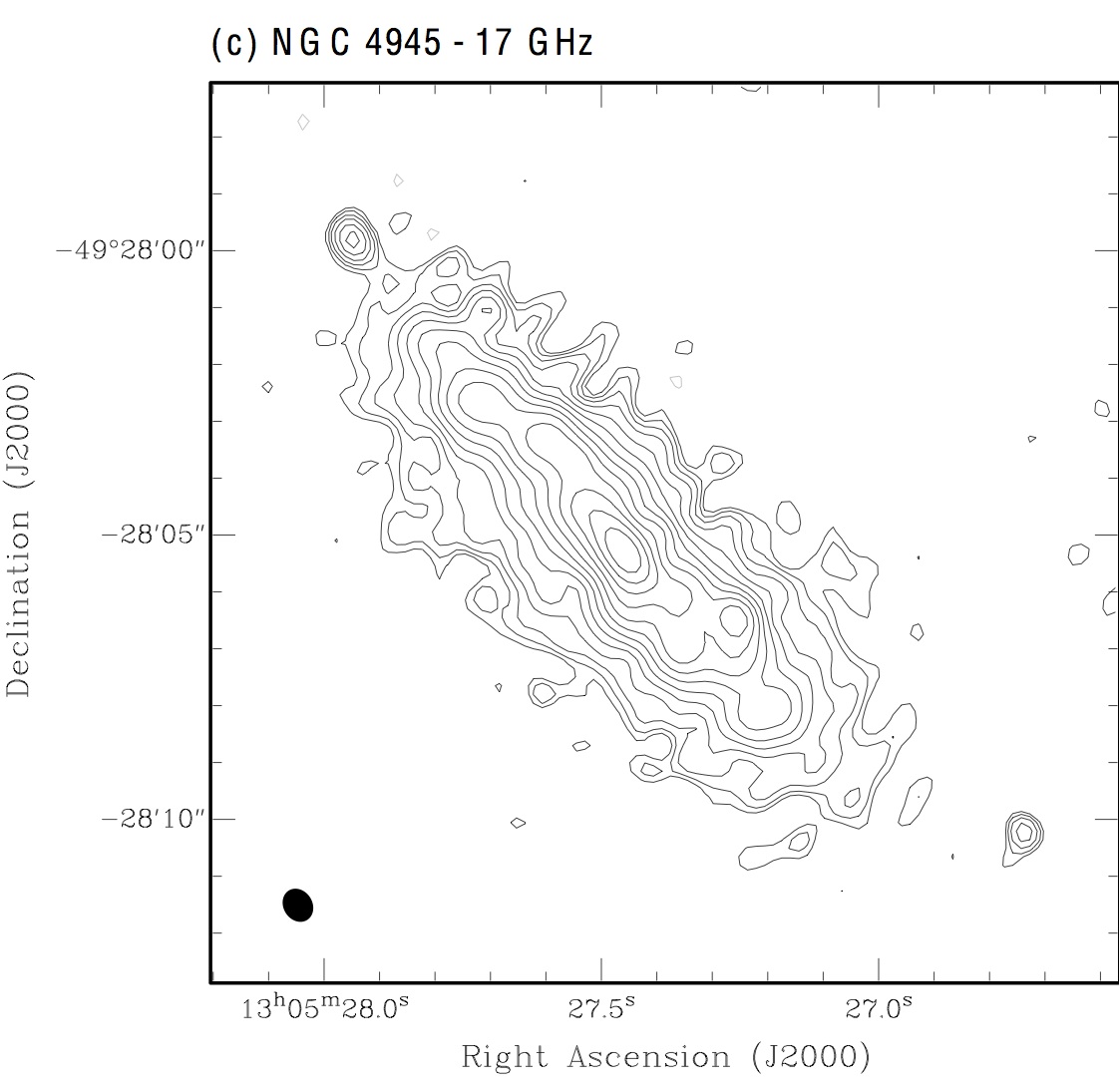} \quad
\plotone{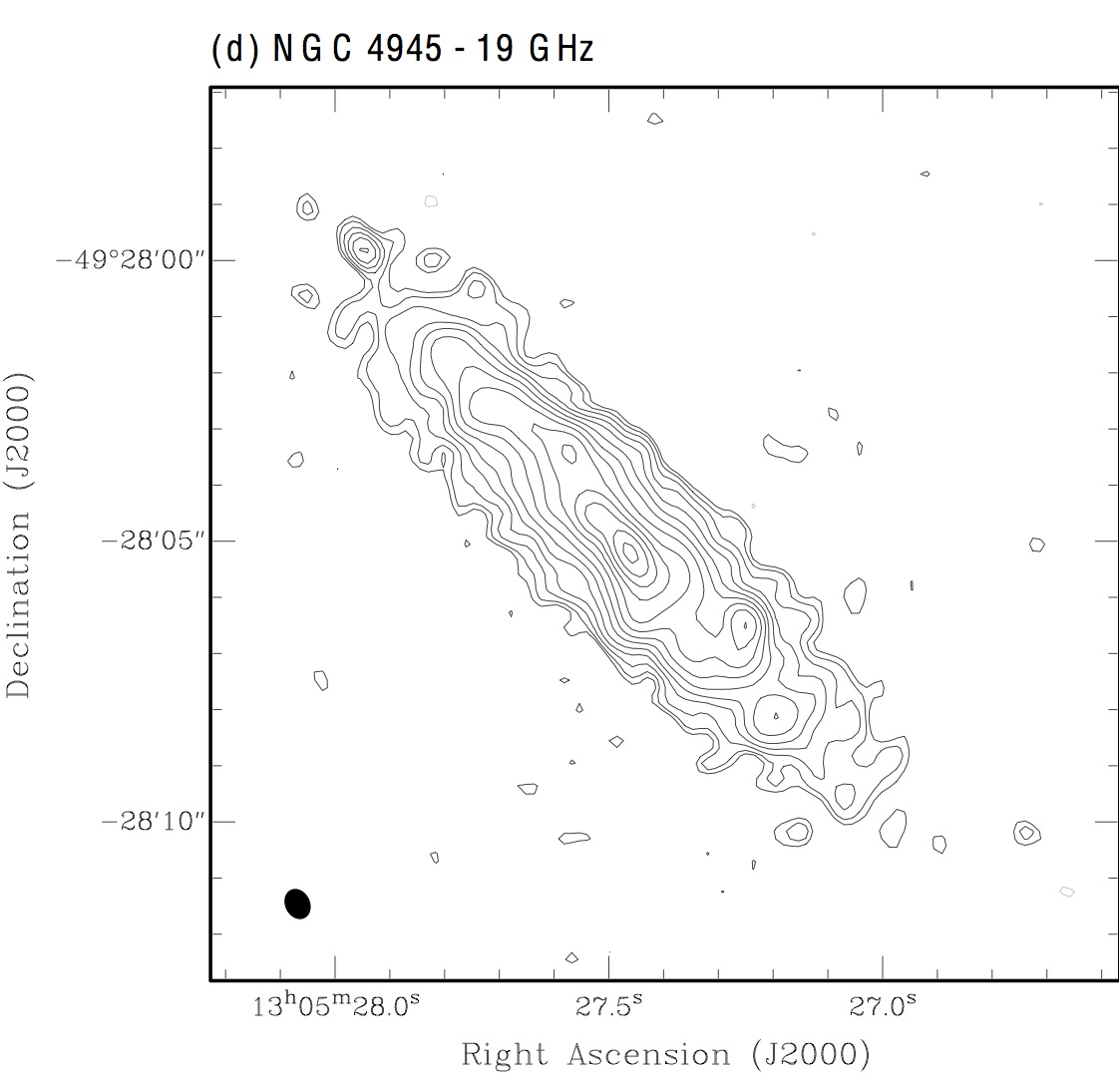}
}
\mbox{
\plotone{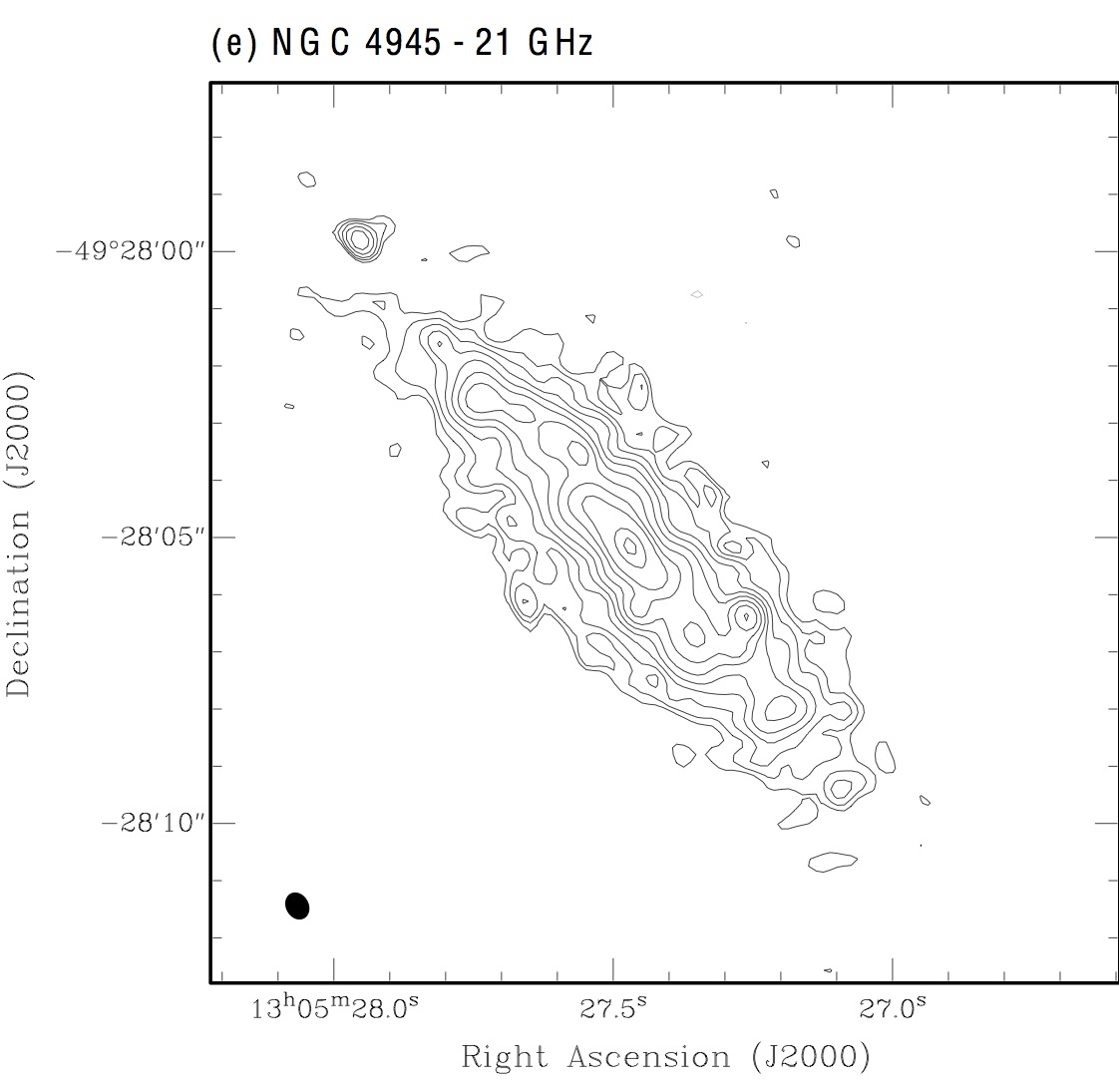} \quad
\plotone{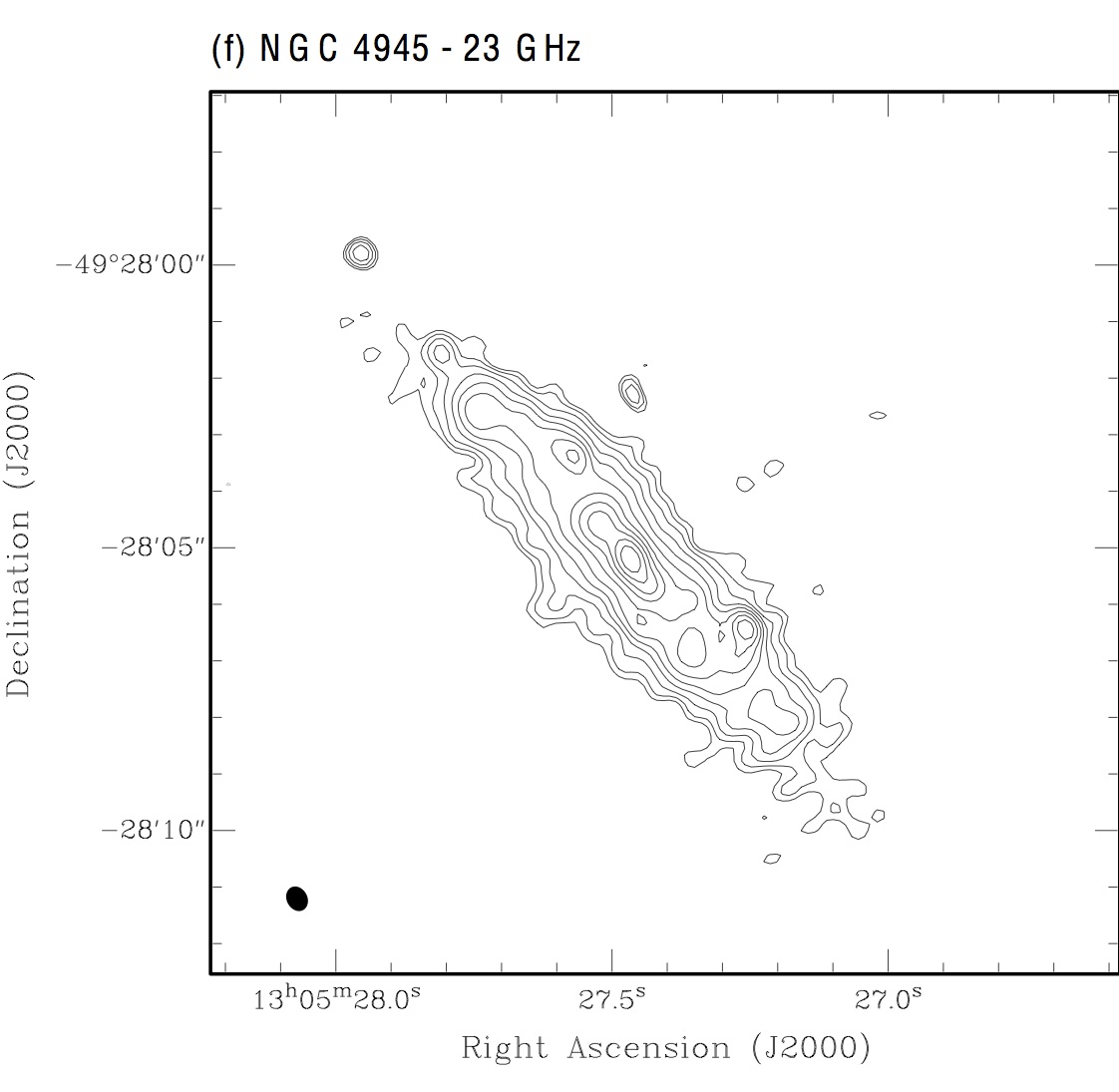}
}
\caption[ATCA images of NGC 4945 at 4.8 GHz, 8.6 GHz, 17 GHz, 19 GHz, 21 GHz and 23 GHz]{Uniformly-weighted total-power maps of NGC 4945 as observed with the ATCA at 4.8 GHz, 8.6 GHz, 17 GHz, 19 GHz, 21 GHz and 23 GHz. Map statistics for the individual maps are shown in Table \ref{tab:p3tabimage}. Contours are drawn at $\pm2^{0}, \pm2^{\frac{1}{2}}, \pm2^{1}, \pm2^{\frac{3}{2}}, \cdots$ times the $3\sigma$ rms noise for all maps except for (a) and (b) where the lowest contour is at $6\sigma$ rms noise.}
\label{fig:p3figATCANGC4945}            
\end{center}
\end{figure}

A list of sources detected above the $6\sigma$ detection threshold in both epochs of the short-baseline VLBI data is given in Table \ref{tab:p3tabfluxes}. The $6\sigma$ threshold provides a conservative false detection rate of less than one in 3000. Flux density errors of $\pm10\%$ are listed due to uncertainties in the absolute flux density scale for Southern Hemisphere VLBI \citep{rey94}. The total flux density of each source was determined by summing its CLEAN model components ($+$ image residuals) in the image plane. The sources were also modelled with elliptical Gaussian components to provide an approximate measure of the full-width half-maximum (FWHM) size, the fitted size and position angle (PA) of each source is also listed in Table \ref{tab:p3tabfluxes}. Two of the detected sources, $27.5291-04.631$ and $27.6164-03.373$, exhibit structure that is significantly larger than the beam width and are shown in greater detail in Figure \ref{fig:p3figlr}. Source $27.5291-04.631$, in particular, appears to exhibit a jet-like morphology and has been divided into three components in Table \ref{tab:p3tabfluxes}: $27.5291-04.631$, $27.5402-04.720$ and $27.5607-04.779$.

\begin{table}[ht]
\begin{center}
{ \tiny
\begin{tabular}{lccccccccc} \hline \hline
Source & FWHM Size\tablenotemark{a} & P.A.\tablenotemark{b} & S$_{2.3} (2005)$  & S$_{2.3} (2007)$ & S$_{8.64}$ & S$_{17}$ & S$_{19}$ & S$_{21}$ & S$_{23}$ \\
       & (mas)                      & (degrees)             & (mJy)             & (mJy)            & (mJy)      & (mJy)    & (mJy)    & (mJy)    & (mJy) \\ \hline \hline
$27.9681-00.044$ & $155\times43$                     & 68  & 6.8    & 5.1    & \nodata & 3.4    & 3.6    & \nodata  & \nodata \\
$27.7631-02.880$ & $91\times44$                      & 73  & 1.6    & 2.9    & 17    & 8.2    & 6.8    & 5.1     & 4.0    \\
$27.6164-03.373$ & $105\times54$                     & 44  & 13    & 12    & 8.4    & 6.0    & 3.9    & 3.0     & 2.7    \\
$27.5734-03.793$ & $87\times56$                      & 83  & 18   & 18    & 13    & 6.3    & 7.2    & 6.6     & 4.9    \\
$27.5607-04.779$\tablenotemark{c} & $97\times62$                      & 45  &9.6    & 8.6    & 20    & 8.1    & 6.6    & 5.6     & 6.7    \\
$27.5402-04.720$\tablenotemark{c} & $156\times12$                     & -55 &9.7    & 15    & \nodata & \nodata & \nodata & \nodata & \nodata \\
$27.5291-04.631$\tablenotemark{c} & $72\times37$                      & 73  &15    & 14    & \nodata & 17    & 16    & 13     & 11    \\
$27.4949-05.063$ & $67\times29$                      & 72  & 8.3    & 8.2    & 38    & 20    & 19    & 14     & 16    \\
$27.4722-04.705$ & $85\times37$                      & 90  & \nodata & 2.9    & 16    & \nodata & \nodata & \nodata  & \nodata \\
$27.4646-05.174$ & $77\times62$                      & 59  & 9.1    & 9.0    & \nodata & \nodata & \nodata & \nodata  & \nodata \\
$27.4646-06.555$ & $79\times33$                      & 77  & 11    & 12    & 15    & 8.9    & 9.0    & 9.7     & 8.8    \\
$27.3861-07.554$ & $84\times39$                      & -84 & 3.4    & 4.0    & 2.8    & 2.3    & 1.3    & 4.5     & 6.1    \\
$27.3659-06.900$ & $79\times35$                      & 72  & 6.3    & 6.4    & 14    & 7.5    & 7.9    & 7.6     & 6.4    \\
$27.2963-07.073$ & $78\times39$                      & 79  & 5.2    & 4.1    & \nodata & 1.3    & 1.7    & 2.4     & 1.2    \\
$27.2710-06.592$ & $54\times24$                      & 83  & 36    & 33    & 34    & 13    & 12    & 12     & 10    \\
\hline
$27.4833-04.570$\tablenotemark{d} & $5764\times982$\tablenotemark{e}  & 42  & \nodata & \nodata & 470     & 300     & 280     & 240      & 230     \\
$27.4602-05.845$\tablenotemark{d} & $422\times422$\tablenotemark{e}   & -51 & \nodata & \nodata & 17      & 19      & 16      & 18       & 18      \\
$27.4525-05.245$\tablenotemark{d} & $566\times311$\tablenotemark{e}   & 34  & \nodata & \nodata & 130     & 66      & 59      & 62       & 57      \\
$27.4371-05.695$\tablenotemark{d} & $9485\times3059$\tablenotemark{e} & 41  &\nodata & \nodata & 600     & 430     & 390     & 390      & 370     \\ \hline
\tablenotetext{a}{Gaussian component sizes measured against short-baseline 2.3 GHz LBA data except where noted otherwise.}
\tablenotetext{b}{Position angle of major axis measured east of north.}
\tablenotetext{c}{$27.5291-04.631$, $27.5402-04.720$ and $27.5607-04.779$ are the apparent source, jet and termination of a jet like source.}
\tablenotetext{d}{Sources used to model diffuse emission observed in ATCA data.}
\tablenotetext{e}{Sizes of large-scale diffuse emission sources measured against 17 GHz ATCA data.}
\end{tabular}
\caption{Flux densities of sources detected in NGC 4945.}
\label{tab:p3tabfluxes}
}
\end{center}
\end{table}

Based on the reduced resolution images, a total of 12 discrete sources fell above the $6\sigma$ threshold (0.91 mJy beam$^{-1}$) of the first epoch observation of NGC 4945, whereas 13 fell above the $6\sigma$ threshold (0.51 mJy beam$^{-1}$) of the second epoch observation. When imaged with the longer baselines included (Hobart and Ceduna), eight of these sources were detected in the first epoch data and nine in the second epoch. While source 27.4646$-$05.174 did not achieve the $6\sigma$ detection threshold in the epoch 1 data-set, an image of the residuals at the source location is included as part of Figure \ref{fig:p3fighra} for completeness sake. No observable position offsets were detected between the sources detected in epoch 1 and those detected in epoch 2.

Comparisons between the VLBI images and the higher frequency ATCA images were complicated by a number of factors. Firstly, source identification in the ATCA images was limited by the large-scale diffuse emission and confusion as a result of the significantly lower resolution of these images compared to that of the VLBI images. Secondly, when compared to the VLBI image, a systematic offset of 900 mas, 160 mas and 205 mas was observed in the 8.6 GHz, 17/19 GHz and 21/23 GHz images, respectively. The 4.8 GHz ATCA image was not used in any further comparisons in this paper as a result of its poor resolution. The large offset observed at 8.6 GHz results from a 945 mas error in the phase calibrator position of the original 1993 observation. The $\sim200$ mas offset observed between the high frequency ATCA images and the VLBI images is well below the ATCA beam size and is typical given atmospheric conditions at these frequencies and the $5\arcdeg$ separation between the phase calibrator and target source.

To aid in discrete source identification, four large elliptical Gaussian components were used to model and effectively subtract the large-scale diffuse emission from each of the ATCA images. The full-width half-maximum (FWHM) size, position angle (PA) and flux density of these components, as determined separately for each frequency observed with the ATCA, are listed in Table \ref{tab:p3tabfluxes}. To compare the resulting ATCA residual sources against the VLBI sources, the elliptical Gaussian components used in the VLBI model were transferred to each of the ATCA data-sets and their positions adjusted accordingly to account for the systematic offsets previously mentioned. Each of the model components were essentially unresolved with respect to even the smallest ATCA beam. With the size, position angle and location fixed, the flux density of each model component was varied to fit the diffuse-emission-subtracted ATCA data. The modelled flux densities for each source and data-set are listed in Table \ref{tab:p3tabfluxes}. No significant residuals were observed in the ATCA image or visibilities after the large-scale structure and source components were subtracted from the image.

\section{Discussion}

\subsection{Radio spectra and free-free absorption modelling of the compact sources}
\label{sec:p3ffmodel}
\citet{Lenc:2006p6695} and \citet{Tingay:2004p778} showed that the spectra of compact radio sources in NGC 253 exhibited a sharp downturn at 1.4 GHz, and to a lesser degree at 2.3 GHz, as a result of free-free absorption. To test a similar scenario quantitatively with the compact radio sources detected in NGC 4945, an analysis was performed using the data tabulated in Table \ref{tab:p3tabfluxes}. In this analysis, the second epoch VLBI data were used to model the low frequency end of the spectrum and the ATCA data were used to model the high frequency end of the spectrum. Archival 8.64 GHz data were also included to model the mid-range spectrum. Observations of compact radio sources in M82 \citep{Ulvestad:1994p1041} and NGC 253 \citep{Ulvestad:1997p907} have shown that for existing sources, little or no change in flux density is observed over periods of order a decade. Here we have assumed that if a source existed in both the archive 8.64 GHz data and our current ATCA or VLBI data then it has not significantly faded or brightened in the intervening years between the observations. Where no 8.64 GHz data exist we do not constrain our models to a lower limit based on the detection threshold at this frequency as we do not know if the source existed at that point in time.

In our analysis we used the same models and techniques described in \citet{Lenc:2006p6695}. However, for sources that were detected in only two observing bands a simple power-law spectrum was used to model the source spectrum, whereas the more highly sampled sources were also modelled with a free-free absorbed-power-law spectrum and a self-absorbed bremsstrahlung (free-free absorbed) spectrum. For the large-scale emission, lower limits were not applied at the low frequency end of the spectrum as this emission is totally resolved by our high-resolution VLBI observations.

\begin{figure}[p]
\epsscale{0.9}
\begin{center}
\plotone{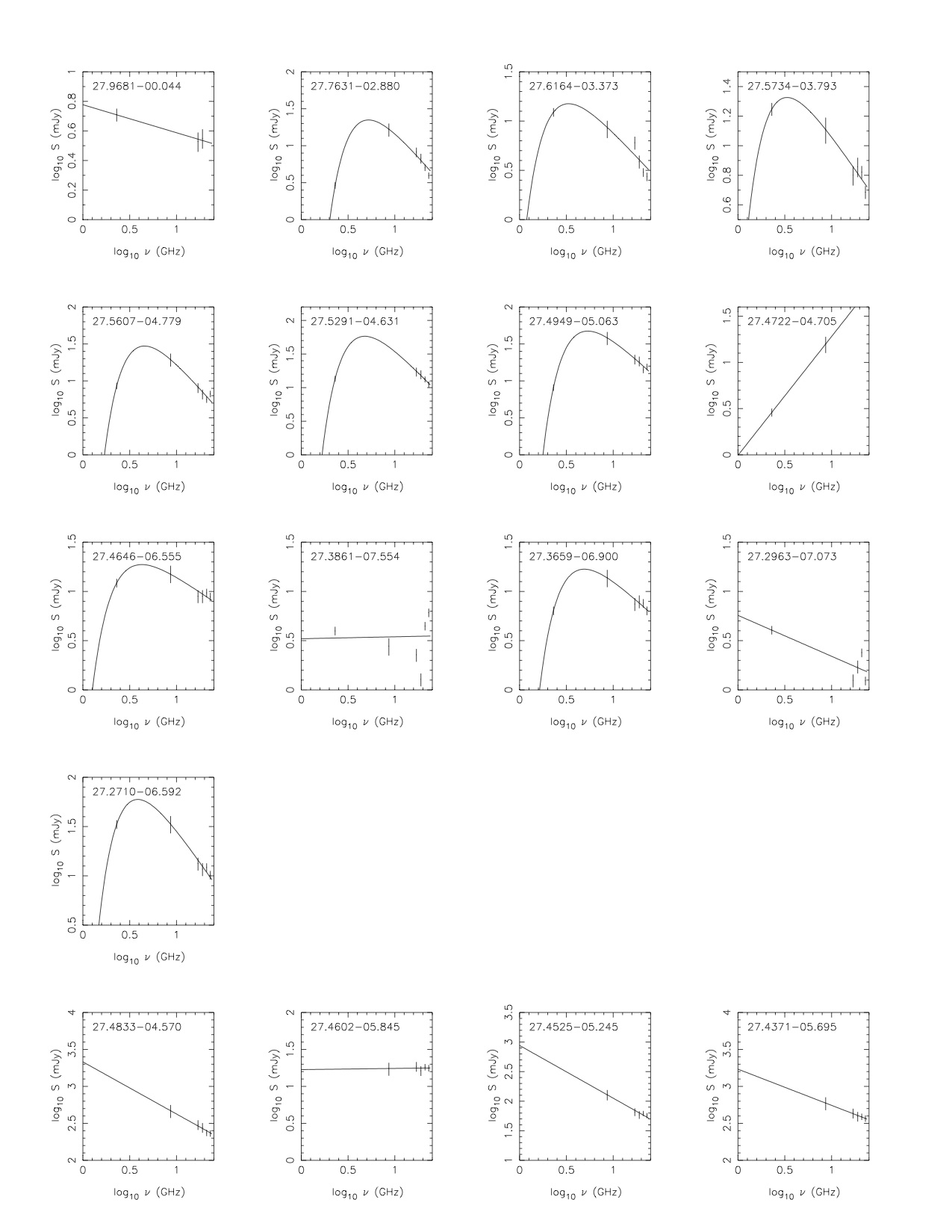}
\caption[Measured flux densities and free-free absorption models for 17 sources in NGC 4945]{Measured flux densities (symbols with error bars) and free-free absorption models (solid lines) for 17 detected sources (upper plots) and four diffuse sources detected with the ATCA data (lower plots).}
\label{fig:p3figff}
\end{center}
\end{figure}

The spectrum for each source was tested against each of the models using a reduced $\chi^{2}$ criterion to determine the best fit. $S_{0}$, $\tau_{0}$ and $\alpha$ from the model fits are listed in Table \ref{tab:p3tabsrcff}. In all cases, the sources were modelled best by either a simple power-law spectrum or a free-free absorbed-power-law spectrum. The model for each source is shown against the corresponding measured spectrum in Figure \ref{fig:p3figff}. Error bars of 10\% are shown for the flux densities at 23 GHz, 21 GHz and 2.3 GHz, whereas slightly larger errors of 15\%, 15\% and 20\% are estimated for the 19 GHz, 17 GHz and 8.64 GHz measures, respectively. The larger errors associated with the lower frequency ATCA observations take into consideration the effects of confusion and the increased levels of diffuse emission. The data shown in Table \ref{tab:p3tabsrcff} and Figure \ref{fig:p3figff} show that at least 9 of the 13 compact radio sources modelled are free-free absorbed.

Using the same criteria as \citet{Lenc:2006p6695}, of the 13 compact sources for which spectral indices could be determined in Table \ref{tab:p3tabsrcff}, three have flat intrinsic power law spectra ($\alpha > -0.4$) indicative of \ion{H}{2} regions dominated by thermal radio emission (T), the 10 remaining sources have steep intrinsic spectra (S), as normally associated with supernova remnants \citep{McDonald:2002p3330}. The proportion of synchrotron sources (77\%) compared to thermal sources (23\%) in NGC 4945 is similar to that found in other nearby starburst galaxies. \citep{McDonald:2002p3330} found 65\% of the compact radio sources in M82 were synchrotron sources and 35\% thermal, whereas \citep{Lenc:2006p6695} found that 60\% of the detections were synchrotron sources and 40\% thermal (after correcting source 5.805$-$38.92 with a spectral index of -1.2). The comparatively lower proportion of thermal sources detected in NGC 4945 compared to M82 and NGC 253 is believed to be a result of the lower resolution of the ATCA compared to the VLA. Thermal sources are more prominent at the high frequency end of the spectrum, however, even at 23 GHz the ATCA does not have sufficiently long baselines to easily resolve thermal sources from each other and from the large-scale diffuse emission.

The red circles in Figures \ref{fig:p3figLBANGC4945}(a) and \ref{fig:p3figLBANGC4945}(b) show how the free-free opacity varies across the field of the nuclear region of NGC 4945. Although there appears to be significantly greater free-free absorption toward the nuclear region of NGC 4945, the ionised medium appears to be clumpy. As we are viewing the galaxy almost edge-on this may be the result of seeing sources in front of the disk, embedded within the disk, or behind the disk, rather than being indicative of variation in the ionised medium. However, it is also possible that the absorption may be localised around the source itself, such as has been observed in young, dense, heavily-embedded clusters \citep{Turner:1998p22446, Kobulnicky:1999p28204, Johnson:2003p28201}.

\begin{table}[ht]
\begin{center}
{ \normalsize
\begin{tabular}{lccccc} \hline \hline
Source & $S_{0}$ & $\alpha$ & $\tau_{0}$ & Type\tablenotemark{a} & $S_{1.4}$\tablenotemark{b} \\
       & (mJy)   &          &            &                       & (mJy)   \\ \hline \hline
$27.9681-00.044$ &  5.99    & -0.19   & \nodata & T       & 5.62    \\
$27.7631-02.880$ & 534     & -1.49   & 22.9    & S       & 0.004   \\
$27.6164-03.373$ &  84.5    & -1.03   & 6.16    & S       & 2.87    \\
$27.5734-03.793$ &  101     & -0.93   & 5.57    & S       & 4.74    \\
$27.5607-04.779$ &  551     & -1.47   & 16.9    & S       & 0.081   \\
$27.5402-04.720$ &  \nodata & \nodata & \nodata & \nodata & \nodata \\
$27.5291-04.631$ &  1110    & -1.45   & 18.4    & S       & 0.079   \\
$27.4949-05.063$ &  608     & -1.19   & 19.1    & S       & 0.033   \\
$27.4722-04.705$ &  0.99    & 1.29    & \nodata & T       & 1.52    \\
$27.4646-05.174$ &  \nodata & \nodata & \nodata & \nodata & \nodata \\
$27.4646-06.555$ & 67.1    & -0.66   & 6.62    & S       & 2.05    \\
$27.3861-07.554$ &  3.31    & -0.02   & \nodata & T       & 3.33    \\
$27.3659-06.900$ &  103     & -0.88   & 11.8    & S       & 0.23    \\
$27.2963-07.073$ &  5.72    & -0.42   & \nodata & S       & 4.97    \\
$27.2710-06.592$ &  745     & -1.38   & 11.2    & S       & 1.84    \\
\hline
$27.4833-04.570$ &  2150    & -0.70   & \nodata & \nodata & 1690    \\
$27.4602-05.845$ &  16.9    &  0.02   & \nodata & \nodata & 17.0    \\
$27.4525-05.245$ &  875     & -0.89   & \nodata & \nodata & 648     \\
$27.4371-05.695$ &  1710    & -0.49   & \nodata & \nodata & 1450    \\ \hline
\tablenotetext{a}{As discussed in the text. Thermal sources are shown as type T, and non-thermal synchrotron sources are shown as type S.}
\tablenotetext{b}{Estimated flux density at 1.4 GHz based on model parameters.}
\end{tabular}
\caption{Parameters of the free-free absorption models for all detected sources in NGC 4945.}
\label{tab:p3tabsrcff}
}
\end{center}
\end{table}

Based on the model parameters, the flux density of each source at 1.4 GHz, $S_{1.4}$, is calculated and listed in Table \ref{tab:p3tabsrcff}. It is expected that up to eight sources may be detected at 1.4 GHz assuming a detection threshold of approximately 1 mJy beam$^{-1}$. Follow-up observations of NGC 4945 have been proposed with the LBA to further constrain the free-free parameters of the sources at lower frequencies. Such an observation would also help to highlight if any of the source flux density measures have been adversely affected by resolution effects since the LBA beam at 1.4 GHz would be substantially larger than at 2.3 GHz. Furthermore, the proposed observation would also highlight any observed structures that may have resulted from structural break-up during deconvolution (Section \S~\ref{sec:p3lbareduction}).

\subsection{Source variation}
\label{sec:fluxvar}

The mean of the individual source flux density ratios between the two VLBI epochs is 1.06, while the median is 0.98. If only relatively strong sources are considered ($S_{2.3}>6$ mJy), to exclude those that are significantly affected by deconvolution errors and measurement uncertainties, we observe a mean flux density ratio of $1.00$ and a median of $0.985$ between epochs. This suggests that there are no significant systematic differences between the flux-density scales or general source fading or brightening between the two epochs. None of the eight sources in this partially restricted sample exhibited flux variations above the 10\% errors of our measures. If we take the median ratio as an upper limit on the median fading of supernova remnants in NGC 4945 then this would equate to a 1.5\% reduction in flux density over the 1.9 yr baseline between the two VLBI epochs, or a fading of $<0.8$\% yr$^{-1}$. Further observations, over a longer period of time, would be required to confirm whether this fading is indeed real.

\subsection{Comments on individual compact radio sources}
\subsubsection{\ion{H}{2} regions}
The sources 27.9681$-$00.044, 27.4722$-$04.705 and 27.3861$-$07.554 are flat-spectrum sources that may be associated with \ion{H}{2} regions. All three sources are comparatively weak at 2.3 GHz and are resolved in the low resolution VLBI images. Only one of the sources, 27.3861-07.554, is detected in the high resolution VLBI images. The source 27.4722-04.705 is detected only in the second, more sensitive, epoch image and is not believed to be a new source.

It should be noted that the classification of the three sources as \ion{H}{2} regions is not entirely consistent with their brightness temperatures ($\sim2\times10^{5}$ K) at the low frequency end of the spectrum. Significant errors in the measured spectral index of these particular sources may have resulted in their misclassification. These errors arise from poor sampling in the spectral energy distribution (SED), scatter in the SED as a result source modelling errors and underestimates in the total flux density of extended sources in VLBI images. Another possibility is that the VLBI detections are of one or more supernova remnants embedded within \ion{H}{2} regions. These possibilities could be tested with further observations at intermediate and lower frequencies by improving the modelling of the SED of these sources.

None of the flat-spectrum sources appear to be significantly free-free absorbed, even those that appear to be relatively close to the nuclear centre. Similar characteristics are also observed in NGC 253 where only two flat-spectrum sources are detected at 2.3 GHz, both of which are only weakly free-free absorbed at this frequency \citep{Lenc:2006p6695}. Any curvature due to free-free absorption is difficult to detect in these sources since the low frequency VLBI observations are only sensitive to high brightness temperature sources, furthermore, the absorption acts in the same direction as the intrinsic spectrum.

\subsubsection{Steep-spectrum sources}

A total of ten steep-spectrum sources have been identified in NGC 4945. Two of the synchrotron sources, $27.5291-04.631$ and $27.5607-04.779$, in combination with $27.5402-04.720$, resemble a jet-like source and may be associated with a relativistic jet near the nucleus of NGC 4945 (Section \S~\ref{sec:jet}). Both of the synchrotron sources are significantly affected by free-free absorption and to a similar degree ($\tau_{0}=17-18$) suggesting that they are in close proximity of each other and deeply buried behind an ionised screen.

Three of the synchrotron sources $27.6164-03.373$, $27.5734-03.793$ and $27.3659-06.900$, have shell-like structures which may be associated with supernova remnants and have shell diameters of 110 mas, 100 mas and 60 mas, respectively. Assuming an average expansion velocity of $v=$10,000 km s$^{-1}$ relative to the shell centre, similar to that measured for the supernova remnant 43.31+592 in M82 \citep{Pedlar:1999p3534}, this implies ages of $\sim100(10^{4}/v)$ yr, $\sim90(10^{4}/v)$ yr and $\sim50(10^{4}/v)$ yr for these remnants, respectively. The source $27.4646-05.174$ also has a shell-like structure in the VLBI image and is assumed to be steep spectrum source since it is not detected at higher frequencies. This source has a diameter of 70 mas and, if it is associated with a supernova remnant, has an age of $\sim60(10^{4}/v)$ yr.

Two of the synchrotron sources $27.7631-02.880$ and $27.2963-07.073$ are comparatively weak and are fully resolved in the high resolution VLBI images. The remaining sources, $27.4646-06.555$, $27.4949-05.063$ and the brightest source, 27.2710-06.592, all are partially resolved and of similar size ($30\times25$ mas) but exhibit no significant structure. These may be young remnants with ages of $\sim25(10^{4}/v)$ yr.

\subsection{A relativistic jet at the nucleus of NGC 4945?}
\label{sec:jet}

\begin{figure}[p]
\epsscale{0.3}
\begin{center}
\mbox{
\plotone{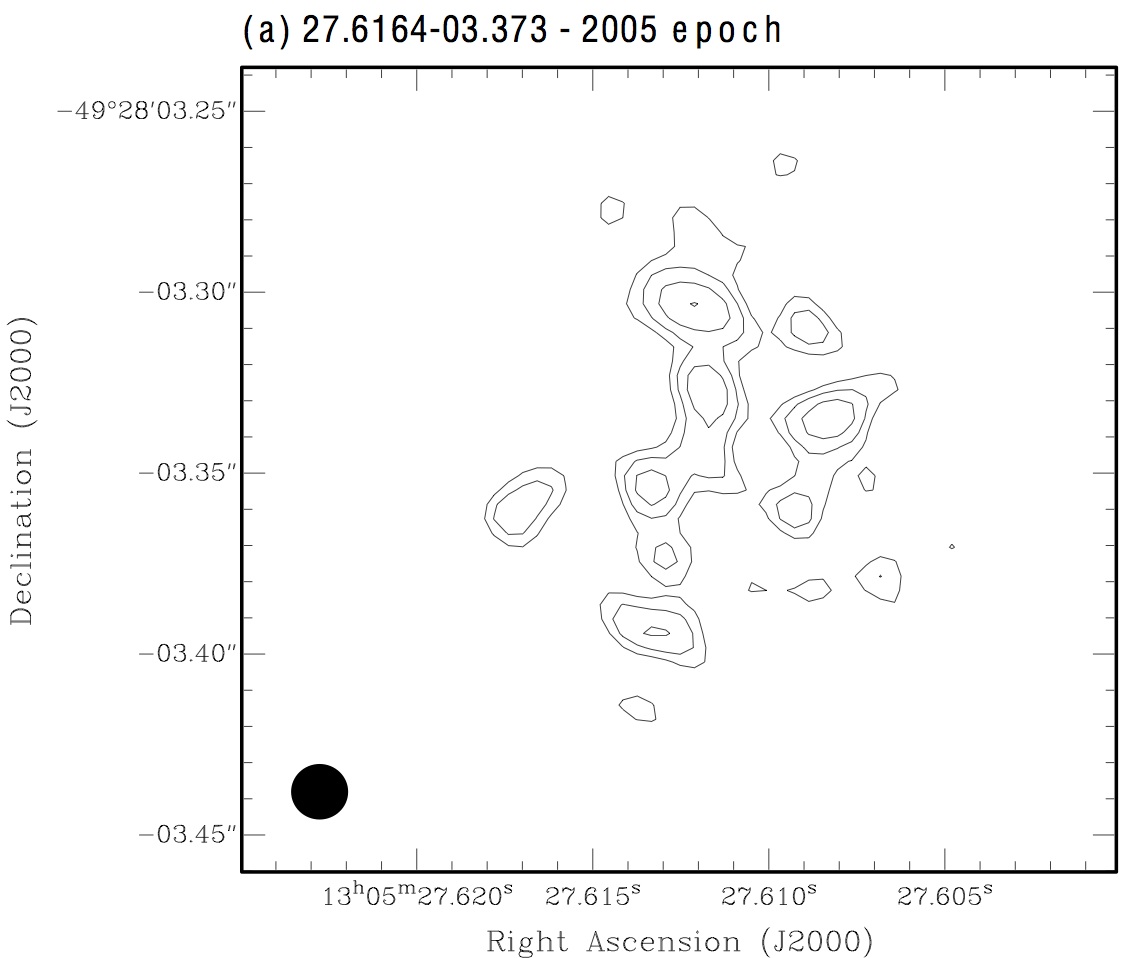} \quad
\plotone{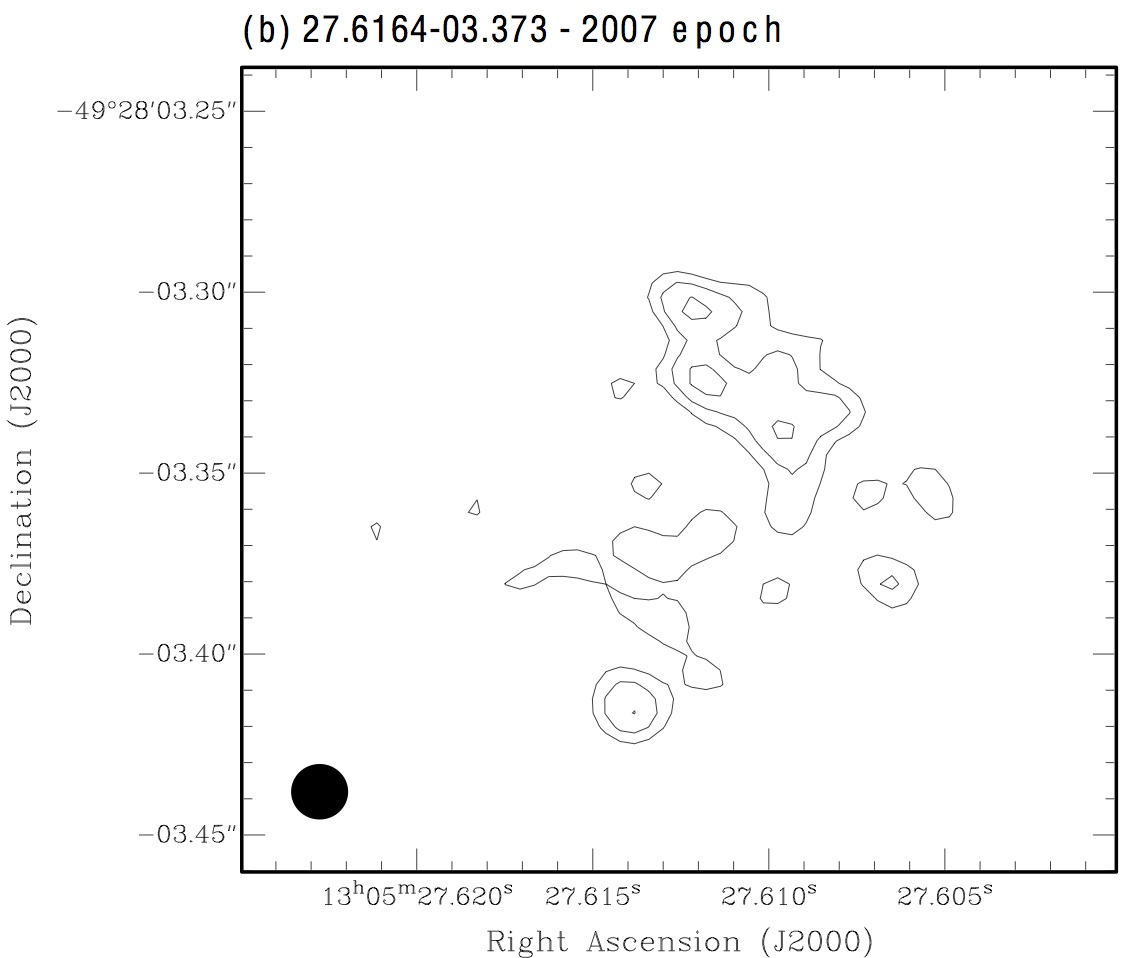} \quad
\plotone{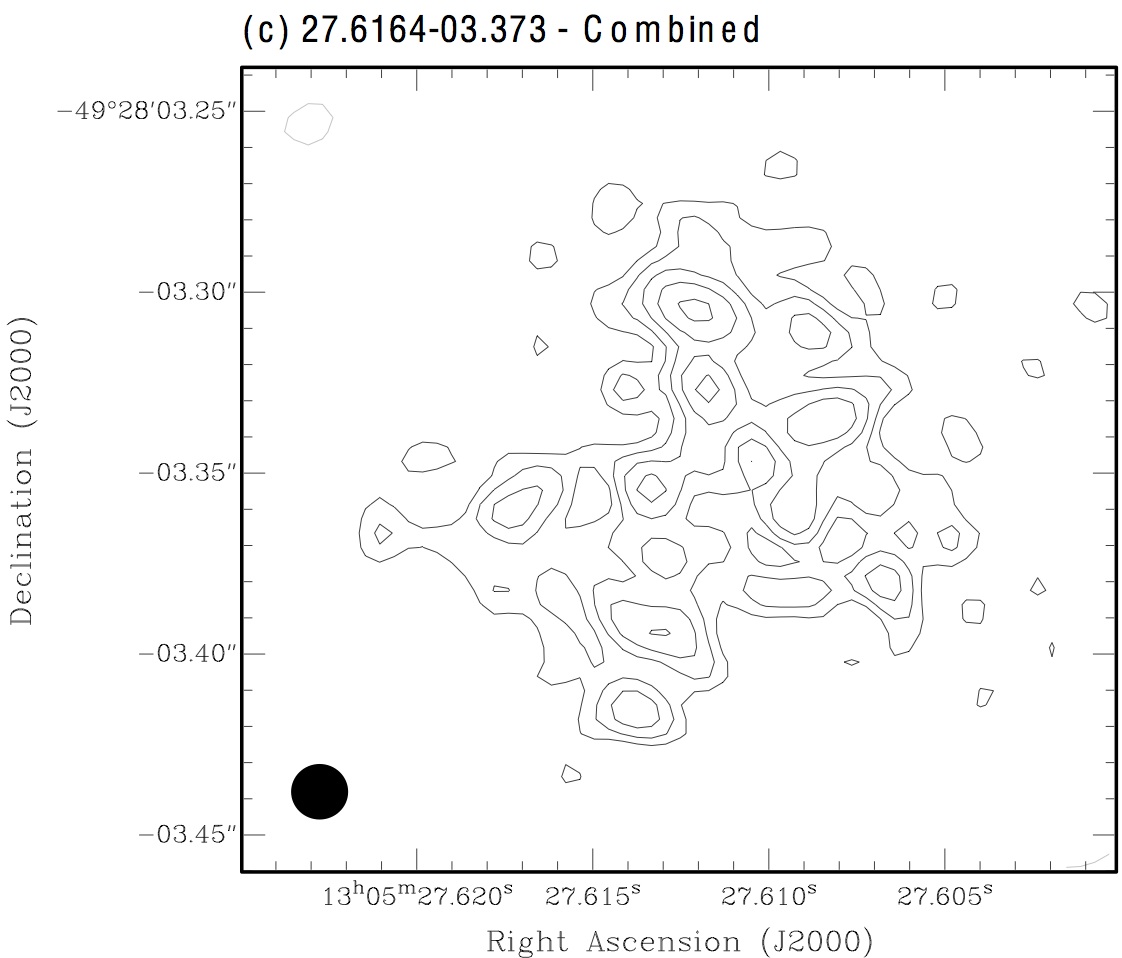}
}
\mbox{
\plotone{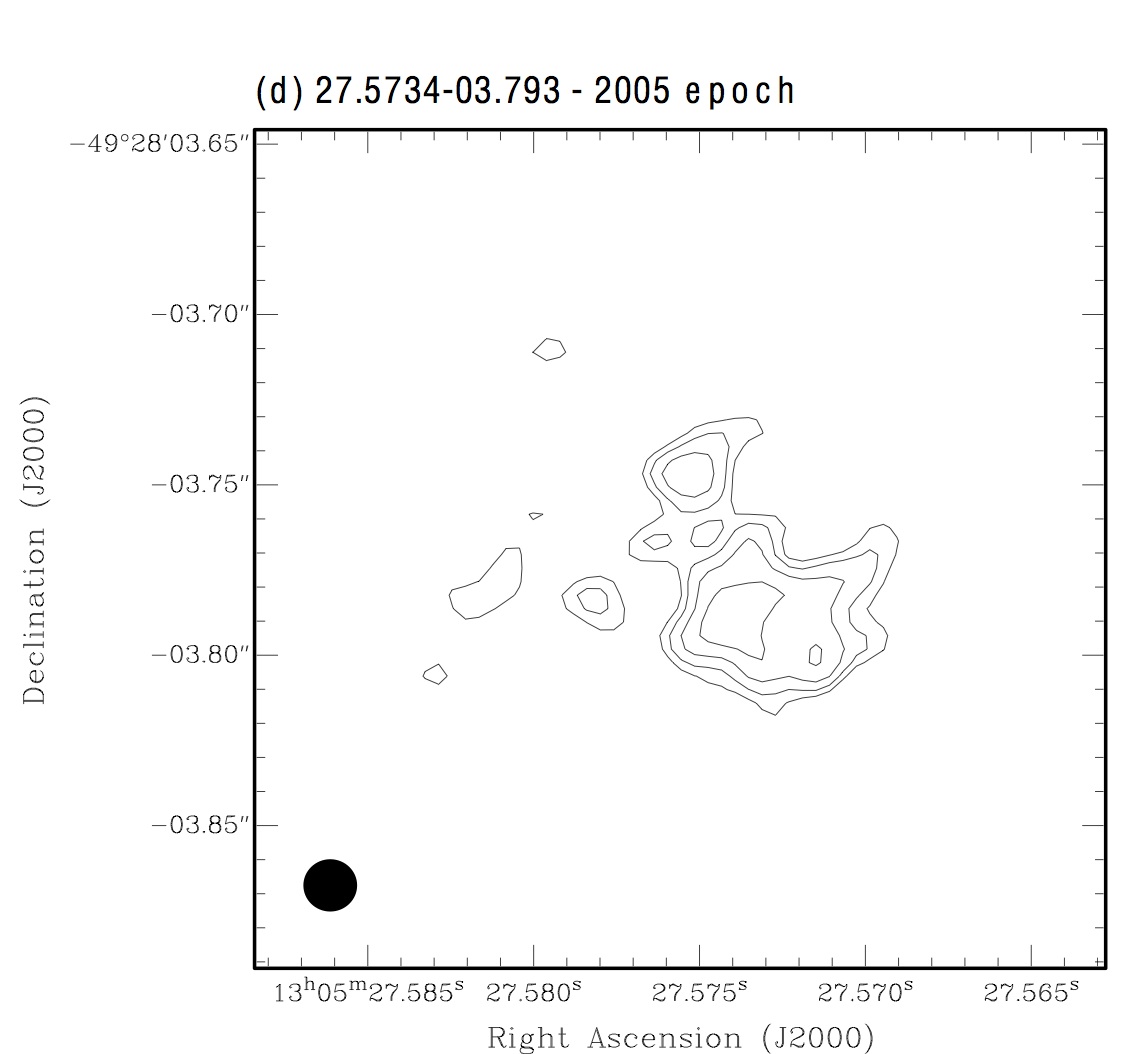} \quad
\plotone{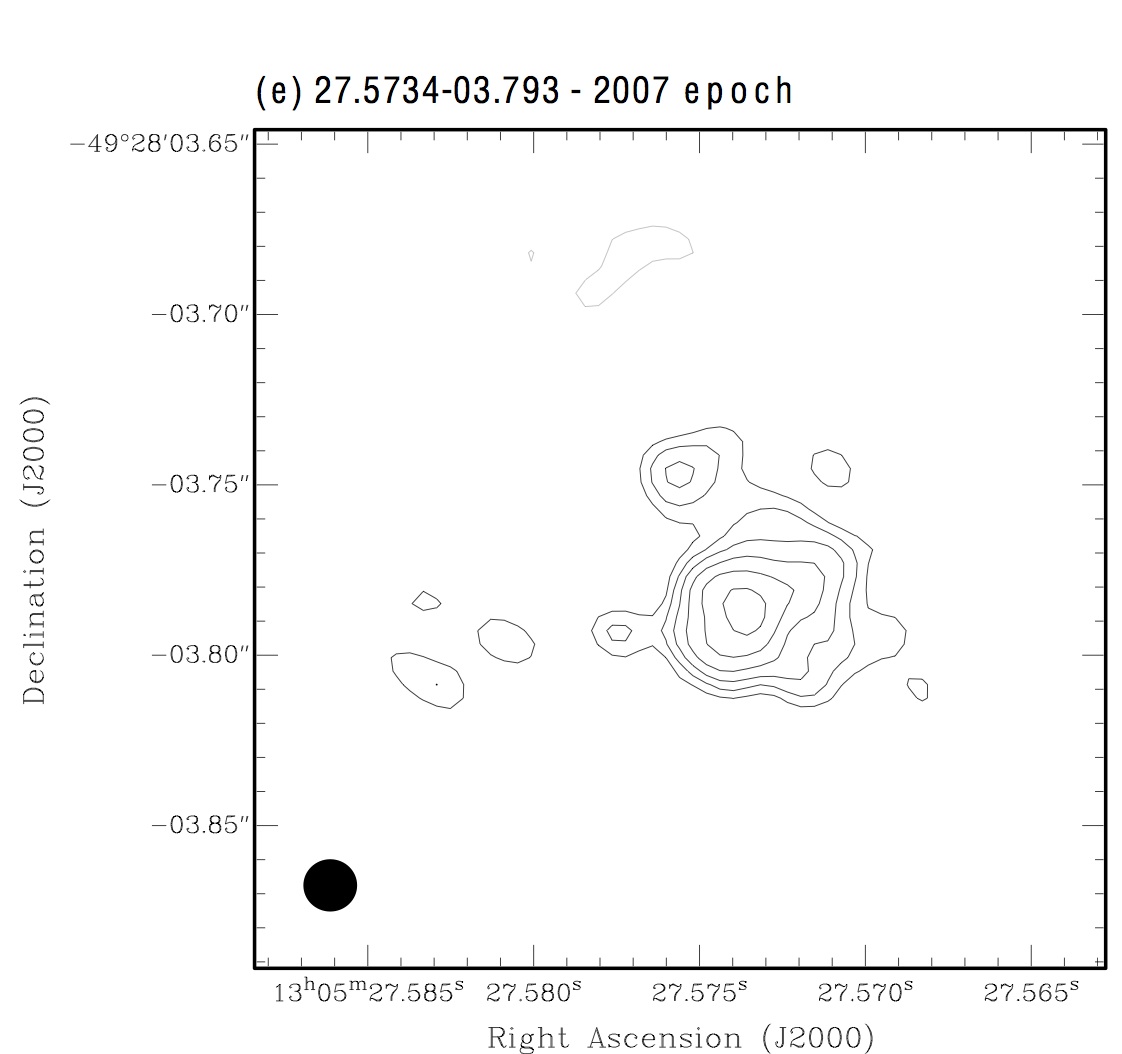} \quad
\plotone{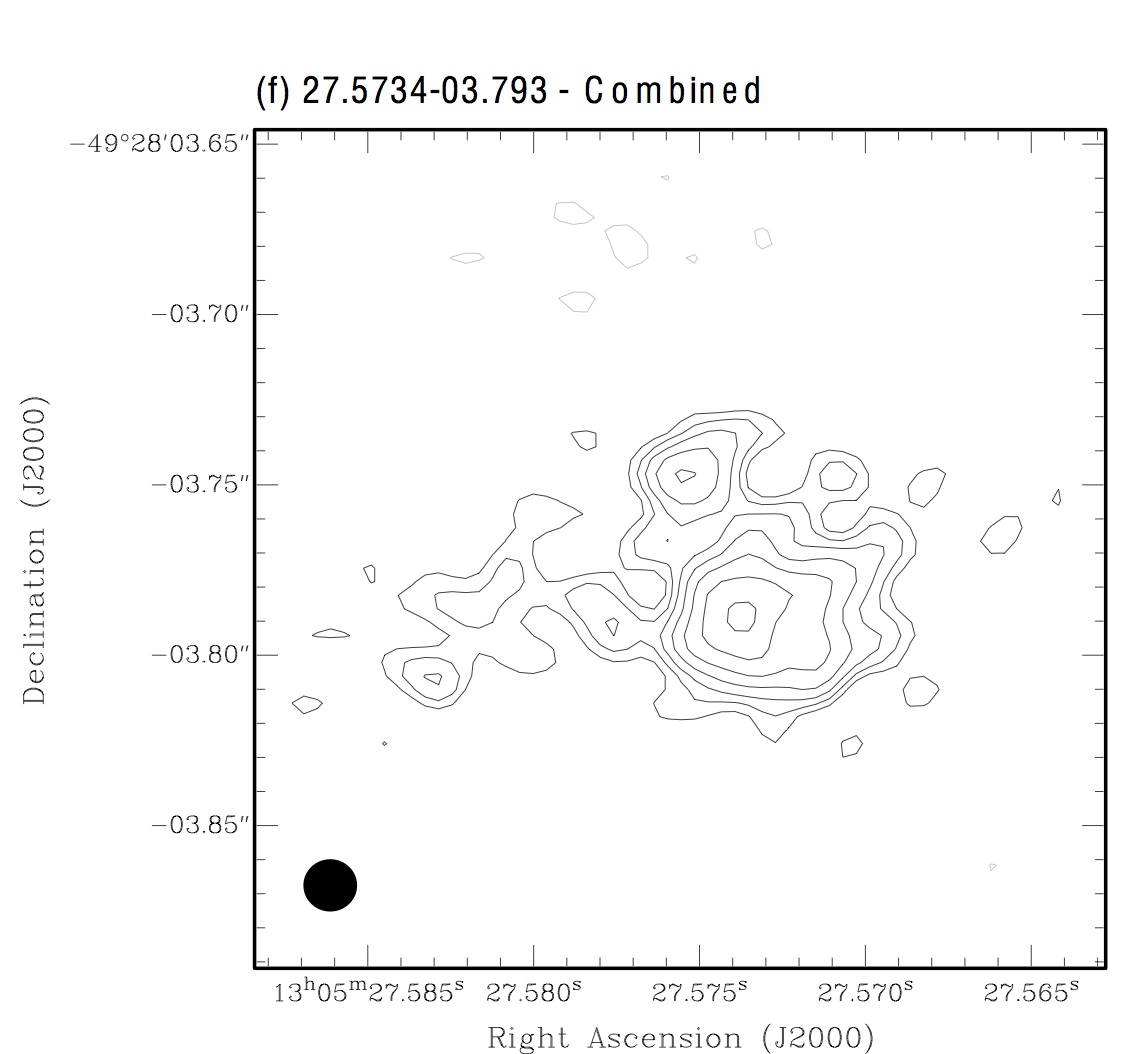}
}
\mbox{
\plotone{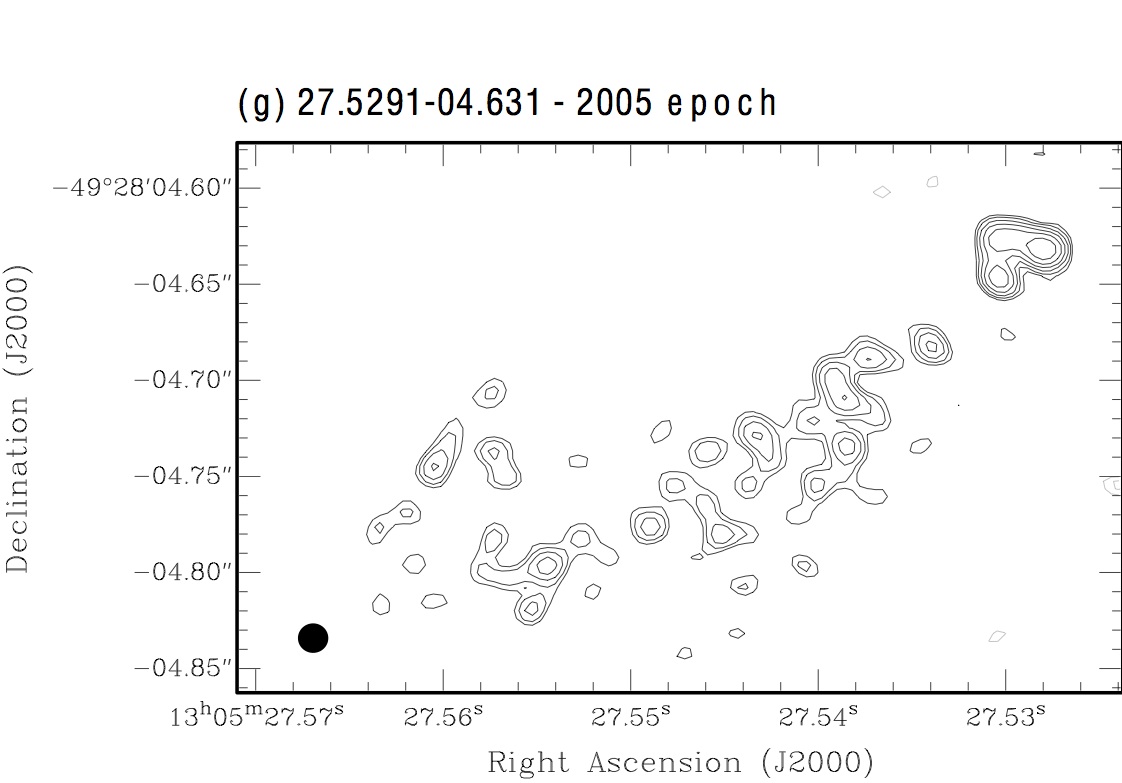} \quad
\plotone{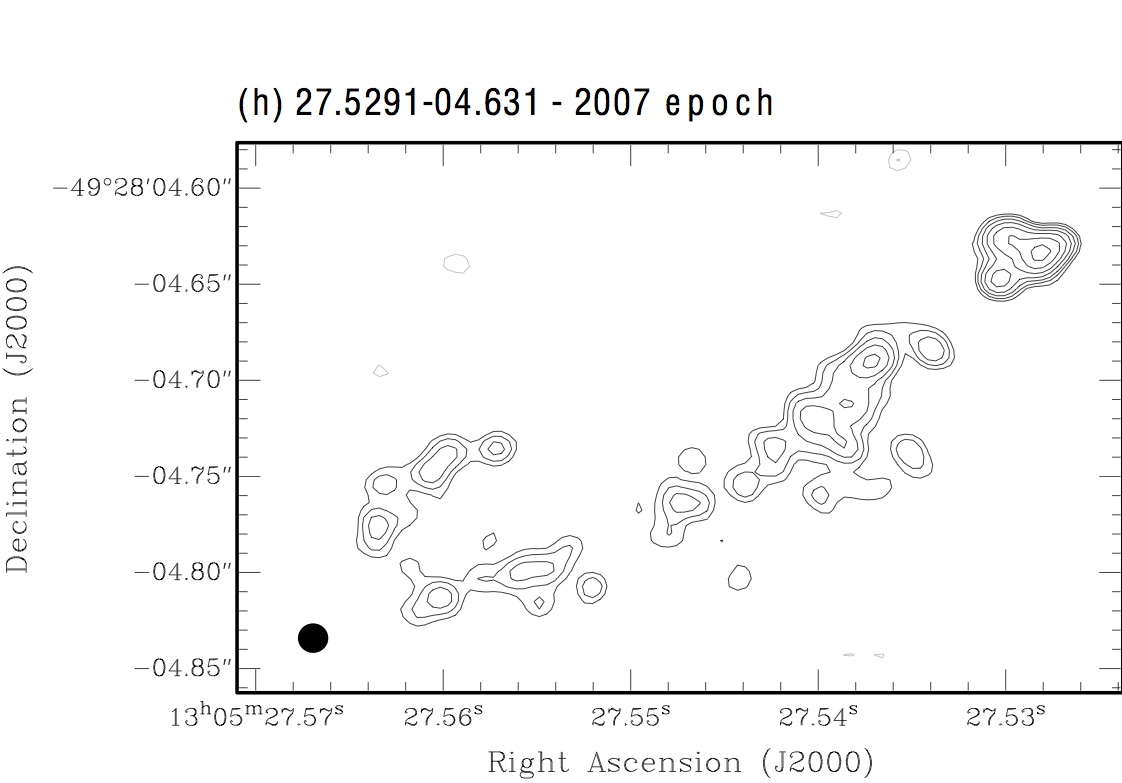} \quad
\plotone{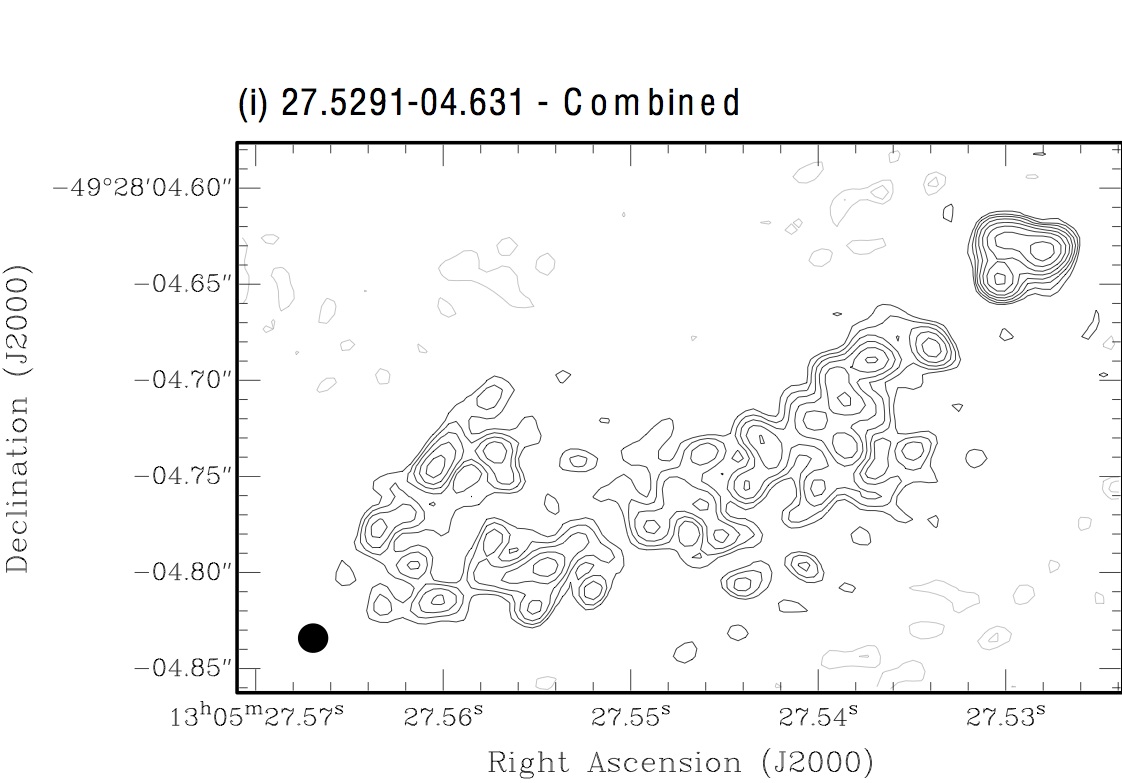}
}
\mbox{
\plotone{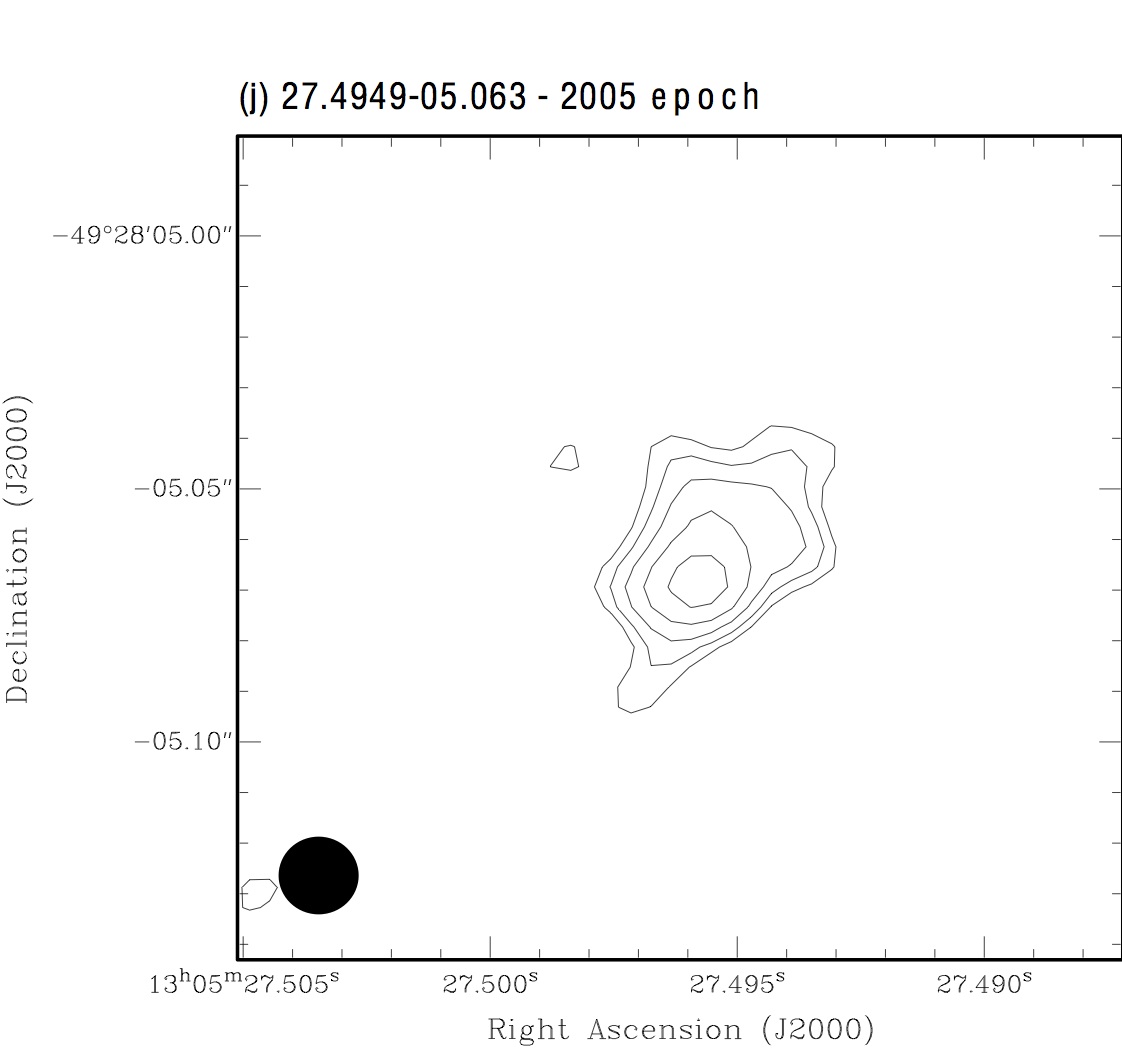} \quad
\plotone{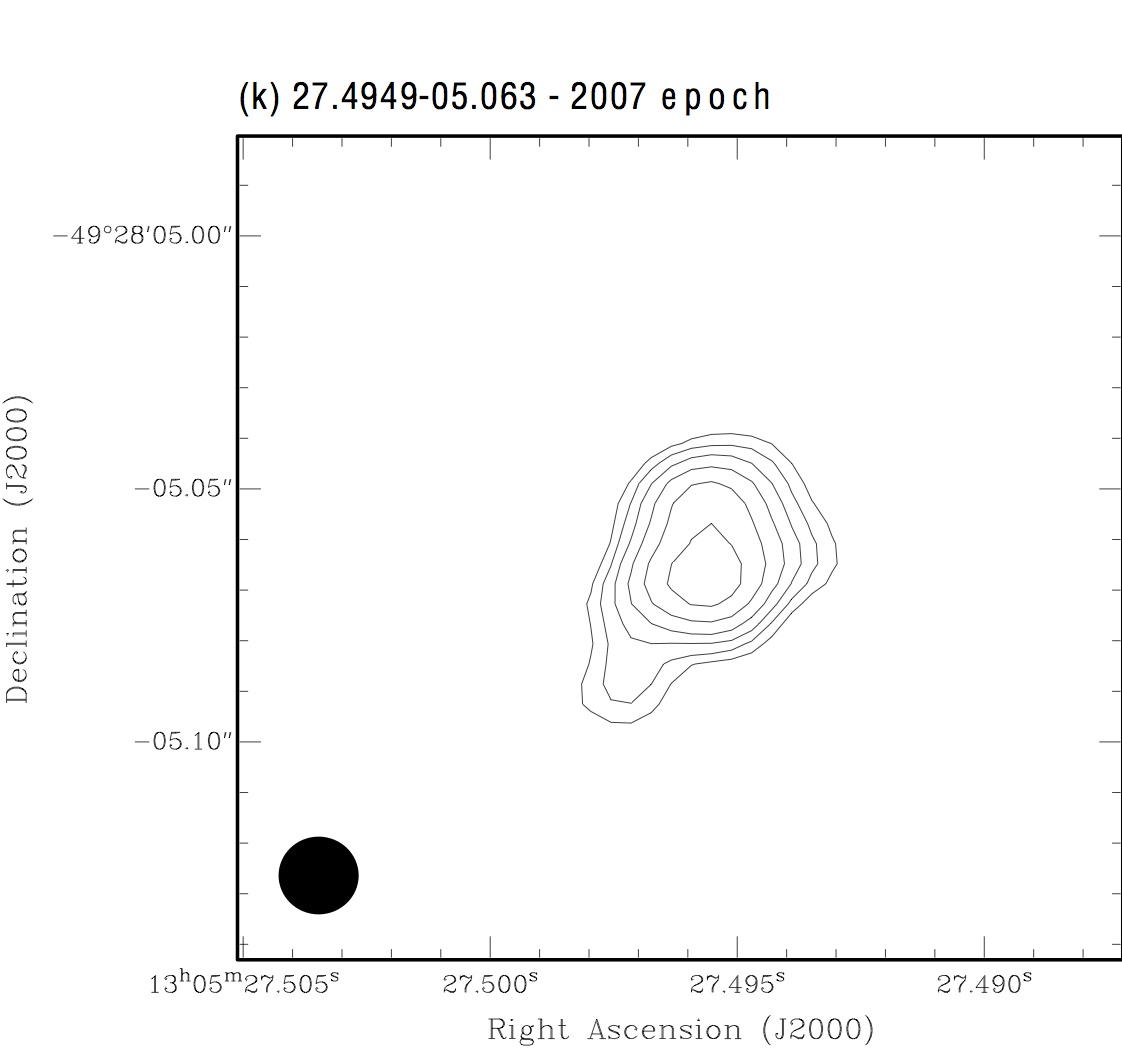} \quad
\plotone{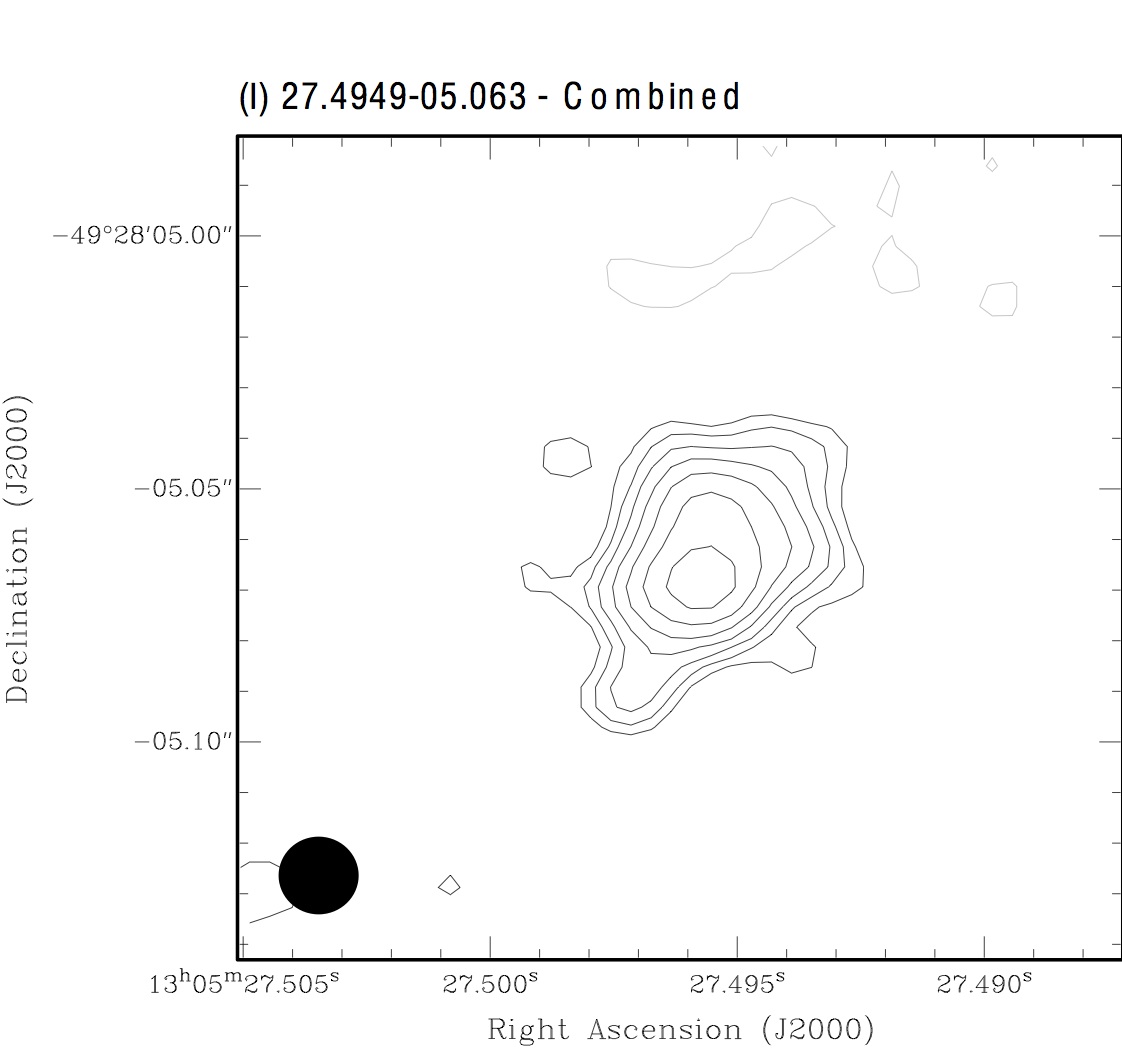}
}
\caption[High resolution Australian LBA images of compact sources detected in NGC 4945 at 2.3 GHz]{Naturally-weighted total-power maps of compact sources detected in NGC 4945 using all six antennas of the LBA at 2.3 GHz. Map statistics for the individual maps are shown in Table \ref{tab:p3tabimage}. Contours are drawn at $\pm2^{0}, \pm2^{\frac{1}{2}}, \pm2^{1}, \pm2^{\frac{3}{2}}, \cdots$ times the $3\sigma$ rms noise for all maps.}
\label{fig:p3fighra}            
\end{center}
\end{figure}

An intriguing feature of our VLBI images of NGC 4945 is the jet-like morphology of $27.5291-04.631$, seen in Figures \ref{fig:p3figlr}(a) and \ref{fig:p3figlr}(b) and, at higher resolution in Figures \ref{fig:p3fighra}(g), \ref{fig:p3fighra}(h) and \ref{fig:p3fighra}(i).  The morphology of this source is unlike that of any of the other radio sources seen in our images of the NGC 4945 starburst.  Rather than being a compact or somewhat resolved discrete source, $27.5291-04.631$ has an appearance strongly suggestive of the core-jet morphologies seen in arcsecond and milli-arcsecond scale radio images of powerful AGN - radio galaxies and quasars. $27.5291-04.631$ consists of a bright and unresolved component, located at the north-west end of an elongated structure of approximately 250 mas angular length and $<80$ mas angular width (5 pc and 1.5 pc, respectively, at the 3.82 Mpc distance of NGC 4945).  The elongated structure consists of at least 4 regions of enhanced brightness when imaged at low resolution (Figures \ref{fig:p3figlr}(a) and \ref{fig:p3figlr}(b)).  In the high resolution images in Figures \ref{fig:p3fighra}(g), \ref{fig:p3fighra}(h) and \ref{fig:p3fighra}(i), the elongated structure appears to be almost more resolved, with no particularly bright, compact components standing out, aside from the compact component at the north-western end of the structure. This indicates a near-uniform surface brightness distribution along the length of the elongated structure.  The images at the two observing epochs are also highly consistent in the features detected.  In particular, high resolution images from the data at both epochs show a highly linear elongated structure close to the compact north-west component, but a clear apparent bend in the structure at the south-east end of the structure.

What is the nature of $27.5291-04.631$?  As noted above, the morphological similarity of the source to a typical AGN core-jet source is striking.  We explore this possibility below.  Another possibility is that $27.5291-04.631$ is the result of a chance alignment along the line of sight of multiple supernova remnants, similar to the other remnants detected in our VLBI image.  While a possibility, this interpretation does not appear overwhelmingly likely.  For example, having detected $\sim10$ remnants in an angular area of 25 sq. arcseconds (Figure \ref{fig:p3figLBANGC4945}), and assuming a uniform distribution of remnants in this area, the chance that 3 remnants (probably the minimum required to explain the $27.5291-04.631$ structure) occur in an area of $\sim0.1$ sq. arcseconds is $\sim0.01$.  Further, the high resolution observations in Figure \ref{fig:p3fighra} provide a further filter on the data, only detecting emission from the highest brightness temperature sources in the field.  Along with $27.6164-03.373$ and $27.5734-03.793$, $27.5291-04.631$ was one of only three high brightness temperature sources detected.  If we revise the above calculation with a total of 5 high brightness temperature sources detected, and again assume a uniform distribution of these sources over the area of the starburst region, the requirement that 3 of these 5 lie together to explain the structure of $27.5291-04.631$ has a probability of only $\sim0.001$.  Even assuming an increase in the density of remnants toward the centre of the starburst does not make the chance alignment argument a plausible one.

The interpretation of $27.5291-04.631$ as a jet implies that the compact component at the north-west end of the elongated structure would coincide with a compact, massive object that provides the engine for the jet, a massive black hole being the likely candidate.  In AGN, the supermassive black holes coincide with the dynamical centroid of the host galaxy.  This does not seem to be the case in NGC 4945, as can be seen from Figure \ref{fig:p3figLBANGC4945}.

The one-sided appearance of the jet suggests that the jet may be relativistic.  We measure the jet to counterjet surface brightness ratio and find $R > 22$.  Assuming the continuous slab jet model \citep{Blandford:1977p21002}, a constraint can be placed on the product $\beta Cos \theta$, where $\beta$ is the speed of the radio emitting jet material (as a fraction of the speed of light) and $\theta$ is the angle that the motion of the jet material makes to our line-of-sight. In this calculation we assume optically thin emission and a jet spectral index of $\alpha=-0.7$ where $S_{\nu}\propto\nu^{\alpha}$. We estimate $\beta Cos \theta > 0.52$, using a value of $R > 22$. Since we only have a limit on $R$, it is not possible to derive hard constraints on $\beta$, given $\theta$, or vice versa.  A value of $\beta=0.9$ implies $\theta<55^{\circ}$ whereas $\beta=0.52$ implies $\theta=0\arcdeg$ (i.e. perfectly aligned with our line of sight).

An alternative explanation of the one-sided appearance of the assumed jet is that the counterjet is heavily free-free absorbed. Our modelling of the jet SED suggests that it is significantly affected by free-free absorption ($\tau=17-18$), the counterjet would presumably be affected to an even greater degree. In our images, the extent of the jet feature ($\sim5$ pc) is significantly shorter than similar features in other Seyfert galaxies, such as NGC 6240 \citep{Gallimore:2004p24805}, Mrk 348 \citep{Peck:2003p24764}, NGC 4151 \citep{Mundell:1995p24903} and NGC 1068 \citep{Gallimore:1996p24849,Gallimore:1996p24868} which have jet features 9 pc, 50 pc, 260 pc and 1.3 kpc in extent, respectively. It is thus feasible that the counterjet is completely obscured by free-free absorption behind a dense ionised torus. Evidence for free-free absorption in gaseous disks has been found in other Seyferts \citep{Gallimore:1999p24742} and would be consistent with the observations in NGC 4945.

If the jet interpretation is correct then this would be the nearest detection of such a feature to date. No such structures are seen in the two prototypical starburst galaxies, M82 \citep{Pedlar:1999p3534,McDonald:2002p3330} and NGC 253 \citep{Lenc:2006p6695}. The next nearest detection of a Seyfert jet is in NGC 4151 at a distance of 13.3 Mpc \citep{Mundell:1995p24903}. This presents the opportunity to study jet physics at the highest possible spatial resolution.

The jet interpretation for 27.5291-04.631 is easily testable with further observations.  If monitored every 1$-$2 years, the motion of components away from the core may be detectable.  In particular, two bright components appear separated from the core by approximately 30 mas, at both epochs shown in Figure \ref{fig:p3fighra}.  We have measured the component separations from the core for these components and can put a limit on their motion, relative to the core, of $\sim10$ mas yr$^{-1}$.  For an intrinsic motion of $\beta=0.9$ and an angle to the line of sight of $45^{\circ}$, we calculate an apparent motion of $\sim25$ mas yr$^{-1}$.  Given this ballpark calculation, several years of monitoring would be required to confidently detect any motion of jet components.  We are continuing to monitor NGC 4945 in order to test the jet hypothesis by directly detecting the transport of energy from the core along the jet.

\subsection{The supernova rate in NGC 4945}
\label{sec:snrate}

A lower limit on the type II supernova rate associated with massive stars can be estimated, based on the number of detected remnants, their size and an assumed expansion rate. We detect of order 10 compact radio sources with sizes limits of $0.5-2$ pc, based on a distance of 3.82 Mpc to NGC 4945, that exhibit steep intrinsic spectra often associated with synchrotron emission from type II supernova remnants. Assuming an expansion rate of 10,000 km s$^{-1}$ (rate of expansion of FWHM of a Gaussian model), we estimate a supernova rate of $>0.1-0.4$ yr$^{-1}$. This is a 
lower limit because sources may have been missed as a result of being intrinsically weak, significantly free-free absorbed, or fully resolved. As our VLBI observations of NGC 4945 currently span a period of less than two years, no significant evolution is detected in the supernova remnants and their mean expansion rate remains unclear. It is therefore appropriate to rewrite the supernova rate limit as $\nu_{SN}>0.1 (v/10^{4})$ yr$^{-1}$, where $v$ is the shell radial expansion velocity in km s$^{-1}$.

It is interesting to note that three times as many VLBI sources are detected in NGC 4945 compared to NGC 253. Yet it is assumed, by considering the effects of confusion at 8.3 GHz in VLA observations, that there may be of order 100 compact radio sources in NGC 253 \citep{Ulvestad:1994p1041}, approximately 10 times the number detected in NGC 4945 at a similar frequency with the ATCA. In Section \S~\ref{sec:p3ffmodel} we discussed the detrimental effects of confusion and diffuse emission on the thermal source counts in NGC 4945. The same problematic effects limit counts of compact steep spectrum sources with the ATCA but even more so as these sources will be significantly weaker at higher frequencies. So it is likely that there are a great deal more sources in NGC 4945 than our observations lead us to believe. How this will effect the lower limit of the supernova rate is unclear and will depend on both the age and number of sources in this population.

An upper limit on the type II supernova rate may be estimated by statistically modelling the effect of non-detections. \citet{Lenc:2006p6695} developed a Monte Carlo simulation based on a hypothetical population of supernova remnants with a uniformly distributed peak luminosity between 5 and 20 times that of Cas A \citep{Weiler:1989p17306,Ulvestad:1991p1008}. This model has been adapted for NGC 4945 by scaling the flux density of the supernova remnants to the distance of this galaxy, 3.82 Mpc. Similarly, simulation parameters were set accordingly for the two epochs of VLBI observations and are summarised in Table \ref{tab:p3tabsnrate}. Our simulations suggest that our VLBI observations should have detected all of the hypothetical supernovae that were simulated for between the two epochs assuming there were no significant absorption effects. Given this simple model we determine a supernova rate of $<1.62$ yr$^{-1}$ at a 95\% confidence level. However, it is known that the nuclear region of NGC 4945 is highly obscured and it has been shown, Section \S~\ref{sec:p3ffmodel}, that the nuclear region is also situated behind a screen of ionized gas. In a second, more complete simulation, the effects of free-free absorption on the upper limit of the supernova rate are taken into consideration. For this test a median value of $\tau_{0}=11.8$ was taken from the free-free modelling in \S~\ref{sec:p3ffmodel}. With this simulation only 11\% of the hypothetical supernovae were detectable in the second epoch observation and the estimated supernova rate increased to $<15.3$ yr$^{-1}$.

\begin{table}[ht]
\begin{center}
{ \normalsize
\begin{tabular}{llllllll} \hline \hline
Epoch & Time & $\nu_{obs}$ & Sensitivity & $\beta_{SN}$\tablenotemark{a} & $\nu_{SN}$\tablenotemark{a} & $\beta_{SN}$\tablenotemark{b} & $\nu_{SN}$\tablenotemark{b} \\
      & (yr) & (GHz)       & (mJy)       &                               & (yr$^{-1}$)                 &                               & (yr$^{-1}$) \\ \hline \hline
1     & $\cdots$ & 2.3         & 0.91         & $\cdots$     & $\cdots$                     & $\cdots$     & $\cdots$     \\  
2     & 1.9      & 2.3         & 0.51         & 1.0          & $<1.62$                      & 0.11         & $<15.3$      \\ \hline
\tablenotetext{a}{Monte Carlo test run with no modelling of free-free absorption.}
\tablenotetext{b}{Monte Carlo test run with a median free-free opacity of $\tau_{0}=11.8$ applied.}
\tablecomments{The supernova rate upper limit based on Monte Carlo simulations run over two observing epochs. The time between epochs, the observing frequency and sensitivity of the observation are listed. At the end of the second epoch the proportion of supernova remnants detected ($\beta_{SN}$) in that epoch is listed together with an estimate of the supernova rate upper limit based on all observations prior to and including that epoch. }
\end{tabular}
\caption{NGC 4945 supernova rate upper limit based on Monte Carlo simulations run over two observing epochs.}
\label{tab:p3tabsnrate}
}
\end{center}
\end{table}

A further method of estimating the supernova rate is to estimate the median age of the supernova remnant population based on the observed flux variation between two epochs. In Section \S~\ref{sec:fluxvar} we observed no significant variation in the detected sources above our error limits. However, if we assume that the median variation in the strongest of sources is real then we find an upper limit of $\sim0.8$\% yr$^{-1}$ to the fading rate in NGC 4945. If the supernovae are assumed to fade in time as the $-0.7$ power of time \citep{Weiler:1986p9393} then we estimate that the median age of the supernova remnant population of $\sim85$ years. This would imply a supernova rate of $\sim0.12$ yr$^{-1}$, a value that falls within the limits determined above.

\subsection{The star formation rate in NGC 4945}
\label{sec:sfrate}

The star-formation rate (SFR) for stars with mass $M\geq5M_{\Sun}$ of a star-forming galaxy can be directly related to its radio luminosity $L_{\nu}$ at wavelength $\nu$ \citep{Condon:1992p10540, Haarsma:2000p10556}. This relation is understood purely from radio considerations by calculating the contribution of synchrotron radio emission from supernova remnants and of thermal emission from \ion{H}{2} regions to the observed radio luminosity. For consistency, we have taken the same form of this relation and the same assumptions described in \citet{Lenc:2006p6695}. Based on the total integrated flux densities measured in the ATCA observations between 8.64 GHz and 23 GHz, we derive a SFR of 14.4$\pm$1.4 $(Q/8.8)$ M$_{\Sun}$ yr$^{-1}$ for the inner $\sim$300 pc of the nuclear region of NGC 4945, where the factor $Q$ accounts for the mass of stars in the interval $0.1-100M_{\sun}$ and has a value of 8.8 if a Saltpeter IMF ($\gamma=2.5$) is assumed \citep{Haarsma:2000p10556}. Similar measurements based on the 24 GHz continuum of NGC 253 give a comparable result of $10.4\pm1.0$ M$_{\Sun}$ yr$^{-1}$ \citep{Lenc:2006p6695}.

Estimates of the SFR can be made based on the measured FIR luminosity $L_{FIR}$ since a large proportion of the bolometric luminosity of a galaxy is absorbed by interstellar dust and reemitted in thermal IR. \citet{Condon:1992p10540} determined the relationship between the SFR of a galaxy and the FIR luminosity by modelling the total energy emission of a massive star and assuming that the contribution of dust heating by old stars was negligible. In NGC 4945 this may be complicated by the presence of an AGN which, based on observations of the relatively high HCN/CO ratio, may contribute as much as 50\% of the bolometric luminosity \citep{Spoon:2000p7502, Curran:2001p10543}. However, more recent observations of CO spectral lines in the nuclear region suggest that star formation, rather than AGN activity, is the primary heating agent \citep{Chou:2007p10537} and so the SFR/FIR relationship should still provide a good estimate of the star formation rate.

For NGC 4945, the far-infrared luminosity, $L_{FIR}$, is $1.69\times10^{10}$ $L_{\sun}$ when adjusted for a distance of 3.82 Mpc \citep{Rice:1988p7793}. Based on \citet{Condon:1992p10540}, this leads to a star formation rate of 1.5 $M_{\sun}$ yr$^{-1}$. This is significantly lower than earlier estimates (8.3 $M_{\sun}$ yr$^{-1}$) based on total FIR luminosity \citep{Dahlem:1993p10544} as a result of the revised distance to NGC 4945. However, the new estimate is comparable to the value $2-8$ $M_{\sun}$ yr$^{-1}$ derived from the mass of ionised gas inferred from radio recombination line observations from the galaxy \citep{2005AIPC..783..303R}. This estimate is also compatible with the lower limit on the star formation rate, $>0.19$ $M_{\sun}$ yr$^{-1}$, derived from H$\alpha$ emission \citep{Strickland:2004p9721}.

For most galaxies it is also possible to directly relate the star formation rate with the supernova rate \citep{Condon:1992p10540}. Using the relation derived by \citet{Condon:1992p10540}, the supernova rate limits of $0.1 (v/10^{4})<\nu_{SN}<15.3$ yr$^{-1}$ (\S~\ref{sec:snrate}) give the limits $2.4 (v/10^{4}) < SFR(M\geq5M_{\Sun})<370$ M$_{\Sun}$ yr$^{-1}$ on the star formation rate. While the upper limit is poorly constrained due to the limited number of epochs thus far observed, it is of the same order of magnitude as the nuclear SFR determined from FIR emissions (1.5 M$_{\Sun}$ yr$^{-1}$) and in agreement with that determined from radio luminosities alone ($14.4\pm1.4$ $(Q/8.8)$ M$_{\Sun}$ yr$^{-1}$). The higher radio estimate of the SFR, compared to the FIR-based estimate, may suggest that the $Q$ factor has been over-estimated. A lower $Q$ would suggest an increased proportion of massive star formation and would also be consistent with the high densities associated with the nuclear region. Nonetheless, it is encouraging that the independent estimates of SFR are only a factor of a few different.

\subsection{Comparisons to the nearby starbursts M82 and NGC 253}

With the recent improvements in distance measurements, it appears that the distances to NGC 4945, M82 and NGC 253 are all within a few percent of 3.9 Mpc. This is a dramatic change, by up to a factor of two, from previous estimates. Such a change in distance has an significant effect on luminosity estimates, which are proportional to the square of the distance, and to a lesser degree on spatial measures, which are linearly proportional to the distance. With the three galaxies now converging on a similar distance it is interesting that the star formation rates and supernova rates are also converging. Based on FIR luminosities, we find that the star formation rates of NGC 4945, NGC 253 and M82 are now 1.5 $M_{\sun}$ yr$^{-1}$, 1.8$-$2.8 M$_{\Sun}$ yr$^{-1}$ \citep{Lenc:2006p6695} and $\sim2.0$ M$_{\Sun}$ yr$^{-1}$ \citep{Pedlar:2003p3309}, respectively. Similarly, the supernova rates are $>0.1 (v/ 10^{4})$ yr$^{-1}$, $>0.14 (v/10^{4})$ yr$^{-1}$ \citep{Lenc:2006p6695} and $\sim0.07$ yr$^{-1}$ \citep{Pedlar:2003p3309}, for NGC 4945, NGC 253 and M82, respectively, based on source counts and sizes.

\section{Summary}

We have imaged NGC 4945 at 2.3 GHz over two epochs using the LBA to produce the highest resolution images of the nuclear starburst region of this galaxy to date. The VLBI observations have been complemented with additional data between 8.64 and 23 GHz observed with the ATCA. We find the following results:
\begin{itemize}
\item Fifteen compact radio sources were detected with the LBA and 13 were identified higher frequency ATCA images.
\item In the highest resolution image (15 mas beam), four supernova remnants are resolved into shell-like structures ranging in size between 60 and 110 mas (1.1 to 2.1 pc) in diameter. Assuming an average radial expansion velocity of $v=10,000$ km s$^{-1}$, the remnants are estimated to be approximately $25 (10^{4}/v)$ to $100 (10^{4}/v)$ years of age.
\item By combining flux density measurements from 2.3 GHz LBA and high frequency ATCA observations, the spectra for 13 compact radio sources were determined. The spectra of 9 sources were found to be consistent with a free-free absorbed power law and 4 with a simple power law spectrum.
\item Ten of the 13 sources have steep intrinsic spectra normally associated with supernova remnants. The three remaining sources have flat intrinsic power law spectra ($\alpha>-0.4$) indicative of \ion{H}{2} regions but may have embedded supernova remnants owing to their high brightness temperature ($\sim2\times10^{5}$ K) at 2.3 GHz.
\item Based on the modelled free-free opacities of 9 sources, the morphology of the ionised medium in the central region of NGC 4945 is complex and clumpy in nature.
\item A type II supernova rate upper limit of 15.3 yr$^{-1}$ in the inner 250 pc region of NGC 4945 was derived from the absence of any new sources between epochs, taking into consideration the improved distance estimate for the galaxy, a median free-free opacity, and the sensitivity limits of two observations over a period of $\sim2$ years. A type II supernova rate of $>0.1 (v/ 10^{4})$ yr$^{-1}$ has been estimated based on an estimate of supernova remnant source counts, their sizes and their expansion rates.
\item A star formation rate of $2.4 (v/10^{4}) < SFR(M\geq5M_{\Sun})<370$ M$_{\Sun}$ yr$^{-1}$ has been estimated directly from supernova rate limits for the inner 250 pc region of the galaxy.
\item The supernova rates and star formation rates determined for NGC 4945 are, to within a factor of two, similar to those observed in NGC 253 and M82.
\item A jet-like source is detected that is offset by $\sim$ 1050 mas from the assumed location of the AGN based on H$_{2}$O megamaser emission, the HNC cloud centroid, the K-band peak and the hard X-ray peak.
\end{itemize}

We need to more stringently constrain the free-free parameters by observing a further epoch using VLBI at 1.4 GHz. Frequent observations will further constrain the upper limit on the supernova rate by increasing the proportion of supernova events that may be captured between epochs. Observations over a longer period of time ($\sim10$ years) will also further constrain the lower limit of the supernova rate by enabling the supernova shell expansion velocity to be measured, and allow us to search for motion in the suggested jet, thereby confirming its nature.

\clearpage
\epsscale{0.3}
\begin{center}
\mbox{
\plotone{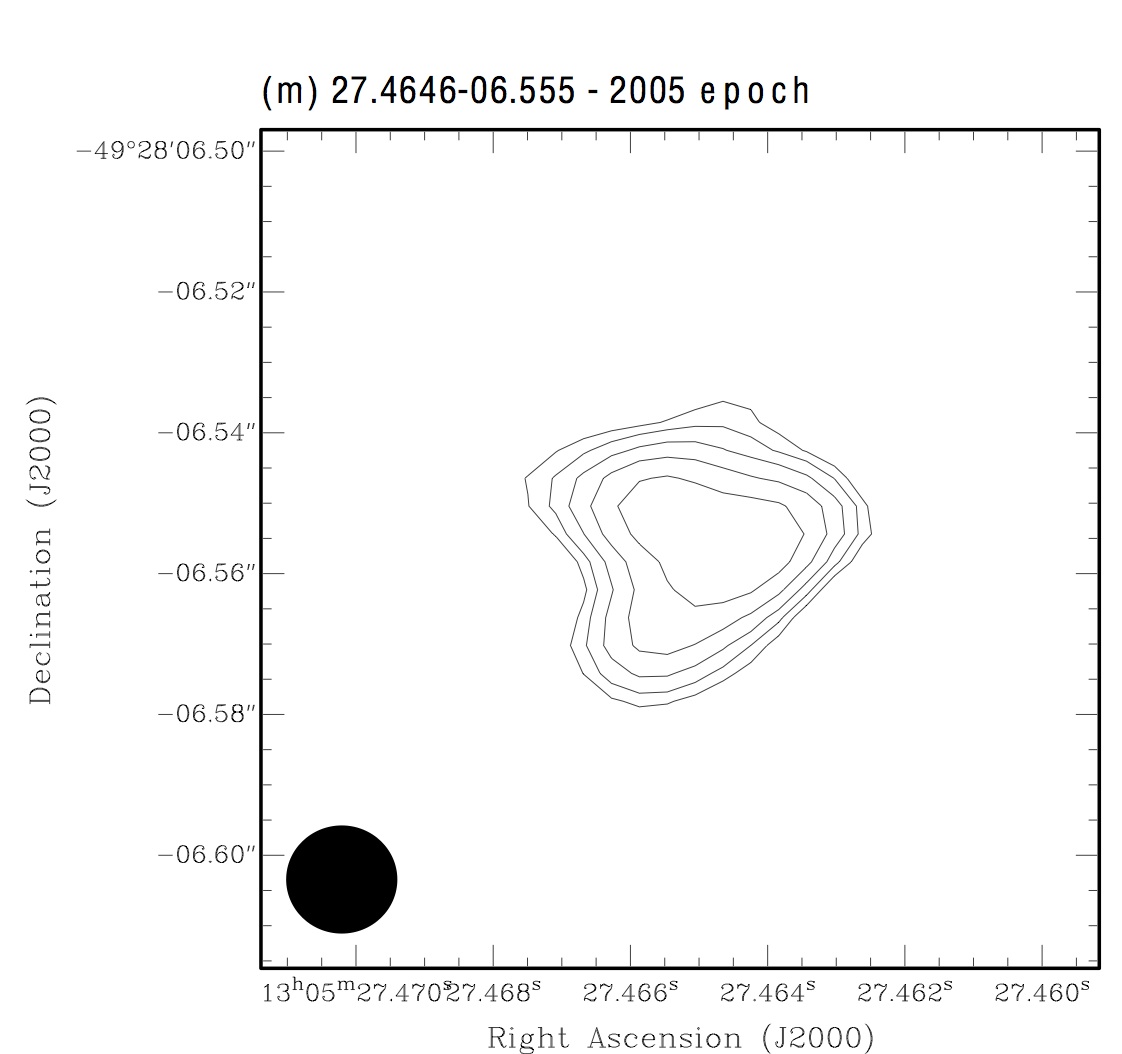} \quad
\plotone{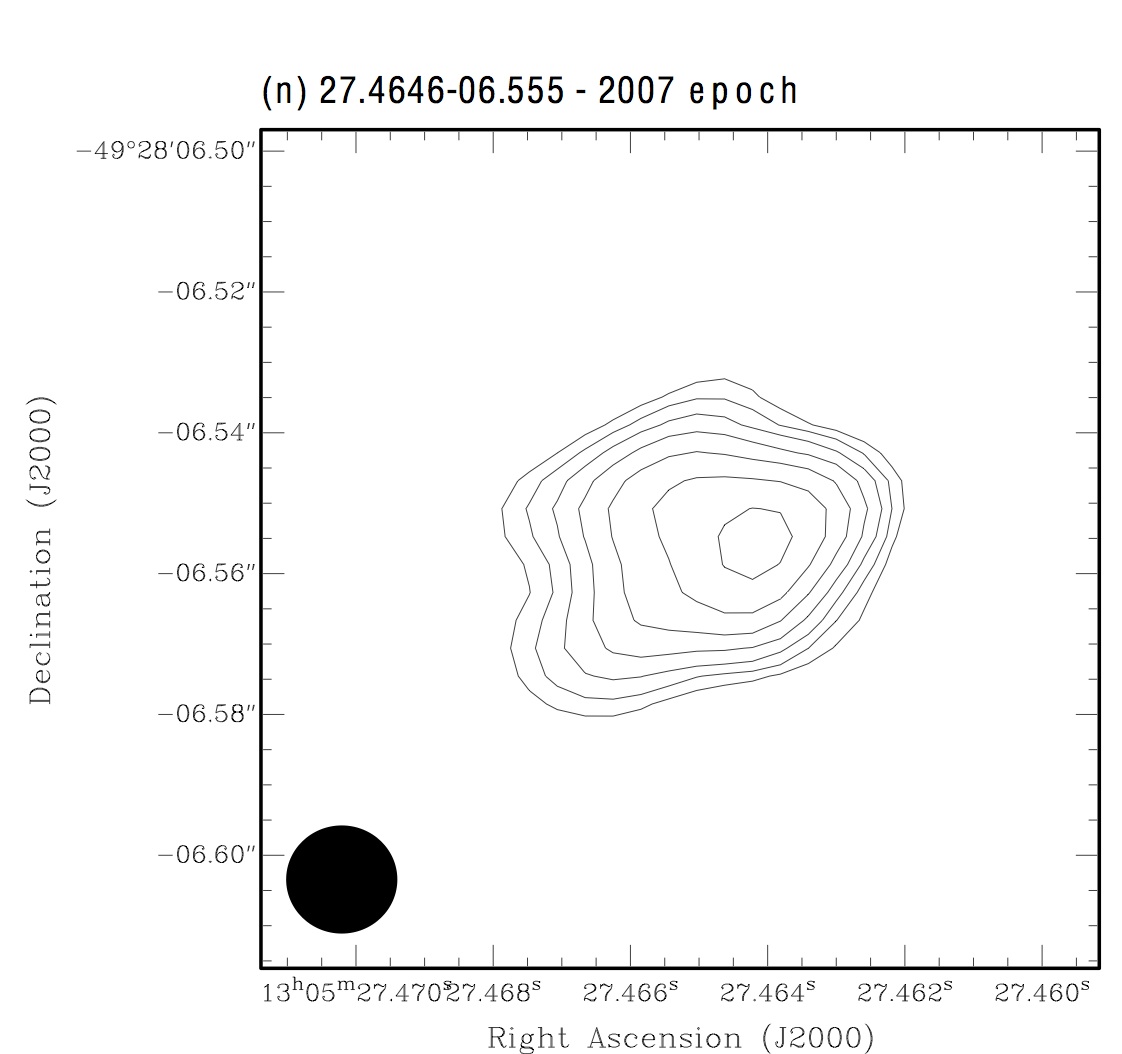} \quad
\plotone{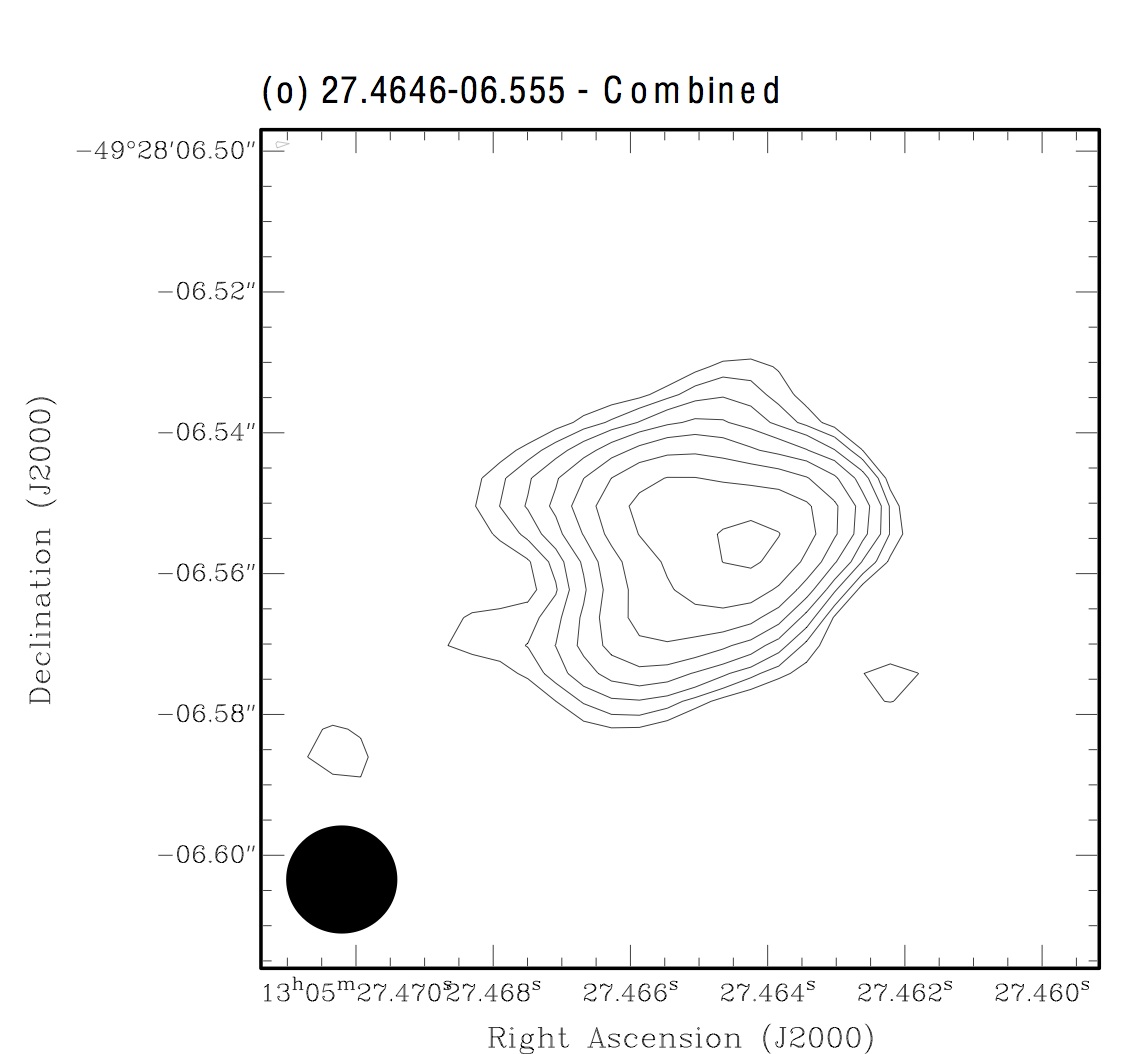}
}
\mbox{
\plotone{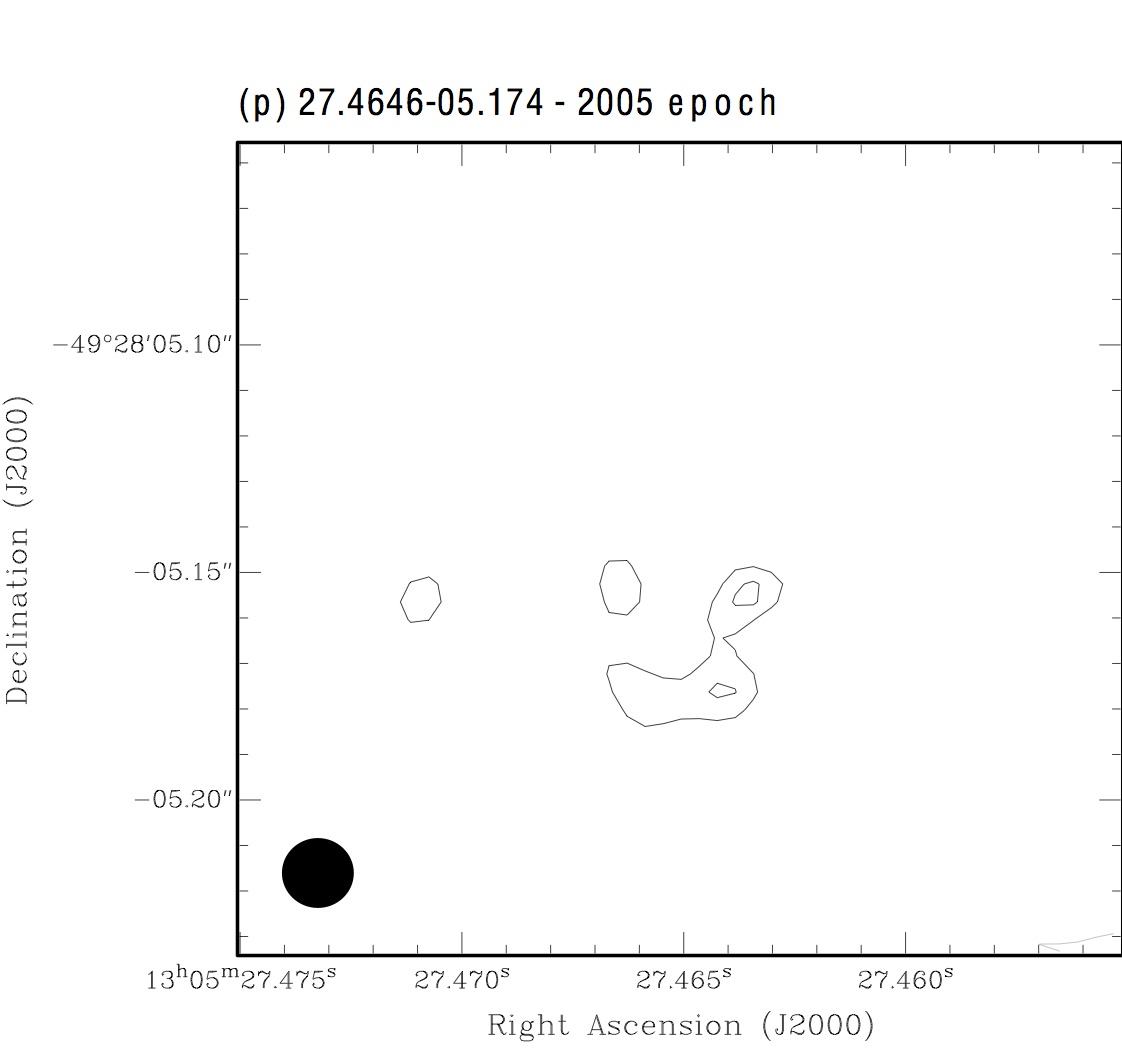} \quad
\plotone{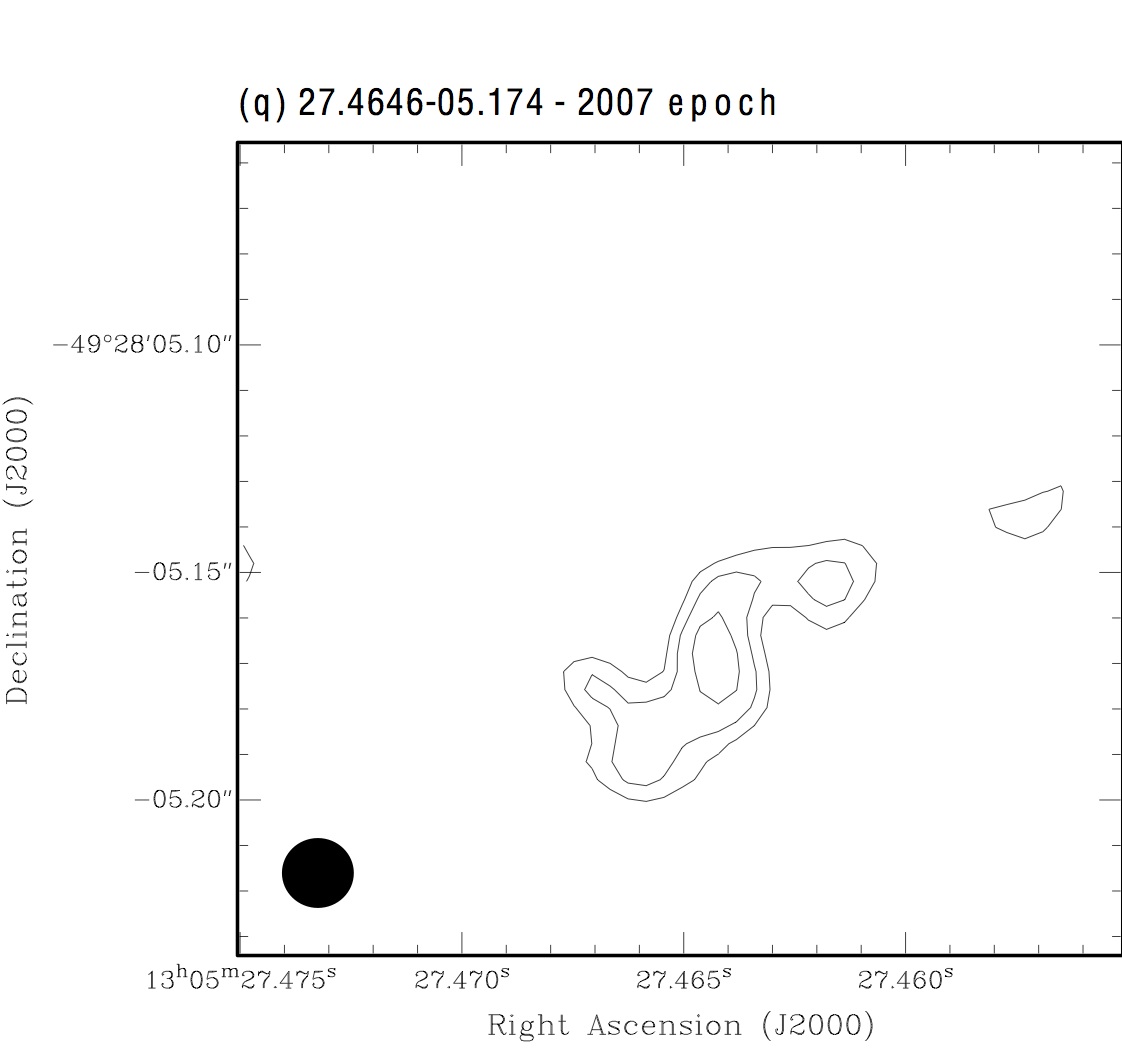} \quad
\plotone{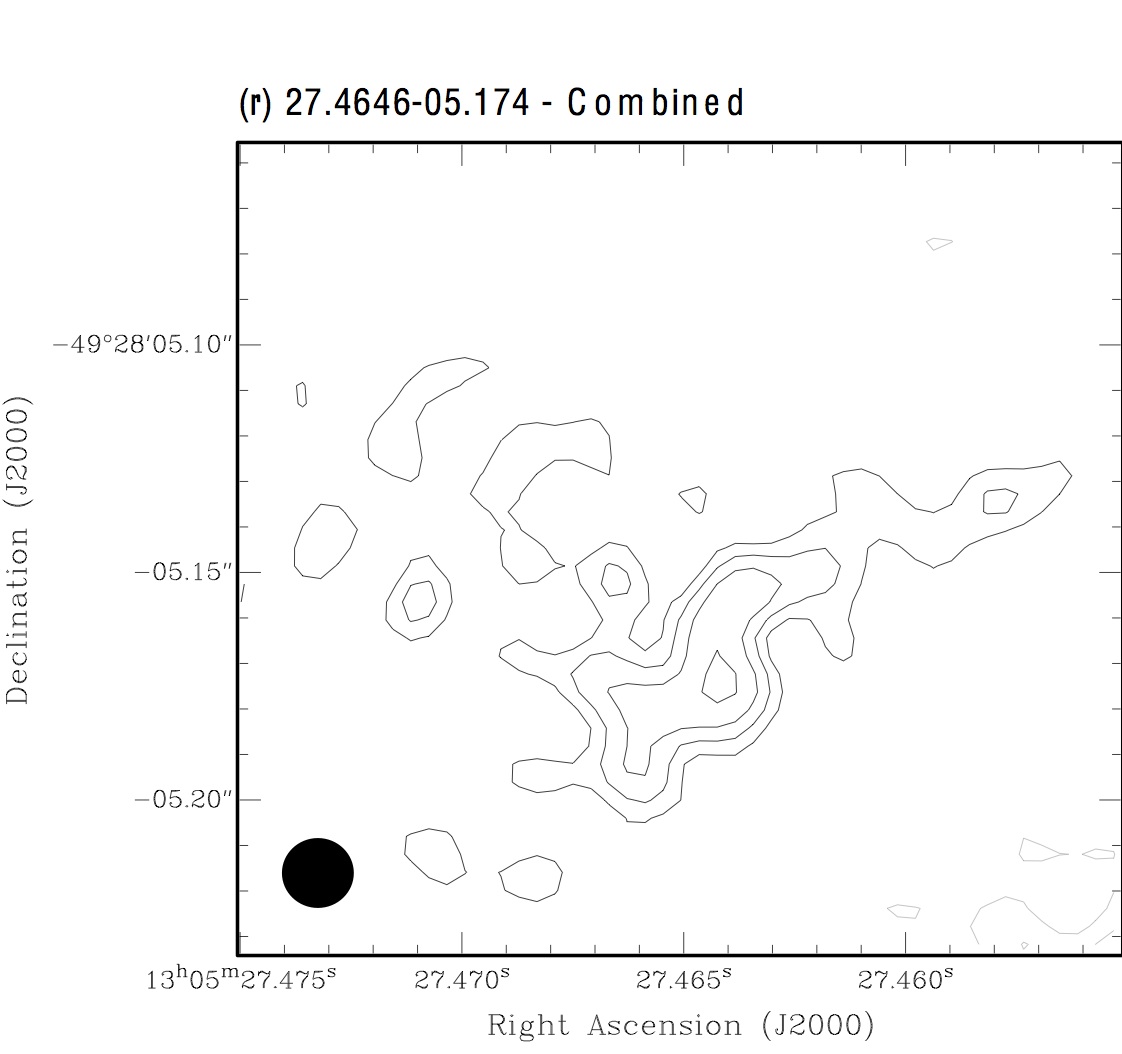}
}
\mbox{
\plotone{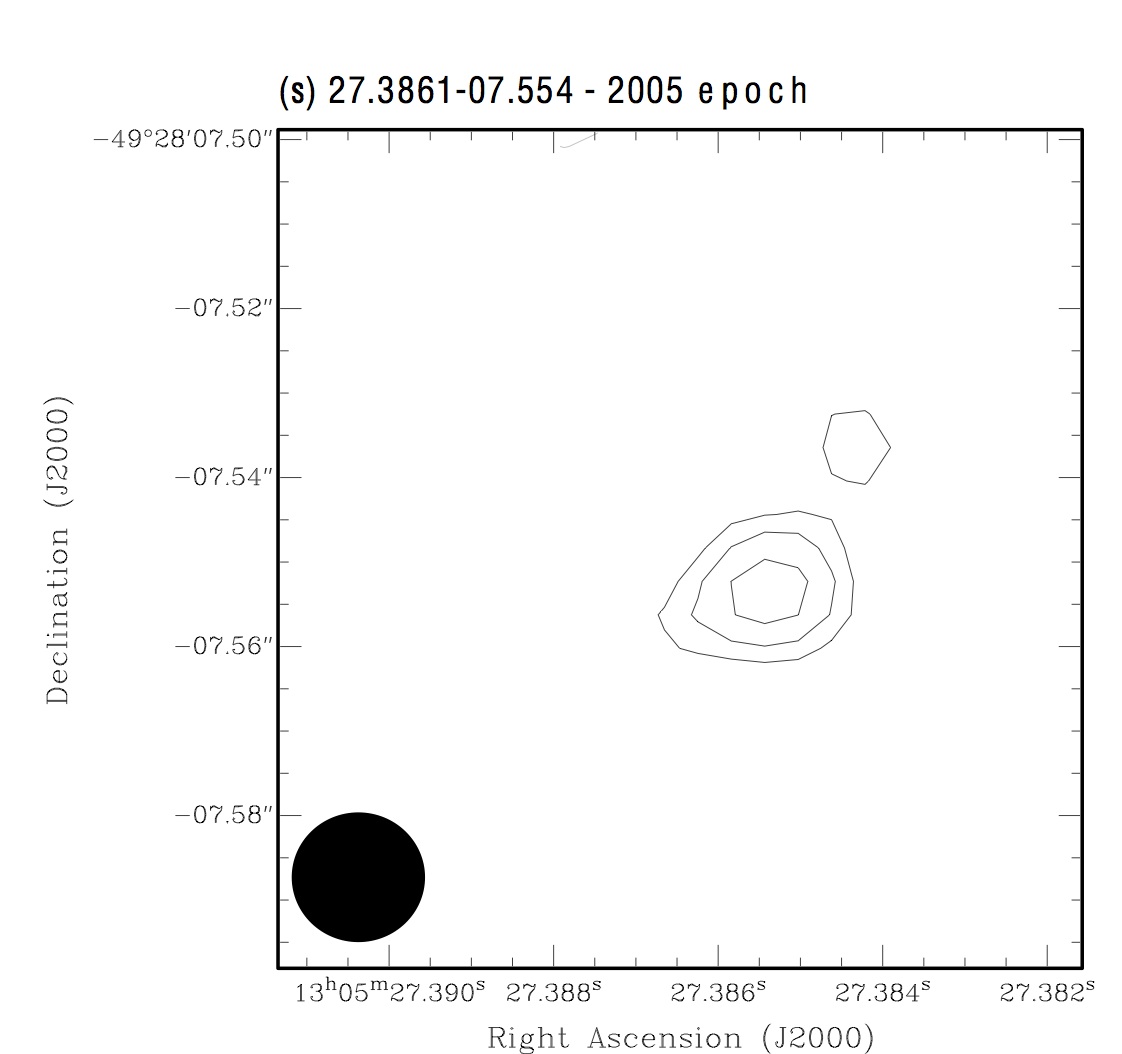} \quad
\plotone{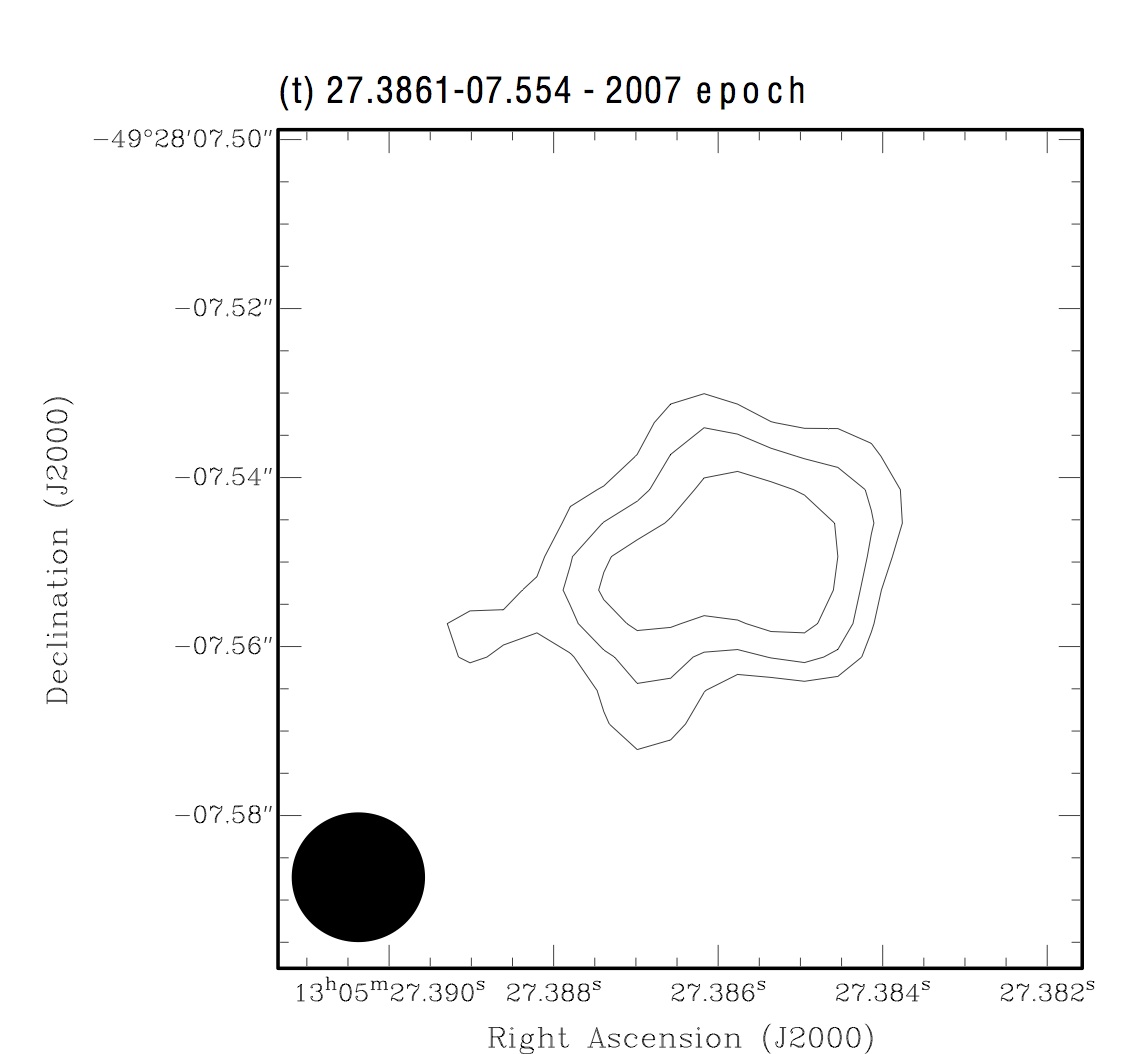} \quad
\plotone{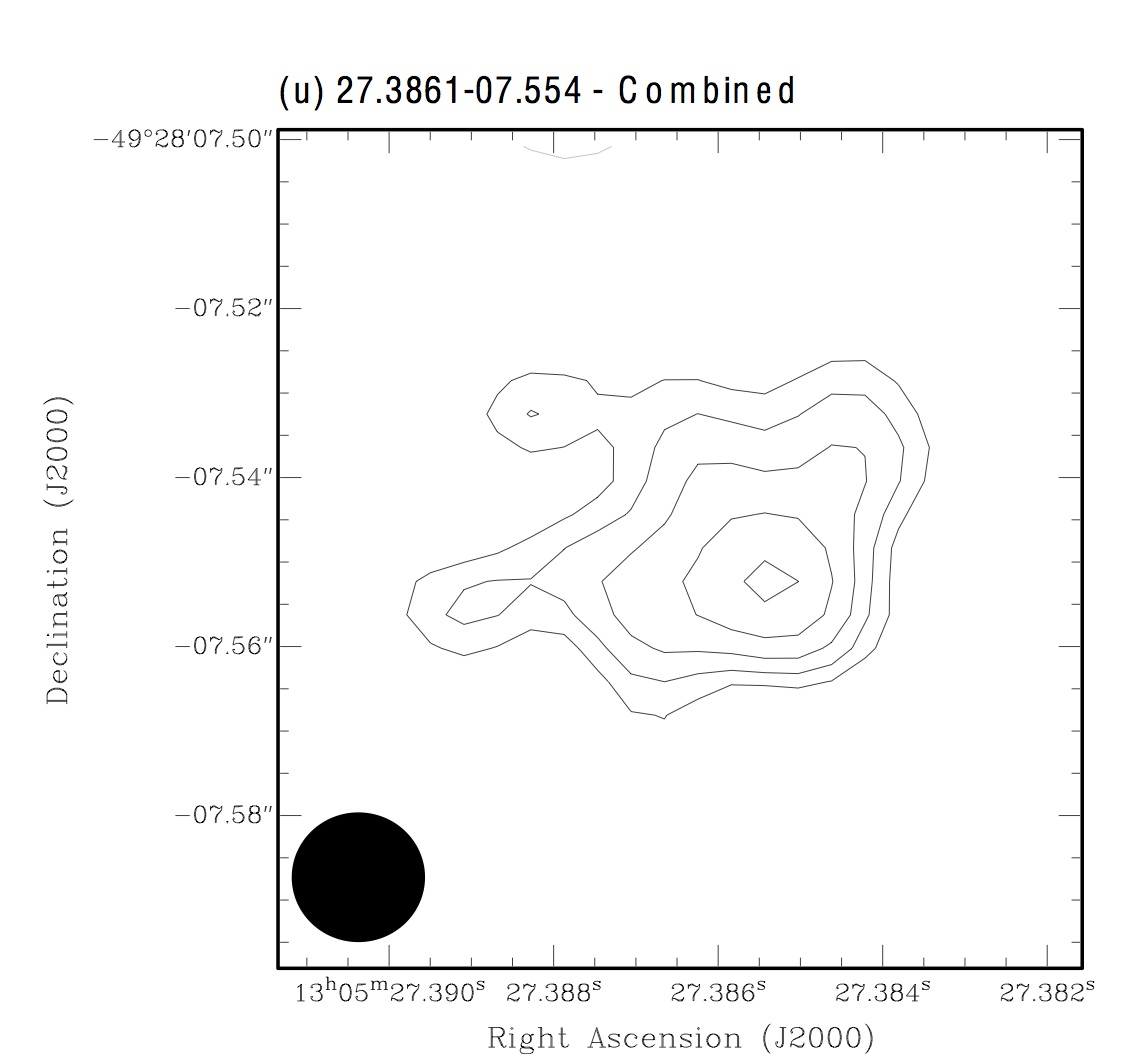}
}
\mbox{
\plotone{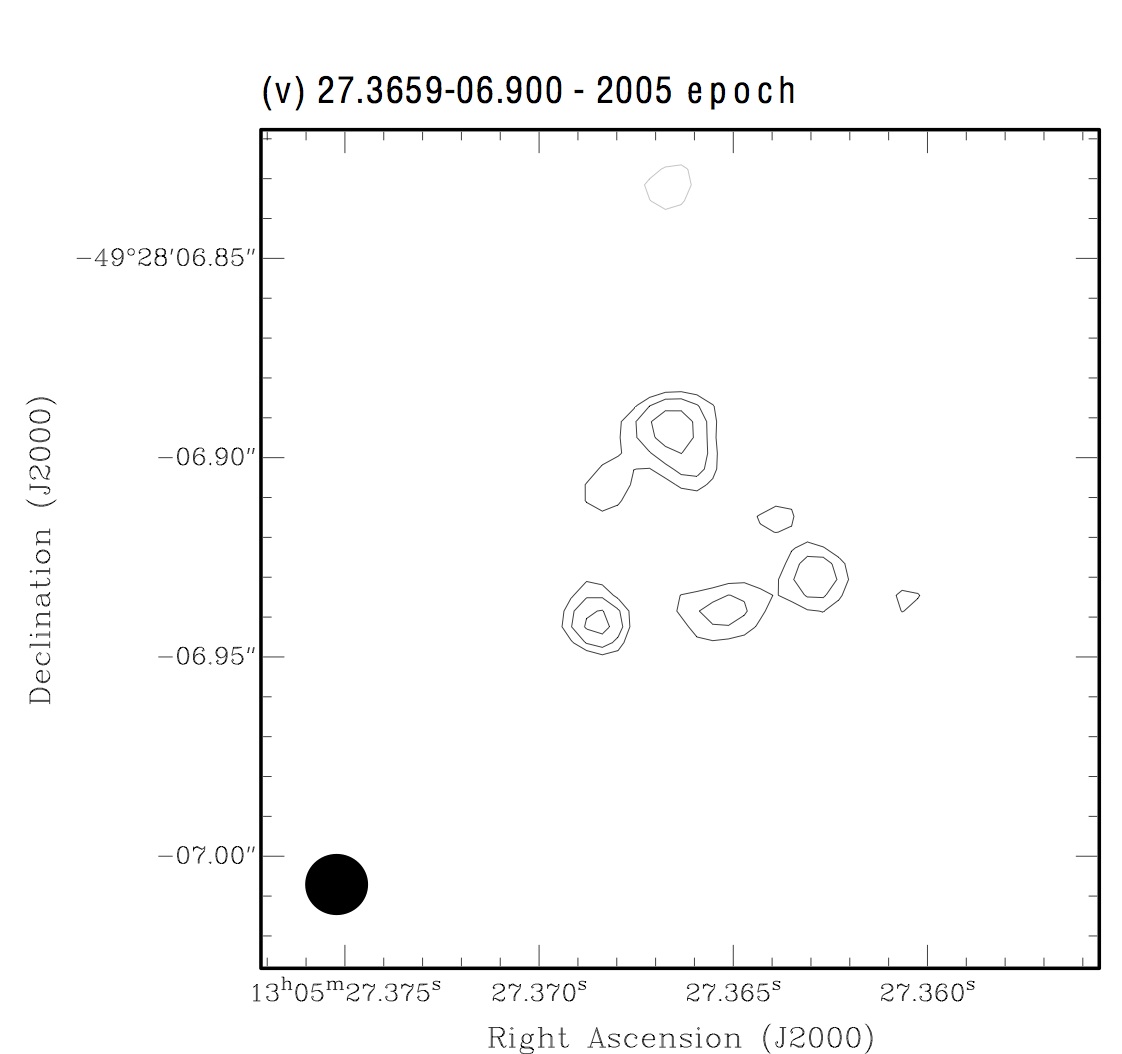} \quad
\plotone{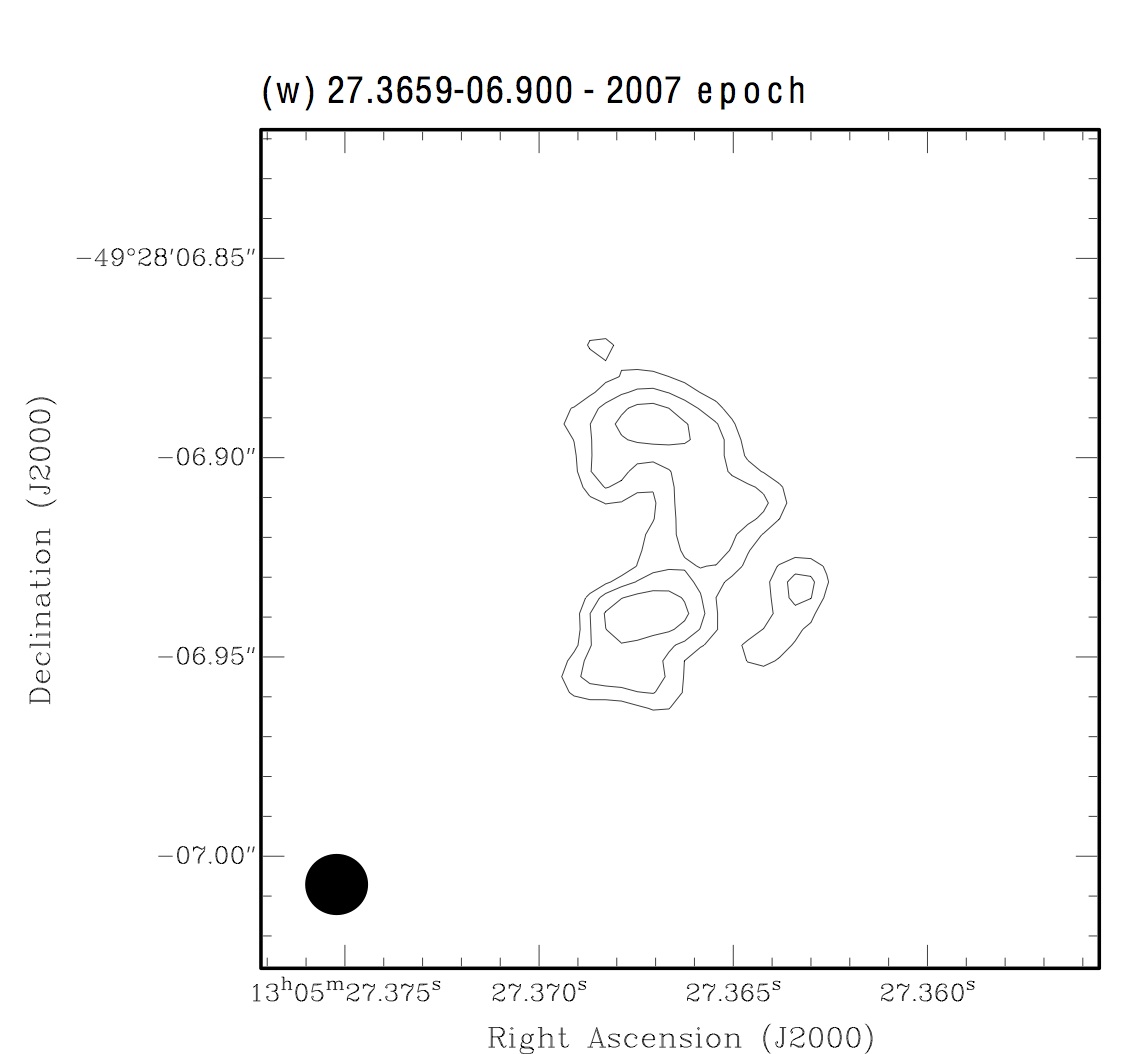} \quad
\plotone{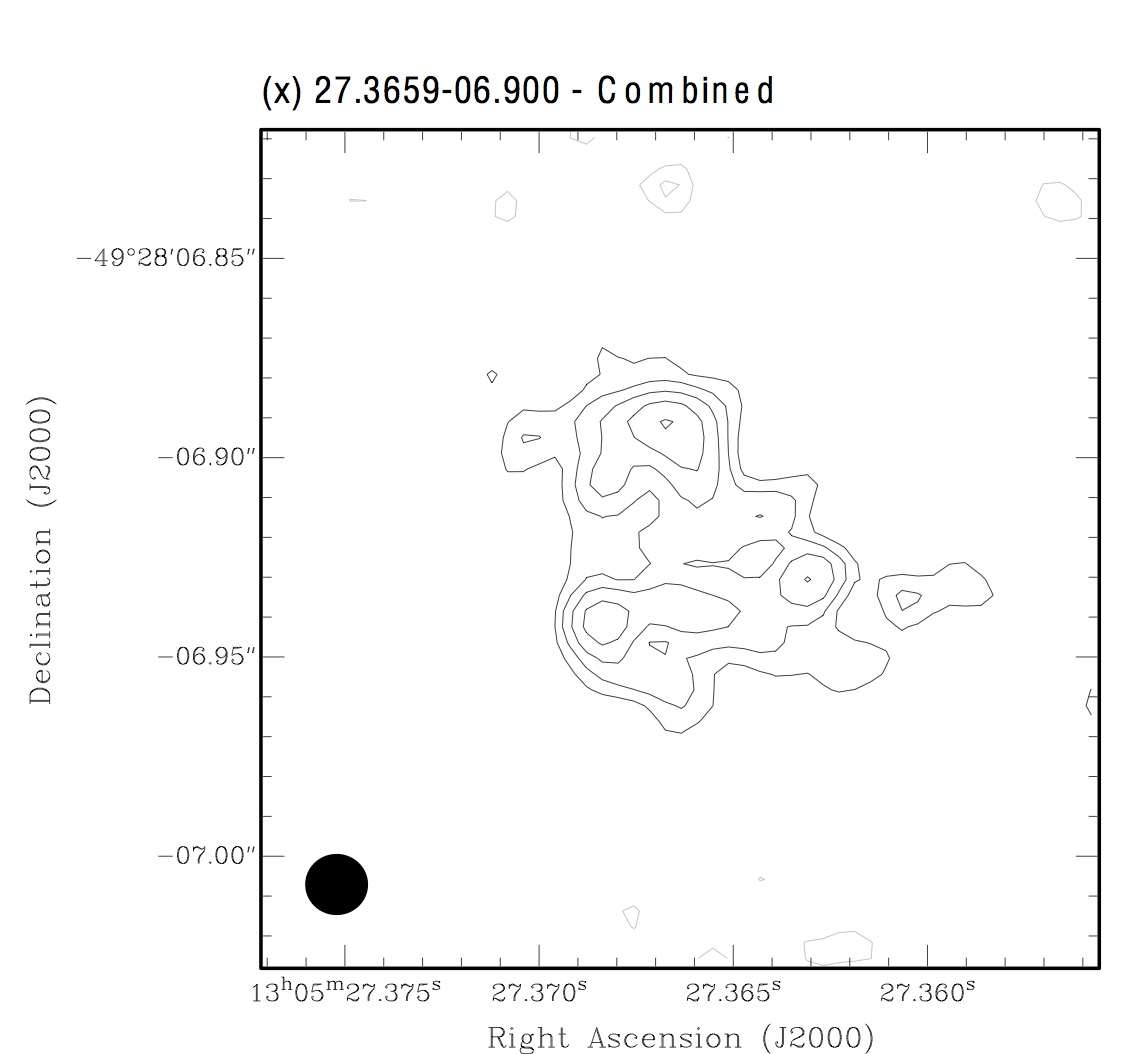}
}
\mbox{
\plotone{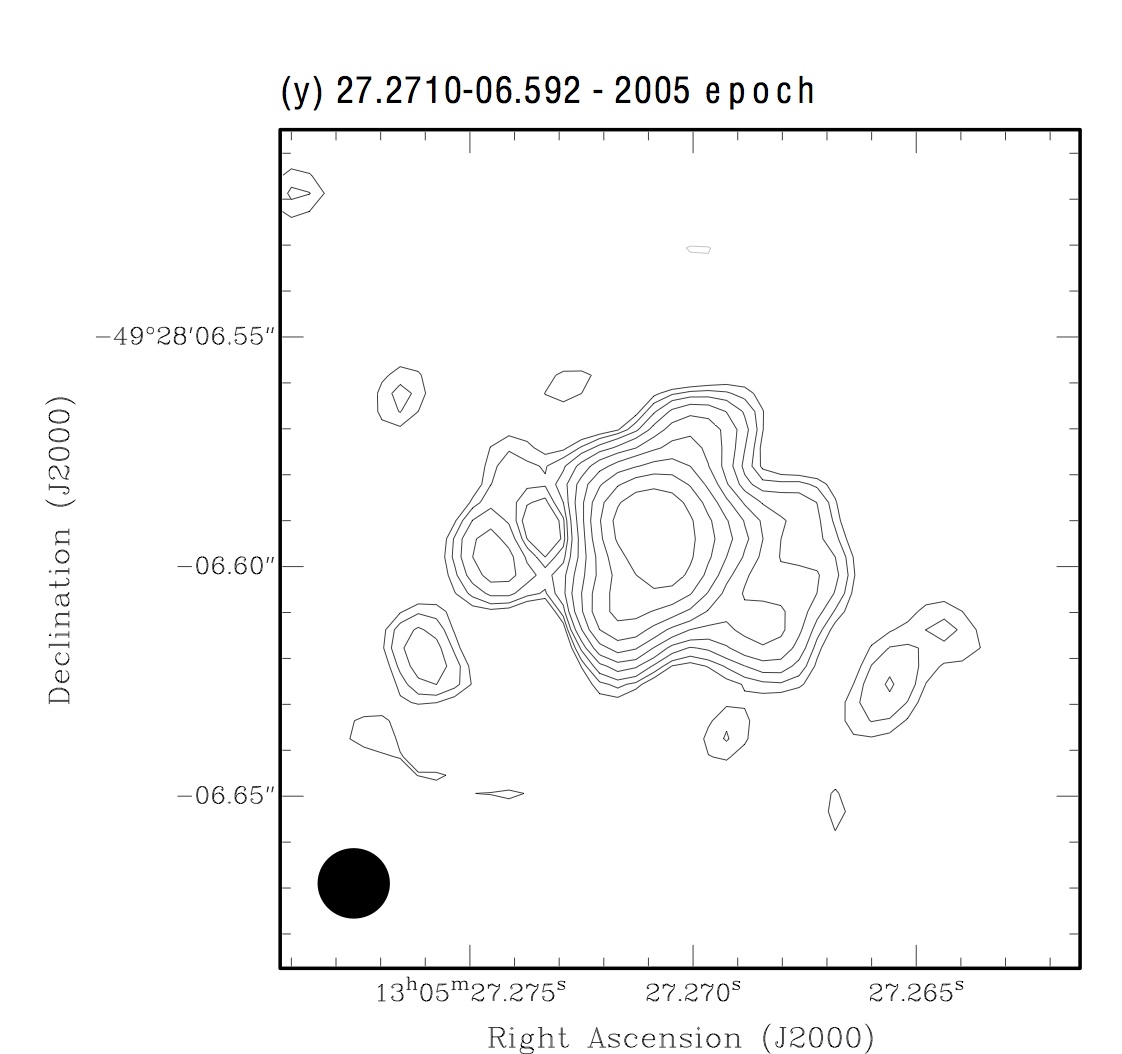} \quad
\plotone{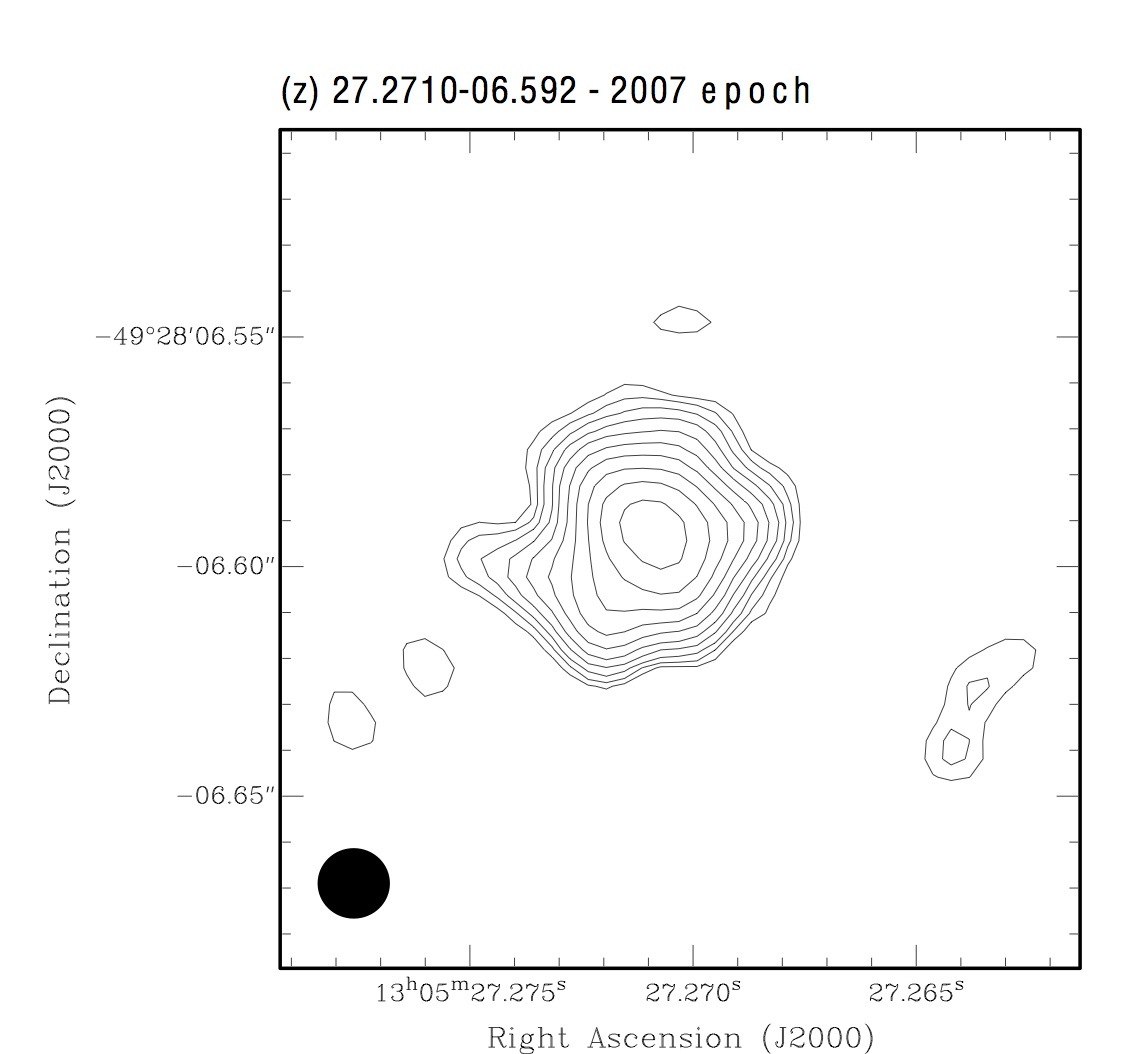} \quad
\plotone{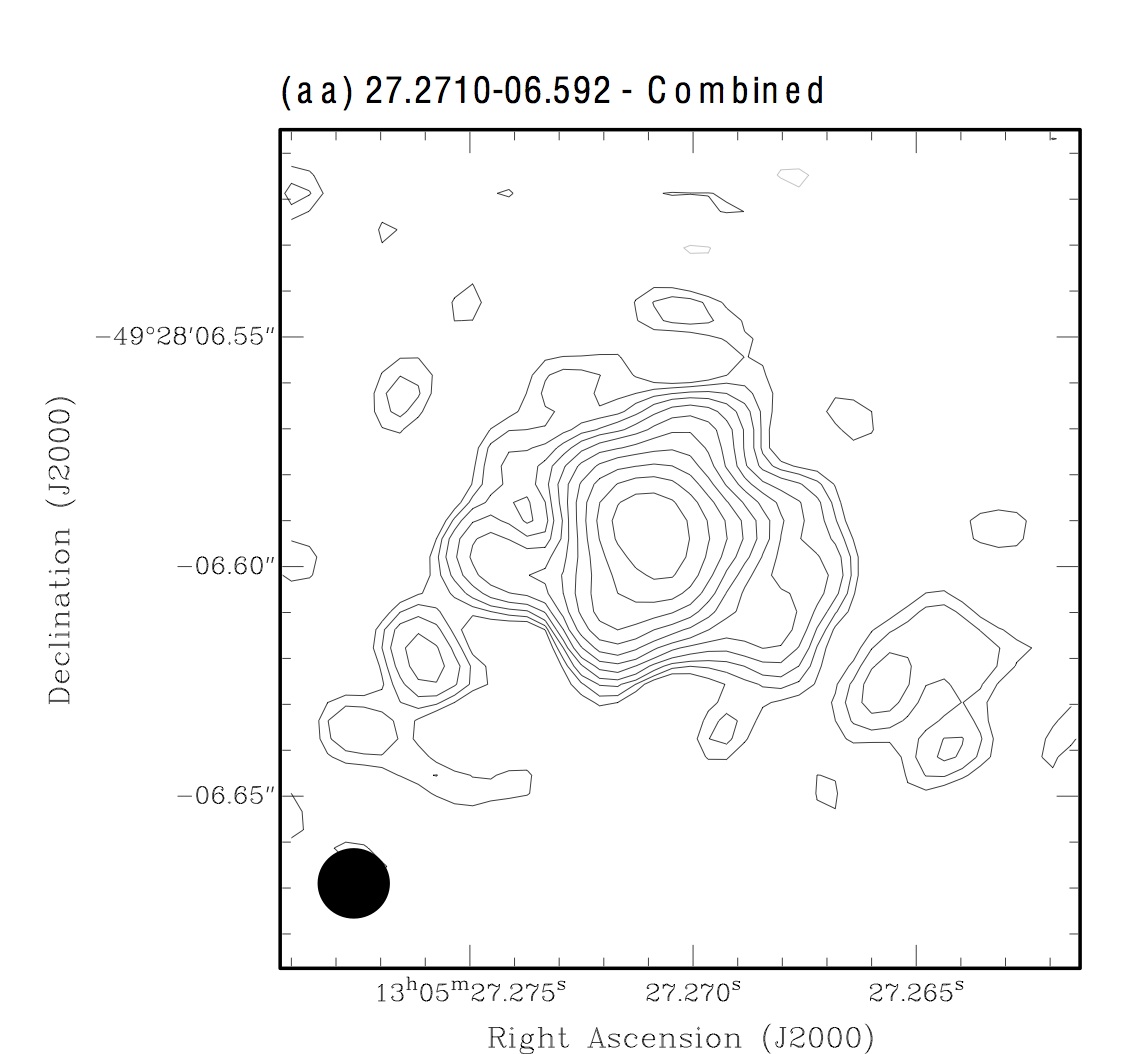}
}\\[5mm]
{Fig. 5.7. --- Continued}
\end{center}

\linespread{1.0}
\normalsize
\begin{savequote}[20pc]
\sffamily
Duct tape is like the force.\\
It has a light side, and a dark side,\\
and it holds the universe together.
\qauthor{Carl Zwanzig}
\end{savequote}

\chapter[Compact Radio Sources in Other Local Starburst Galaxies]{The compact radio source population of NGC 55, NGC 1313, NGC 5236 and NGC 5253}
\label{chap:starbursts}
\begin{center}
{\it Adapted from:}

E. Lenc \& S.J. Tingay

in preparation
\end{center}
\small
Wide-field, very long baseline interferometry (VLBI) observations of the nearby starburst galaxies NGC 55, NGC 5236 and NGC 5253, obtained with the Australian Long Baseline Array (LBA), have detected no compact radio sources associated with these galaxies at 2.3 GHz. Similar observations of NGC 1313 detected two compact radio sources, one of which is the supernova remnant SN 1978K. ATCA observations of these starburst galaxies between 17 and 23 GHz have resulted in the detection of weak and extended thermal emission from each of NGC 55, NGC 5236 and NGC 5253. No emission was detected in the ATCA observations of NGC 1313, at 17 and 23 GHz. The VLBI non-detections in NGC 55 and NGC 5253 and the small number of detections in NGC 1313 are consistent with the low star formation rates (SFRs) implied from the far-infrared flux density of these galaxies when compared against the number of detections verses SFR for prototypical starbursts such as NGC 253 and NGC 4945. Furthermore, weak $17-23$ GHz emission in all four galaxies is suggestive of low density nuclear environments, resulting in weak and short-lived supernova emission that rapidly fades below the VLBI detection limit.

\clearpage
\linespread{1.3}
\normalsize
\section{Introduction}
\label{sec:starbursts.introduction}

This paper is the third of a series of papers that completes the study of the sub-parsec scale properties of local ($D<10$ Mpc), bright ($S_{1.4}>10$ mJy) southern ($\delta<-20\arcdeg$) starburst galaxies. Starburst galaxies are the result of processes acting on a wide range of scales and themselves drive energetic activity on a wide range of scales. In the first two papers of this series we discussed the properties of the nuclear starburst region of NGC 253 \citep{Lenc:2006p6695} and NGC 4945 (Lenc \& Tingay, submitted). Direct observations of supernova remnants enabled an investigation of the star-formation and supernova history in these galaxies. Furthermore, high-resolution radio observations were used to map the ionized gaseous environments of the starburst regions. These observations provide a link between the large-scale dynamical effects that feed the starburst regions, the activity in the starburst regions, and the energetic phenomenon that in turn is driven by the starbursts. In the present work (Paper III of the series) we discuss the sub-parsec scale properties of NGC 55, NGC 1313, NGC 5236 (M83) and NGC 5253, and implications on the star-formation and supernova rates in these galaxies.





\section{Observations, data reduction, and results}

\subsection{LBA Observations}
\label{sec:starbursts.reduction}

VLBI observations of NGC 55, NGC 1313, NGC 5236 and NGC 5253 were made between 13 May 2005 and 25 March 2006, using a number of the Long Baseline Array (LBA) telescopes: the 70 m NASA Deep Space Network (DSN) antenna at Tidbinbilla; the 64 m antenna of the Australia Telescope National Facility (ATNF) near Parkes; the ATNF Australia Telescope Compact Array (ATCA) near Narrabri; the ATNF Mopra 22 m antenna near Coonabarabran; the University of Tasmania's 26 m antenna near Hobart; and the University of Tasmania's 30 m antenna near Ceduna. For observations of NGC 1313 and NGC 5236, $5\times22$ m antennas of the ATCA were used as a phased array, whereas for observations of NGC 55 and NGC 5253 only $2\times22$ m antennas were used as a phased array in order to maintain a compact configuration suitable for wide-field imaging. For the observations of NGC 55, NGC 1313, NGC 5236 and NGC 5253, Parkes was unavailable for 1, 3, 1 and 2 hours, respectively. For the observation of NGC 5253, Tidbinbilla was unavailable for 3 hours. All observations utilised the S2 recording system \citep{Cannon:1997p10533} to record 2 $\times$ 16 MHz bands (digitally filtered 2-bit samples) in the frequency ranges: 2252 - 2268 MHz and 2268 - 2284 MHz.  Both bands were upper side band and right circular polarisation.

During each of the VLBI observations, three minute scans of the target galaxy were scheduled, alternating with three minute scans of a phase reference calibration source. PKS B0010$-$401 ($\alpha=00\rah12\ram59\fs90986$; $\delta=-39\arcdeg54\arcmin26\farcs0553$ [J2000]), PKS B0302$-$623 ($\alpha=03\rah03\ram50\fs631332$; $\delta=-62\arcdeg11\arcmin25\farcs54983$ [J2000]), PKS B1339$-$287 ($\alpha=13\rah42\ram15\fs345611$; $\delta=-29\arcdeg00\arcmin41\farcs83126$ [J2000]) and PKS B1339$-$287 were used as phase reference calibration sources for NGC 55, NGC 1313, NGC 5236 and NGC 5253, respectively. Observing parameters associated with each of the LBA observations are shown in Table \ref{tab:sbtabobs}.

\begin{sidewaystable}[!p]
\begin{center}
{ \footnotesize
\begin{tabular}{llcccccccc} \hline \hline
Source   & Observatory & Frequency & $\alpha$ & $\delta$ & Date & Config. & Duration & Bandwidth & $\Delta$t \\
         &             & (MHz)     & (J2000)  & (J2000)  &      &         & (h)      & (MHz)     & (s) \\ [0.5ex] \hline \hline
NGC 55   & ATCA    & 17000.0  & $00\rah14\ram53\fs6$ & $-39\arcdeg11\arcmin48$         & 24/25 MAR 2006 & 6C      & 3  & 128 & 10 \\
         & \nodata & 19000.0  & \nodata              & \nodata                         & \nodata        & \nodata & 3  & 128 & 10 \\
         & \nodata & 21000.0  & \nodata              & \nodata                         & \nodata        & \nodata & 3  & 128 & 10 \\
         & \nodata & 23000.0  & \nodata              & \nodata                         & \nodata        & \nodata & 3  & 128 & 10 \\
\hline
NGC 55   & LBA & 2252.0  & $00\rah14\ram53\fs6$    & $-39\arcdeg11\arcmin48$  & 11/12 MAR 2006 & \nodata & 11   & 16  & 2  \\
         & \nodata & 2268.0  & \nodata              & \nodata                         & \nodata        & \nodata & 11   & 16  & 2  \\
\hline
NGC 1313 & ATCA    & 17000.0 & $03\rah18\ram16\fs0$  & $-66\arcdeg29\arcmin54$        & 24/25 MAR 2006    & 6C   & 3 & 128 & 10 \\
         & \nodata & 19000.0 & \nodata              & \nodata                         & \nodata        & \nodata & 3 & 128 & 10 \\
         & \nodata & 21000.0 & \nodata              & \nodata                         & \nodata        & \nodata & 3 & 128 & 10 \\
         & \nodata & 23000.0 & \nodata              & \nodata                         & \nodata        & \nodata & 3 & 128 & 10 \\
\hline
NGC 1313 & LBA & 2252.0  & $03\rah18\ram16\fs0$  & $-66\arcdeg29\arcmin54$  & 13/14 MAY 2005 & \nodata & 12  & 16  & 2  \\
         & \nodata & 2268.0  & \nodata              & \nodata                         & \nodata        & \nodata & 12  & 16  & 2  \\
\hline
NGC 5236 & ATCA    & 17000.0 & $13\rah37\ram00\fs9$ & $-29\arcdeg51\arcmin57$  & 21 MAR 2006    & 6C      & 1.5 & 128 & 10 \\
(M83)    & \nodata & 19000.0 & \nodata              & \nodata                         & \nodata        & \nodata & 1.5 & 128 & 10 \\
         & \nodata & 21000.0 & \nodata              & \nodata                         & \nodata        & \nodata & 1.5 & 128 & 10 \\
         & \nodata & 23000.0 & \nodata              & \nodata                         & \nodata        & \nodata & 1.5 & 128 & 10 \\
\hline
NGC 5236 & LBA & 2252.0  & $13\rah37\ram00\fs9$ & $-29\arcdeg51\arcmin57$  & 13/14 MAY 2005 & \nodata & 9   & 16  & 2  \\
(M83)    & \nodata & 2268.0  & \nodata              & \nodata                         & \nodata        & \nodata & 9   & 16  & 2  \\
\hline
NGC 5253 & ATCA    & 17000.0 & $13\rah39\ram55\fs9631$ & $-31\arcdeg38\arcmin24\farcs388$ & 21 MAR 2006    & 6C      & 1.5 & 128 & 10 \\
         & \nodata & 19000.0 & \nodata              & \nodata                         & \nodata        & \nodata & 1.5 & 128 & 10 \\
         & \nodata & 21000.0 & \nodata              & \nodata                         & \nodata        & \nodata & 1.5 & 128 & 10 \\
         & \nodata & 23000.0 & \nodata              & \nodata                         & \nodata        & \nodata & 1.5 & 128 & 10 \\
\hline
NGC 5253 & LBA & 2252.0  & $13\rah39\ram55\fs9631$ & $-31\arcdeg38\arcmin24\farcs388$  & 16 MAR 2006 & \nodata & 11  & 16  & 2  \\
         & \nodata & 2268.0  & \nodata              & \nodata                         & \nodata        & \nodata & 11  & 16  & 2  \\ \hline
\end{tabular}
\caption{Summary of remaining nearby southern starburst galaxy observations.}
\label{tab:sbtabobs}
}
\end{center}
\end{sidewaystable}

\subsection{LBA correlation}
The data were correlated using the ATNF Long Baseline Array (LBA) processor at ATNF headquarters in Sydney \citep{Wilson:1992p9290}, using an integration time of 2 seconds and with 32 frequency channels across each 16 MHz band (channel widths of 0.5 MHz). In the observations of NGC 1313 and NGC 5236, the ATCA primary beam limits the field of view to a half-width half-maximum (HWHM) of $\sim2\arcmin$. At the HWHM point, bandwidth smearing and time-averaging smearing losses are estimated to be approximately 42\% and 20\%, respectively (9\% and 2\%, respectively, if the longer baselines associated with Ceduna and Hobart are removed). In the observations of NGC 55 and NGC 5253, the field of view is primarily limited by bandwidth smearing effects to approximately $1\arcmin$ for 10\% losses in measured peak flux densities ($\sim2.3\arcmin$ if the longer baselines associated with Ceduna and Hobart are removed).

\subsection{LBA Data Reduction}

The correlated data were reduced using the techniques developed and described in Section \S~\ref{sec:p3lbareduction} to correct for any structure that may exist in the phase calibrator sources. For NGC 55 and NGC 5253 observations the theoretical thermal noise in the images is 0.12 mJy beam$^{-1}$, while for the NGC 1313 and NGC 5236 observations the theoretical thermal noise in the images is 0.10 mJy beam$^{-1}$. The calibrated data resulting from the data reduction process achieve a one sigma image noise of 0.12 mJy beam$^{-1}$, 0.11 mJy beam$^{-1}$, 0.14 mJy beam$^{-1}$, 0.14 mJy beam$^{-1}$ for NGC 55, NGC 1313, NGC 5236 and NGC 5253, respectively.

\begin{figure}[ht]
\epsscale{0.45}
\begin{center}
\mbox{
\plotone{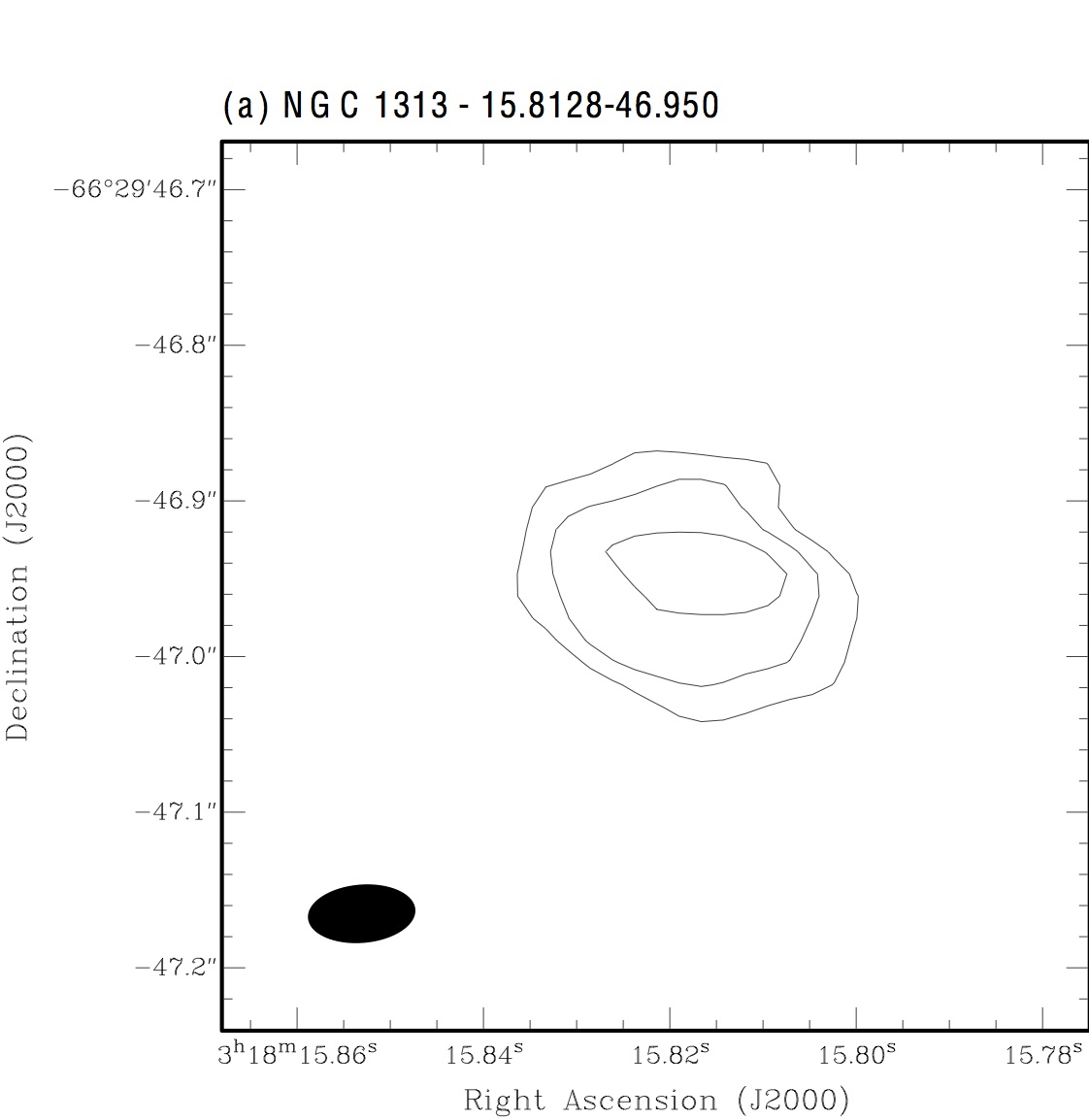} \quad
\plotone{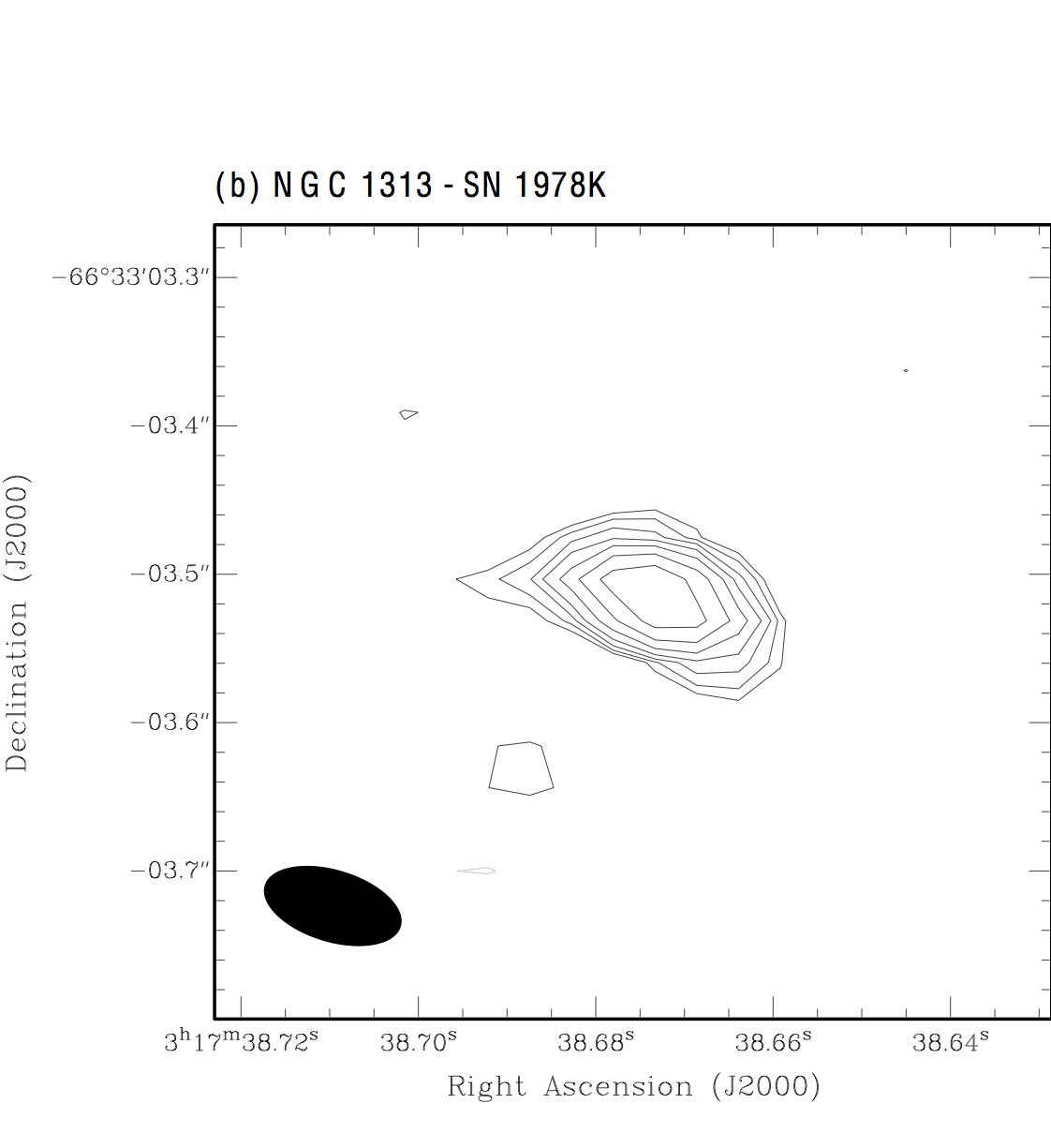}
}
\caption[LBA images of NGC 1313 at 2.3 GHz]{(a) Naturally-weighted total-power map of the source 15.813-46.95. (b) Naturally-weighted total-power map of SN 1978K in NGC 1313 as observed with the LBA at 2.3 GHz using only a single baseline between Parkes and Tidbinbilla. Map statistics for the individual maps are shown in Table \ref{tab:sbtabimage}. Contours are drawn at $\pm2^{0}, \pm2^{\frac{1}{2}}, \pm2^{1}, \pm2^{\frac{3}{2}}, \cdots$ times the $3\sigma$ rms noise.}
\label{fig:sbngc1313}            
\end{center}
\end{figure}

No detection of compact radio sources from the VLBI data were made above a 6 sigma threshold in the NGC 55, NGC 1313, NGC 5236 and NGC 5253 naturally-weighted images. To emphasise larger-scale structure that may exist, the data were re-imaged using steadily more restricted $(u,v)$ ranges. This yielded a weak ($8\sigma$) detection of a radio source (designated 15.813$-$46.95) in NGC 1313 when the longer baselines associated with the antennas at Ceduna and Hobart were removed. The source has a peak flux density of 0.87 mJy beam$^{-1}$ and has a position of $\alpha=03\rah18\ram15\fs813$; $\delta=-66\arcdeg29\arcmin46\farcs95$ (J2000). An image of the source is shown in Figure \ref{fig:sbngc1313}(a).

As a confirmation of the quality of the data reduction process an attempt was made to detect the supernova remnant SN 1978K in NGC 1313. The source is situated approximately $5\arcmin$ from the phase centre of the observation and so is well outside of the nominal field of view of the observation. To reduce non-coplanar array distortion \citep{Perley:1999p10576}, the AIPS task UVFIX was used to shift to the known location of SN 1978K at $\alpha=03\rah17\ram38\fs62$ and $\delta=-66\arcdeg33\arcmin03\farcs4$ (J2000), as determined by \citet{Smith:2007p7796}. Furthermore, to reduce the effects of bandwidth and time-averaging smearing \citep{Bridle:1999p10564} and primary beam effects, only the shortest and most sensitive baseline was used (i.e. between Parkes and Tidbinbilla). This resulted in a strong detection ($42\sigma$) of the supernova remnant. Model-fitting to the rotated and averaged single baseline data was performed to effectively image the source with the limited available data, Figure \ref{fig:sbngc1313}(b). The detected remnant is unresolved, as is further supported by the observed flux density (Section \S~\ref{sec:starbursts.galaxies.ngc1313}), and is offset from the nominal position of the source by approximately 300 mas ($\alpha=03\rah17\ram38\fs67$; $\delta=-66\arcdeg33\arcmin03\farcs5$). The observed position error is largely attributable to the differences and uncertainties of the primary beam shape out to $\sim5\arcmin$ (2.3 GHz) at Parkes and Tidbinbilla and to a lesser degree the large restoring beam of the resulting image map ($96\times49$ mas). Following this detection, a subsequent survey of the south-west star-forming region of the galaxy was performed but no further detections were made. The image parameters for both of the sources detected in NGC 1313 are shown in Table \ref{tab:sbtabimage}.

\begin{figure}[ht]
\epsscale{0.45}
\begin{center}
\plotone{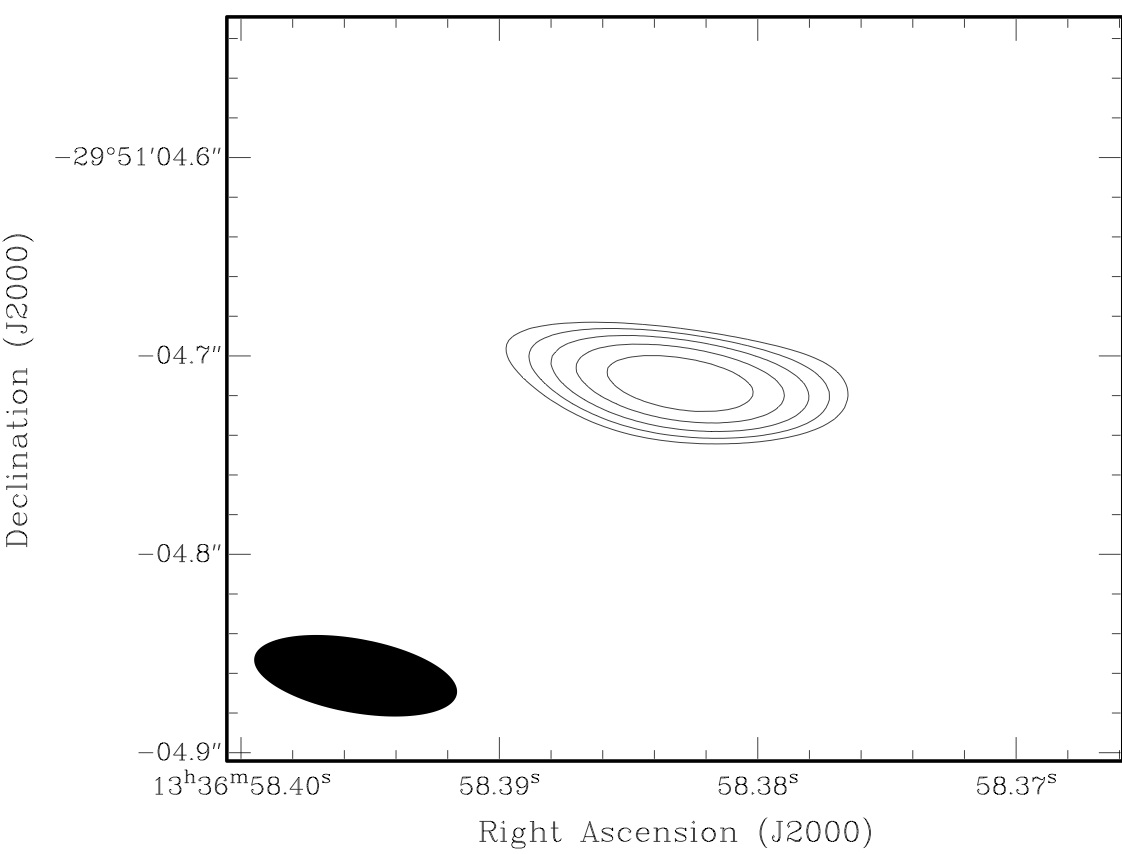}
\caption[LBA image of source 58.383$-$04.71 in NGC 5236 at 2.3 GHz]{Naturally-weighted total-power map of 58.383$-$04.71 in NGC 5236 as observed with the LBA at 2.3 GHz using only a single baseline between Parkes and Tidbinbilla. Map statistics for the map is shown in Table \ref{tab:sbtabimage}. Contours are drawn at $\pm2^{0}, \pm2^{\frac{1}{2}}, \pm2^{1}, \pm2^{\frac{3}{2}}, \cdots$ times the $3\sigma$ rms noise.}
\label{fig:sbfigs28}            
\end{center}
\end{figure}

A similar technique (using only the Parkes-Tidbinbilla baseline to maximise the observable field of view) was used to survey a larger region of NGC 5236 (extending out to $2\arcmin$ from the centre of the galaxy) to include locations in which supernova remnants and \ion{H}{2} regions were known to exist. The search resulted in the $17\sigma$ detection of a single unresolved source ($\alpha=13\rah36\ram58\fs383$; $\delta=-29\arcdeg51\arcmin04\farcs71$) approximately $62\arcsec$ from the phase centre. A map of the source, which we designate 58.383$-$04.71, was produced by model-fitting to the rotated and averaged single baseline data and is shown in Figure \ref{fig:sbfigs28}. The image parameters for the source are shown in Table \ref{tab:sbtabimage}.

\subsection{ATCA Observations and Data Reduction}

Observations of each of the four starburst galaxies were carried out using all six antennas in the 6C configuration of the Australia Telescope Compact Array (ATCA) on 2006 March 12, 24 and 25. The flux scale was set using PKS B1934$-$638 and the phase calibrators PKS B0010$-$401, PKS B0302$-$623, PKS B1313$-$333 and PKS B1313$-$333 were used for NGC 55, NGC 1313, NGC 5236 and NGC 5253, respectively. During each observing run, we alternated between two setups every 6.5 minutes. The first setup observed at 17 GHz and 19 GHz and the second setup at 21 GHz and 23 GHz. A list of all ATCA observations reported in this paper and their associated observing parameters are tabulated in Table \ref{tab:sbtabobs}.

\begin{table}[ht]
\begin{center}
{ \scriptsize
\begin{tabular}{lcccccccc} \hline \hline
Figure                     & Source  & Frequency & Synthesized Beam & $\sigma$ & $S_{p}$ & $S_{int}$ \\
                           &          & (GHz)     & (mas)           & (mJy beam$^{-1}$) & (mJy beam$^{-1}$) & (mJy) \\ [0.5ex] \hline \hline
\ref{fig:sbngc1313}(a)     & 15.813$-$46.95 & 2.3 & $69\times38$  & 0.110  & 0.87 & 9.8 \\
\ref{fig:sbngc1313}(b)     & SN 1978K & 2.3 & $96\times49$          & 0.26  & 8.4\tablenotemark{a}  & 9.8\tablenotemark{a} \\ \hline
\ref{fig:sbfigs28}         & 58.382$-$04.71 & 2.3 & $104\times37$ & 0.04 & 0.67 & 0.65 \\ \hline
\ref{fig:sbfigNGC55}(a)    & NGC 55   & 17 & $5400\times4200$       & 0.240 & 2.1  & 6.7 \\
\ref{fig:sbfigNGC55}(b)    & NGC 55   & 19 & $5500\times4000$       & 0.280 & 2.8  & 5.5 \\
\ref{fig:sbfigNGC55}(c)    & NGC 55   & 21 & $5900\times3900$       & 0.450 & 2.3  & 4.5 \\ \hline
\ref{fig:sbfigNGC5236}(a)  & NGC 5236 (M83) & 17 & $2420\times1270$       & 0.540 & 8   & 42 \\
\ref{fig:sbfigNGC5236}(b)  & NGC 5236 (M83) & 19 & $2550\times1230$       & 0.580 & 8   & 34 \\
\ref{fig:sbfigNGC5236}(c)  & NGC 5236 (M83) & 21 & $3180\times1290$       & 1.20  & 11   & 31 \\
\ref{fig:sbfigNGC5236}(d)  & NGC 5236 (M83) & 23 & $3020\times1320$       & 1.40  & 13   & 27 \\ \hline
\ref{fig:sbfigNGC5253}(a)  & NGC 5253 & 17 & $880\times480$         & 0.330 & 19   & 27 \\
\ref{fig:sbfigNGC5253}(b)  & NGC 5253 & 19 & $820\times430$         & 0.390 & 19   & 26 \\
\ref{fig:sbfigNGC5253}(c)  & NGC 5253 & 21 & $720\times390$         & 0.810 & 21   & 28 \\
\ref{fig:sbfigNGC5253}(d)  & NGC 5253 & 23 & $640\times340$         & 1.15  & 21   & 30 \\ \hline
\tablenotetext{a}{Uncorrected peak and integrated flux density - see text for details.}
\end{tabular}
\caption{Map Statistics for starburst galaxy images from the LBA and ATCA.}
\label{tab:sbtabimage}
}
\end{center}
\end{table}

\begin{figure}[ht]
\epsscale{0.3}
\begin{center}
\mbox{
\plotone{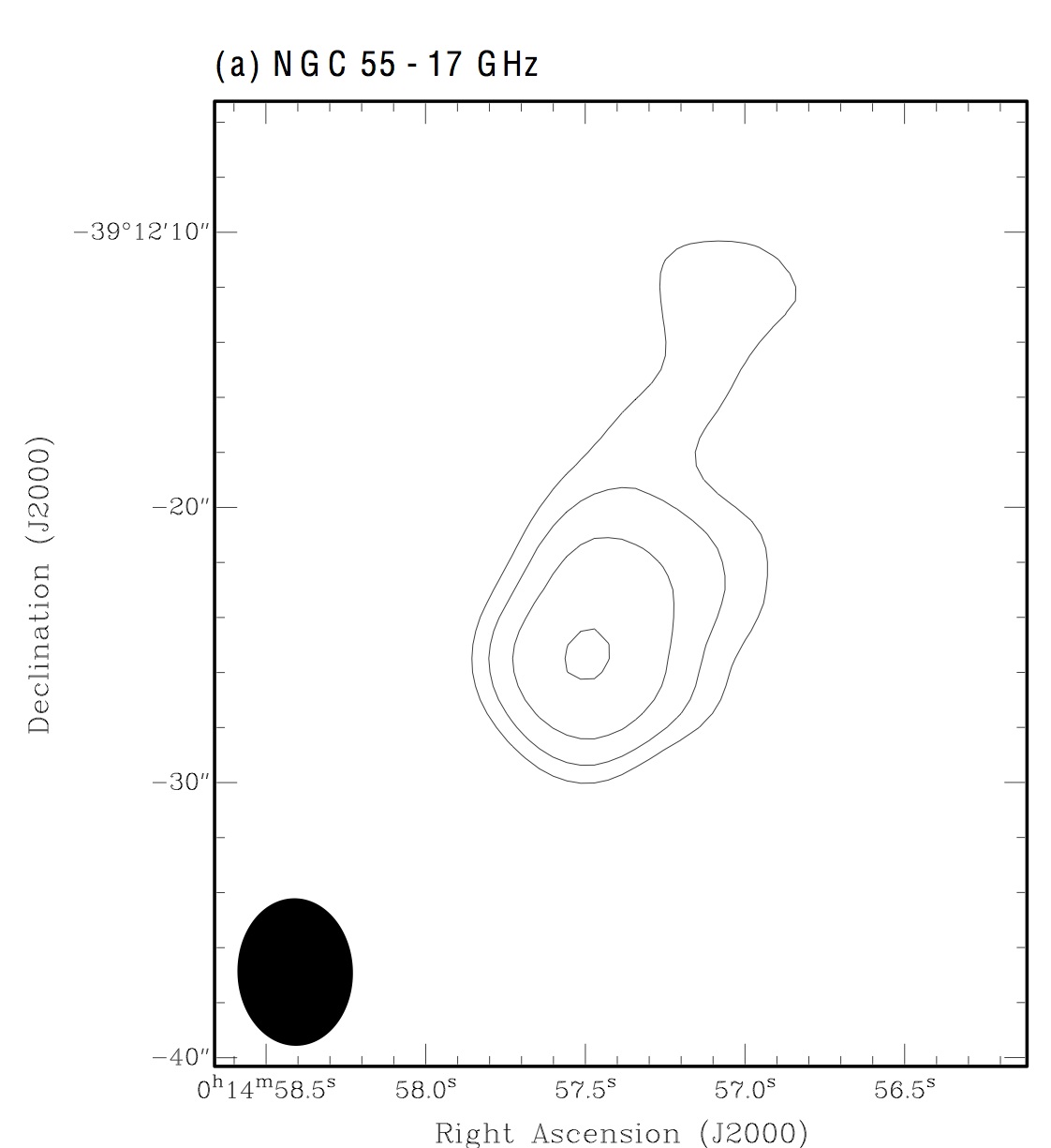} \quad
\plotone{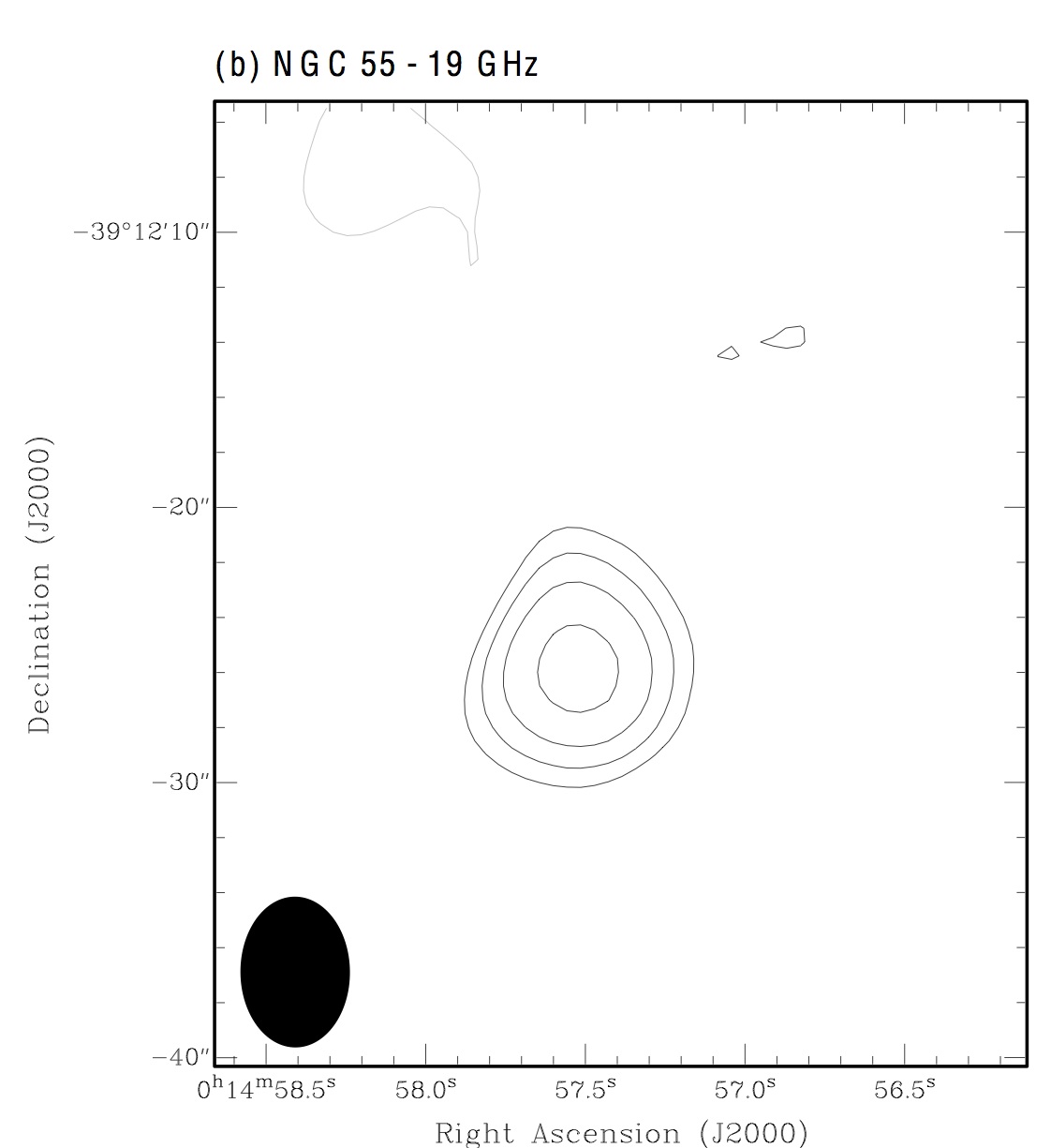} \quad
\plotone{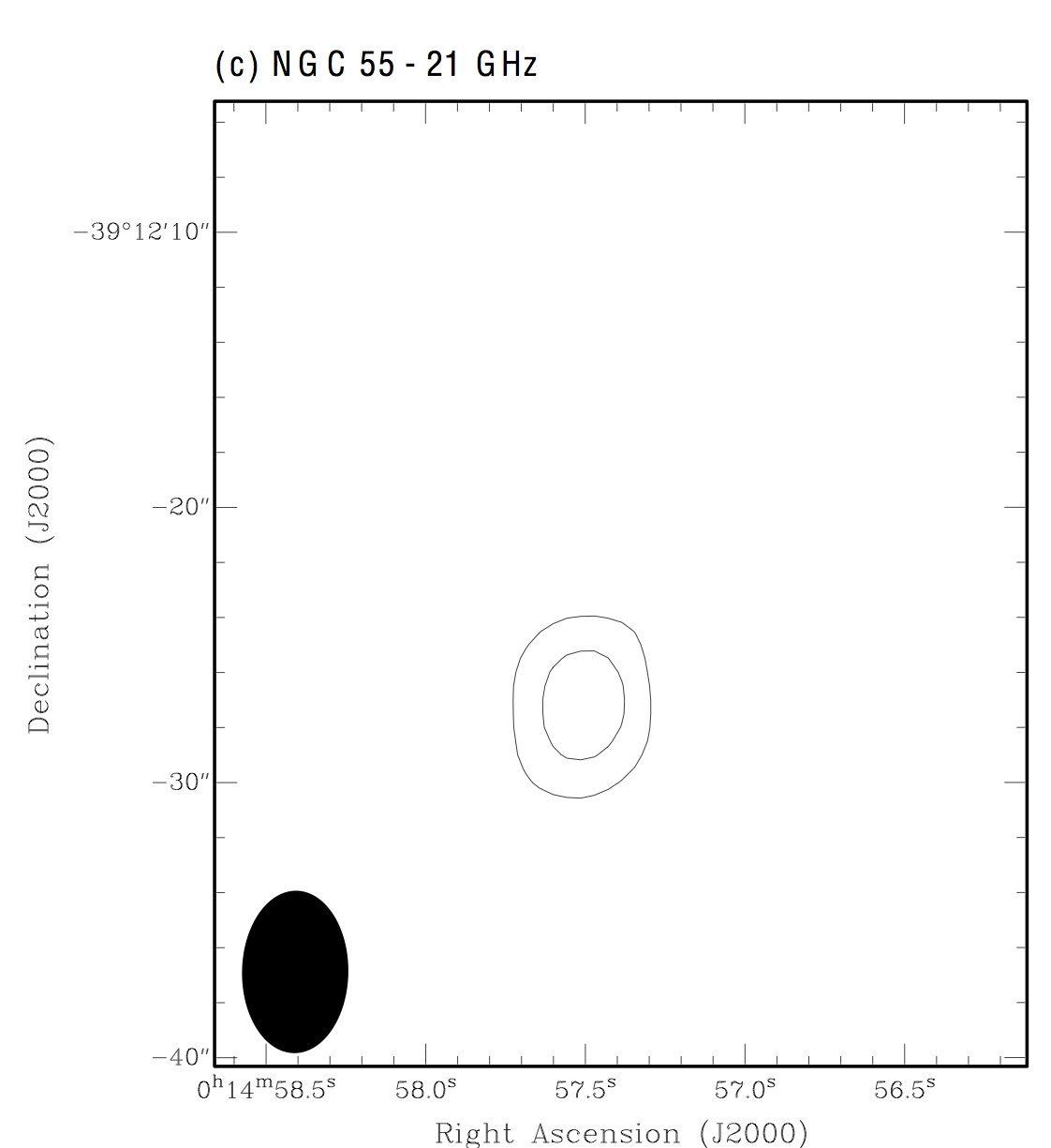}
}
\caption[ATCA image of NGC 55 at 17, 19 GHz and 21 GHz]{Uniformly-weighted total-power map of NGC 55 as observed with the ATCA at 17 GHz, 19 GHz and 21 GHz. Map statistics for the individual maps are shown in Table \ref{tab:sbtabimage}. Contours are drawn at $\pm2^{0}, \pm2^{\frac{1}{2}}, \pm2^{1}, \pm2^{\frac{3}{2}}, \cdots$ times the $3\sigma$ rms noise.}
\label{fig:sbfigNGC55}            
\end{center}
\end{figure}

\begin{figure}[ht]
\epsscale{0.22}
\begin{center}
\mbox{
\plotone{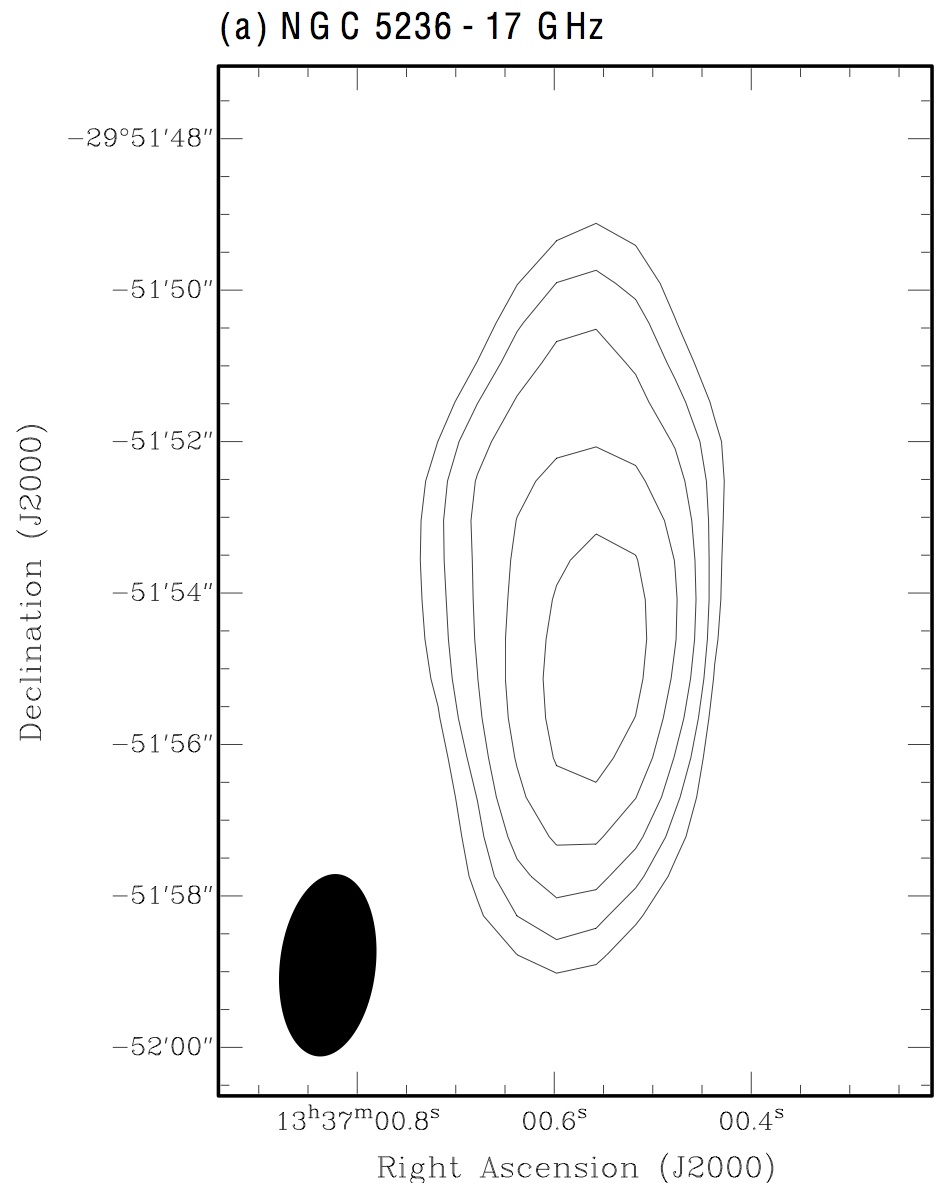} \quad
\plotone{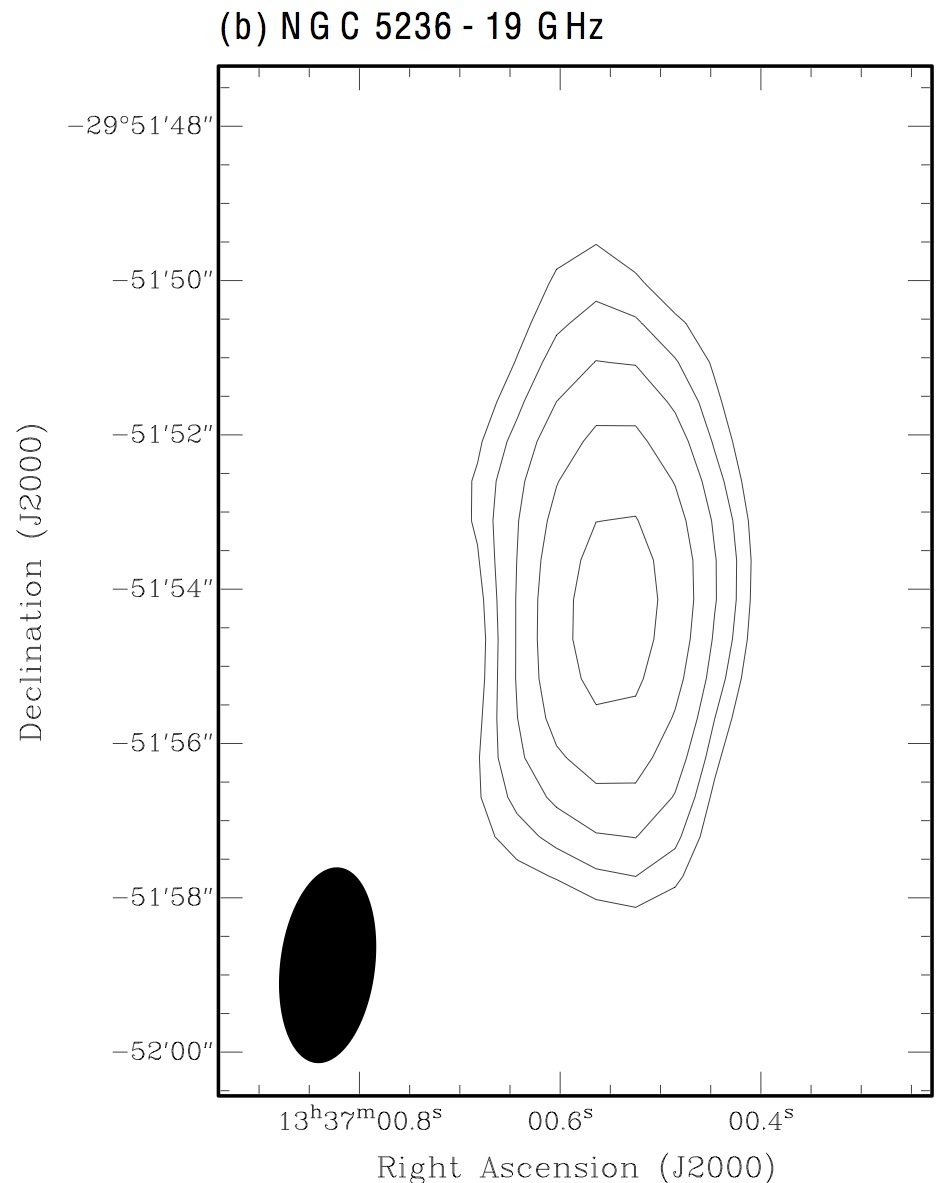} \quad
\plotone{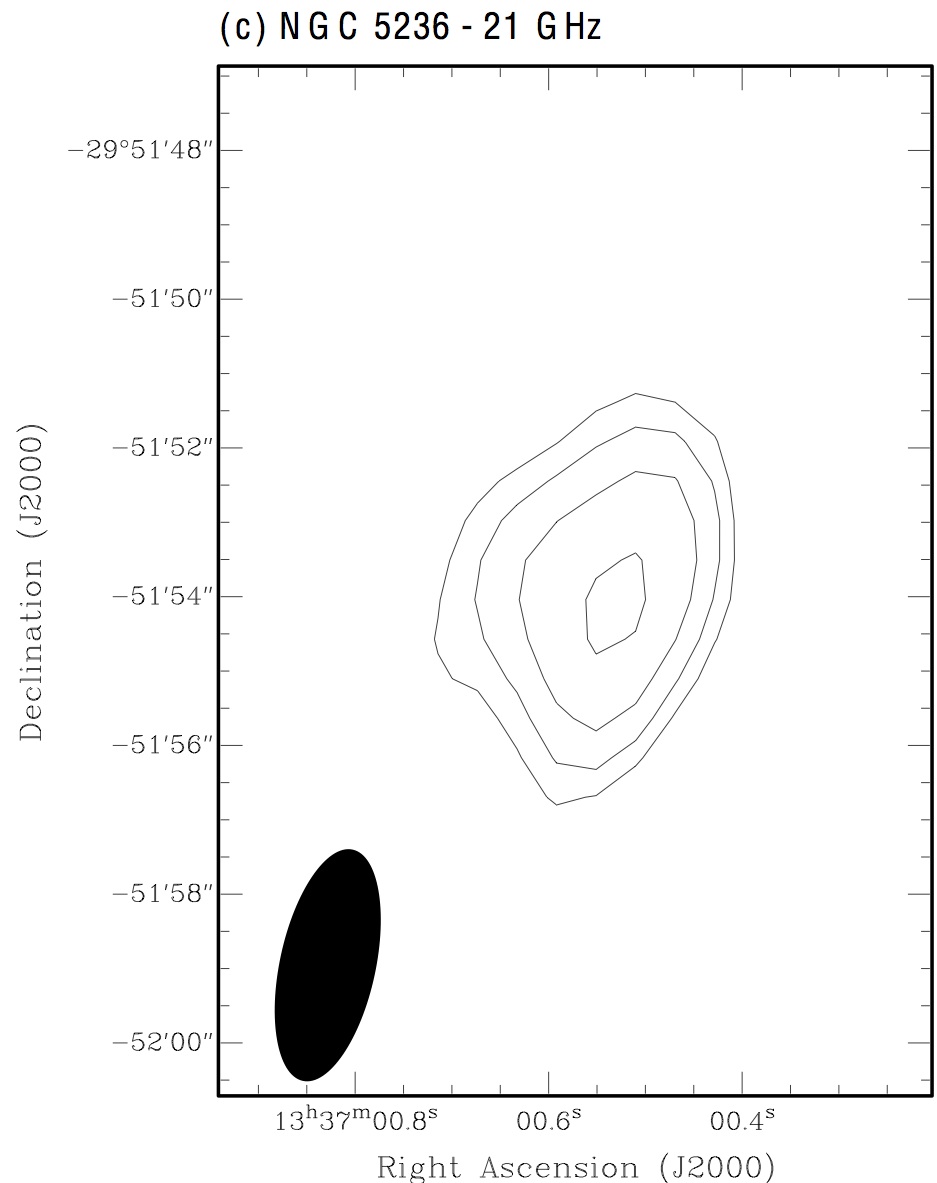} \quad
\plotone{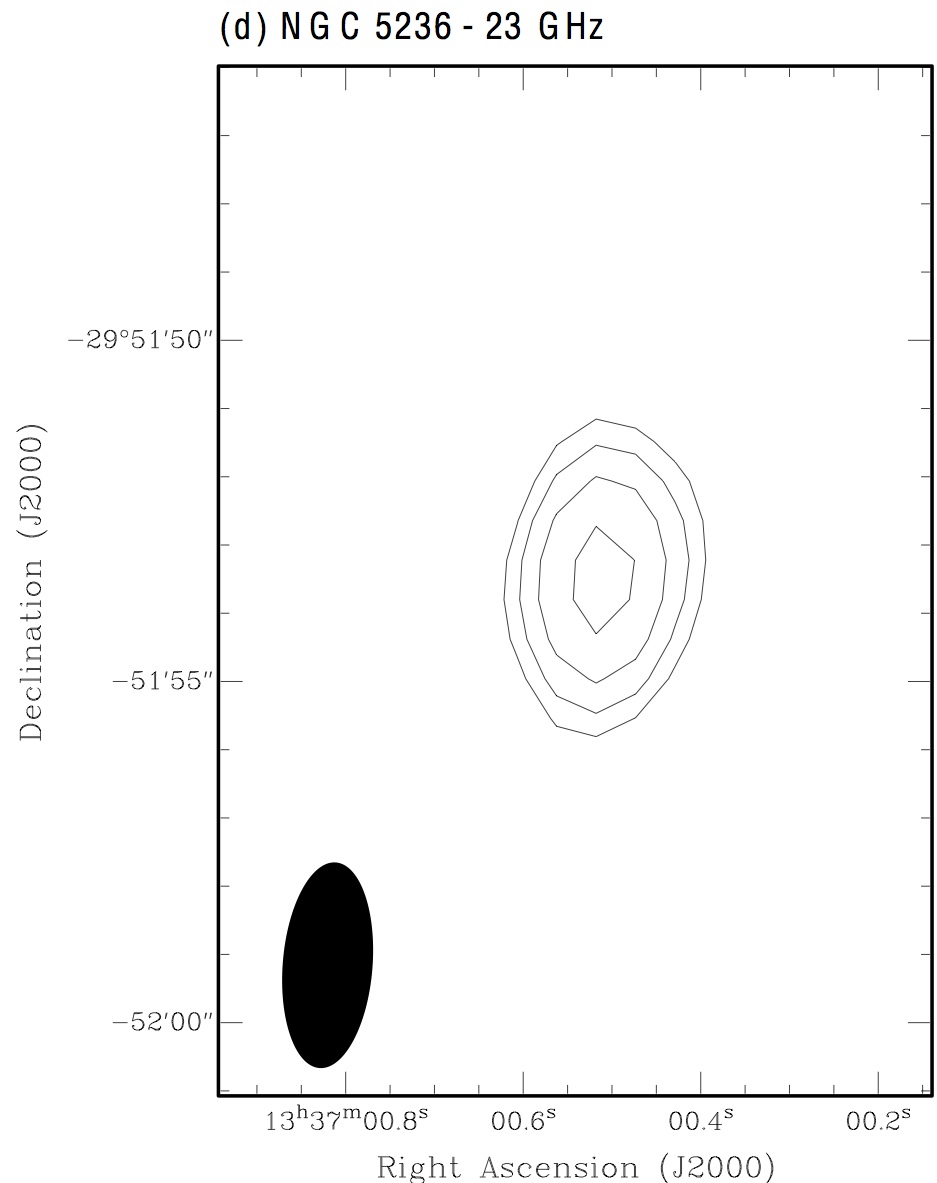}
}
\caption[ATCA images of NGC 5236 (M83) between 17 and 23 GHz]{Naturally-weighted total-power map of NGC 5236 (M83) as observed with the ATCA at 17 GHz, 19 GHz, 21 GHz and 23 GHz. Map statistics for the individual maps are shown in Table \ref{tab:sbtabimage}. Contours are drawn at $\pm2^{0}, \pm2^{\frac{1}{2}}, \pm2^{1}, \pm2^{\frac{3}{2}}, \cdots$ times the $3\sigma$ rms noise.}
\label{fig:sbfigNGC5236}            
\end{center}
\end{figure}

\begin{figure}[ht]
\epsscale{0.22}
\begin{center}
\mbox{
\plotone{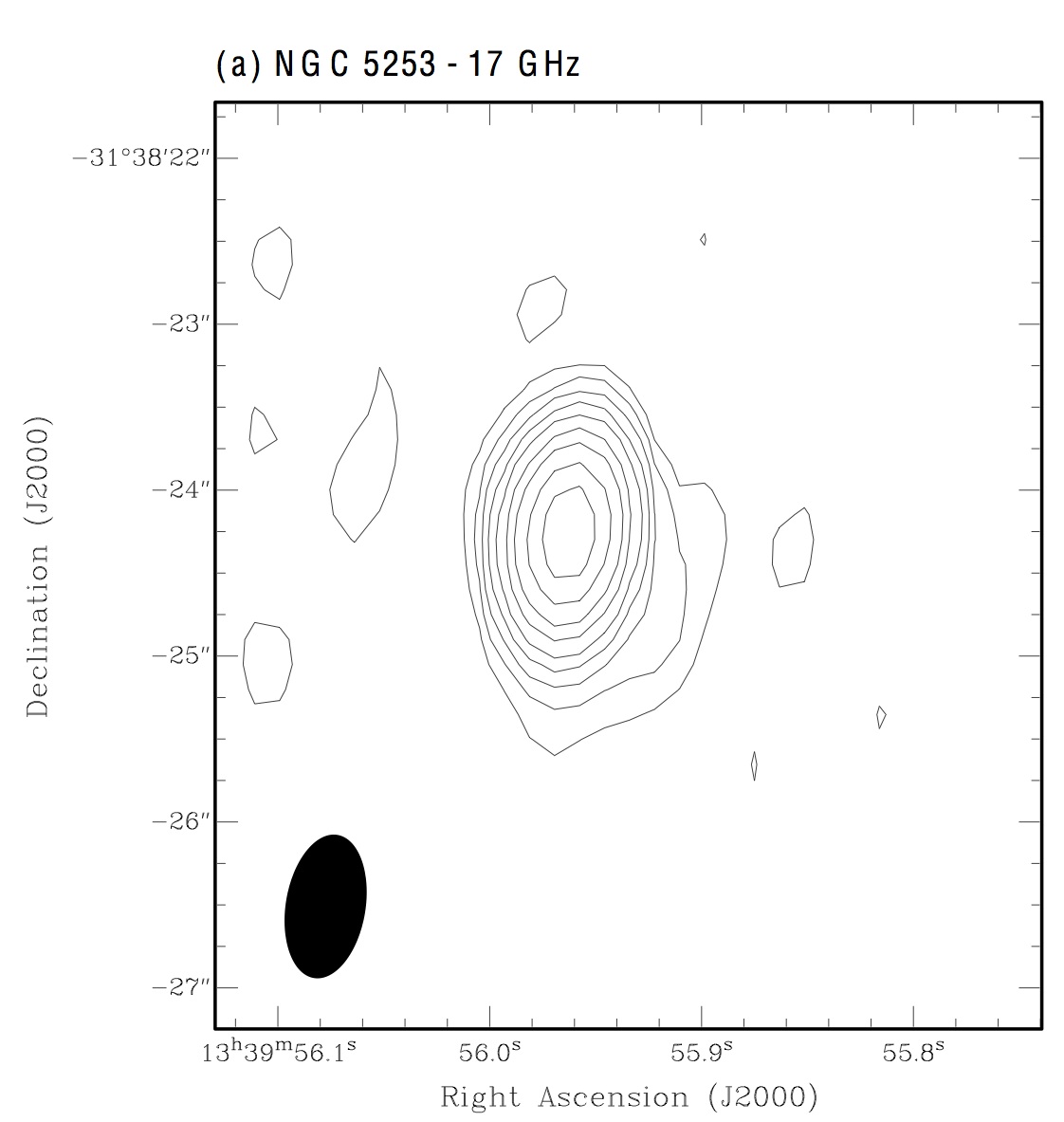} \quad
\plotone{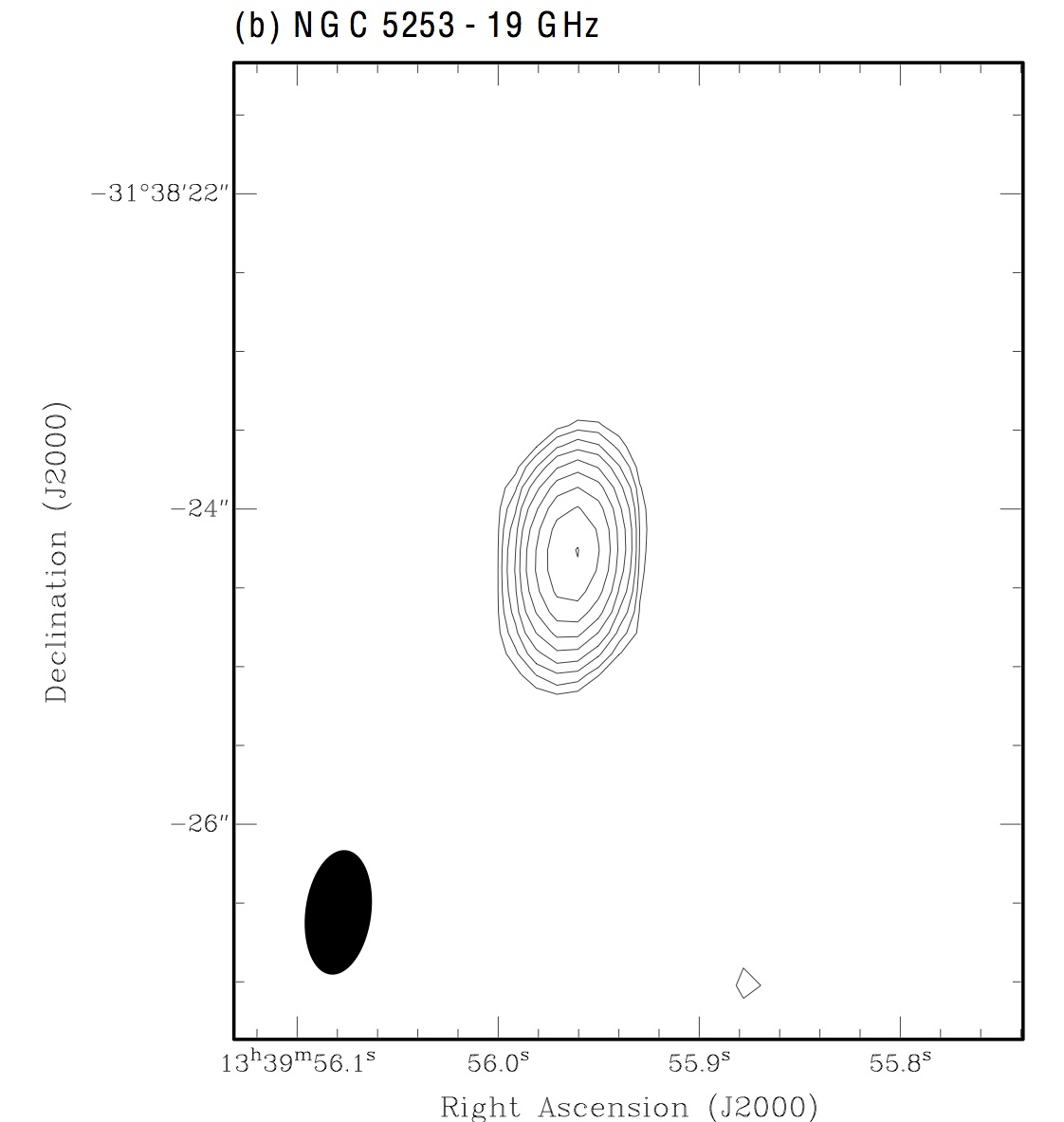} \quad
\plotone{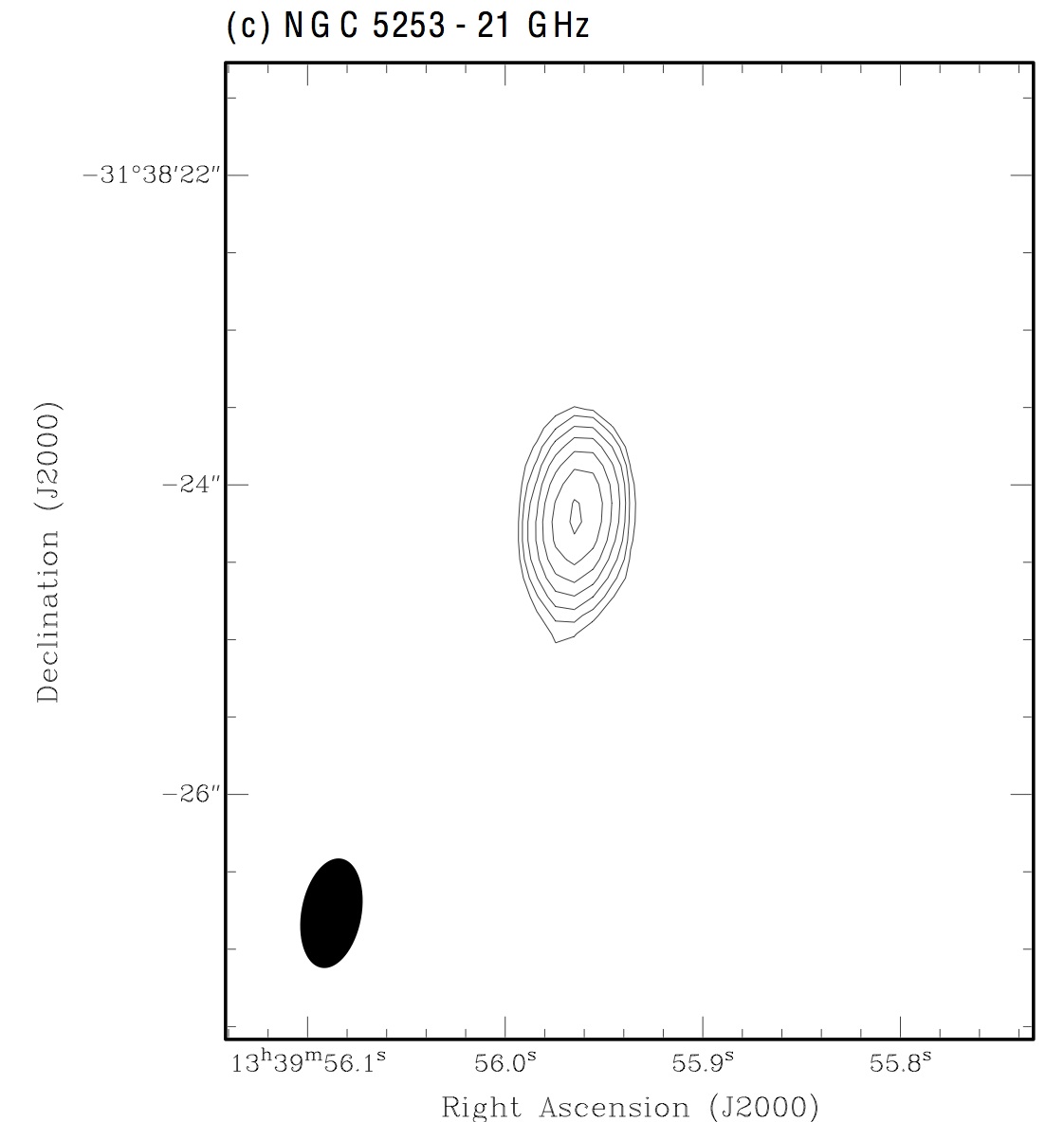} \quad
\plotone{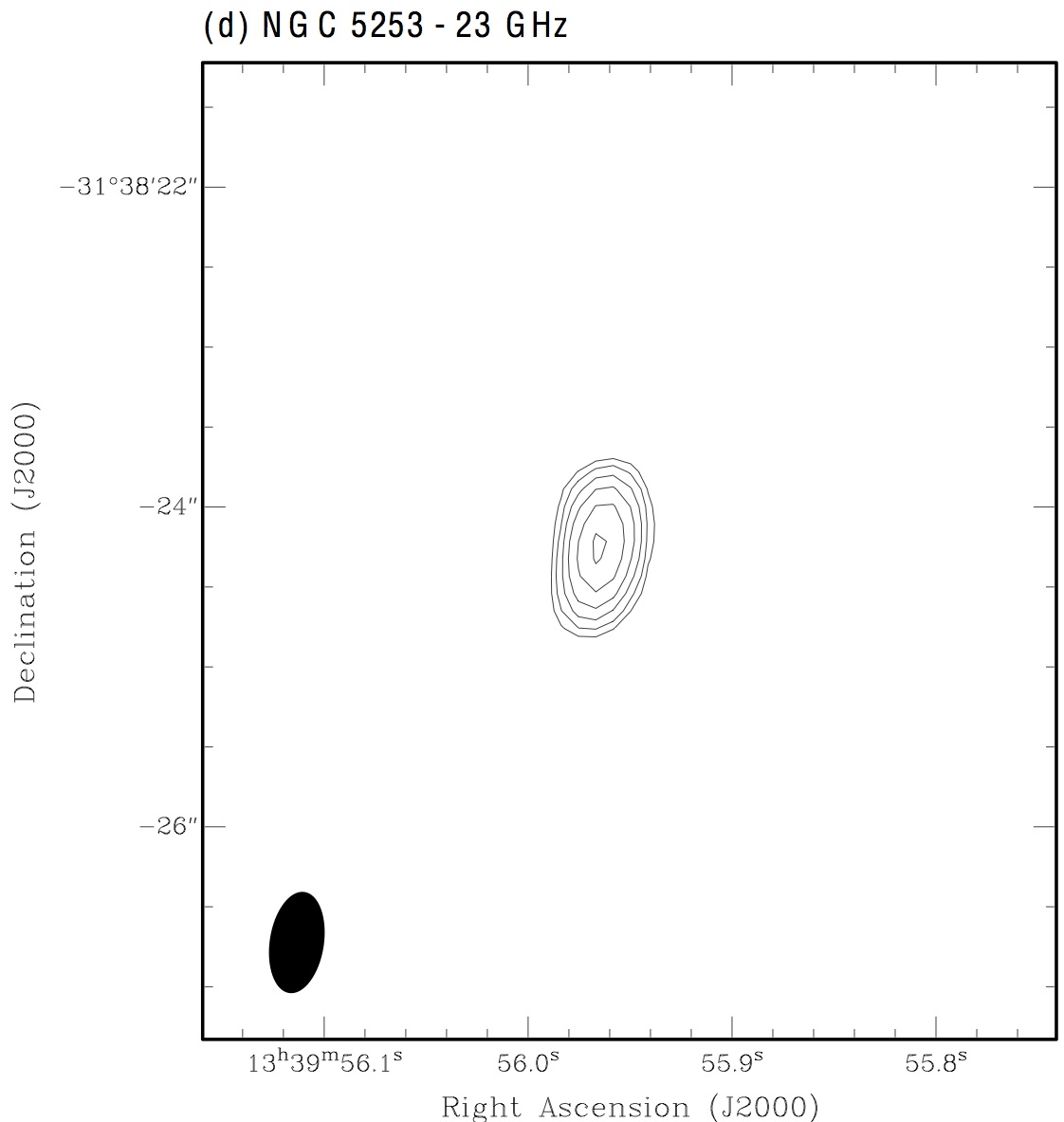}
}
\caption[ATCA images of NGC 5253 between 17 and 23 GHz]{Uniformly-weighted total-power map of NGC 5253 as observed with the ATCA at 17 GHz, 19 GHz, 21 GHz and 23 GHz. Map statistics for the individual maps are shown in Table \ref{tab:sbtabimage}. Contours are drawn at $\pm2^{0}, \pm2^{\frac{1}{2}}, \pm2^{1}, \pm2^{\frac{3}{2}}, \cdots$ times the $3\sigma$ rms noise.}
\label{fig:sbfigNGC5253}            
\end{center}
\end{figure}

All ATCA data were initially calibrated using the MIRIAD software package \citep{Sault:1995p10582}. Subsequent calibration, deconvolution and imaging was performed with the DIFMAP \citep{Shepherd:1994p10583} software package and final images were created using the KARMA software package \citep{Gooch:1996p7263}. No compact or large-scale structure was detected in NGC 1313. No compact structure was detected in NGC 55 and NGC 5236, however, larger-scale structure was detected when the $(u,v)$ range of the data-sets were restricted to 0.05 and 0.1 M$\lambda$, respectively. A single compact radio source was detected in NGC 5253. The resulting contour maps of the detected sources are shown in Figure \ref{fig:sbfigNGC55}, Figure \ref{fig:sbfigNGC5236}, and Figure \ref{fig:sbfigNGC5253} for NGC 55, NGC 5236 and NGC 5253, respectively. All associated beam parameters are tabulated in Table \ref{tab:sbtabimage}. The measured one sigma noise in the images is approximately $2-3$ times that of the estimated thermal noise of the observations. The excess noise results from phase errors that scatter substantial amounts of radio flux throughout the image and is typical for similar data-sets in which self-calibration is difficult to perform.

\section{Individual Galaxies}
\label{sec:starbursts.galaxies}

\subsection{NGC 55}
NGC 55 is classified as an SB(s)m galaxy \citet{deVaucouleurs:1961p21474} and is one of the closest members of the Sculptor group at a distance of 2 Mpc \citep{Davidge:2005p11573}. It is highly inclined with an angle near $81\arcdeg$ \citep{KiszkurnoKoziej:1988p24221}. The irregular structure of NGC 55 is similar to the LMC and it is often referred to as a dwarf irregular.

At visible wavelengths the main concentration of the galaxy is offset from the geometric centre, \citet{deVaucouleurs:1961p21474} concluding that this is a bar seen end-on. \ion{H}{1} observations indicate that 20\% of the total mass resides in the bar, while the remainder is distributed throughout the disk \citep{Hummel:1986p21216}. Radio continuum observations at 6 cm reveal a concentration of active star formation slightly offset from the bar centre but coincident with a discrete \ion{H}{1} cloud and H$\alpha$ emission \citep{Hummel:1986p21216}. Recent \emph{Spitzer} far-infrared observations suggests this is a young ($<2$ Myr) star formation complex \citep{Engelbracht:2004p15104}. Multi-wavelength observations of this galaxy have not clearly revealed an obvious nucleus.

While star formation is concentrated mostly on the bar region, there is active star formation throughout much of the disk of NGC 55. The effects of this disk-based star formation can be seen through the ejection of gas from the disk into the halo \citep{Davidge:2005p11573}. Further evidence is provided by shell structures and chimneys in the extra-planar regions \citep{Ferguson:1996p21235, Otte:1999p21501} as these are the signatures of supernovae explosions and stellar winds - the inevitable result of active star formation. \citet{Stobbart:2006p21260} detect five supernova remnants spread $10\arcmin$ around the central part of the galaxy based on hardness ratios measured with \emph{XMM-Newton}.

Our 17 GHz, 19 GHz and 21 GHz ATCA observations of NGC 55 reveal only a single extended component (FWHM$\sim4.5\arcsec$, $\alpha=00\rah14\ram57\fs51$; $\delta=-39\arcdeg12\arcmin25\farcs6$ [J2000] with a position error of $\sim0.5\arcsec$) that appears to be associated with the concentrated active star forming region. No detections are made at 23 GHz owing to the increased noise levels at that frequency. Although the detected source appears to have a steep spectral index ($\alpha=-1.6\pm1.0$, where $S\propto\nu^{\alpha}$), no detections of this or any other source were made with the LBA at 2.3 GHz. This is not totally unexpected considering the extended nature of the source. Furthermore, the concentrated star forming region, in which detections would be more probable, is situated at the edge of the LBA field of view (approximately 1$\arcmin$ from the phase centre of the observation) where sensitivity is limited.

The total power emission of NGC 55 follows a simple power law with spectral index $\alpha=-0.9\pm0.1$ \citep{Hummel:1986p21216}. Assuming this trend continues to 1.3 cm wavelengths then an integrated flux density of 45 mJy and 41 mJy would be expected at 17 and 19 GHz, respectively. While the spectral index of the detected source is consistent with that of the total power emission observed at lower frequencies, it is clear that there is substantially more emission at larger scales that is resolved out with our 1.3 cm observations. There are hints of larger scale emission at the noise level in the 17 GHz data but not in the higher frequency ATCA observations where the restoring beam is smaller and the image noise is greater.


\subsection{NGC 1313}
\label{sec:starbursts.galaxies.ngc1313}

NGC 1313 is a late-type barred spiral galaxy at a distance of 4.13 Mpc \citep{Mendez:2002p21191}. The galaxy has patchy optical morphology with a disk inclined at $48\arcdeg$ to the line of sight \citep{Ryder:1995p21602}. Although the galaxy appears to be isolated, historically it has been catalogued as an interaction remnant or as a multiple system. However, a number of \ion{H}{1} supershells have been imaged in the disk of NGC 1313, suggesting an internal origin for its visual appearance \citep{Ryder:1995p21602}. The largest of these supershells spans approximately one-fifth of the entire \ion{H}{1} extent of the galaxy and is one of the largest supershells so far discovered.

The chaotic optical morphology, the detection of \ion{H}{1} regions and the richness of young clusters \citep{Larsen:1999p21661} indicate that the galaxy has undergone vigourous irregular star formation. This star formation is believed to be driven by a mechanism similar to that of galaxy harassment \citep{Moore:1996p22088}, in which nearby gas clouds fall into, or orbit, the galaxy and prompt localised starbursts. Three large patches of star-forming activity are observed to the south-west of the galaxy and do not appear to be associated with the bar or the spiral arms of the galaxy \citep{Ryder:1995p21602}. The larger size of the \ion{H}{2} regions here, compared to those in the main part of the galaxy, are believed to be as a result of the low gas density and the presence of very massive young clusters \citep{Marcelin:1983p21571}.

Our observations of NGC 1313 were concentrated on the central region of the galaxy where the radio continuum, far-infrared emission and H$\alpha$ emission peak. The 1.3 cm ATCA observations of NGC 1313 did not detect any sources above the $6\sigma$ detection limit within the FWHM of the telescope primary beam. A brief search was performed in the large star-forming regions to the south-west of the galaxy, in which there is a known supernova remnant (SN 1978K), however as the region fell just outside the ATCA field-of-view ($\sim3.5\arcmin$ at 1.3 cm) image sensitivity was limited and resulted in no further detections.

At 2.3 GHz, our VLBI observation resulted in a weak detection ($8\sigma$) of a single extended ($280\times130$ mas) source, 15.813$-$46.95, approximately $7\arcsec$ from the phase centre. The lack of a detection at 1.3 cm would suggest that this source is non-thermal and that it may be an aged supernova remnant. Given the FWHM of the source, it has an age of $500\pm180(4\times10^{3}/\nu)$ years assuming an average expansion velocity of $\nu=$4,000 km s$^{-1}$ relative to the remnant centre\footnote{The average expansion velocity is based on the upper limit of the expansion velocity for the supernova remnant SN 1978K \citep{Smith:2007p7796} in the same galaxy}.

A VLBI search of the south-west star-forming region revealed only a single source above the $6\sigma$ threshold, the previously known supernova remnant SN 1978K. Our observation of SN 1978K at the periphery of the VLBI field measured an uncorrected integrated flux density of $9.8\pm1.0$ mJy for this source. At $5\arcmin$ from the phase centre of the observation, we estimate a loss of approximately $67\%$ as a result of the primary beam shape (assuming a simple Gaussian beam). Bandwidth smearing and time-averaging effects are estimated to be at the few percent level. Thus the corrected integrated flux density of the source is estimated to be $30\pm3$ mJy. An earlier epoch VLBI observation of this source measured an integrated flux density of $33\pm3$ mJy \citep{Smith:2007p7796}. If it is assumed that the rate of fading at VLBI resolution is consistent with that observed at lower resolution with the ATCA \citep{Smith:2007p7796,Schlegel:1999p17016}\footnote{The decline in radio luminosity is taken to follow $S\propto(t-t_{0})^{\beta}$ from \citet{Schlegel:1999p17016}, where $\beta=-1.53\pm0.13$ and $(t-t_{0})$ is the number of days since the explosion ($t_{0}$ is assumed to be 1978 May 22 UT).}, then an integrated flux density of $29\pm3$ mJy would be expected at our second epoch observation - a value that is consistent with our estimate.

\subsection{NGC 5236 (M83)}

NGC 5236 is a nearby SAB(s)c \citep{deVaucouleurs:1991p23821} galaxy that is nearly face-on with an inclination of $24\arcdeg$ \citep{Talbot:1979p22773} and located at $5.16\pm0.41$ Mpc \citep{Karachentsev:2007p7477}. The presence of significant quantities of gas and dust suggest enhanced levels of star formation. Previous studies of the nuclear region have also noted vigourous star formation in a double circumnuclear ring composed of an inner and outer ring with a radius $2.8\arcsec$ ($\sim70$ pc) and $8.6\arcsec$ ($\sim220$ pc), respectively \citep[e.g.][]{Elmegreen:1998p22784}. Optical observations of the galaxy provide further evidence of high star formation in the form of \ion{H}{2} regions \citep{Rumstay:1983p22186} and supernova remnants \citep{Blair:2004p22952}.

NGC 5236 has a close dynamical companion, NGC 5253, which also exhibits recent star formation but on a smaller scale. It is unlikely that the enhanced levels of star formation in these two galaxies are as a result of direct interaction since the last close passage occurred $1-2$ GYrs ago \citep{Rogstad:1974p25446} whereas the clusters in these galaxies are mostly $1-10$ Myrs in age \citep{Harris:2004p15625}. It is more likely that the activity in these galaxies is driven by bar-induced gas dynamics. Near-infrared \citep{Elmegreen:1998p22784}, H$\alpha$ \citep{Harris:2001p25769} and CO $(J=3-2)$ molecular-line observations \citep{Muraoka:2007p25850} of NGC 5236 highlight the bar driving the outer spiral arms as well as inflow from its inner Lindblad resonance (ILR) toward the inner ILR radius. This process is seen in other, similarly driven, starbursts such as M82 \citep{ForsterSchreiber:2000p8653}, NGC 253 \citep{Lenc:2006p6695} and NGC 4945 (see Section \S~\ref{sec:p3introduction}). However, the near face-on orientation of NGC 5236, compared to the near edge-on orientation of these classic starbursts, provides a unique perspective of this interaction.

The central $80\times80$ pc$^{2}$ region of the galaxy has been forming stars at a rate of $\sim0.04$ $M_{\sun}$ yr$^{-1}$ ($\sim6\times10^{-6}$ pc$^{-2}$yr$^{-1}$) for at least $10^7$ years, with evidence of prolonged \citep{Harris:2001p25769} and episodic \citep{Fathi:2008p21332} star formation. While the star formation rate is a factor of 30 higher than that in the outer disk \citep{Fathi:2008p21332}, it is an order of magnitude lower than the peak star formation rate per unit area found in other local starbursts \citep{Meurer:1997p26557}.

Six supernovae have been observed in NGC 5236 over the past 85 years, three type II supernova (SN 1923A, SN 1950B and SN 1957D), a suspected type Ib (SN 1983N) and two that remain unclassified (SN 1945B and SN 1968L). Four of these have been observed at radio wavelengths with the VLA: SN 1923A, SN 1950B, SN 1957D and SN 1983N \citep{Cowan:1982p23466, Cowan:1985p23064, Cowan:1994p3926, Sramek:1984p25951}, and have been monitored at 6 cm and 20 cm wavelengths over three epochs spanning a total of 15 years \citet{Eck:1998p23474,Eck:2002p16551,Stockdale:2006p3851, Maddox:2006p3835}. SN 1983N and SN 1957D have been observed to fade rapidly, whereas no significant change in flux density has been observed in SN 1923A and SN 1950B. Aside from the four supernovae already discussed, a further 51 compact radio sources have been identified in the galaxy \citep{Maddox:2006p3835}, at least five of which have been associated with supernova remnants and over 20 of which have been associated with \ion{H}{2} regions.

In addition to the optical and radio surveys, several X-ray studies have been performed on NGC 5236 \citep[e.g.][]{Kilgard:2002p22895, Kilgard:2005p23511}. \emph{Chandra} observations of this galaxy have identified 127 discrete sources \citep{Soria:2003p3786}, at least 15 of which are within the inner $16\arcsec$ region of the galaxy \citep{Soria:2002p22964}. \citet{Maddox:2006p3835} found radio counterparts to 10 X-ray sources listed by \citet{Soria:2003p3786} and \citet{Kilgard:2005p23511}. Approximately half of these sources are associated with supernova remnants, with the remainder being identified with \ion{H}{2} regions, candidate X-ray binaries and a background galaxy.

The nucleus of NGC 5236 has traditionally been associated with the optical and near-infrared peak.  After studying the central region of the galaxy with near-infrared long-slit kinematics, \citet{Thatte:2000p23959} suggested the possibility of a second nucleus or mass concentration $\sim3\arcsec$ south-west of the primary nucleus. The mass concentration was subsequently linked with a strong velocity gradient in the H$\alpha$ velocity map \citep{Mast:2006p13139}. More recently there has been some controversy with regards to the position and nature of the nucleus and it has been postulated that the ``second nucleus'' of \citet{Thatte:2000p23959} is in fact the true, albeit heavily absorbed, nucleus, whereas the optical/near-IR peak is a massive super star cluster (Sharp, private communication). 

Our 1.3 cm ATCA observations, shown in Figures \ref{fig:sbfigNGC5236}(a)$-$\ref{fig:sbfigNGC5236}(d), reveal a single steep spectrum source ($\alpha=-1.5\pm0.7$) at $\alpha=13\rah37\ram00\fs54$; $\delta=-29\arcdeg51\arcmin53\farcs5$ (J2000) with a position error of $\sim0.5\arcsec$. The source is coincident with the location of the ``second nucleus'' of \citet{Thatte:2000p23959} and the true nucleus proposed by Sharp. At 17 GHz there is evidence of diffuse emission which is progressively resolved out in the higher frequency images. At the 23 GHz the source has a FWHM of $\sim1.8\arcsec$ which corresponds to a size of 45 pc.

At 2.3 GHz, our VLBI observations detect a single source, 58.383$-$04.71 in Figure \ref{fig:sbfigs28}, which coincides with source 28 of \citet{Maddox:2006p3835} in their 20 cm and 6 cm VLA observations of the galaxy. Radio and X-ray spectral analysis of this source indicates that it is likely associated with the nucleus of a background galaxy \citep{Stockdale:2001p26568,Soria:2003p3786}. The background galaxy is also associated with sources 27 and 29 of \citet{Cowan:1994p3926}, which form the radio lobes of an FR II radio galaxy that lies along the line of sight \citep{Stockdale:2001p26568}. Our VLBI observations reveal no compact radio sources associated with NGC 5236 itself.




\subsection{NGC 5253}

NGC 5253 is a blue dwarf irregular galaxy located at $3.60\pm0.20$ Mpc \citep{Karachentsev:2007p7477}. The galaxy hosts several groups of young ($<5$ Myr) star clusters, called super star clusters or SSCs \citep{Gorjian:2001p22497, Calzetti:1997p22602}. Approximately half of the young star clusters are located in a tight grouping near the centre of the galaxy \citep{Harris:2004p15625}. A nearly flat radio continuum spectrum suggests that free-free emission is the dominant mechanism and thus rules out supernova remnants and radio supernovae as a major contributing factor \citep{Turner:2000p5783} - a finding consistent with the young age of the star clusters.

VLA radio observations of the galaxy at 6 cm, 3.6 cm and 2 cm reveal a complex structure in the central $20\arcsec\times40\arcsec$ region that is dominated by sources \citep{Turner:1998p22446,Beck:1996p22664} on the sub-arcsecond-scale. Higher angular resolution observations with the VLA at 1.3 cm and 2 cm reveal two compact sources \citep{Turner:2000p5783} and may be related to a double cluster observed with NICMOS observations in the galactic nucleus \citep{AlonsoHerrero:2004p3984}. The main radio source, first referred to as a ``the supernebula'' by \citet{Turner:1998p22446}, is compact but partially resolved in 7 mm observations on scales of $0.05\arcsec$\citep{Turner:2004p16373}. Coincident with the location of the supernebula is an unresolved cluster that dominates the emission in mid-infrared \emph{Spizter} observations of the galaxy \citep{Beirao:2006p21432}.

Two historical supernovae have been observed in NGC 5253 over a period of 112 yr, SN 1895B \citep{Campbell:1897p26586} and SN 1972E \citep{Kowal:1972p26591}. Both supernovae are Type I which, owing to their rapid turn-on and turn-off at radio wavelengths, have not been detected in recent radio observations of the galaxy \citep{Ulvestad:2007p9194}. Based on H$\alpha$ observations, \citet{Calzetti:1997p22602} deduce a total star formation rate averaged over the central $6\arcsec$ of $0.1-1.1$ M$_{\sun}$ yr$^{-1}$. This corresponds to a supernova rate of $5\times10^{-3}-5\times10^{-2}$ yr$^{-1}$ \citep{Condon:1992p10540}. \citet{Ulvestad:2007p9194} estimate an upper limit on the supernova rate of $<10^{-2}$ yr$^{-1}$ based on VLBI non-detections, assuming an upper limit of one supernova above $10^{18}$ W Hz$^{-1}$, and scaling by the supernova rate in Arp 299.

Our 1.3 cm ATCA observations, shown in Figures \ref{fig:sbfigNGC5253}(a)-\ref{fig:sbfigNGC5253}(d), reveal a single source, at $\alpha=13\rah39\ram55\fs96$; $\delta=-31\arcdeg38\arcmin24\farcs3$ (J2000) with a position error of $\sim0.1\arcsec$, that coincides with the supernebula observed with the VLA \citep{Turner:2000p5783,Turner:2004p16373}. The flat spectrum of the source ($\alpha=0.2\pm0.6$) at 1.3 cm is consistent with that observed between 2 cm and 20 cm \citet{Beck:1996p22664}. The total flux density measured for this source lies between that measured in the higher resolution VLA observations of \citet{Turner:2000p5783} (20 mJy at 23 GHz) and the lower resolution VLA observations of \citet{Beck:1996p22664} and \citet{Turner:1998p22446} (50 mJy at 15 GHz).

We find no detections in our LBA image of the galaxy. This is consistent with higher resolution observations with the VLBA$+$VLA$+$GBT at 5 GHz where no detections where made \citep{Ulvestad:2007p9194}.





\subsection{Star formation and supernova rates}
\begin{figure}[ht]
\epsscale{1.0}
\begin{center}
\plotone{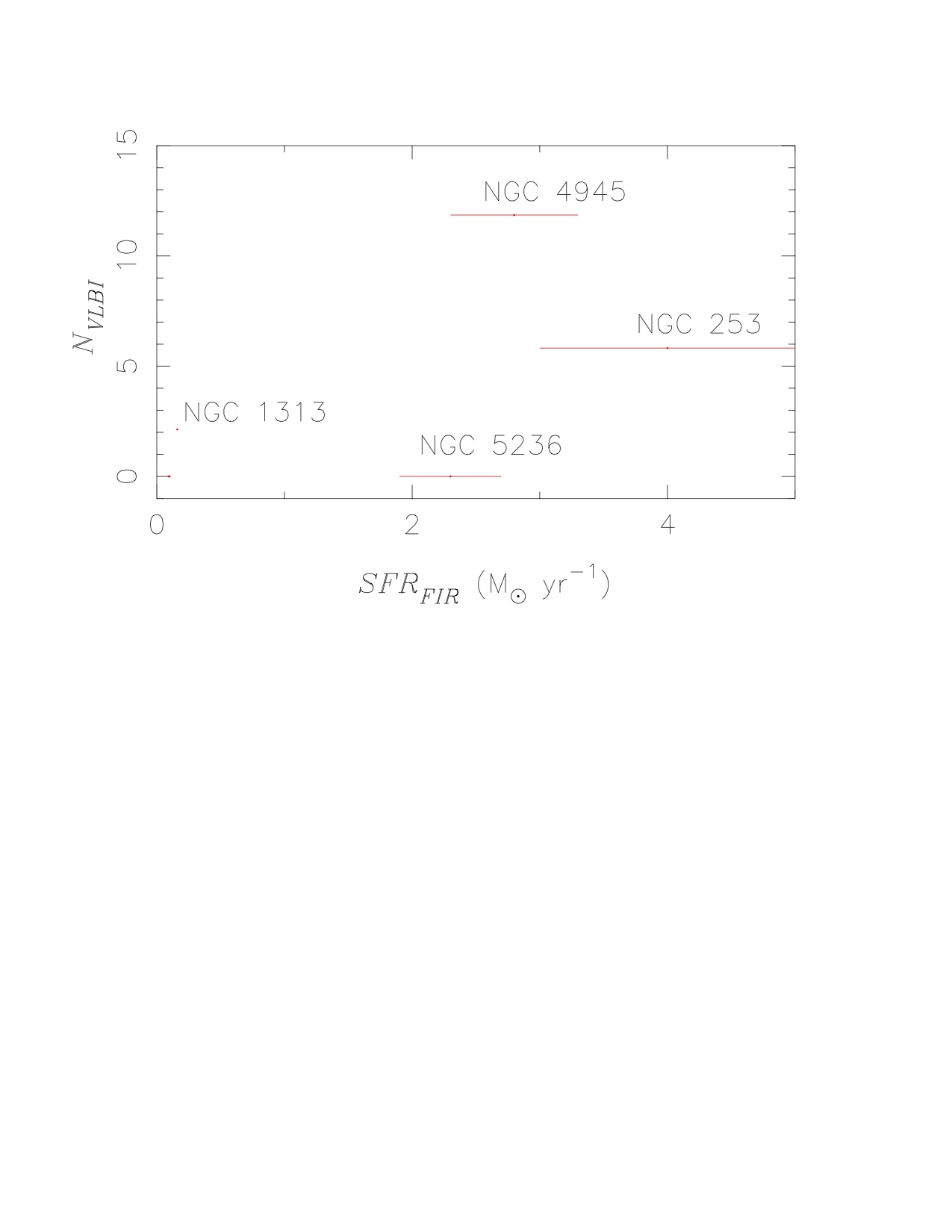}
\caption[Number of VLBI detections versus SFR]{Number of VLBI detections versus SFR determined from FIR measures for the sample of galaxies surveyed. Data points (red) for all sources labelled except for NGC 55 and NGC 5253 which are at $SFR_{FIR}\sim0.1$ M$_{\sun}$ yr$^{-1}$ and $N_{VLBI}=0$.}
\label{fig:sbNvsSFR}            
\end{center}
\end{figure}

Compared to our VLBI observations of NGC 253 and NGC 4945, as described in Chapters \ref{chap:ngc253} and \ref{chap:ngc4945}, it is clear that the starburst activity in NGC 55, NGC 1313, NGC 5236 and NGC 5253 is either significantly lower or the nature of the starburst environment is different. The total far-infrared flux can be used as an approximate measure of the star formation rate (SFR) within each galaxy and this can also be used to estimate the supernova rate, see Section \S~\ref{sec:p1sfrate}. The far-infrared flux, SFR and supernova rate for the entire set of starburst galaxies observed in our sample is shown in Table \ref{tab:sbsummary}. The number of compact sources detected in each galaxy with our VLBI observations is plotted against the SFR implied from the far-infrared flux in Figure \ref{fig:sbNvsSFR}. In this figure, it is clear that the star-formation rate in NGC 55, NGC 1313 and NGC 5253, and consequently the supernova rate, is well over an order of magnitude lower than that in NGC 253 and NGC 4945. As a result, one would expect a similarly reduced fraction of compact radio sources in these galaxies, or $\lesssim1$ compact source in the low SFR galaxies - a finding that is consistent with our observations.

\begin{table}[ht]
\begin{center}
{ \tiny
\begin{tabular}{lllrrlll} \hline \hline
Source         & D             & Ref.\tablenotemark{a}           & FIR\tablenotemark{b}  & $SFR_{IR}$\tablenotemark{c}      & $\nu_{SN}$\tablenotemark{d} \\
               & (Mpc)         &                                 & (W m$^{-2}$)          & (M$_{\Sun}$ yr$^{-1}$) & yr$^{-1}$         \\ [0.5ex] \hline \hline
NGC 55         & $2.0$         & \citet{Davidge:2005p11573}      & $4.697\times10^{-12}$ & 0.1             & $4.1\times10^{-3}$       \\
NGC 253        & $3.94\pm0.5$  & \citet{Karachentsev:2003p10562} & $4.769\times10^{-11}$ & $4.0\pm1.0$     & $1.6\pm0.4\times10^{-1}$ \\
NGC 1313       & $4.13$        & \citet{Mendez:2002p21191}       & $1.794\times10^{-12}$ & 0.16            & $6.8\times10^{-3}$       \\
NGC 4945       & $3.82\pm0.31$ & \citet{Karachentsev:2007p7477}  & $3.709\times10^{-11}$ & $2.8\pm0.5$     & $1.2\pm0.2\times10^{-1}$ \\
NGC 5236 (M83) & $5.16\pm0.41$ & \citet{Karachentsev:2007p7477}  & $1.525\times10^{-11}$ & $2.3\pm0.4$     & $9.3\pm1.5\times10^{-2}$ \\
NGC 5253       & $3.6\pm0.20$  & \citet{Karachentsev:2007p7477}  & $1.349\times10^{-12}$ & $0.095\pm0.011$ & $3.9\pm0.5\times10^{-3}$ \\ \hline
\tablenotetext{a}{References for distances based on the tip of the red giant branch.}
\tablenotetext{b}{Estimate of total far-infrared flux between 42.5 and 122.5 $\um$, defined as FIR$=1.26(2.58S_{60}+S_{100})\times10^{-14}$ \citep{Helou:1988p26551}. $S_{60}$ and $S_{100}$ are taken from \citet{Sanders:2003p7880}.}
\tablenotetext{c}{Star formation rate implied from infrared luminosity, using equation (4) of \citet{Kennicutt:1998p8678}.}
\tablenotetext{d}{Supernova rate implied from star formation rate, using equation \ref{eq:sfr2snr}.}
\end{tabular}
\caption{Summary of southern starburst galaxies.}
\label{tab:sbsummary}
}
\end{center}
\end{table}

The situation in NGC 5236 is not so easily explained. While this galaxy has a SFR that is approximately the same as that of NGC 4945, we detect no compact radio sources in it, see Table \ref{tab:sbsummary} and Figure \ref{fig:sbNvsSFR}. The greater distance to the galaxy causes some weaker sources to fall below the detection threshold of the observation, assuming they have similar intrinsic luminosity, but $\sim5-7$ sources should still be detectable. Interestingly, our ATCA observations of NGC 5236 measure 20 times less flux between $17-23$ GHz than that of NGC 4945, an order of magnitude lower than what would be expected from distance differences alone. This suggests that star formation in the nuclear region is not as great and that the environment is not as dense. \citet{Meurer:1997p26557} found that the peak star formation rate per unit area in NGC 5236 was an order of magnitude lower than that found in other starbursts and this is consistent with our ATCA observations. Furthermore, five of the six historically observed supernovae in NGC 5236 are situated along a spiral ring approximately $3\arcmin$ from the nucleus. It is apparent that the nuclear region does not overly dominate the star formation (and subsequent supernovae), with much of the star formation occurring in the less dense regions surrounding it.

In high density nuclear regions, such as those observed in NGC 253 and NGC 4945, the supernova ejecta will be slowed dramatically causing both powerful and long-lived radio emission \citep{Chevalier:2001p4204}, consistent with the compact radio sources observed in these galaxies. In NGC 5236, where the comparative density of the nuclear region is lower, the radio emission of young Type II supernovae is dominated by the interaction of the supernova blast wave with the mass-loss shell from the parent star \citep{Chevalier:1982p26861}, resulting in weaker emission that fades rapidly with time $t$ at rates between $t^{-0.7}$ to $t^{-1.8}$ \citep{Weiler:2002p26875}. This behaviour has been observed with SN 1987A in the Large Magellanic Cloud where a sudden brightening of radio emission occurred $\sim15$ yr after the supernova event as a result of interaction between the expanding shock and the swept-up circumstellar medium \citep{StaveleySmith:1992p27067}. This is certainly consistent with VLA radio observations of known supernovae in NGC 5236 which have been found to be both weak and fading \citep{Stockdale:2006p3851}. Thus the apparent lack of compact radio sources in this galaxy appears to be a result of the star formation occurring in less dense regions than those in prototypical starbursts such as NGC 253, NGC 4945 and M82.

\begin{figure}[ht]
\epsscale{1.0}
\begin{center}
\plotone{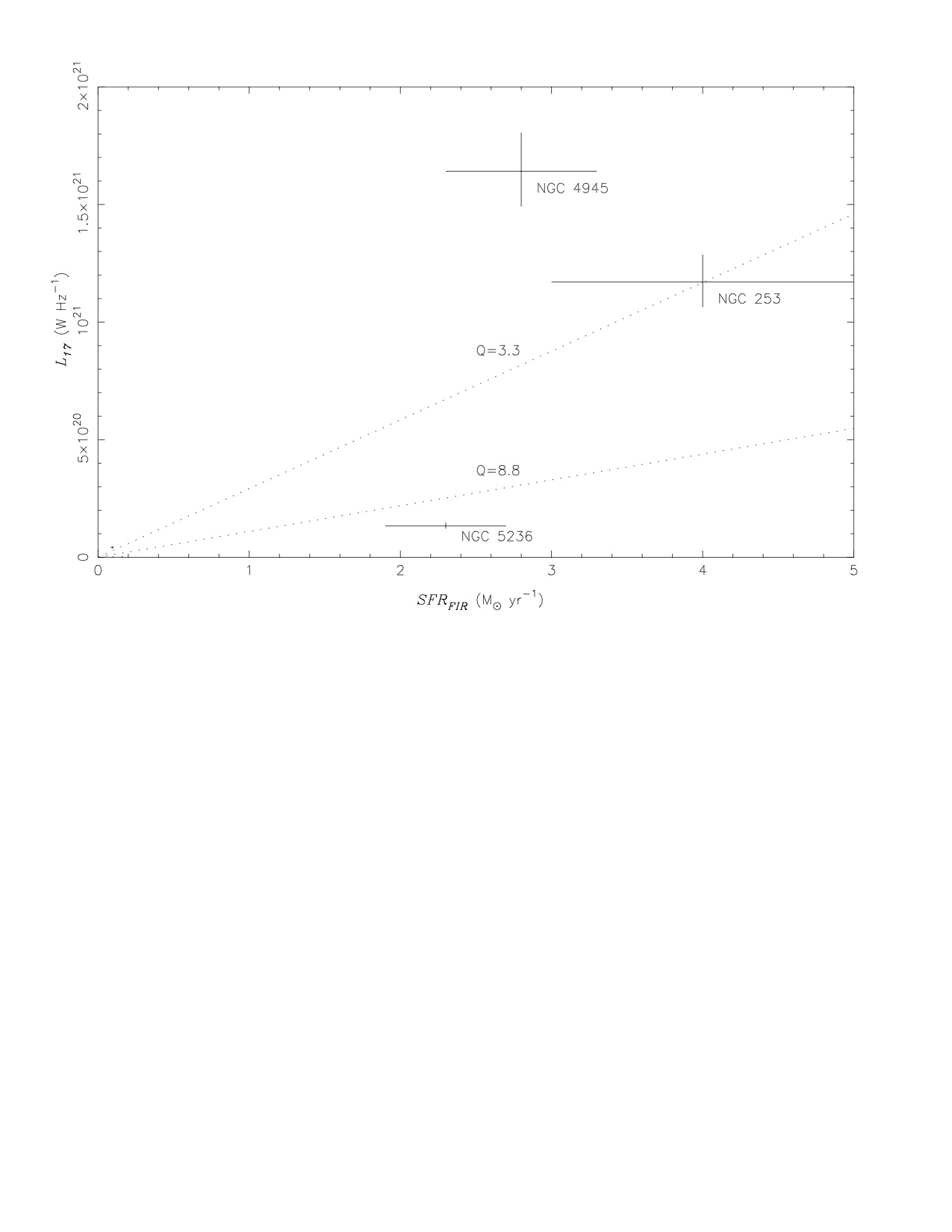}
\caption[Luminosity at 17 GHz versus SFR]{Luminosity at 17 GHz versus SFR determined from FIR measures for the sample of galaxies surveyed. Dotted lines show expected relationship between the luminosity and the FIR-based SFR for Q=8.8 and Q=3.3 \citep{Condon:1992p10540,Haarsma:2000p10556}. Data points for NGC 253, NGC 4945 and NGC 5236 are labelled, the remaining three sources are located at $SFR_{FIR}<0.2$ M$_{\sun}$ yr$^{-1}$ and $L_{17}<5.0\times10^{19}$ W Hz$^{-1}$.}
\label{fig:sbLvsSFR}            
\end{center}
\end{figure}

A plot of 17 GHz luminosity against the SFR also suggests that the nuclear environment of NGC 5236 differs from NGC 4945 and NGC 253 (Figure \ref{fig:sbLvsSFR}). As has been discussed, in Section \S~\ref{sec:p1sfrate}, the SFR is linearly proportional to the radio flux density at a given frequency. A factor $Q$ is also included to account for the mass of all stars in the interval $0.1-100$ $M_{\sun}$,

\begin{equation}
Q=\frac{\int_{0.1 M_{\Sun}}^{100 M_{\Sun}} M\psi(M)dM}{\int_{5 M_{\Sun}}^{100 M_{\Sun}} M\psi(M)dM},
\end{equation}

where $\psi(M)\propto M^{-\gamma}$ is the IMF. Regions that are more conducive to high-mass star formation (and often associated with dense nuclear regions) have a smaller $Q$ than those that are not. Figure \ref{fig:sbLvsSFR} shows the expected relationship between the SFR and the 17 GHz flux density for two different environments, one with a $Q$ of 3.3 and another with a $Q$ of 8.8. It is clear that NGC 253 and NGC 4945 are associated with low $Q$ values which imply a larger proportion of high-mass stars, whereas NGC 5236 is associated with a high $Q$ value and thus has a lower proportion of high-mass stars.

Given that NGC 55, NGC 1313 and NGC 5253 also show very weak emission in our $17$ GHz ATCA observations, see Figure \ref{fig:sbLvsSFR}, it is possible that these galaxies also contain environments which result in weak and short-lived supernovae. This, combined with the significantly lower SFR implied from their far-infrared flux, Figure \ref{fig:sbNvsSFR}, would make VLBI supernovae detection in these galaxies difficult at current sensitivity levels.

\section{Summary}
\label{sec:starbursts.summary}

We have imaged NGC 55, NGC 1313, NGC 5236 and NGC 5253 with the LBA at 2.3 GHz to search for compact radio sources. The VLBI observations have been complemented with additional data between 17 GHz and 23 GHz observed with the ATCA. We find the following results:

\begin{itemize}
\item Two compact radio sources were detected in NGC 1313 with the LBA and one was identified with a previously known supernova remnant SN 1978K.
\item No compact radio sources were detected in NGC 55, NGC 5236 and NGC 5253.
\item No diffuse emission was detected in $17-23$ GHz observations of NGC 1313 with the ATCA and only weak emission was detected in the remaining galaxies. The observations are suggestive of low density nuclear environments in these galaxies.
\item The VLBI non-detections in NGC 55 and NGC 5253 and the small number of detections in NGC 1313 are consistent with the low star formation rates and supernova rates implied from the far-infrared flux density of these galaxies.
\item The VLBI non-detections in NGC 5236 are explained as a result of supernovae occurring in low density environments, as implied from weak emission at $17-23$ GHz, resulting in weak and short-lived supernova emission that falls below our detection limits. Weak $17-23$ GHz emission in the NGC 55, NGC 1313 and NGC 5253 suggest that low density environments may also contribute to the lack of VLBI detections in these galaxies.
\end{itemize}

\linespread{1.0}
\normalsize
\begin{savequote}[20pc]
\sffamily
I have yet to see any problem, however complicated, which,\\
when you looked at it in the right way,\\
did not become still more complicated.
\qauthor{Poul Anderson}
\end{savequote}

\chapter{The Jet Termination Shock in Pictor A}
\label{chap:pictora}
\begin{center}
{\it Adapted from:}

S.J. Tingay, E. Lenc, G. Brunetti \& M. Bondi

Astronomical Journal, 136, 2473 (2008)
\end{center}
\small
Images made with the Very Long Baseline Array have resolved the region in a nearby ($z=0.035$) radio galaxy, Pictor A, where the relativistic jet that originates at the nucleus terminates in an interaction with the intergalactic medium, a so-called radio galaxy hot spot.  This image provides the highest spatial resolution view of such an object to date, the maximum angular resolution of 23 mas corresponding to a spatial resolution of 16 pc, more than three times better than previous VLBI observations of similar objects.  The north-west Pictor A hot spot is resolved into a complex set of compact components, seen to coincide with the bright part of the hot spot imaged at arcsecond-scale resolution with the VLA.  In addition to a comparison with VLA data, we compare our VLBA results with data from the \emph{HST} and \emph{Chandra} telescopes, as well as new Spitzer data.  The presence of pc-scale components in the hot spot, identifying regions containing strong shocks in the fluid flow, leads us to explore the suggestion that they represent sites of synchrotron X-ray production, contributing to the integrated X-ray flux of the hot spot, along with X-rays from synchrotron self-Compton scattering.  This scenario provides a natural explanation for the radio morphology of the hot spot and its integrated X-ray emission, leading to very different predictions for the higher energy X-ray spectrum compared to previous studies.  From the sizes of the individual pc-scale components and their angular spread, we estimate that the jet width at the hot spot is in the range 70 - 700 pc, which is comparable to similar estimates in PKS 2153$-$69 ($z=0.028$),  3C 205 ($z=1.534$), and 4C 41.17 ($z=3.8$).  The lower limit in this range arises from the suggestion that the jet may dither in its direction as it passes through hot spot backflow material close to the jet termination point, creating a $``$dentist drill" effect on the inside of a cavity 700 pc in diameter.
\clearpage
\linespread{1.3}
\normalsize
\section{Introduction}
\label{sec:radiojets.introduction}

The hot spots of powerful radio galaxies, where relativistic jets originating at the active galactic nucleus (AGN) terminate in an interaction with the intergalactic medium (IGM), are known to radiate strongly over the radio to X-ray wavelength range (\citet{Georganopoulos:2003p10756} list some of the best studied objects).  At radio wavelengths the emission mechanism is widely agreed to be due to the synchrotron process.  The details of the emission mechanisms responsible for the production of the X-rays in the jets and hot spots of powerful radio galaxies are a matter for more debate \citep{Hardcastle:2006p2321,Georganopoulos:2003p10756,Tavecchio:2000p10780}, with the relative contributions of synchrotron emission and the various flavours of inverse Compton emission (synchrotron self-Compton and external Compton) often difficult to determine from the models.  Complicating factors include the orientation of the jet relative to the observer and the degree of Doppler boosting from the different regions of the jet and hot spot, caused by possibly relativistic flows in these regions.

Although our knowledge of the high energy emission from radio galaxy hot spots has greatly improved since the launch of \emph{Chandra} \citep{Schwartz:2000p10823}, and knowledge of the synchrotron emission at low photon energies has been provided for many years from instruments such as the VLA and large optical/infrared telescopes, little is still understood of the structure of the hot spots at the highest possible spatial resolutions.  As the hot spot emission is widely considered to originate in strong shocks produced when the jet interacts with the IGM, high resolution observations have the potential to determine the structure of these shocked regions.  Very long baseline interferometry (VLBI) is the highest resolution direct imaging technique in astronomy and can produce milliarcsecond resolution images of the radio emission from radio galaxy hot spots.  Important parameters in models for the X-ray emission from the hot spots can be estimated from these observations, such as the volume and energy density of the shocked regions.  

Despite this, the literature contains few reports of VLBI observations of radio galaxy hot spots.  \citet{Kapahi:1979p10893} presented single-baseline VLBI observations of hot spots in 35 high luminosity, distant ($z>0.5$) type II quasars or radio galaxies, showing that structures more compact than 150 mas can exist.  \citet{Lonsdale:1998p176}, \citet{Lonsdale:1989p10912}, and \citet{Lonsdale:1984p10888} report on VLBI observations of a hot spot in 3C 205 at a redshift of 1.534, finding a compact and complex structure $\sim$300 mas in extent (corresponding to $\sim$2.5 kpc\footnote{In this paper we adopt a cosmology with $H_{0}=71$ km/s/Mpc, $\Omega_{m}=0.27$, and $\Omega_{\Lambda}=0.73$ \citep{Spergel:2003p9720}.  We calculate all parameters for Pictor A using this cosmology and recompute previously published results in the literature also according to this cosmology, for comparison.}), observing with an angular resolution of 6.9 $\times$ 8.8 mas (corresponding to a spatial resolution of approximately 60 pc).  The comprehensive investigation of 3C 205 led \citet{Lonsdale:1998p176} to the conclusion that the complex structure seen in the hot spot is due to a continuous jet flow around a bend caused by interaction with a dense medium.  

\citet{Gurvits:1997p10768} report on VLBI observations of the hot spot in 4C 41.17 ($z = 3.8$) that show compact structure with not more than 15 mas angular extent ($\sim$110 pc at this redshift), accounting for $\sim$30\% of the integrated flux density of the hot spot as imaged with the VLA.  \citet{Gurvits:1997p10768} conclude that the structure detected with VLBI is the location of an interaction with a massive clump in the interstellar medium, causing a deflection in the jet direction.

Most recently, \citet{Young:2005p5449} report on VLBI observations of the southern lobe hot spot of the nearby radio galaxy PKS 2153-69 ($z=0.028$), showing that, at 90 $\times$ 150 mas resolution, the hot spot is marginally resolved into three circular Gaussian components of 100 - 220 mas FWHM and 10 - 65 mJy, within a 400 mas diameter area.  The spatial resolution of the PKS 2153-69 observations were comparable to the 3C 205 observations of \citet{Lonsdale:1998p176}, approximately 50 pc.

Pictor A is the closest powerful Fanaroff-Riley type II (FR-II) radio source, at a redshift of 0.03498(5) \citep{Eracleous:2004p10752}.  By virtue of its proximity, Pictor A is bright over a wide wavelength range, from radio to X-ray, and offers good spatial resolution return for high angular resolution (1 milli-arcsecond $=$ 0.7 pc).  Previously, a number of authors have taken advantage of Pictor A (and its host galaxy) as a target for detailed observations on a variety of spatial scales, at radio \citep{Perley:1997p6689,Tingay:2000p6116,Simkin:1999p462}, optical \citep{Simkin:1999p462}, and X-ray \citep{Wilson:2001p536} wavelengths.  \citet{Wilson:2001p536} undertook an exploration of the emission mechanisms possible for the hot spot X-rays, arriving at a number of plausible explanations, involving several distinct physical interpretations.  Their analysis highlights the general uncertainties in models for X-ray emission from hot spots noted above.  Better knowledge of the high resolution structure of the Pictor A hot spots may help our understanding of the X-ray emission mechanisms at play, which may be useful for our general understanding of radio galaxies.

In this paper, we present VLBA observations of the north-west Pictor A hot spot that reveal complex and compact structure on an angular scale of 23 milli-arcseconds, corresponding to a spatial scale of 16 pc, resolving the termination shock of the jet at the IGM.  We make a comparison of the VLBA data to the VLA data of \citet{Perley:1997p6689}, our ATCA data, the \emph{Chandra} data of \citet{Wilson:2001p536}, \emph{HST} data of \citet{Meisenheimer:1997p19596} and new Spitzer infrared data, discussing possible physical explanations for the high resolution structure of the hot spot and implications for synchrotron and inverse Compton modelling of radio galaxy hot spots.  The VLBA image of the Pictor A hot spot is the highest spatial resolution image of a radio galaxy hot spot to date.

\section{Observations and results}
\label{sec:radiojets.observation}

\subsection{ATCA observations}

Observations of Pictor A were carried out using all six antennas in five different configurations of the Australia Telescope Compact Array (ATCA) between 2000 October 1 and 2001 August 27. The flux density scale was set using PKS B1934$-$638 and the phase calibrator was PKS B0537$-$441. Five minute scans of the phase calibrator were performed after each 30 minute scan of the target source. Dual 128 MHz bands were recorded at 6 cm (full polarisation) during each observation. A list of all ATCA observations reported in this paper and their associated observing parameters are tabulated in Table \ref{tab:p4tabobs}.

\begin{sidewaystable}[!p]
\begin{center}
{ \footnotesize
\begin{tabular}{lcccccccc} \hline \hline
Observatory & Frequency & $\alpha$ & $\delta$ & Date & Config. & Duration & Bandwidth & $\Delta$t \\
            & (MHz)     & (J2000)  & (J2000)  &      &         & (h)      & (MHz)     & (s) \\ \hline \hline
ATCA    & 4800.0  & $5\rah19\ram49\fs75$ & $-45\arcdeg46\arcmin43\farcs80$ & 01 OCT 2000 & 6A      & 14  & 128 & 30 \\
\nodata & 4928.0  & \nodata              & \nodata                         & \nodata        & \nodata & 14  & 128 & 30 \\
\hline
ATCA    & 4800.0  & $5\rah19\ram49\fs75$ & $-45\arcdeg46\arcmin43\farcs80$ & 19 OCT 2000 & 6C      & 14  & 128 & 30 \\
\nodata & 4928.0  & \nodata              & \nodata                         & \nodata        & \nodata & 14  & 128 & 30 \\
\hline
ATCA    & 4800.0  & $5\rah19\ram49\fs75$ & $-45\arcdeg46\arcmin43\farcs80$ & 05 APR 2001 & 6E      & 14  & 128 & 30 \\
\nodata & 4928.0  & \nodata              & \nodata                         & \nodata        & \nodata & 14  & 128 & 30 \\
\hline
ATCA    & 4800.0  & $5\rah19\ram49\fs75$ & $-45\arcdeg46\arcmin43\farcs80$ & 16 JUN 2001 & 375     & 14  & 128 & 30 \\
\nodata & 4928.0  & \nodata              & \nodata                         & \nodata        & \nodata & 14  & 128 & 30 \\
\hline
ATCA    & 4800.0  & $5\rah19\ram49\fs75$ & $-45\arcdeg46\arcmin43\farcs80$ & 27 AUG 2001 & 6B     & 14  & 128 & 30 \\
\nodata & 4928.0  & \nodata              & \nodata                         & \nodata        & \nodata & 14  & 128 & 30 \\
\hline
VLBA    & 1653.48  & $5\rah19\ram49\fs702$ & $-45\arcdeg46\arcmin43\farcs795$ & 27 AUG 2001 & \nodata  & 5  & 4 & 0.26 \\
\nodata & 1661.48  & \nodata              & \nodata                         & \nodata        & \nodata & 5  & 4 & 0.26 \\
\nodata & 1669.48  & \nodata              & \nodata                         & \nodata        & \nodata & 5  & 4 & 0.26 \\
\nodata & 1677.48  & \nodata              & \nodata                         & \nodata        & \nodata & 5  & 4 & 0.26 \\
\hline
VLBA    & 2257.49  & $5\rah19\ram49\fs702$ & $-45\arcdeg46\arcmin43\farcs795$ & 27 AUG 2001 & \nodata  & 5  & 4 & 0.26 \\
\nodata & 2265.49  & \nodata              & \nodata                         & \nodata        & \nodata & 5  & 4 & 0.26 \\
\nodata & 2273.49  & \nodata              & \nodata                         & \nodata        & \nodata & 5  & 4 & 0.26 \\
\nodata & 2281.49  & \nodata              & \nodata                         & \nodata        & \nodata & 5  & 4 & 0.26 \\ \hline
\end{tabular}
\caption{Summary of Pictor A observations.}
\label{tab:p4tabobs}
}
\end{center}
\end{sidewaystable}

All ATCA data were initially calibrated using the MIRIAD software package \citep{Sault:1995p10582}. The data were exported from MIRIAD and individual frequency channels were converted into sub-arrays to reduce the effects of bandwidth smearing \citep{Bridle:1999p10564} during imaging. Subsequent calibration, deconvolution and imaging was performed with the DIFMAP \citep{Shepherd:1994p10583} software package. Gaussian model components were used to model the large-scale structure of Pictor A, whereas smaller-scale structure was modelled with CLEAN components. Final images were created using the KARMA software package \citep{Gooch:1996p7263}.

The uniformly weighted images of the north-west and south-east hot spot of Pictor A are shown in Figure \ref{fig:p4fig3}. The measured rms noise of both images is 170 $\mu$Jy beam$^{-1}$ and the beam size is approximately $2.1\arcsec\times1.6\arcsec$ at a position angle of $-0.5\arcdeg$. The north-west hot spot has a peak flux density of 1.07 Jy beam$^{-1}$ with the lowest contour at $\pm0.16$\% of the peak and subsequent contours increasing by a factor of $\sqrt{2}$ up to a maximum of 82\% of the peak.  The south-east hot spot has a flux density of 97.8 mJy beam$^{-1}$ with the lowest contour at $\pm1.7\%$ of the peak and subsequent contours increasing by a factor of $\sqrt{2}$ up to a maximum of 78.6\% of the peak.

The measured rms image noise of our ATCA images are a factor of $\sim20$ times greater than the theoretical thermal noise. While deconvolution and calibration errors are a contributing factor to this, owing to the complex structure of Pictor A on all scales, imaging of the phase calibrator has revealed similar limits on the achievable dynamic range. These limits have become more apparent as ATCA is pushed to create deeper images \citep[e.g.][]{Norris:2006p6530} and appears to be a problem that is unique to the observatory itself. Investigations are currently underway to identify and resolve factors that may be limiting the dynamic range achievable with the ATCA \citep[e.g.][]{Middelberg:2006p26778}.

\subsection{VLBA observations}

Pictor A was observed at 18 cm and 13 cm using the NRAO Very Long Baseline Array (VLBA) on 2005 November 20 and 21, respectively. Due to the low declination of the target, $-45\arcdeg$, only the southern antennas of the array at Los Alamos, Fort Davis, Pie Town, Kitt Peak, and Owens Valley were used. The observation spanned approximately 5 hours with 10 minute scans of the fringe finder PKS 0537$-$441 made on an hourly basis. Dual circular polarisation data were recorded across four 4 MHz intermediate frequencies (IFs) centred on 1653.48, 1661.48, 1669.48 and 1677.48 MHz for the 18 cm observation and 2257.49, 2265.49, 2273.49 and 2281.49 MHz for the 13 cm observation. The data were correlated at the NRAO VLBA correlator (Socorro, NM, USA) with the AGN core as the phase centre. To reduce the effects of bandwidth smearing and time averaging smearing at the north-west and south-east hot spots, in order to make a wide field of view available for imaging, the correlator generated data with 128 spectral points per baseline/IF and an integration time of 0.26 s. The one sigma theoretical thermal noise of the 18 cm and 13 cm data-sets is 170 $\mathrm{\mu}$Jy beam$^{-1}$ and 190 $\mathrm{\mu}$Jy beam$^{-1}$, respectively. Observing parameters associated with the two VLBA observations are shown in Table \ref{tab:p4tabobs}. The $uv$-coverage of the 18 cm VLBA observation is shown in Figure \ref{fig:p4uv}.

\clearpage
\begin{figure}[p]
\center
\epsscale{0.7}
\plotone{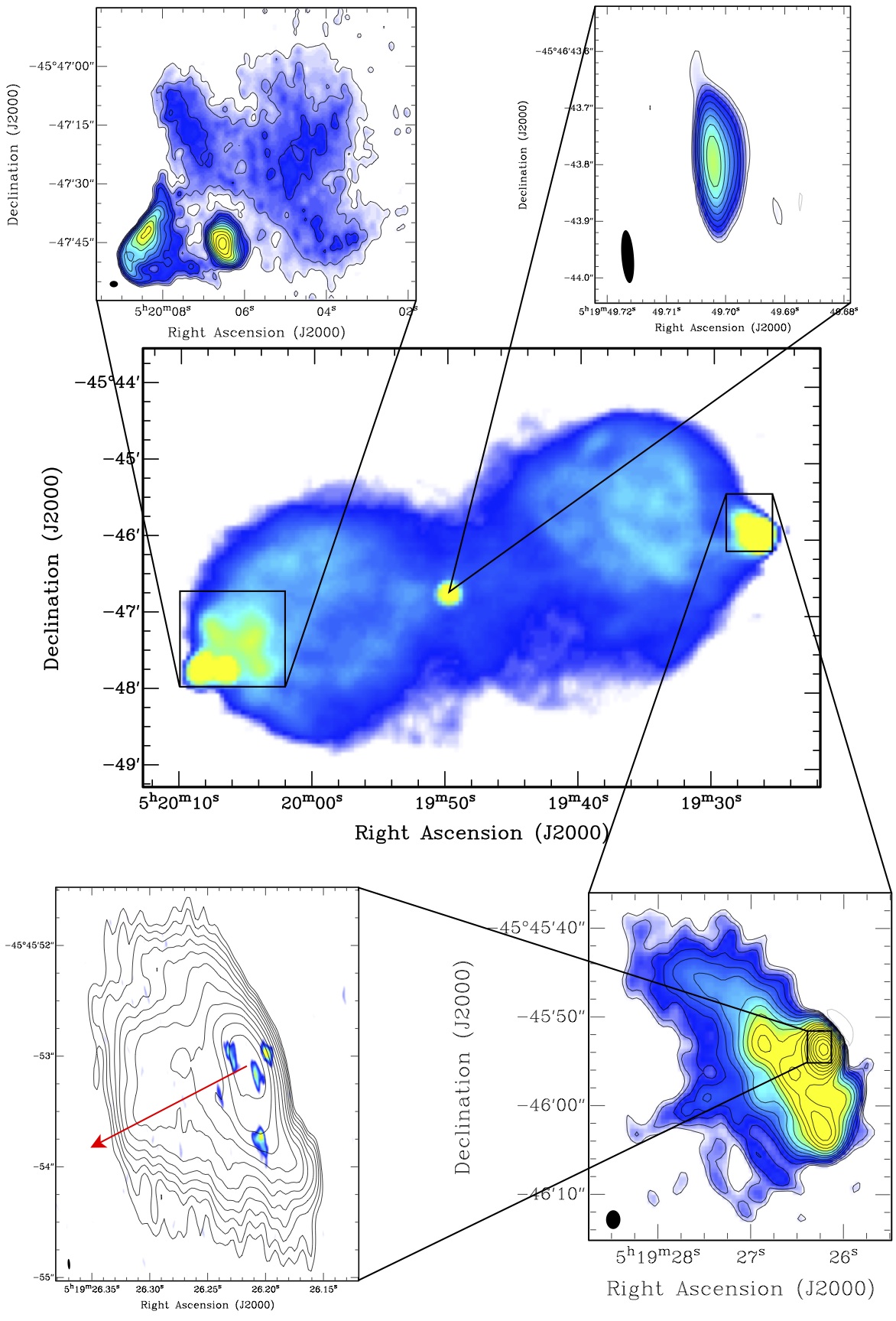}
\caption[Context of the VLBA observations in the overall radio galaxy structure of Pictor A]{Context of the VLBA observations in the overall radio galaxy structure.  The 20 cm VLA image of \citet{Perley:1997p6689} is shown centrally, with zooms into the two hot spots (6 cm ATCA data) and the 18 cm core (this paper).  A further zoom indicates the high resolution 18 cm VLBA false-colour image of the north-west hot spot, overlaid on the 2 cm VLA contour map of the hot spot from \citet{Perley:1997p6689}.  The red arrow indicates the direction to the nucleus from the location of the peak brightness in the 2 cm VLA image.}
\label{fig:p4fig3}
\end{figure}

Initial calibration of the data was performed using standard VLBA data calibration techniques in AIPS\footnote{The Astronomical Image Processing System (AIPS) was developed and is maintained by the National Radio Astronomy Observatory, which is operated by Associated Universities, Inc., under co-operative agreement with the National Science Foundation}. 

The calibration was refined by using the nucleus of Pictor A as an in-beam calibrator. To account for structure in the nucleus (which we use as a phase reference source) a new DIFMAP \citep{Shepherd:1994p10583} task, \emph{cordump}\footnote{The \emph{cordump} patch is available for DIFMAP at \url{http://astronomy.swin.edu.au/~elenc/DifmapPatches/}} \citep{Lenc:2006p32}, was developed to enable the transfer of all phase and amplitude corrections made in DIFMAP during the imaging process to an AIPS compatible solution table. The \emph{cordump} task greatly simplified the calibration of the data-set. 

First, the data were averaged in frequency and exported to DIFMAP where several iterations of modelling and self-calibration of both phases and amplitudes were performed. \emph{cordump} was then used to transfer the resulting phase and amplitude corrections back to the unaveraged AIPS data-set. The bandpass for the data was calibrated against observations of PKS $0537-441$. After application of these corrections, the DIFMAP model of the nucleus was subtracted from the unaveraged $(u,v)$ data-set. The 18 cm and 13 cm images of the nucleus had a measured RMS noise of 240 and 500 $\mathrm{\mu}$Jy beam$^{-1}$, respectively. The higher than theoretical noise in the 13 cm data is attributed to substantial levels of RFI observed at Kitt Peak and Los Alamos during the observation.

At 18 cm, the nuclear core total flux density is 734 mJy and is composed of a 526 mJy ($27\times11$ mas at a position angle of $-8\arcdeg$) component and a 209 mJy ($62\times38$ mas at a position angle of $-61\arcdeg$) extension to the north-west.  At 13 cm, the nuclear core total flux density is 885 mJy and is composed of a 624 mJy ($18\times5$ mas at a position angle of $-36\arcdeg$) component and a 260 mJy ($110\times62$ mas at a position angle of $10\arcdeg$) extension to the north-west.  Figure \ref{fig:p4fig1} shows the 18 and 13 cm images of the AGN in Pictor A.  The resolution of these images is significantly worse than the 3 cm images in \citet{Tingay:2000p6116}, so a detailed comparison with those previous results is not possible.  The unaveraged, calibrated, core-subtracted data-sets were then used to image the regions around the north-west and south-east hot spots in Pictor A.

\begin{figure}[ht]
\epsscale{0.6}
\plotone{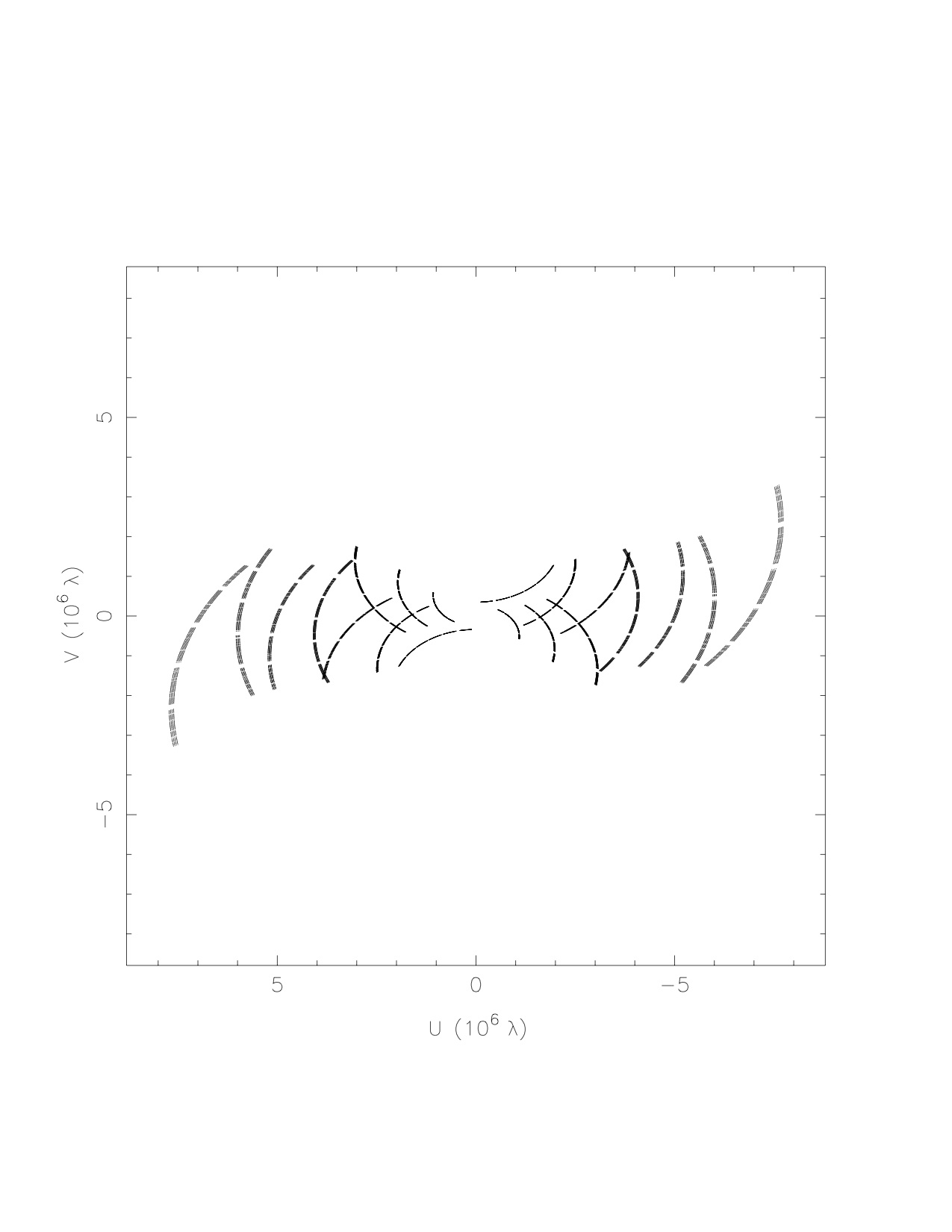}
\caption[Plot of $uv$-coverage for VLBA observation of Pictor A at 18 cm]{Plot of $uv$-coverage for VLBA observation of Pictor A at 18 cm.}
\label{fig:p4uv}
\end{figure}

\begin{figure}[ht]
\epsscale{0.45}
\mbox{
\plotone{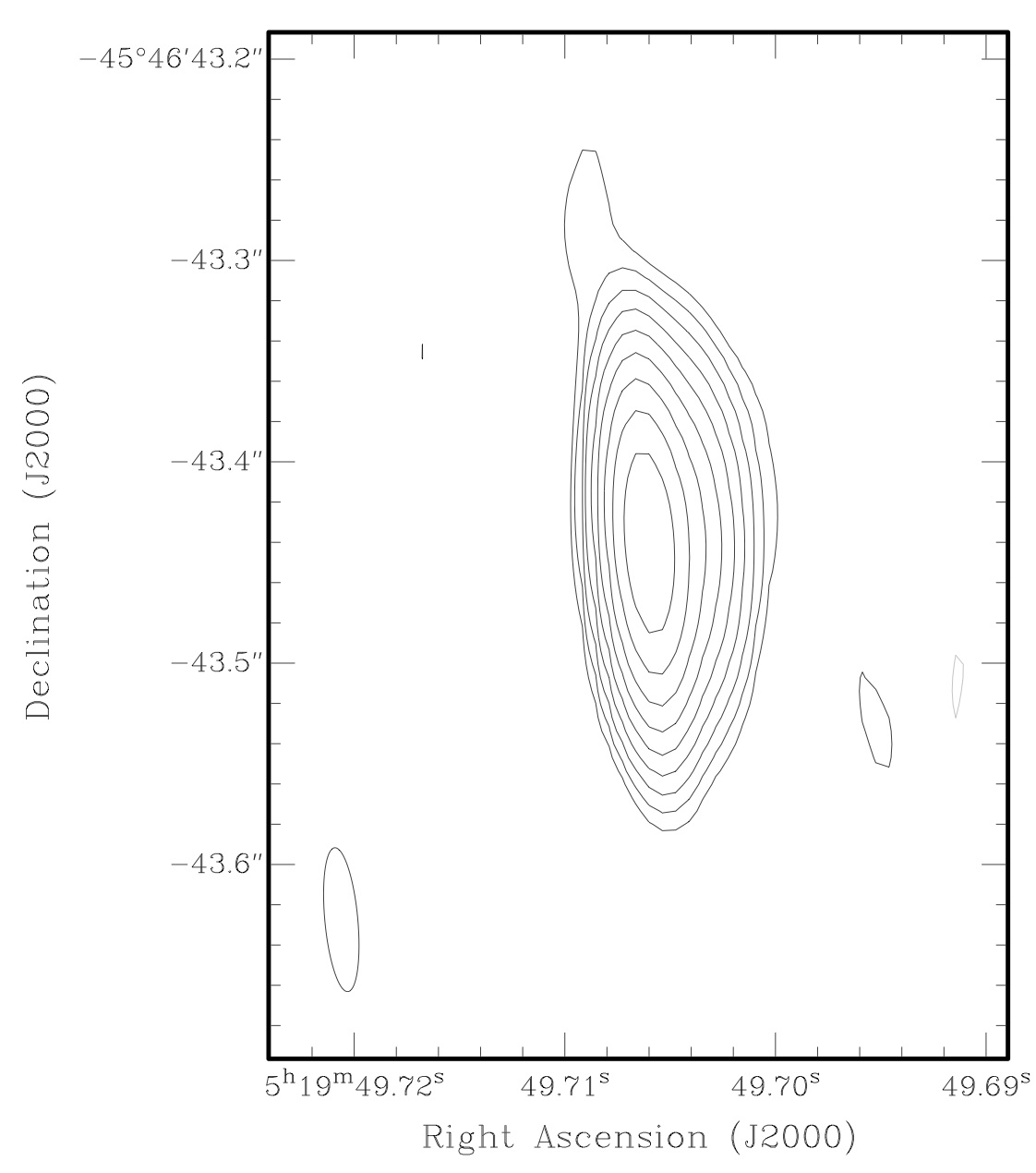} \quad
\plotone{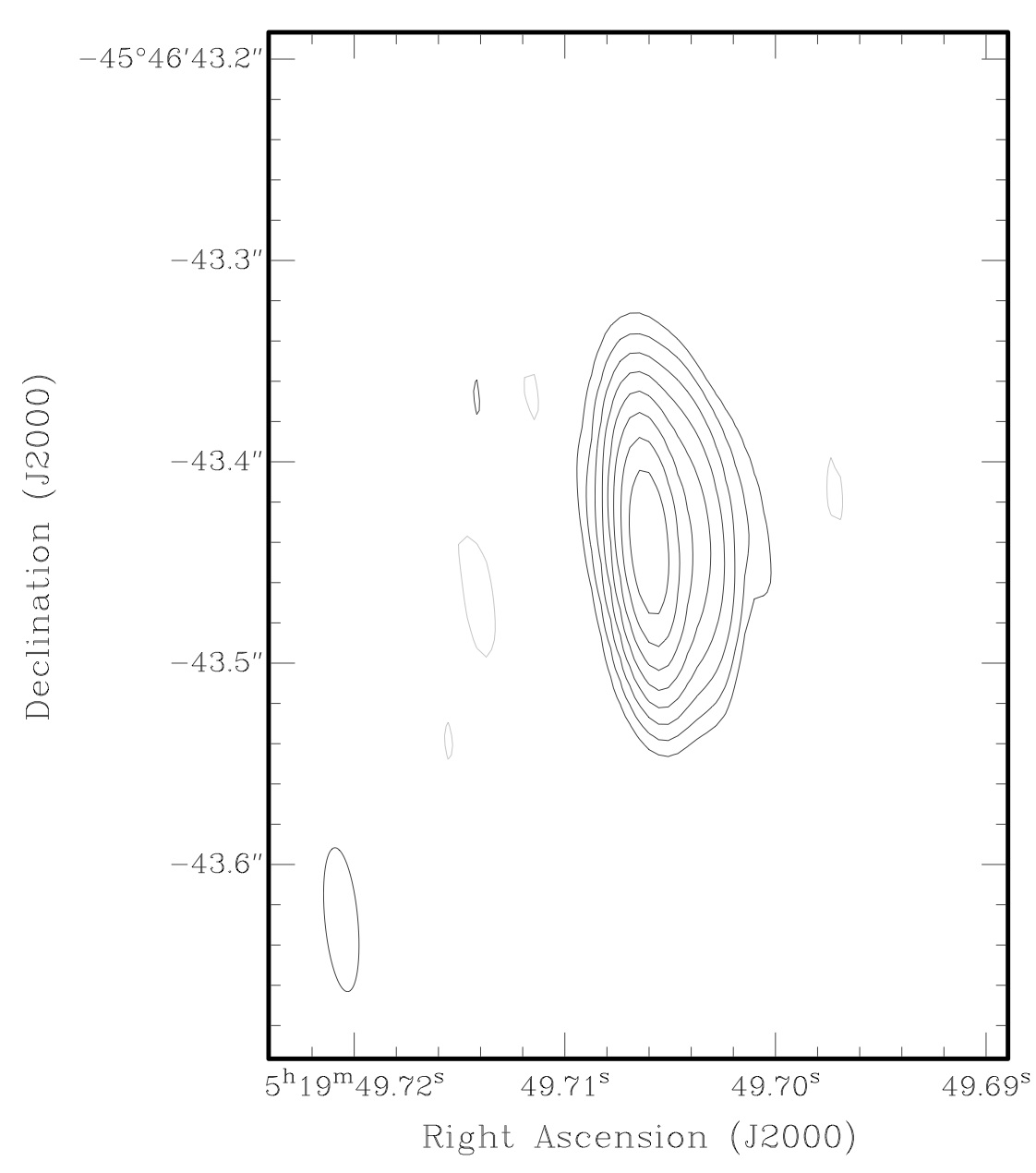}
}
\caption[VLBA images of the Pictor A nucleus at 18 cm and 13 cm]{VLBA images of the Pictor A nucleus at 18 cm (left) and 13 cm (right).  At 18 cm the image has an RMS noise level of approximately 240 $\mathrm{\mu}$Jy beam$^{-1}$. The peak flux density is 592 mJy beam$^{-1}$ and contours are set at -0.2\%, 0.2\%, 0.4\%, 0.8\%, 1.6\%, 3.2\%, 6.4\%, 12.8\%, 25.6\% and 51.2\% of the peak. The beam size is approximately $94\times22$ mas at a position angle of $5\arcdeg$.  At 13 cm the image has an RMS noise level of approximately 500 $\mathrm{\mu}$Jy beam$^{-1}$. The peak flux density is 654 mJy/beam and contours are set at -0.4, 0.4\%, 0.8\%, 1.6\%, 3.2\%, 6.4\%, 12.8\%, 25.6\% and 51.2\% of the peak. The beam size is approximately $72\times16$ at a position angle of $5\arcdeg$.}
\label{fig:p4fig1}
\end{figure}

In the first phase of the hot spot imaging process, the AIPS task IMAGR was used to make naturally weighted dirty images and beams of regions covering both the south-east and north-west hot spots. A $3\times3$ grid of $\sim12\arcsec$ square dirty maps were imaged simultaneously using the multi-field option within IMAGR.  The DO3D option, in combination with the gridded imaging, was used to reduce non-coplanar array distortion. Each grid was centred about the brightest components of each hot spot, as determined from lower resolution 6 cm ATCA (Australia Telescope Compact Array) images. Since each dirty map contains $\sim10^5$ synthesized-beam areas, a conservative $6\sigma$ detection threshold was imposed to avoid spurious detections. Furthermore, only the inner 75\% of each dirty map was searched for candidate detections to avoid erroneous detections as a result of map edge effects.

The second phase of the imaging process involved creating a $uv$-shifted data-set for each of the positive detections, using the AIPS task UVFIX. The shifted data-sets were averaged in frequency, effectively reducing the field of view of each of the targeted sources to approximately $10\arcsec$, and then exported to DIFMAP. In DIFMAP, the visibilities were averaged over 10 second intervals to reduce the size of the data-set and to speed up the imaging process. Each target was imaged in DIFMAP, with natural weighting applied, using several iterations of CLEAN.

Earlier observations of Pictor A with the VLA \citep{Perley:1997p6689} determined the position of the AGN core as $\alpha=05\rah18\ram23\fs590$ and $\delta=-45\arcdeg49\arcmin41\farcs40$ (B1950 epoch 1979.9) or $\alpha=05\rah19\ram49\fs706$ $\delta=-45\arcdeg46\arcmin43\farcs44$ (J2000). To provide a direct comparison with the VLA data we have re-referenced our data so that our nuclear core coincides with the VLA position.


No sources were detected in the hot spot regions at 13 cm. Only the north-west hot spot was detected at 18 cm. Given the peak flux density we observe at 18 cm, the expected hot spot spectral index leading to weaker emission at 13 cm, the smaller beam size at 13 cm, and the less sensitive VLBA image at 13 cm, we calculate that the expected 13 cm peak flux density would lie below our detection limit.

\begin{figure}[p]
\epsscale{1.0}
\plotone{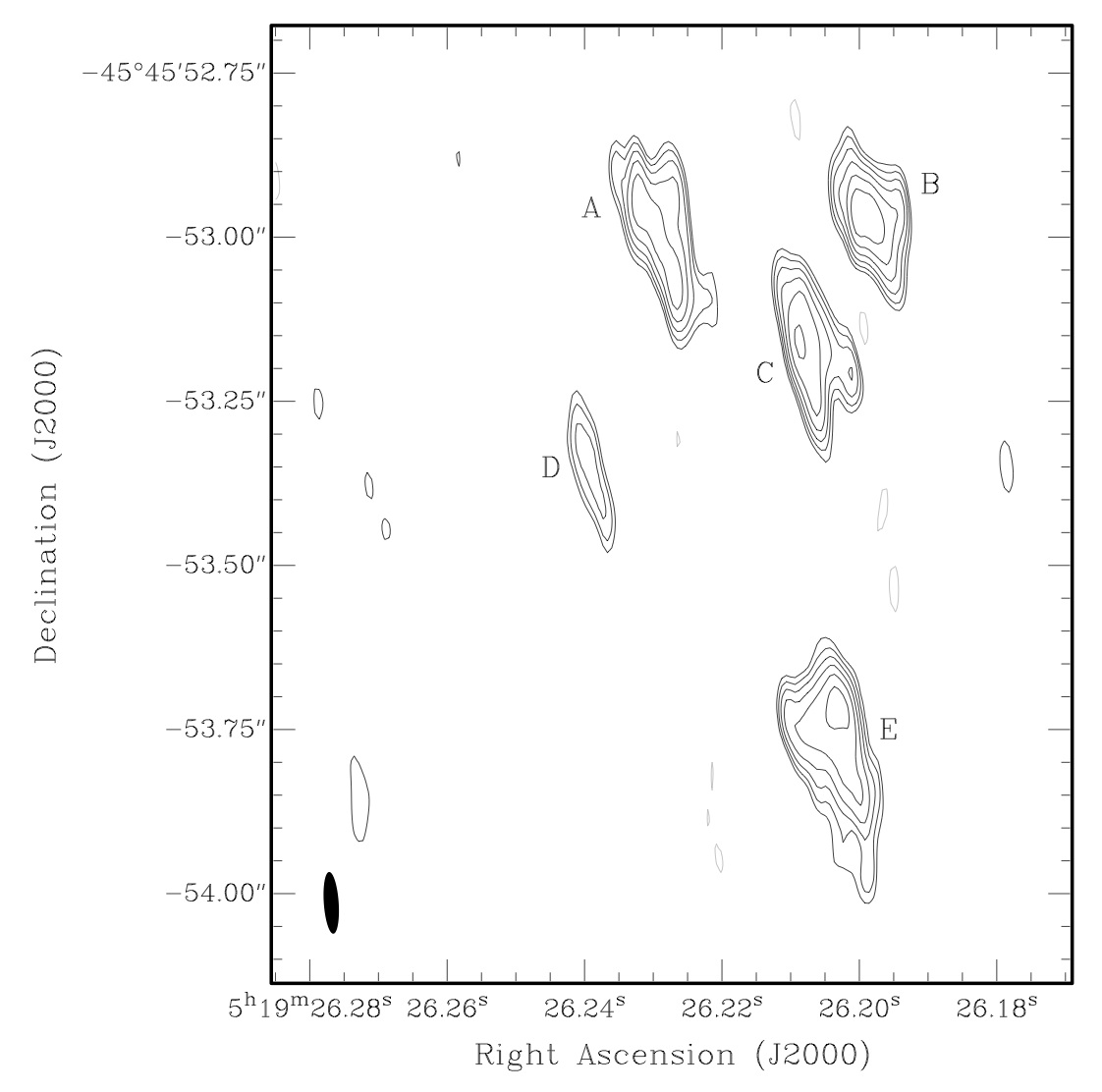}
\caption[VLBA image of the Pictor A hot spot at 18 cm]{VLBA image of the Pictor A hot spot at 18 cm.  The components referred to as A, B, C, D, \& E have their properties tabulated in Table \ref{tab:p4tabcomponents}.  The image has an RMS noise level of 260 $\mathrm{\mu}$Jy beam$^{-1}$. The peak flux density is 6 mJy beam$^{-1}$ and contours are set at $\pm3$ times the RMS noise level and increase by factors of $\sqrt{2}$. The beam size is approximately $94\times23$ at a position angle of $4\arcdeg$.
\label{fig:p4fig2}}
\end{figure}

Figure \ref{fig:p4fig2} shows the VLBA image of the north-west Pictor A hot spot at 18 cm.  The image consists of a number of components.   Figure \ref{fig:p4fig3} shows the final VLBA image of the hot spot, overlaid on a VLA image at 15 GHz (Figure 18 from \citet{Perley:1997p6689}), showing the hot spot within the context of the overall structure of the Pictor A radio source, using VLA data from \citet{Perley:1997p6689} and ATCA data.

In our 18 cm observation, sources approximately 4$\arcmin$ from the phase centre, the approximate distance to the south-east and north-west hot spots, exhibit a 6.4\% loss in peak flux density as a result of the cumulative effect of bandwidth smearing (1.6\%), time averaging smearing (0.8\%) and primary beam effects (4.1\%).  Our measured flux densities were corrected for primary beam effects (4.1\% adjustment to flux densities), but not corrected for smearing effects, due to the intrinsic extended nature of the components seen in the hot spots.  The combined 1.6\% and 0.8\% (2.4\%) effect of smearing has been included in the error estimates for the hot spot flux densities.  The errors on the peak and integrated flux densities are therefore the quadrature combination of the smearing effects and the absolute error in the calibration of the flux density scale, $\sim$10\%.

The parameters of the components (as annotated in Figure \ref{fig:p4fig2}) are listed in Table \ref{tab:p4tabcomponents}, adjusted for the primary beam effects as outlined above.

The inherent positional uncertainty in both the VLA and VLBA images is approximately 50 mas.  The nucleus in the VLBA image is manually aligned with the VLA coordinates of the nucleus, which have approximately 50 mas uncertainty \citep{Perley:1997p6689}.  Although the positional accuracy of the VLA image of the hot spot at 15 GHz is not stated in \citet{Perley:1997p6689}, since the hot spot image was referenced to the same calibrator and self-calibrated, presumably the positional accuracy of VLA image of the hot spot is also of order 50 mas.  The alignment between the VLA and VLBA images is therefore probably good at the 100 mas level. 

\begin{table}[ht]
\begin{center}
{ \tiny
\begin{tabular}{lccccccccc} \hline \hline
          &        &         &                   &         &   \multicolumn{3}{c}{FWHM Size}    &   \multicolumn{2}{c}{FWHM Size}    \\
Component & R.A.   & Decl.   & $S_{P}$           & $S_{I}$ & major & minor & P.A.  & major  & minor  \\
          &        &         & (mJy beam$^{-1}$) & (mJy)   & (mas) & (mas) & (deg) & (pc)   & (pc)   \\ [0.5ex] \hline \hline
		A   &   05 19 26.2291   &   -45 45 52.993   &   4.3$\pm$0.4   &   28.3$\pm$2.8   &   247   &   62   &   72   &   170   &   42   \\
		B   &   05 19 26.1986   &   -45 45 52.967   &   6.3$\pm$0.6   &   25.9$\pm$2.6   &   126   &   67   &   -113   &   87   &   46   \\
		C   &   05 19 26.2075   &   -45 45 53.159   &   5.0$\pm$0.5   &   24.9$\pm$2.5   &   185   &   62   &   -100   &   127   &   43   \\
		D   &   05 19 26.2392   &   -45 45 53.352   &   1.9$\pm$0.2   &   6.0$\pm$0.6   &   201   &   40   &   -103   &   138   &   28   \\
		E   &   05 19 26.2036   &   -45 45 53.751   &   5.7$\pm$0.6   &   36.2$\pm$3.6   &   178   &   85   &   67   &   122   &   58   \\ \hline
\end{tabular}
\caption{Parameters of components in VLBA image of Pictor A NW hot spot.}
\label{tab:p4tabcomponents}
}
\end{center}
\end{table}

The positional uncertainties, combined with the difference in angular resolution and observing frequency between the VLA data and the VLBA data, make a detailed registration of the two data-sets difficult.  The VLBI components trace the bright part of the VLA image but no one component marks the peak of the brightness distribution in the VLA image.  The brightest of the VLBA components lies furthest from the core, beyond the peak in the VLA image, in the jet direction.  It appears that the structure seen at lower resolution with the VLA may therefore be a combination of the compact components seen with the VLBA, combined with diffuse emission to which the VLBA observations are not sensitive, since the sum of the flux densities of the components seen with the VLBA are a small fraction of the VLA flux density in the hot spot at the same frequency.  The integrated flux density in the VLBA image is approximately 120 mJy, whereas the integrated flux density at 1.7 GHz within the 1\arcsec~ region encompassing the emission seen with the VLBA is close to 8 Jy (based on an interpolation of the radio data in \citet{Perley:1997p6689}).  The unseen diffuse emission must trace the compact components however, as the compact components follow the higher contours of the VLA image.  This implies that the emission seen with the VLBA represents the high brightness temperature $``$cores" of the hot spot sub-components.

\begin{figure}[ht]
\center
\epsscale{0.8}
\plotone{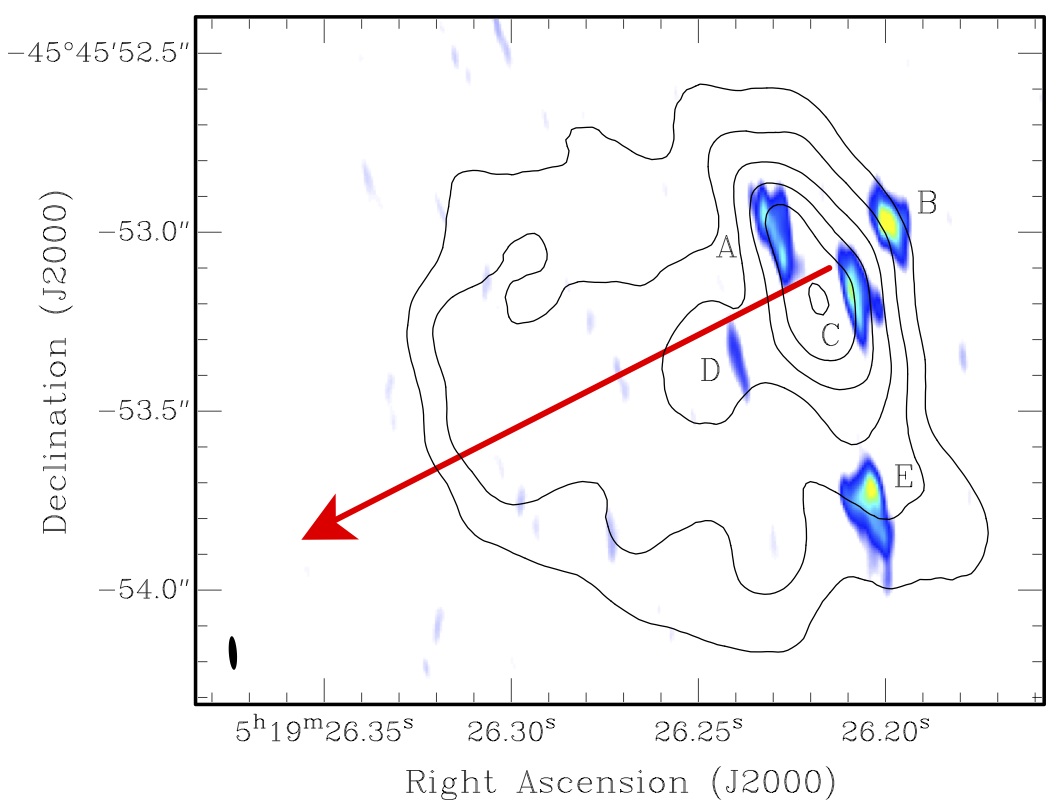}
\caption[\emph{HST} and VLBA overlay of Pictor A north-west hot spot]{\emph{HST} (contours) and VLBA (pseudo-colour image) hot spot overlay.  The red arrow indicates the direction to the nucleus as in Figure \ref{fig:p4fig3}.  Contours are at 28, 42, 57, 71, \& 85\% of the peak intensity (which has not been calibrated).}
\label{fig:p4fig4}
\end{figure}

Figure \ref{fig:p4fig4} shows a comparison between the VLBA data and the optical data of \citet{Thomson:1995p552}.    The raw optical image was retrieved from the \emph{HST} archive and a Gaussian smoothing function (with a kernel radius of 10) was applied to the image using the SAOImage DS9 \citep{Joye:2003p10892}, an astronomical imaging and data visualisation application, before comparing to the VLBA image.  The peak of the optical image (uncalibrated intensity) was aligned with the peak of the 15 GHz VLA image of \citet{Perley:1997p6689}, in order to register against the VLBA data.  Again, significant positional uncertainties exist between the \emph{HST} image and the VLBA image, at least of order 100 mas, due to the alignment of the peaks in the \emph{HST} image and the VLA 15 GHz image.  Likely this \emph{HST}-VLBA registration is significantly more uncertain than at the 100 mas level because of the frequency difference between the \emph{HST} and VLA 15 GHz images.

\subsection{SPITZER observations}

The north-west hot spot in Pictor A was observed by Spitzer with IRAC (Infrared Array Camera) and MIPS (Multiband Imaging Photometer for SIRTF).  The IRAC observations were made on 2004 November 26 with a total time of 48 s. The MIPS (24 $\mu$m only) observations were made on 2004 September 21 with an observation time of 100 s. The images were produced using the data analysis tool MOPEX \citep{Makovoz:2006p24713}.  The IRAC and MIPS fluxes were obtained from circular regions with radius of 3.6 arcsec and 6 arcsec, respectively, centred on the radio peak of the hot spot. These values were corrected applying the appropriate aperture correction for the adopted extraction region.  Table \ref{tab:p4ir} lists the Spitzer data.

\begin{table}[ht]
\begin{center}
{ \scriptsize
\begin{tabular}{ccc} \hline \hline
Log$_{10}$($\nu$) & Log$_{10}$(flux erg/s/Hz) & Error \%  \\ [0.5ex] \hline \hline
13.10     &      -25.08       &          10 \\
13.57      &     -25.46       &           5 \\
13.71      &     -25.60       &           7 \\
13.82     &      -25.72       &           5 \\
13.92      &     -25.89       &           5 \\ \hline
\end{tabular}
\caption{Spitzer infrared data for Pictor A north-west hot spot.}
\label{tab:p4ir}
}
\end{center}
\end{table}

\section{Discussion}
\label{sec:radiojets.discussion}

We have detected only the north-west hot spot in Pictor A at radio wavelengths with the VLBA, thus this hot spot contains higher brightness temperature components than the south-east hot spot.  This is consistent with the ideas put forward by \citet{Georganopoulos:2003p10756}, that radio galaxies with somewhat aligned jets (they count Pictor A among these objects) have brighter hot spots on the side of the approaching jet, due to Doppler boosting of the emission (assuming that the hot spots are identical).  The north-west hot spot brightness temperature may therefore be enhanced by Doppler boosting compared to the south-east hot spot, making it easier to detect with VLBI.  This is consistent with the jet sidedness as seen in VLBI images of the AGN core \citep{Tingay:2000p6116} and the \emph{Chandra} X-ray data of \citet{Wilson:2001p536}.  Alternatively or additionally, the south-east hot spot of Pictor A has a significantly different morphology to the north-west hot spot, a double hot spot morphology similar to structures observed in a number of other FR-II radio galaxies e.g. \citet{Lonsdale:1989p10912}.  In the south-east hot spot, the jet may release energy in multiple weaker interactions, rather than in a single strong interaction as appears in the north-west hot spot.  The nature of the south-east hot spot is briefly discussed below.

\subsection{Implications for X-ray emission models}

\subsubsection{Previous models - balanced synchrotron and SSC emission}

In their exploration of various models for the X-ray emission from the north-west hot spot in Pictor A, \citet{Wilson:2001p536} consistently assume a radius for the emission region of 250 pc = 0.4\arcsec, based on the VLA data.  As is usual in such models, the assumption is made of a homogenous, spherical emission region, which does not appear to be a bad assumption based on the VLA data, alone.  The VLBI data, however, show the danger in such an assumption, revealing significant structure on scales more than an order of magnitude smaller than the assumed size.  Further modelling is therefore required to assess the X-ray emission from a combination of compact and diffuse structures, to determine deviations from the models presented in \citet{Wilson:2001p536}, and we explore these models in detail below.  It is worth noting that, although a relatively small percentage of the integrated flux density of the hot spot is contained in the VLBI components, significant structure is likely to be present on spatial scales that fall between the VLA and VLBA resolutions, to which neither array is sensitive.  It is therefore possible that the majority of the flux density of the hot spot may be contained in sub-components that are a factor of 2 or 3 smaller than assumed by \citet{Wilson:2001p536}. Given the resolved nature of the structures we detect with the VLBA, it is unlikely that a significant portion of the detected flux density contained in these structures exists in 
unresolved components. We estimate conservatively that less than 10\% of the detected flux density could be contained in unresolved components. Observations on longer and more sensitive VLBI baselines would be required to quantify this statement further.

\citet{Wilson:2001p536} note that the X-rays could be due to synchrotron radiation and find that their synchrotron model is consistent with particle acceleration in strong shocks, the electrons requiring re-acceleration in shocks on pc scales.  The components seen in our VLBI images are on the right spatial scales to be consistent with this picture.

Therefore, both the possibilities of a smaller emission region for the X-rays, and the generation of synchrotron X-rays could be explored further in models for the north-west Pictor A hot spot.  Interestingly, \citet{Wilson:2001p536} find that a composite synchrotron plus synchrotron self-Compton (SSC) model can match the \emph{Chandra} observations, but requires similar contributions from both processes in the \emph{Chandra} band.

However, we believe that these models are not strongly predictive, as \citet{Wilson:2001p536} model the synchrotron peak as being due to a broken power-law electron distribution, giving the synchrotron tail required in the \emph{Chandra} band, with the SSC peak derived from an electron energy distribution that cuts off sharply at a maximum energy.  The addition of the synchrotron and SSC peaks is therefore based on two different electron energy distribution models that are qualitatively similar, but strictly inconsistent.

\subsubsection{New models - synchrotron dominated emission}

\begin{figure}[p]
\center
\epsscale{1.0}
\plotone{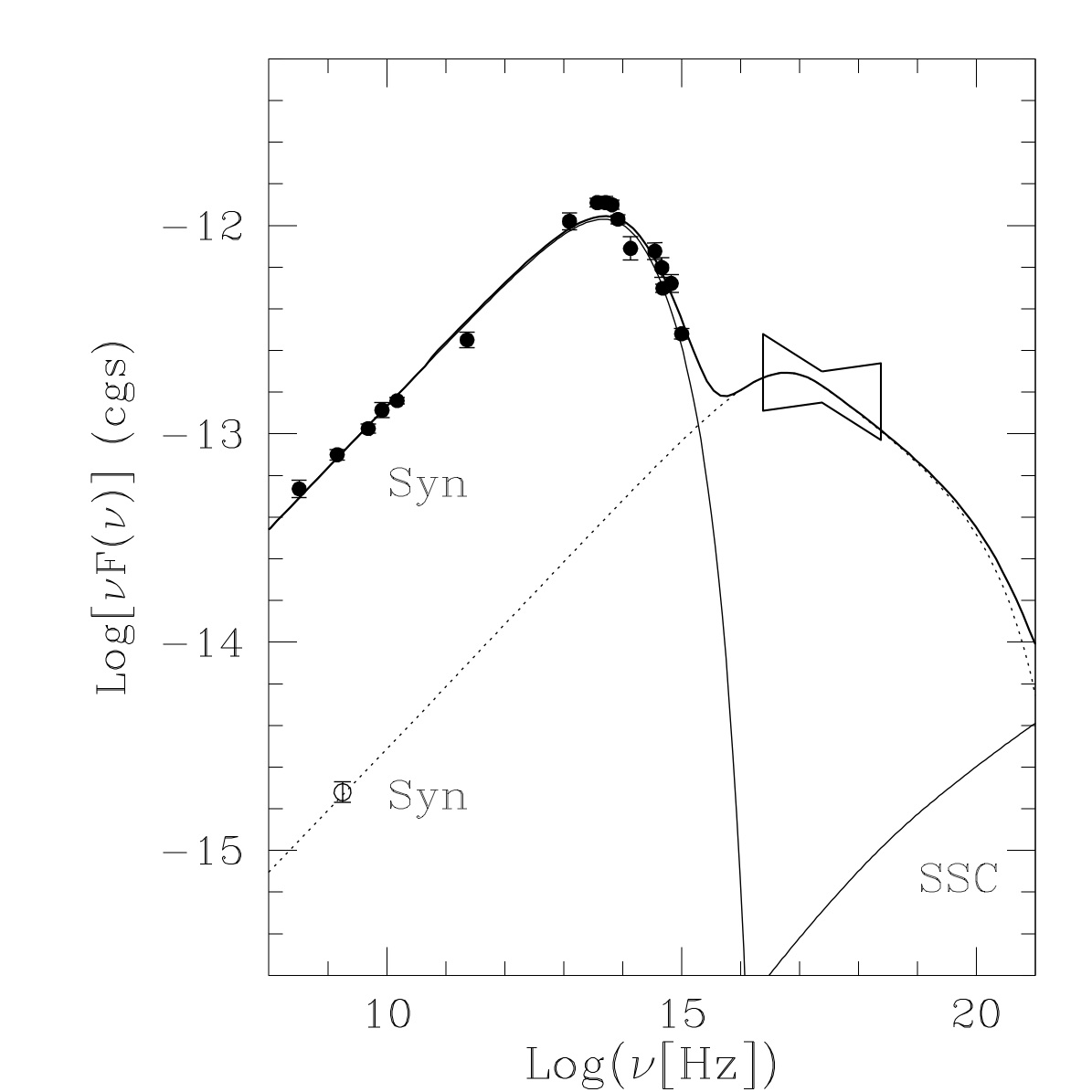}
\caption[Expected synchrotron and inverse Compton spectrum for north-west hot spot of Pictor A]{The expected synchrotron and inverse Compton spectrum for north-west hot spot of Pictor A is compared with the radio, IR, optical and X-ray data.  The synchrotron spectrum is emitted by two components : diffuse hot spot (thin solid line) and pc--scale VLBA regions (dotted line).  The SSC components (thin solid line) is contributed from the hot spot and pc--scale VLBA regions. SSC calculations are reported assuming a reference value of $B=350~\mu$G in both hot spot and compact VLBA regions and adopting the measured size of the components.  The total emission is reported in thick--solid line.}
\label{fig:p4fig5}
\end{figure}

With this in mind, and the new information available from the VLBA observations, we have endeavoured to find a more natural explanation for the radio to X-ray spectrum for the Pictor A hot spot.  The pc--scale structures seen in the hot spot are compact regions with energy density 2--3 times larger than that of the (average) hot spot, thus they may trace the region where electrons are accelerated very recently via shocks and turbulence.  These should be transient regions that are expected to expand reaching pressure equilibrium with the surrounding plasma. Their maximum dynamical time--scale can be estimated by the Alfvenic crossing time or by the time required (with an advance speed $\approx$ 0.05--0.1 $c$) to cross these regions, and this comes out to be between $\sim$5 years and a few decades.  This is much smaller than the radiative time scale of the hot spot that is about 100-700 years, assuming a break frequency of the synchrotron spectrum, $\nu_b \approx 10^{14}$ Hz (resulting from the frequency of the peak of the synchrotron emission originating in the diffuse region surrounding the pc--scale hot spot structures: Figure \ref{fig:p4fig5}), and a magnetic field in the range $100-400~\mu$G (this range being consistent with the expected equipartition magnetic field strength).  Thus the frequency of the radiative break in the synchrotron spectrum of the dynamically young regions discovered by our VLBA observations, that scales with $\nu_b \propto B^{-3} \tau^{-2}$, is expected to be $\approx$100--10000 times higher than that of the overall hot spot.  Therefore the synchrotron emission contribution from the pc--scale components would extend into the \emph{Chandra} band.

Based on these considerations we adopt a model of the hot spot region assuming two emitting components : the diffuse contribution to the hot spot, that dominates in the radio and optical band, and the pc--scale components embedded within the diffuse emission.  As an approximation, each component (diffuse and each pc--scale component) is described by a homogenous sphere with constant magnetic field and fixed properties of the relativistic electrons.  We model the spectral energy distributions of the emitting electrons in both the hot spot and the pc--scale components by means of the formalism described in \citet{Brunetti:2002p24680} that accounts for the acceleration of electrons at shocks and for the effect due to synchrotron and inverse Compton losses on the shape of the spectrum.

First we fit the synchrotron emission of the diffuse emission using the radio, IR and optical data points (using same data as used by \citet{Wilson:2001p536}, plus our new Spitzer data) and derive the relevant parameters of the synchrotron spectrum (injection spectrum $\alpha$, break frequency $\nu_b$, cut-off $\nu_c$) and the slope of the spectrum of the electrons as injected at the shock ($\delta = 2 \alpha +1$).  For a given value of the magnetic field strength this allows us to fix the spectrum of the emitting electrons (normalization, break and cut-off energy), and to calculate synchrotron self-Compton (SSC) emission from the hot spot region, following \citet{Brunetti:2002p24680}.

The synchrotron spectrum of the pc--scale components is normalised to the fluxes at 18 cm derived from our VLBA observations assuming the same $\alpha$ measured for the emission from the diffuse region and a synchrotron break at higher frequencies.  For a given value of the magnetic field strength in the region of the pc--scale components the spectrum of the electrons is fixed and the inverse Compton emission from these electrons is calculated by taking into account both SSC and the scattering of the external radio photons from the diffuse region in which the pc--scale components are embedded.

In Figure \ref{fig:p4fig5} we show the radio to hard X-ray emission expected from our modelling of the hot spot region of Pictor A.  Synchrotron emission is responsible for the observed properties of the hot spot, while the inverse Compton emission is expected to give appreciable contribution only at very high energies.  In Figure \ref{fig:p4fig5} the SSC is calculated assuming a reference value B=350 $\mu$G in both the diffuse and pc--scale components\footnote{this value is compatible (within a factor of 2) with the equipartition field in both these regions calculated with equipartition formulae given in \citet{Brunetti:1997p24676} with $\gamma_{min}=100$ and F=1.} with the two components giving a similar contribution to the total SSC spectrum.

As noted in \citet{Wilson:2001p536}, an assumed magnetic field an order of magnitude lower than the equipartition field is required to reproduce the observed X-ray luminosity, but fails to reproduce the X-ray spectral slope.

Thus, our new observations have given a basis for a synchrotron X-ray component which, when self-consistently calculated along with a SSC X-ray component, shows that the \emph{Chandra} band is dominated by synchrotron emission from the pc--scale components, with a relatively small SSC contribution.  The implication of this is that at energies higher than observed with \emph{Chandra}, the new model predictions stand in stark contrast to those of \citet{Wilson:2001p536}, providing a clear test of the two models in the future.

The new synchrotron-dominated model provides a much more natural explanation for the Pictor A hot spot radio to X-ray emission, taking into account the new VLBA observations of significant pc--scale internal structure.  The model calculation is also fully self-consistent, compared to the previous models, so we believe that there is good reason to suppose that our analysis reflects closely on the true physical situation for the hot spots.  Further higher energy observations with good angular resolution and sensitivity are required to ultimately discriminate between the very different predictions of our model and those of \citet{Wilson:2001p536}.

\subsubsection{Other considerations}

\citet{Georganopoulos:2003p10756} raise the possibility that the bulk flow of the material in radio galaxy hot spots, in the lab frame, may be at least mildly relativistic and therefore allow inverse Compton scattering of the cosmic microwave background as a viable model for the X-ray emission.  The post-shock flow may be relativistic \citep{Georganopoulos:2003p10756} and the hot spot advance into the IGM may be mildly relativistic (\citet{Wilson:2001p536}; \citet{Georganopoulos:2003p10756}).  Of particular interest, therefore, is any measurement of the hot spot advance speed in Pictor A, which could strengthen or weaken this argument.  The mismatch in resolution between the VLA and VLBA images in Figure \ref{fig:p4fig3}, and the sum of the positional uncertainties in both images, do not allow any meaningful constraint on the hot spot advance speed.  Further VLBI observations are required at much later epochs, to determine any hot spot advance.  \citet{Arshakian:2000p4487} find an average value of $v_{adv} \sim 0.1c$ for powerful radio galaxies.  Given this value at the distance of Pictor A, $\sim$0.05 mas/yr could be expected for the hot spot advance.  Even over a 20 year period, only of order a milli-arcsecond of difference in angular position could be expected.  At the resolution and sensitivity of current VLBI observations, such a detection of motion would be very difficult.  Any changes in the $\sim$15 years between the VLA and VLBA observations are therefore insignificant compared to the resolution of either observation.

\subsection{Terminal shock morphology and comparison to 3C 205, PKS 2153$-$69, and 4C 41.17}

How does the distribution of presumably shocked jet material seen in the VLBA image of the Pictor A hot spot relate to the termination of the jet at the IGM?  The $``$dentist drill" scenario explored by \citet{Scheuer:1982p10884} and \citet{Cox:1991p10746} means that the average jet power released in the jet termination is spread over a larger area than the cross-section area of the jet.  Although there would only be a single termination shock at any one time, the locations of previous termination shocks would also be sites of radio emission, until their electron populations lost enough energy to be undetectable at radio wavelengths.  However, the $``$dentist drill" model is used to explain hot spot morphologies on much larger scales ($\sim$kpc) than probed with our VLBA data.  Additionally, the $``$dentist drill" model, as simulated by \citet{Cox:1991p10746} was caused by a substantial precession of the jet ($5-10\arcdeg$ half-angle variations of the jet direction), manually introduced into the simulation.  No such precession of the north-west jet in Pictor A can be seen.  The VLBI, VLA, and \emph{Chandra} data only indicate a very straight jet that terminates at the hot spot.

A small scale analog of the $``$dentist drill" effect may be possible.  As the straight jet approaches the termination point, traversing the relatively high density backflow from the termination point, the direction of the jet may be altered, causing the termination point to vary over a working surface larger than the cross-section of the jet.  Assuming that this is the case, and each of the components seen in Figure \ref{fig:p4fig2} are current or previous interaction sites, it implies that the jet width is of order 100 mas $=$ 70 pc, the observed size of the components in the VLBI image.  Optical observations of the hot spots in 3C 445 showing clumps of emission proposed as ``local accelerators" could also be interpreted under a small-scale dentist drill scenario \citep{Prieto:2002p24132}. A 70 pc jet width at the hot spot implies a very small jet opening angle, $<0.2\arcdeg$.

Alternatively, the shocks may simply be the highest brightness temperature regions of an extended termination shock front, which is spread over a larger area than indicated by each of the compact components.  In this case, the spread of components may be a better indication of the cross-section of the jet at the hot spot, 1000 mas $=$ 700 pc.  In this case, the distribution of compact components may indicate the structure of the IGM that the jet is interacting most strongly with.

Comparisons of the VLBA image to both the VLA image and the \emph{HST} image of the hot spot shows that the compact structures are distributed in a symmetric fashion around the jet direction.  The direction of the nucleus at the hot spot is indicated in both Figures \ref{fig:p4fig3} and \ref{fig:p4fig4} and the two alternatives discussed above cannot be distinguished on the basis of the distribution of the compact emission.

We note, in comparison, that the overall size of the hot spot detected with VLBI in 3C 205, by \citet{Lonsdale:1998p176}, is 1400 pc, although the morphology is quite different to that seen in Pictor A.  The brightest, most compact part of the 3C 205 hot spot (Component 3 in the \citet{Lonsdale:1998p176} nomenclature) appears to be of order 250 pc in size and is much more luminous than the hot spot in Pictor A.  The conclusion that \citet{Lonsdale:1998p176} apply to the 3C 205 data, of a continuously bending jet in reaction to the jet interaction with a dense IGM, does not seem to immediately apply to the Pictor A data.  There is no clear connection between the components in Figure \ref{fig:p4fig2} that would indicate a flow of material from one component to another.  In this sense, the small-scale $``$dentist drill" model may be more plausible.

In 4C 41.17, the observed compact structure in the hot spot at the suggested site of a jet interaction has a size of $\sim$110 pc, comparable to that seen for the individual components in the VLBA image of the Pictor A hot spot.

The only other well observed example of a radio galaxy hot spot at VLBI resolution is PKS 2153$-$69 \citep{Young:2005p5449}.  In PKS 2153$-$69, the overall hot spot size is approximately 200 pc, containing structures as small as 50 pc in extent, comparable to the size of the structures seen in 3C 205 and Pictor A.   Interestingly, in PKS 2153$-$69, the hot spot structure on scales of $\sim$5 kpc shows evidence for a similar morphology as seen in 3C 205, a curved structure with a strong surface brightness gradient.  In PKS 2153$-$69, the overall radio morphology, along with X-ray and optical data are interpreted to support a model of jet precession, where the curved radio structure in the hot spot traces the varying position of the jet interaction region with time, rather than an {\it in situ} bend of the flow due to the jet interaction.  This is the type of radio galaxy structure that the classical $``$dentist drill" model of \citet{Scheuer:1982p10884} seeks to explain.  Again, the current data for Pictor A do not seem to easily fit such an interpretation.

\subsection{Comparison to jet simulations}

\citet{Saxton:2002p557} present two-dimensional, axisymetric, non-relativistic, hydrodynamic simulations of the north-west Pictor A hot spot, in an effort to reproduce the distinctive hot spot structure on arcsecond scales (in particular the filament behind the hot spot that appears perpendicular to the jet direction).  Although these simulations are simplistic relative to the physical situation in Pictor A, they show that the hot spot advance proceeds in an episodic and ephemeral manner, driven by the formation of a channel or cavity evacuated by the jet, the temporary frustration of the jet as the channel periodically closes, and a rapid advance of the hot spot as it breaks through the temporary obstacle, traversing the channel rapidly and reforming the working surface at the IGM.  

The simulations show that a rich variety of complex structures are possible at the terminal shock of the jet, although the simulation resolution does not reach the equivalent of the mas scales of Figure \ref{fig:p4fig2}.  The estimated time-scales for these predicted temporal variations in the hot spot position ($10^{4} - 10^{5}$ years, depending on the simulation) are much longer than the period between the VLA and VLBA images for Pictor A and the observations would not be expected to detect these changes in the hot spot position with time.

\subsection{The south-east hot spot in Pictor A}

Although the south-east hot spot in Pictor A was not detected with our VLBA observations, it provides an interesting contrast to the north-west hot spot.  The south-east hot spot appears to be a double hot spot, as noted in a number of other radio galaxies.

The classical $``$dentist drill" model, as simulated by \citet{Cox:1991p10746} invokes jet precession to provide position angle changes of the jet, allowing the termination point of the jet to vary across an area larger than the width of the jet.  In Pictor A, there is no evidence for jet precession in the north-west jet, and therefore not likely to be any precession of the south-east jet which feeds the south-east lobe.  We therefore find that the $``$dentist drill" model is not likely to explain the morphology of the south-east lobe.  

The jet bending hypothesis put forward by \citet{Lonsdale:1998p176} to explain the double hot spot in 3C 205 is supported by the evidence that the jet from the nucleus in this object connects with a curved, bright hot spot that subsequently appears to feed a weaker hot spot further from the nucleus.

\begin{figure}[ht]
\center
\epsscale{1.0}
\plotone{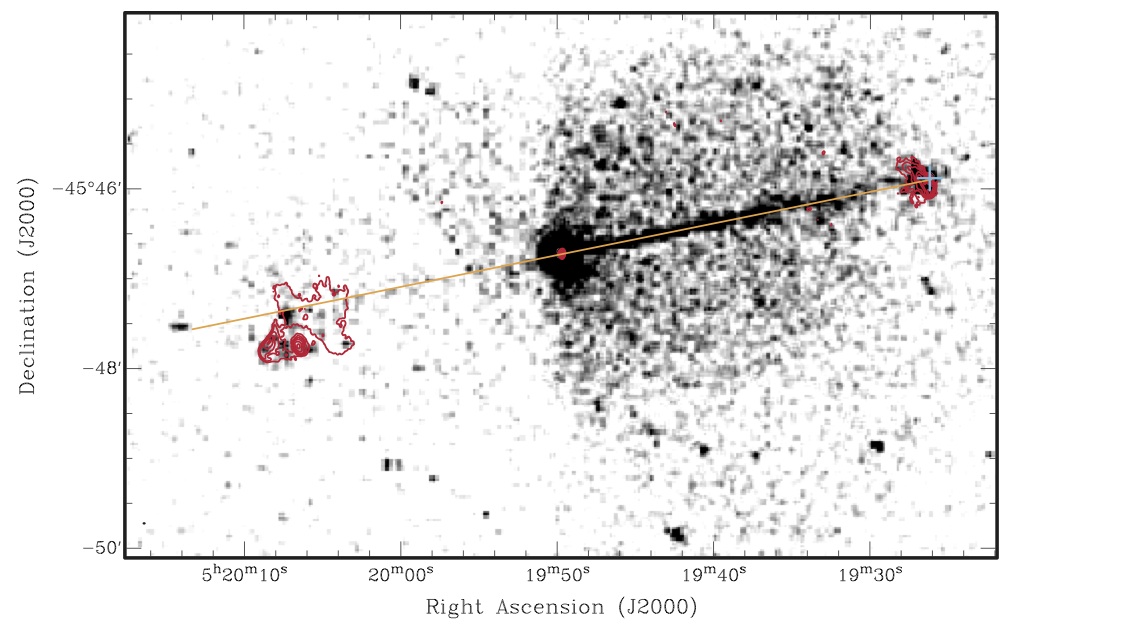}
\caption[\emph{Chandra} and ATCA overlay of Pictor A north-west hot spot]{Overlay of \emph{Chandra} archival data (gray scale) and ATCA data (red contours).  Indicated on the overlay is the jet direction as defined by the jet seen at X-ray wavelengths and the VLBI data of \citet{Tingay:2000p6116}, which agree well and both point to the structures seen with the VLBA (indicated by the blue cross).  The jet direction is extrapolated to the south-east, to show the relationship of the south-east hot spot to the expected jet direction, revealing an offset in position angle, as discussed in the text.}
\label{fig:p4fig6}
\end{figure}

In Pictor A, such a situation is not apparent.  Figure \ref{fig:p4fig6} shows an overlay of radio contours from our ATCA image with co-added images from the \emph{Chandra} archives (also published in \citet{Hardcastle:2005p534}).  The yellow line in the figure indicates the direction of the X-ray and nuclear VLBI jets, which shows they are highly aligned with the north-west hot spot that we have detected with VLBI.  This jet direction, extrapolated to the counterjet side, is misaligned with the south-east double hot spot, by approximately 20$^{\circ}$.  There is no point along this extrapolated jet direction that provides obvious evidence of an interaction that would bend the jet and deflect it to the location of the observed double hot spots.  A relatively strong X-ray source appears aligned with the extrapolated jet direction, immediately before the double hot spots, however, there is no radio counterpart to the X-ray component (c.f. Figure \ref{fig:p4fig3}).  The observations therefore do not obviously support the jet bending hypothesis \citet{Lonsdale:1998p176} used for 3C 205.

The nature of the south-east hot spots in Pictor A, and the apparent misalignment with the jet, relative to the north-west jet and hot spot is therefore an interesting aspect of this galaxy, requiring further observation and interpretation, in particular very deep radio and X-ray imaging in order to trace the jet on its path to the south-east hot spots.  The \emph{Chandra} co-added image shown in Figure \ref{fig:p4fig6} is more sensitive on the north-west side of the galaxy than on the south-east, as this is where strong emission was known to exist and the observations concentrated.  Further \emph{Chandra} observations of the region of the south-east jet, hot spots, and lobes would be very interesting.

\section{Conclusions}

Of the four radio galaxy hot spots that have been imaged in detail using VLBI techniques, all have similar gross properties in that they possess structure on scales between 20 and 100 pc, despite being at redshifts of 0.028, 0.035, 1.534, and 3.8.  The details of the structures observed in the different hot spots are rather different, however, and different interpretations have been applied to these structures, from the bent jet model of \citet{Lonsdale:1998p176} to the precessing jet model of \citet{Young:2005p5449}, and the jet interaction conclusion of \citet{Gurvits:1997p10768}.  A small-scale analog of the $``$dentist drill" model may be a viable scenario for the north-west hot spot Pictor A, but other scenarios are also possible.  

The detection of parsec-scale structures in radio galaxy hot spots has implications for the models used to explain the X-ray emission from the hot spots.  In particular we find that a natural explanation for the radio to X-ray spectrum is that the pc--scale components in the hot spot represent recently accelerated regions of electrons, with higher break frequencies than the remainder of the hot spot electrons, on larger spatial scales.  A consequence of this is that the X-rays in the \emph{Chandra} band are dominated by synchrotron X-rays, with a relatively weak SSC contribution.  This is in stark contrast to the previous modelling of \citet{Wilson:2001p536}, who predict a balanced SSC and synchrotron emission in the \emph{Chandra} band and strong SSC emission above the \emph{Chandra} band.  Future high energy observations may readily distinguish between these models.

The south-east hot spot has a substantially different morphology to the north-west hot spot, which may have to do with the apparent misalignment between the jet and the south-east hot spot location.  Further observations, particularly at X-ray wavelengths would help to determine the nature of the south-east hot spots.

Obviously, further systematic VLBI observations of hot spots in other nearby powerful radio galaxies would be highly useful, to determine if similar high resolution structures exist in those sources.  FR-II radio galaxies at low redshift, where high spatial resolution can potentially be obtained, are rare.  We have therefore embarked on a program to observe the hot spots in a number of powerful radio galaxies with $z<0.1$ in the Southern Hemisphere.  

A further potentially interesting high angular resolution observation of the Pictor A hot spot has been proposed by \citet{Wilson:2001p536}.  Their modelling hints at the presence of a low energy electron population in the hot spot that is not visible at cm wavelengths.  To detect such a population, low frequency observations with high resolution would be required.  At some point in the future, the Murchison Widefield Array (MWA), being built by an international consortium at the candidate Square Kilometre Array (SKA) site in Western Australia, may be able to make such an observation.  Also possible would be to observe the Pictor A hot spots with the 90 cm system on the VLBA, to obtain very high resolution at long wavelengths.  Although too far north to observe Pictor A, the eLOFAR array will have good angular resolution at very low radio frequencies and will be able to observe the hot spots of many Northern Hemisphere radio galaxies to search for low energy electron populations, such as suggested by \citet{Wilson:2001p536} for Pictor A.

\linespread{1.0}
\normalsize
\begin{savequote}[20pc]
\sffamily
The Universe is full of magical things,\\
patiently waiting for our wits to grow sharper.
\qauthor{Eden Philpotts}
\end{savequote}

\chapter{Radio Jet Interactions in PKS $0344-345$ and PKS $0521-365$}
\label{chap:jets}
\small
Wide-field, very long baseline interferometry (VLBI) observations of jet interactions in the galaxy PKS $0344-345$ ($z=0.0538$), obtained with the Australian Long Baseline Array (LBA), have not detected any emission associated with the jet termination hot spot at 2.3 GHz. Low resolution ATCA radio images of the interaction region suggest that the region is extended ($\sim6.8$ kpc) and has a spectral index of $\alpha=-0.84$ that is comparable to a similar interaction region in PKS $2152-699$. A 1-KeV X-ray flux density of $\sim8$ nJy is predicted for this interaction region based on a simple power-law model for the spectral energy distribution. Similar observations of PKS $0521-365$ ($z=0.05534$) reveal two components associated with the south-east hot spot of the galaxy at 1.6 GHz and 2.3 GHz. A simple power-law, with spectral index $\alpha=-1.06$, provides a good fit between our VLBI data and \emph{Chandra} X-ray data. The multi-component nature, overall extent and size of individual components of the hot spot is similar to that observed in Pictor A ($z=0.035$). The similarities suggest that the same emission mechanisms may be at work in the two galaxies, however, further multi-wavelength observations will be required to verify this.

\clearpage
\linespread{1.3}
\normalsize
\section{Introduction}
\label{sec:jets.introduction}

\subsection{PKS $0344-345$}

\begin{figure}
\epsscale{1.0}
\plotone{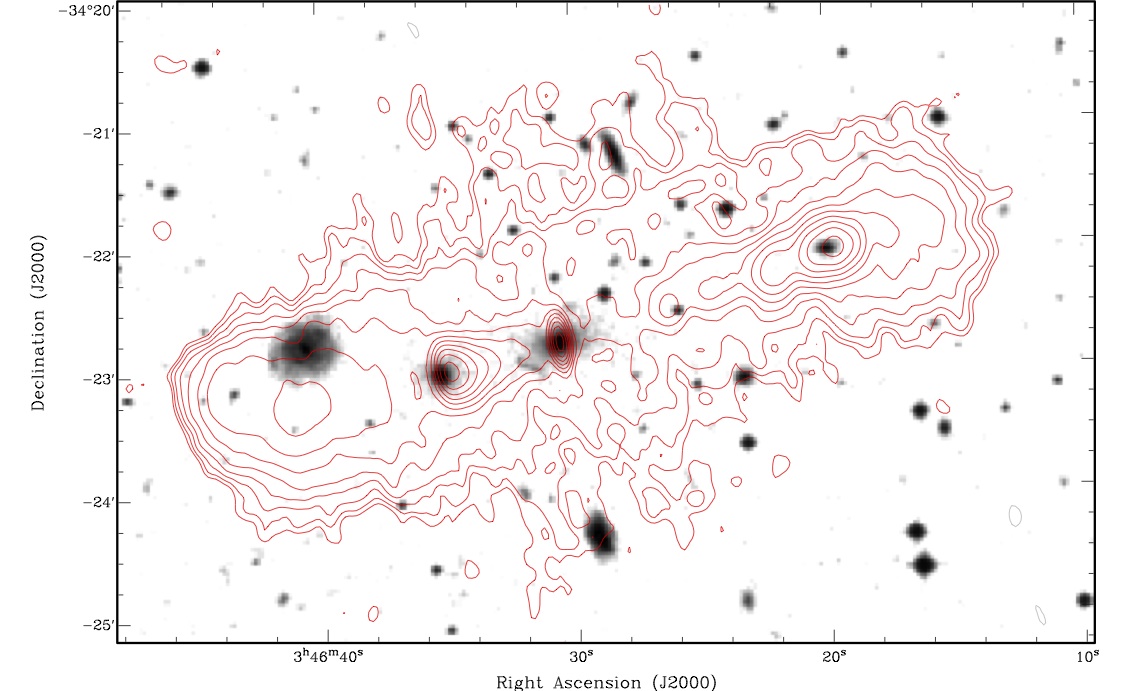}
\caption[ATCA 1.384 GHz image of PKS $0344-345$ overlaid with the Digitised Sky Survey 2 optical image.]{ATCA 1.384 GHz image of PKS $0344-345$ overlaid with the Digitised Sky Survey 2 optical blue image.  Contours are drawn at $\pm2^{\frac{1}{2}}, \pm2^{1}, \pm2^{\frac{3}{2}}, \cdots$ times the $3\sigma$ rms noise. Restoring beam is $15.9\arcsec\times6.1\arcsec$ with a PA of $9\arcdeg$ and rms image noise is 0.35 mJy beam$^{-1}$.}
\label{fig:fig0344ovl}
\end{figure}

PKS $0344-345$ is a radio galaxy which is a member of a low redshift (z=0.05380) group \citep{Jones:1992p681}.  The radio structure was imaged with the VLA by \citet{Ekers:1989p17319} at a frequency of 1.5 GHz, revealing a complex source with at least three extended components and diffuse bridge emission.  Our 1.4 GHz and 2.5 GHz ATCA images, shown in Figures \ref{fig:figself3}(s) and \ref{fig:figself3}(t), reveal similar structure with an interaction region approximately $55\arcsec$ west of the core component and diffuse lobe emission to the east and west. When overlaid with a digital sky survey image (Figure \ref{fig:fig0344ovl}) it can be seen that the eastern jet emerges from the host galaxy and impacts another nearby galaxy in the group.  At the site of interaction the jet is disrupted, forming a bright and compact radio hot spot.  The western jet from the nucleus is unimpeded on its way to the western lobe.  At the western lobe, there is evidence of optical emission from the lobe.  There is also evidence from the optical image of diffuse emission from the jet, close to the jet interaction.

The lobes each have a compact hot spot that is resolved in higher resolution ATCA images at 4.8 GHz (Figure \ref{fig:figsel0344}(a)) and completely resolved at 8.64 GHz and 18 GHz (Figures \ref{fig:figsel0344}(b) and \ref{fig:figsel0344}(c), respectively). The interaction region is detected at 18 GHz but is significantly resolved with a FWHM size of approximately $9.8\arcsec\times6.2\arcsec$ and a total flux density of 17 mJy.

\subsection{PKS $0521-365$}

PKS $0521-365$ has a redshift of z=0.05534 \citep{Keel:1986p10557}, placing it at a distance of approximately 240 Mpc. It has a prominent radio jet \citep{Danziger:1979p26828,Keel:1986p10557,Macchetto:1991p26838,Falomo:1994p26842} which like M87 is also detectable at optical wavelengths; the detection of an optical jet being rare in such objects \citep[and references therein]{ODea:1999p26847,Scarpa:1999p17053}. VLA images at 5 GHz reveal a knot in the jet approximately $2\arcsec$ north-west from the nucleus \citep{Keel:1986p10557} which appears resolved at 15 GHz and is perpendicular to the jet suggesting an internal shock.

Although no counter-jet has been observed, a strong hot spot with complex structure in both intensity and polarisation appears $8\arcsec$ south-east of the core on the opposite side of the jet and is resolved at 15 GHz. \emph{Chandra} observations of the south-east hot spot detected emission believed to arise from synchrotron self-Compton emission from a near-equipartition plasma \citep{Birkinshaw:2002p4973}. \emph{HST} observations of PKS $0521-365$ detected the core and jet but found no emission associated with the south-east hot spot \citep{Scarpa:1999p17053}.

\citet{Tingay:1996p5577} observed PKS $0521-365$ over three epochs at 4.8 GHz and one observation at 8.4 GHz using the SHEVE (Southern Hemisphere VLBI Experiment) array of telescopes and found no apparent motion of a jet component relative to the core over a period of almost a year. This together with lower limits on the jet-to-counterjet surface brightness ratio are consistent with mild relativistic speeds and moderate beaming. Multi-epoch observations with the VLBA and VOSP, \citet{Tingay:2002p5468} found no evidence of super-luminal motion but found evidence for evolution internal to a component in the parsec-scale jet.

Our recent J-Band observations with the Magellan telescope, Figure \ref{fig:fig0521}, reveal the core, jet, a knot associated with the jet and features associated with both hot spots.

\section{Observations, data reduction, and results}
\label{sec:jets.reduction}

\subsection{PKS $0344-345$ Observation}
VLBI observations of PKS $0344-345$ were made on 16/17 May 2006, using a number of the Long Baseline Array (LBA) telescopes : the 64 m antenna of the Australia Telescope National Facility (ATNF) near Parkes (8 hours); the ATNF Australia Telescope Compact Array (ATCA) near Narrabri (12 hours); the ATNF Mopra 22 m antenna near Coonabarabran (12 hours); the University of Tasmania's 26 m antenna near Hobart (12 hours); and the University of Tasmania's 30 m antenna near Ceduna (12 hours). For these observations, $5\times22$ m antennas of the ATCA were used as a phased array in order to maintain a compact configuration suitable for wide-field imaging. The observations utilised the S2 recording system \citep{Cannon:1997p10533} to record 2 $\times$ 16 MHz bands (digitally filtered 2-bit samples) in the frequency ranges: 2252 - 2268 MHz and 2268 - 2284 MHz.  Both bands were upper side band and right circular polarisation.

During the VLBI observation, three minute scans of PKS $0344-345$ were scheduled, alternating with three minute scans of a phase reference calibration source $0335-364$ ($\alpha=03\rah36\ram54\fs0235$; $\delta=-36\arcdeg16\arcmin06\farcs224$ [J2000]). Observing parameters associated with each of the LBA observations are shown in Table \ref{tab:tabjetobs}.

\begin{sidewaystable}[!p]
\begin{center}
{ \footnotesize
\begin{tabular}{llcccccccc} \hline \hline
Source     & Observatory          & Frequency & $\alpha$              & $\delta$                         & Date & Duration & Bandwidth & $\Delta$t \\
           &                      & (MHz)   & (J2000)                 & (J2000)                          &      & (h)      & (MHz)     & (s) \\ [0.5ex] \hline \hline
$0344-345$ & LBA & 2252.0  & $03\rah46\ram30\fs58$   & $-34\arcdeg22\arcmin46\farcs3$   & 16/17 MAY 2006 & 12   & 16  & 2  \\
           & \nodata              & 2268.0  & \nodata                 & \nodata                          & \nodata        & 12   & 16  & 2  \\
\hline
$0521-365$ & LBA & 1634.0  & $05\rah22\ram57\fs9846$ & $-36\arcdeg27\arcmin30\farcs848$ & 15 NOV 2006    & 12  & 16  & 1  \\
           & \nodata              & 1650.0  & \nodata                 & \nodata                          & \nodata        & 12  & 16  & 1  \\
           & \nodata              & 1666.0  & \nodata                 & \nodata                          & \nodata        & 12  & 16  & 1  \\
           & \nodata              & 1682.0  & \nodata                 & \nodata                          & \nodata        & 12  & 16  & 1  \\
\hline
$0521-365$ & LBA & 2268.0  & $05\rah22\ram57\fs9846$ & $-36\arcdeg27\arcmin30\farcs848$ & 20/21 JUN 2007 & 12  & 16  & 2  \\
           & \nodata              & 2284.0  & \nodata                 & \nodata                          & \nodata        & 12  & 16  & 2  \\
           & \nodata              & 2300.0  & \nodata                 & \nodata                          & \nodata        & 12  & 16  & 2  \\
           & \nodata              & 2316.0  & \nodata                 & \nodata                          & \nodata        & 12  & 16  & 2  \\ \hline
\end{tabular}
\caption{Summary of $0344-345$ and $0521-365$ observations.}
\label{tab:tabjetobs}
}
\end{center}
\end{sidewaystable}

\subsection{PKS $0521-365$ Observations}

VLBI observations of PKS $0521-365$ were made at 1.6 GHz on 15 November 2006 and at 2.3 GHz on 20/21 June 2007. The 1.6 GHz observation used a number of the Long Baseline Array (LBA) telescopes : the 70 m NASA Deep Space Network (DSN) antenna at Tidbinbilla (8 hours); the 64 m antenna of the Australia Telescope National Facility (ATNF) near Parkes (10 hours); the ATNF Australia Telescope Compact Array (ATCA) near Narrabri (12 hours); and the ATNF Mopra 22 m antenna near Coonabarabran (6 hours). The 2.3 GHz observation used the 70 m Tidbinbilla antenna (4 hours); the 64 m Parkes antenna (5 hours); the ATCA (6 hours); and the Mopra antenna (2.5 hours). For the 1.6 GHz observation, $3\times22$ m antennas of the ATCA were used as a phased array (with a maximum baseline of 490 m), whereas for 2.3 GHz observation $5\times22$ m antennas were used as a phased array (with a maximum baseline of 252 m) in order to maintain a compact configuration suitable for wide-field imaging. All observations were correlated using the DiFX Software Correlator \citep{Deller:2007p10545}. The 1.6 GHz data were correlated using an integration time of 1 second and with 128 frequency channels across each 16 MHz band (channel widths of 0.125 MHz), the 2.3 GHz data were correlated using an integration time of 2 seconds and with 64 frequency channels across each 16 MHz band (channel widths of 0.25 MHz). Observing parameters associated with each of the LBA observations are shown in Table \ref{tab:tabjetobs}.

\subsection{Data Reduction and Results}

\begin{figure}[ht]
\epsscale{0.8}
\begin{center}
\plotone{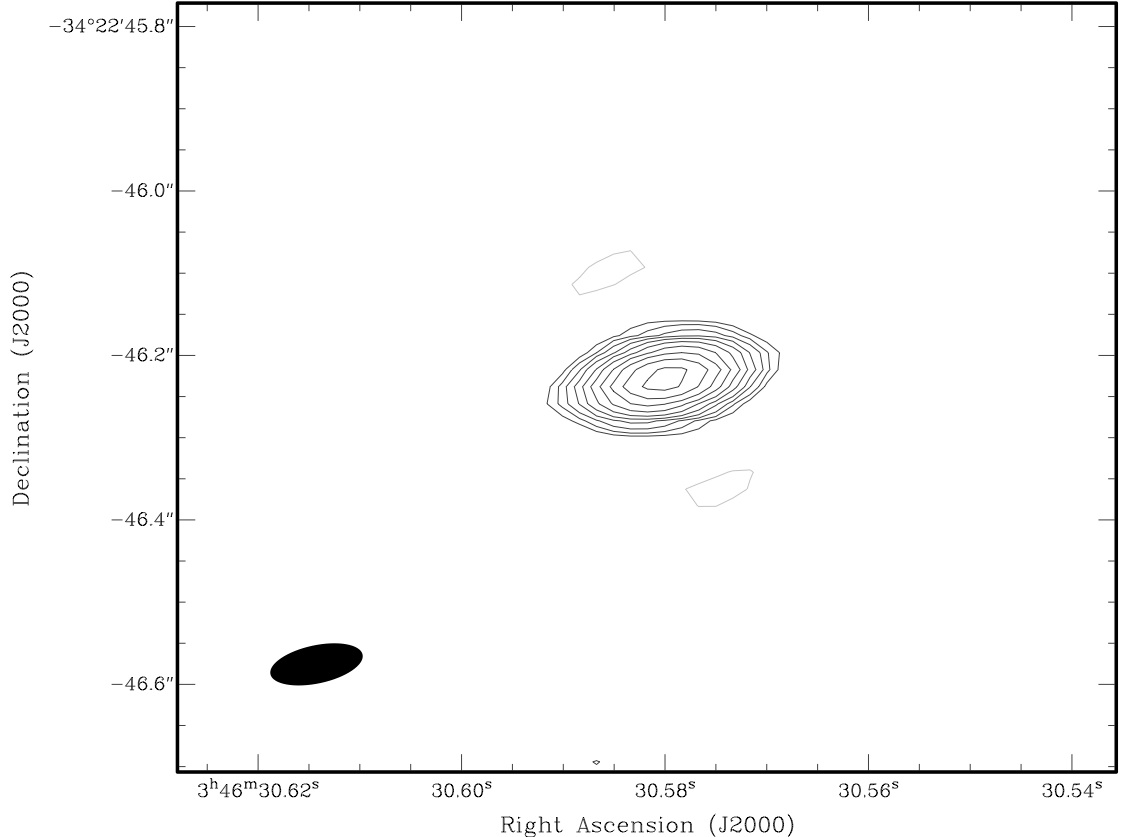}
\caption[LBA image of PKS $0344-345$ at 2.3 GHz]{Naturally-weighted total-power map of the source PKS $0344-345$ as observed with the LBA at 2.3 GHz. Map statistics are shown in Table \ref{tab:tabjetimage}. Contours are drawn at $\pm2^{0}, \pm2^{\frac{1}{2}}, \pm2^{1}, \pm2^{\frac{3}{2}}, \cdots$ times the $3\sigma$ rms noise.}
\label{fig:jet0344}            
\end{center}
\end{figure}

For the observation of PKS $0344-345$, the correlated data were reduced using the techniques developed and described in Section \S~\ref{sec:p3lbareduction} to correct for any structure that may exist in the phase calibrator source. The calibrated data resulting from the data reduction process achieved a one sigma RMS noise of 0.387 mJy beam$^{-1}$, compared to the expected theoretical thermal image noise of $\sim0.25$ mJy beam$^{-1}$. With the longest baselines included the core component was weak and highly resolved, making it difficult to refine calibration of the source any further. With a 6 sigma detection threshold, there was no detection of the interaction region or any of the hot spots, suggesting that the sources were completely resolved by the VLBI observation. A restricted $(u,v)$ range was applied to exclude the longest baselines to Ceduna and Hobart. While the core component was clearly detected, Figure \ref{fig:jet0344}, no detections were made of the interaction region or the hot spots.

In both observations of PKS $0521-365$, the bright and compact core component was used as an in-beam calibrator. The data were modelled using iterations of Gaussian component fitting followed by phase self-calibration. The resulting 1.6 GHz and 2.3 GHz images achieve a one sigma RMS noise of 0.21 mJy beam$^{-1}$ and 0.74 mJy beam$^{-1}$, respectively. The 1.6 GHz one sigma noise is approximately 6 times greater than the expected theoretical thermal noise and at 2.3 GHz it is approximately 3 times greater, the higher than expected image noise is as a result of significant telescope down-time and insufficient overlap between between telescopes to allow for amplitude self-calibration.

\begin{table}[ht]
\begin{center}
{ \scriptsize
\begin{tabular}{lcccccccc} \hline \hline
Figure                 & Source     & Frequency & Synthesized Beam & $\sigma$ & Peak Flux & Integrated Flux \\
                       &            & (GHz)     & (mas)            & (mJy beam$^{-1}$) & (mJy beam$^{-1}$) & (mJy) \\ [0.5ex] \hline \hline
\ref{fig:jet0344}      & PKS $0344-345$ & 2.3 & $115\times46$          & 0.387 & 29.6  & 37.3 \\
\ref{fig:jet0521}(a)   & PKS $0521-365$ & 1.6 & $109\times45$          & 0.210 & 1090  & 1780 \\
\ref{fig:jet0521}(b)   & PKS $0521-365$ & 2.3 & $155\times49$          & 0.930 & 1430  & 1770 \\
\ref{fig:jet0521se}(a) & PKS $0521-365$ & 1.6 & $109\times45$          & 0.210 & 27.7  & 194  \\
\ref{fig:jet0521se}(b) & PKS $0521-365$ & 2.3 & $155\times49$          & 0.930 & 34.4  & 153  \\
\ref{fig:jet0521agn2cm}& PKS $0521-365$ & 2.3 & $14\times12$           & 0.740 & 1047 & 1699 \\
\ref{fig:jet0521agn2cm}& PKS $0521-365$ & 15  & $3.4\times0.7$         & 0.530 & 1520  & 2150  \\ \hline
\end{tabular}
\caption{Map statistics for PKS $0344-345$ and PKS $0521-365$ images.}
\label{tab:tabjetimage}
}
\end{center}
\end{table}

\begin{figure}[ht]
\epsscale{0.8}
\begin{center}
\plotone{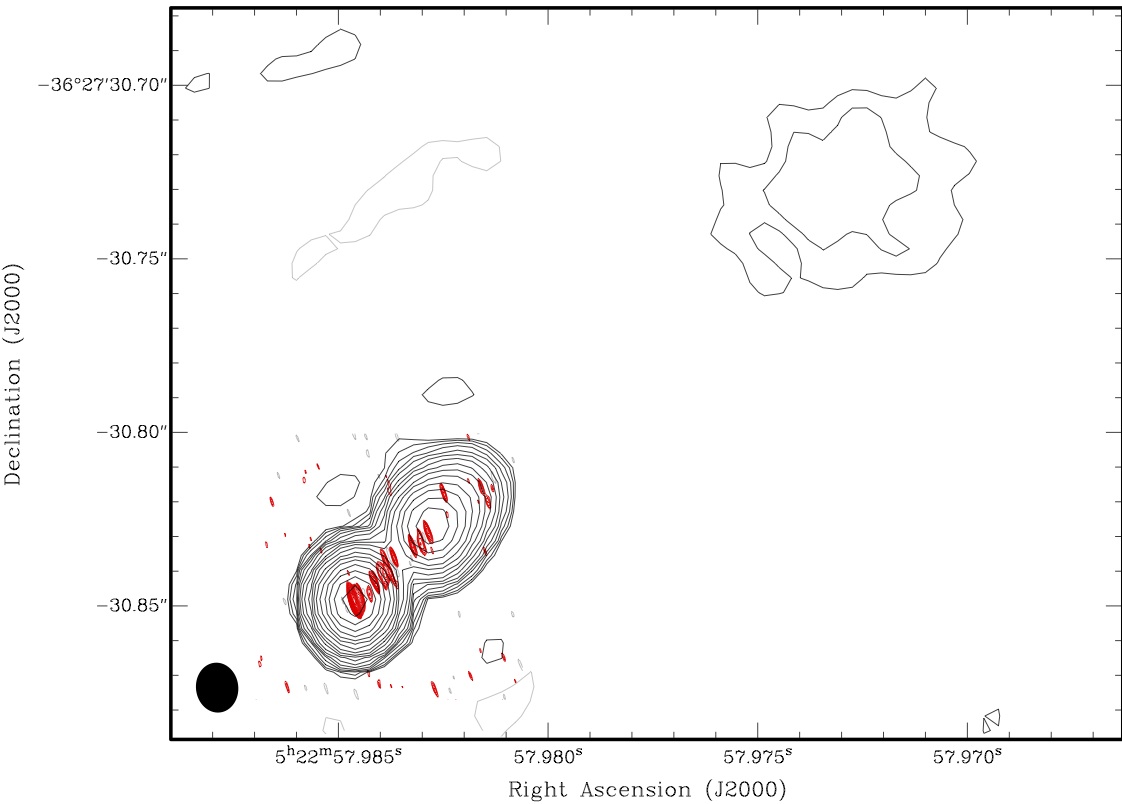}
\caption[LBA image of PKS $0521-365$ at 2.3 GHz overlaid with VLBA image at 15 GHz]{Black contours: Naturally-weighted total-power map of the source PKS $0521-365$ as observed with the LBA at 2.3 GHz. Red contours: VLBA image at 15 GHz \citep{Kellermann:1998p18091}. Map statistics are shown in Table \ref{tab:tabjetimage}. Contours are drawn at $\pm2^{0}, \pm2^{\frac{1}{2}}, \pm2^{1}, \pm2^{\frac{3}{2}}, \cdots$ times the $3\sigma$ rms noise.}
\label{fig:jet0521agn2cm}            
\end{center}
\end{figure}

In the highest resolution image at 2.3 GHz, Figure \ref{fig:jet0521agn2cm}, the nucleus is resolved into multiple components, which is in line with expectations for the compact jet components detected with the VLBA at 15 GHz \citep{Kellermann:1998p18091}. An additional extended component is observed $\sim180$ mas from the core lying outside of the field of view of the original VLBA observation. In the full resolution image, neither the knot nor the hot spots are detected above a 6 sigma threshold.

\begin{figure}[p]
\epsscale{0.8}
\begin{center}
\plotone{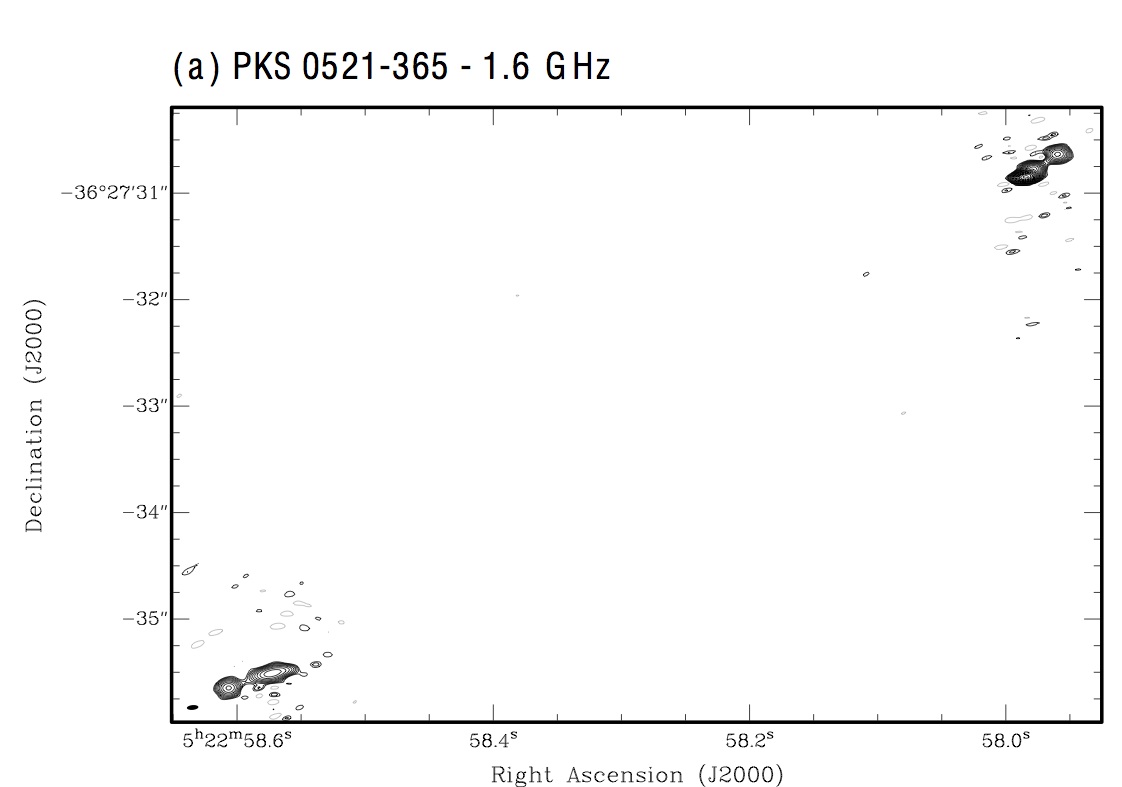}
\plotone{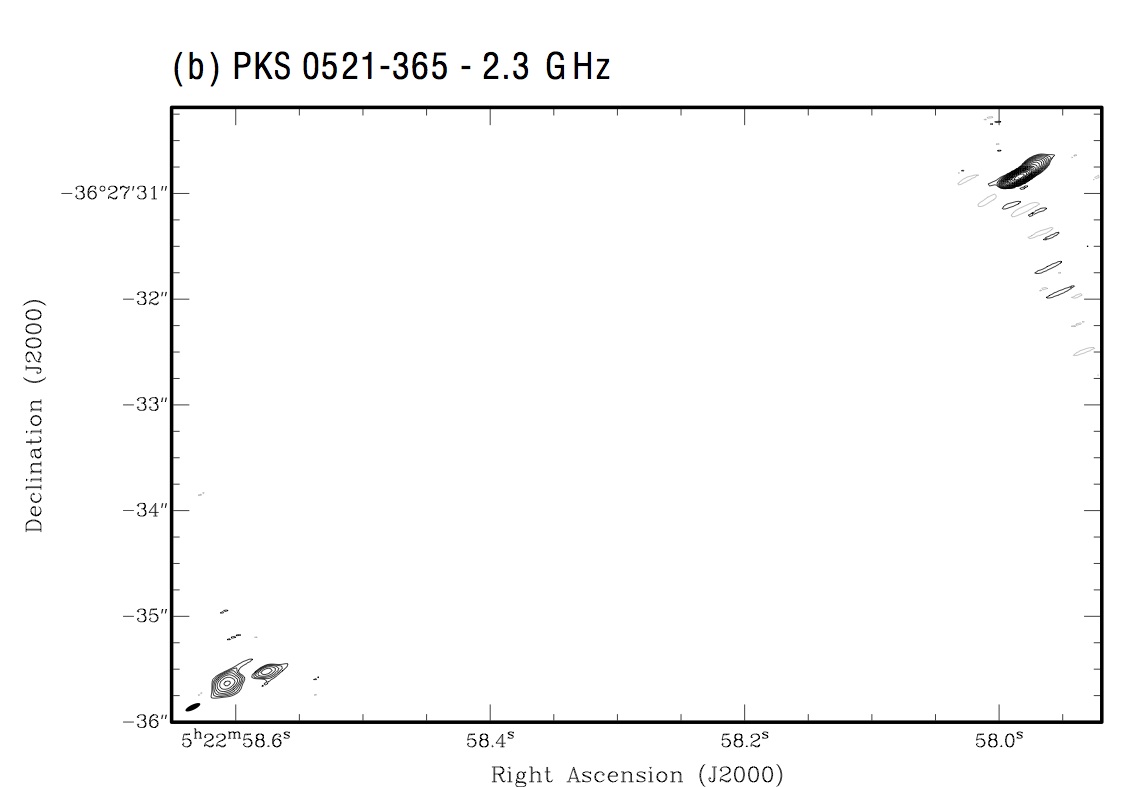}
\caption[LBA images of PKS $0521-365$ at 1.6 GHz and 2.3 GHz]{Naturally-weighted total-power map of the source PKS $0521-365$ as observed with the LBA at 1.6 GHz and 2.3 GHz. Map statistics for the individual maps are shown in Table \ref{tab:tabjetimage}. Contours are drawn at $\pm2^{1}, \pm2^{\frac{3}{2}}, \cdots$ times the $6\sigma$ rms noise for image (a) and at $\pm2^{0}, \pm2^{\frac{1}{2}}, \pm2^{1}, \pm2^{\frac{3}{2}}, \cdots$ times the $3\sigma$ rms noise for image (b).}
\label{fig:jet0521}            
\end{center}
\end{figure}

\begin{figure}[p]
\epsscale{0.8}
\begin{center}
\plotone{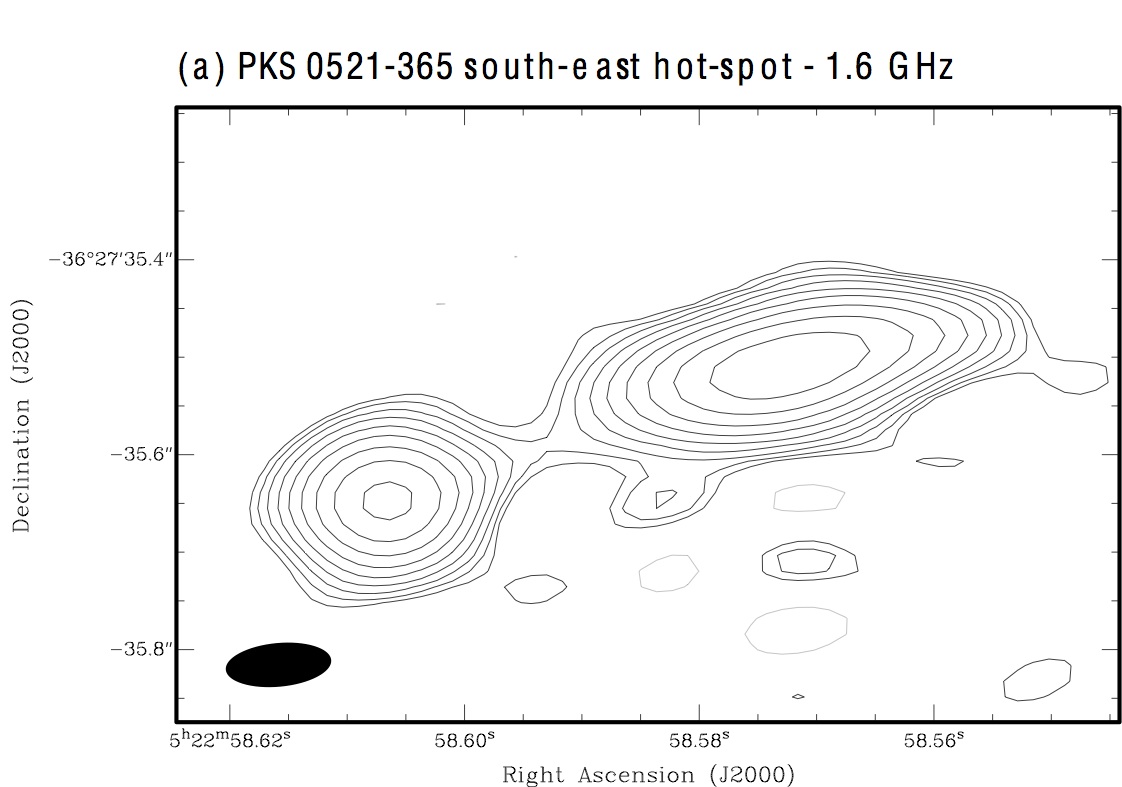}
\plotone{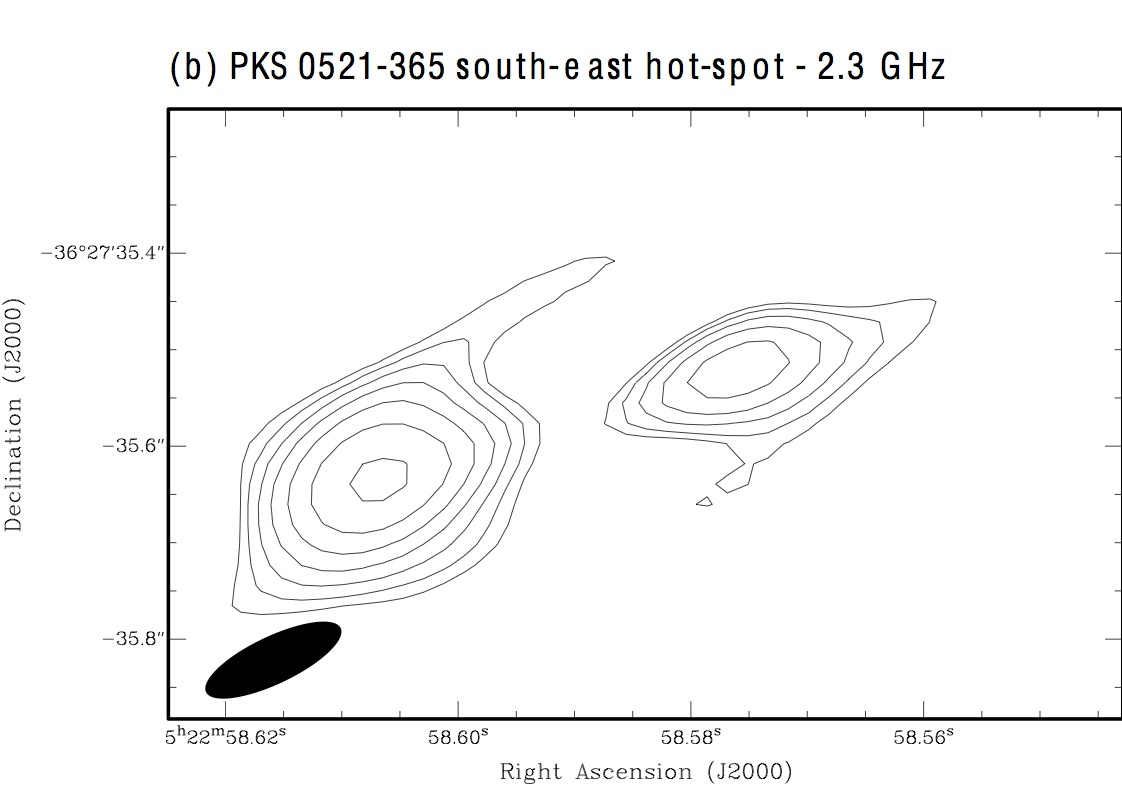}
\caption[LBA images of south-east hot spot in PKS $0521-365$ at 1.6 GHz and 2.3 GHz]{Naturally-weighted total-power map of the south-east hot spot in PKS $0521-365$ as observed with the LBA at 1.6 GHz and 2.3 GHz. Map statistics for the individual maps are shown in Table \ref{tab:tabjetimage}. Contours are drawn at $\pm2^{1}, \pm2^{\frac{3}{2}}, \cdots$ times the $6\sigma$ rms noise for image (a) and at $\pm2^{0}, \pm2^{\frac{1}{2}}, \pm2^{1}, \pm2^{\frac{3}{2}}, \cdots$ times the $3\sigma$ rms noise for image (b).}
\label{fig:jet0521se}            
\end{center}
\end{figure}

At 1.6 GHz, only the short baselines associated with Parkes, Tidbinbilla, ATCA and Mopra were available resulting in a larger synthesized beam of $109\times45$ mas. At this reduced resolution two components are detected associated with the south-east hot spot, see Figures \ref{fig:jet0521}(a) and \ref{fig:jet0521se}(a). These hot spot components are also detected at 2.3 GHz when the longer baselines associated with Hobart and Ceduna are removed, see Figures \ref{fig:jet0521}(b) and \ref{fig:jet0521se}(b), and are coincident with the 1.6 GHz components.

\section{Discussion}
\label{sec:jets.discussion}

\subsection{PKS $0344-345$}
While there was no detection of the interaction region in PKS $0344-345$, an estimate of the size and spectral index of the region can be made from our low resolution ATCA images (Section \S~\ref{sec:selection.agn}). From these images we estimate that the interaction region has a spectral index of $\alpha=-0.84$ ($S\propto\nu^{\alpha}$) and a FWHM size of $\sim6.4\arcsec$ (6.8 kpc). The spectral index of this region is comparable to that observed in the northern hot spot of PKS $2152-699$ which has a measured spectral index of $\alpha=-0.87$ \citep{Young:2005p5449}. The size of the region, however, is an order of magnitude larger than the $\sim200$ pc extent measured in PKS $2152-699$. If a simple power-law is assumed to describe the spectral energy distribution of the interaction region then an X-ray flux density of $\sim8$ nJy is estimated at 1-KeV. Further observations at optical, infrared and X-ray wavelengths will be required to improve the modelling of the interaction region.

\subsection{PKS $0521-365$}
A jet-knot is observed in ATCA radio images \citep{Birkinshaw:2002p4973}, \emph{HST} optical images \citep{Scarpa:1999p17053}, and is partially resolved with a 300 mas FWHM beam at 15 GHz with the VLA \citep{Keel:1986p10557}. Our LBA observation does not detect any emission from the knot location above the $6\sigma$ detection threshold. This suggests that the knot is completely resolved by the significantly smaller beam of the LBA and that the knot does not contain any significant flux at $\sim50$ pc scales.

Our LBA observations of the south-east hot spot reveals two components at both 1.6 GHz and 2.3 GHz. The components are separated by $\sim450$ mas ($\sim500$ pc) and have FWHM sizes of $93\times80$ mas and $224\times76$ mas at 1.6 GHz. The total flux density of the components is $0.19\pm0.02$ Jy at 1.6 GHz and $0.15\pm0.02$ Jy at 2.3 GHz, and has a spectral index of $\alpha=-0.65\pm0.5$ (where $S\propto\nu^{\alpha}$).

\begin{figure}[ht]
\epsscale{1.0}
\begin{center}
\plotone{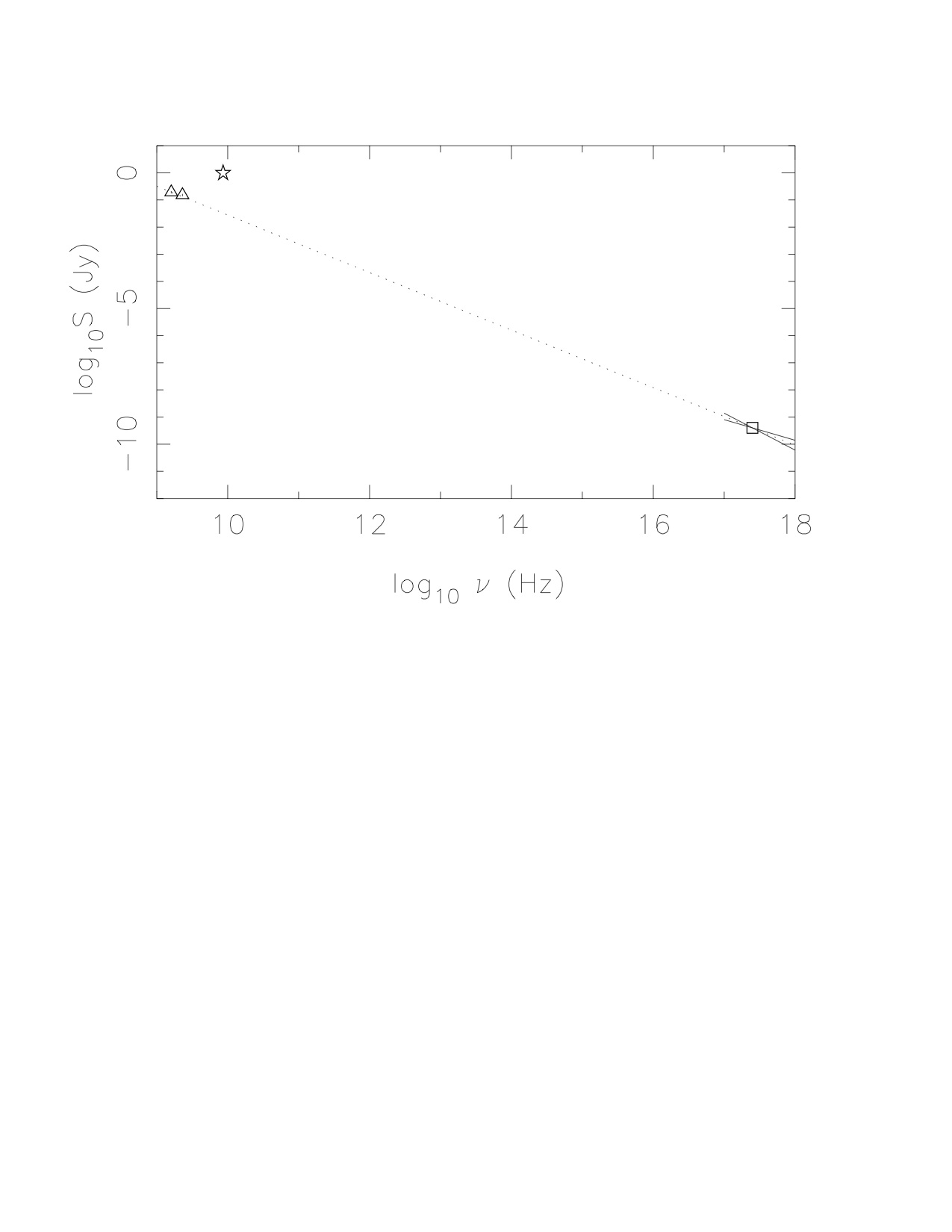}
\caption[Spectral energy distribution of the south-eastern hot spot in PKS $0521-365$]{Spectral energy distribution of the south-eastern hot spot in PKS $0521-365$ showing the 1.6 GHz and 2.3 GHz radio flux (triangles), the 8.6 GHz flux of the more extended emission (star), and the X-ray spectrum (bow-tie) associated with the $3\sigma$ of \citet{Birkinshaw:2002p4973} (square). The dotted line shows a power law $S_{\nu}\propto\nu^{-1.06}$.}
\label{fig:jet0521spec}            
\end{center}
\end{figure}

\citet{Birkinshaw:2002p4973} observed the south-east hot spot with the ATCA at 8.64 GHz and estimated a FWHM size of approximately 400 mas - a value that is comparable to the separation of the two components detected with the LBA. When modelled using a 400 mas sphere, completely filled with a uniform magnetic field and a relativistic electron-proton plasma, the energetics of which are dominated by the electrons, and assuming equipartition and electron energy limits 600 MeV$-$400 GeV, they predicted a synchrotron self-Compton X-ray flux density from the hot spot of about 0.2 nJy. This is a factor of two less than their $3\sigma$ detection of the hot spot in \emph{Chandra} observations at 1-KeV. If a simple power-law model is assumed for the spectrum of the south-east hot spot, then a spectral index of $\alpha=-1.06$ would provide a good fit between the radio points and the $3\sigma$ x-ray detection of \citet{Birkinshaw:2002p4973} (Figure \ref{fig:jet0521spec}).

The south-east hot spot is similar, in many respects, to the north-west hot spot in Pictor A (Chapter \ref{chap:pictora}). They have a similar overall extent (500 pc in PKS $0521-365$ compared to 700 pc in Pictor A), and are composed of multiple components of comparable size (80 pc in PKS $0521-365$ compared to 70 pc in Pictor A). In Pictor A, it is believed that the small-scale components are sites of synchrotron X-ray production as a result of strong shocks. These components contribute to the overall X-ray flux along with X-rays from synchrotron self-Compton scattering. Given the similarities between the two sources, it is possible that the same mechanism may be at work in PKS $0521-365$. Currently, there is insufficient data to test this possibility. However, new observations of the hot spot at optical and infrared wavelengths, together with high frequency VLBI observations will allow a more detailed study of the emission mechanisms at play. 

\section{Summary}
\label{sec:jets.summary}

We have imaged PKS $0344-345$ with the LBA at 2.3 GHz and PKS $0521-365$ with the LBA at 1.6 GHz and 2.3 GHz in an attempt to detect jet interaction regions in these galaxies. We find the following results:

\begin{itemize}
\item No detection of the jet interaction region in PKS $0344-345$.
\item Based on low resolution ATCA data, the interaction region in PKS $0344-345$ has an extent of $\sim6.8$ kpc and a spectral index of $\alpha=-0.84$ that is comparable to that of a similar interaction region in PKS $2152-699$.
\item A 1-KeV X-ray flux density of $\sim8$ nJy is predicted for the interaction region in PKS $0344-345$ assuming a simple power-law model for the spectral energy distribution.
\item No detection of the jet-knot in PKS $0521-365$.
\item A detection of two compact sources in the south-east hot spot of PKS $0521-365$.
\item A simple power-law model, with $\alpha=-1.06$, provides a good fit for the spectral energy distribution of the south-east hot spot in PKS $0521-365$ between on our VLBI data and previously published X-ray data.
\item The multi-component nature, overall extent and size of individual components in the hot spot of PKS $0521-365$ is similar to that of the north-west hot spot in Pictor A ($z=0.035$).
\item Further multi-wavelength observations of the south-east hot spot are required to determine if the emission mechanisms at work there are similar to that of the north-west hot spot in Pictor A.
\end{itemize}

\linespread{1.0}
\normalsize
\begin{savequote}[20pc] \sffamily
Not only is the universe stranger than we imagine,\\
it is stranger than we can imagine.
\qauthor{Sir Arthur Eddington}
\end{savequote}

\chapter{A Deep, High Resolution Survey of the Low Frequency Radio Sky}
\label{chap:lfwfvlbi}
\begin{center}
{\it Adapted from:}

E. Lenc,  M.A. Garrett, O. Wucknitz, J.M. Anderson, \& S.J. Tingay

Astrophysical Journal, 673, 78--95 (2008)
\end{center}
\small
We report on the first wide-field, very long baseline interferometry (VLBI) survey at 90 cm.  The survey area consists of two overlapping 28 deg$^{2}$ fields centred on the quasar J0226$+$3421 and the gravitational lens B0218$+$357. A total of 618 sources were targeted in these fields, based on identifications from Westerbork Northern Sky Survey (WENSS) data. Of these sources, 272 had flux densities that, if unresolved, would fall above the sensitivity limit of the VLBI observations. A total of 27 sources were detected as far as $2\arcdeg$ from the phase centre. The results of the survey suggest that at least $10\%$ of moderately faint (S$\sim100$ mJy) sources found at 90 cm contain compact components smaller than $\sim0.1$ to $0.3$ arcsec and stronger than $10\%$ of their total flux densities. A $\sim90$ mJy source was detected in the VLBI data that was not seen in the WENSS and NRAO VLA Sky Survey (NVSS) data and may be a transient or highly variable source that has been serendipitously detected. This survey is the first systematic (and non-biased), deep, high-resolution survey of the low-frequency radio sky. It is also the widest field of view VLBI survey with a single pointing to date, exceeding the total survey area of previous higher frequency surveys by two orders of magnitude. These initial results suggest that new low frequency telescopes, such as LOFAR, should detect many compact radio sources and that plans to extend these arrays to baselines of several thousand kilometres are warranted.
\clearpage
\linespread{1.3}
\normalsize
\section{Introduction}
\label{sec:p2introduction}

The general properties of the 90 cm sky are not very well known and even less is known at VLBI resolution. Previous snapshot surveys at these wavelengths have only targeted the brightest sources and were plagued by poor sensitivity, radio interference and limited coherence times. Furthermore, the field of view that could be imaged was typically limited by the poor spectral and temporal resolution of early generation hardware correlators and the available data storage and computing performance at the time. As a result, although several hundred 90 cm VLBI observations have been made over the past two decades, images of only a few tens of sources have been published e.g. \citet{Altschuler:1995p9758}; \citet{Lazio:1998p10559}; \citet{Chuprikov:1999p10566}; \citet{Cai:2002p10532}. With such a small sample it is difficult to quantify the total population and nature of these sources. In particular, the sub-arcsecond and sub-Jansky population of 90 cm sources is largely unexplored. 

Recent improvements to the EVN hardware correlator at JIVE \citep{vanLangevelde:2004p9712}, have enabled significantly finer temporal and spectral resolution. Combined with vast improvements in storage and computing facilities, it is now possible to image fields as wide as, or even wider than, the FWHM of the primary beam of the observing instrument. To complement the hardware improvements, new approaches to calibration and imaging have been developed to better utilise the available data and processing platforms. For example, \cite{Garrett:2005p10555} performed a deep VLBI survey at 20 cm of a $36\arcmin$ wide field by using a central bright source as an in-beam calibrator. The approach was ideal for survey work as it permitted the imaging of many potential target sources simultaneously by taking advantage of the full sensitivity of the observation across the entire field of view. We have applied a similar technique at 90 cm by piggybacking on an existing VLBI observation of the gravitational lens B0218$+$357 and the nearby quasar J0226$+$3421, with the aim of surveying a 28 deg$^{2}$ field around each of the sources. The results provide an important indication of what may be seen by future low-frequency instruments such as the Low Frequency Array (LOFAR), European LOFAR (E-LOFAR) and the Square Kilometre Array (SKA).

In this paper, we present the results of a 90 cm wide-field VLBI survey that covers two partially overlapping regions of 28 deg$^{2}$ each, surveying 618 radio source targets at angular resolutions ranging between 30 and 300 mas. For sources located at a redshift of $z=1$, the linear resolution corresponding to 30 mas is 230 pc. A \emph{WMAP} cosmology \citep{Spergel:2003p9720} with a flat Universe, $H_{0}=72$ km s$^{-1}$ Mpc$^{-1}$ and $\Omega_{m}=0.29$ is assumed throughout this paper.

\section{Observations and Correlation}

A VLBI observation of the gravitational lens B0218$+$357 was made on 11 November 2005 using all ten NRAO Very Long Baseline Array (VLBA) antennas, the Westerbork Synthesis Radio Telescope (WSRT) as a phased array and the Jodrell Bank, 76$-$m Lovell Telescope (JB). The primary aim of this observation was to investigate, in detail, propagation effects in the lensing galaxy and the substructure in the lens. The secondary aim, as investigated in this paper, was to serve as a wide-field test observation to study the faint source population at 90 cm over a good fraction of the primary beam.

The observation spanned 14 hours with approximately 6 hours of data recorded at JB and WSRT and 13 hours at the VLBA stations. Ten minutes scans of the target source B0218$+$357 ($\alpha=02\rah21\ram05\fs4733$ and $\delta=35\arcdeg56\arcmin13\farcs791$) were interleaved with three minute scans of the nearby quasar J0226$+$3421 ($\alpha=02\rah26\ram10\fs3332$ and $\delta=34\arcdeg21\arcmin30\farcs286$). Five minute scans of the fringe finder 3C84 were made approximately every four hours. Dual circular and cross polarisation data were recorded across four 4 MHz IFs centred on 322.49, 326.49, 330.49 and 610.99 MHz respectively, resulting in a total data recording rate of 128 Mbits s$^{-1}$. The 610.99 MHz data were only recorded at the VLBA antennas. The data were correlated at the European VLBI Network (EVN) correlator at the Joint Institute for VLBI in Europe (JIVE, Dwingeloo, the Netherlands) in multiple passes to create a single-IF, single polarisation, wide-field data-set and a multi-IF, dual circular polarisation, narrow-field data-set. The narrow-field data-set concentrated on the B0218$+$357 with greater sensitivity and the results of this observation will be presented elsewhere (Wucknitz et al., in preparation). The wide-field data consisted of a single polarisation (LL), single IF with a 4 MHz band centred on 322.49 MHz. A third correlator pass centred on another source was used to create a second wide-field data-set with a single IF and RR polarisation but was not used in our data reduction process.

To reduce the effects of bandwidth smearing and time averaging smearing, and thus image the largest possible field, the EVN correlator generated data with 512 spectral points per baseline and an integration time of 0.25 s. The spectral and temporal resolution exercised the current physical limits of the JIVE hardware correlator and resulted in a final data-set size of 77.5 Gbytes. The wide-field data-set has a one sigma theoretical thermal noise of $\sim$1.2 and $\sim$0.7 mJy/beam for the quasar and gravitational lens, respectively.

\section{VLBI Calibration and Imaging}
\label{sec:p2vlbical}
The data from the narrow-field data-set were used to perform the initial editing and calibration of the phase reference to take advantage of the increased sensitivity available with the additional bands and polarisations. Nominal corrections to counter the effects of the total electron content (TEC) of the ionosphere were applied with the AIPS\footnote{The Astronomical Image Processing System (AIPS) was developed and is maintained by the National Radio Astronomy Observatory, which is operated by Associated Universities, Inc., under co-operative agreement with the National Science Foundation} task TECOR. An amplitude calibration table was derived from measures of the system temperature of each antenna throughout the observation using the AIPS task APCAL, and applied to the data-set.

Delays across the IF bands, which were assumed to be constant throughout the observation, were calibrated by fringe fitting on 3C84. A multi-band fringe fit was then performed on the quasar. The flagging and calibration tables of the narrow-field data-set were transferred to the wide-field data-set using a ParselTongue\footnote{A Python scripting tool for AIPS. ParselTongue was developed in the context of the ALBUS project, which has benefited from research funding from the European Community's sixth Framework Programme under RadioNet R113CT 2003 5058187. ParselTongue is available for download at \url{http://www.radionet-eu.org/rnwiki/ParselTongue}} script. Further editing was applied to the wide-field data-set to remove the frequency band edges and frequency channels adversely affected by RFI. The bandpass for the observation was calibrated against observations of 3C84.

As both the phase reference and the gravitational lens were to be used as in-beam calibrators for their respective fields, accurate calibration of the amplitudes and phases of both fields was essential. The calibration was complicated by the complex structure of both sources. To account for this structure a new DIFMAP \citep{Shepherd:1994p10583} task, \emph{cordump}\footnote{The \emph{cordump} patch is available for DIFMAP at \url{http://astronomy.swin.edu.au/~elenc/DifmapPatches/}} \citep{Lenc:2006p32}, was developed to enable the transfer of all phase and amplitude corrections made in DIFMAP during the imaging process to an AIPS compatible SN table. The \emph{cordump} task greatly simplified the calibration of both of the fields. First, the phase reference data were averaged in frequency and exported to DIFMAP where several iterations of modelling and self-calibration of both phases and amplitudes were performed. \emph{cordump} was then used to transfer the resulting phase and amplitude corrections back to the unaveraged AIPS data-set. After application of these corrections, the DIFMAP model of the phase reference source was subtracted from the AIPS $(u,v)$ data-set. The quasar self-calibration solutions were then applied to the lens field as an initial calibration for that field. The calibration for the lens field was further refined using the same approach as for the quasar and upon completion the DIFMAP model of the lens was subtracted from the calibrated field. The images of the phase reference and the lens had measured RMS noise of 1.8 and 1.0 mJy beam$^{-1}$, respectively. The higher than theoretical noise is attributed to substantial levels of RFI observed on some baselines and the shorter time available with the WSRT and JB observations.

In the first phase of the imaging process, the AIPS task IMAGR was used to make naturally weighted dirty images and beams of regions selected from WENSS data \citep{Rengelink:1997p10578} of the two fields being surveyed, the source selection criteria are described in detail in \S~\ref{sec:p2selection}. Targets falling within a certain annulus around each field were imaged simultaneously using the multi-field option within IMAGR, the DO3D option was used to reduce non-coplanar array distortion. The data from both fields were kept in an unaveraged form to prevent smearing effects during imaging. For each target, the dirty image subtended a square of approximately $51\arcsec$ on each side, an area that covers approximately half that of the WENSS beam at the observation declination ($54\arcsec\times92\arcsec$). Since the dirty image of each target source contains $\sim2\times10^5$ synthesized-beam areas, a conservative $6\sigma$ detection threshold was imposed to avoid spurious detections. Furthermore, only the inner 75\% of each dirty image was searched for candidate detections to avoid erroneous detections as a result of map edge effects. For each positive detection the co-ordinate of the VLBI peak flux density was recorded. This first imaging step was used to determine whether the target source had been detected with the VLBI observation.  Based on our detection criteria, we estimate a false detection rate of approximately one in every 3300 images. While we expect the majority of unresolved WENSS sources to have peaks that fall within the imaged areas, approximately 9.5\% of the WENSS sources exhibit resolved structure. For these sources, we would not detect bright compact components that may exist outside of the central region that was imaged.

During the calibration process, it was noted that the lens had significantly weaker signal on the longer baselines compared to that of the quasar. To test the effectiveness of the refined lens field self-calibration solutions, we re-imaged one of the B0218$+$357 field sources, B0215.1$+$3710, with only the phase reference self-calibration solutions applied. B0215.1$+$3710 is located on the side of the lens field that is furthest from the phase reference ($\sim3.45\arcdeg$) and so is most sensitive to changes from the nominal conditions that were corrected for. The target-calibrated dirty image for this source has an rms noise of 4.7 mJy beam$^{-1}$ and a peak of 42 mJy beam$^{-1}$. With only the phase reference self-calibration solutions applied, the rms noise is 6.6 mJy beam$^{-1}$ and the peak 21 mJy beam$^{-1}$. It is clear that without the refined calibration, B0215.1$+$3710 would not have been detected above the $6\sigma$ threshold.

The second phase of the imaging process involved creating a $(u,v)$ shifted data-set for each of the positive detections, using the AIPS task UVFIX, such that the new image centre coincided with the co-ordinate of the image peak recorded in the first phase. The shifted data-sets were averaged in frequency, effectively reducing the field of view of each of the targeted sources to approximately $0.5\arcmin$, and then exported to DIFMAP. In DIFMAP, the visibilities were averaged over 10 second intervals to reduce the size of the data-set and to speed up the imaging process. Each target was imaged in DIFMAP, with natural weighting applied, using several iterations of model fitting. Phase self-calibration was performed between iterations to adjust for the varying effects of the ionosphere across the field.

During the imaging process it was noted that the self-calibration phase corrections varied significantly between fields and even among sources within each field. It is believed that these were due to ionospheric variations that occurred across the survey field. Observations at 90 cm will invariably suffer degradation as a result of ionospheric variations and the nominal TEC corrections made in the initial calibration stages assumed that these corrections would be valid across the entire field. This is not a valid assumption when imaging extremely large fields. To provide position dependent corrections within a field, ParselTongue scripts were developed to implement two alternate methods that could be tested against the data.

The first method calculated ionospheric corrections based on TEC measures to each source of interest in the field of view and applied differential corrections, based on the TEC correction already made at the phase centre, prior to imaging. This allowed a differential correction to be applied after self calibrating on the bright central sources in each field of view of these observations. The corrections did not result in any significant improvement in the resulting images. We suspect that the currently available TEC solutions may be too coarse, both spatially and temporally, to account for the ionospheric variations across the field.

The second method used a parameterized ionospheric model to determine the corrections to each source of interest in the field of view. As with the TEC corrections, these were applied after the central source in each field had been self calibrated. Preliminary tests of these corrections indicated that they performed better than the differential TEC solutions with improvements of the order of a few percent in flux density observed in approximately 70\% of sources tested.

The testing of these libraries is not yet complete and only the 11 wide-field sources of \citet{Lenc:2006p32} were used in our initial tests. Further tests will be required to more robustly analyse the performance of the two approaches to ionospheric calibration.

Following our first attempt to survey the inner $0\arcdeg-1\arcdeg$ region of each field \citep{Lenc:2006p32} we discovered that the positional accuracy of the detected sources degraded significantly with radial distance from the phase centre when compared to the positions derived from observations with other instruments. While most sources observed with other instruments only had a positional accuracy of $\sim1\arcsec$ it was still clear that our fields were being scaled by a factor of $0.99871\pm7\times10^{-5}$, a factor that corresponds to an offset of $53\pm3$ frequency channels in our data-set. Interestingly, this appeared to corresponded to the 50 lower-band channels that were flagged during editing in the 31DEC05 version of AIPS that was used for the processing of the data. When the processing was repeated in the 31DEC06 version of AIPS, the positional discrepancies disappeared so we suspected that there may have been a software issue with the earlier version of AIPS. However, as wider fields were imaged we discovered four sources that were common to both of the fields. Each of these sources should have been well aligned between the two fields, however, discrepancies of $0.41\arcsec-0.65\arcsec$ were being observed in approximately the same position angle. This was also indicative of a radial scaling but to a lesser degree, $0.999927\pm1.4\times10^{-5}$, corresponding to an offset of $3\pm0.6$ channels. The source of this error has not yet been identified, however all of the sources positions and images in this paper have been corrected to account for this effect\footnote{Following the publication of this paper the source of the scaling error was identified as a software bug in AIPS and has since been resolved.}.

\section{Survey annuli, survey depths, and source selection}
\label{sec:p2selection}
We split the survey of each field into six annuli based on the radial distance from the correlation phase centre of that field. These are referred to as the $0\arcdeg-0.25\arcdeg$, $0.25\arcdeg-0.5\arcdeg$, $0.5\arcdeg-1\arcdeg$, $1\arcdeg-1.5\arcdeg$, $1.5\arcdeg-2\arcdeg$, and $2\arcdeg-3\arcdeg$ annuli in field 1 (centred on J0226$+$3421) and field 2 (centred on B0218$+$357). Our survey attempts to detect sources at large radial distances from the antenna pointing position and the correlation phase centre. To reduce the effect of bandwidth and time-averaging smearing, increasingly restrictive $(u,v)$ ranges are employed in the outer annuli. Furthermore, the fall-off of the response of the primary beam is an effect that significantly limits the sensitivity within each annulus. In particular, WSRT and JB have significantly narrower primary beams ($\sim1\arcmin$ and $\sim0.5\arcdeg$ HWHM respectively), owing to their larger effective aperture, compared to the VLBA ($\sim1.3\arcdeg$ HWHM). As such, the WSRT data were only used to image the source directly at the phase centre, whilst the JB observatory data was only used in the $0\arcdeg-0.25\arcdeg$ annulus (restrictions in $(u,v)$ range effectively excludes the JB data from the $0.25\arcdeg-0.5\arcdeg$ annulus even though it has a significant response within this annulus).

Since only the VLBA antennas were used outside the $0\arcdeg-0.25\arcdeg$ annulus, the reduced response in the other annuli is composed of only three independent components: the VLBA primary beam response ($R_{VLBA}$) and the reduced response due to bandwidth and time-averaging smearing ($R_{bw}$, $R_{t}$). The combined reduced response, $R$, is given by $R=R_{bw}R_{t}R_{VLBA}$. We have estimated $R_{bw}$ and $R_{t}$ following \citet{Bridle:1999p10564} and have adopted a fitted function used to model the VLA antennas, as documented in the AIPS task PBCOR, to model the primary beam response of the VLBA under the assumption that the 25 m antennas in these arrays have a similar response \citep{Garrett:2005p10555}. In Table \ref{tab:tabp2t1}, we calculate the total response and estimated $1\sigma$ rms noise at the outer edge of each annulus of the two surveyed fields. The $(u,v)$ range has been restricted to limit the effects of bandwidth and time-averaging smearing to at most a few percent.

\begin{table}[ht]
\begin{center}
{ \tiny
\begin{tabular}{lcccccccc} \hline \hline
Survey                 & Annulus  & Maximum       & Response           & $1\sigma$ rms          & Survey     & $S_{P}$\tablenotemark{b} & $N_{WENSS}$ & $\langle N_{VLBI}\rangle$\tablenotemark{c} \\
Field\tablenotemark{a} & Range    & $(u,v)$ Range & R\tablenotemark{b} & Noise\tablenotemark{b} & Resolution &                          &             & \\
                       & (arcdeg) & (M$\lambda$)  &                    & (mJy beam$^{-1}$)      & (mas, mas) & (mJy beam$^{-1}$)        &             & \\ \hline \hline
1 & $0.00-0.25$  & 2.50   & 0.94  & 3.7   & $40\times20$    & $>24$  &  4    &     3     \\
1 & $0.25-0.50$  & 1.00   & 0.89  & 4.8   & $140\times130$  & $>32$  &  15   &     10    \\
1 & $0.50-1.00$  & 0.75   & 0.65  & 6.8   & $180\times170$  & $>63$  &  55   &     27    \\
1 & $1.00-1.50$  & 0.75   & 0.36  & 8.9   & $230\times220$  & $>147$ &  91   &     28    \\
1 & $1.50-2.00$  & 0.50   & 0.17  & 11.0  & $350\times290$  & $>378$ &  128  &     16    \\
1 & $2.00-3.00$  & 0.50   & 0.06  & 15.1  & $360\times290$  & $>1603$&  411  &     11    \\ \hline
2 & $0.00-0.25$  & 2.50   & 0.94  & 1.9   & $90\times80$    & $>12$  &  2    &     2     \\
2 & $0.25-0.50$  & 1.00   & 0.89  & 2.3   & $140\times130$  & $>16$  &  15   &     15    \\
2 & $0.50-1.00$  & 0.75   & 0.65  & 3.2   & $180\times170$  & $>30$  &  50   &     44    \\
2 & $1.00-1.50$  & 0.75   & 0.36  & 4.1   & $230\times220$  & $>68$  &  97   &     52    \\
2 & $1.50-2.00$  & 0.50   & 0.17  & 5.1   & $350\times290$  & $>174$ &  128  &     41    \\
2 & $2.00-3.00$  & 0.50   & 0.06  & 6.9   & $360\times290$  & $>749$ &  375  &     23    \\ \hline
\tablenotetext{a}{Field 1 is centred about J0226$+$3421 and field 2 is centred about B0218$+$357.}
\tablenotetext{b}{The estimated worst case values at the edge of the annulus.}
\tablenotetext{c}{The number of WENSS sources that, if unresolved, have a peak flux that would fall above the estimated VLBI sensitivity limit.}
\end{tabular}
\caption{Survey fields, depths and source counts at 324 MHz.}
\label{tab:tabp2t1}
}
\end{center}
\end{table}

The WENSS catalogue was used as a guide for potential targets in the survey. Table \ref{tab:tabp2t1} lists the total number of WENSS sources that exist within each annulus of each field. For a WENSS source to be detected by our survey it must have a peak flux density, $S_{P}($WENSS$)$, that satisfies the constraint $S_{P}($WENSS$)>6\sigma R^{-1}$. Estimates of this limit at the edge of each annulus, $S_{P}$, and the number of WENSS sources that meet this constraint, $\langle N_{VLBI}\rangle$, are listed in Table \ref{tab:tabp2t1}. Even though it was estimated that many of the WENSS sources would fall below our detection limits, for completeness, we targeted all WENSS sources in the $0\arcdeg-2\arcdeg$ annuli of each field given the possibility that some sources might exhibit strong variability. Between $2\arcdeg-3\arcdeg$ only candidate sources that were within our sensitivity limits were targeted.

\section{Results}

\begin{figure}[p]
\epsscale{1.0}
\plotone{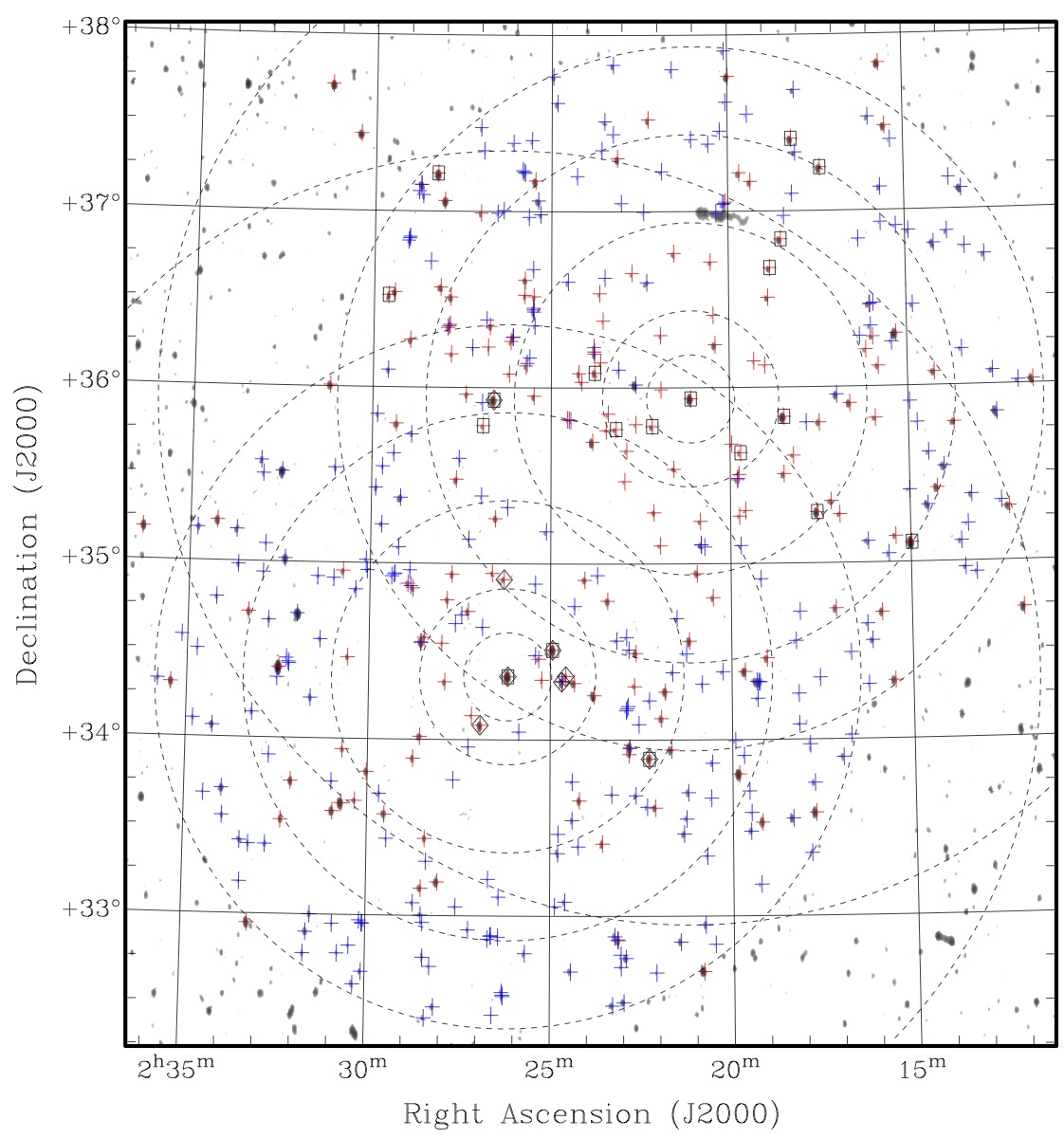}
\caption[Gray-scale WENSS image of the two fields surveyed by the VLBA, JB and WSRT VLBI observation.]{Gray-scale WENSS image of the two fields surveyed by the VLBA, JB and WSRT VLBI observation. The dashed circles define the six sub-fields that are co-located on the phase centre of each field of the VLBI observation. The crosses located across the image denote all WENSS sources that were targeted by the VLBI observation, red crosses mark unresolved WENSS sources with peak flux densities above the VLBI sensitivity limit and green crosses mark sources that were resolved in WENSS or fell below the VLBI sensitivity limit. Targets that are boxed identify VLBI sources detections in the B0218$+$357 field and those that are contained within a diamond identify VLBI detections in the J0226$+$3421 field. The four sources that were commonly detected within both of the fields and are marked within a box and diamond.}
\label{fig:figp2f1}            
\end{figure}

A total of 618 WENSS sources were targeted by the 90 cm wide-field VLBI survey at radial distances of up to $2.89\arcdeg$ from the phase centre of the survey fields (Figure \ref{fig:figp2f1}). The complete survey of all WENSS sources within the inner $0\arcdeg-2\arcdeg$ annulus of each field did not detect any source that had peak flux below our sensitivity limit, or that was partially resolved in WENSS. The combined total area imaged around each of the targeted sources in these fields represents $\sim0.5\%$ of the area surveyed.

Of all of the WENSS sources targeted, a total of 272 sources, 95 in the J0226$+$3421 field (field 1) and 177 in the more sensitive B0218$+$357 field (field 2), would have peak flux densities above our VLBI detection limits, if unresolved by our VLBI observations. The WENSS characteristics of these sources (position, peak flux density per solid beam angle, integrated flux and where available, the WENSS/NVSS spectral index $\alpha$ where $S_{\nu}\propto\nu^{\alpha}$) are listed in Tables \ref{tab:tabf1} and \ref{tab:tabf2} for fields 1 and 2, respectively. Where the target source is also detected by the VLBI survey, the corrected source position (see \S~\ref{sec:p2vlbical}), VLBI peak flux density per solid beam angle and integrated flux are listed in the following row. The VLBI peak and integrated flux density have been corrected for the primary beam response but not for bandwidth and time-averaging smearing losses. Based on these losses and uncertainties in the amplitude calibration, we estimate the absolute flux density scales of the VLBI observations to be better than $10\%$ in the inner $0.5\arcdeg$ of each field and better than $20\%$ elsewhere.

A total of 27 sources were detected and imaged by the survey, eight of these sources detected in the J0226$+$3421 field (field 1) and the remaining 19 detected in the more sensitive B0218$+$357 field (field 2). Nine of the sources were detected outside of the half-power point of the VLBA primary beam (HWHM $\sim1.3\arcdeg$). Four sources, B0223.1$+$3408, B0221.9$+$3417, B0219.2$+$3339 and B0223.5$+$3542, were detected in both fields in the region of overlap. B0219.2+3339, in particular, was detected at $2.06\arcdeg$ from the phase centre of the second field and is well past the quarter power point of the VLBA primary beam, it is also the only source detected in the outer annulus of the survey. Table \ref{tab:tabp2t2} lists all of the detected sources, the detection field, the distance from the phase centre of that field, the restoring beam size and the one sigma residual RMS noise. A positional comparison of our corrected source positions is also made with respect to the best known radio positions, where $d_{E}$ and $\theta_{E}$ are the observed offset and position angle from this position, and $d_{\sigma}$ is the offset in terms of the combined one sigma position error of the two compared positions.

The VLBI positions in the $0\arcdeg-0.25\arcdeg$ and $0.25\arcdeg-0.5\arcdeg$ annuli are limited by errors introduced by the ionosphere.  We estimate a one sigma error of $\sim3-12$ mas in each co-ordinate. The positional accuracy of the outer, heavily tapered, annuli are further limited by RMS noise errors. We estimate the one sigma error in each co-ordinate of these outer fields to be better than 15 mas, 20 mas and 30 mas for the $0.5\arcdeg-1\arcdeg$, $1\arcdeg-1.5\arcdeg$ and $1.5\arcdeg-2\arcdeg$ annuli, respectively. The positional accuracy of B0219.2+3339 is also expected to be $\sim30$ mas as it is located quite close to the $2\arcdeg$ boundary. After a correction was applied for an apparent scaling effect in the image (see \S~\ref{sec:p2vlbical}), residual errors of 40 mas, 100 mas, 200 mas and 180 mas were measured for the cross-field detections of sources B0223.1$+$3408, B0221.9$+$3417, B0219.2$+$3339 and B0223.5$+$3542 respectively. These extremely wide-field sources exhibited significant ionospheric phase fluctuations that distorted their initial dirty images. While the fluctuations were partially corrected for with phase self-calibration, it is believed that this may have adversely affected the position accuracy.

Contour maps for each of the sources detected in field 1 are shown in Figure \ref{fig:figp2f2} and those detected in field 2 are shown in Figure \ref{fig:figp2f3}. While not formally part of the wide-field survey, the fringe finder 3C84 has also been imaged and is shown in Figure \ref{fig:figp2f4}.

The total survey required nearly six weeks of processing on a single 2 GHz computer, required $\sim200$ gigabytes of workspace, and generated a total of $\sim20$ gigabytes of image data. We estimate that six years of processing would be required to completely image the FWHM beam of the VLBA at 320 MHz using similar techniques. Fortunately, the problem can be easily broken down to run efficiently in the parallel environment of a supercomputing cluster. With a basic 100-node cluster, the entire FWHM beam of the VLBA could be imaged within three weeks using a simple brute force method that would image targets on individual processors. For a single field, the cluster would generate mosaic of the primary beam comprising of $\sim1$ terabyte of image data. More elaborate algorithms may be employed to improve the efficiency of this processing further, for example, by using a recursive approach that creates successively smaller sub-fields by performing a combination of (u-v) shifting and data averaging.

\subsection{Comments on individual sources}

\subsubsection{3C84}
For 3C84, we measure a VLBI peak flux density of 2.33 Jy beam$^{-1}$ and an integrated flux of 6.07 Jy, whereas the WENSS peak flux density and integrated flux is 19.396 Jy beam$^{-1}$ and 42.8 Jy respectively. This is the only source in our sample of imaged sources with extended structure in WENSS, where it has an estimated size of $115\arcsec\times84\arcsec$ at a position angle of $115\arcdeg$. In our VLBI image we have recovered $\sim14\%$ of the WENSS flux. Similar VLBI observations at 327 MHz, with a larger synthesised beam, measure a slightly greater integrated flux of 7.47 Jy \citep{Ananthakrishnan:1989p9842} suggesting that the missing flux is most likely related to the larger-scale structure that is resolved out by VLBI observations. We measure a Largest Angular Size (LAS) of 150 mas which corresponds to a Largest Linear Size (LLS) of $\sim50$ pc at its measured redshift of $z=0.017559\pm0.000037$ \citep{Strauss:1992p9719}. A 15 GHz VLBA contour map \citep{Lister:2005p10561} is shown overlaid with our 90 cm image of 3C84 in Figure \ref{fig:figp2f4}. The smaller scale structures within this image appear to align with the jet-like feature that appears in our 90 cm image and extends 100 mas to the south of the core.

\subsubsection{B0223.1+3408 (J0226$+$3421, 4C$+$34.07)}
The quasar, J0226$+$3421, was imaged with the full $(u,v)$ range and recovers approximately 80\% of the WENSS flux. The contour map of this source is shown in Figure \ref{fig:figp2f2}(a) and is overlaid with a naturally weighted 2 cm A$-$configuration VLA with Pie Town contour map (Wucknitz et al., in preparation). The source is dominated by a bright core ($\sim0.85$ Jy) and an extended lobe to the west ($\sim1.6$ Jy). A weaker lobe appears to the north ($\sim0.28$ Jy) and a partially resolved hot spot ($\sim0.14$ Jy) approximately mid-way between the core and the western lobe. The source has a LAS of $1.15\arcsec$ which corresponds to a LLS of $\sim9$ kpc at its measured redshift of $z=2.91\pm0.002$ \citep{Willott:1998p9363}. All of the large-scale structures observed at 90 cm with VLBI are also clearly detected at 2 cm with the VLA and Pie Town. MERLIN$+$VLBI images at 18 cm \citep{Dallacasa:1995p10568} detect the core and western lobe with large-scale structure and positions that are consistent with our image, however, their observations do not detect the northern lobe or hot spot.   The source was also detected in the second field and is shown in Figure \ref{fig:figp2f3}(q) with an image of the field 1 source restored using the same beam. An offset of 40 mas is observed between the field 1 and field 2 source after correcting for the larger-scale offset described in \S~\ref{sec:p2vlbical}.

\subsubsection{B0218.0+3542 (B0218$+$357)}
B0218.0$+$3542 is a gravitational lens that has been mapped at higher frequencies \citep[e.g.][]{Biggs:2001p10528,Wucknitz:2004p9221}, with VLBI \citep[e.g.][]{Biggs:2003p10529}, and at various wavelengths by \citet{Mittal:2006p10563}. The source is the smallest known Einstein radio ring \citep{Patnaik:1993p10575}. As this source was the main target of the original observation, it is placed in the most sensitive field and annulus of the survey and has been imaged with the full $(u,v)$ range. Our image of the source, Figure \ref{fig:figp2f3}(a), has been restored with a beam that is $\sim4$ times larger than normal to highlight the large-scale structure within the source. The source has a LAS of 690 mas and is dominated by the A and B lensed images to the west and east, respectively. The two images are separated by $\sim340$ mas and appear to have weaker components that are mirrored on either side of the lens. These weaker components may be a small portion of a lensed jet that is tangentially stretched. Our observations recover $\sim54\%$ of the WENSS flux suggesting the presence of structures that are fully resolved out even with our shortest baselines. L-Band VLA images of the source seem to suggest that there is indeed a larger-scale emission surrounding the source \citep{ODea:1992p10554}. The measured redshift of the lensing galaxy $z=0.68466\pm0.00004$ \citep{Browne:1993p10565} and that of the lensed object is $z=0.944\pm0.002$ \citep{Cohen:2003p10538}.   B0218.0+3542 will be studied in greater detail with the high sensitivity, narrow-field observations of this source at 327 MHz and at 610 MHz by Wucknitz et al. (in preparation).


\subsubsection{B0221.9$+$3417}
The VLBI source we detect within the B0221.9$+$3417 field, Figure \ref{fig:figp2f2}(b), is offset by $9.94\arcsec$ and at a position angle of $146\arcdeg$ compared to the WENSS position. The source is also detected in field 2, Figure \ref{fig:figp2f3}(o). The separation between both detections is within 100 mas, after correcting for the larger-scale offset described in \S~\ref{sec:p2vlbical}, and both have a similar flux density confirming that the detected source is indeed at this position. The position offset also exists when compared against the same source in NVSS, the 365 MHz Texas survey \citep{Douglas:1996p10548} and VLSS \citep{Cohen:2007p10539}. As only 11\% of the WENSS flux was recovered by the VLBI observation, the position offset hints at a larger component $\sim10\arcsec$ to the north-west of the VLBI source. Furthermore, NVSS lists a fitted source size with a major axis of $16\arcsec$ and the Texas survey categorises the source as a symmetric double with a component separation of $13\pm2\arcsec$ at a position angle of $155\pm11\arcdeg$. These observations are consistent with the VLBI observation if we assume a compact south-eastern component has been detected. The VLBI source has a weaker component 360 mas to the north-west that is directly in line with the WENSS source. We measure a LAS of 360 mas for the VLBI source and an LLS 2.8 kpc at its measured redshift of $0.852\pm0.002$ \citep{Willott:2002p9292}.

\subsubsection{B0221.6$+$3406B}
B0221.6$+$3406B, as shown in Figure \ref{fig:figp2f2}(c), has a complex morphology. The source has an LAS of 830 mas which corresponds to a LLS of 6.7 kpc at a measured redshift of $z=2.195\pm0.003$ \citep{Willott:1998p9363}. Our VLBI observations have recovered $\sim90\%$ of the WENSS flux suggesting that there is little or no extended structure above what has already been imaged. Based on the integrated flux densities in WENSS and NVSS, the source has a spectral index of $-0.93$. 

\subsubsection{B0223.9$+$3351}
B0223.9$+$3351, as shown in Figure \ref{fig:figp2f2}(d), appears to be an AGN with a 180 mas jet extension to the north. The source has a LAS of 420 mas which corresponds to a LLS of 3.5 kpc at a measured redshift of $z=1.245\pm0.004$ \citep{Willott:2002p9292}. Our VLBI observations have recovered $\sim50\%$ of the WENSS flux. 




\subsubsection{B0215.4$+$3536}
B0215.4$+$3536 is an ultra-steep spectrum source, a characteristic that is an excellent tracer of galaxies at redshifts $z\geq2$ \citep[e.g.][and references therein]{Roettgering:1994p10579}. The source has a LAS of $7.1\arcsec$ which corresponds to a LLS of $\geq60$ kpc for a redshift of $z\geq2$. Figure \ref{fig:figp2f3}(e) shows our 90 cm VLBI image overlaid with an L-Band VLA image \citep{Roettgering:1994p10579}. The core and peaks within the two lobes align very closely, to within $0.1\sigma$, with the VLA image. A compact component of emission, possibly a jet interaction region, is detected by our observations approximately mid-way between the core and the south-west lobe and appears to align with extended edge of that lobe. Approximately 50\% of the WENSS flux is recovered by our observation, the remaining flux is likely associated with the extended lobes and is resolved out by our observation.




\subsubsection{B0219.2$+$3339}
\begin{figure}[p]
\epsscale{0.45}
\begin{center}
\mbox{
\plotone{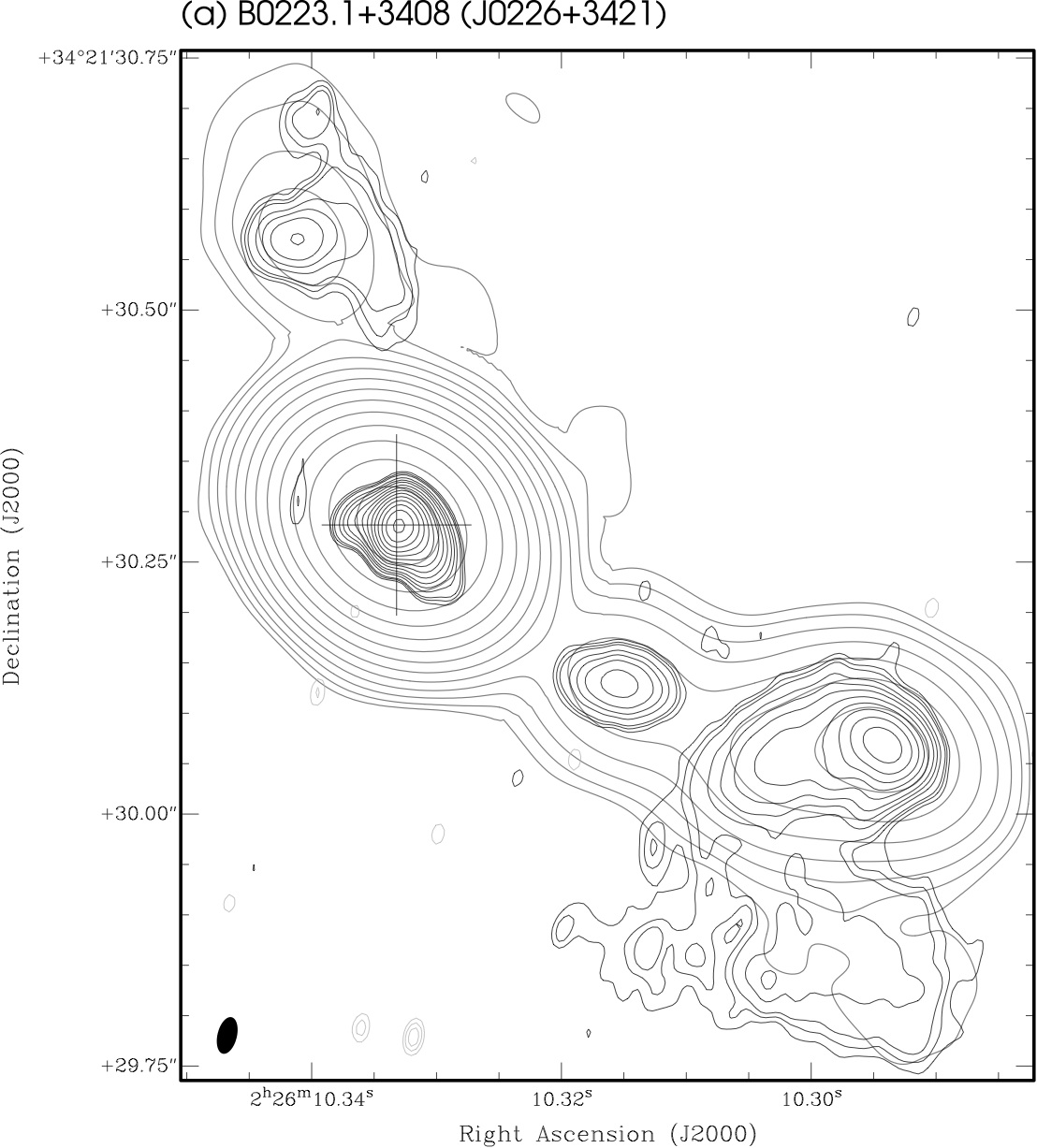} \quad
\plotone{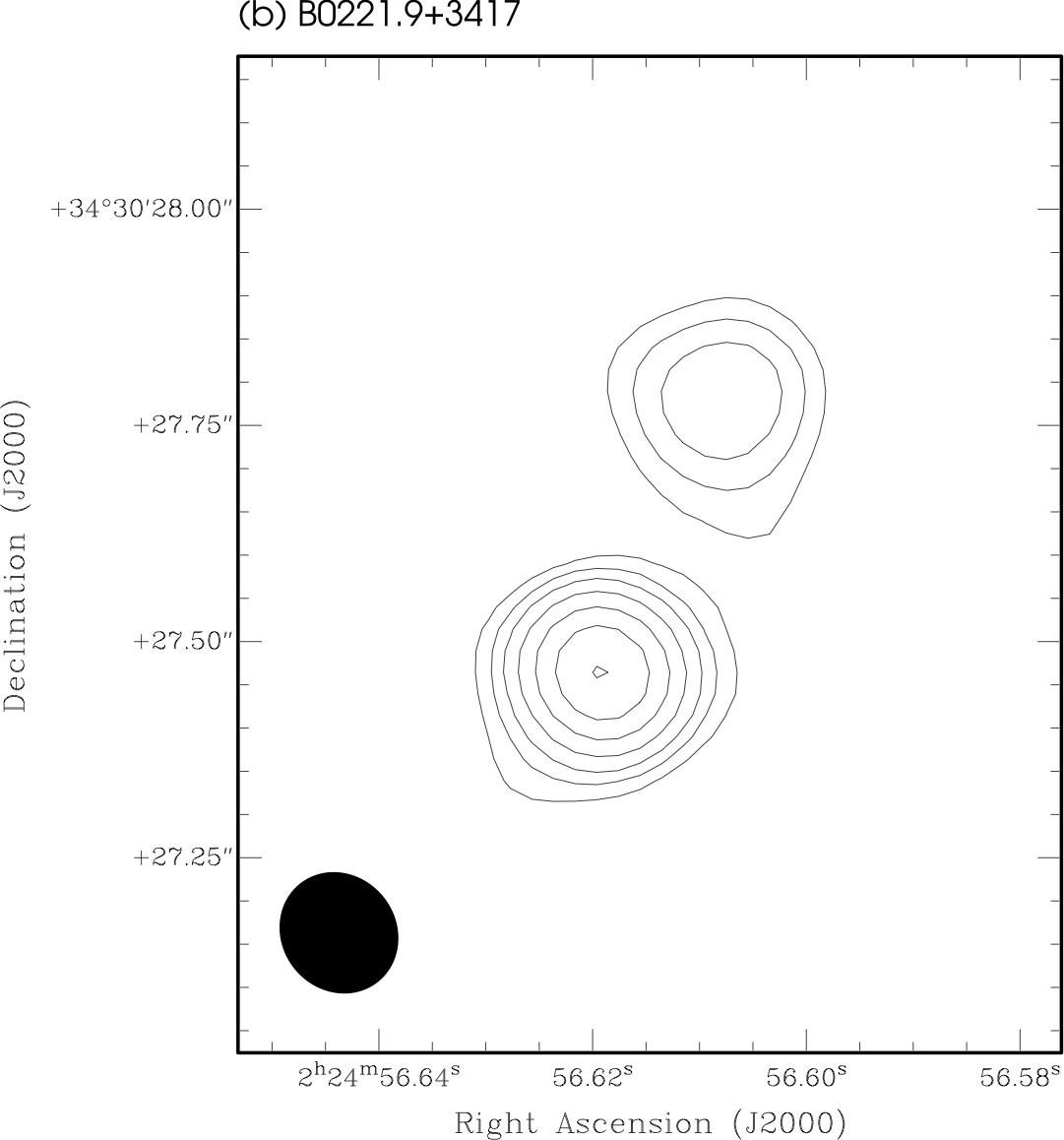}
}
\mbox{
\plotone{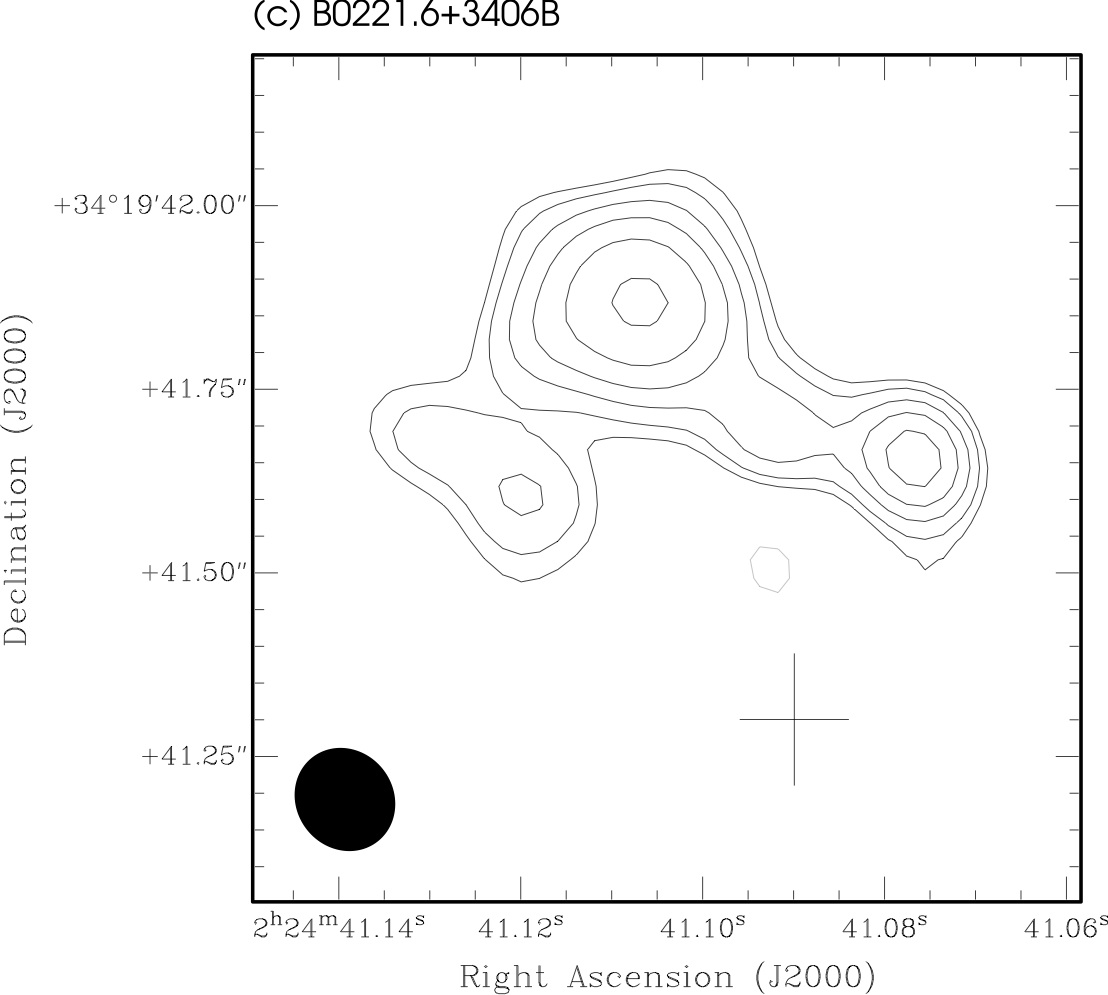} \quad
\plotone{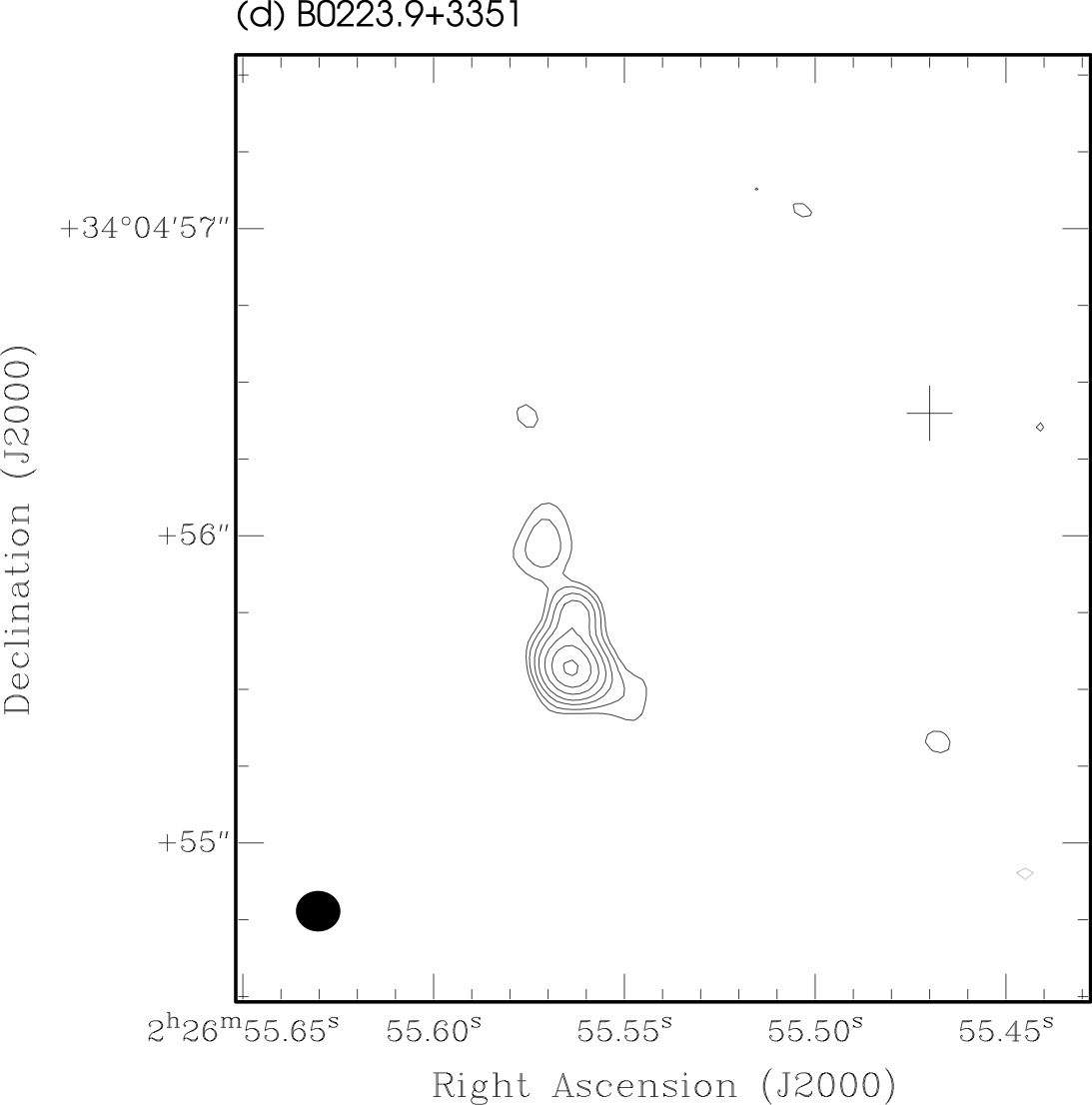}
}
\caption[Naturally weighted VLBI images of field 1 sources.]{Naturally weighted VLBI images of field 1 sources. Contours are drawn at $\pm2^{0}, \pm2^{\frac{1}{2}}, \pm2^{1}, \pm2^{\frac{3}{2}}, \cdots$ times the $3\sigma$ rms noise except for B0223.1$+$3408 where the lowest contour is at $1.5\sigma$ rms noise. Restoring beam and rms image noise for all images can be found in Table \ref{tab:tabp2t2}. Crosses mark the best known radio positions (see Table \ref{tab:tabp2t2}). Notes for sub-figures (a) Grey contours: 2 cm VLA $+$ Pie Town contour map; $1\sigma$ RMS noise is 0.053 mJy beam$^{-1}$; contours at $\pm1, \pm2, \pm4, \pm8, \cdots$ times the $3\sigma$ rms noise; beam size is $118\times96$ mas at P.A.$=51\arcdeg$; (h) Grey contours: Field 2 detection of source restored using the same beam as the field 1 source.}
\label{fig:figp2f2}
\end{center}
\end{figure}

\begin{figure}[ht]
\epsscale{0.45}
\begin{center}
\mbox{
\plotone{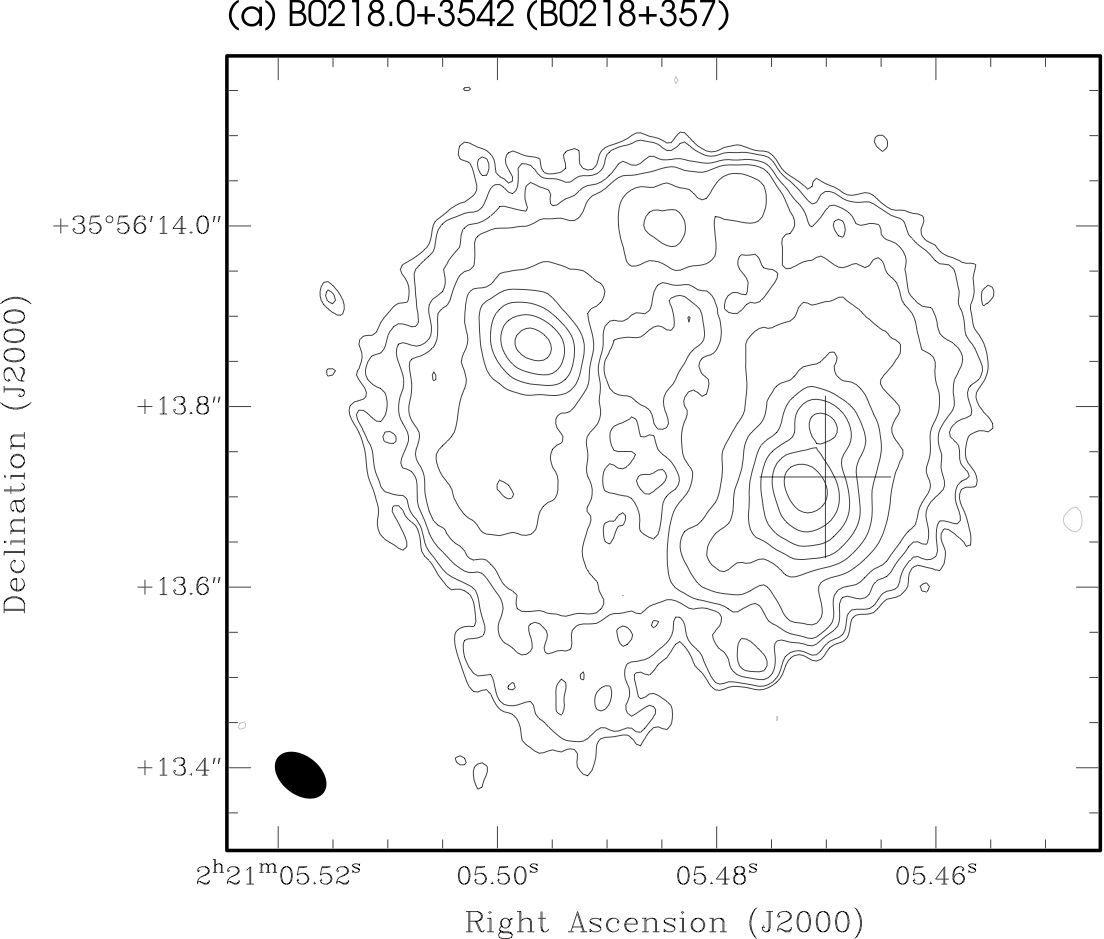} \quad
\plotone{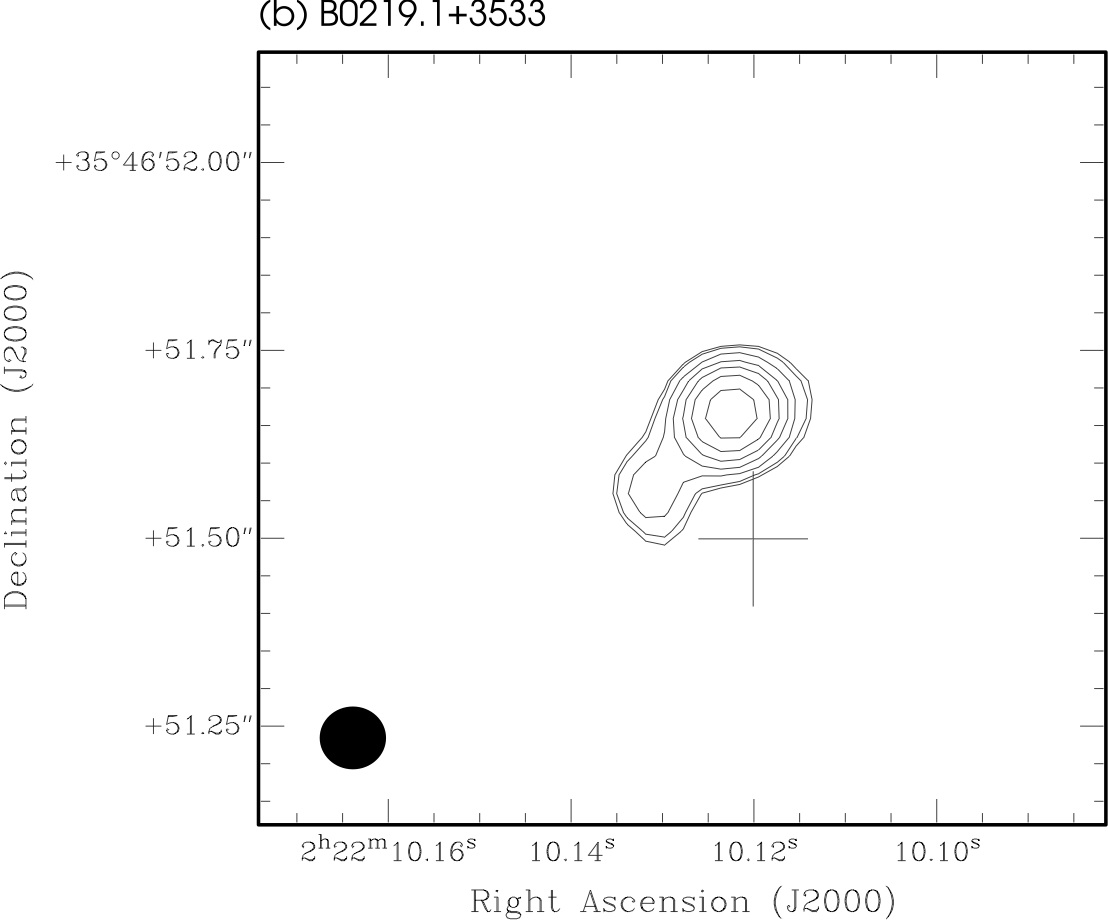}
}
\mbox{
\plotone{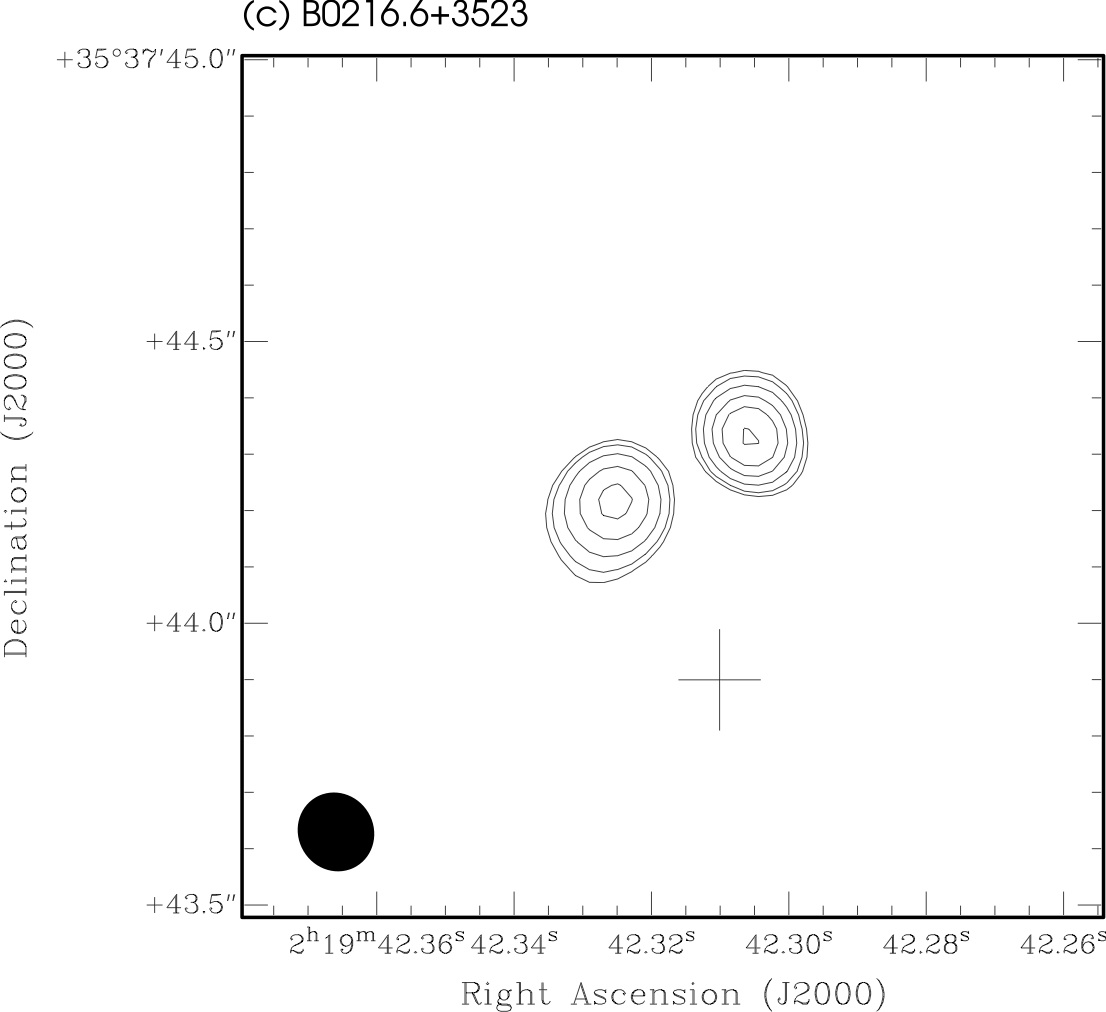} \quad
\plotone{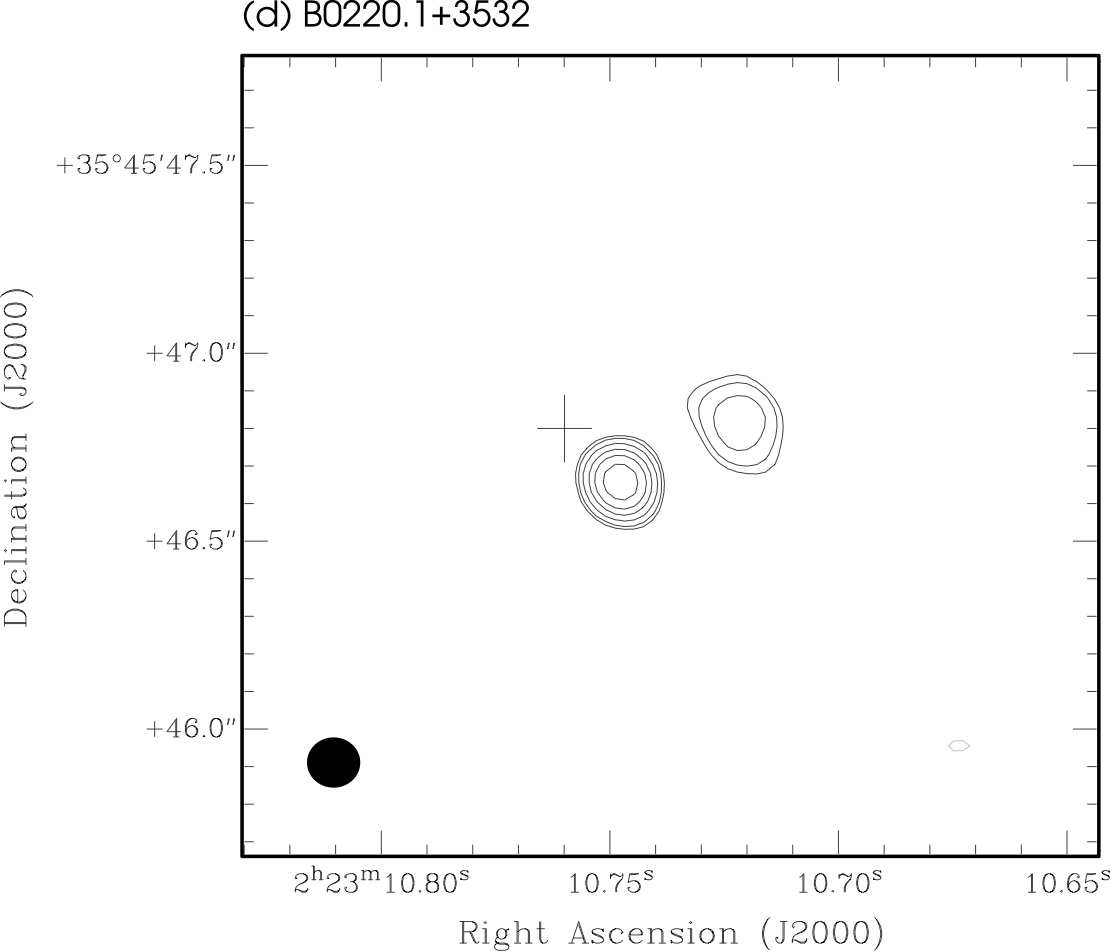}
}
\caption[Naturally weighted VLBI images of field 2 sources.]{Naturally weighted VLBI images of field 2 sources. Contours are drawn at $\pm2^{0}, \pm2^{\frac{1}{2}}, \pm2^{1}, \pm2^{\frac{3}{2}}, \cdots$ times the $3\sigma$ rms noise. Restoring beam and rms image noise for all images can be found in Table \ref{tab:tabp2t2}. Crosses mark the best known radio positions (see Table \ref{tab:tabp2t2}). Notes for sub-figures (e) and (h) Grey contours: 1465 MHz VLA contour map \citep{Roettgering:1994p10579}; (o), (q) and (s) Grey contours: Field 1 detection of source restored using the same beam as the field 2 source.}
\label{fig:figp2f3}            
\end{center}
\end{figure}

\begin{figure}[ht]
\epsscale{0.4}
\plotone{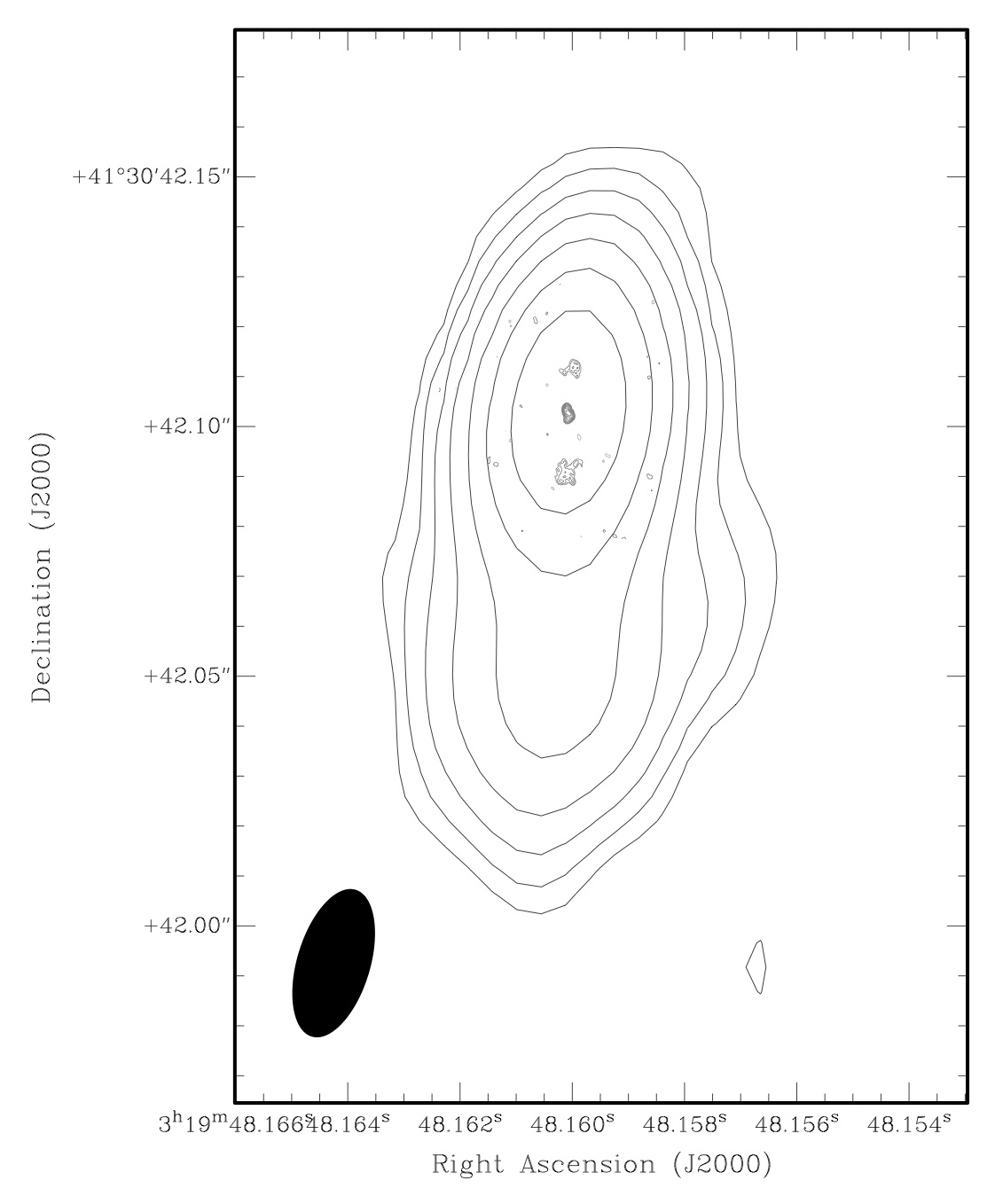}
\caption[Naturally weighted VLBI image of 3C84.]{Naturally weighted VLBI image of 3C84. Contours drawn at $\pm1, \pm2, \pm4, \cdots$ times the $3\sigma$ rms noise. Restoring beam and rms image noise for all images can be found in Table \ref{tab:tabp2t2}. A 15 GHz VLBA contour map is shown in grey \citep{Lister:2005p10561} with a peak flux density of 2.65 Jy beam$^{-1}$ and an integrated flux of 10.64 Jy. Contours are shown at 10, 20, 40, 80,$\cdots$, 2560 mJy beam$^{-1}$ and the restoring beam is $0.69\times0.55$ mas at a position angle of $2\arcdeg$.}
\label{fig:figp2f4}
\end{figure}

B0219.2$+$3339 is detected in both field 1 and field 2 and is shown in Figures \ref{fig:figp2f2}(g) and \ref{fig:figp2f3}(s), respectively. A significant residual offset of 200 mas exists between these two independent detections even after correcting for the larger-scale offset described in \S~\ref{sec:p2vlbical}. The phases of this source in the J0226$+$3421 field appeared to be more heavily affected by the ionosphere than those in the B0218$+$357 field and it is believed that the offset may have been introduced by the phase self-calibration process. The source has a LAS of 900 mas which corresponds to a LLS of 6.6 kpc for a measured redshift of $z=0.752\pm0.002$ \citep{Willott:2002p9292}.

\clearpage
\clearpage
\epsscale{0.45}
\begin{center}
\mbox{
\plotone{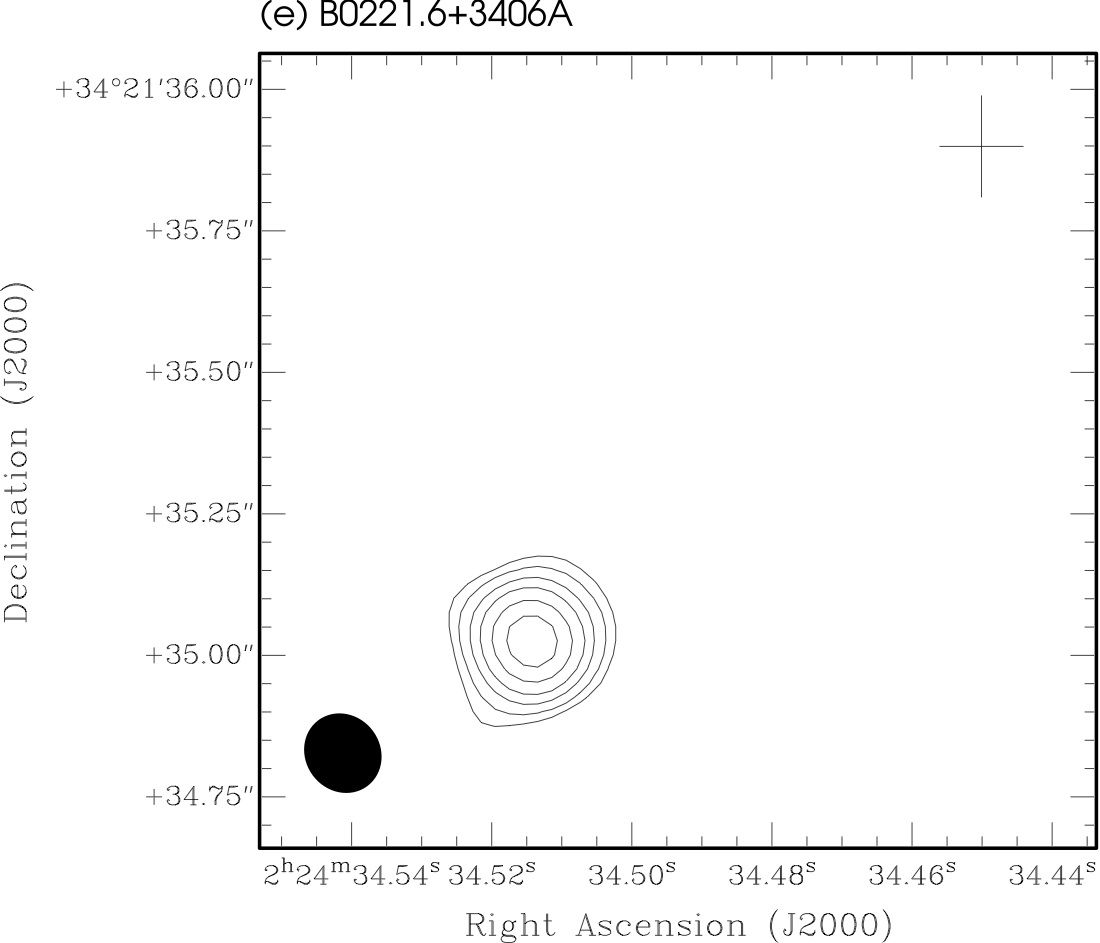} \quad
\plotone{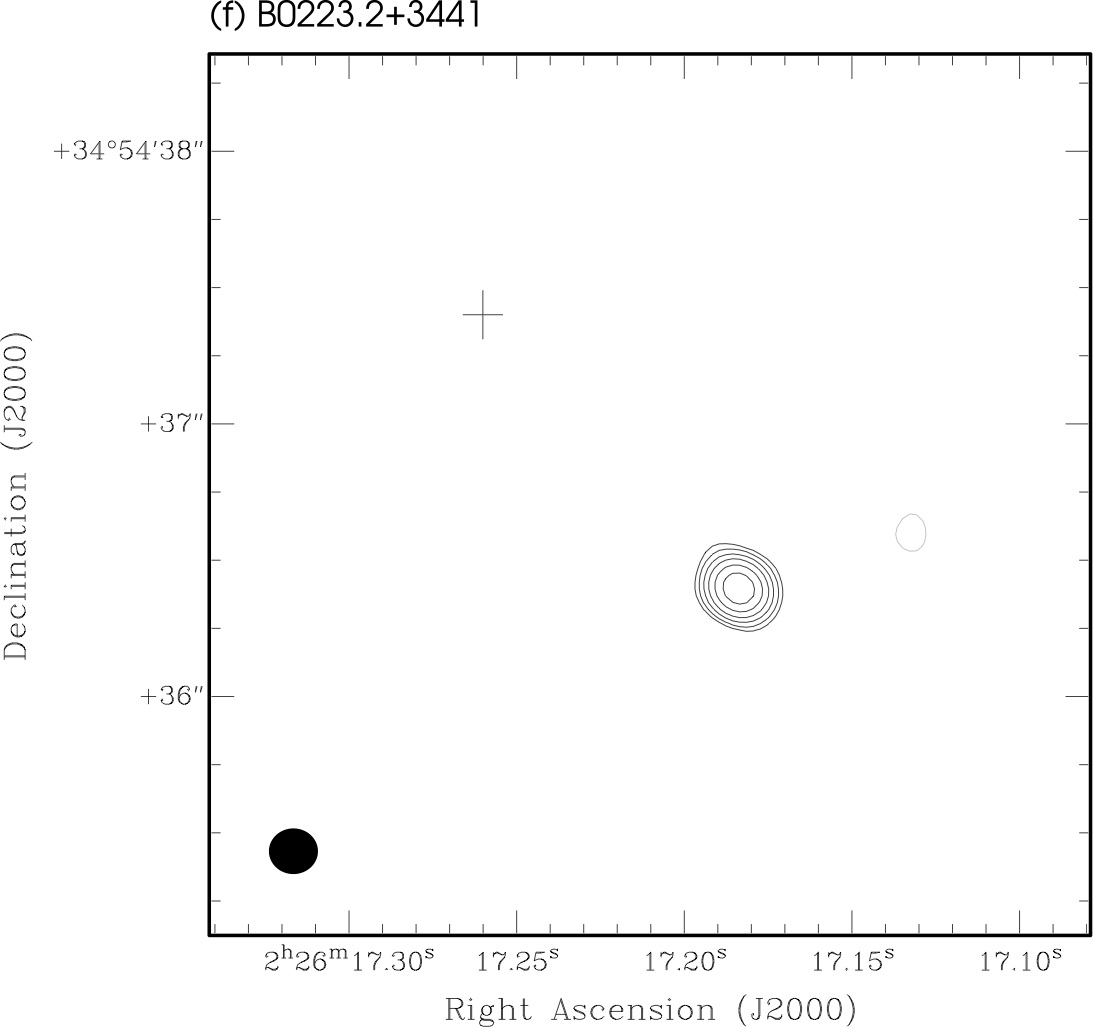}
}\\[5mm]
\mbox{
\plotone{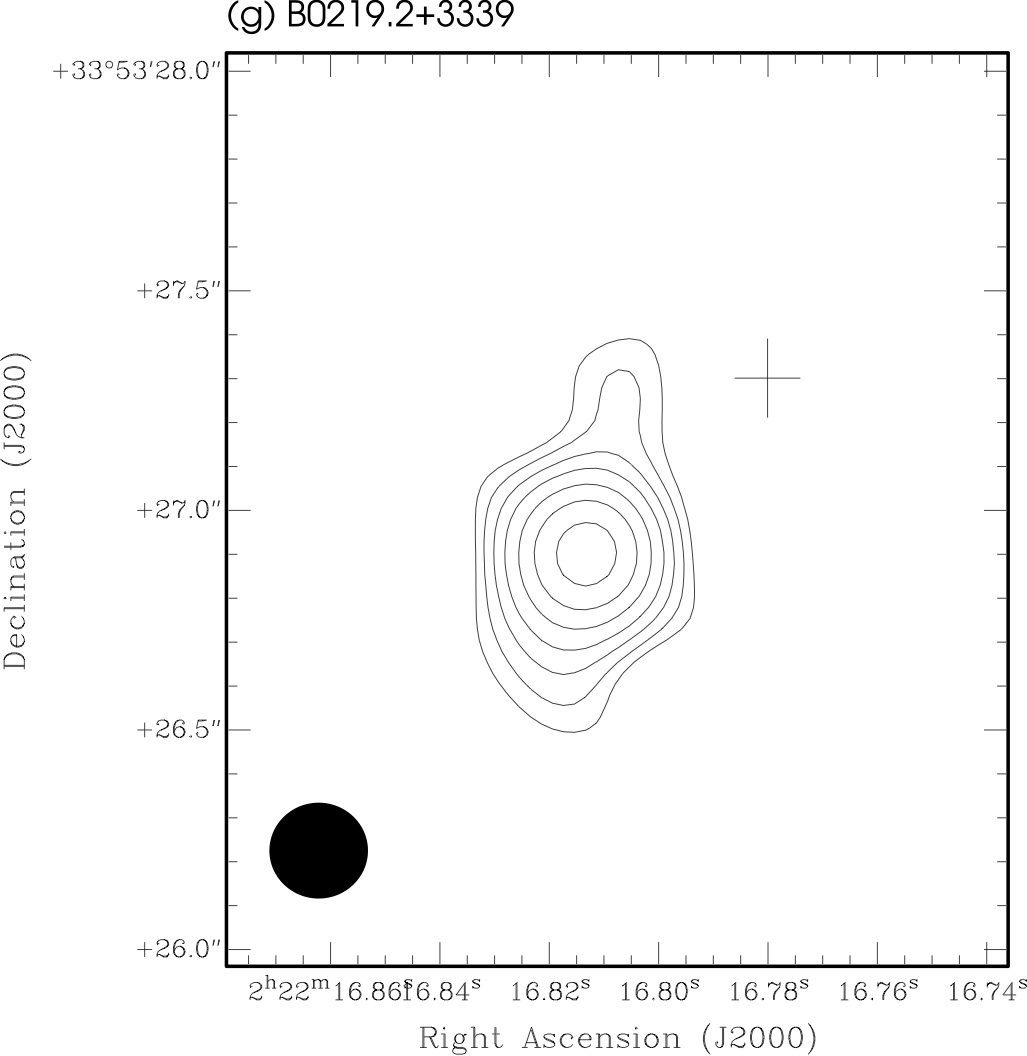} \quad
\plotone{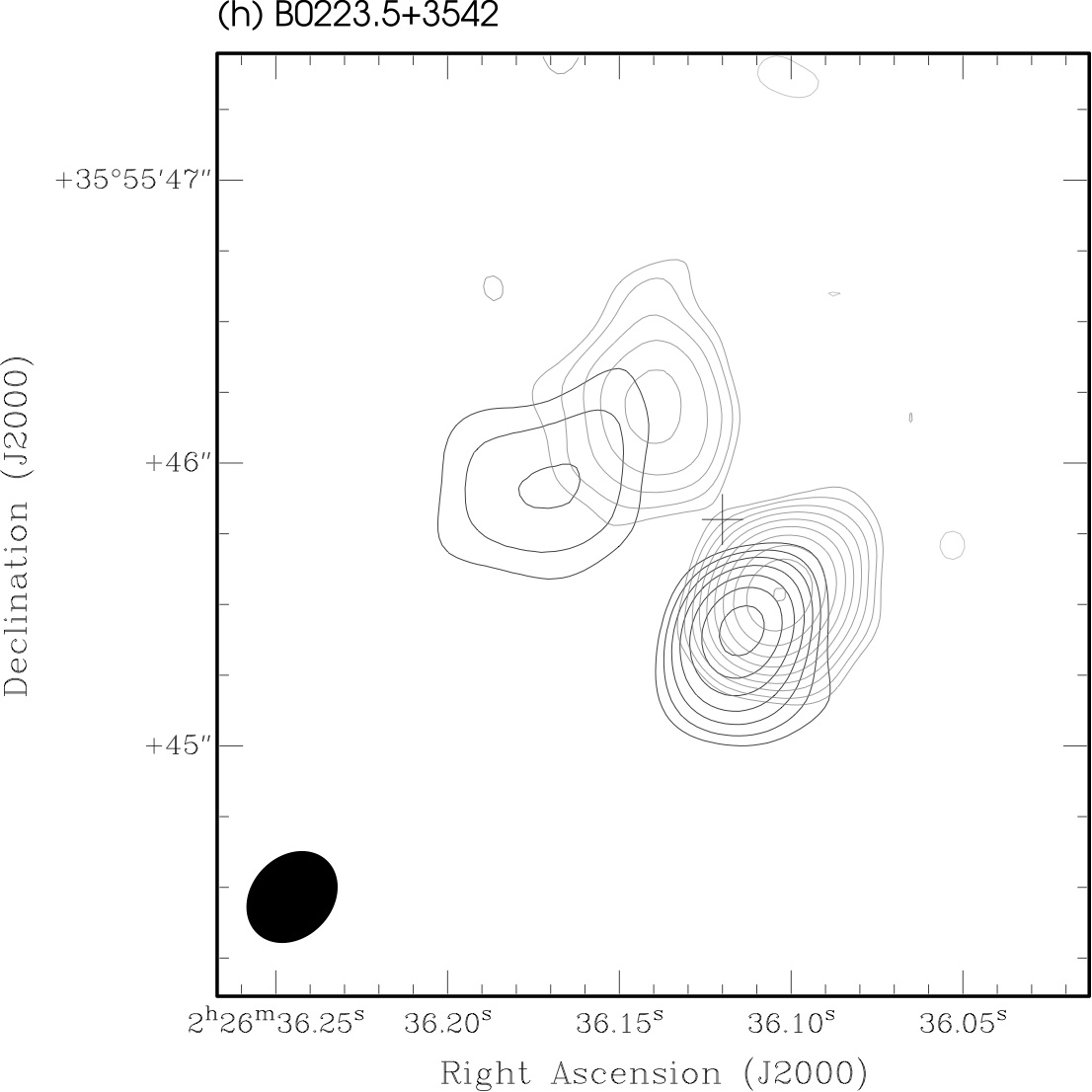}
}\\[5mm]
{Fig. 9.2. --- Continued}
\end{center}
\clearpage

\clearpage
\epsscale{0.45}
\begin{center}
\mbox{
\plotone{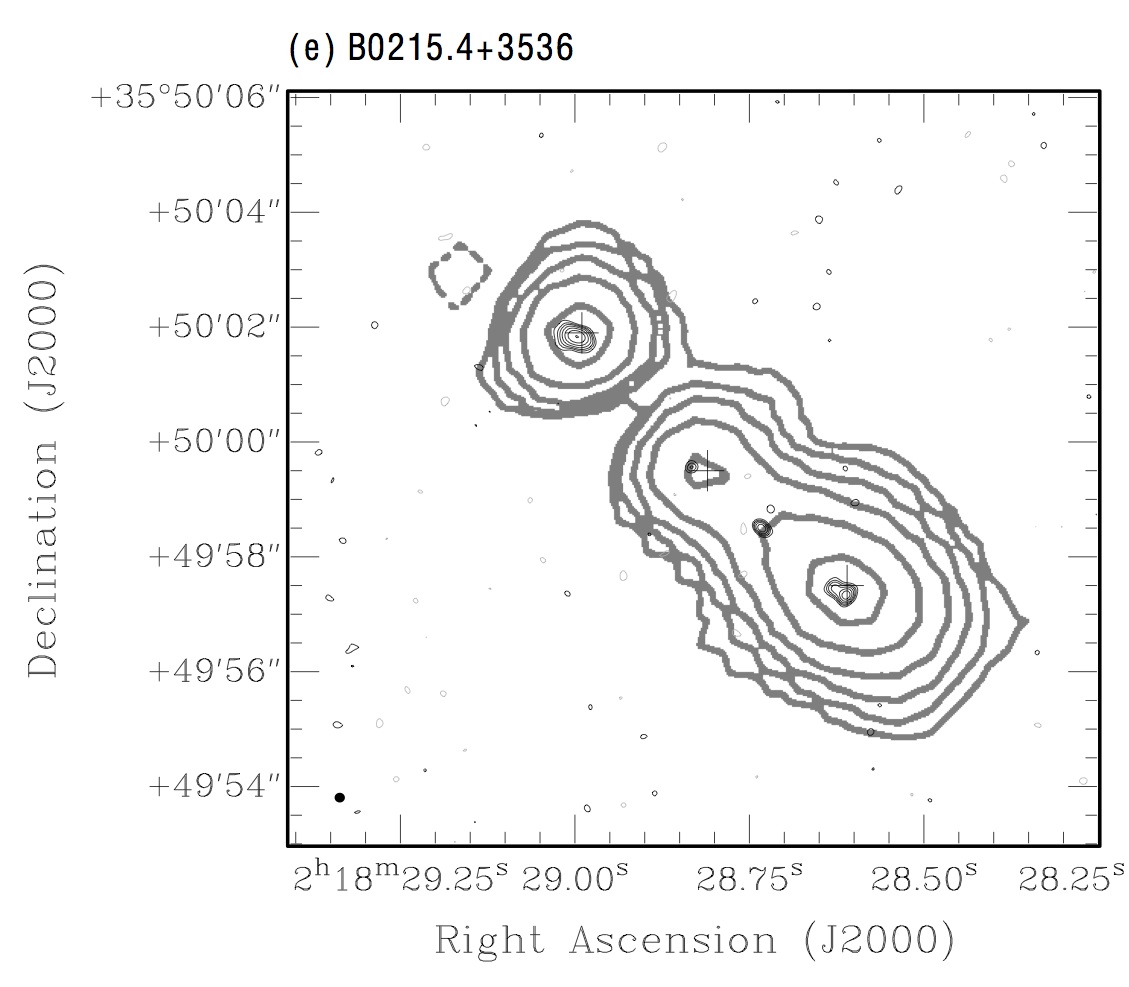} \quad
\plotone{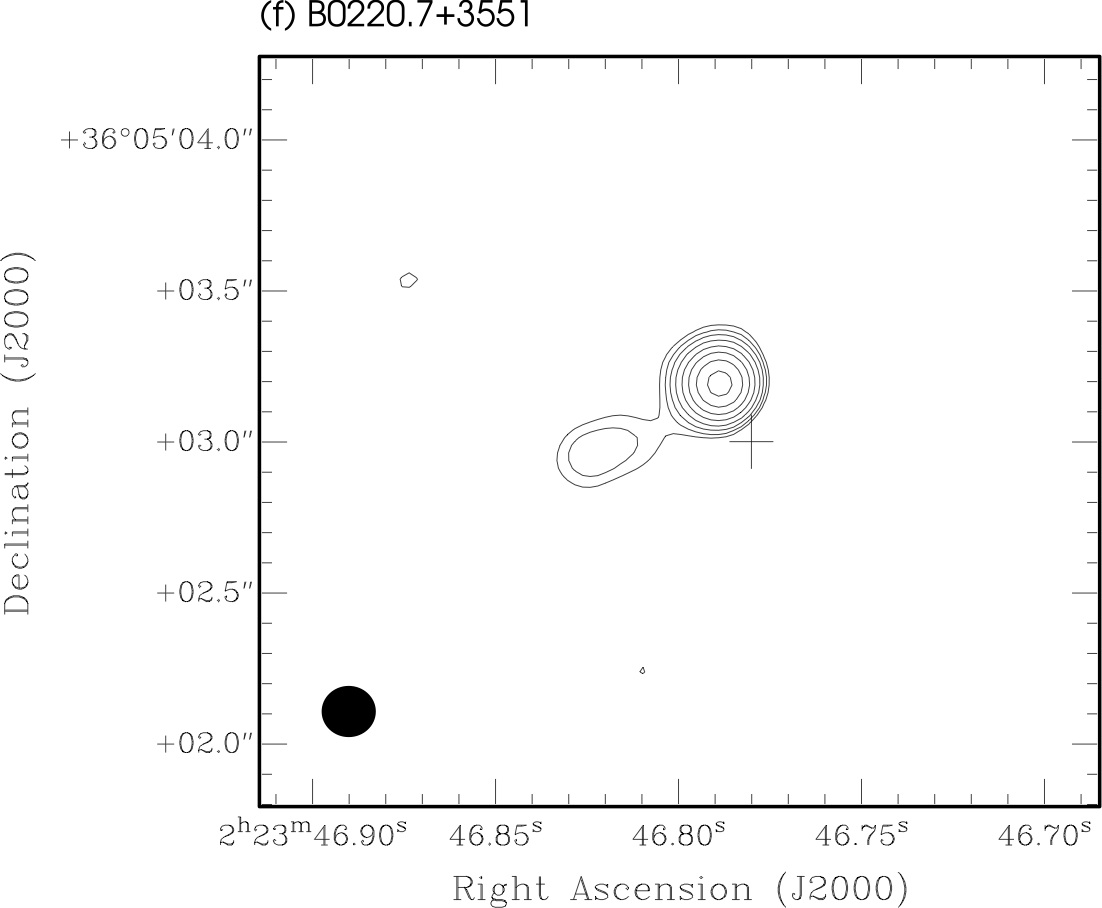}
}\\[5mm]
\mbox{
\plotone{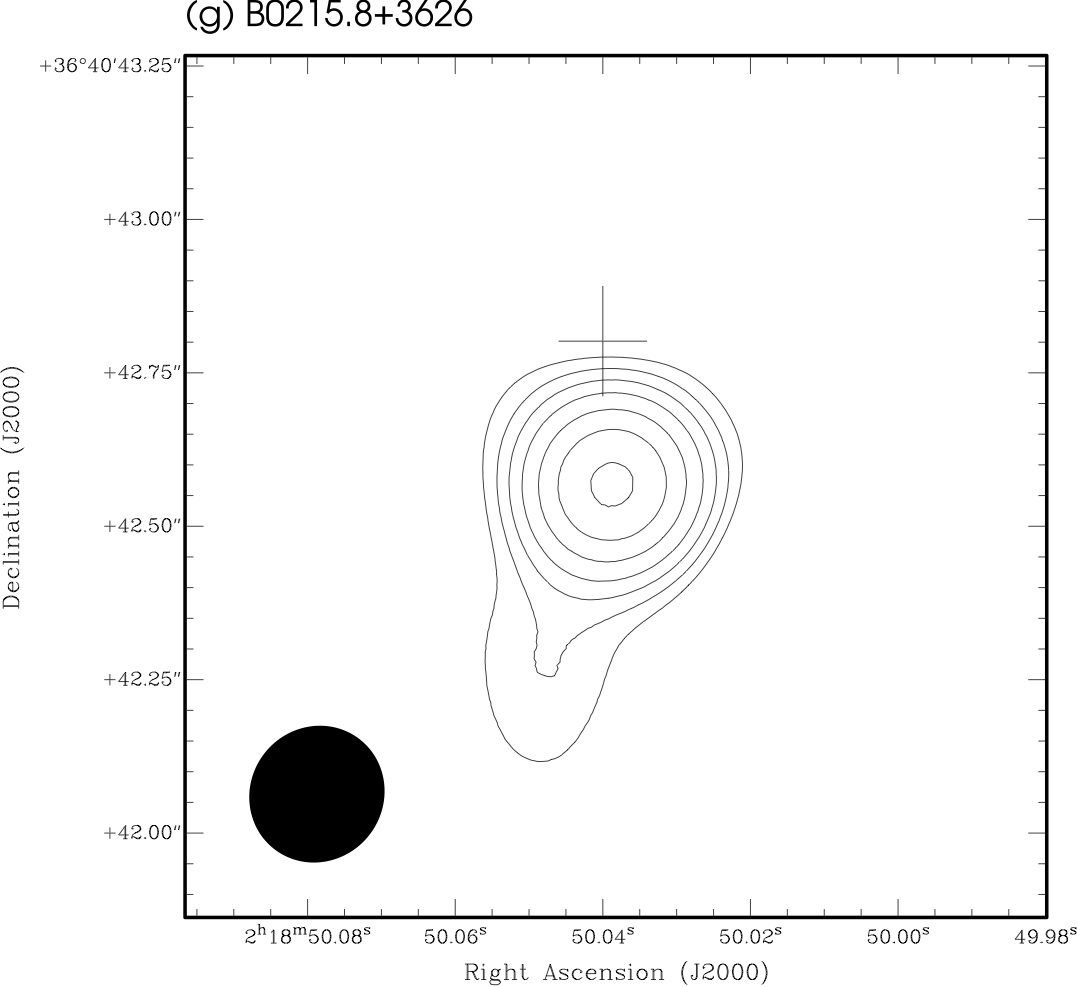} \quad
\plotone{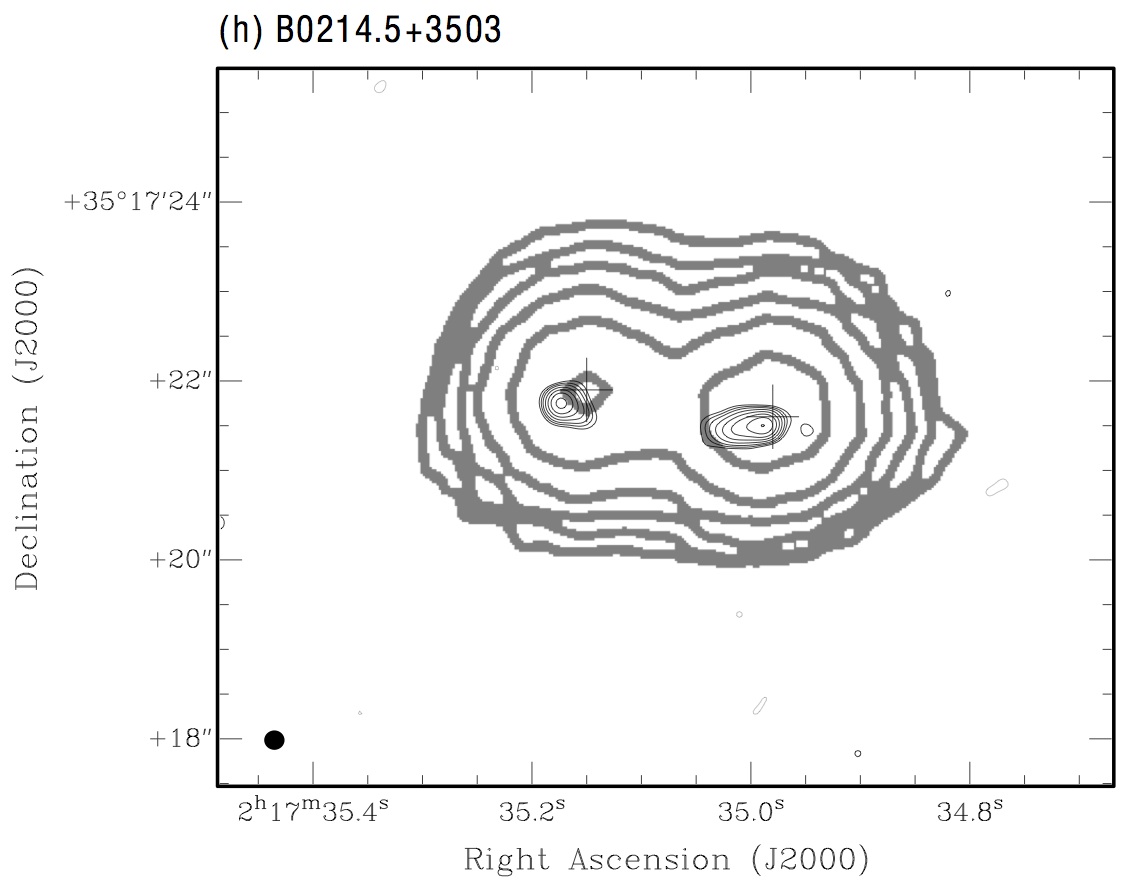}
}\\[5mm]
\mbox{
\plotone{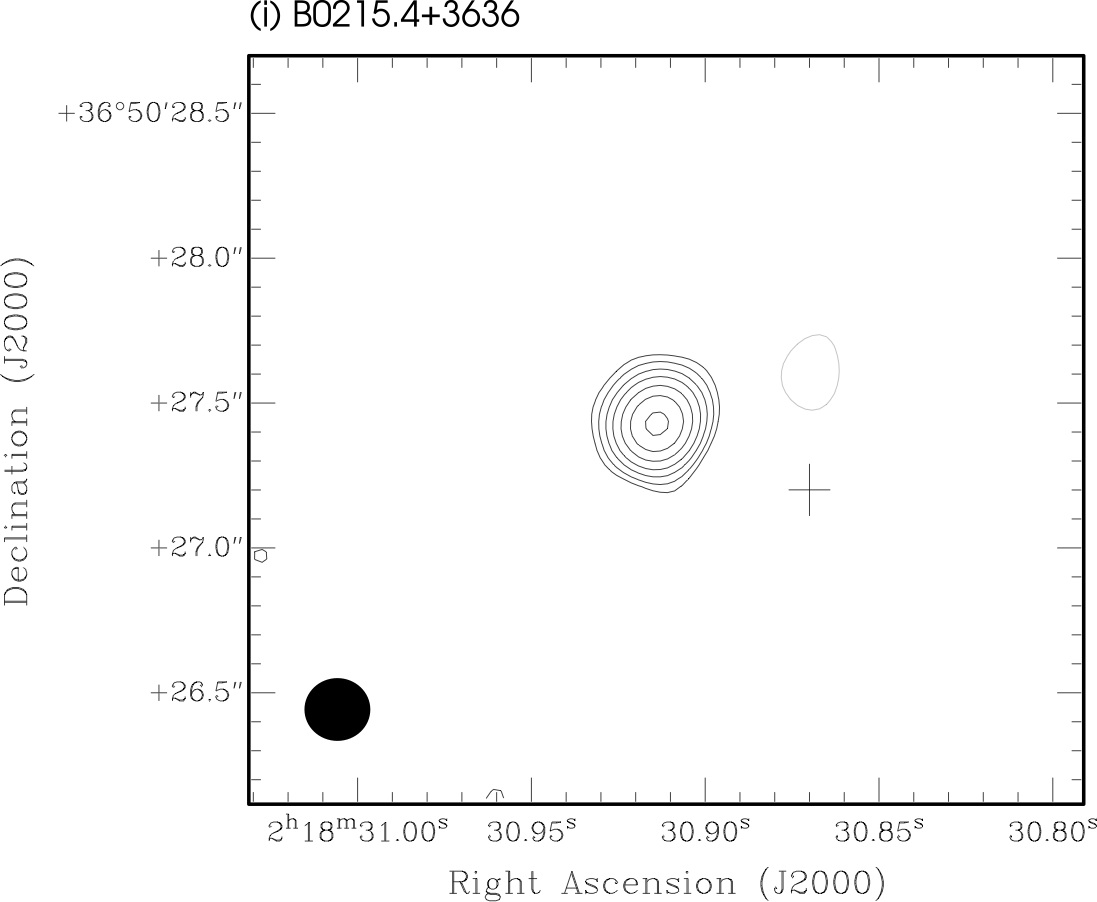} \quad
\plotone{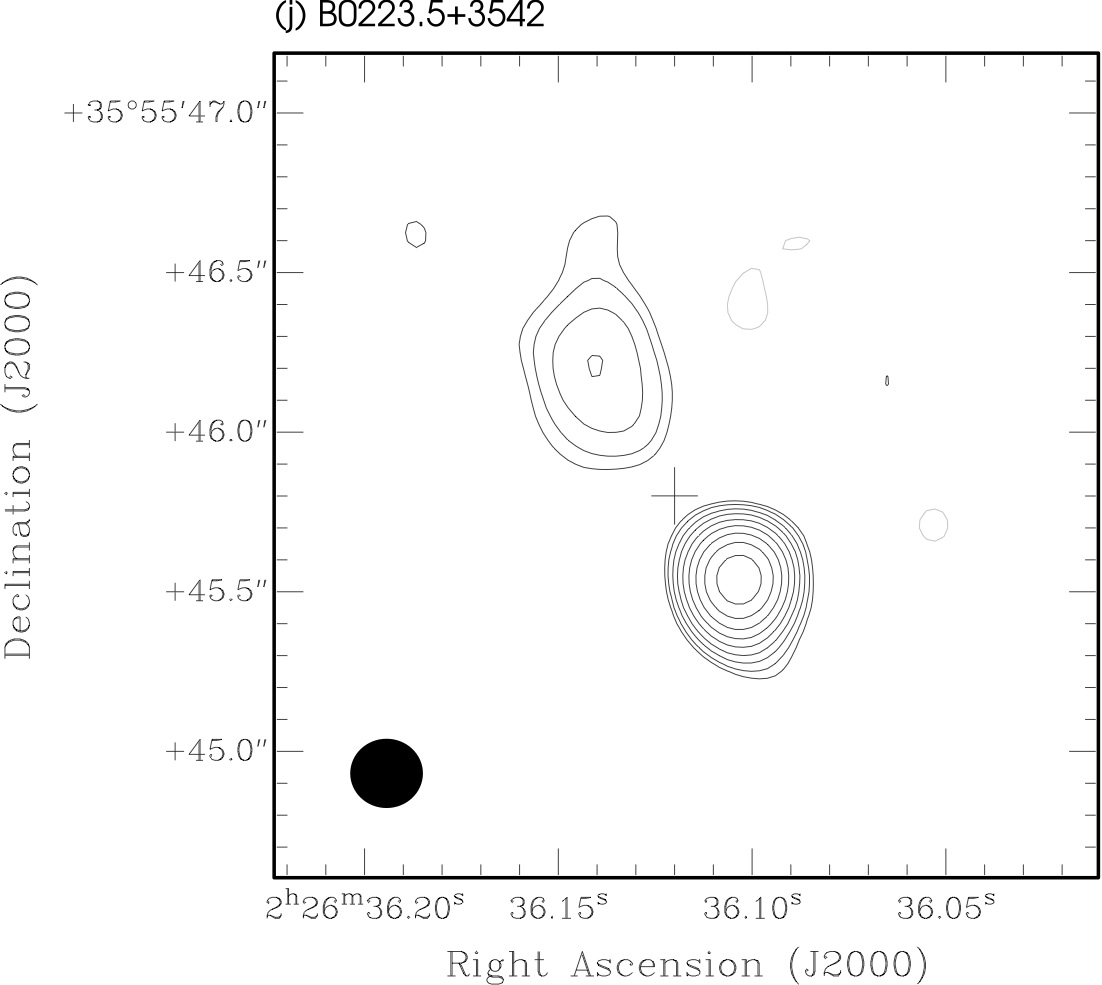}
}\\[5mm]
{Fig. 9.3. --- Continued}
\end{center}
\clearpage
\epsscale{0.40}
\begin{center}
\mbox{
\plotone{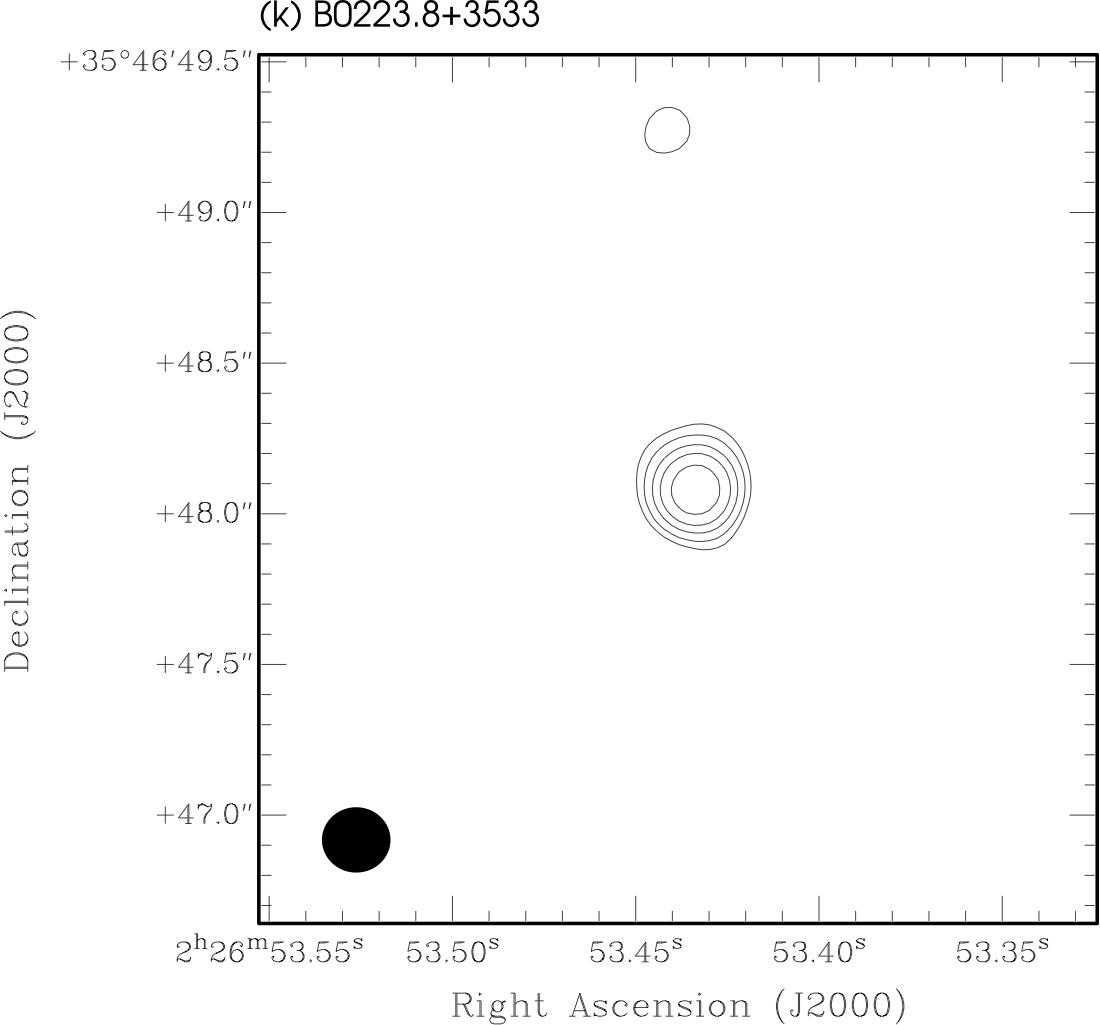} \quad
\plotone{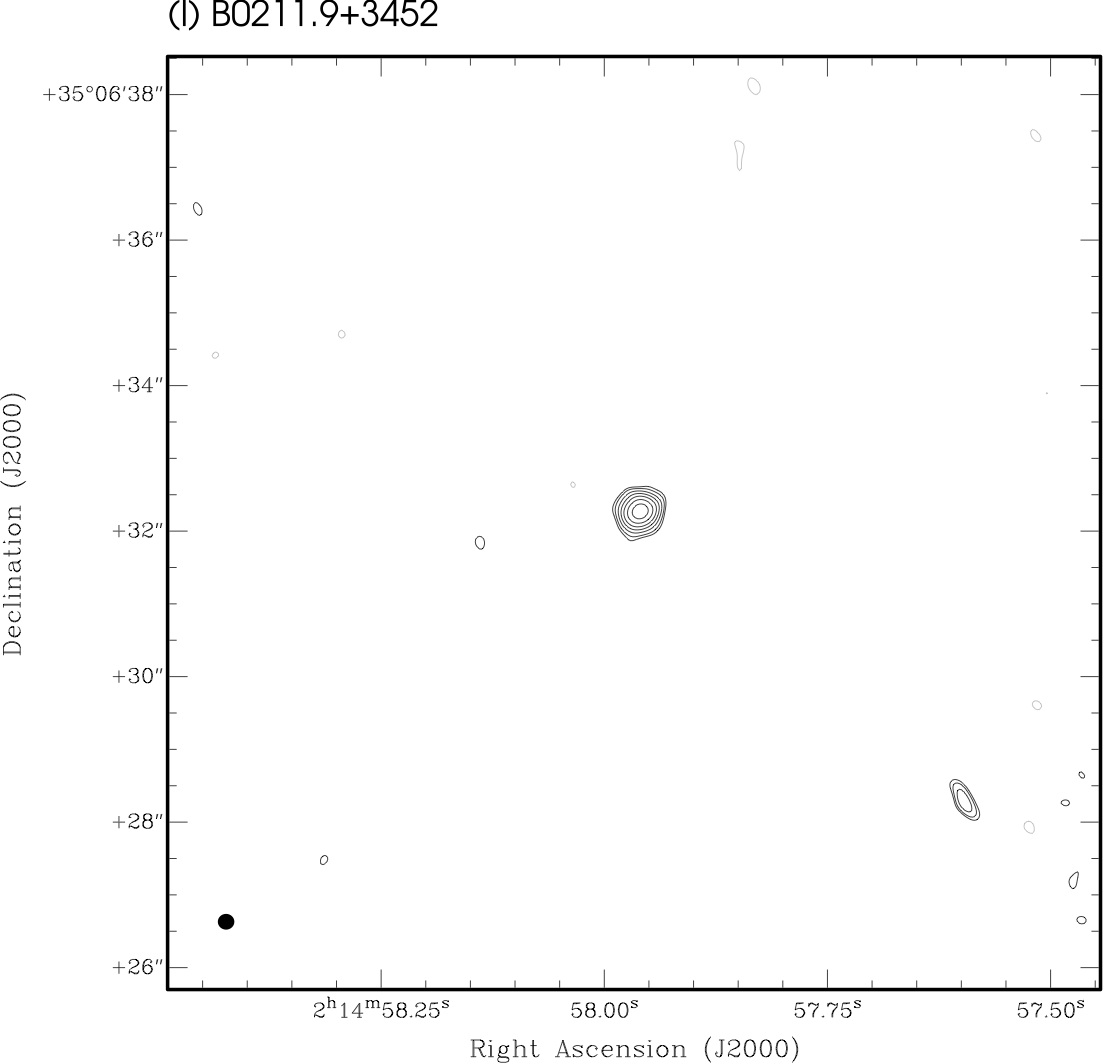}
}\\[5mm]
\mbox{
\plotone{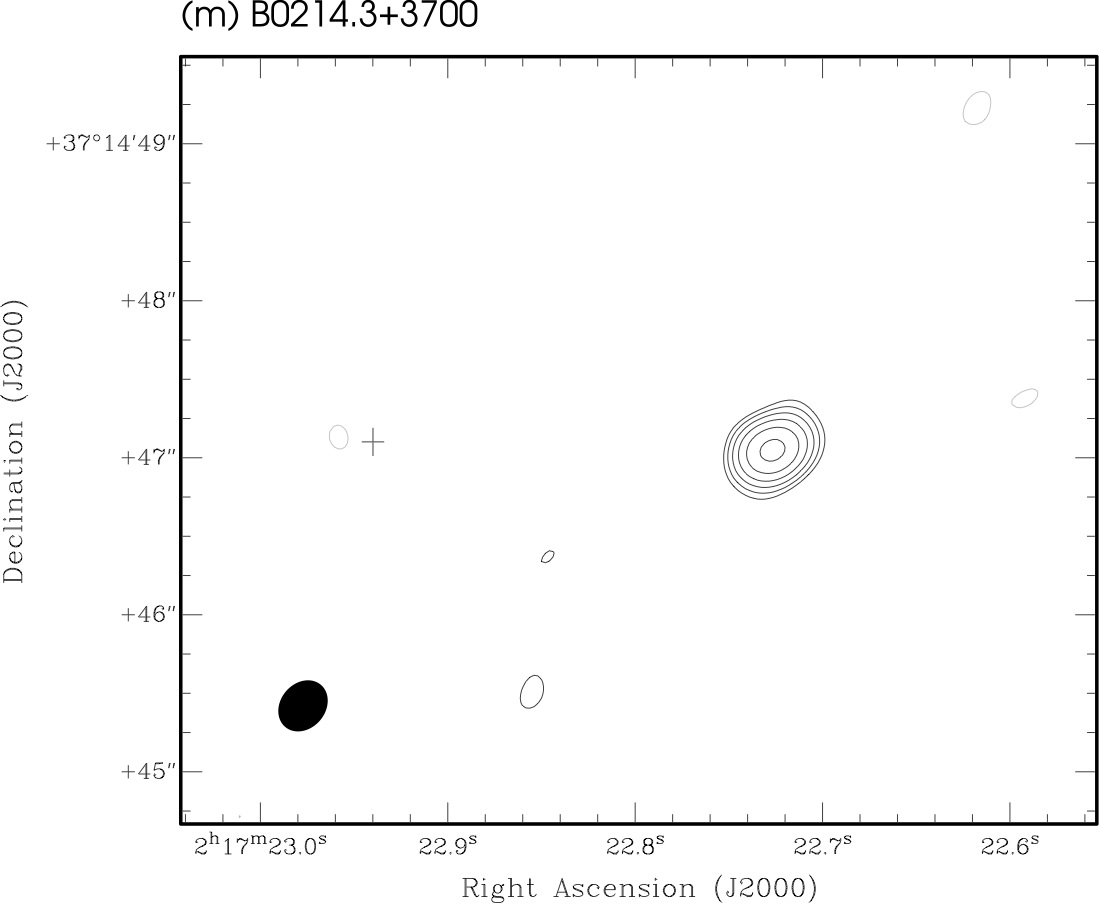} \quad
\plotone{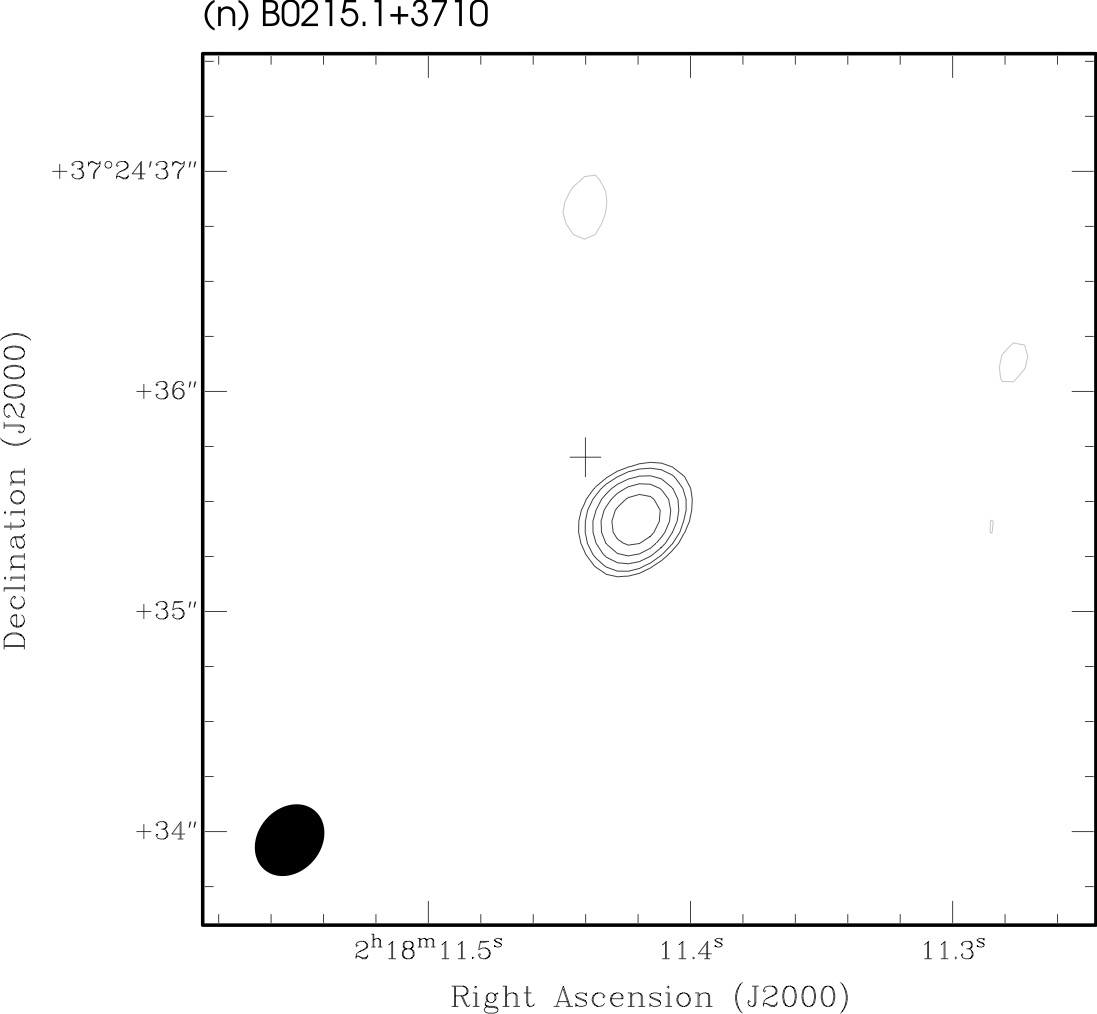}
}\\[5mm]
\mbox{
\plotone{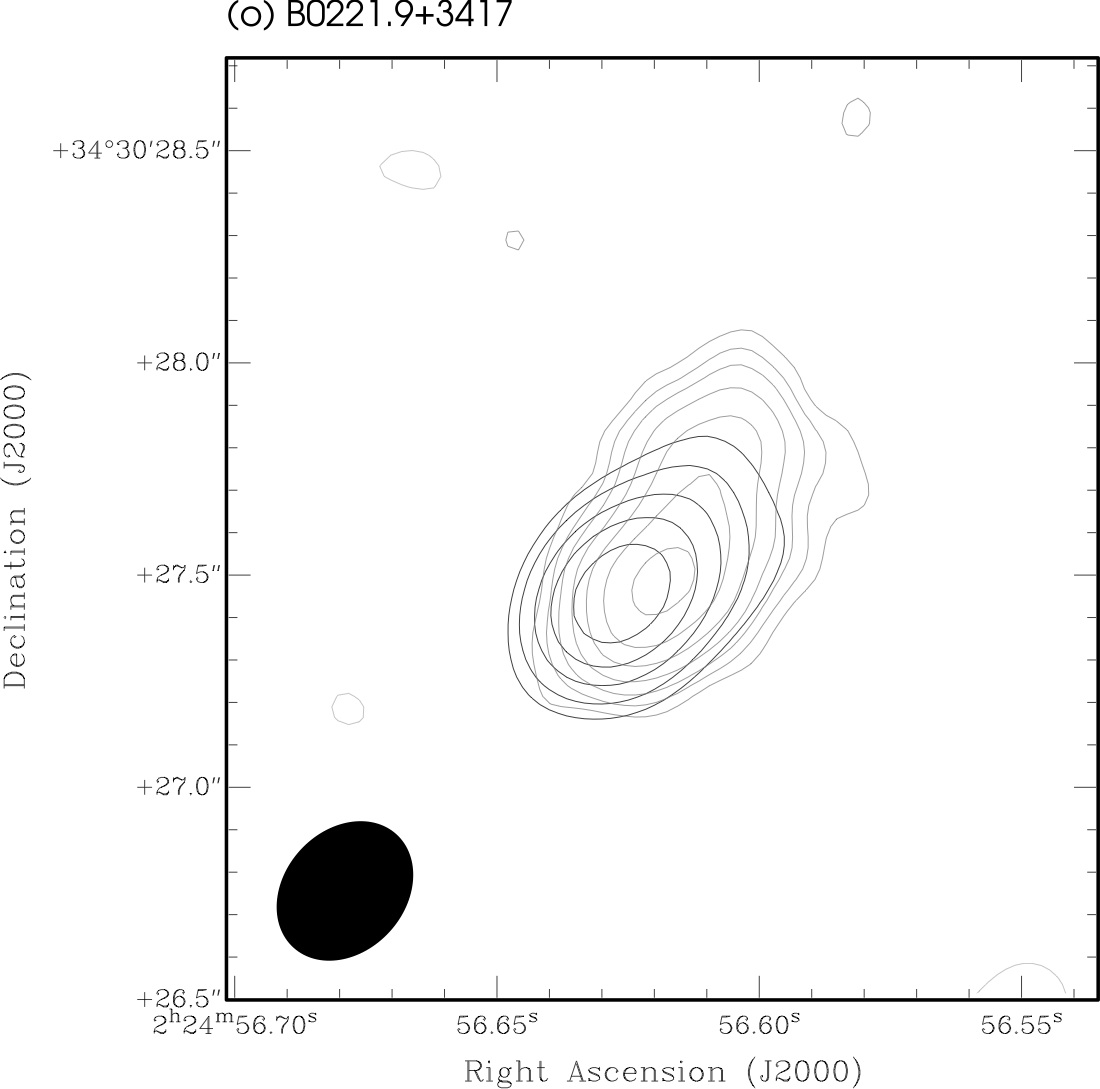} \quad
\plotone{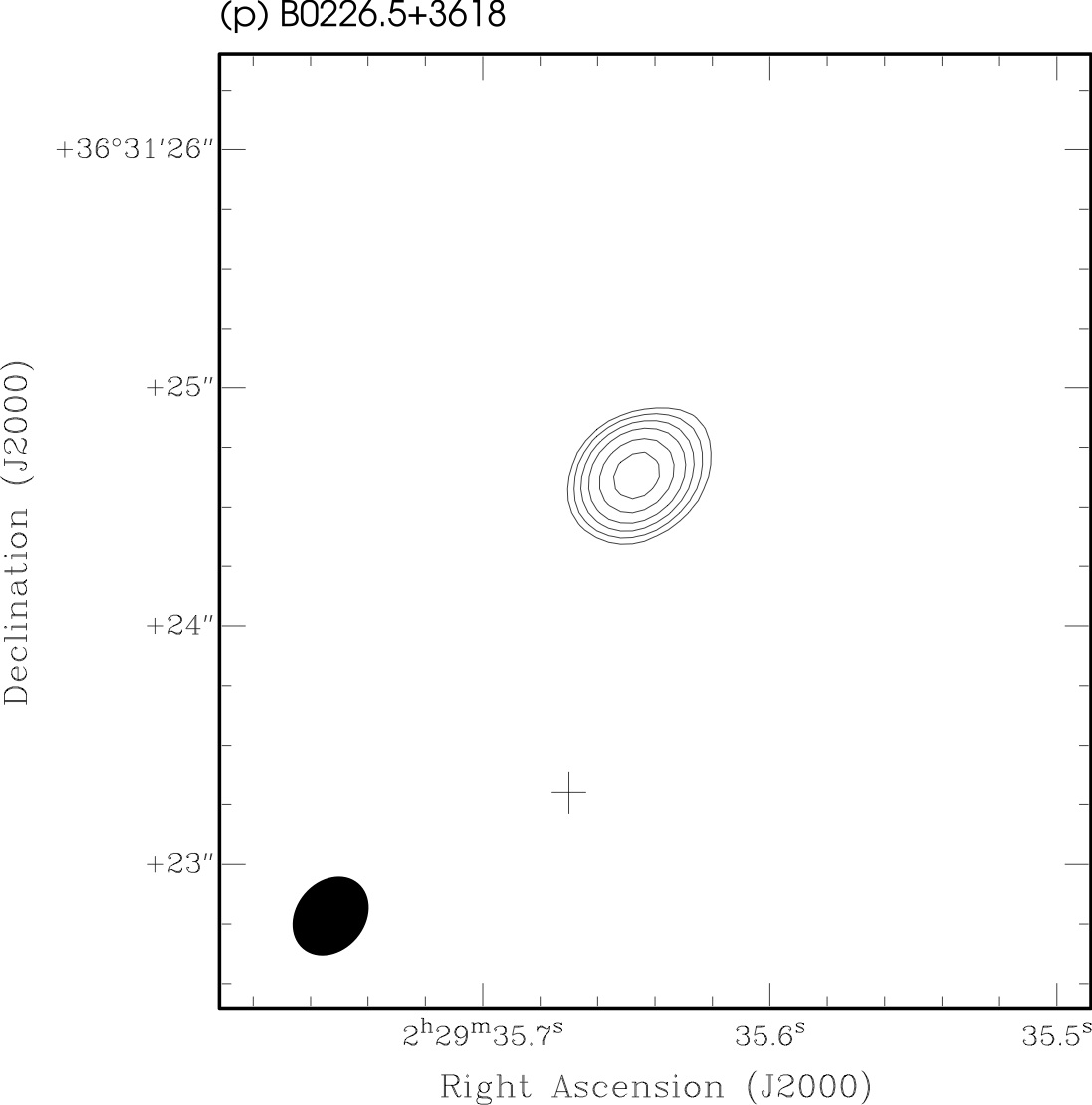}
}\\[5mm]
{Fig. 9.3. --- Continued}
\end{center}
\clearpage
\epsscale{0.45}
\begin{center}
\mbox{
\plotone{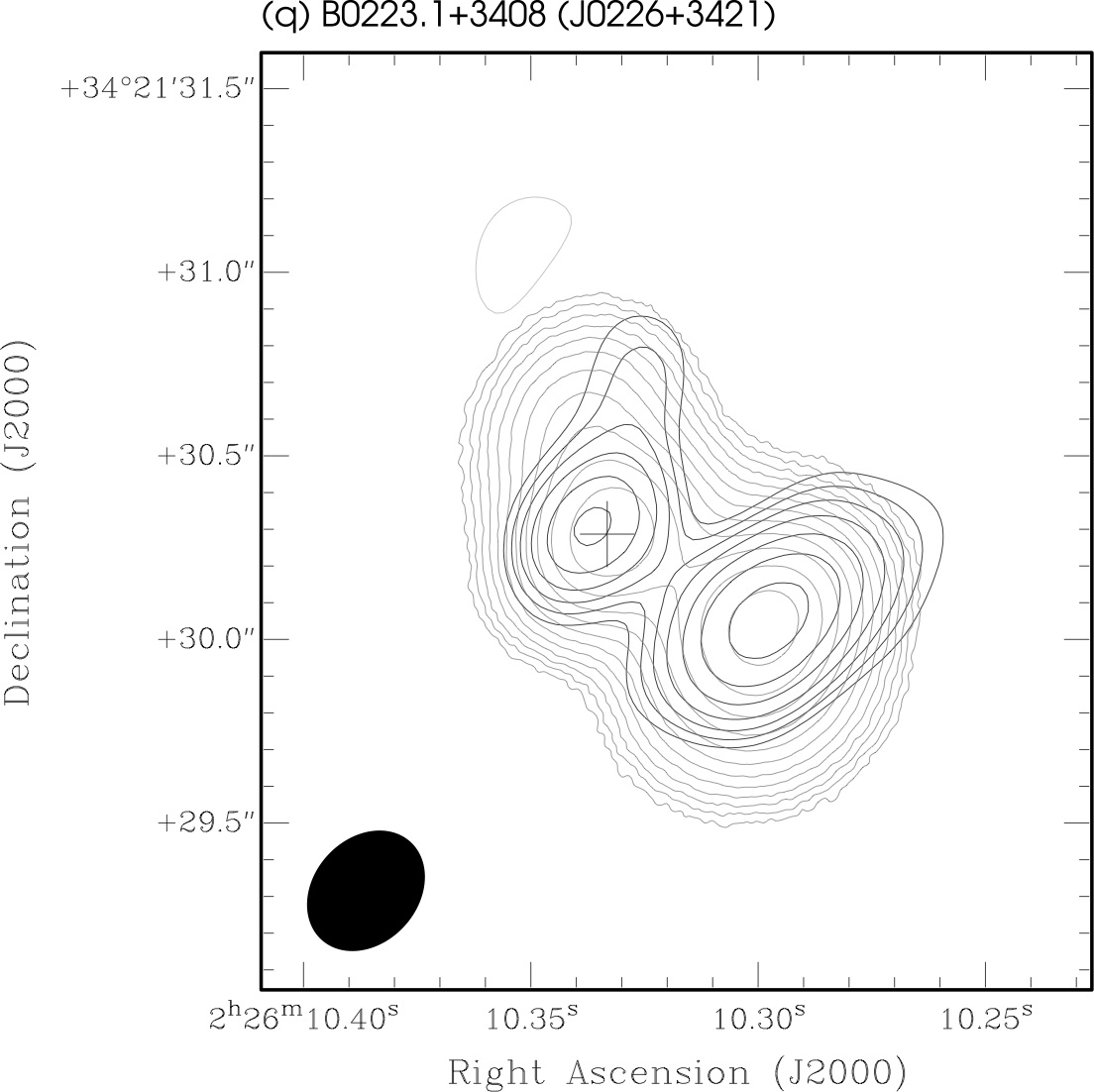} \quad
\plotone{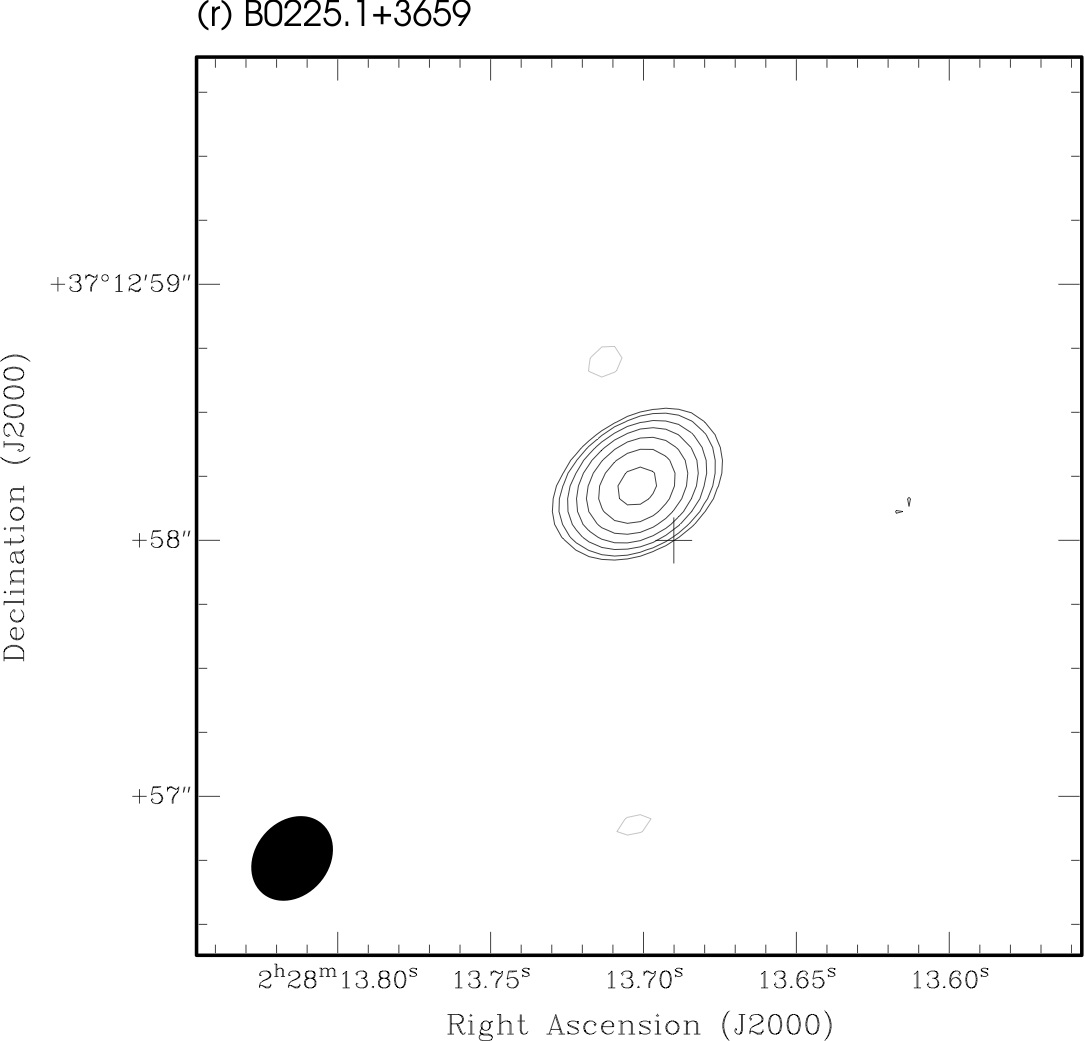}
}
\plotone{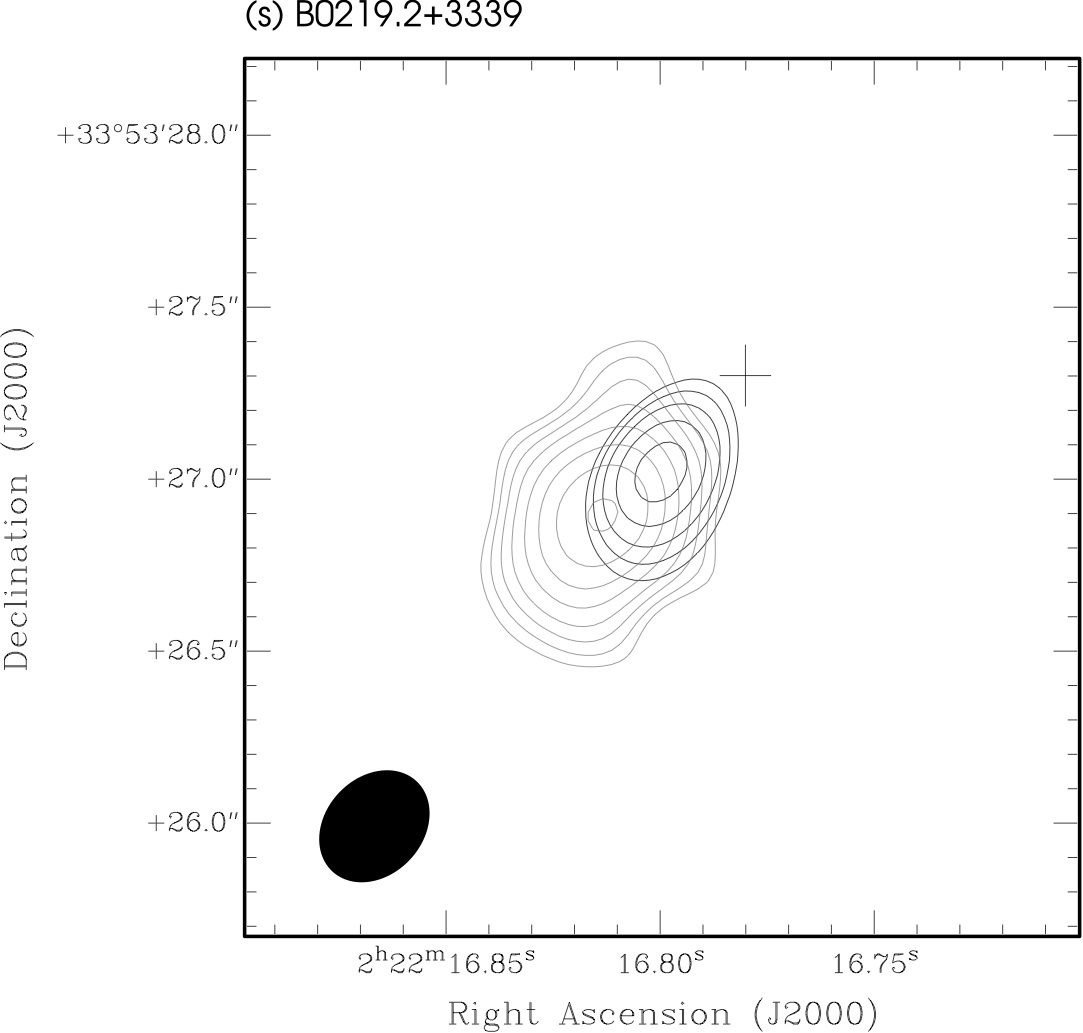}\\[5mm]
{Fig. 9.3. --- Continued}
\end{center}

\clearpage

\subsubsection{B0214.5$+$3503}
B0215.4$+$3503 has been previously imaged by \citet{Roettgering:1994p10579} at L-Band with the VLA and is classified as an ultra-steep spectrum source. Contours of the L-Band VLA image are shown overlaid with the 90 cm VLBI observation in Figure \ref{fig:figp2f3}(h). The source has a LAS of $2.8\arcsec$ which corresponds to a LLS of $\geq24$ kpc if a redshift of $z\geq2$ is assumed. The two components of the VLBI source align very closely, to within $0.2\sigma$, with the VLA image. Approximately 75\% of the WENSS flux is recovered by our observation, the remaining flux is likely associated with the extended structure observed in the VLA image.


\subsubsection{B0223.5$+$3542}
B0223.5$+$3542 is detected in both field 1 and field 2 and is shown in Figures \ref{fig:figp2f2}(h) and \ref{fig:figp2f3}(j), respectively. A significant residual offset of 180 mas exists between these two independent detections even after correcting for the larger-scale offset described in \S~\ref{sec:p2vlbical}. The phases of the source in the J0226$+$3421 field appeared to be more heavily affected by the ionosphere than those in the B0218$+$357 field and it is believed that the offset may have been introduced by the phase self-calibration process. This may have also adversely affected the measure of the integrated flux density which is approximately 60\% greater in the J0226$+$3421 field compared to the B0218$+$357 field. The two components of the source are separated by 830 mas. Approximately 90\% of the WENSS flux is recovered in the B0218$+$357 field observation of this source.

\subsubsection{B0223.8$+$3533}
Observations of this field revealed an unresolved compact source, Figure \ref{fig:figp2f3}(k), approximately $17\arcsec$ from the position reported by WENSS and over $25\arcsec$ from the more accurate position reported by NVSS, these are by far the greatest offsets observed in all of our detected sources. NVSS also places a limit on the fitted major axis of this source at less than $15\arcsec$, a size that does not encompass our observed source. Furthermore, our measure of integrated flux of 90 mJy is more than twice that measured by WENSS. The size, position and integrated flux density of our VLBI source suggest that it may be an unrelated transient or highly variable source that has been serendipitously detected within the target field. The existence of this source was tested by splitting the VLBI data into four equal length periods and independently imaging each of these. The source was detected in all four data-sets with an integrated flux of $92\pm11$ mJy beam$^{-1}$. Furthermore, as would be expected for an unresolved source, scalar averaging of the visibility amplitudes over 30 minute intervals revealed amplitudes that were approximately equal on all baselines.

\subsubsection{B0211.9$+$3452}
We observe an almost $7\arcsec$ offset at a position angle of $13\arcdeg$ between our VLBI detection of this source, Figure \ref{fig:figp2f3}(l), and the best known position \citep{Condon:1998p10541}. NVSS places a limit on the fitted HWHM model of this source at less than $28.6\arcsec$ and also notes the presence of large residual errors which are indicative of a complex source.  As only 30\% of the WENSS flux was recovered by the VLBI observation there is a suggestion that we have detected a compact component of a larger source. This is also supported by the 365 MHz Texas survey which classifies the source as an asymmetric double with a separation of $27\pm1\arcsec$ at a position angle of $25\pm2\arcdeg$.




\section{Discussion}

Our survey results indicate that at least $10\%$ of moderately faint (S$\sim100$ mJy) sources found at 90 cm contain compact components smaller than $\sim0.1$ to $0.3$ arcsec and stronger than $10\%$ of their total flux densities. This is a strict lower limit as the sensitivity of our observation was limited by the primary beam at the edge of the survey fields. None of the surveyed sources that were even slightly resolved by WENSS were detected. Similarly, none of the WENSS sources that were below the sensitivity limits of the VLBI observation were detected either, suggesting that none of these sources had significantly increased in brightness since the WENSS observations were carried out.  

The apparent lack of sources varying above our detection threshold must at least in part be due to resolution effects.  As 90\% of the WENSS sources above the VLBI detection threshold are not detected, they must be at least partially resolved at the VLBI resolution with the compact component of radio emission only a fraction of the WENSS flux density.  For the compact component of these sources to vary enough to be detected with VLBI, they must increase in strength by factors of perhaps at least a few (the reciprocal of the ratio of compact flux to WENSS flux) to be detectable with VLBI; the compact component of the flux needs to increase above the VLBI sensitivity limit.  Resolution effects are masking variability in these sources.  As discussed below, the detection of one apparent highly variable source in the VLBI data is rather remarkable.

The interpretation of our detection statistics is complicated, in that the survey has a non-uniform sensitivity over both fields, due to the primary beam response of the VLBA antennas and the fact that we are imaging objects well beyond the half-power points of the primary beam.  In addition, due to time and bandwidth smearing effects, as one images objects further from the phase centre, data on the long baselines is discarded, since the smearing effects make imaging difficult.  A consequence of this is that the angular resolution is also non-uniform across the surveyed fields, with low resolution far from the phase centre.  Not only is the flux limit variable across the field, the brightness temperature sensitivity also varies.

It is possible to estimate the detection statistics of our survey for a uniform flux density and brightness temperature limit by considering sources not too far from the phase centre and for a flux sensitivity between the extremes at the phase centre and field edge.  For example, if a sensitivity limit of 30 mJy beam$^{-1}$ is considered (achieved in the $0.25 - 0.5$ degree annulus of field 1 and in the $0.5 - 1$ degree annulus of field 2, and exceeded in the lower radius annuli in each field), 11 out of 55 possible sources are detected, a detection rate of 20\%, higher than the strict lower limit of 10\% estimated above for all sources at all annuli.

\citet{Garrett:2005p10555} performed a similar survey of the NOAO Bootes field at 1.4 GHz, using the NRAO VLBA and 100 m Green Bank Telescope. The survey covered a total of 0.28 deg$^{2}$, one hundredth of the area covered by our survey, and detected a total of 9 sources. The survey achieved sensitivities of $0.074-1.2$ mJy beam$^{-1}$ that enabled the detection of both weak and extended sources, whereas our 90 cm observations detected mainly compact sources or slightly resolved bright sources. Nonetheless, we can estimate the number of detections in this region that could be achieved using the 90 cm survey techniques described in this paper. The 0.28 deg$^2$ NOAO Bootes field contains a total of 13 WENSS sources, 6 of which have integrated flux densities $>30$ mJy. Based on our detection rate of 20\% for such sources we would expect to detect one WENSS source at 90 cm. Assuming a median spectral index of -0.77, only two of the \citet{Garrett:2005p10555} sources have integrated flux densities above our 30 mJy beam$^{-1}$ limit at 90 cm, however, one of these is extended and would have a VLBI peak flux density that falls below our limit. Thus the observations of \citet{Garrett:2005p10555} are consistent with our 90 cm VLBI results for sources with a peak flux density above 30 mJy beam$^{-1}$.

Estimates of the percentage of sources detected with VLBI gives an estimate of the relative contribution of AGN (that contain compact radio emission and are detectable with VLBI) and starburst galaxies (which contain low brightness temperature radio emission not detectable with VLBI).  Analysis of the ratio of starburst galaxies to AGN as a function of redshift (at high redshifts) can help to determine the initial sources of ionising radiation early in the Universe.  As very little redshift data for our surveyed sources are available, such an analysis is not currently possible with this data-set.  In practice, VLBI data at an additional frequency is also required, to confirm that the compact radio emission attributed to AGN has plausible spectral indices.

The distribution of morphologies in the detected survey sources are typical of AGN.  10/27 sources are unresolved point sources, consistent with core-dominated AGN.  A further 8/27 are clearly resolved into double component sources, consistent with being core-jet AGN or double-lobed radio galaxies.  7/27 sources have complex or extended structures, not obviously clear double components.  Again, these sources may be core-jet AGN or radio galaxies.  The remaining 2/27 sources are the gravitational lens and the quasar at the phase centres of the two fields.

The serendipitous detection of a likely highly variable, very compact source near the target WENSS source B0223.8$+$3533 is intriguing. The total area imaged by this survey represents $\sim0.5\%$ of the area within the $0\arcdeg-2\arcdeg$ annulus and is equivalent to $\sim2.2\%$ of the FWHM of the VLBA primary beam. While it is difficult to place any limits on the real population of variable sources based on this one observation, it does highlight the importance of imaging wide-fields completely, in order to improve our understanding of such sources.

\subsection{Future Prospects}

The observations presented here demonstrate that extremely wide-field surveys can now be piggybacked on current and future VLBI observations at 90 cm. While this survey has mainly concentrated on detecting and imaging sources already detected by other surveys, we find tantalising evidence of a transient or highly variable source. We were fortunate to have found one that appeared in close proximity to one of our target sources but this may not always be the case. This provides a motive to take on a more ambitious survey of the entire field. Such a survey is not beyond the reach of current technology, it would require at most $\sim45$ times more processing compared to the project presented here, in order to image the entire primary beam of the VLBA using a similar faceted approach. While this is not the most efficient means of detecting transients, it will help progress the development of algorithms and techniques needed for next generation, survey-class instruments that operate at wavelengths or sensitivities not matched by current instruments.

The observations presented in this paper were limited by the spectral and temporal resolution of the EVN correlator at the time of the observation. To minimise the effects of bandwidth and time-averaging smearing it was necessary to compromise resolution and image noise. Future technical developments in the capabilities of correlators will allow wide-field, global VLBI studies to be conducted without such restrictions. In particular, software correlators can provide extremely high temporal and spectral resolution, limited only by the time it takes to process the data \citep{Deller:2007p10545}. They also allow for some pre-processing to be applied during the correlation process to, for example, mitigate the effects of radio interference or to correlate against multiple phase centres simultaneously.

\subsection{Implications for LOFAR and SKA}

The results of these observations provide important information on the nature and incidence of compact, low-frequency radio sources, with consequences for next generation, low-frequency instruments such as LOFAR and the SKA. LOFAR is currently being deployed across The Netherlands but remote stations are already under construction in neighbouring countries, in particular Germany. Other countries (e.g. UK, France, Sweden, Italy and Poland) are also expected to join this European expansion of LOFAR (E-LOFAR), extending the longest baseline from a few hundred, to a few thousand km. This development will provide LOFAR with sub-arcsecond resolution at its highest observing frequency (the $120-240$ MHz high-band). One concern associated with extending LOFAR to much longer baselines is whether enough cosmic sources will remain unresolved - this characteristic is required in order to ensure there are enough calibrator sources in the sky in order to calibrate the instrument across its full, very wide, field-of-view.  The observations presented here suggest that at least one tenth of all radio sources (at the several tens of mJy level) are likely to exhibit compact VLBI radio structure in the LOFAR high-band. In all likelihood, an even larger fraction of the E-LOFAR source population will therefore be bright and compact enough to form a grid of calibrator sources across the sky. From our results, we estimate the spatial number density of relatively bright (S$>10$ mJy) and compact (LAS$<200$ mas) sources at 240 MHz to be $\sim3$ deg$^{-2}$. The aggregate total of these compact sources within a beam should serve as a good calibrator for E-LOFAR and enable most of the low-frequency radio sky to be imaged with excellent sub-arcsecond resolution and high dynamic range. Extrapolation to LOFAR's low-band (10$-$80 MHz) is probably very dangerous, but there is every reason to believe that a large number of these sources will remain compact.

In order to assess the relative numbers of starburst galaxies and AGN as a function of redshift, obviously large redshift surveys need to take place for these radio continuum objects.  Such a survey could be conducted using the redshifted HI signal from these galaxies, using the SKA.

\renewcommand{\thefootnote}{\alph{footnote}}
\scriptsize
\begin{center}
\begin{longtable}{llllccc}
\caption[Astrometry and photometry at 324 MHz for field 1 targets.]{Astrometry and photometry at 324 MHz for field 1 targets.}
\label{tab:tabf1}
 \\
\hline \hline \\[-2ex]
   Annulus & WENSS Source\tablenotemark{a} & R.A.\tablenotemark{b}      & Decl.\tablenotemark{c}     & $S_{P}$           & $S_{I}$ & $\alpha$\tablenotemark{d} \\
                    &                       & (J2000.0) & (J2000.0) & (mJy beam$^{-1}$) & (mJy)   & \\[0.5ex] \hline
   \\[-1.8ex]
\endfirsthead

\multicolumn{7}{c}{{\tablename} \thetable{} -- Continued} \\[0.5ex]
\hline \hline \\[-2ex]
   Annulus & WENSS Source\tablenotemark{a} & R.A.\tablenotemark{b}      & Decl.\tablenotemark{c}     & $S_{P}$           & $S_{I}$ & $\alpha$\tablenotemark{d} \\
                    &                       & (J2000.0) & (J2000.0) & (mJy beam$^{-1}$) & (mJy)   & \\[0.5ex] \hline
   \\[-1.8ex]
\endhead

\multicolumn{7}{l}{{Continued on Next Page\ldots}} \\
\endfoot

\\[-1.8ex] \hline \hline
\endlastfoot

$0\arcdeg-0.25\arcdeg$ & B0223.1+3408 & 02 26 10.15   & +34 21 30.7   & 3793 & 3678 & $-0.16$   \\
& \nodata      & 02 26 10.3332 & +34 21 30.286 & 375  & 2880 & \nodata \\
& B0222.2+3406 & 02 25 14.12   & +34 20 22.4   & 32   & 33   & $-1.09$   \\
& B0222.3+3414 & 02 25 19.18   & +34 27 31.7   & 41   & 50   & \nodata \\
\hline
$0.25\arcdeg-0.5\arcdeg$ & B0221.9+3417  & 02 24 56.18  & +34 30 35.8   & 1657 & 1638 & $-0.91$   \\
& \nodata       & 02 24 56.620 & +34 30 27.47  & 111  & 186  & \nodata \\
& B0224.1+3354  & 02 27 09.72  & +34 08 11.4   & 35   & 17   & $-0.94$   \\
& B0221.6+3406B & 02 24 41.08  & +34 19 43.6   & 459  & 433  & $-0.93$   \\
& \nodata       & 02 24 41.106 & +34 19 41.88  & 78   & 382  & \nodata \\
& B0221.6+3406  & 02 24 39.96  & +34 20 03.1   & 459  & 543  & \nodata \\
& B0223.9+3351  & 02 26 55.50  & +34 04 55.7   & 532  & 525  & $-1.02$   \\
& \nodata       & 02 26 55.564 & +34 04 55.57  & 114  & 269  & \nodata \\
& B0221.6+3406A & 02 24 34.48  & +34 21 39.5   & 104  & 109  & $-1.11$   \\
& \nodata       & 02 24 34.514 & +34 21 35.03  & 97   & 84   & \nodata \\
& B0224.8+3406  & 02 27 54.18  & +34 19 39.5   & 83   & 73   & $-0.88$   \\
& B0221.3+3405  & 02 24 20.16  & +34 19 17.5   & 195  & 193  & $-0.84$   \\
& B0224.9+3419  & 02 27 59.75  & +34 32 46.5   & 50   & 33   & $0.15$    \\
& B0224.2+3430  & 02 27 16.40  & +34 43 42.9   & 129  & 128  & $-0.96$   \\
\hline
$0.5\arcdeg-1\arcdeg$ & B0220.7+3401  & 02 23 47.53  & +34 15 08.3   & 403  & 421 & $-0.80$   \\
& B0225.5+3419A & 02 28 29.29  & +34 34 43.2   & 89   & 75  & $-0.33$   \\
& B0225.5+3419  & 02 28 34.05  & +34 33 09.5   & 504  & 568 & $-1.06$   \\
& B0225.5+3419B & 02 28 34.63  & +34 32 58.5   & 504  & 492 & $-0.97$   \\
& B0223.2+3441  & 02 26 17.12  & +34 54 36.5   & 88   & 85  & $-0.71$   \\
& \nodata       & 02 26 17.184 & +34 54 36.40  & 126  & 119 & \nodata \\
& B0224.8+3434  & 02 27 51.46  & +34 47 36.1   & 100  & 99  & $-0.65$   \\
& B0223.5+3443  & 02 26 36.80  & +34 56 42.9   & 38   & 33  & $-0.39$   \\
& B0225.5+3347  & 02 28 34.76  & +34 01 00.2   & 148  & 139 & $-0.71$   \\
& B0224.7+3442  & 02 27 43.89  & +34 56 23.1   & 100  & 85  & $-0.90$   \\
& B0221.0+3440  & 02 24 03.11  & +34 54 26.6   & 107  & 102 & $-0.41$   \\
& B0225.7+3340  & 02 28 45.59  & +33 53 31.1   & 95   & 84  & $-0.66$   \\
& B0220.4+3433  & 02 23 25.14  & +34 47 20.1   & 134  & 145 & \nodata \\
& B0219.6+3404  & 02 22 40.67  & +34 18 10.7   & 54   & 41  & $-0.87$   \\
& B0225.8+3438A & 02 28 48.50  & +34 51 32.8   & 87   & 86  & $-0.85$   \\
& B0219.6+3415  & 02 22 38.27  & +34 29 29.1   & 188  & 174 & $-0.81$   \\
& B0225.8+3438  & 02 28 52.37  & +34 52 17.3   & 87   & 161 & \nodata   \\
& B0225.8+3438B & 02 28 56.96  & +34 53 04.5   & 81   & 74  & $-0.92$   \\
& B0219.8+3343A & 02 22 48.79  & +33 57 18.9   & 638  & 638 & $-0.83$   \\
& B0219.8+3343  & 02 22 48.85  & +33 57 11.3   & 638  & 688 & $-0.88$   \\
& B0221.2+3325  & 02 24 11.73  & +33 39 18.1   & 185  & 178 & $-0.78$   \\
& B0219.8+3343B & 02 22 49.91  & +33 55 04.3   & 49   & 49  & $-0.84$   \\
& B0218.8+3402  & 02 21 50.01  & +34 16 19.1   & 359  & 347 & $-0.88$   \\
& B0223.5+3501  & 02 26 32.36  & +35 15 21.8   & 122  & 102 & $-0.81$   \\
& B0218.9+3353  & 02 21 56.58  & +34 07 17.5   & 127  & 122 & $-0.92$   \\
& B0227.5+3414  & 02 30 35.36  & +34 27 45.3   & 74   & 67  & $-0.86$   \\
& B0219.2+3339  & 02 22 16.75  & +33 53 27.3   & 516  & 499 & $-0.67$   \\
& \nodata       & 02 22 16.812 & +33 53 26.90  & 251  & 417 & \nodata \\
& B0227.0+3335  & 02 30 01.40  & +33 48 48.3   & 142  & 138 & \nodata \\
\hline
$1\arcdeg-1.5\arcdeg$ & B0218.6+3342  & 02 21 40.00 & +33 56 36.1 & 195  & 204  & $-0.84$   \\
& B0227.6+3343  & 02 30 41.94 & +33 56 23.3 & 68   & 59   & $-0.35$   \\
& B0225.4+3312  & 02 28 25.15 & +33 26 14.0 & 128  & 111  & $-0.63$   \\
& B0226.5+3321  & 02 29 31.62 & +33 34 23.8 & 215  & 207  & $-0.79$   \\
& B0218.1+3419  & 02 21 09.69 & +34 33 37.8 & 252  & 252  & $-0.54$   \\
& B0220.5+3310  & 02 23 33.90 & +33 24 35.5 & 179  & 175  & $-0.80$   \\
& B0227.7+3443  & 02 30 44.68 & +34 57 14.6 & 98   & 96   & $-0.53$   \\
& B0227.3+3325  & 02 30 19.97 & +33 38 53.8 & 76   & 68   & $-0.80$   \\
& B0219.1+3323  & 02 22 06.57 & +33 36 49.0 & 123  & 112  & $-0.88$   \\
& B0224.6+3515  & 02 27 38.33 & +35 28 43.8 & 164  & 147  & $-0.82$   \\
& B0227.7+3324  & 02 30 42.60 & +33 37 59.8 & 1029 & 996  & $-0.77$   \\
& B0226.1+3509  & 02 29 11.30 & +35 22 22.6 & 160  & 178  & \nodata   \\
& B0217.9+3442  & 02 20 56.23 & +34 56 36.2 & 94   & 104  & $-0.72$   \\
& B0225.0+3258  & 02 28 03.91 & +33 11 27.5 & 399  & 388  & $-0.96$   \\
& B0228.9+3429  & 02 31 59.53 & +34 42 26.7 & 2361 & 2556 & $-1.02$   \\
& B0227.9+3322  & 02 30 56.80 & +33 35 16.5 & 477  & 433  & $-0.77$   \\
& B0225.5+3255  & 02 28 30.73 & +33 09 16.4 & 161  & 131  & $-0.68$   \\
& B0229.4+3410  & 02 32 28.67 & +34 24 03.0 & 9334 & 9459 & $-0.88$   \\
& B0217.8+3452B & 02 20 49.19 & +35 06 28.3 & 107  & 98   & $-1.30$   \\ 
& B0217.8+3452  & 02 20 48.92 & +35 06 25.2 & 107  & 109  & $-1.37$   \\
& B0216.6+3409  & 02 19 37.82 & +34 23 10.5 & 555  & 515  & $-1.06$   \\
& B0229.0+3332  & 02 32 05.82 & +33 45 22.2 & 321  & 305  & $-1.24$   \\
& B0220.8+3527  & 02 23 50.17 & +35 41 31.7 & 398  & 393  & $-0.83$   \\
& B0216.2+3405B & 02 19 18.40 & +34 19 49.5 & 273  & 338  & \nodata   \\
& B0216.2+3405  & 02 19 15.49 & +34 19 42.1 & 412  & 802  & \nodata   \\
& B0216.8+3334  & 02 19 48.77 & +33 48 16.0 & 1352 & 1357 & \nodata   \\
& B0229.3+3447  & 02 32 21.61 & +35 01 06.8 & 352  & 381  & $-1.30$   \\
& B0216.2+3405A & 02 19 13.44 & +34 19 37.2 & 412  & 463  & \nodata   \\
\hline
$1.5\arcdeg-2\arcdeg$ & B0230.3+3429  & 02 33 20.42  & +34 42 54.1   & 170  & 172  & $0.06$    \\
& B0229.3+3318  & 02 32 21.28  & +33 32 08.4   & 168  & 150  & $-0.77$   \\
& B0223.5+3542  & 02 26 36.14  & +35 55 46.2   & 1758 & 1730 & $-0.59$   \\
& \nodata       & 02 26 36.114 & +35 55 45.44  & 675  & 1491 & \nodata   \\
& B0226.2+3534  & 02 29 20.00  & +35 47 40.2   & 167  & 163  & $-0.80$   \\
& B0220.1+3238  & 02 23 09.03  & +32 51 58.0   & 206  & 257  & \nodata   \\
& B0220.1+3238A & 02 23 08.00  & +32 51 43.3   & 206  & 191  & $-0.77$   \\
& B0216.2+3318  & 02 19 11.54  & +33 31 49.9   & 302  & 284  & $-0.53$   \\
& B0229.4+3517  & 02 32 29.32  & +35 30 53.6   & 2149 & 2359 & $-0.86$   \\
& B0231.2+3500  & 02 34 15.22  & +35 13 53.0   & 294  & 300  & $-0.95$   \\
& B0219.6+3547  & 02 22 37.70  & +36 01 01.1   & 487  & 515  & $-0.93$   \\
& B0218.0+3542  & 02 21 05.40  & +35 56 13.0   & 2534 & 2460 & $-0.25$   \\
& B0232.4+3405  & 02 35 26.28  & +34 18 30.3   & 391  & 384  & $-0.55$   \\
& B0214.7+3321  & 02 17 43.31  & +33 35 03.0   & 683  & 657  & $-0.85$   \\
& B0223.0+3603  & 02 26 04.82  & +36 17 10.0   & 337  & 355  & $-1.11$   \\
& B0223.0+3603A & 02 26 04.43  & +36 17 19.8   & 337  & 308  & $-1.02$   \\
& B0214.5+3503  & 02 17 35.04  & +35 17 22.6   & 1146 & 1092 & \nodata   \\
\hline
$2\arcdeg-3\arcdeg$ & B0217.8+3227  & 02 20 48.04 & +32 41 07.0 & 1415 & 1393 & $-0.28$   \\
& B0230.1+3243  & 02 33 12.09 & +32 56 48.5 & 594  & 606  & $-0.98$   \\
& B0230.1+3243B & 02 33 12.37 & +32 56 51.2 & 594  & 572  & $-0.94$   \\
& B0215.4+3536  & 02 18 28.69 & +35 49 58.6 & 1445 & 1401 & $-0.89$   \\
& B0227.2+3206B & 02 30 16.48 & +32 20 27.1 & 642  & 685  & \nodata   \\
& B0227.2+3206  & 02 30 14.54 & +32 19 48.0 & 713  & 1473 & \nodata   \\
& B0227.2+3206A & 02 30 12.82 & +32 19 12.7 & 713  & 788  & $-1.06$   \\
& B0233.2+3458  & 02 36 19.28 & +35 11 15.2 & 871  & 843  & $-0.97$   \\
& B0211.9+3452  & 02 14 58.14 & +35 06 39.5 & 3011 & 3084 & $-0.83$   \\
& B0224.9+3650  & 02 28 01.68 & +37 03 36.3 & 1409 & 1455 & $-0.90$   \\
& B0225.1+3659  & 02 28 13.67 & +37 12 57.1 & 1566 & 1519 & $-1.09$   
\footnotetext[1]{The first data entry for each target refers to the WENSS target source, and the second entry, when present, refers to the VLBI detection.}
\footnotetext[2]{Units of right ascension are hours, minutes and seconds.}
\footnotetext[3]{Units of declination are degree, arcminutes and arcseconds.}
\footnotetext[4]{The spectral index $\alpha$ is defined as $S_{\nu}\propto\nu^{\alpha}$ and is estimated from WENSS and NVSS integrated flux densities where available.}
\end{longtable}
\end{center}

\clearpage

\begin{center}
\begin{longtable}{llllccc}
\caption[Astrometry and photometry at 324 MHz for field 2 targets.]{Astrometry and photometry at 324 MHz for field 2 targets.}
\label{tab:tabf2}
 \\
\hline \hline \\[-2ex]
   Annulus & WENSS Source\tablenotemark{a} & R.A.\tablenotemark{b}      & Decl.\tablenotemark{c}     & $S_{P}$           & $S_{I}$ & $\alpha$\tablenotemark{d} \\
                    &                       & (J2000.0) & (J2000.0) & (mJy beam$^{-1}$) & (mJy)   & \\[0.5ex] \hline
   \\[-1.8ex]
\endfirsthead

\multicolumn{7}{c}{{\tablename} \thetable{} -- Continued} \\[0.5ex]
\hline \hline \\[-2ex]
   Annulus & WENSS Source\tablenotemark{a} & R.A.\tablenotemark{b}      & Decl.\tablenotemark{c}     & $S_{P}$           & $S_{I}$ & $\alpha$\tablenotemark{d} \\
                    &                       & (J2000.0) & (J2000.0) & (mJy beam$^{-1}$) & (mJy)   & \\[0.5ex] \hline
   \\[-1.8ex]
\endhead

\multicolumn{7}{l}{{Continued on Next Page\ldots}} \\
\endfoot

\\[-1.8ex] \hline \hline
\endlastfoot

$0\arcdeg-0.25\arcdeg$ & B0218.0+3542 & 02 21 05.40   & +35 56 13.0   & 2534 & 2460 & $-0.25$   \\
& \nodata      & 02 21 05.4720 & +35 56 13.716 & 90   & 1320 & \nodata \\
& B0218.8+3545 & 02 21 55.21   & +35 59 20.4   & 20   & 22   & $-0.90$ \\
\tableline
$0.25\arcdeg-0.5\arcdeg$ & B0219.1+3533 & 02 22 10.16  & +35 46 48.3   & 89  & 87  & $-1.08$   \\
& \nodata      & 02 22 10.123 & +35 46 51.67  & 48  & 85  & \nodata   \\
& B0219.6+3547 & 02 22 37.70  & +36 01 01.1   & 487 & 515 & $-0.93$   \\
& B0217.3+3600 & 02 20 23.66  & +36 14 36.8   & 174 & 178 & $-0.96$   \\
& B0219.6+3533 & 02 22 37.61  & +35 47 30.9   & 20  & 19  & $-1.02$   \\
& B0216.9+3526 & 02 19 57.00  & +35 40 41.9   & 31  & 32  & $-1.33$   \\
& B0218.8+3604 & 02 21 55.81  & +36 17 49.5   & 32  & 29  & $-1.18$   \\
& B0218.5+3518 & 02 21 34.18  & +35 32 12.5   & 58  & 44  & $-0.78$   \\
& B0216.6+3523 & 02 19 42.13  & +35 37 43.9   & 48  & 59  & $-0.24$   \\
& \nodata      & 02 19 42.305 & +35 37 44.33  & 41  & 84  & \nodata   \\
& B0216.2+3555 & 02 19 18.61  & +36 08 59.4   & 23  & 21  & \nodata   \\
& B0220.0+3552 & 02 23 06.20  & +36 06 06.5   & 22  & 36  & \nodata   \\
& B0220.1+3532 & 02 23 10.75  & +35 45 50.0   & 80  & 76  & $-0.75$   \\
& \nodata      & 02 23 10.747 & +35 45 46.65  & 53  & 97  & \nodata   \\
& B0215.9+3553 & 02 19 00.18  & +36 07 34.5   & 22  & 14  & $-0.12$   \\
& B0219.8+3524 & 02 22 52.98  & +35 38 26.2   & 29  & 29  & $-0.64$   \\
& B0220.3+3537 & 02 23 23.98  & +35 51 05.8   & 35  & 33  & $-0.50$   \\
& B0217.4+3610 & 02 20 26.67  & +36 24 39.8   & 22  & 29  & $-0.88$   \\
\tableline
$0.5\arcdeg-1\arcdeg$ & B0220.4+3531  & 02 23 27.04  & +35 45 29.4   & 25   & 23   & $-0.32$   \\
& B0216.7+3515A & 02 19 44.91  & +35 30 18.8   & 18   & 18   & \nodata   \\
& B0216.7+3515  & 02 19 47.22  & +35 29 00.0   & 48   & 60   & \nodata   \\
& B0216.7+3515B & 02 19 47.92  & +35 28 32.8   & 48   & 41   & $-1.02$   \\
& B0215.4+3536  & 02 18 28.69  & +35 49 58.6   & 1445 & 1401 & $-0.89$   \\
& \nodata       & 02 18 28.996 & +35 50 01.84  & 93   & 670  & \nodata   \\
& B0220.5+3554  & 02 23 36.81  & +36 08 34.1   & 28   & 24   & $-0.83$   \\
& B0220.7+3551  & 02 23 46.60  & +36 05 01.3   & 176  & 173  & $-0.97$   \\
& \nodata       & 02 23 46.790 & +36 05 03.20  & 104  & 149  & \nodata   \\
& B0219.9+3514  & 02 22 56.36  & +35 28 05.4   & 26   & 23   & $-0.73$   \\
& B0220.7+3558A & 02 23 47.14  & +36 11 34.5   & 85   & 89   & $-0.79$   \\
& B0220.7+3558  & 02 23 47.37  & +36 12 17.3   & 85   & 137  & \nodata   \\
& B0220.8+3527  & 02 23 50.17  & +35 41 31.7   & 398  & 393  & $-0.83$   \\
& B0220.7+3558B & 02 23 47.92  & +36 13 49.8   & 45   & 47   & $-0.93$   \\
& B0221.1+3548  & 02 24 08.54  & +36 01 55.2   & 40   & 34   & $-1.06$   \\
& B0221.1+3551  & 02 24 13.51  & +36 04 51.9   & 73   & 67   & $-0.86$   \\
& B0220.5+3609  & 02 23 32.65  & +36 22 47.7   & 22   & 22   & \nodata   \\
& B0215.2+3522  & 02 18 13.64  & +35 36 42.5   & 42   & 39   & $-0.80$   \\
& B0215.4+3516  & 02 18 29.78  & +35 30 27.7   & 110  & 101  & $-1.02$   \\
& B0219.1+3503  & 02 22 07.70  & +35 17 26.9   & 73   & 66   & $-0.95$   \\
& B0221.4+3535A & 02 24 26.91  & +35 48 50.9   & 26   & 19   & $-0.94$   \\
& B0217.8+3500  & 02 20 50.30  & +35 14 26.9   & 52   & 38   & $-0.41$   \\
& B0221.4+3535  & 02 24 29.69  & +35 49 01.3   & 26   & 38   & \nodata   \\
& B0219.2+3622  & 02 22 18.62  & +36 35 49.7   & 61   & 89   & \nodata   \\
& B0216.5+3504  & 02 19 34.57  & +35 18 02.7   & 29   & 25   & $-1.08$   \\
& B0221.4+3535B & 02 24 32.69  & +35 49 11.6   & 26   & 18   & $-0.72$   \\
& B0216.7+3502  & 02 19 44.76  & +35 16 08.4   & 26   & 31   & $-1.06$   \\
& B0215.8+3616  & 02 18 53.75  & +36 30 35.9   & 126  & 119  & $-0.81$   \\
& B0214.4+3534  & 02 17 29.52  & +35 47 56.3   & 105  & 111  & $-0.74$   \\
& B0217.4+3629  & 02 20 30.75  & +36 42 48.3   & 58   & 48   & $-0.52$   \\
& B0219.6+3625  & 02 22 43.74  & +36 39 02.3   & 34   & 24   & $-0.88$   \\
& B0220.5+3618  & 02 23 38.39  & +36 32 08.0   & 30   & 21   & $-0.19$   \\
& B0213.9+3543  & 02 16 59.68  & +35 57 21.6   & 162  & 179  & $-0.72$   \\
& B0218.5+3632  & 02 21 32.74  & +36 45 45.5   & 54   & 63   & $-0.75$   \\
& B0217.8+3452B & 02 20 49.19  & +35 06 28.3   & 107  & 98   & \nodata   \\
& B0217.8+3452  & 02 20 48.92  & +35 06 25.2   & 107  & 109  & \nodata   \\
& B0218.9+3452  & 02 21 57.11  & +35 06 09.7   & 37   & 30   & $-0.6$    \\
& B0215.8+3626  & 02 18 49.99  & +36 40 41.6   & 156  & 144  & $-0.07$   \\
& \nodata       & 02 18 50.038 & +36 40 42.58  & 101  & 126  & \nodata   \\
& B0222.4+3543  & 02 25 28.81  & +35 57 10.2   & 108  & 89   & $-0.68$   \\
& B0213.6+3540  & 02 16 37.63  & +35 54 21.1   & 224  & 234  & $-1.03$   \\
& B0222.6+3554  & 02 25 40.74  & +36 08 19.9   & 48   & 74   & \nodata   \\
& B0222.6+3554B & 02 25 42.54  & +36 07 28.8   & 48   & 45   & $-0.54$   \\
& B0221.4+3622  & 02 24 31.52  & +36 36 02.1   & 33   & 51   & $-0.60$   \\
& B0214.5+3503  & 02 17 35.04  & +35 17 22.6   & 1146 & 1092 & $-0.90$   \\
& \nodata       & 02 17 34.989 & +35 17 21.49  & 187  & 808  & \nodata   \\
& B0214.1+3507  & 02 17 11.78  & +35 21 05.3   & 81   & 81   & $-0.73$   \\
& B0217.9+3442  & 02 20 56.23  & +34 56 36.2   & 94   & 104  & $-0.72$   \\
\tableline
$1\arcdeg-1.5\arcdeg$ & B0222.4+3613A & 02 25 29.65  & +36 25 50.9   & 60   & 91   & \nodata   \\
& B0222.4+3613  & 02 25 30.21  & +36 26 34.7   & 99   & 185  & \nodata   \\
& B0222.4+3613B & 02 25 30.65  & +36 27 08.9   & 99   & 93   & $-0.77$   \\
& B0213.1+3558  & 02 16 09.27  & +36 12 34.7   & 44   & 37   & $-0.67$   \\
& B0223.1+3551  & 02 26 10.41  & +36 04 41.9   & 33   & 29   & \nodata   \\
& B0215.4+3636  & 02 18 30.87  & +36 50 27.2   & 339  & 321  & $-0.55$   \\
& \nodata       & 02 18 30.913 & +36 50 27.45  & 127  & 136  & \nodata   \\
& B0212.9+3535  & 02 15 55.24  & +35 49 12.0   & 61   & 64   & $-0.68$   \\
& B0222.4+3617  & 02 25 29.56  & +36 31 03.3   & 64   & 57   & $-0.85$   \\
& B0217.3+3645  & 02 20 20.42  & +36 59 16.7   & 367  & 2686 & \nodata   \\
& B0223.0+3603A & 02 26 04.43  & +36 17 19.8   & 337  & 308  & $-1.02$   \\
& B0223.0+3603  & 02 26 04.82  & +36 17 10.0   & 337  & 355  & $-1.11$   \\
& B0213.9+3502  & 02 16 57.40  & +35 16 33.5   & 61   & 64   & $-0.77$   \\
& B0223.0+3603B & 02 26 08.11  & +36 15 47.6   & 50   & 47   & \nodata   \\
& B0212.8+3553  & 02 15 48.66  & +36 06 56.9   & 83   & 79   & $-0.80$   \\
& B0212.9+3603  & 02 15 57.29  & +36 16 55.4   & 93   & 91   & $-0.82$   \\
& B0222.7+3618  & 02 25 45.12  & +36 31 32.5   & 57   & 57   & $-0.81$   \\
& B0223.5+3542  & 02 26 36.14  & +35 55 46.2   & 1758 & 1730 & $-0.59$   \\
& \nodata       & 02 26 36.104 & +35 55 45.53  & 518  & 954  & \nodata   \\
& B0217.0+3648B & 02 20 08.92  & +37 02 21.4   & 42   & 54   & $-1.11$   \\
& B0217.0+3648  & 02 20 06.68  & +37 02 41.2   & 45   & 95   & \nodata   \\
& B0217.0+3648A & 02 20 04.01  & +37 03 08.9   & 45   & 40   & $-0.69$   \\
& B0217.4+3434  & 02 20 29.93  & +34 48 25.1   & 62   & 57   & $-1.08$   \\
& B0213.0+3614B & 02 16 02.26  & +36 27 54.2   & 98   & 94   & $-0.81$   \\
& B0213.0+3614  & 02 16 01.21  & +36 27 59.1   & 98   & 117  & \nodata   \\
& B0222.6+3623  & 02 25 44.67  & +36 36 36.7   & 171  & 158  & $-0.55$   \\
& B0223.7+3600  & 02 26 46.21  & +36 13 50.2   & 45   & 36   & $-0.53$   \\
& B0223.8+3533  & 02 26 54.06  & +35 47 03.2   & 48   & 45   & $-0.89$   \\
& \nodata       & 02 26 53.435 & +35 46 48.08  & 86   & 90   & \nodata    \\
& B0221.0+3440  & 02 24 03.11  & +34 54 26.6   & 107  & 102  & $-0.41$   \\
& B0223.6+3607  & 02 26 43.34  & +36 20 55.7   & 109  & 100  & $-0.57$   \\
& B0212.2+3604  & 02 15 18.14  & +36 17 57.2   & 1278 & 1303 & $-0.96$   \\
& B0220.4+3433  & 02 23 25.14  & +34 47 20.1   & 134  & 145  & $-0.98$   \\
& B0224.3+3544  & 02 27 23.05  & +35 57 42.5   & 72   & 75   & $-0.67$   \\
& B0216.3+3656  & 02 19 21.87  & +37 10 05.7   & 106  & 98   & $-0.77$   \\
& B0223.5+3501  & 02 26 32.36  & +35 15 21.8   & 122  & 102  & $-0.81$   \\
& B0216.6+3659  & 02 19 40.76  & +37 13 05.2   & 82   & 89   & $-0.61$   \\
& B0218.1+3419  & 02 21 09.69  & +34 33 37.8   & 252  & 252  & $-0.54$   \\
& B0224.7+3558  & 02 27 48.73  & +36 11 25.1   & 61   & 47   & $-0.02$   \\
& B0211.2+3550  & 02 14 14.41  & +36 04 40.5   & 182  & 186  & $-0.84$   \\
& B0212.4+3454  & 02 15 24.71  & +35 08 50.6   & 78   & 66   & $-0.86$   \\
& B0224.6+3515  & 02 27 38.33  & +35 28 43.8   & 164  & 147  & $-0.82$   \\
& B0222.3+3650  & 02 25 22.43  & +37 03 40.9   & 146  & 172  & $-0.95$   \\
& B0220.0+3704  & 02 23 07.86  & +37 18 03.6   & 200  & 198  & $-0.82$   \\
& B0224.8+3607A & 02 27 51.15  & +36 21 29.6   & 97   & 98   & $-1.23$   \\
& B0224.8+3607  & 02 27 52.76  & +36 21 01.3   & 97   & 168  & \nodata   \\
& B0224.8+3607B & 02 27 54.99  & +36 20 24.2   & 88   & 69   & $-0.79$   \\
& B0214.1+3430  & 02 17 05.91  & +34 44 30.4   & 128  & 136  & $-0.82$   \\
& B0211.5+3505  & 02 14 29.85  & +35 19 20.0   & 213  & 232  & \nodata   \\
& B0223.2+3441  & 02 26 17.12  & +34 54 36.5   & 88   & 85   & $-0.71$   \\
& B0224.7+3617  & 02 27 50.29  & +36 30 47.1   & 200  & 178  & $-0.82$   \\
& B0219.6+3415  & 02 22 38.27  & +34 29 29.1   & 188  & 174  & $-0.81$   \\
& B0211.2+3511  & 02 14 13.61  & +35 25 07.2   & 357  & 359  & $-0.91$   \\
& B0211.0+3518  & 02 14 01.38  & +35 32 28.7   & 174  & 189  & $-0.73$   \\
& B0211.9+3452  & 02 14 58.14  & +35 06 39.5   & 3011 & 3084 & $-0.83$   \\
& \nodata       & 02 14 57.959 & +35 06 32.27  & 259  & 869  & \nodata   \\
\tableline
$1.5\arcdeg-2\arcdeg$ & B0210.7+3533  & 02 13 43.25  & +35 47 29.5   & 149  & 141  & $-0.97$   \\
& B0214.3+3700  & 02 17 22.91  & +37 14 47.8   & 669  & 650  & $-1.18$   \\
& \nodata       & 02 17 22.726 & +37 14 47.08  & 207  & 248  & \nodata   \\
& B0222.3+3656  & 02 25 27.31  & +37 10 27.9   & 373  & 380  & $-0.45$   \\
& B0216.0+3413  & 02 19 01.20  & +34 27 45.7   & 131  & 136  & $-0.66$   \\
& B0225.0+3620  & 02 28 07.42  & +36 34 09.5   & 227  & 235  & $-0.83$   \\
& B0216.6+3409  & 02 19 37.82  & +34 23 10.5   & 555  & 515  & $-1.06$   \\
& B0215.1+3710  & 02 18 11.43  & +37 24 36.5   & 481  & 473  & $-0.70$   \\
& \nodata       & 02 18 11.421 & +37 24 35.44  & 184  & 192  & \nodata   \\
& B0223.9+3646  & 02 26 59.91  & +36 59 27.9   & 76   & 60   & $-0.75$   \\
& B0219.2+3717  & 02 22 15.42  & +37 31 16.7   & 144  & 123  & $-0.08$   \\
& B0225.9+3602  & 02 28 57.74  & +36 16 13.8   & 125  & 122  & $-0.80$   \\
& B0212.8+3429  & 02 15 49.40  & +34 43 02.8   & 213  & 199  & $-0.19$   \\
& B0221.9+3417  & 02 24 56.18  & +34 30 35.8   & 1657 & 1638 & $-0.91$   \\
& \nodata       & 02 24 56.629 & +34 30 27.43  & 184  & 222  & \nodata   \\
& B0211.1+3634  & 02 14 11.25  & +36 48 24.9   & 228  & 244  & $-1.20$   \\
& B0216.2+3405B & 02 19 18.40  & +34 19 49.5   & 273  & 338  & $-0.96$   \\
& B0216.2+3405  & 02 19 15.49  & +34 19 42.1   & 412  & 802  & \nodata   \\
& B0216.2+3405A & 02 19 13.44  & +34 19 37.2   & 412  & 463  & $-0.86$   \\
& B0218.8+3402  & 02 21 50.01  & +34 16 19.1   & 359  & 347  & $-0.88$   \\
& B0226.2+3534  & 02 29 20.00  & +35 47 40.2   & 167  & 163  & $-0.80$   \\
& B0224.7+3442  & 02 27 43.89  & +34 56 23.1   & 100  & 85   & $-0.90$   \\
& B0221.6+3406A & 02 24 34.48  & +34 21 39.5   & 104  & 109  & $-1.11$   \\
& B0209.5+3536  & 02 12 31.09  & +35 50 40.1   & 381  & 482  & \nodata   \\
& B0226.1+3509  & 02 29 11.30  & +35 22 22.6   & 160  & 178  & \nodata   \\
& B0224.2+3430  & 02 27 16.40  & +34 43 42.9   & 129  & 128  & $-0.96$   \\
& B0221.3+3405  & 02 24 20.16  & +34 19 17.5   & 195  & 193  & $-0.84$   \\
& B0221.6+3406  & 02 24 39.96  & +34 20 03.1   & 459  & 543  & \nodata   \\
& B0221.6+3406B & 02 24 41.08  & +34 19 43.6   & 459  & 433  & $-0.93$   \\
& B0220.7+3401  & 02 23 47.53  & +34 15 08.3   & 403  & 421  & $-0.80$   \\
& B0226.3+3619  & 02 29 24.87  & +36 32 26.6   & 191  & 174  & $-0.85$   \\
& B0224.9+3650  & 02 28 01.68  & +37 03 36.3   & 1409 & 1455 & $-0.90$   \\
& B0226.5+3618  & 02 29 35.60  & +36 31 22.9   & 417  & 401  & $-0.67$   \\
& \nodata       & 02 29 35.650 & +36 31 24.65  & 294  & 335  & \nodata   \\
& B0218.9+3353  & 02 21 56.58  & +34 07 17.5   & 127  & 122  & $-0.92$   \\
& B0216.9+3732  & 02 19 59.02  & +37 45 56.1   & 373  & 362  & $-0.78$   \\
& B0223.1+3408  & 02 26 10.15  & +34 21 30.7   & 3793 & 3678 & $-0.16$   \\
& \nodata       & 02 26 10.338 & +34 21 30.28  & 700  & 1834 & \nodata   \\
& B0212.4+3714  & 02 15 31.32  & +37 28 43.3   & 333  & 323  & $-0.71$   \\
& B0209.2+3504  & 02 12 14.34  & +35 18 24.7   & 253  & 230  & $-0.85$   \\
& B0225.1+3659  & 02 28 13.67  & +37 12 57.1   & 1566 & 1519 & $-1.09$   \\
& \nodata       & 02 28 13.706 & +37 12 58.23  & 643  & 692  & \nodata   \\
& B0208.4+3547  & 02 11 28.80  & +36 01 26.2   & 446  & 431  & $-0.94$   \\
& B0210.3+3652  & 02 13 22.53  & +37 07 00.0   & 303  & 320  & \nodata   \\
& B0225.6+3655  & 02 28 42.20  & +37 08 58.0   & 306  & 308  & $-0.85$   \\
& B0225.6+3655A & 02 28 42.07  & +37 09 03.6   & 306  & 286  & $-0.80$   \\
& B0212.5+3405  & 02 15 29.49  & +34 19 47.5   & 337  & 328  & $-0.86$   \\
\tableline
$2\arcdeg-3\arcdeg$ & B0218.6+3342  & 02 21 40.00  & +33 56 36.1   & 195  & 204  & $-0.84$   \\
& B0219.8+3343A & 02 22 48.79  & +33 57 18.9   & 638  & 638  & $-0.83$   \\
& B0219.8+3343  & 02 22 48.85  & +33 57 11.3   & 638  & 688  & $-0.88$   \\
& B0228.1+3547  & 02 31 11.57  & +36 00 28.4   & 321  & 303  & $-1.33$   \\
& B0219.2+3339  & 02 22 16.75  & +33 53 27.3   & 516  & 499  & $-0.67$   \\
& \nodata       & 02 22 16.801 & +33 53 26.98  & 331  & 337  & \nodata   \\
& B0225.5+3419  & 02 28 34.05  & +34 33 09.5   & 504  & 568  & $-1.06$   \\
& B0225.5+3419B & 02 28 34.63  & +34 32 58.5   & 504  & 492  & $-0.97$   \\
& B0216.8+3334  & 02 19 48.77  & +33 48 16.0   & 1352 & 1357 & $-0.81$   \\
& B0212.6+3736  & 02 15 40.01  & +37 50 09.3   & 691  & 655  & $-0.54$   \\
& B0223.9+3351  & 02 26 55.50  & +34 04 55.7   & 532  & 525  & $-1.02$   \\
& B0208.9+3429  & 02 11 54.12  & +34 44 01.7   & 569  & 551  & $-1.08$   \\
& B0210.5+3404  & 02 13 28.47  & +34 18 20.0   & 402  & 482  & $-0.90$   \\
& B0216.0+3756B & 02 19 09.03  & +38 10 00.7   & 439  & 645  & $-0.84$   \\
& B0216.0+3756  & 02 19 08.14  & +38 10 15.6   & 439  & 739  & \nodata   \\
& B0206.6+3533  & 02 09 38.91  & +35 47 48.9   & 4259 & 5519 & $-0.67$   \\
& B0206.6+3533A & 02 09 38.87  & +35 47 48.9   & 4259 & 5489 & $-0.66$   \\
& B0213.2+3750  & 02 16 19.76  & +38 04 39.4   & 709  & 782  & $-0.89$   \\
& B0229.4+3517  & 02 32 29.32  & +35 30 53.6   & 2149 & 2359 & $-0.86$   \\
& B0227.3+3713  & 02 30 25.65  & +37 26 18.0   & 1034 & 1013 & $-1.10$   \\
& B0214.7+3321  & 02 17 43.31  & +33 35 03.0   & 683  & 657  & $-0.85$   \\
& B0228.9+3429  & 02 31 59.53  & +34 42 26.7   & 2361 & 2556 & $-1.02$   \\
& B0228.1+3729  & 02 31 14.98  & +37 42 57.0   & 1426 & 1397 & $-1.03$   \\
& B0229.4+3410  & 02 32 28.67  & +34 24 03.0   & 9334 & 9459 & $-0.88$   
\footnotetext[1]{The first data entry for each target refers to the WENSS target source, and the second entry, when present, refers to the VLBI detection.}
\footnotetext[2]{Units of right ascension are hours, minutes and seconds.}
\footnotetext[3]{Units of declination are degree, arcminutes and arcseconds.}
\footnotetext[4]{The spectral index $\alpha$ is defined as $S_{\nu}\propto\nu^{\alpha}$ and is estimated from WENSS and NVSS integrated flux densities where available.}
\end{longtable}
\end{center}

\renewcommand{\thefootnote}{\arabic{footnote}}
\normalsize

\begin{table}[ht]
\begin{center}
{ \tiny
\begin{tabular}{lccccccccc} \hline \hline
WENSS  & \multicolumn{2}{c}{Source Location}                & \multicolumn{3}{c}{Image Characteristics}                & \multicolumn{4}{c}{Position Comparison} \\
Source & Field\tablenotemark{a} & $d_{PC}$\tablenotemark{b} & Beam Size & Beam P.A. & $1\sigma$ Noise\tablenotemark{c} & Ref.\tablenotemark{d} & $d_{E}$\tablenotemark{e} &  $\theta_{E}$\tablenotemark{f} & $d_{\sigma}$\tablenotemark{g} \\
       &                        & (arcdeg)                  & (mas,mas) & (degrees) & (mJy beam$^{-1}$)                &                       & (arcsec) & (degrees) & ($\sigma_{p}$) \\ \hline \hline
3C84      & \nodata & \nodata & $31\times15$   & $-17$  & 6.1 & BE02     & 0.002  & 228  & 0.7   \\
B0223.1+3408  &  1  &  0.00   & $37\times19$   & $-13$  & 1.8 & BE02     & 0.002  & 270  & 0.7   \\
\nodata       &  2  &  1.89   & $360\times285$ & $-42$  & 5.8 & BE02     & 0.042  & 125  & 1.4   \\
B0218.0+3542  &  2  &  0.00   & $64\times43$   & $52$   & 1.0 & PA92     & 0.024  & 75   & 0.9   \\
B0219.1+3533  &  2  &  0.27   & $88\times84$   & $80$   & 1.5 & CO98     & 0.173  & 168  & 0.1   \\
B0221.9+3417  &  1  &  0.29   & $145\times132$ & $38$   & 4.4 & DO96     & 9.938  & 34   & 13.2  \\
\nodata       &  2  &  1.63   & $361\times285$ & $-42$  & 4.1 & DO96     & 10.003 & 34   & 13.0  \\
B0221.6+3406B &  1  &  0.31   & $145\times132$ & $37$   & 4.1 & CO98     & 0.613  & 160  & 0.6   \\
B0223.9+3351  &  1  &  0.32   & $145\times132$ & $38$   & 4.3 & CO98     & 1.432  & 55   & 1.5   \\
B0221.6+3406A &  1  &  0.33   & $145\times132$ & $37$   & 4.3 & CO98     & 1.178  & 42   & 0.8   \\
B0216.6+3523  &  2  &  0.42   & $142\times133$ & $31$   & 2.2 & CO98     & 0.437  & 188  & 0.4   \\
B0220.1+3532  &  2  &  0.46   & $142\times134$ & $30$   & 2.0 & CO98     & 0.215  & 313  & 0.2   \\
B0215.4+3536  &  2  &  0.54   & $180\times170$ & $23$   & 2.5 & R\"{O}94 & 0.103  & 56   & 0.1   \\
B0223.2+3441  &  1  &  0.55   & $180\times168$ & $38$   & 5.0 & CO98     & 1.382  & 317  & 1.1   \\
B0220.7+3551  &  2  &  0.56   & $180\times170$ & $23$   & 2.4 & CO98     & 0.231  & 148  & 0.2   \\
B0215.8+3626  &  2  &  0.87   & $227\times217$ & $-38$  & 3.0 & CO98     & 0.226  & 354  & 0.2   \\
B0219.2+3339  &  1  &  0.93   & $224\times218$ & $-62$  & 6.3 & CO98     & 0.572  & 46   & 0.6   \\
\nodata       &  2  &  2.06   & $359\times284$ & $-43$  & 4.2 & CO98     & 0.374  & 42   & 0.4   \\
B0214.5+3503  &  2  &  0.96   & $227\times217$ & $-36$  & 3.9 & R\"{O}94 & 0.162  & 49   & 0.2   \\
B0215.4+3636  &  2  &  1.04   & $228\times217$ & $-36$  & 3.3 & CO98     & 0.579  & 114  & 0.6   \\
B0223.5+3542  &  2  &  1.12   & $228\times217$ & $-37$  & 5.3 & DO96     & 0.833  & 295  & 1.8   \\
\nodata       &  1  &  1.57   & $357\times287$ & $-44$  & 9.1 & DO96     & 0.783  & 308  & 1.6   \\
B0223.8+3533  &  2  &  1.19   & $228\times217$ & $-35$  & 3.1 & CO98     & 25.146 & 344  & 11.0  \\
B0211.9+3452  &  2  &  1.50   & $227\times217$ & $-37$  & 3.6 & CO98     & 6.936  & 344  & 10.6  \\
B0214.3+3700  &  2  &  1.51   & $352\times286$ & $-40$  & 4.6 & CO98     & 2.544  & 271  & 2.6   \\
B0215.1+3710  &  2  &  1.58   & $353\times287$ & $-41$  & 4.1 & CO98     & 0.368  & 322  & 0.4   \\
B0226.5+3618  &  2  &  1.81   & $360\times288$ & $-40$  & 3.8 & CO98     & 1.362  & 192  & 1.4   \\
B0225.1+3659  &  2  &  1.92   & $358\times288$ & $-40$  & 5.3 & CO98     & 0.256  & 143  & 0.3   \\ \hline
\tablenotetext{a}{Field 1 is centred about J0226+3421 and field 2 is centred about B0218$+$357.}
\tablenotetext{b}{Distance of the source from the phase centre of the observed field.}
\tablenotetext{c}{The $1\sigma$ residual noise after the source model has been subtracted from the image.}
\tablenotetext{d}{References for positions.~
  BE02 = 2.3/8.3 GHz VLBA \citep{Beasley:2002p10526};
  CO98 = 1.4 GHz VLA \citep{Condon:1998p10541};
  DO96 = 365 MHz Texas interferometer \citep{Douglas:1996p10548};
  PA92 = 8.4 GHz VLA \citep{Patnaik:1992p10574};
  R\"{O}94 = 1.4 GHz VLA \citep{Roettgering:1994p10579}
}
\tablenotetext{e}{The measured offset of the VLBI source peak from the position listed in reference.}
\tablenotetext{f}{The position angle of the VLBI source peak in relation to the position listed in reference.}
\tablenotetext{g}{The measured offset of the VLBI source peak in relation to the position listed in reference in terms of the $1\sigma$ astrometric precision $\sigma_{p}$.}
\end{tabular}
\caption{Source and image characteristics and astrometric errors.}
\label{tab:tabp2t2}
}
\end{center}
\end{table}

\linespread{1.0}
\normalsize
\begin{savequote}[20pc]
\sffamily
Nothing exists except atoms and empty space;\\
everything else is opinion.
\qauthor{Democritus}
\end{savequote}

\linespread{1.3}
\normalsize
\chapter{Conclusion}
\label{chap:conclusion}

The research reported in this thesis has, for the first time, applied new wide-field VLBI imaging techniques to real astronomical sources. This has been made possible through recent improvements in data storage and processing capacity, improvements in instrument sensitivity, the development of new processing algorithms and advances in hardware and software correlators. Through these advances and the improved calibration techniques described in this thesis, wide-field VLBI has matured into a powerful tool that has increased the imageable field-of-view from traditional sub-arcsecond fields to fields spanning arc-minutes and even arc-degrees at low frequencies. Wide-field VLBI observations have been used to successfully target three main science areas: (i) a study of compact radio sources in local southern starburst galaxies, (ii) a study of jet interactions in southern AGN, and (iii) a 28 deg$^{2}$ unbiased survey of the 90 cm sky at VLBI resolution. The results of these studies are briefly summarised in the following sections.

\section{Starbursts}

\subsection{NGC 253}

Wide-field, 2.3 GHz VLBI observations of the prominent starburst galaxy NGC 253, obtained with the LBA, have produced the highest angular resolution ($\sim15$ mas) image of this source to date. The image revealed six sources, all of which corresponded to sources identified in higher frequency ($>5$ GHz) VLA images. One of the sources, supernova remnant 5.48$-$43.3, is resolved into a shell-like structure approximately 90 mas (1.7 pc) in diameter and is estimated to have an age of $80(10^{4}/v)$ yr, based on an assumed radial expansion velocity of $v=$10,000 km s$^{-1}$. From these data and data from the literature, the spectra of 20 compact radio sources in NGC 253 were modelled and found to be consistent with free-free absorbed power laws. Broadly, the free-free opacity is highest toward the nucleus but varies significantly throughout the nuclear region ($\tau_0\sim 1->20$), implying that the overall structure of the ionised medium is clumpy or that the sources are situated at differing depths within the dense nuclear region.

Multi-wavelength comparisons of the free-free opacity against optical and non-optical tracers of ionised gas failed to show any significant correlation. For the optical tracers, the lack of correlation is most likely due to the characteristically high levels of interstellar extinction associated with starbursts. For the non-optical tracers, the lack of correlation here is attributed to the substantially lower resolution of these images compared to the LBA images. A comparison with radio recombination line images show that four of the modelled sources have free-free optical depths expected by RRL models.

Of the 20 sources, eight have flat intrinsic spectra associated with thermal radio emission and the remaining 12 have steep intrinsic spectra, associated with synchrotron emission from supernova remnants. Based on the lack of detection of new sources over a period spanning 17 years, a supernova rate upper limit of 2.4 yr$^{-1}$ is determined for the inner 320 pc region of the galaxy at the 95\% confidence level. Similarly, estimates of supernova remnant source counts, sizes and expansion rates were used to derive an upper limit to the supernova rate of $>0.14 (v/ 10^{4})$ yr$^{-1}$, where $v$ is the radial expansion velocity of the supernova remnant in km s$^{-1}$. Based on these limits, the star formation rate is estimated to be $3.4 (v/10^{4}) < SFR(M\geq5M_{\Sun})<59$ M$_{\Sun}$ yr$^{-1}$ and is of the same order of magnitude as rates determined from integrated FIR and radio luminosities. Both upper and lower limits on the supernova rate could be further constrained with more frequent, high sensitivity observations with the LBA or the VLBA.

\subsection{NGC 4945}

Wide-field, 2.3 GHz LBA observations of the nearby starburst galaxy NGC 4945 over two epochs, have resulted in the first high resolution ($\sim15$ mas) images of this galaxy. The observations were complemented by ATCA observations between 17 GHz and 23 GHz. A total of 15 compact radio sources were observed in the VLBI images, 13 of which correspond to sources identified in the ATCA images. Four of the sources are resolved into shell-like structures ranging between 60 and 110 mas (1.1 to 2.1 pc) in diameter and have ages of $25 (10^{4}/v)$ to $100 (10^{4}/v)$ yr, assuming an average radial expansion velocity of $v=10,000$ km s$^{-1}$.

The spectral energy distribution of the detected sources were consistent with free-free absorbed power laws in nine of the sources and a simple power law spectrum in the remaining four. As with NGC 253, the free-free opacity was found to be highest toward the nucleus but varying significantly throughout the nuclear region ($\tau_0\sim 6-23$). Of the 13 sources, 10 have steep intrinsic spectra, associated with synchrotron emission from supernova remnants. Three sources have flat intrinsic spectra which may be associated with thermal radio emission, however, as the sources have brightness temperatures of $>2\times10^{5}$ K, it is likely that these may be \ion{H}{2} regions with an embedded supernova remnant.

Based on the non-detection of new sources in observations spanning a relatively short period of 1.9 years between the two observing epochs, a supernova rate upper limit of 15.3 yr$^{-1}$ is determined for the inner 250 pc region of the galaxy at the 95\% confidence level.  Similarly, based on estimates of supernova remnant source counts, sizes and expansion rates, the lower limit of the supernova rate is estimated to be $>0.1 (v/ 10^{4})$ yr$^{-1}$, where $v$ is the radial expansion velocity of the supernova remnant in km s$^{-1}$. From the supernova rate limits, a star formation rate of $2.4 (v/10^{4}) < SFR(M\geq5M_{\Sun})<370$ M$_{\Sun}$ yr$^{-1}$ is estimated and is of the same order of magnitude as rates determined from integrated FIR (1.5 M$_{\Sun}$ yr$^{-1}$) and radio luminosities ($14.4\pm1.4$ M$_{\Sun}$ yr$^{-1}$). The supernova rates and star formation rates determined for NGC 4945 are, to within a factor of two, similar to those measured in NGC 253 and M82.

A non-thermal source with a jet-like morphology is detected in the nuclear region of the galaxy. The source is offset by $\sim1050$ mas from the assumed location of the AGN based on H$_{2}$O megamaser emission, the HNC cloud centroid, the K-band peak and the hard X-ray peak.

\subsection{NGC 55, NGC 1313, NGC 5236 and NGC 5253}

An attempt was made to image the nearby starburst galaxies NGC 55, NGC 5236 and NGC 5253, with the LBA at 2.3 GHz. No compact radio sources have been found to be associated with these galaxies. Similar observations of NGC 1313 detected two compact radio sources, one of which is the supernova remnant SN 1978K. The low frequency VLBI observations were complemented with ATCA observations between 17 and 23 GHz. Weak and extended thermal emission was detected in NGC 55, NGC 5236 and NGC 5253, however, no emission was detected in NGC 1313.

The absence of compact radio sources in NGC 55 and NGC 5253 and the small number of detections in NGC 1313 is consistent with the low star formation rates (SFRs) implied from the far-infrared flux density of these galaxies when compared against the number of detections verses SFR for prototypical starbursts such as NGC 253 and NGC 4945. Furthermore, the weak diffuse emission observed with the ATCA in all four galaxies is suggestive of low density nuclear environments, resulting in weak and short-lived supernova emission that rapidly fades below the VLBI detection limit. This would explain the absence of detections in NGC 5236 where the SFR implied from FIR observations is comparable to that of NGC 4945 and fleeting supernova events have been observed in the past.

\section{AGN jet interactions}

\subsection{Pictor A}

The first VLBI image of the north-west hot spot of Pictor A ($z=0.035$) has been made using wide-field VLBI techniques. The image is the highest spatial resolution image (16 pc), by a factor of three, of such an object to date. The north-west hot spot of Pictor A is resolved into a complex set of components that coincide with the bright part of the hot spot imaged at arcsecond-scales with the VLA at radio wavelengths and with the \emph{HST} at optical wavelengths.

The detection of parsec-scale structures in radio galaxy hot spots has implications for the models used to explain the X-ray emission from the hot spot. Previous modelling by \cite{Wilson:2001p536} predicted a balanced SSC and synchrotron emission in the \emph{Chandra} band and strong SSC emission above the \emph{Chandra} band. In contrast, our observations suggest that the small-scale structures indicate regions of strong shocks in the fluid flow and so represent regions of recently accelerated electrons, which have higher break frequencies than the hot spot electrons on larger spatial scales.  A consequence of this is that the X-rays in the \emph{Chandra} band are dominated by synchrotron X-rays, with a relatively weak SSC contribution. Future high energy observations may readily distinguish between the model we have put forward and that of \citet{Wilson:2001p536}.

Based on the sizes of the individual small-scale components of the hot spot and their angular spread, we estimate that the jet width at the hot spot is in the range 70 - 700 pc. The lower limit arises from the suggestion that the jet may dither in its direction as it passes through hot spot backflow material close to the jet termination point, creating a $``$dentist drill" effect on the inside of a cavity 700 pc in diameter. The estimated jet width is comparable to similar estimates in PKS 2153$-$69 ($z=0.028$),  3C 205 ($z=1.534$), and 4C 41.17 ($z=3.8$).

\subsection{PKS $0344-345$}

Wide-field VLBI observations of jet interactions in the galaxy PKS $0344-345$ ($z=0.0538$), obtained with the LBA, have detected no emission associated with the hot spot at 2.3 GHz. Low resolution ATCA radio images of the interaction region suggest that the region is extended ($\sim6.8$ kpc) and has a spectral index of $\alpha=-0.84$ that is comparable to a similar interaction region in PKS $2152-699$. A 1-KeV X-ray flux density of $\sim8$ nJy is predicted for this interaction region based on a simple power-law model for the spectral energy distribution.

\subsection{PKS $0521-365$}

Wide-field VLBI observations of PKS $0521-365$ ($z=0.05534$), obtained with the LBA, reveal two components associated with the south-east hot spot of the galaxy at 1.6 GHz and 2.3 GHz. A simple power-law, with spectral index $\alpha=-1.06$, provides a good fit between our VLBI data and \emph{Chandra} X-ray data. The size of the individual components and the overall extent of the hot spot is similar to that observed in Pictor A using similar methods.. The similarities suggest that the same emission mechanisms may be at work in the two galaxies, however, further multi-wavelength observations at optical and infrared wavelengths will be required to verify this.

\section{Wide-field surveys}

We have completed the first wide-field, deep, unbiased, VLBI survey at 90 cm. The survey area consisted of two overlapping 28 deg$^{2}$ fields centred on the quasar J0226$+$3421 and the gravitational lens B0218$+$357. This is the widest field of view VLBI survey with a single pointing to date, exceeding the total survey area of previous higher frequency surveys by two orders of magnitude. A total of 618 sources were targeted within these fields, based on identifications from Westerbork Northern Sky Survey (WENSS) data. Of these sources, 272 had flux densities that, if unresolved, would fall above the sensitivity limit of the VLBI observations.

A total of 27 sources were detected as far as $2\arcdeg$ from the phase centre and their distribution of morphologies are typical of AGN.  Of these, 10 are unresolved (consistent with core-dominated AGN), 8 are clearly resolved into double component sources (consistent with being core-jet AGN or double-lobed radio galaxies), 7 have complex or extended structures (these may be core-jet AGN or radio galaxies) and the remaining 2 sources are the gravitational lens and the quasar at the phase centres of the two fields.

The results of the survey suggest that at least $10\%$ of moderately faint (S$\sim100$ mJy) sources found at 90 cm contain compact components smaller than $\sim0.1$ to $0.3$ arcsec and stronger than $10\%$ of their total flux densities.  This is a strict lower limit as the sensitivity of our observation was limited by the primary beam at the edge of the survey fields. These initial results suggest that new low frequency telescopes, such as LOFAR, should detect many compact radio sources and that plans to extend these arrays to baselines of several thousand kilometres are warranted.

None of the surveyed sources that were even slightly resolved by WENSS were detected. Similarly, none of the WENSS sources that were below the sensitivity limits of the VLBI observation were detected either, suggesting that none of these sources had varied significantly since the WENSS observations were carried out. Only one source appears to be a serendipitous detection of a likely highly variable, very compact source near the target WENSS source B0223.8$+$3533.

\section{Future work}

\subsection{Starbursts}

Multi-epoch and multi-wavelength VLBI observations of the two most prominent galaxies in our sample, NGC 253 and NGC 4945, should be continued in the future. For NGC 4945, additional observations at 1.4 GHz would aid in more stringently defining the free-free parameters for the detected sources. Frequent observations of these galaxies will help constrain the upper limit on the supernova rate by increasing the proportion of supernova events that may be captured between epochs. Observations over a longer period of time ($\sim10$ years) will also further constrain the lower limit of the supernova rate by enabling the supernova shell expansion velocity to be measured.

Our understanding of the jet-like object discovered in NGC 4945 would benefit from multi-wavelength and multi-epoch observations. Multi-wavelength observations would enable us to better characterise the spectral energy distribution of the source. Whereas multi-epoch observations would allow us to search for motion in the suggested jet, thereby confirming its nature.

In terms of additional sources for study, Circinus may provide an interesting target. The galaxy is host to nuclear starburst activity \citep{Harnett:1990p27197} and shows evidence of an AGN \citep{Gardner:1982p27219} but was initially left off our selection list as it is highly obscured by our own galaxy, thus complicating multi-wavelength studies. Recently however, a supernova has been discovered near the nuclear region of the galaxy \citep{Bauer:2007p10525} and so is worthy of study at radio and infrared wavelengths.

A brief literature survey of northern-sky starbursts reveals that, apart from the prominent starburst M82 \citep{McDonald:2001p3426}, few have been studied with wide-field VLBI. This presents an opportunity to undertake and complete an all-sky survey of all local starburst galaxies. Such an endeavour can be easily achieved with instruments such as the VLBA and the EVN and would provide a richer data-set with which the characteristics of these galaxies could be compared against.

\subsection{AGN jet interactions}

Further high-resolution observations of Pictor A at the low frequency end of the spectrum may yield interesting science results. Models from \cite{Wilson:2001p536} hint at the presence of a low energy electron population at the hot spot that is not visible at cm wavelengths. In the near future, observations with upcoming instruments such as the Murchison Widefield Array (MWA), being built by an international consortium at the candidate Square Kilometre Array (SKA) site in Western Australia, may be able to make such an observation.  Also possible would be to observe the Pictor A hot spots with the 90 cm system on the VLBA, to obtain very high resolution at long wavelengths.

During the time-frame of this research, it was not possible to observe one of our selected sources, PKS $2152-699$, using the full sensitivity available with the LBA. We believe that such an observation may detect the jet-interaction region in this galaxy and possibly the northern hot spot. Such observation would allow a detailed study of the emission mechanisms at play in these regions and may indicate if they are still interacting with the jet.

Another source of interest that was not included in our sample of sources is Centaurus A. VLA observations of this source have revealed a number of radio knots associated with the jet and an apparent bifurcation of the jet \citep{Hardcastle:2003p2376}. There are some questions raised as to whether or not they are associated with supernovae or supernova remnants that have been triggered by the jet \citep{Capetti:2002p27194}. Thus, Centaurus A provides a fascinating link between both the starburst and jet-interaction investigations conducted in this research and is worthy of further investigation.

\subsection{Wide-field surveys}
The observations presented in Chapter \ref{chap:lfwfvlbi} were limited by the spectral and temporal resolution of the EVN correlator at the time of the observation. To minimise the effects of bandwidth and time-averaging smearing it was necessary to compromise resolution and image noise. With the advent of software correlators \citep[e.g.][]{Deller:2007p10545}, it is now possible to achieve improvements in both temporal and spectral resolution. They can also correlate against multiple phase centres simultaneously and allow for some pre-processing to be applied during the correlation process to, for example, mitigate the effects of radio interference. This opens up new possibilities to not only continue the wide-field survey work presented here but to extend it to enable some SKA science now with available facilities.

For example, the study of the evolution of radio AGN populations and the star formation history of the early Universe figure highly in the science case for the SKA \citep{Jackson:2004p27172}. An improved understanding of the star-formation history of the Universe can be obtained with three concurrent surveys. The first is a targeted survey of ``famous'' fields, the second is a targeted low frequency survey (which also includes ``famous'' fields) and the third is an unbiased survey. Results from the targeted surveys may be compared against the unbiased survey as a control.

A targeted wide-field VLBI survey of “famous” fields, such as the Chandra Deep Field South and the associated ATLAS (Australia Telescope Large Area Survey) – CDFS deep field \citep{Norris:2006p6530}, would enable a multi-wavelength analysis of sources detected within these fields. Using wide-field techniques it would be possible to map out the milli-arcsecond-scale morphologies of the sources, which in combination with the multi-wavelength analysis would enable an estimate of the proportion of AGN to starburst galaxies. Such a survey has now been proposed for the LBA and is currently underway with the VLBA.

The wide-field VLBI surveys initiated at 1.4 GHz \citep{Garrett:2005p10555} and extended at 90 cm (Chapter \ref{chap:lfwfvlbi}), should be continued. This would allow a comparison between the high and low frequency data and help to distinguish between starburst and AGN populations, based on spectral index and morphological considerations. The low frequency observations, in particular, will also provide important information on the nature and incidence of compact, low-frequency radio sources, with consequences for next generation, low-frequency instruments such as LOFAR and the SKA. In particular, whether there are sufficient calibrator sources in the sky in order to calibrate the instrument across its full, very wide, field-of-view. The observations presented in Chapter \ref{chap:lfwfvlbi} suggest that, at least for the two fields surveyed, this is the case. However, a larger survey of the sky will be needed to provide a more definite statement with regard to the population of these sources.

To complement the targeted and low frequency surveys and to act as a control, an unbiased survey may be performed by “piggybacking” on top of any phase-referenced observations made with the LBA. By using the software correlator currently available at this facility, observed data can be re-correlated with finer frequency and temporal outputs allowing them to be imaged in wide-field mode. The wider field of view will allow the detection and imaging of weak sources within the observed field that would not have been recovered using traditional LBA observations. Using this method, up to $\sim15$ fields each year could be imaged in wide-field mode without requiring additional LBA resources. A processing pipeline can be developed for the LBA to automate the wide-field correlation and imaging of data-sets as they become available.

\subsection{Wide-field techniques}

While wide-field VLBI is now possible with current computing platforms it can still be a painfully slow process for large data-sets, such as those generated from wide-field surveys. Surveying, lends itself particularly nicely to parallel processing as the problem space may be easily divided across computing nodes of a Beowulf cluster or cores in a multi-core workstation. The feasibility of this form of processing has been demonstrated using scripting languages such as ParselTongue \citep{Kettenis:2006p27101} on an extremely wide-field low frequency VLBI data-set (Chapter \ref{chap:lfwfvlbi}) by dividing the main field into many smaller target fields -- these may then be processed concurrently on several computing nodes.

Further advances may be achieved by reworking some of the core processing algorithms of existing software packages to work more effectively with multi-core processors. Even today this would provide performance gains of a factor of $2-8$, with 80-core processors planned to be available in commercial numbers by 2010, performance gains of almost 2 orders of magnitude should be possible within the next few years. Furthermore, networking many of these within a Beowulf cluster would result in total performance gains of 3-4 orders of magnitude over a single-core processor. Such performance gains are critical if the processing of wide-field VLBI data from the SKA is to become feasible.

\newpage
\addcontentsline{toc}{chapter}{Bibliography}
\bibliographystyle{thesis}
\bibliography{thesis}
\newpage

\clearpage
\addcontentsline{toc}{chapter}{Publications}
\chapter*{Publications}
\label{chap:publications}

The following publications report on results derived from or supplemented
by the work presented in this thesis. \\

\noindent
Lenc, E., and Tingay, S.J.
2009,
The Sub-parsec Scale Radio Properties of Southern Starburst Galaxies. II. Supernova Remnants, the Supernova Rate, and the Ionised Medium in the NGC 4945 Starburst.
{\it Astronomical J.}, {\bf 137}, 537--553  \\ [-2mm]

\noindent
Tingay, S.J., Lenc, E., Brunetti, G., and Bondi M.
2008,
A high resolution view of the jet termination shock in a hot spot of the nearby radio galaxy Pictor A.
{\it Astronomical J.}, {\bf 136}, 2473--2482  \\ [-2mm]

\noindent
Lenc, E., Garrett, M.A., Wucknitz, O., Anderson, J.M., and Tingay, S.J.
2008,
A deep, high resolution survey of the low frequency radio sky.
{\it Astrophys. J.}, {\bf 673}, 78--95  \\ [-2mm]

\noindent
Johnston, S., et al.
2007,
Science with the Australian Square Kilometre Array Pathfinder.
{\it Pub. of the Astron. Soc. of Aust.}, {\bf 24}, 174--188  \\ [-2mm]

\noindent
Lenc, E.
2006,
Living life on the edge - Widefield VLBI at 90 cm!
in {\it Proceedings of the 8th European VLBI Network Symposium.},
(eds W. Baan, R. Bachiller, R. Booth, P. Charlot, P. Diamond, M. Garrett, X. Hong, J. Jonas, A. Kus, F. Mantovani, A. Marecki, H. Olofsson, W. Schlueter, M. Tornikoski, N. Wang, and A. Zensus)
{\bf 8}, 79--82  \\ [-2mm]

\noindent
Lenc, E., and Tingay, S.J.
2006,
The Sub-parsec Scale Radio Properties of Southern Starburst Galaxies. I. Supernova Remnants, the Supernova Rate, and the Ionised Medium in the NGC 253 Starburst.
{\it Astronomical J.}, {\bf 132}, 1333--1345  \\ [-2mm]

\end{document}